\documentclass[a4paper,11pt,openleft]{book}
\usepackage[latin1]{inputenc}
\usepackage[english]{babel}
\usepackage{amssymb}
\usepackage{amsfonts}
\usepackage{amsmath}
\usepackage{mathrsfs}
\usepackage{eufrak}
\usepackage{fancyhdr}
\usepackage{verbatim}
\usepackage[dvips]{graphicx}
\DeclareMathAlphabet{\mathpzc}{OT1}{pzc}{m}{it}
\newcommand{\sectioncount}[1]{\section{#1}\setcounter{equation}{0}}
\newenvironment{enumeratea}{\renewcommand{\theenumi}{\alph{enumi}}\renewcommand{\labelenumi}{(\theenumi)}\setcounter{enumi}{0}\begin{enumerate}}{\end{enumerate}}
\newenvironment{enumeratei}{\renewcommand{\theenumi}{\roman{enumi}}\renewcommand{\labelenumi}{(\theenumi)}\setcounter{enumi}{0}\begin{enumerate}}{\end{enumerate}}
\newcommand{\Zgr}{\mathbb Z}
\newcommand{\cep}{\mathcal{C}_\epsilon}
\newcommand{\dep}{\mathcal{D}_\epsilon}

\newcommand{\oneover}[1]{\frac{1}{#1}}
\newcommand{\lagra}{\mathcal{L}}

\newcommand{\intd}{\mathrm{d}}
\newcommand{\intD}{\mathcal{D}}
\newcommand{\ano}{\mathcal{A}_{(6)}}
\newcommand{\sano}{\mathcal{A}_{(6)\star}}
\newcommand{\kb}{\frac{\k}{b^2}}

\newcommand{\ka}{\frac{k}{a^2}}

\newcommand{\sT}{T_\star}
\newcommand{\sH}{H_\star}
\newcommand{\sqq}{q_\star}
\newcommand{\sF}{F_\star}

\newcommand{\sB}{B_\star}

\newcommand{\sR}{R_\star}
\newcommand{\sbb}{b_\star}
\newcommand{\hg}{\hat g}
\newcommand{\srho}{\rho_\star}
\newcommand{\ssig}{\s_\star}
\newcommand{\ssigp}{\s_{p\star}}
\newcommand{\ssigpp}{\s_{\p\star}}
\newcommand{\schi}{\chi_\star}
\newcommand{\rhofo}{\varrho}
\newcommand{\chifo}{\mathpzc{x}}
\newcommand{\rhofz}{\varrho_0}
\newcommand{\chifz}{\mathpzc{x}_0}
\newcommand{\cHH}{\check H}
\newcommand{\cTT}{\check T}
\newcommand{\cAA}{\check A}
\newcommand{\Afo}{\mathpzc{A}}
\newcommand{\cAAfo}{\check{\mathpzc{A}}}

\newcommand{\ck}{\check\k}

\newcommand{\cqq}{\check q}

\newcommand{\crho}{\check\rho}
\newcommand{\cchi}{\check\chi}
\newcommand{\crhofo}{\check\varrho}
\newcommand{\cchifo}{\check{\mathpzc{x}}}
\newcommand{\csH}{\check H_\star}

\newcommand{\csrho}{\check\rho_\star}
\newcommand{\cschi}{\check\chi_\star}

\newcommand{\stano}{\mathcal{\tilde A}_{(6)\star}}
\newcommand{\weyl}{\hat{\mathcal{W}}}
\newcommand{\mani}{\mathcal{M}}
\newcommand{\hx}{\hat{x}}
\newcommand{\hD}{\hat{\Delta}}
\newcommand{\sY}{{Y}_{\star}}
\newcommand{\stY}{{\tilde Y}_{\star}}
\newcommand{\tn}{\tilde\nu}

\newcommand{\tcreq}{\tilde c_{\rho\scriptscriptstyle{(eq)}}}
\newcommand{\tcrst}{\tilde c_{\rho\scriptscriptstyle{(st)}}}
\newcommand{\tcrx}{\tilde c_{\rho,\scriptscriptstyle\x}}
\newcommand{\tcrd}{\tilde c_{\rho,\scriptscriptstyle d}}
\newcommand{\tcrsix}{\tilde c_{\rho,\scriptscriptstyle{d=6}}}
\newcommand{\tcv}{\tilde c_V}
\newcommand{\tcveq}{\tilde c_{V\scriptscriptstyle{(eq)}}}
\newcommand{\tcvst}{\tilde c_{V\scriptscriptstyle{(st)}}}
\newcommand{\tcvx}{\tilde c_{V,\scriptscriptstyle\x}}
\newcommand{\tcvd}{\tilde c_{V,\scriptscriptstyle d}}
\newcommand{\tcvsix}{\tilde c_{V,\scriptscriptstyle{d=6}}}

\newcommand{\Mpl}{M_{Pl}^4}
\newcommand{\mpl}{M_{Pl}^2}
\newcommand{\MPl}{M_{Pl}^6}
\newcommand{\sgn}{\mathrm{sgn}}
\newcommand{\traccia}{\mathrm{tr}}

\newcommand{\diag}{\mathrm{diag}}

\newcommand{\ex}{\mathrm{e}}
\newcommand{\imm}{\mathrm{i}}

\newcommand{\ogr}{\mathcal{O}}

\newcommand{\im}{\imath}
\newcommand{\cd}{\cdot}
\newcommand{\pa}{\partial}

\newcommand{\bea}{\begin{eqnarray}}
\newcommand{\ena}{\end{eqnarray}}
\newcommand {\non}{\nonumber}

\newcommand{\refeq}[1]{(\ref{#1})}
\renewcommand{\a}{\alpha}
\renewcommand{\b}{\beta}
\renewcommand{\d}{\delta}
\renewcommand{\th}{\theta}

\newcommand{\g}{\gamma}
\newcommand{\G}{\Gamma}

\newcommand{\D}{\Delta}
\newcommand{\e}{\epsilon}
\newcommand{\eps}{\epsilon}
\newcommand{\z}{\zeta}
\renewcommand{\k}{\kappa}

\renewcommand{\l}{\lambda}
\renewcommand{\L}{\Lambda}
\newcommand{\m}{\mu}
\newcommand{\M}{\Mu}
\newcommand{\n}{\nu}

\newcommand{\x}{\xi}

\newcommand{\p}{\pi}

\newcommand{\s}{\sigma}
\newcommand{\pab}{{\bar{\partial}}}
\renewcommand{\S}{\Sigma}
\renewcommand{\t}{\tau}

\newcommand{\bps}{\bar\psi}
\renewcommand{\o}{\omega}
\renewcommand{\O}{\Omega}
\def\bseq{\begin{subequation}}  % = 1a 1b
\def\eseq{\end{subequation}}
\def\bsea{\begin{subeqnarray}}  % = 1.1a 1.1b
\def\esea{\end{subeqnarray}}
\newcommand{\bbox}{\lower.2ex\hbox{$\Box$}}

\newcommand{\beq}{\begin{equation}}
\newcommand{\eeq}{\end{equation}}
\newcommand{\eea}{\end{eqnarray}}

\def\Ups{\Upsilon}

\newcommand{\ad}{{\dot{\alpha}}}
\newcommand{\bd}{{\dot{\beta}}}

\def\sfrac#1#2{{\textstyle\frac{#1}{#2}}}
\def\rd#1{\buildrel{_{_{\hskip 0.01in}\rightarrow}}\over{#1}}
\def\ld#1{\buildrel{_{_{\hskip 0.01in}\leftarrow}}\over{#1}}
\def\+{\dagger}
\def\={\ =\ }
\def\res{\mathrm{res}}
\def\ch{\mathrm{ch}}

\def\th{\mathrm{th}}

\newcommand{\mbf}[1]{{\boldsymbol{#1}}}
\newcommand{\diff}{\mathrm{d}}
\newcommand{\R}{{\mathbb{R}}}
\newcommand{\C}{{\mathbb{C}}}

\newcommand{\Ecal}{{\cal E}}
\newcommand{\mb}{\bar{\mu}}
\newcommand{\Lb}{\bar{L}^{A,B}}
\newcommand{\Lbt}{\widetilde{\bar{L}}^{A,B}}
\newcommand{\Pht}{\widetilde{\Phi}}
\newcommand{\Pt}{\widetilde{P}}
\newcommand{\St}{\widetilde{S}}
\newcommand{\Tt}{\widetilde{T}}
\newcommand{\gt}{\widetilde{g}}
\newcommand{\Psh}{\widehat{\Psi}}
\newcommand{\Sh}{\widehat{S}}

\begin{document}
\fancyhf{}
\thispagestyle{empty}
\begin{center}
\includegraphics[height=1.5cm]{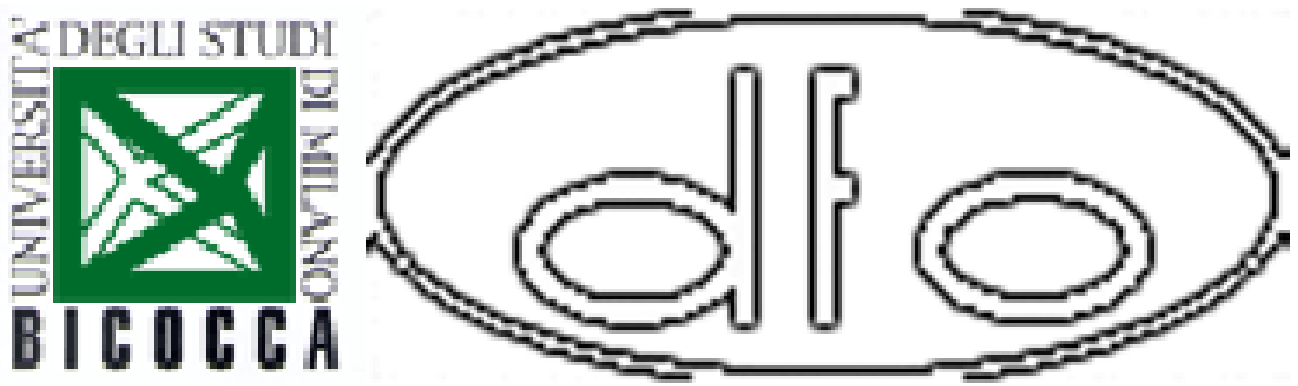}
\hspace{\stretch{1}}
\includegraphics[height=1.5cm]{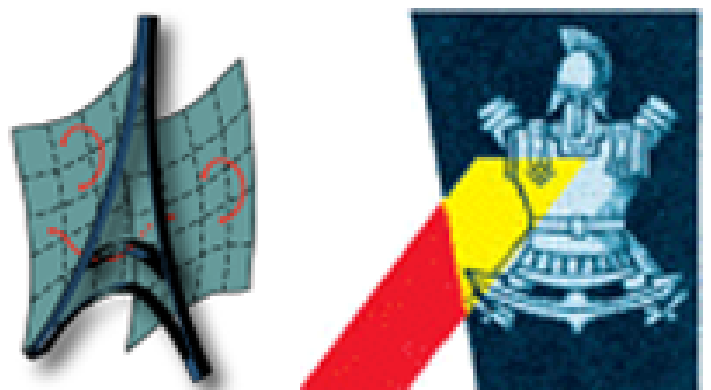}
\\
~\\
\large{\sf{Universit\`a di Milano-Bicocca - Dip.~di Fisica G.~Occhialini\\\'Ecole Polytechnique - CPhT\\}}
\vspace{2.5cm}\hrulefill
\LARGE{\\\textsc{Topics in noncommutative integrable theories\\and holographic brane--world cosmology}}
\\\hrulefill\end{center}
\vspace{2.5cm}
\begin{flushright}\texttt{Doctoral Thesis of}: Liuba Mazzanti\\\footnotesize{\tt{(international Ph.D. project)}}\end{flushright}
\vspace{1cm}
\texttt{Supervisor at University of Milano-Bicocca}: Prof.~Silvia Penati\\
\texttt{Supervisor at \'Ecole Polytechnique}: Prof.~Elias Kiritsis\\

\noindent
\texttt{External Referees}: Prof.~Gianluca Grignani, Prof.~David Langlois and Prof.~Boris Pioline\\
\vspace{\stretch{2}}
\begin{flushright}\small{Dottorato di Ricerca in Fisica ed Astronomia (XIX ciclo)\\Doctorat en Physique Th\'eorique}\end{flushright}
\newpage\thispagestyle{empty}\cleardoublepage
\thispagestyle{empty}
\vspace*{\stretch{1}}
\begin{flushright}
{\LARGE{\bf{Acknowledgements}}}
% acknowledgments

\vspace{0.3cm}

I am grateful to all who somehow shared with me these 3-4 years of Ph. D.\dots\\
\vspace{0.3cm}

First of all, it is my duty and pleasure to thank my two advisors Silvia Penati and Elias Kiritsis.\\
\vspace{0.1cm}

Then the whole two string groups in Milano and Paris; in particular Roberto (if he can hear me, lost in baby's cries), Angel, Francesco, Umut (I'll have to thank you two for some more time ahead) and Alberto
(MOA, for once), Gabri (see you in Perth), Marco, Ago, Alberto (a kiss).\\
\vspace{0.3cm}

Finally my family for supporting. Marcello (dad) and Claudia for constructive and healthy criticisms, appreciations and all their love. Muriel (mum) and Andrea for everything I could need and more. A huge
kiss to my sister Arianne, always giving me the sweetest encouragements. Alberto, for making sun shine, most of the times.

\end{flushright}
\vspace{\stretch{1}}
\newpage\thispagestyle{empty}\cleardoublepage
\chapter*{Abstract}
\thispagestyle{empty}
\fancyhf{}
\fancyhead[RO]{\bf{Abstract}}
\fancyhead[LE]{\bf{Abstract}}
This thesis follows two main lines of research, both related to relevant aspects of string theory and its phenomenological/cosmological applications. 

On the one hand, noncommutative (NC) geometry is a natural consequence of the presence of branes and fluxes in string theory. Here we study two different generalizations of the integrable sine--Gordon (SG)
model to NC geometry, after discussing general properties and issues of integrable theories and NC field theories --- mentioning their role in string theory. The question is whether we can obtain an
integrable NC SG theory, according to the theorem  relating integrability in two dimensions to the factorization of the S matrix and absence of particle production. The first NC SG model is characterized by a
non factorized S matrix, spoiling the validity of the aforementioned theorem in this NC set--up. However, the second model exhibits the good properties of S matrix (at tree level) required by integrability in
2D.  The model is derived by dimensional reduction from the stringy NC self--dual Yang--Mills (SDYM) theory in $(2+2)$ dimensions, describing ${\cal N}=2$ fermionic strings.  Since the higher dimensional NC
SDYM is integrable, due to integrability of the underlying string theory, intergability of this NC SG model seems to be ensured. A method to build multi--soliton solutions is also given.

On the other hand, branes play a fundamental role in the brane--world context as well. A particular brane--world model is here analyzed both from the cosmological point of view and in the spirit of holography
(via the AdS/CFT correspondence). An introduction is given on conventional cosmology, brane--worlds, and AdS/CFT. Randall--Sundrum (RS) models are thoroughly illustrated, together with their dual theory. The
7D RS set--up with brane--bulk energy exchange we propose leads to a non conventional cosmological evolution possibly characterized by late time acceleration and an intermediate decelerated phase. All fixed
points have positive acceleration factor and are found to be stable for a wide range of choices for the parameters. We construct the holographic dual theory, represented by a renormalized 6D CFT coupled to
6D gravity (at low energies we would get an effective 4D theory, due to compactification on a 2D internal manifold). The matching of parameters on the two sides of the duality is then achieved in specific
approximations.

\newpage\thispagestyle{empty}\cleardoublepage
\chapter*{R\'esum\'e de la th\`ese}
\thispagestyle{empty}
\fancyhf{}
\fancyhead[RO]{\bf{R\'esum\'e de la th\`ese}}
\fancyhead[LE]{\bf{R\'esum\'e de la th\`ese}}
Ma th\`ese se d\'eroule suivant deux principales lignes de recherche. Les deux arguments trait\'es constituent une relation entre la th\'eorie des cordes et les aspects ph\'enom\'enologiques/cosmologiques.
D'une part, la g\'eom\'etrie noncommutative (NC) est une cons\'equence naturelle de la pr\'esence de branes et flux dans la th\'eorie des cordes. La noncommutativit\'e d\'eforme certaines propri\'et\'es
fondamentales des th\'eories ordinaires d\'ecrivant par exemple les interactions \'electro--faibles et fortes ou les mod\`eles statistiques. C'est dans ce sens que la g\'eom\'etrie NC repr\'esente une
application \`a la ph\'enom\'enologie des cordes. D'autre part, les branes sont l'ingr\'edient cl\'e des mod\`eles d'univers branaires. Le mod\`ele de Randall--Sundrum (RS) en particulier offre de nouvelles
perspectives tant du point de vue de la cosmologie, ouvrant des sc\'enarios d'\'evolution cosmologique non conventionnelle, que du point de vue de l'holographie.

La premi\`ere partie de la th\`ese est d\'edi\'ee \`a la g\'eom\'etrie NC et, en particulier, aux th\'eories de champs NC int\'egrables. Le but principal du travail a \'et\'e d'\'etudier les cons\'equences de
la noncommutativit\'e par rapport \`a l'int\'egrabilit\'e. Plus pr\'ecis\'ement, on a voulu v\'erifier ou r\'efuter dans un contexte NC le th\'eor\`eme qui lie, en deux dimensions, l'int\'egrabilit\'e \`a la
factorisation de la matrice S. Avec int\'egrabilit\'e on parle de l'existence d'un nombre infini de courants locaux conserv\'es, associ\'es aux sym\'etries de la th\'eorie de champs. Le point de d\'epart a
donc \'et\'e de garantir la pr\'esence de tels courants, au moyen du formalisme du bicomplexe. Cette m\'ethode permet d'obtenir les \'equations du mouvement en tant que conditions d'int\'egrabilit\'e d'un
syst\`eme d'\'equations diff\'erentielles lin\'eaires. \`a partir des solutions du m\^eme syst\`eme lin\'eaire suivent les courants conserv\'es.
En exploitant le formalisme de Weyl, la proc\'edure est imm\'ediatement g\'en\'eralisable \`a la g\'eom\'etrie NC. Une alg\`ebre de fonctions (op\'erateurs de Weyl) d\'efinie sur un espace NC est associ\'ee
\`a une alg\`ebre NC de fonctions où la multiplication est ex\'ecut\'ee au moyen d'un produit NC de Moyal: le produit $\star$. En introduisant le produit $\star$ au niveau du syst\`eme lin\'eaire et en en
d\'eduisant les \'equations du mouvement NC, on obtient la g\'en\'eralisation NC du bicomplexe.
On a inf\'er\'e le premier mod\`ele en g\'en\'eralisant le bicomplexe du mod\`ele de sine--Gordon (SG) \`a la g\'eom\'etrie NC. Nous avons d\'eduit (en collaboration avec Grisaru, Penati, Tamassia) l'action
correspondante aux \'equations du mouvement pr\'ec\'edemment \'etablies par Grisaru et Penati. Le calcul des amplitudes de diffusion et production a d\'etermin\'e les caract\'eristiques de la matrice S du
mod\`ele. Des comportements acausaux ont \'et\'e relev\'es pour les processus de diffusion. En outre, les processus de production poss\`edent une amplitudes non nulle: d'où la non validit\'e du th\'eor\`eme
d'int\'egrabilit\'e vs. factorisation pour cette version NC du mod\`ele de SG. D'autres propri\'et\'es ont \'et\'e mises en \'evidence, comme la relation avec la th\'eorie des cordes et la bosonisation.
Le deuxi\`eme mod\`ele de SG NC a \'et\'e propos\'e en collaboration avec Lechtenfeld, Penati, Popov, Tamassia. Les \'equations du mouvement ont \'et\'e tir\'ees de la r\'eduction dimensionnelle du mod\`ele
sigma NC en 2+1 dimensions, qui \`a son tour est la r\'eduction de la th\'eorie de self--dual Yang--Mills NC en 2+2 dimensions (d\'ecrivant les supercordes $\mathcal N=2$ avec champs $B$). L'action a \'et\'e
calcul\'ee de m\^eme que les amplitudes. Les processus de production poss\'edant des amplitudes nulles et ceux de diffusion ne d\'ependant pas du param\`etre de NC, entra\^inent ainsi un comportement causal.
Le deuxi\`eme mod\`ele de SG NC semble donc ob\'eir \`a l'\'equivalence entre int\'egrabilit\'e et factorisation de la matrice S. La r\'eduction de la th\'eorie des cordes garde sa validit\'e m\^eme au niveau
de l'action, contrairement au mod\`ele pr\'ec\'edent.

La deuxi\`eme partie de ma th\`ese traite des mod\`eles d'univers branaires, ou plus pr\'ecis\'ement des mod\`eles de RS. Le mod\`ele propos\'e par Randall et Sundrum se situe dans un bulk 5--dimensionnel,
caract\'eris\'e per une sym\'etrie d'orbifold $\Zgr_2$ par rapport \`a la position de la brane 4--dimensionnelle. Gr\^ace au facteur de warp qui multiplie le sous--espace 4--dimensionnel parall\`ele \`a la
brane, on obtient la localisation des modes du graviton. Par cons\'equent, le potentiel gravitationnel efficace est newtonien aux \'energies inf\'erieures \`a la masse de Planck. En introduisant en outre un
terme de mati\`ere dans le bulk et en consid\'erant l'\'echange d'\'energie entre brane et bulk, une vari\'et\'e de nouvelles cosmologies en d\'erive.
Dans la premi\`ere partie de mon travail sur RS nous avons propos\'e un mod\`ele analogue situ\'e dans un bulk 7--dimensionnel. La brane 6--dimensionnelle --- ayant compactifi\'e deux dimensions --- est
plac\'ee au point fixe de l'orbifold $\Zgr_2$. Afin d'\'etudier l'\'evolution cosmologique en nous mettant en relation avec les observations, nous avons introduit l'\'echange d'\'energie entre brane et bulk.
Les sc\'enarios possibles sont nombreux et d\'ependent de la forme explicite du param\`etre d'\'echange d'\'energie. Entre autres, les points fixes poss\`edent une acc\'el\'eration positive, pouvant ainsi
repr\'esenter la r\'ecente acc\'el\'eration de l'univers. Il sont \'egalement stables pour un large ensemble des valeurs des param\`etres. Finalement, on peut tracer des sc\'enarios qui partent d'une phase
initiale acc\'el\'er\'ee, en passant successivement \`a une \`ere de d\'ec\'el\'eration, pour terminer sur un point fixe stable d'inflation.
Les mod\`eles d'univers branaires \`a la RS poss\`edent un dual holographique via AdS/CFT. La correspondance AdS/CFT \'etablit qu'une th\'eorie de supergravit\'e (ou, plus g\'en\'eralement, de cordes) dans un
champ de fond d'anti de Sitter (AdS) en $d+1$ dimensions est duale \`a une th\'eorie de champs conforme (CFT) en $d$ dimensions. Tenant compte des divergences pr\'esentes dans les deux descriptions, cette
correspondance \`a \'et\'e rendue plus pr\'ecise par la formulation de la renormalisation holographique. Si l'espace de AdS est r\'egularis\'e au moyen d'un cutoff infrarouge, la correspondante CFT r\'esulte
r\'egularis\'ee par un cutoff ultraviolet et coupl\'ee \`a la gravit\'e $d$--dimensionnelle.
En analogie \`a l'analyse effectu\'ee en cinq dimensions par Kiritsis, nous avons construit la th\'eorie duale au mod\`ele cosmologique de RS en sept dimensions. Pour capturer les dynamiques dict\'ees par
l'\'echange d'\'energie entre brane et bulk, la th\'eorie holographique en six dimensions a \'et\'e g\'en\'eralis\'ee au cas interagissant (entre mati\`ere et CFT) et non conforme. Le r\'esultat sont les
relations entre les param\`etres de masse appartenant aux deux descriptions et entre l'\'echange d'\'energie, d'un c\^ot\'e, et le param\`etre d'interaction, de l'autre. De plus, le param\`etre de rupture
conforme est associ\'e au param\`etre d'auto--interaction du bulk dans la description de supergravit\'e 7--dimensionnelle.

Le travail de recherche inclut donc des r\'esultats pouvant trouver leur application dans la ph\'e\-no\-m\'e\-no\-lo\-gie et cosmologie des cordes. D'une part on a enqu\^eter sur l'influence de la
noncommutativit\'e li\'ee \`a l'int\'egrabilit\'e du mod\`ele de SG. D'autre part, les cons\'equences cosmologiques de l'emplacement du mod\`ele de RS en sept dimensions ont \'et\'e \'etudi\'ees et la
correspondance AdS/CFT a \'et\'e appliqu\'ee afin d'en tirer des informations sur la th\'eorie duale, coupl\'ee \`a la gravit\'e.

\newpage\thispagestyle{empty}\cleardoublepage
\chapter*{Riassunto della tesi}
\thispagestyle{empty}
\fancyhf{}
\fancyhead[RO]{\bf{Riassunto della tesi}}
\fancyhead[LE]{\bf{Riassunto della tesi}}
Il mio lavoro di tesi si sviluppa seguendo due principali linee di ricerca. Entrambi gli argomenti affrontati costituiscono una relazione tra la teoria delle stringhe e aspetti fenomenologici/cosmologici. Da
un lato, la geometria noncommutativa (NC) \`e una naturale conseguenza della presenza di brane e flussi nella teoria di stringa. La noncommutativit\`a deforma alcune propriet\`a fondamentali delle teorie
ordinarie che ad esempio descrivono le interazioni elettro--debole e forte o modelli statistici. In tal senso, la geometria NC rappresenta un'applicazione alla fenomenologia di stringa.  D'altro canto, 
le brane rappresentano un ingrediente chiave nei modelli di brane--world. Il modello di Randall--Sundrum (RS), in particolare, offre nuove prospettive sia dal punto di vista della cosmologia, aprendo
scenari di evoluzione cosmologica non convenzionale, sia dell'olografia.

La prima parte della tesi \`e dedicata alla geometria NC ed, in particolare, a teorie di campo NC integrabili. Il principale scopo del lavoro di ricerca \`e stato studiare le conseguenze della
noncommutativit\`a sull'integrabilit\`a. Pi\`u esplicitamente, si \`e voluto verificare o confutare in un contesto noncommutativo il teorema che lega, in due dimensioni, l'integrabilit\`a alla fattorizzazione
della matrice S. Per integrabilit\`a si intende l'esistenza di un infinito numero di correnti locali conservate, associate alle simmetrie della teoria di campo. Il punto di partenza \`e stato dunque garantire
la presenza di tali correnti attraverso il formalismo del bicomplex. Questo metodo consente di ottenere le equazioni del moto come condizioni di integrabilit\`a di un sistema di equazioni differenziali
lineari. Dalle soluzioni dello stesso sistema lineare \`e possibile ricavare le infinite correnti conservate. 
Sfruttando il formalismo di Weyl, il procedimento \`e immediatamente generalizzabile alla geometria NC. Un algebra di funzioni (operatori di Weyl) definite sullo spazio NC viene associata ad un algebra NC di
funzioni in cui la moltiplicazione \`e implementata attraverso un prodotto NC di Moyal: il prodotto $\star$. Introducendo nel sistema differenziale lineare il prodotto $\star$ e deducendone le equazioni del
moto NC, si ottiene la generalizzazione NC del metodo del bicomplex.
Il primo modello considerato \`e stato ricavato generalizzando il bicomplex per il modello di sine--Gordon (SG) alla geometria NC. Dalle equazioni del moto ottenute in precedenza da Grisaru e Penati abbiamo
dedotto l'azione corrispondente (in collaborazione con Grisaru, Penati, Tamassia). Il calcolo delle ampiezze di scattering e produzione ha determinato le caratteristiche della matrice S del modello.  Sono
risultati comportamenti acausali per i processi di scattering. Inoltre, poich\'e i processi di produzione di particelle non possiedono ampiezza nulla, il teorema integrabilit\`a vs. fattorizzazione non rimane
valido per tale generalizzazione NC del modello di SG. Altre propriet\`a sono state evidenziate, come la relazione con la teoria di stringa e con la bosonizzazione.
Il secondo modello di SG NC \`e stato proposto in collaborazione con Lechtenfeld, Penati, Popov, Tamassia. Le equazioni del moto sono state derivate dalla riduzione dimensionale del modello sigma NC in 2+1
dimensioni, che a sua volta \`e la riduzione dimensionale della teoria di self--dual Yang--Mills NC in 2+2 dimensioni (che descrive le superstringhe $\mathcal N=2$ con campo $B$). Anche in questo caso \`e
stata dedotta l'azione ed \`e stato effettuato il calcolo delle ampiezze ad albero. I processi di produzione risultano possedere ampiezza nulla e le ampiezze di scattering non dipendono dal parametro di NC,
implicando un comportamento causale. Perci\`o questo secondo modello di SG NC sembra obbedire all'equivalenza tra integrabilit\`a e fattorizzazione della matrice S. La riduzione dalla teoria di stringa \`e
valida anche a livello dell'azione, al contrario di quanto accade per il primo modello analizzato.

La seconda parte della tesi tratta di modelli di brane--world, o pi\`u specificatamente di modelli di RS. Il modello proposto da Randall e Sundrum \`e ambientato in un bulk 5--dimensionale, caratterizzato da
una simmetria di orbifold $\Zgr_2$ rispetto alla collocazione della brana 4--dimensionale. Grazie al fattore di warp che moltiplica il sottospazio 4--dimensionale parallelo alla brana, si ottiene la
localizzazione dei modi gravitonici. Conseguentemente, il potenziale gravitazionale efficace \`e newtoniano per energie inferiori alla massa di Planck. Introducendo un termine di materia nel bulk e
considerando lo scambio di energia tra brana e bulk, si ottiene una variet\`a di nuove possibili cosmologie. 
Nella prima parte del mio lavoro su RS \`e stato proposto un modello analogo, ambientato in un bulk 7--dimensionale. La brana 6--dimensionale --- di cui due dimensioni sono compattificate --- \`e posta nel
punto fisso dell'orbifold $\Zgr_2$. Al fine di studiare l'evoluzione cosmologica ponendoci in relazione con le osservazioni abbiamo introdotto lo scambio di energia tra brana e bulk. I possibili scenari sono
numerosi e dipendono dalla forma esplicita del parametro di scambio di energia. In particolare, tutti i punti fissi possiedono accelerazione positiva, sono stabili per appropriati valori dei parametri e
potrebbero dunque rappresentare la presente era accelerata. \`E possibile inoltre ipotizzare scenari in cui, partendo da una fase iniziale con grande accelerazione positiva, si passi da una era decelerata,
per terminare sul punto fisso inflazionario e stabile. 
Modelli di brane--world \`a la RS possiedono un duale olografico via AdS/CFT. La corrispondenza AdS/CFT stabilisce che una teoria di supergravit\`a (o, pi\`u in generale, di stringa) in un background di anti
de Sitter (AdS) in $d+1$ dimensioni \`e duale ad una teoria di campo conforme (CFT) in $d$ dimensioni. Tale corrispondenza \`e stata resa pi\`u precisa mediante la formulazione della rinormalizzazione
olografica, tenendo conto delle divergenze presenti in entrambe le descrizioni. Se lo spazio di AdS viene regolarizzato tramite un cutoff infrarosso, la corrispondente CFT risulta regolarizzata da un cutoff
ultravioletto e accoppiata alla gravit\`a $d$--dimensionale.
In analogia all'analisi effettuata in cinque dimensioni da Kiritsis, abbiamo costruito la teoria duale al modello cosmologico di RS in sette dimensioni. Per catturare le dinamiche dettate dallo scambio di
energia tra brana e bulk, la teoria olografica in sei dimensioni \`e stata generalizzata al caso in cui materia e settore nascosto (appartenente alla CFT) interagiscano e l'invarianza conforme sia rotta. Come
risultato sono state trovate le relazioni tra i parametri di massa nelle due descrizioni e tra scambio di energia, da una parte, e parametro di interazione, dall'altra.  Inoltre, il parametro di rottura
conforme risulta associato al parametro di auto--interazione del bulk nella descrizione di supergravit\`a 7--dimensionale.

Il lavoro di tesi comprende dunque risultati che possono trovare applicazione nella fenomenologia o cosmologia di stringa. Da un lato si \`e investigata l'influenza della noncommutativit\`a
sull'integrabilit\`a del modello di SG. Dall'altro, sono state studiate le conseguenze cosmologiche dell'am\-bien\-ta\-zio\-ne del modello di RS in sette dimensioni ed \`e stata applicata la corrispondenza
AdS/CFT per ricavare informazioni sulla teoria duale, accoppiata alla gravit\`a.

\newpage\thispagestyle{empty}\cleardoublepage
\begin{frontmatter}
\fancyhf{}
\fancyhead[RO]{\rightmark}
\fancyhead[LE]{\leftmark}
\fancyfoot[C]{\thepage}
\pagenumbering{roman}
\addcontentsline{toc}{chapter}{Contents}
\tableofcontents
\newpage\thispagestyle{empty}\cleardoublepage
\addcontentsline{toc}{chapter}{List of figures}
\listoffigures
\newpage\thispagestyle{empty}\cleardoublepage
\addcontentsline{toc}{chapter}{Introduction and outline}
\chapter*{Introduction and outline}
\fancyhf{}
\fancyhead[RO]{\bf{Introduction and outline}}
\fancyhead[LE]{\bf{Introduction and outline}}
\fancyfoot[C]{\thepage}
String theory is a wide web of interlacing theories which encloses gauge theories and gravity in some low energy limits. By now, string theory in its supersymmetric version, provides a consistent description
of quantum gravity. However, it is still not completely clear how to merge real (hence non supersymmetric) fundamental interactions in the strings framework, despite the fact that much work has been 
recently devoted to this purpose. The unifying theories of strings (which in turn may be argued to be incorporated in the larger M theory) contain degrees of freedom which cannot be described by ordinary
gauge theories. This is why new features arise in this context and new mathematical techniques as well as new objects must be studied. My thesis essentially tackles two of these stringy issues:
noncommutative geometry and brane--worlds.

Both topics deal with stringy effects on some aspects of (hopefully) realistic description of known physics. On the one hand, noncommutative (NC) geometry emerges in relation to particular string
configurations involving branes and fluxes. Gauge theories arise in the low energy limit of open string dynamics, with string ends attached on the branes. When non trivial fluxes are turned on, ordinary field
theories get deformed by noncommutativity. On the other hand, brane--worlds originated from the intuition that matter fields can be localized on branes, while gravity propagates in the whole string target
space.  Brane--worlds can thus yield effectively four dimensional gauge theories with obvious phenomenological implications, despite the existence of extra dimensions. Furthermore, non staticity of the brane
worldvolume produces cosmological evolution, opening the issue of the brane--world cosmology. Branes turn out to be key ingredients for both topics. They indeed represent at present the main motivation to
study noncommutative geometry and create a link between string theory and phenomenology/cosmology.  

\subsubsection{Noncommutative geometry} 
Independently of string theory successes, NC geometry was initially formulated with the hope that it could milden ultraviolet divergences in quantum field theories \cite{Snyder}. However, noncommutativity
usually does not qualitatively modify renormalization properties, except for the mixing of infrared and ultraviolet divergences --- IR/UV mixing ---, which on the contrary generally spoils renormalization.
Noncommutative relation among space--time coordinates may also be interpreted as a possible deformation of geometry beyond the Planck scale. In fact, we can imagine that space--time can be no more endowed
with a point--like structure. Indeed, this is a consequence of noncommutativity.  Points would be subtituted by space cells with Planck length dimension, so that ordinary geometry is recovered at energies
lower than the Planck scale.  It is also true that noncommutativity arises in the large magnetic field limit of quantum Hall effect \cite{Szabo,Bigatti}. There, space coordinates are forced not to commute due
to the very large magnetic field, or equivalently to the very small particle mass.  However, the strongest motivation is string theory, since it naturally describes noncommutative embeddings.

Field theories in NC geometry represent the low energy limit of dynamics of open strings ending on branes with appropriate non zero fluxes. A paradigmatic configuration is that of D3--branes in IIB string
theory with constant Neveu--Schwarz--Neveu--Schwarz (NS-NS) form $B_{\m\n}$, yielding noncommutative four dimensional super Yang--Mills theory (SYM), with instanton solutions described by self--dual
Yang--Mills (SDYM) equations \cite{SW}.  Noncommutative versions of well known field theories have been studied over the last ten years \cite{list1}--\cite{hamanaka}, which single out the interesting
results ensuing from noncommutativity. For instance, noncommutative relation among space--time coordinates imply a correlation between infrared and ultraviolet divergences in the field theory
\cite{UVIR1,UVIR2}.  This follows intuitively from the uncertainty principle involving the coordinates, which connects small distances to large distances dynamics, just as quantum mechanic uncertainty
principle connects large momenta to small distances and viceversa.  Hence, field theory renormalization also depends on the IR behavior. Although in most cases renormalizability doesn't change going to NC
geometry --- except for UV/IR mixing ---, it can be explicitly destroyed in some particular models by noncommutativity, as I will show. 

Besides renormalization, there has been much interest in studying integrability properties of noncommutative generalizations. Integrable theories share very nice features, in particular restricting to two
dimensions \cite{ZZ}. Their S--matrix has to be factorized in simple two particle processes and can be explicitly calculated in some cases. Furthermore, no particle production or annihilation occurs.  Momenta
of initial states must be mapped in the final states, precisely. Solitons, i.e. localized classical solutions preserving their shape and velocity in scattering processes, are usually present. The origin of
these nice properties is the presence of an infinite number of conserved currents, which are indeed responsible for the integrability of the theory. It is interesting to note that most of the known integrable
bidimensional models come from dimensional reduction of four dimensional SDYM. In turn, $(2+2)$--dimensional SDYM is the effective field theory for ${\cal N}=2$ open superstrings on D3--branes, whose
noncommutative version is obtained by turning on a constant NS-NS two form \cite{Ooguri:1990ww,LPS}. We may now wonder if noncommutativity influences integrability of known models. This is basically the
question I tried to answer with my collaborators, restricting to a special integrable model, namely the two dimensional sine--Gordon theory.

Sine--Gordon equations of motion are related to the integrability of a system with an infinite number of degrees of freedom, giving the infinite number of conserved currents. The gauged bicomplex approach
guarantees the existence of the local currents as solutions to an infinite chain of conservation equations, for any integrable theory. These come from solving a matrix valued equation, order by order in an
expansion parameter --- a Lax pair of differential operators (guaranteeing integrability) can also be found in some cases related to the bicomplex formulation. Furthermore, the compatibility condition of the
matrix equation yields the equations of motion, from which an action can in some cases be derived --- for sine--Gordon, for instance. Soliton solutions can also be constructed via the dressing method
\cite{dressing} in integrable theories, exploiting the solutions to the integrable linear system of equations.

Using the gauged bicomplex formalism, S. Penati and M.T. Grisaru wrote the equations of motion for a noncommutative version of sine--Gordon, introducing noncommutativity in the two dimensional matrix equation
\cite{GP}.  Successively, we found the corresponding noncommutative action and studied properties of the S--matrix at tree level \cite{GMPT}.  Noncommutativity entailed acausal behaviors and non factorization
of the S--matrix. Acausality is actually a typical problem in NC field theories when noncommutativity involves the time coordinate. It has been shown that also unitarity is broken by time/space
noncommutativity. Indeed, in two dimensions, a noncommuting time is unavoidable.

However, NC generalizations are not unique, since different deformations can yield the same ordinary theory in the commutative limit. Indeed, a second noncommutative sine--Gordon model was proposed in my
publication \cite{LMPPT} in collaboration with O.~Lechtenfeld, S.~Penati, A.~D.~Popov, L.~Tamassia, where noncommutativity has been implemented at an intermediate step in the dimensional reduction from SDYM.
The action and tree level S--matrix were computed. Scattering processes displayed the nice properties expected in integrable models and causality was not violated. Moreover, we also provided a general method
to calculate multi--soliton solutions in this integrable NC sine--Gordon model.

As I anticipated, string theory suggests a deformation of space--time, leading to noncommutative field theories. The relation to specific string configurations pass through dimensional reduction of higher
dimensional integrable theories --- namely 4-- or $(2+2)$--dimensional SDYM --- describing the open string dynamics. Phenomenological consequences other that integrability in two dimensions can be
investigated. Most of related literature focuses on Lorentz violation in Standard Model noncommutative generalizations \cite{ncsm}. In fact, in other than two dimensions, Lorentz invariance is broken, due to
the non tensorial nature of noncommutativity parameter (which I assume to be constant --- non constant generalizations have been considered, though \cite{Kontse}). Cosmological issues, such as noncommutative
inflation, have also been subjects of research \cite{Kar:2006mt,Alexander:2001dr}. Thorough studies have been devoted to non(anti)commutative generalizations of supersymmetric theories, which imply a non
trivial extension of noncommutative relations to fermionic variables \cite{nonanti}. Summarizing, NC geometry has its modern origin in string theory and its implications can be analyzed in the perspective of
finding phenomenological indications of strings.

\subsubsection{Brane--worlds} 
Conversely, taking as an input the low energy physics as we know it --- Standard Model, General Relativity --- we may wish to find its description inside the string theory framework. A very successful
intuition going in this direction is the brane--world idea. As I mentioned, the low energy effective field theories living on the brane worldvolume are gauge theories. Thus, we may hope to describe
electroweak interactions and QCD in a brane--world picture, allowing large and eventually non compact extra dimensions. However, going towards realistic theories implies for instance that supersymmetry and
conformal invariance, as they appear in string theory, have to be broken. Some literature is devoted to the search of branes configurations \cite{Blumen,Kiritsis:2003mc,Cascales:2003wn} (intersecting branes,
for example) which realize Standard Model features in string theory. 

A great improvement in the subject of brane--world models is represented by the AdS/CFT correspondence. In early times, it was already pointed out that large $N$ gauge theories --- $N$ is the rank of the
gauge group --- displayed stringy characteristics. The large $N$ expansion can indeed be related to the closed string loop expansion if the string coupling is identified with $1/N$ \cite{thooft:1973}.
Furthermore, gauge theories naturally arise in string theory in the presence of branes, more precisely D--branes. On the other hand, so called black brane solutions in supergravity were argued to describe
D--branes in the classical limit \cite{Polchibrane}. This was a hint going towards the formulation of a duality connecting gauge theories on the D--branes to supergravity in the black brane backgrounds.
Stronger indications came from the counting of BPS states and absorption cross sections calculations in D--branes configurations compared to entropy and absorption processes in the supergravity description.
In particular, the $(1+1)$--dimensional CFT living in the intersection of the D1-D5 system was suggested to be dual to a charged black hole supergravity solution, whose near horizon geometry yields
$AdS_3\times S^3$ \cite{Witten:1995,D1D5,Aharony:1999ti}. Analogously, the D3 configuration in type IIB string theory, giving an effective $SU(N)$ ${\cal N}=4$ SYM theory ($N$ is the number of coincident
D--branes), was compared to the black 3--brane supergravity solution where the near horizon geometry is $AdS_5\times S^5$ ($N$ units of five form flux are present and $N$ also determines the $AdS_5$ and $S^5$
radii) \cite{D3,Aharony:1999ti}. 

Finally, Maldacena formulated his conjecture \cite{Maldacena:1997re}, stating that the large $N$ field theory describing the dynamics of opens strings on D--branes (or M--branes, if we consider M theory) is
dual to the full string theory in the corresponding AdS background. Such a duality is holographic in the sense that the dynamics in the supergravity bulk is determined only in terms of boundary degrees of
freedom.  Furthermore, the boundary conditions are exactly identified with the sources of the CFT operators. In this spirit, the gauge theory can be thought to live on the boundary of the AdS space. It is
particularly interesting to note that the Bekenstein--Hawking formula for entropy already suggested the existence of an holographic principle, relating gravity solutions to the dynamics of the background
boundary. The matching of global symmetries also supported Maldacena's idea. The highly non trivial content of AdS/CFT correspondence is that perturbative approximations in the two descriptions hold in
opposite regimes for the effective string coupling constant $g_sN$. The correspondence instead relates the two full theories.

From the time the conjecture was formulated, many checks (mainly on protected quantities) and improvements have been worked out. A rigorous treatment of the divergences that plague the two sides of the
duality is provided by holographic renormalization \cite{Henningson:1998gx,Skenderis:1999nb,deHaro:2000xn}. It has hence been used to perform correlation functions calculations on the gravity side, using a
covariant regularization, and to compare them with the CFT results. An important consequence of holographic renormalization is its application to AdS/CFT duals of supergravity solutions with cutoff
space--times. Such backgrounds appear in Randall--Sundrum models (RS) \cite{Randall:1999ee}, where space--time is a slice of AdS with a brane placed at the fixed point of a $\Zgr_2$ orbifold, playing the role
of a IR cutoff. It has been argued \cite{Hawking:2000kj,Arkani-Hamed:2000ds,Perez-Victoria:2001pa,Nojiri:2000eb} that the holographic dual theory is a cutoff CFT living on the boundary of AdS, coupled to
gravity and higher order corrections.  Indeed, it can be shown that Einstein--Hilbert action and the higher order corrections are the boundary covariant counterterms appearing in the regularization procedure.
The presence of gravity is intriguing since we expect the brane--world to display gravitational interaction if it has to describe real universe (using General Relativity as a theoretical instrument). 
%Since gravity is present on the field theory side, the issue of cosmological evolution for these brane--worlds can be addressed both from the supergravity and holographic point of view.

Brane--world cosmology is a rather broad subject, including applications of stringy models to different issues of cosmology. Among the mostly investigated scenarios, brane induced gravity --- proposed by
Dvali, Gabadadze and Porrati, also called the DGP model \cite{Dvali:1998pa} --- and Randall--Sundrum model \cite{Randall:1999ee} are two alternative ways to obtain 4D gravity in a background with an infinite
extra dimension. DGP and RS models display Newtonian gravity in opposite regimes. Namely, gravity induced on the brane yields effective 4D behaviors at high energies, while in this same regime 5D effects
appear in RS. If the two models are merged, getting induced gravity on a RS brane \cite{Kiritsis:2002ca}, Newton's potential can be recovered at all energies if the brane induced gravity term is strong
compared to the RS scale.  On the other hand, we may modify gravity to get non conventional cosmological features --- primordial inflation, late time acceleration --- including higher order corrections such
as Gauss--Bonnet terms. Gauss--Bonnet brane--worlds \cite{Nojiri:2000gv}--\cite{Sami:2005zc} admit a 4D gravity description in the low energy regime, as RS models. Moreover, the issue of primordial
inflation has been addressed in further string theory contexts. Brane/antibrane inflation \cite{Kachru:2003sx,Blanco-Pillado:2006he,Conlon:2005jm,Burgess:2001fx,Alexander:2001ks,Buchel:2006em}, for instance,
is a thoroughly investigated subject embodying the initial inflationary era in string compactification. Related topics include brane inflation \cite{Dvali:1998pa}, further works on brane induced gravity
\cite{Gabadadze:2006tf}--\cite{Maartens:2003tw}, particular examples with varying speed of light \cite{Alexander:1999cb}, cosmological evolution induced by the rolling tachyon \cite{Gibbons:2002md} and
recent brane--world models \cite{Setare:2006pj}--\cite{Maartens:2006qf}. Brane--world cosmology \cite{Langlois:2002bb}--\cite{Kiritsis:2005bm} can also be analyzed, in a rather general way, following the
mirage cosmology approach \cite{Kehagias:1999vr,Kiritsis:1999tj}, where evolution is driven by a mirage energy density, which encodes bulk effects --- no matter term is there from the beginning. 

Non conventional cosmology \cite{Binetruy:1999ut} at late times can be obtained in RS brane--worlds by considering the interaction between brane and bulk. Indeed, models with brane--bulk energy exchange have
been discussed in literature \cite{Kiritsis:2002zf}--\cite{Cai:2005qm}. It has been shown that a rich variety of cosmologies are produced in the original 5D RS model with the presence of energy exchange. The
brane motion is driven both by the matter energy density and by a mirage radiation, which takes account of the bulk dynamics. Some of the features that these models exhibit can fit into the cosmological
observational data. For instance, eternally accelerating solutions can be found. Furthermore, an holographic cosmology analysis has been performed in \cite{Kiritsis:2005bm}, exploiting gauge/gravity duality
specified to 5D RS model. Non conventional cosmology results are found in the 4D picture and compared to the 5D description, yielding the matching of dimensionful parameters on the two sides.

My work is inspired to the cosmological analysis in RS brane--worlds, both from the bulk gravitational point of view and in the holographic description.  A RS model in seven dimensions was considered in
\cite{Bao:2005ni}, tracing it back to the M5-M2 configuration in M theory. Bao and Lykken concentrated on the graviton mode spectrum analysis.  They found new features with respect to the 5D picture. New
Kaluza--Klein (KK) and winding modes appear due to the additional compactification on the internal two dimensional compact manifold.  Whether the two additional extra dimensions also lead to new properties
for the cosmological evolution is the question I address in the second part of my thesis.

I proposed a 7D RS set--up, with a codimension--one 5--brane and matter on the brane, as well as in the bulk \cite{Mazzanti:2007dq}. In order to make contact with our four dimensional universe, I further
compactify space--time on a two dimensional internal manifold, without necessarily impose homogeneity --- in the sense that evolution in the 3D and 2D spaces may in general be different. The detailed study
of the brane cosmological evolution from the 7D gravity viewpoint is carried, yielding accelerating solutions at late times, among the other possibilities. New features with respect to the 5D set--up appear.
I moreover constructed the holographic dual theory and compared it to the 7D description, generalizing the 6D set--up to the non conformal and interacting case, in analogy to the 4D model.  \vspace{0.5cm}

The structure of this thesis is composed by two parts. The first part is devoted to noncommutative integrable field theories and to my results on noncommutative integrable sine--Gordon. The second part is
dedicated to brane--world holographic cosmology and 7D RS results.

An introduction to integrable systems is given in the first chapter. In chapter \ref{chapter ncft}, I review the Weyl--Moyal formalism for noncommutative geometry, its application to noncommutative quantum
field theories and the relation to string theory. The third and fourth chapters contain the two generalizations to noncommutative geometry of sine--Gordon model that I proposed with my collaborators. The
first theory exhibits acausality and non factorization of the S--matrix, which is calculated at tree level, as shown in chapter \ref{sgI}. As a result, the connection to NC 4--dimensional self--dual
Yang--Mills and to NC Thirring model are also illustrated. The second theory, examined in chapter \ref{sgII} displays integrability properties of the S--matrix. It is shown how this model comes from
dimensional reduction from NC $(2+2)$--dimensional self--dual Yang--Mills, via the intermediate $(2+1)$--dimensional modified sigma model. I give the procedure allowing to construct the noncommutative
multi--soliton solutions and calculate tree level amplitudes.

Chapter \ref{chapter adscft} is a review of AdS/CFT correspondence, particularly focusing on holographic renormalization and RS dual. It is followed by a summary of conventional cosmology issues and by an
introduction on brane--worlds in chapter \ref{cosmo braneworld}. Cosmological evolution in the 5D RS brane--world and the comparison to the holographic dual scenario is also reviewed in chapter \ref{cosmo
braneworld}. The new results on 7D RS brane--world cosmology and holography are illustrated in the two following chapters. In particular, chapter \ref{7D RS cosmo} is devoted to the critical point analysis
and brane cosmological evolution on the 7D gravity side. I construct the 6D dual theory in chapter \ref{7D RS dual}, deriving the Friedmann--like equations and the matching with the 7D description. A summary
on results concludes the thesis.

\end{frontmatter}
\begin{mainmatter}
\fancyhf{}
\fancyhead[RO]{\rightmark}
\fancyhead[LE]{\leftmark}
\fancyfoot[RO]{\thepage}
\fancyfoot[LE]{\thepage}
\part{Noncommutative integrable theories}
\chapter{Integrable systems and the sine-Gordon model}
It is well known that integrable theories can be related to statistical models in their continuous limit. Statistical systems are of high interest in physics literature, for the study of correlation
functions, critical exponents, and other physical measurable quantities. Integrable models are of interest on their own since they are by definition endowed with a number of conserved currents equal to the
number of degrees of freedom. In the case of integrable field theories this number is infinite. Due to this property, integrable models are exactly solvable and in many cases the exact mass spectrum and
S--matrix are calculable. The presence of an infinite number of local conserved currents is a consequence of the equations of motions of the theory and do not need to be generated by a specific action for the
fields. Nevertheless, in some cases of particular interest the action leading to the equations of motion is known and classical and quantum characteristics of the theory may be derived. One of these models is
the sine--Gordon model that I will briefly review in section \ref{sine gordon}. The first part of my thesis is based on the sine--Gordon generalization to noncommutative geometry. Before facing the
sine--Gordon quantum theory, I will clarify its relation to the statistical XY model in subsection \ref{XY} and to the fermionic Thirring theory in subsection \ref{thirring}. I will then sketch some
properties of the soliton solutions and their construction in the last section of this chapter.

I will focus on the properties possessed by the S--matrix for a two dimensional integrable theory. This issue has been studied in the noncommutative generalizations of the sine--Gordon
models constructed in my first two publications \cite{GMPT,LMPPT}. Also noncommutative solitons solutions have been systematically produced, being another important aspect of integrable systems. 
\sectioncount{Sine--Gordon and relations to other models}\label{relations to other models}
I will sketch in this section a couple of interesting links between the sine--Gordon theory on one hand and apparently different models on the other hand. The first topic shows how sine--Gordon field can
describe a 2D Coulomb gas via XY model in the continuous limit. The second correspondence is bosonizations, which relates sine--Gordon to massive Thirring fermionic theory. 
\subsection{The XY model as the sine--Gordon theory}\label{XY}
The XY model describes two dimensional spin variables $\mathbf S$ on a $N$ site lattice of dimension $L$ and step $a$. The components of $\mathbf S$ are $(S_x,S_y)=(\cos\theta,\sin\theta)$, since $\mathbf S$
is normalized to be $|\mathbf S|^2=1$. The system partition function is given by
\bea\label{Z XY}
Z=\int\prod_i\frac{\intd\theta_i}{2\p}\ex^{K\sum_{<i,j>}\cos(\theta_i-\theta_j)}
\ena
where $K=\frac{J}{k_B T}$ ($J$ is the spin coupling, $k_B$ is the Boltzman constant) and $<i,j>$ indicates first neighbors.

This two dimensional spin variable model gives a nice description of a 2D classical Coulomb gas (but also has applications for thin films and fluctuating surfaces). It is thus interesting to note that, due
to the equivalence (in the continuous limit) to the sine--Gordon theory, renormalization group (RG) equation of the quantum field theory describes the dynamics of such statistical systems. 

The spin lattice displays two different behaviors --- and hence a phase transition --- in the high and low temperature regimes. At high temperatures, the correlation function for two spins located at two
different sites of the lattice is exponentially decreasing with the distance between the two sites, while at low temperatures one gets a power dependence. In terms of the Coulomb gas this is interpreted as a
transition between a plasma phase at high temperatures and a neutral gas with coupled charges at low temperatures, where the effect of vortices can be neglected. The critical temperature can be evaluated as
the temperature for which vortices are no more negligible. This gives $2 k_B T_c=\p J$ (renormalization group analysis have also been performed and gives as a result the RG flow of the sine--Gordon model,
that I will shortly sketch in section \ref{sine gordon}).

The equivalence with the sine--Gordon theory is derived by rewriting the partition function \refeq{Z XY} on the dual lattice in the continuous limit $a\to0$ (integrating over the angular spin variables)
\bea\label{Z V}
Z_{V}&=&\sum_n\frac{f^{2n}}{(2n)!}\int\intD\phi\exp\left\{-\int\intd^2 x\left[\oneover{2}\pa^\m\phi(x)\pa_\m\phi(x)\right]\right\}\non\\
&&\int\intd^2 x\left[\ex^{2\imm\p\sqrt K \phi(x)}+\ex^{-2\imm\p\sqrt K \phi(x)}\right]^{2n}
\ena
Here $\phi(x)$ is the continuous limit of the dual lattice variables and $f\equiv\ex^{-\p K\log\frac{a}{r_0}}$ represents the fugacity of vortices in the Coulomb gas language ($r_0$ regulates the UV).
Identifying the sine--Gordon coupling constants by
\bea
\b=2\p\sqrt K \;,\qquad \g=2f=2\ex^{-\p K\log\frac{a}{r_0}}
\ena
one obtains the exact (euclidean) sine--Gordon partition function once summed over $n$ in \refeq{Z V}
\bea
Z_{SG}=\int\intD\phi\exp\left\{-\int\intd^2 x\left[\oneover{2}\pa^\m\phi(x)\pa_\m\phi(x)-\g\cos\b\phi(x)\right]\right\}
\ena

The critical point for the temperature phase transition is thus translated for the sine--Gordon parameters to $\b^2=8\p$. In the short overview of the renormalization results of section \ref{sine gordon} it
will be clear indeed why this phase transition occurs in the quantum field theory.
\subsection{Massive Thirring model and bosonization}\label{thirring}
A second interesting duality relates the massive Thirring model to sine--Gordon via bosonization. Bosonization acts by means of an integration over the fermionic fields in the Thirring partition function,
leaving as a result a path integral over a scalar field, which becomes the sine--Gordon field \cite{coleman,witten,bosonization}.  Explicitly, the partition function for the massive Thirring lagrangian
\bea
\lagra_{MT}=\bps\left(\imm\g^\m\pa_\m-m\right)\psi-\frac{g_{MT}}{2}\bps\g^\m\psi\bps\g_\m\psi
\ena
cab be put in the following form (up to an overall normalization coefficient and gauge fixing)
\bea\label{Z MT}
Z=\int\intD\phi\intD A_\m\intD\phi\,\Delta\left[\pa A\right]\exp\left\{\lagra_{MT}+\imm\bps\g^\m\psi A_\m+\oneover{2}\phi\epsilon^{\m\n}F_{\m\n}\right\}
\ena
The Lagrange multiplier $\phi$ has been introduced imposing a null gauge for the $U(1)$ field strength $F_{\m\n}$ associated to the gauge field $A_\m$ coupled to the fermions. The Faddeev--Popov determinant
is given by $\Delta\left[\pa A\right]=\prod_{x,t}\d\left[\pa^\m A_\m(x,t)\right]$. Fixing the residue gauge to be $A_\m=0$, \refeq{Z MT} becomes the partition function for the Thirring model.

Roughly%
\footnote{The complete calculation needs the introduction of an additional vector field $h_\m$ that appears quadratically in the action coupled to the $U(1)$ current, whose integration immediately gives
\refeq{Z MT}. To be more precise the Thirring lagrangian without mass term is rewritten as $\lagra_{T}=\imm\bps\g^\m\pa_\m\psi-\oneover{2}h^\m h_\m+\imm g_{MT}h^\m\bps\g_\m\psi$.} %
performing in the first place the integration over the fermions, then over the gauge fields, one obtains --- disregarding for the moment the mass term --- that the partition function \refeq{Z MT} corresponds
the the bosonic theory governed by the lagrangian
\bea
\lagra=\oneover{2}\pa^\m\phi\pa_\m\phi
\ena
($\phi$ has been rescaled to $\phi/\sqrt{\p+g_{MT}^2}$). 

The mass term contribution to the bosonic theory is computed exploiting the properties of chiral symmetry breaking. In order to cancel the chiral anomaly when no mass is present one has to impose that the
scalar field $\phi$ transforms under infinitesimal chiral transformations $\psi_L\to\ex^{\imm\a}\psi_L$, $\psi_R\to\ex^{-\imm\a}\psi_R$, parametrized by $\a$ as $\phi\to\phi+\frac{\a}{\p}$. Obviously, the
mass term $m\bps\psi=m\left(\bar\psi_R\psi_L+\bar\psi_L\psi_R\right)\equiv m\left(\mathcal M[\phi]+\mathcal M^\dag[\phi]\right)$ breaks chiral symmetry explicitly. Its transformation rules give a direct
evaluation of the term appearing in the bosonic lagrangian $\mathcal M[\phi]$ and $\mathcal M^\dag[\phi]$ once integrated over fermions. In fact, one gets the following equations
\bea
\mathcal M\left[\phi+\frac{\a}{\p}\right]=\ex^{-2\imm\a}\mathcal M\left[\phi\right] \;,\quad \mathcal M^\dag\left[\phi+\frac{\a}{\p}\right]=\ex^{2\imm\a}\mathcal M^\dag\left[\phi\right]
\ena
yielding $\mathcal M[\phi]\propto\ex^{-2\imm\p\phi}$ and $\mathcal M^\dag[\phi]\propto\ex^{2\imm\p\phi}$. The mass term thus gives a cosine potential for the rescaled field $\phi$.

Putting all together, the bosonic theory is a sine--Gordon
\bea\label{lagra sine gordon}
\lagra_{SG}=\oneover{2}\pa^\m\phi\pa_\m\phi-\g\cos\b\phi
\ena
where $\g=Cm$ ($C$ is a constant) and $\b=\frac{2\p}{\sqrt{\p+g_{MT}^2}}$. From the expression for $\b$ it is clear that bosonization is a strong/weak coupling duality.
\sectioncount{Sine--Gordon at classical and quantum level}\label{sine gordon}
I will briefly mention some important characteristics of the sine--Gordon model. In particular, in the next subsection I will derive the classical action from the equations of motion that one obtains using
the bicomplex formalism. Following \cite{DMH}, this ensures the integrability of the theory. It will be the starting point for the noncommutative generalizations constructed and studied in \cite{GMPT,LMPPT}.
The reduction from self--dual Yang--Mills is also outlined in subsection \ref{ordinary bicomplex}, while quantum properties are described in subsection \ref{commutative renormalization}. Finally, subsection
\ref{ordinary S matrix} contains some general remarks on the S--matrix.
\subsection{Gauging the bicomplex}\label{ordinary bicomplex}
The bicomplex technique guarantees to supply the theory, whose equations of motion can be derived from a matrix valued equation, with an infinite number of local conserved currents. Hence it
offers a systematic method to build integrable field theories. Since this procedure acts at the level of the equations of motion it is not assured that an action can be found. 

In two euclidean dimensions the bicomplex technique is illustrated as follows. Space is spanned by complex coordinates
\beq
z =\frac{1}{\sqrt{2}} (x^0 +\imm x^1) \;,\qquad \bar{z} =\frac{1}{\sqrt{2}} (x^0 -\imm x^1)
\eeq
The bicomplex is a triple $({\mathcal M}, \intd, \d)$ where ${\mathcal M}=\otimes_{r\ge 0}{\mathcal M}^r$ is an $N_0$--graded associative (not necessarily commutative) algebra, ${\mathcal M}^0$ is the algebra
of functions on $\R^2$ and $\intd,\d:{\mathcal M}^r\rightarrow {\mathcal M}^{r+1}$ are two linear maps satisfying the conditions $\intd^2=\d^2=\{\intd,\d\}=0$. ${\mathcal M}^r$ is therefore a space of
$r$--forms. The linear equation characterizing the bicomplex is
\beq\label{linsis}
\d \xi=\l \intd\xi
\eeq 
where $\l$ is a real parameter and $\xi \in {\mathcal M}^s$ for a given $s$. If a non trivial solution $\tilde \xi$ exists, we wish to expand it in powers of the parameter $\l$ as 
\beq
\tilde\xi=\sum_{i=0}^\infty \l^i \xi^{(i)}
\eeq 
The components $\xi^{(i)}\in {\mathcal M}^s$ are then related by an infinite set of linear equations
\bea
\d\xi^{(0)}&=&0\non\\
\d\xi^{(i)}&=&\intd \xi^{(i-1)}\;,\quad i\ge1
\label{claws}
\ena
which give us the desired chain of $\d$--closed and $\d$--exact forms
\beq\label{closed exact forms}
\Xi^{(i+1)}\equiv \intd\xi^{(i)}=\d\xi^{(i+1)}\;,\quad i\ge 0
\eeq
We remark that for the chain not to be trivial $\xi^{(0)}$ must not be $\d$--exact. Now, the equations of motion of the theory should come from the conditions $\intd^2=\d^2=\{\intd,\d\}=0$. When the two
differential maps $d$ and $\d$ are defined in terms of ordinary derivatives in $\R^2$, these conditions are trivial and the chain of conserved currents \refeq{claws} is not associated to
any second order differential equation. 

To have non trivial equations the bicomplex must be gauged. The procedure gets modified by introducing a connection such that
\beq
D_\intd=\intd+ A \;,\qquad D_\d=\d + B
\eeq 
The flatness conditions now amount to $D^2_\intd=D^2_\d=\{D_\intd,D_\d\}=0$ and are non trivial. In fact, the gauged bicomplex provides the differential equations
\bea
&&{\mathcal F}(A)\equiv \intd A +A^2=0\cr
&& {\mathcal F}(B)\equiv\d B + B^2 =0\cr
&& {\mathcal G}(A,B)\equiv \intd B+\d A +\{A,B\}=0 
\label{nonlin}
\ena
In analogy to the trivial set--up, the theory is equipped with an infinite number of conserved currents originating from the solution to the linear differential equation
\beq\label{lin2}
{\mathcal D}\xi \equiv (D_\d - \l D_\intd)\xi=0
\eeq
The nonlinear equations \refeq{nonlin} play the role of the compatibility conditions for \refeq{lin2}
\beq
0={\mathcal D}^2\xi=\left[{\mathcal F}(B)+\l^2{\mathcal F}(A) -\l{\mathcal G}(A,B)\right]\xi
\eeq
A solution $\tilde\xi\in {\mathcal M}^s$ to \refeq{lin2} can be expanded as $\tilde\xi=\sum_{i=0}^\infty \l^i\xi^{(i)}$, giving as a consequence the possibly infinite chain of relations 
\bea\label{chain}
D_\d \xi^{(0)}&=&0\non\\
D_\d \xi^{(i)}&=&D_\intd \xi^{(i-1)} \;,\qquad i\ge 1
\ena
In analogy to \refeq{closed exact forms}, the $D_\d$--closed and $D_\d$--exact forms $\Xi^{(i)}$ can be constructed when $\xi^{(0)}$ is not $D_\d$--exact.

For suitable connections $A$ and $B$ we obtain an infinite number of local%
\footnote{The $\xi^{(i)}$ currents may in general not be local functions of the coordinates. However it is possible to define local conserved currents in terms of the $\xi^{(i)}$ which will have physical
meaning.} % conserved currents associated to non trivial second order differential equations, which are interpreted as the equations of motion of the corresponding integrable model.
\subsubsection{Sine--Gordon from gauged bicomplex}
We define the elements of the bicomplex to be ${\mathcal M}={\mathcal M}_0 \otimes \L$, where ${\mathcal M}_0$ is the space of $2\times 2$ matrices with entries in the algebra of smooth functions on ordinary
$\R^2$ and $\L=\otimes_{i=0}^{2} \L^i$ is a two dimensional graded vector space. We call the $\L^1$ basis $(\tau,\s)$ and impose $\tau^2=\s^2=\{\tau,\s\}=0$. Finally, we define the non gauged differential
maps
\beq
\d \x=\pab \x\tau -R\x\s \;,\qquad \intd\x=-Sf\tau + \pa \x \s
\eeq
in terms of the commuting constant matrices $R$ and $S$. The flatness conditions $\intd^2=\d^2=\{\intd,\d\}=0$ are trivially satisfied in this case. But when gauging the bicomplex --- dressing the $\intd$ map
---
\beq
D \x\equiv G^{-1} \intd(G\x) 
\eeq
by means of a generic invertible matrix $G\in{\mathcal M}_0$, the condition $D ^2=0$ is trivially satisfied, while $\{\d,D\}=0$ yields the nontrivial second order differential equation 
\beq\label{matrixsgo}
\pab\left(G^{-1}\pa G\right)=\left[R,G^{-1} S G\right]
\eeq
In order to specify to the sine--Gordon equation, we choose $R,S$ to be
\beq
R \= S \= \sqrt \g\, \Bigl(\begin{matrix} 0 & 0 \\ 0 & 1 \end{matrix}\Bigr) 
\eeq
and $G\in SU(2)$ as 
\beq
G \= e^{\frac{\im}{2} \s_2 \phi} \= \biggl( \begin{matrix} 
\phantom{-}\cos{\frac{\phi}{2}} & \ \sin{\frac{\phi}{2}} 
\\[4pt]
-\sin{\frac{\phi}{2}} & \ \cos{\frac{\phi}{2}} 
\end{matrix} \biggr)
\eeq
The sine-Gordon equation then follows from the off--diagonal part of the matrix equation (\ref{matrixsgo})
\beq\label{sinegordon}
\pab\pa\phi=\g \sin\phi
\eeq
while the diagonal part gives a trivially satisfied identity. 

The bicomplex approach is straightforward generalizable to noncommutative geometry, inducing noncommutative equations of motion for the theory under exam --- sine--Gordon for our purposes. We note that
deriving the action is a non trivial calculation. In \cite{GMPT} we constructed such an action starting from the deformed equations of motion obtained by generalizing the bicomplex with the introduction of
the noncommutative $\star$--product $D_\intd=\intd+A\star$ and $D_\d=\d +B\star$. This will be explained in chapter \ref{sgI}. Moreover, noncommutativity implies an extension of the $SU(2)$ symmetry group,
which is no longer closed. It will be necessary to consider $U(2)$, rather that $SU(2)$, leading to an extra $U(1)$ factor. The extension of the symmetry group must be carried carefully, as we have shown in
\cite{LMPPT} (see chapter \ref{sgII}).
\subsubsection{From self--dual Yang--Mills to sine--Gordon}
Self--dual Yang--Mills was conjectured by Ward to give origin to all integrable equations in two dimensions via dimensional reduction. The conjecture has been tested over the last years for the most
important known integrable systems. Indeed, sine--Gordon can be obtained both from euclidean ($\mathbb R^4$) and kleinian ($\mathbb R^{(2,2)}$) signature for the four dimensional Yang--Mills equations. In
the first case the dimensional reduction leads to euclidean sine--Gordon, while in the second one gets minkowskian signature for the two dimensional metric.

\paragraph{Self--dual Yang--Mills in short}
Let me summarize the relation between the Yang--Mills self--duality equations and the associated matrix valued equation. Four dimensional Yang--Mills theory with signature $(++--)$ has a stringy origin since
it describes $\mathcal N=2$ strings, which indeed live in a real $(2+2)$--dimensional target space \cite{Ooguri:1990ww}. Self-duality of Yang--Mills models in $\mathbb R^4$ or $\mathbb R^{(2,2)}$ is expressed
by the following equation \cite{Y} 
\bea\label{sdeq}
\frac{1}{2}\e_{\m\n\rho\s}F^{\rho\s}=F_{\m\n}
\ena
where $F_{\m\n}$ is the field strength of the $A_\m$ gauge field $F_{\m\n}=\pa_\m A_\n-\pa_\n A_\m+[A_\m,A_\n]$. Equation \refeq{sdeq} is integrable. In fact, taking for instance the gauge group to be $SU(N)$,
the self--duality equation can be rewritten in terms of complex coordinates $y,\bar y,z,\bar z$ in four dimensions performing an analytical continuation on $A_\m$
\beq\label{complexsd}
F_{yz}=F_{\bar y \bar z}=0 \;,\qquad F_{y \bar y}\pm F_{z \bar z}=0
\eeq
The sign in the second equation depends on the signature of the metric, being $+$ is the euclidean case and $-$ in the kleinian case. The zero value for the mixed $yz$ and $\bar y\bar z$ components of the
field strength makes the fields $A_y,A_z$ (for fixed $\bar y$ and $\bar z$) and $A_{\bar y},A_{\bar z}$ (for fixed $y$ and $z$) pure gauges. Gauge fields may thus be expressed in terms of two $N\times N$
complex matrices $B$ and $\bar B$
\bea
&&A_y=B^{-1}\pa_y B \;,\qquad A_z=B^{-1}\pa_z B\cr
&&A_{\bar y}={\bar B}^{-1}\pa_{\bar y}{\bar B} \;,\qquad A_{\bar z}={\bar B}^{-1}\pa_{\bar z}{\bar B}
\ena
Finally, the Yang formulation of Yang--Mills theory in light--cone gauge is obtained defining a complex $N\times N$ matrix $J=B\bar B^{-1}$. In terms of $J$ the self--duality equations read
\beq\label{sdymyang}
\pa_{\bar y}(J^{-1}\pa_y J)\pm\pa_{\bar z}(J^{-1}\pa_z J)=0
\eeq
The action whose variation leads to such equations of motion is
\bea
S&=&\int \intd^2 y\intd^2z\,{\rm tr}(\pa_y J\pa_{\bar y}J^{-1})-\int\intd^2y\intd^2z\int_0^1\intd\rho\,{\rm tr}\left(\hat J^{-1}\pa_\rho\hat J[\hat J^{-1}\pa_{\bar y}\hat J,\hat J^{-1}\pa_y\hat J]\right)\cr
&&+\int\intd^2 y\intd^2 z\,{\rm tr}(\pa_z J\pa_{\bar z}J^{-1})-\int\intd^2y\intd^2z\int_0^1\intd\rho\,{\rm tr}\left(\hat J^{-1}\pa_\rho\hat J[\hat J^{-1}\pa_{\bar z}\hat J,\hat J^{-1}\pa_z \hat J]\right)
\ena% num ali
where $\hat J(y,\bar y, z,\bar z,\rho)$ is a homotopy path satisfying $\hat J(\rho=0)=1$ and $\hat J(\rho=1)=J$.

A Leznov formulation \cite{L} of the same equations \refeq{complexsd} has also been derived. It corresponds to a different choice of light--cone gauge and is ruled by a cubic action in terms of an algebra
valued field.  I will tell more about this formulation in its noncommutative version in the reduction procedure through the three dimensional modified non linear sigma model in chapter \ref{sgII}. It is
interesting to note that Yang--Mills in the Leznov gauge completely describes $\mathcal N=2$ strings at tree level, while Yang formulation is related to the zero instanton sector of the same theory
\cite{LPS}.

\paragraph{Through dimensional reduction}
The dimensional reduction from four dimensional self--dual Yang--Mills in order to get two dimensional integrable theories must satisfy a requirement about group invariance. More precisely, the theory must be
invariant under any arbitrary subgroup of the group of conformal transformations in four dimensional space--time. The dependence on the disregarded coordinates is eliminated through an algebraic constraint on
the arbitrary matrices involved in the reduction.

As I anticipated, sine--Gordon equations of motion, both in their euclidean and lorentzian version, can arise from the self--duality equations of Yang--Mills theory in four dimensions. It is not trivial that
the action of the bidimensional models in general can be obtained from the Yang--Mills action via the dimensional reduction operated at the level of the equations of motion (as an example we will discuss two
different cases in the noncommutative generalizations, in chapters \ref{sgI} and \ref{sgII}).

Euclidean sine--Gordon comes from the euclidean version of Yang equation (\ref{sdymyang}) when the $B$ and $\bar B$ matrices are chosen to be
\beq
B=\ex^{\frac{z}{2}\s_1} \ex^{\imm\frac{\phi}{2}\s_3} \;,\qquad \bar B=\ex^{\frac{\bar z}{2}\s_1}
\eeq
where $\phi=\phi(y,\bar y)$. In fact, it immediately turns out that the field $\phi$ satisfies (\ref{sinegordon}) with $4\g=-1$.

Kleinian self--dual Yang-Mills equations instead lead to lorentzian sine--Gordon through a two--step reduction procedure. Yang equation (\ref{sdymyang}) is required to have no dependence on one of
the real coordinates $x^i,i=1,2,3,4$, let's say $x^4$. This first step brings to the $(2+1)$--dimensional sigma model equations
\beq
(\eta^{\m\n}+ V_\a \e^{\a\m\n})\pa_\m (J^{-1}\pa_\n J)=0
\eeq
with $V_\a$ defined to be a constant vector in space--time. Non zero $V_\a$ implies the breaking of Lorentz invariance but guarantees integrability if it is chosen to be a space--like vector with unit length
(nonlinear sigma models in $(2+1)$ dimensions can be either Lorentz invariant or integrable but cannot share both these properties \cite{ward}). Once we fix the value for $V_\a$ as $V_\a=(0,1,0)$, the second
step consists in performing a reduction on the matrix $J$ \cite{leese} factorizing the dependence on the third coordinate $x\equiv x^3$
\beq
J=\left(\begin{matrix}\cos \frac{\phi}{2} & \ex^{-\frac{i}{2}x} \sin\frac{\phi}{2}\\
 -\ex^{\frac{i}{2}x} \sin\frac{\phi}{2} & \cos \frac{\phi}{2} 
\end{matrix}\right)\in SU(2)
\eeq
Here $\phi$ is a function of two coordinates only $\phi=\phi(t,y)$, $t\equiv x^1,y\equiv x^2$ (not to be confused with the complex coordinates $y,\bar y$) with different signature. The field $\phi$ satisfies
the sine--Gordon equations of motion in $(1+1)$ dimensions.
\subsection{Quantum properties of ordinary sine--Gordon}\label{commutative renormalization}
We already expect from the discussion about the XY model and its relation to sine--Gordon to get a phase transition for the critical value of the $\b$ coupling constant $\b^2=8\p$. However, this estimate is
naive, since it doesn't take account of the running of the $\b$ coupling. The sine--Gordon theory undergoes a change of regime from super--renormalizability, for $\b<8\p$, to non--renormalizability, for
$\b\ge8\p$. In the non--renormalizable regime, the theory may still be finite pertubatively in $\d\equiv\b^2-8\p$, but the renormalization of $\d$ (or $\b$ equivalently) is needed in addition to $\g$
renormalization (the lagrangian of the theory is precisely given by \refeq{lagra sine gordon}).
\paragraph{Super--renormalizability regime: renormalizing $\mathbf\g$}
Renormalization in the $\b^2<8\p$ regime has to cure divergences which only come from tadpoles with multiple legs (multitadpoles). The value of such Feynmann diagrams is the same for any arbitrary number of
external legs (more precisely, the coefficient depending on the number $M$ of external propagators factors out for every $M$). The implication of this feature is that all correlation functions can be
renormalized at the same time. Moreover, the series over $N$ of (IR and UV) regulated multitadpoles with $N$ internal propagators ($N$ tadpoles) sums to an exponential (up to overall factors) $\propto
\exp\left\{\frac{\b^2}{8\p}\log\frac{m^2}{\L^2}\right\}$, where $m$ and $\L$ are respectively the IR and UV cutoff. Renormalization of $\g$ alone is thus needed and amounts to define the renormalized coupling
constant at a scale $\m$ according to
\bea
\g_R=\m^{-2}\g\left(\frac{\m^2}{\L^2}\right)^{\frac{\b^2}{8\p}}
\ena
at all orders in $\b$.

The renormalization group flow driven from the beta--function $\b_\g=2\g_R\left(\frac{\b^2}{8\p}-1\right)$ displays UV fixed points for $\b^2<8\p$ and would have IR fixed points in the $\b^2>8\p$ regime (this
will hold in the $\b^2\gtrsim8\p$ analysis). Trajectories in the $\g_R/\b^2$ phase space are straight lines parallel to the $\g_R$ axe, since $\b$ doesn't get renormalized in this regime.
\paragraph{Non--renormalizability regime: renormalizing $\mathbf\b$}
New divergences appear in the non--re\-nor\-ma\-li\-za\-bi\-li\-ty regime. They emerge in the two vertices correlation functions, i.e. in the $\g$ second order contribution to the effective action. Individual
Feynmann diagrams are convergent, but their sum over all orders in the $\b$ (or field) expansion diverges. The divergence has a different nature distinguishing the two super-- or non--renormalizable phases:
it is IR for $\b^2<8\p$ and turns to UV for larger values of $\b^2$, $\b^2>8\p$ (the critical value $\b^2=8\p$ gives logarithmical UV divergences). Renormalization of the $\b$ coupling constant is needed. We
consider to be in the proximity of the naive critical point $\d\equiv\b^2-8\p\simeq0$. The renormalized coupling constants (and renormalized field --- since $\b$ renormalization also implies field $\phi$
renormalization not to get the cosine potential renormalized) read
\bea
\g_R=Z^{-1}_\g\m^-2\g &\quad\Rightarrow\quad& Z\g=\left(\frac{\L^2}{\m^2}\right)^{\frac{\b^2}{8\p}}\non\\
\phi_R=Z^{-1/2}_\phi\phi\;,\;\b_R=Z^{1/2}_\phi\b &\quad\Rightarrow\quad& Z_\phi=1-\frac{\p}{2}\b^2\g^2\log\frac{\L^2}{\m^2}
\ena

The renormalization group flow equations are obtained from the beta--functions $\b_\g=2\g\d$ and $\b_\d=8\p^2\g^2$. Noting that $\k^2\equiv\d^2-\g^2$ is a RG invariant, we can divide the RG phase space
$2\p\g_R/\d_R$ according to the sing of $\k$. For $\k>0$ ($\k<0$) one gets IR (UV) fixed points where the theory becomes asymptotically free, but in opposition to the super--renormalizability analysis
trajectories are hyperboloids with axes $\k^2=0$. For imaginary $\k$, $\k^2<0$, the hyperbolic trajectories intersect the $\g$ axe and flow to large negative $\d$ in the IR. 

The transition from dipole to plasma phase happens at $\g=\d$ (intersected with the equation giving $\d$ in terms of $\g$ in the XY model of section \ref{XY}). For $\g<\d$ (dielectric) the trajectories go
towards the region of validity of our small $\d$ approximation, while for $\g>\d$ (conductor) they rapidly flow away.
\subsection{The S--matrix and its features for 2D integrable theories}\label{ordinary S matrix}
Here I make some comments on the S--matrix of integrable two dimensional theories and in particular of the sine--Gordon model. 

\paragraph{The theorem: integrability vs. factorization} It is important to note that the property of integrability for a system in two dimensions is equivalent to the property of factorization of the
S--matrix. More in detail, if a theory possesses an infinite number of local conserved currents --- and hence an infinite number of conserved charges that must be components of Lorentz tensors of increasing
rank --- it
follows that the S--matrix is constrained to be elastic and factorized in two particle scattering \cite{raja,ZZ}. The number of particles involved in a process with given mass and momentum is thus always
conserved and all processes can be described only by some number of two particle scattering, i.e. no production or annihilation occurs. Moreover the two particle S--matrix is associated to a cubic
equation. the solution to this cubic equation can give in most cases the exact form of the S--matrix. For example this is the case for the sine--Gordon theory.

It is worthwhile to note some unavoidable restrictions that must be applied to the integrable theory in order to prove the integrability vs. factorization theorem just stated. This restrictions play an
important role in the generalization to noncommutative geometry, since they fail in noncommutative theories. Precisely, we must have locality and unitarity in the theory. Both these two properties are
typically absent in noncommutative generalizations of quantum field theories, as I will point out in subsection \ref{nc problems}. So, we don't expect the theorem on integrable two dimensional models to be
valid in general. Indeed, I will show two noncommutative examples: the first \cite{GMPT} gives a non factorized S--matrix at tree level, non vanishing production processes and acausality while the second
\cite{LMPPT} displays nice properties such as factorization, absence of production, causality.
\sectioncount{Generalities on solitons}\label{solitons}
I now move to illustrate a very peculiar and useful characteristic of integrable models: the solitonic solutions. Solitons are widely studied in literature (see \cite{raja} for a review). They are defined as
localized solutions of the non linear equations of motion carrying a finite amount of energy. They were originally thought of as a kind of solitonic wave that doesn't change shape and velocity in time or
after scattering processes with other such waves. Since at infinity they have to approach a constant value labeled by an integer $\phi_{sol}\stackrel{\scriptscriptstyle{x\to\pm\infty}}{\sim}2\p n_\pm$, they
come along with an associated (integer) topological charge defined by
\bea
Q_{sol}=\oneover{2\p}\int_{-\infty}^{+\infty}\intd x\;\frac{\pa\phi_{sol}}{\pa x}=n_+-n_-
\ena
Topological charges $Q_{sol}$ of $\pm1$ are associated the the simplest solutions: one--(anti)soliton. 

For the ordinary euclidean sine--Gordon system such a classical solutions of the equations of motion is known to be
\bea
\phi_{sol}(x^0,x^1)=4\arctan \ex^{\sqrt{2\g}\frac{x_1-x^1_0-\imm v x^0}{\sqrt{1-v^2}}}=-\phi_{antisol}(x^0,x^1)
\ena
where $v$ is the velocity parameter of the soliton and $\g$ is the sine--Gordon coupling constant.
% Derrick argument?
\paragraph{Dressing the solitons} Multi--soliton solutions can be constructed using a recursive procedure that I will refer to as the dressing method \cite{dressing}, inspired by the original work by Belavin
and Zakharov \cite{Belavin:1977cz}. Babelon and Bernard worked out multi--soliton solutions recursively from the one--soliton \cite{Babelon:1993yd}. In our paper \cite{LMPPT} sine--Gordon noncommutative
solitons are obtained by reduction from self--dual Yang--Mills and $(2+1)$--dimensional sigma model. Since most of the known integrable theories descend from self--dual Yang--Mills, as I already pointed out,
I will briefly mention the procedure for deriving its multi--soliton solutions. 

The key observation is that self--duality equations \refeq{sdeq} can be interpreted as the integrability conditions of a linear system containing the spectral parameter $\z\in\mathbb{CP}^1$ (in complex
coordinates)
\bea\label{psi eq}
\left(\bar D_1-\z D_2\right)\psi=0 \;,\quad \left(\bar D_2-\z D_1\right)\psi=0 \;,\quad \pab_\z\psi=0
\ena
Covariant derivatives are defined as $D_i=\pa_i+A_i$ (and analogously for $\bar D_i$) where $A_i$ are the complex gauge fields $A_1=\oneover{2}\left(A_{x^1}-\imm A_{x^2}\right)$,
$A_2=\oneover{2}\left(A_{x^3}-\imm A_{x^4}\right)$ and their complex conjugate $\bar A_1,\bar A_2$, follows from the definition. Equations \refeq{psi eq} must be solved for the arbitrary field $\psi$. Gauge
fields are then deduced by reverting \refeq{psi eq} as
\bea\label{gauge fields eq}
\bar A_1-\z A_2=\psi\left(\pab_1-\z\pa_2\right)\psi^{-1} \;,\quad \bar A_2-\z A_1=\psi\left(\pab_2-\z\pa_1\right)\psi^{-1}
\ena
and imposing the reality condition on $\psi$
\bea
\psi^{-1}(x,\z)=\left[\psi\left(x,-\bar\z^{-1}\right)\right]^\dagger
\ena
This comes from noticing that $\left\{\left[\psi\left(x,-\bar\z^{-1}\right)\right]^\dagger\right\}^{-1}$ also solves equations \refeq{psi eq}. The field $\psi$ is assumed to be meromorphic in the spectral
parameter $\z$, so that it solves \refeq{gauge fields eq} only if the residues vanish. In fact, the l.h.s. of \refeq{gauge fields eq} is linear in $\z$ since the gauge fields are $\z$--independent. Hence no
poles exist. The requirement of vanishing residues allows to calculate simple soliton solutions for $A_i$ and $\bar A_i$ \cite{Belavin:1977cz} such as the BPST one--instanton \cite{BPST}. 

The dressing procedure generates new solutions $\psi$ to \refeq{psi eq} from a known solution $\tilde\psi$, multiplying it by a dressing factor $\chi$ on the left $\psi(x,\z)=\chi(x,\z)\tilde\psi(x,\z)$.
The factor $\chi$ is a function of the complex coordinates. Being meromorphic in $\z$, it can be expanded as %
\footnote{Here I use the notations of \cite{Horvath}.}%
\bea\label{chi expansion}
\chi=\z R_{-1}+R_0+\sum_{i=1}^r\frac{R_i}{\m_i\z+\n_i}
\ena
where $R_n$ are some $\z$--independent complex matrices.  Equations \refeq{gauge fields eq} written in terms of the dressing factor read
\bea\label{new sol eq}
\chi\left(\tilde{\bar D_1}-\z \tilde D_2\right)\chi^\dagger={{\bar A}}_1-\z A_2 \;,\quad \chi\left(\tilde{\bar D_2}-\z\tilde D_1\right)\chi^\dagger={{\bar A}}_2-\z A_1 
\ena
where covariant derivatives are though in terms of the old $\tilde A_i,\tilde{\bar A_i}$ gauge fields. Again, since $ A_i, {{\bar A}}_i$ are $\z$--independent, the solutions to \refeq{new sol eq} are found
imposing zero value for all the residues associated to the poles appearing in $\chi$ \refeq{chi expansion}. This yields a set of differential equations for the $\z$--independent coefficients $R_{-1}$, $R_0$
and $R_i$ for $i=1,\dots,r$. In turn, using \refeq{new sol eq} and substituting the derived expression for $\chi$, we get a set of differential equations for the new solutions $A_i$ and ${\bar A}_i$.

Specific multi--soliton solutions for self--dual Yang--Mills were constructed in \cite{Belavin:1977cz}, while solutions by dressing method for noncommutative self--dual Yang--Mills are described in
$(1+1)$--dimensional NC sine--Gordon multi--solitons from $(2+1)$--dimensional NC sigma model solutions, which is itself derived by dimensional reduction from $(2+2)$--dimensional self--dual Yang--Mills.

\chapter{Basics and origins of noncommutative field theories}\label{chapter ncft}
The aim of this chapter is to motivate and introduce the study of noncommutative geometry and noncommutative field theories. The interest in noncommutativity has grown over the years thanks to the important
improvements in understanding string theories and the consequent emergence of noncommutative backgrounds in its context. Some examples of space--time coordinate noncommutativity originating from string theory
will be mentioned in subsection \ref{nc strings}. Nonetheless, there exists a famous example of quantum mechanics which already introduces noncommutativity relations among coordinates: the quantum Hall
effect, which I will sketch in subsection \ref{hall}. Even earlier, motivations to noncommutative geometry applied to quantum field theories were adopted, such as the novelty of an intrinsic UV cutoff
furnished to the theory, due to the noncommutation relations among space coordinates. However, the intrinsic cutoff doesn't seem to give better renormalization results in comparison to the usual
regularization schemes. In addition it shows a typical feature in noncommutative theories mixing UV with IR divergences, in such a way that the two high and low energy limits don't commute (more
about this will be discussed together with the main common problems and properties of NC field theories in subsection \ref{nc problems}).

Noncommutative geometry formalism has hence been developed for more than twenty years \cite{Connes,Szabo} and has been recently understood in terms of strings and branes \cite{SW}.
Extensive studies have since then been performed on noncommutative generalization of quantum field theories. The fundamental relation characterizing noncommutative geometry is the non vanishing commutator
\bea
\left[x^\m,x^\n\right]=\imm\theta^{\m\n}\non
\ena
which is determined by the noncommutativity (antisymmetric) parameter $\theta^{\m\n}$ (the constant value of $\theta^{\m\n}$ will be justified in the next section). How an algebra of functions (fields) can be
defined in noncommutative geometry is the subject of the next section, while concrete construction of NC FT are discussed in section \ref{ncft} and, in particular, integrable NC deformations are described in
the last section.
\sectioncount{Weyl formalism and Moyal product}\label{star product section}
Non vanishing commutation relation among coordinates remind of the quantum phase space for particles, which is described by a non trivial commutator between momenta and positions
\bea\label{qm}
&&\left[\hat{x}^i,\hat{p}_j\right]=\imm\hbar\d^i_{j}\cr
&&\left[\hat{x}^i,\hat{x}^j\right]=\left[\hat{p}_i,\hat{p}_j\right]=0
\ena
In the same way as the quantum mechanics commutators imply the well--known uncertainty principle $\D\hat x^i\D\hat p_j\ge\frac{\hbar}{2}\d^i_j$, the noncommutative geometry coordinate algebra
\bea\label{nc geo}
\left[\hx^\m,\hx^\n\right]=\imm\theta^{\m\n}
\ena
--- generally time/space noncommutativity $\theta^{0i}\ne0$ can be considered and related problems will be illustrated in section \ref{nc problems} --- gives rise to the space--time uncertainty relations
\beq
\D \hx^\m\D \hx^\n\geq\frac{1}{2}\vert\theta^{\m\n}\vert 
\eeq
Hence, at distances lower that the order of the noncommutativity parameter $\sqrt{\vert\theta^{\m\n}\vert}$, ordinary geometry can no longer be used to describe space--time. In fact there are reasons to
believe that at very short distances (i.e. shorter than the Planck length) known geometry should be replaced by some new physics, since quantum effects of gravity could arise. When the NC parameter vanishes,
ordinary geometry is recovered. 

The parallel between quantum mechanics phase space and noncommutative geometry can be pushed further. Analogously to the correspondence between functions of the phase space variables $x^i,p_j$ and the
associated operators expressed in terms of the quantum momentum and position operators $\hat x^i,\hat p_j$, one can construct a map going from the commutative algebra of functions over $\R^d$ (where a
noncommutative product is implemented) to the noncommutative algebra of operators generated by the coordinate operators obeying to \refeq{nc geo}. This is formally achieved by the Weyl transform.
\subsection{Moyal product arising from Weyl transform}
The case of my interest is two dimensional space with variables $x^1,x^2$ associated to noncommuting operators $\hat{x}^1,\hat{x}^2$ that satisfy $\left[\hat{x}^1,\hat{x}^2\right]=\imm\theta^{12}$ with
constant $\theta^{12}\equiv\theta$. The Weyl transform associates an operator $\weyl_f(\hx^1,\hx^2)$ to a function $f(x^1,x^2)$ of the coordinates --- provided with the usual pointwise product. The Weyl
operator $\weyl_f(\hx^1,\hx^2)$ is defined via the Fourier transform of the function $f$
\bea
\weyl_f(\hx^1,\hx^2)=\int\frac{\intd^2x}{(2\p)^2}\tilde f(k_1,k_2)\,\ex^{\imm k_i\hx^i}=\int\intd^2xf(x^1,x^2)\hD(x^1,x^2)
\ena
where the Fourier transform is as usual
\bea
\tilde f(k_1,k_2)=\int\intd^2x\,\ex^{\imm k_ix^i}f(x^1,x^2)
\ena
The map $\hD$ is hermitian and equal to
\bea
\hD(x^1,x^2)=\int\frac{\intd^2k}{(2\p)^2}\,\ex^{\imm k_i\hx^i}\ex^{-\imm k_ix^i}
\ena
It is interpreted as a mixed basis for operators and fields on the two dimensional space.  Moreover, it reduces to $\hD_0(x)=\d^2(\hx-x)$ when commutativity is restored $\theta\to0$. The trace of the Weyl
operator gives an integral of the associated function over the space
\bea
\traccia\left[\weyl_f(\hx^1,\hx^2)\right]=\int\intd^2x\,f(x^1,x^2)
\ena
if we normalize $\traccia\,\hD(x)=1$. From the trace normalization 
\bea
\traccia\left[\hD(x^1,x^2)\hD(y^1,y^2)\right]=\d^{(2)}(x-y)
\ena
and the product of two maps $\hD$ (the Baker--Campbell--Hausdorff formula should be used) we can deduce that the map between functions $f(x^1,x^2)$ and operators $\weyl_f(\hx^1,\hx^2)$ via $\hD(x^1,x^2)$ is
invertible and represents a one--to--one correspondence between Weyl operators and Wigner distribution functions. Indeed, these functions are obtained by means of the following inverted relation
\bea
f(x^1,x^2)=\traccia\left[\weyl_f(\hx^1,\hx^2)\hD(x^1,x^2)\right]
\ena 

A noncommutative product among functions belonging to the commutative algebra is introduced as the image via the inverse $\hD$ map of the product of Weyl operators. In fact, using the expression for the
product of two $\hD$ operators
\bea
\hD(x^1,x^2)\hD(y^1,y^2)=\oneover{\p^2\vert\det\theta\vert}\int\intd^2z\hD(z^1,z^2)\,\ex^{-2\imm\left(\theta^{-1}\right)_{ij}(x-z)^i(y-z)^j}
\ena
it follows that
\bea
\traccia\left[\weyl_f(\hx)\weyl_g(\hx)\hD(x)\right]=\oneover{\p^2\vert\det\theta\vert}\int\intd^2y\intd^2z\,f(y)g(z)\ex^{-2\imm\left(\theta^{-1}\right)_{ij}(x-y)^i(y-z)^j}
\ena
The product between two Weyl operators is thus mapped to a noncommutative product between functions
\bea
\weyl_f(\hx^1,\hx^2)\weyl_g(\hx^1,\hx^2)=\weyl_{f\star g}(\hx^1,\hx^2)
\ena
where the noncommutative $\star$\emph{--product} is defined by
\bea\label{star prod}
\left(f\star g\right)(x)=\ex^{\frac{\imm}{2}\theta^{ij}\pa_i\pa_j'}\,f(x)g(x')\bigg|_{x=x'}=f(x)\ex^{\frac{\imm}{2}\overleftarrow{\pa}_{i}\theta^{ij}\overrightarrow{\pa}_{j}}g(x)
\ena
Obviously, the usual product is recovered in \refeq{star prod} when $\theta\to0$. For $n$ functions the $\star$--product formula is straightforward generalized to
\bea
f_1(x_1)\star\dots\star f_n(x_n)=\prod_{a<b}\ex^{\frac{\imm}{2}\theta^{ij}\pa_{x^i_a}\pa_{x^j-b}}f_1(x_1)\cdots f_n(x_n)
\ena
We note that the commutator of the commutative algebra, evaluated substituting the $\star$--product to the usual one, reproduces the quantum commutation relation of the noncommutative operator
algebra
\bea\label{star brackets}
[x^i,x^j]_\star=\imm\theta^{ij}
\ena
The Weyl transform is extendible to any number of space--time dimensions and may also be generalized to more complicated quantized algebras where the $\hx^i$'s commutators are not only c--numbers
\cite{Kontse}.
\paragraph{Properties $\star$--product} There are three main properties of $\star$--product, which are fundamental for perturbative calculations in noncommutative field theories.
\begin{enumeratei}
\item
Associativity remains a property of this noncommutative product. In fact, the $\star$--product defined in \refeq{star prod} is a special example of the associative products that arise in the deformation
quantization \cite{Castellani}. The deformation of an algebra is defined by a formal power series expansion in the deformation parameter $\l$, such that the $\l^0$ order restores the algebra itself. The
multiplication rule between elements of the algebra $f,g$ is defined as
\bea\label{deform product}
f\star_\l g=fg+\sum_{n=1}^{\infty}\l^n C_n(f,g)
\ena
Equality \refeq{star prod} hence defines a unique deformation of the algebra of function to a noncommutative associative algebra (up to local redefinitions of the elements of the algebra), since it can be
rewritten in the following form
\bea
f\star g=fg+\sum_{n=1}^{\infty}\left(\frac{\imm}{2}\right)^n\oneover{n!}\theta^{i_1j_1}\cdots\theta^{i_nj_n}\pa_{i_1}\cdots\pa_{i_n}f(x)\,\pa_{j_1}\cdots\pa_{j_n}g(x)
\ena
I will further discuss algebra deformations in the next subsection.
%More precisely, the deformation quantization just described defines a Poisson structure for the manifold over which the algebra of functions $f,g,\dots$ acts.
\item 
The $\star$--product is closed under complex conjugation. For complex valued functions we get $\left(f\star g\right)^*=g^*\star f^*$.
\item
Ciclic invariance under integration is a very important property of $\star$--product, which directly comes from the ciclicity of trace of Weyl operators
\bea
\int\intd^2xf_1\star\dots\star f_n=\traccia\left[\weyl_{f_1}\cdots\weyl_{f_n}\right]
\ena
In particular, the $\star$--product of two functions is equivalent to the usual product when integrated.
\end{enumeratei}
\subsection{Moyal product defined by translation covariance and associativity}
As I mentioned above, the $\star$--product appearing in the Weyl formalism can be interpreted as coming from a special algebra deformation where the product \refeq{deform product} is defined by the Poisson
bracket of functions. I will now formulate in a more rigorous way how the $\star$--product we chose can be uniquely derived by imposing the properties of associativity and translation invariance to a Poisson
structure product over a generic manifold $\mathcal M$.

The manifold $\mani$ is endowed with a Poisson structure $P$ if for any two functions $f,g\in\mathcal A$ in the algebra $\mathcal A$ (in general a $C^*$--algebra) defined over $\mani$ we specify the Poisson
bracket
\bea
P(f,g)=\left\{f,g\right\}_P=P^{\m\n}\nabla_{\m}\nabla'_{\n}f(x)g(x')\bigg|_{x=x'}=f\overleftarrow{\nabla}_{\m}P^{\m\n}\overrightarrow{\nabla}_{\n}g
\label{poisson}
\ena
A generic product on $\mani$ is then defined through the derivatives $\nabla_\m$ defined on the manifold itself (with vanishing torsion and curvature) by
\bea\label{svil1}
f\star_\l g =\sum_{n=0}^{\infty}\l^r\frac{a_n}{n!}P^n(f,g)
\ena
where the coefficients $a_0$ and $a_1$ must be chosen to be $a_0=a_1=1$ and
\bea\label{svil2}
P^n(f,g)=P^{\m_1 \n_1}\cdots P^{\m_n \n_n}\nabla_{\m_1}\dots\nabla_{\m_n}f\nabla_{\n_1}\dots\nabla_{\n_n}g
\ena
Comparing \refeq{svil1} to \refeq{deform product} we can immediately relate the deformation coefficients $C_n(f,g)$ to the Poisson structure $P(f,g)$ by $C_n(f,g)=\frac{a_n}{n!}P^n(f,g)$.  A necessary
hypothesis on $P$ in order to have associativity for the Moyal product \refeq{svil1} is $\nabla_\m P=0$, i.e. the Poisson structure must be constant. However this is not the only constraint that associativity
implies. The property
\bea\label{moyalasso}
(f\star g)\star h=f\star (g\star h)
\ena
gives order by order equations for the coefficients $a_n$. As a result, we get that all the $a_n$ must be equal to $a_n=1$. It is essential for the Poisson structure to be constant and for the derivatives to
be
curvature and torsion free. If any of these assumptions on $P$ and $\nabla_\m$ is dropped, the Moyal product defined by \refeq{svil1} is no longer associative and \refeq{moyalasso} doesn't hold anymore. We
can see the complete analogy with the $\star$--product of \refeq{star prod} by rewriting \refeq{svil1}. Substituing the results from associativity constraints, we obtain
\bea
f\star_\l g=f\ex^{\l\overleftarrow{\nabla}_{\m}P^{\m\n}\overrightarrow{\nabla}_{\n}}g
\ena
Moreover, if $\mani$ is flat with coordinates $x^\m$ and $\nabla_\m$ are ordinary derivatives, the Moyal commutator between coordinates exactly gives
\bea
\left[x^{\m}, x^{\n}\right]_{\star_\l}=2\l P^{\m\n}.
\ena
Renaming $2\l P^{\m\n}=\imm\theta^{\m\n}$ we obtain our $\star$--product \refeq{star prod} and the Moyal brackets \refeq{star brackets}.
% note on Konstevich product
\paragraph{Poicaré invariance}
A constant value for the Poisson structure is not only necessary for Moyal product to be associative but also for translations to be symmetries of the theory defined by the Moyal deformation (i.e.
substituing usual product with the Moyal deformation rule). Let's impose Poincar\'e invariance for a theory of fields over the noncommutative algebra $[x^\m,x^\n]=\imm\theta^{\m\n}(x)$ including coordinate
dependence (I drop the hat operator symbol to simplify notation). Supposing that the $\theta^{\m\n}$ matrix doesn't transform, we get that $x\to x'$ implies
\bea
[x^\m,x^\n]=\imm\theta^{\m\n}(x)\longrightarrow[x'^\m,x'^\n]=\imm\theta^{\m\n}(x')
\ena
and not $\theta'^{\m\n}(x')$. For translations $x^\m\to (x^\m+a^\m)$, the commutation relation becomes
\bea\label{invariance}
[x'^{\m},x'^{\n}]=[x^{\m}+a^{\m},x^{\n}+a^{\n}]=[x^{\m},x^{\n}]
\ena
As a consequence $\theta$ must satisfy the constraint
\bea\label{invariance2}
\theta^{\m\n}(x+a)=\theta^{\m\n}(x)
\ena
It has to be constant as a local function of the coordinates. I now turn to Lorentz transformations $x^\m\to\L^\m_\n x^\n$%
~\footnote{This are the \emph{particle} Lorentz transformation. \emph{Observer} Lorentz transformations (i.e. static particle in moving frame) would imply only covariance for the $\theta^{\m\n}$ tensor.}%
. The commutator changes according to
\bea
[x'^\m,x'^\n]=[\L^\m_\rho x^\rho,\L^\n_\s x^\s]=\L^\m_\rho[x^\rho,x^\s]\L^\n_\s
\ena
Since generally
\bea
\L^{\m}_{\rho}\theta^{\rho\s}\L^{\n}_{\s}\,\stackrel{d>2}{\neq}\,\theta^{\m\n}
\ena
we conclude that Lorentz invariance is not preserved in theories deformed with the noncommutative $\star$--product \refeq{nc geo}. Nevertheless, two dimensional space plays a particular role, since
two dimensional Lorentz transformations represented by the $\L^{\m\n}$ matrices commute with the antisymmetric $\theta^{\m\n}$ and are both multiples of the Ricci tensor $\epsilon^{\m\n}$.
\sectioncount{From quantum Hall to strings and branes}\label{hall and strings}
As I remarked at the beginning of last section, the main motivation to study field theories in noncommutative geometry is the natural appearance of noncommutative backgrounds in string theory. Noncommutative
Yang--Mills theory arises in type IIB string theory when a constant NS--NS two form $B^{\m\n}$ is turned on. The commutator of stringy coordinates is analogue to \refeq{nc geo} and the effective action
constructed from vertex operators contains a $\star$--product with the form \refeq{star prod} as multiplication among fields. In the small energy limit, the noncommutative parameter is inversely proportional
to the NS--NS two form $\theta\propto B^{-1}$. This same relation arises in the analysis of the quantum Hall effect, where a strong magnetic field $B$ is considered. I will show in the following subsection
how noncommutativity emerges in this simple case. The superstring context will be then illustrated in subsection \ref{nc strings}.
\subsection{Noncommutativity in strong magnetic field}\label{hall}
The dynamics of particles of mass $m$ moving on a two dimensional surface in a magnetic field $B$ with potential $A_i=-\frac{B}{2}\epsilon_{ij}x^j$ is governed by the Hamiltonian
\bea
\mathcal H=\oneover{2m}\mathfrak p^2
\ena
where $\mathfrak p$ is the physical gauge invariant momentum $\mathfrak p=m\dot{\mathbf x}=\mathbf p+\mathbf A$ ($\mathbf p$ is the conjugate momentum to $\mathbf x$). The commutation relation for $\mathbf
p_i$'s is immediately derived
\bea
[\mathbf p_i,\mathbf p_j]=\imm B\epsilon_{ij}
\ena
Noncommutativity of the coordinates $x^i$'s emerges when the spectrum of the quantum system is projected over the lowest of Landau levels, which are separated by an energy $\D E\propto\frac{B}{m}$. In the
strong magnetic field limit, $\mathfrak p\to0$ and the Lagrangian is
\bea
\lagra=-\frac{B}{2}\dot x^i\epsilon_{ij}x^j
\ena
From canonical quantum commutation rules $[x^i,p_j]=\imm\d^i_j$ and using $p_i=-\frac{B}{2}\epsilon_{ij}x^j$, we finally obtain
\bea
[x^i,x^j]=\frac{\imm}{B}\epsilon^{ij}
\ena
which is precisely \refeq{nc geo} in two dimensions with $\theta=B^{-1}$. Since the Hamiltonian vanishes in the strong magnetic field limit, the theory becomes topological. Moreover, every function of space
coordinates $x^1,x^2$ is a function of momenta $p_1,p_2$. This dependence is encoded in a phase factor of the form $\exp\left\{\frac{2\imm}{B}k_i\epsilon^{ij}p_j\right\}$ appearing in the Fourier
(anti)transform
\bea
f(\mathbf x)=\int\intd^2k\tilde f(\mathbf k)\,\ex^{\frac{2\imm}{B}k_i\epsilon^{ij}p_j}
\ena
The analogy with open strings in a constant NS--NS two form field will become clear in the next subsection.
\subsection{String backgrounds and noncommutative geometry}\label{nc strings}
Different string theories can lead to different non(anti)commutative deformations of space--time. It is known that turning on the Neveu--Schwarz $B$--field in a background with D--branes generally amounts to a
noncommutative deformation of the space--time geometry seen by the open strings ending on the coincident D--branes. The resulting effective small energy gauge theories living on the branes ($\mathcal N=4$
four dimensional SYM for D3 in IIB strings, for instance) acquire a noncommutative $\star$--product. I will illustrate a couple of examples of my interest (for applying subsequent dimensional reductions and
obtaining the two dimensional integrable theories examined in my papers) on how the deformation works. The discussion will mainly follow \cite{SW} for bosonic and type IIB strings and \cite{LPS} for $\mathcal
N=2$ superstring theory.

Another non(anti)commutative model arising in string theories is worth being mentioned. In pure spinor formalism it has been shown that Ramond--Ramond field backgrounds are related to non trivial
anticommutator of fermionic coordinates \cite{nonanti}, in analogy to the Neveu--Schwarz--Neveu--Schwarz field, which is connected to noncommutativity of bosonic coordinates.
\subsubsection{A constant $B$ field and noncommutative (S)YM}
A famous example of noncommutative field theory emerging in string theory is described in \cite{SW}. The set--up is the bosonic sector of string theory, in ten dimensional flat background given by the metric
$g_{\m\n}$, the constant NS--NS two form $B_{\m\n}$ and some D$p$--branes. The electric $B_{0i}\neq0$ and magnetic $B_{0i}=0$ cases must be analyzed separately, since an electric $B$ field displays some
essential differences with respect to magnetic backgrounds, in particular in the small energy decoupling limit.

Let us start with a magnetic $B_{\m\n}$. If its rank is equal to $r$ we assume that $B_{ij}\neq0$ only for $i,j=1,\dots,r$ and that the D$p$--branes have dimensions $p+1\ge r$. The metric $g_{ij}$ has
vanishing components for $i=1,\dots,r$ and $j\neq1,\dots,r$. The string worldsheet action is separated in two contributions from the worldsheet bulk $\S$ and boundary $\pa\S$
\bea\label{string}
S&=&\frac{1}{4\p\a'}\int_{\S}\left(g_{\m\n}\pa_a x^{\m}\pa^a x^{\n}-2\p \imm\a' B_{ij}\e^{ab}\pa_a x^i \pa_b x^j\right)=\cr
&=&\frac{1}{4\p\a'}\int_{\S}g_{\m\n}\pa_a x^{\m}\pa^a x^{\n}-\frac{\imm}{2}\int_{\pa \S}B_{ij}x^{i}\pa_t x^{j}
\ena
Here $\a'$ is as usual related to the string length, $\pa_t$ is the tangent derivative w.r.t. the boundary $\pa\S$. The boundary term is a consequence of the presence of the D$p$--branes and thus modifies the
boundary conditions for the open strings along the branes
\bea\label{BC}
g_{ij}\pa_n x^{j}+2\p \imm\a' B_{ij}\pa_t x^{j}\big|_{\pa\S}=0
\ena
The derivative transverse to the boundary is $\pa_n$ and $i,j$ run over directions along the brane. Boundary conditions \refeq{BC} interpolate between the Neumann and Dirichlet ones, for $B=0$ and
$B\to\infty$ with maximal rank, respectively. In the classical approximation of string theory, we can parametrize $\S$ by the complex coordinates $z,\bar z$ as a disc conformally mapped into the upper half
plane, with $\Im (z)\ge0$ and the boundary located at $z=\bar z$. Equation \refeq{BC} reads
\bea\label{BC2}
g_{ij}(\pa-\bar{\pa})x^{j}+2\p\a'B_{ij}(\pa+\bar{\pa})x^{j}\big|_{z=\bar{z}}=0
\ena
In order to write the commutator for the space--time coordinates $x^i$'s, we first have to calculate the corresponding correlator. The OPE from the action \refeq{string} yields
\bea\label{propag}
\langle x^{i}(z) x^{j}(z')\rangle&=&-\a'\bigg[g^{ij}\log\vert z-z'\vert- g^{ij}\log\vert z-\bar{z}'\vert+\cr
&&+G^{ij} \log\vert z-\bar{z}'\vert^2+\frac{1}{2\p\a'}\theta^{ij}\log\frac{z-\bar{z}'}{\bar{z}-z'}+D^{ij}\bigg]
\ena
Here
\bea\label{param}
G^{ij}&=&\left(\frac{1}{g+2\p\a' B}\right)^{ij}_S=\left(\frac{1}{g+2\p\a' B}g\frac{1}{g-2\p\a' B}\right)^{ij}\cr
G_{ij}&=&g_{ij}-\left(2\p\a'\right)^2\left(Bg^{-1}B\right)_{ij}\cr
\theta^{ij}&=&2\p\a'\left(\frac{1}{g+2\p\a' B}\right)^{ij}_A=-\left(2\p\a'\right)^2\left(\frac{1}{g+2\p\a' B}B \frac{1}{g-2\p\a' B}\right)^{ij}
\ena
where $(~~)_S$, $(~~)_A$ denote the symmetric and antisymmetric part of the matrix in brackets, respectively. The last term in the propagator \refeq{propag}, $D^{ij}$, is a constant. It is independent of
$z,\bar z$, but may in general depend on $B_{ij}$. It can thus be fixed to some specific value given the arbitrariness of $B_{ij}$. We can restrict to the real axe $z\to\t\in\R$ since open string vertex
operators are always inserted on the boundary of the string worldsheet. The propagator evaluated as a function of the real variables $\t,\t'$ yields
\bea\label{open}
\langle x^{i}(\t) x^{j}(\t')\rangle=-\a'G^{ij}\log(\t-\t')^2+\frac{\imm}{2}\theta^{ij}\e(\t-\t')
\ena
where $\epsilon(\t)$ denotes the sign of $\t$. From \refeq{open} we deduce that $G^{ij}$ is the metric seen by the open strings, just as $g^{ij}$ is the closed string one. Moreover, $G^{ij}$ reduces to
$g^{ij}$ in the $\theta^{ij}\to0$ limit. We can already interpret $\theta^{ij}$ as a noncommutativity parameter for the coordinates along the brane, since the commutator can be derived as
time--ordered product.  As a result we indeed obtain
\bea\label{nc boso strings}
\left[x^i(\t),x^j(\t)\right]=T\left(x^i(\t)~x^j(\t^-)-x^i(\t)~x^j(\t^+)\right)=\imm\theta^{ij}
\ena
The known relation \refeq{nc geo} has been explicitly obtained for the space--time directions parallel to the D$p$--branes in the $B$ field background of critical bosonic strings. 

Correlation functions exhibit a phase factor that corresponds to the introduction of $\star$--product in the $(p+1)$--dimensional space--time string effective action on the brane, in the $\a'\to0$ limit. If
we start by evaluating for instance the product $\mathcal O(\t)\mathcal O(\t')$ of two tachyon operator vertices $\mathcal O(\t)=\ex^{\imm p\cd x}(\t)$ we expect that the leading short distance behavior of
the operator product is independent of $(\t-\t')$ in the limit $\a'\to0$. The product would also have to be associative, to preserve translation invariance. It furthermore has to be given by \refeq{nc boso
strings}. These conditions uniquely determine the $\star$--product, as I observed in section \ref{star product section}. In formulae, the two--tachyon OPE reads
\bea
\ex^{\imm p\cdot x}(\t)\cdot\ex^{\imm q\cdot x}(\t')\sim(\t-\t')^{2\a'G^{ij}p_i q_j}\ex^{-\frac{\imm}{2}\theta^{ij}p_i q_j}\ex^{\imm (p+q)\cdot x}(\t')+...
\ena
which reduces to the $\star$--product multiplication
\bea\label{openc}
e^{\imm p\cdot x}(\t) e^{\imm q\cdot x}(\t')\sim e^{\imm p\cdot x}\star e^{\imm q\cdot x}(\t') 
\ena
in the zero slope limit $\a'\to0$, i.e. when the term $(\t-\t')^{2\a'G^{ij}p_i q_j}$ can be ignored. It is possible to write all correlation functions evaluated for a non vanishing noncommutativity parameter
$\theta^{ij}$ in terms of the correlation functions themselves evaluated at zero $\theta_{ij}$. Moreover, the factor relating one to the other is a phase factor that can be traced back to the $\star$--product
phase factor (that appears in the quantum Hall analysis as well). It is understood that we here use a description of the open string theory on branes in terms of the open string parameters $G^{ij}$ and
$\theta^{ij}$, which represent the effective metric seen by the open strings and the noncommutativity parameter. We instead started with a dependence on the closed string metric and two form, $g^{ij}$ and
$B^{ij}$. Noncommutativity is explicit when $\theta^{ij}$ enters in the description. Generalizing the formula for product of two tachyon vertex operators, we obtain the expression for the expectation
value of products of $k$ polynomial operators of the form $P(\pa x,\pa x^2,\dots)\ex^{\imm p\cd x}$ ($x$'s are directions along the brane)
\bea\label{corr func star}
&&\left\langle\prod_{n=1}^k P_n\left(\pa x(\t_n),\pa^2 x(\t_n),\dots\right)e^{ip^n\cd x(\t_n)}\right\rangle_{G,\theta}\non\\
&=&\ex^{-\frac{i}{2}\sum_{n>m}p_i^n\theta^{ij}p_j^m\e(\t_n-\t_m)}\left\langle\prod_{n=1}^k P_n\left(\pa x(\t_n),\pa^2 x(\t_n),\dots\right)e^{ip^n\cdot x(\t_n)}\right\rangle_{G,\theta=0}% num ali
\ena
The path integrals of vertex operators associated to the massless spectrum of strings allow to write the effective action of the theory, order by order in $\a'$. The effective action for a $k$--point function
is expressed in terms of $N\times N$ matrix--valued fields $\Phi_i$ ($N\times N$ is the number of Chan--Paton factors and hence $N$ is the rank of the gauge group living on the brane). In the $B^{ij}=0$ case
(i.e. $\theta^{ij}=0$) we get
\bea
\int \intd^{p+1}x\,\sqrt{\det G}\,\traccia\left(\pa^{n_1}\Phi_1\pa^{n_2}\Phi_2\dots\pa^{n_k}\Phi_k\right)
\ena
where $\intd^{p+1}x$ is a measure along the brane directions. When the $B$--field is turned on, we can encode the dependence on this NS field simply by implementing the $\star$--product in the effective
action
\bea
\int \intd^{p+1}x\,\sqrt{\det G}\,\traccia\left(\pa^{n_1}\Phi_1\star\pa^{n_2}\Phi_2\star\dots\star\pa^{n_k}\Phi_k\right)
\ena
This modification in the fields product leads to the correlation functions of the form \refeq{corr func star}. To see how noncommutative effective gauge theory describing string theory with constant
$B$--field are generated, we have to take the $\a'\to0$ limit. To consistently keep $G^{ij}$ and $\theta^{ij}$ finite while $\a'$ vanishes we choose
\bea
&&\a'\sim\e^{\frac{1}{2}}\longrightarrow0\cr
&&g_{ij}\sim\e\longrightarrow0 \quad i,j=1,...,r
\ena
In this regime the effective metric and noncommutativity parameter become
\bea\label{paramalpha}
&&G^{ij}=\left\{
\begin{array}{ll}
-\frac{1}{(2\p\a')^2}\left(\frac{1}{B}g\frac{1}{B}\right)^{ij}&{\rm{for}}\quad i,j=1,...,r\cr
g^{ij}&\rm{otherwise}
\end{array}\right.\cr
&&G_{ij}=\left\{
\begin{array}{ll}
-(2\p\a')^2(Bg^{-1}B)_{ij}&{\rm{for}}\quad i,j=1,...,r\cr
g_{ij}&\rm{otherwise}
\end{array}\right.\cr
&&\theta^{ij}=\left\{
\begin{array}{ll}
\left(\frac{1}{B}\right)^{ij}&{\rm{for}}\quad i,j=1,...,r\cr
0&\rm{otherwise}
\end{array}\right.
\ena
The propagator simplifies to
\bea\label{propalpha}
\langle x^i(\t)x^j(0)\rangle=\frac{i}{2}\theta^{ij}\e(\t)
\ena
From the most singular behavior of the OPE of two generic functions
%:f(x(\t)):~:g(x(0)):~=~:e^{\frac{i}{2}\e(\t)\theta^{ij}\frac{\pa}{\pa x^i(\t)}\frac{\pa}{\pa x^j(0)}}f(x(\t))g(x(0)):
\bea\label{funstar}
\lim_{\t\to 0^+}:f(x(\t)):\,:g(x(0)):\,=\,:f(x(0))\star g(x(0)):
\ena
we obtain the correlation functions of $k$ generic operators on the worldsheet boundary
\bea\label{correlators}
\left\langle\prod_n f_n(x(\t_n))\right\rangle=\int\intd x f_1(x)\star\dots\star f_k(x)
\ena
This shows that the effective theory correlation functions turn out to contain $\star$--products just as in the noncommutative \emph{natural} field theory, in the sense of the definition of chapter
\ref{chapter ncft}.
\paragraph{(Non)commutative Yang--Mills from gauge fields in the $\a'\to0$ limit}
The effective YM description of the open string theory is obtained when gauge field are added on the brane, leading to an additional boundary contribution to the worldsheet action \refeq{string}
\bea\label{actiongauge}
-i\int d\t A_i(x)\pa_{\t}x^i
\ena
This is an example of rank one gauge fields, which I will assume for simplicity to be the case. The dependence on the constant $B$ field in the string action \refeq{string} may be written in terms of a gauge
field $A_i=-\oneover{2}B_{ij}x^j$ with field strength $F_{ij}=B_{ij}$. In the $B$ field description we get an invariance of the bosonic string theory under the following gauge symmetry
\bea
\d B_{\m\n}=\pa_{[\m} \L_{\n]}
\ena
combined with a transformation changing the boundary term
\bea
\d A_\m = \L_\m
\ena
This tells us that the gauge invariant, physically meaningful, combination of NS two form and one form gauge field is $(F_{ij}+B_{ij})$. On the other hand, the gauge field contribution \refeq{actiongauge}
to the worldsheet action is naively invariant under gauge transformations
\bea\label{trasf}
\d A_i=\pa_i \l
\ena
However, this invariance may be modified by the necessary regularization prescription. Indeed, noncommutativity pops out when we use point splitting regularization, where the product of operators at the same
point never appears. The variation of the exponential of the action \refeq{actiongauge} under the transformations \refeq{trasf}, evaluated eliminating the region $|\t-\t'|<\d$, yields 
\bea
-\int\intd\t A_i(x)\pa_{\t}x^i\cdot\int\intd\t'\pa_{\t'}\l
\ena
at first order in $A_i$. Integrating and taking then the $\d\to0$ limit one finds that non vanishing boundary contributions survive and break invariance for naive gauge transformations if we use
\refeq{funstar}.  These terms can be written as a $\star$--product commutator
\bea\label{termine}
&&-\int\intd\t:A_i(x(\t))\pa_{\t}x^i(\t):\;:\left[\l(x(\t^-))-\l(x(\t^+))\right]:\cr
&&=-\int\intd\t:\left(A_i(x)\star\l-\l\star A_i(x)\right)\pa_{\t}x^i:
\ena
The true gauge symmetry in the point splitting regularization scheme is thus given by
\bea\label{noncomm trasf}
\d \hat{A}_i=\pa_i \hat{\l}+i\hat{\l}\star\hat{A}_i-i\hat{A}_i\star\hat{\l}
\ena
This result is valid at each order in the $A_i$ expansion of the action exponential. However, point splitting is not the only regularization that one may choose. Pauli--Villars regulated action remains
invariant under ordinary transformations \refeq{trasf}. The existence of two descriptions depending on the choice of regularization means that the two different theories associated to the gauge symmetries
\refeq{trasf} and \refeq{noncomm trasf} has to be related by some duality. More precisely, Seiberg and Witten argued that, on the one hand, point splitting regularization gives noncommutative YM. In fact the
expectation value of gauge vertex operators can be traced back to the noncommutative effective action
\bea
S_{\rm{eff}}\propto \int \sqrt{\det G}~G^{ii'}G^{jj'}Tr\left( \hat{F}_{ij}\ast \hat{F}_{i'j'}\right)
\ena
On the other hand, Pauli--Villars procedure leads instead to ordinary YM. The duality relating the two descriptions involves the change of variables $\hat A=\hat A(A)$ and $\hat\l=\hat\l(\l,A)$, if we denote
with hatted notation the fields belonging to the noncommutative theory. The precise form of the duality has been worked out in \cite{SW}.
%\EQ
%\hat{A}(A)+\hat{\d}_{\hat{\l}}\hat{A}(A)=\hat{A}(A+\d_{\l}A)
%\label{regola}
%\EN

At the beginning of this paragraph, I assumed that the gauge field to be a rank one field. Nevertheless, taking $N$ coincident D$p$--branes rather than only one, immediately changes the gauge group into
$U(N)$.  Noncommutative $U(N)$ YM can thus emerge. $SU(N)$ YM cannot be described in a NC context, since the $U(1)$ degree of freedom doesn't decouple in noncommutative embedding --- in other words, $SU(N)$
is not closed under NC product. Other gauge groups can also be obtained by orientifolding, such as $SO(N)$ and $Sp(N)$ \cite{matsubara}.
\paragraph{A note on electric $B$ field}
I here draw some crucial conclusions about the electric $B$--field case, i.e. when we drop the assumption $B^{0i}=0$. The main point that I will discuss is that field theories with time/space noncommutativity
do not consistently emerge from string theory, in the effective field theory description.

First of all, stability problems arise. An electric field $E\equiv B^{01}$ extending along a spatial direction $i$ --- for instance we choose $i=1$ --- and along the time direction, should be bounded by an
upper critical value $E_c$ 
\bea
E\leq E_c\,\qquad  {\rm where}\; E_c=\frac{g}{2\p\a'}
\ena
Moreover, the open string parameters $G\equiv G^{01}$ and $\theta\equiv\theta^{01}$ satisfy the relation \cite{SST}
\bea
\a' \left(G^{-1}\right)= \frac{1}{2\p}\frac{E}{E_c}\theta
\ena
This equation shows that the zero slope limit $\a'\to0$ together with the request of finite effective open string metric $G^{01}$ must also entail the commutative limit represented by the vanishing of the
noncommutative parameter $\theta\to0$. Hence, the effective field theory description in the decoupling limit is necessarily commutative, while if we want to keep the noncommutativity parameter turned on, the
full string theory with finite $\a'$ must be considered.

On the other hand, a consistent limit can be taken \cite{SST} where
\bea
\frac{E}{E_c}\longrightarrow 1 \,, \qquad g\sim\frac{1}{1-\left(\frac{E}{E_c}\right)^2} \,,\qquad \theta\longrightarrow2\p\a'G^{-1}
\ena
The noncommutative parameter is finite and we get an open string theory on noncommutative space--time where open strings decouple from closed strings. It can be argued that this open string theory on
time/space noncommutative background is dual --- via $S$--duality --- to strongly coupled $\mathcal N=4$ YM with space noncommutativity \cite{Gopakumar:2000na}. More details about string time/space
noncommutative string theory can be found in \cite{Klebanov:2000pp}. The absence of a decoupled field theory limit of string theory with constant electric field reflects the problems of quantum field theories
where time/space components of the noncommutativity parameter are assumed not to vanish. All stringy massive modes are necessary to get a consistent theory. On the other hand, these problems involve causality
and unitarity that I will discuss in subsection \ref{nc problems}.
\paragraph{Susy Ramond--Neveu--Schwarz and Green--Schwarz strings}
I quickly mention how a constant NS--NS field modifies the superstring effective field theory description with respect to the zero $B$ field background set--up. Noncommutativity still appears among the
bosonic coordinates alone, while fermionic ones are not involved.

Including the fermionic sector in the context of string theories with manifest $\mathcal N=1$ worldsheet supersymmetry and turning on the Neveu--Schwarz--Neuveu--Schwarz two form $B^{ij}$, gives the following
action
\bea\label{actionRNS}
S&=&\frac{1}{4\p\a'}\int\intd^2 z\left[\bar{\pa} x^{\m}\pa x_{\m}+\imm\psi^{\m}\bar{\pa}\psi_{\m}+\imm\bar{\psi}^{\m}\pa\bar{\psi}_{\m}+\right.\non\\
&&\left.-2\p \imm\a'B_{ij}\left(\e^{ab}\pa_a x^i\pa_b x^j+\bar\psi^i\rho^a\pa_a\psi^j\right)\right]
\ena
Again $i,j$ denote the directions along the D$p$--brane. Boundary conditions take the form
\bea\label{BCRNS}
\left.g_{ij}(\pa-\bar{\pa})x^{j}+2\p\a'B_{ij}(\pa+\bar{\pa})x^{j}\right|_{z=\bar{z}}&=&0 \cr
\left.g_{ij}(\psi^j-\bar{\psi}^j)+2\p\a'B_{ij}(\psi^j+\bar{\psi}^j)\right|_{z=\bar{z}}&=&0
\ena
The last term in the action \refeq{actionRNS} is to be added so that \refeq{BCRNS} are true supersymmetric invariant boundary contributions to the field equations of motion. As a consequence, the action and
boundary conditions are invariant under the following supersymmetry transformations, written in terms of the supersymmetry parameter $\eta$
\bea\label{superconf}
&&\d x^i=-\imm\eta(\psi^i+\bar{\psi}^i)\cr
&&\d\psi^i=\eta\pa x^i\cr
&&\d\bar{\psi}^i=\eta\bar{\pa}x^i
\ena
Here $\eta$ is a worldsheet spinor and space--time scalar.  The YM description can be obtained in analogy to the bosonic string case, coupling the string to a gauge field $A_i$ --- a $U(1)$ gauge field for
simplicity. The full action has to be modified with the addition of an extra boundary term
\bea\label{accopp}
-\imm\int d\t\left[A_i(x)\pa_{\t}x^i-\imm F_{ij}\Psi^i\Psi^j\right]
\ena
where the ordinary field strength $F_{ij}$ is coupled to the fermionic combination 
\bea
\Psi^i=\frac{1}{2}\left(\psi^i+\bar{\psi}^i\right)
\ena
Under supersymmetry \refeq{superconf} the gauge field action \refeq{accopp} transforms by a total derivative, as in the bosonic string theory. However, using point splitting regularization, surface
terms don't cancel. Instead, the variation of the exponential of the action, expanded at first order in $A_i$, gives
\bea\label{point split boundary}
\imm\int d\t\int d \t'\left[A_i\pa_{\t}x^i(\t)-\imm F_{ij}\Psi^i\Psi^j(\t)\right]\left[-2\imm \eta\pa_{\t'}A_k\Psi^k(\t')\right]
\ena
So, when $\t'$ approaches $\t$ we need some extra boundary term in the action in order to compensate for the contribution coming from \refeq{point split boundary}. Extending the analysis to all orders in
$A_i$, Seiberg and Witten came to the conclusion that the correct gauge field action that accounts for point splitting regularization is \refeq{accopp} where usual product gets replaced by noncommutative
$\star$--product, i.e.
\bea
F_{ij}\longrightarrow\hat{F}_{ij}=\pa_i\hat A_j-\pa_j\hat A_i-\imm\hat A_i\star\hat A_j+\imm\hat A_j\star\hat A_i
\ena
Hatted fields denotes noncommutative functions of space--time coordinates. The gauge theory that can be derived in the zero slope limit of RNS open strings ending on a D$p$--brane with constant $B$
field through point splitting regularization is a noncommutative YM theory, where noncommutativity is introduced among bosonic coordinates spanning the directions along the D$p$--brane. On the other hand,
Pauli--Villars regularization leads to an ordinary commutative gauge theory. The two theories are related by the Seiberg--Witten duality.

Supersymmetry in target space is realized by the Green--Schwarz (GS) string. Fermionic coordinates $\theta^{\a i}$, $i=1,2$ are included in the set of coordinates $(x^\m,\theta^{\a i})$ for $\mathcal N=2$
supersymmetric ten dimensional space--time. The authors of \cite{Chu} showed that for GS string with D$p$--branes and constant NS--NS two form the fermionic coordinates are not involved in the typical
noncommutativity emerging from the appropriate constant $B$ field background. Constant non trivial Ramond--Ramond backgrounds are instead necessary in order to get noncommuting $\theta^{\a i}$'s
\cite{nonanti} (I'm not going to discuss this kind of non(anti)commutativity, though). The analysis for the NS--NS background is performed in \cite{SW}. Boundary conditions for the bosonic coordinates are
equal to those in the bosonic string theory \refeq{BC}, while also the fermionic coordinates must obey to some $B$--dependent relation on the boundary. These boundary conditions are constraints on the phase
space and imply a noncommutative algebra for the $x^i$'s, while the $\theta^{\a i}$'s surviving to the gauge fixing exhibit the usual commutation rules.
\paragraph{Noncommutative self--dual Yang--Mills from $\mathcal N=2$ strings}
Extended $\mathcal N=2$ supersymmetry in the open string worldsheet leads to ordinary self--dual YM (SDYM) when no NS--NS two form is present in the background. It was shown in \cite{SW} that a constant $B$
field, just as in the bosonic case, causes noncommutativity among the space coordinates along the $N$ D3--branes, which are space--filling. In fact, the target space for such a supersymmetric string theory
has to be $(2+2)$--dimensional. It is spanned by the real coordinates $x^\m$, $\m=1,\dots,4$. The RNS Majorana spinors $\psi^\m$, arising as superpartners of bosonic coordinates, as well as the $x^\m$ are
coupled to the $\mathcal N=2$ supergravity multiplet.

By gauging away all the gravity degrees of freedom and remaining with a residual $\mathcal N=2$ supersymmetry out of the full superconformal group, the space--time action for open $\mathcal N=2$ string theory
in the superconformal gauge reads
\bea
S=-\frac{1}{4\pi\a'}\int_\Sigma \intd^2 \xi \,\eta^{\a\b}\left(\pa_\a x^\m \pa_\b x^\n + i \bar \psi ^\m \rho_\a \pa_\b \psi^\n\right)g_{\m\n}
\ena
The $\mathcal N=2$ supersymmetry transformations under which the action is invariant are
\bea\label{2susy}
\d x^\m &=&\bar \e_1 \psi^\m + J^\m_\n \bar \e_2 \psi^\n \cr
\d\psi^\m &=& -i \rho^\a \pa_\a x^\m \e_1 + i J^\m_\n \rho^\a \pa_\a x^\n \e_2
\ena
where $J_\m^\n$ is a complex structure compatible with the metric
\bea\label{J condition}
g_{\m\n}J^\n_\l + J_\m^\n g_{\l\n}=0
\ena
(in flat metric $J_2^1=-J_1^2=J_4^3=-J_3^4=1$). The spectrum of this open string theory only includes one massless scalar (just as in the $\mathcal N=2$ closed string theory) and the tree level string
amplitudes beyond three point functions vanish. Zero value for an infinite number of tree level amplitudes should imply conservation for an infinite number of currents, i.e. an infinite number of symmetries.
This suggests that the theory is integrable. In fact, an alternative definition for integrability in four dimensions was argued to be the vanishing of all amplitudes beyond the three point functions.
Moreover, the three point function is shown \cite{LPS} to come from the self--dual Yang--Mills action in $(2+2)$ dimensions. A stack of $N$ D3--branes is hence described by an $SU(N)$ SDYM theory.

Coupling the $\mathcal N=2$ superstrings with the NS--NS two form field $B^{\m\n}$ yields additional boundary terms in the space--time action
\bea
S&=&-\frac{1}{4\pi\a'}\int_\Sigma \intd^2 \xi\left[\left(\eta^{\a\b}g_{\m\n} + \e^{\a\b}2\pi\a' B_{\m\n}\right)\pa_\a x^\m \pa_\b x^\n \right.\cr
&&\left.+ \left(g_{\m\n}+ 2\pi\a' B_{\m\n}\right) i \bar \psi ^\m \rho^\a \pa_\a\psi^\n\right]
\ena
A boundary $B$--dependent term has been added just as in the $\mathcal N=1$ string theory, in order to get boundary conditions consistent with supersymmetry transformations. Thus, we obtain
\bea
\left.\left[\left(g_{\n\m}+2\p\a'B_{\m\n}\right)J^\n_\l\pa_+x^\l-\left(g_{\m\n}+2\p\a'B_{\m\n}\right)J^\n_\l\pa_-x^\l\right]\right|_{\pa\Sigma}=0
\ena
where $\pa_{\pm}=\pa_0\pm\pa_1$. The $B_{\m\n}\wedge\intd x^\m\wedge\intd x^\n$ background two form field is K\"ahler on $\R^{2,2}$, closed and constrained by the compatibility condition with respect to the
complex structure $J^\n_\m$
\bea
B_{\m\n}J^\n_\l-J^\n_\m B_{\l\n}=0
\ena
where $J^\n_\n$ still satisfies \refeq{J condition}.

Noncommutativity arises first through the evaluation of the open string correlators, in analogy to the bosonic and ten dimensional $\mathcal N=1$ theory,
\bea\label{propa}
\left\langle x^\m(\tau)\,x^\n(\tau^{\prime})\right\rangle&=&-\a' G^{\m\n}\,\log(\tau-\tau^{\prime})^2+\frac{\imm}{2}\theta^{\m\n}\,\epsilon(\tau-\tau^{\prime})\quad\\
\left\langle\psi^\m(\tau)\,\psi^\n(\tau^{\prime})\right\rangle&=&\frac{G^{\m\n}}{\tau-\tau^{\prime}}
\ena
with $\tau,\tau^{\prime}\in\pa\Sigma$ and $[(g+2\pi\a^{\prime}B)^{-1}]^{\mu\nu}=G^{\m\n}+\frac{1}{2\p\a^{\prime}}\theta^{\mu\nu}$. The effective metric seen by the open string is as usual referred to as
$G^{\m\n}$, while the noncommutative parameter is $\theta^{\m\n}$. Choosing specific generators of the $SO(2,2)$ symmetry group, we can write $J$ and $B$ in the basis of $U(1)\times U(1)\subset SO(2,2)$
generators
\bea
J_{12}=-J_{21}=J_{34} =-J_{43}\equiv1\,,\quad B_{12}=-B_{21}\equiv B_1\,, \quad B_{34} =-B_{43}\equiv B_2
\ena
Hence
\bea
G^{11}=G^{22}\equiv\frac{\z}{\z^2+(2\p\a'B_1)^2}\,, \quad G^{33}=G^{44}\equiv-\frac{\z}{\z^2+(2\p\a'B_2)^2}\non\\
\theta^{12}=-\theta^{21}=-\frac{(2\p\a')^2B_1}{\z^2+(2\p\a'B_1)^2}\,, \quad \theta^{34}=-\theta^{43}=-\frac{(2\p\a')^2B_2}{\z^2+(2\p\a'B_2)^2}
\ena

In the zero slope limit $\a'\to0$ (and $\z\sim(\a')^2\to0$, $g^{\m\n}\to0$), keeping the open string parameters finite, it is shown that the three string amplitudes at tree level lead to noncommutative
self--dual Yang--Mills in the Leznov gauge (also Yang gauge can be obtained) \cite{Ooguri:1990ww}. The effective theory yields a cubic lagrangian
\bea
{\cal L}=\frac{1}{2} G^{\m\n}{\rm tr}\left(\pa_\m \Phi\star\pa_\n \Phi\right) + \frac{1}{3}\e^{\ad\bd}{\rm tr}\left(\Phi\star\hat\pa_{0\ad} \Phi\star\hat\pa_{0\bd}\Phi\right)
\ena
where $\hat{\pa}_{0\dot{0}}\equiv\hat{\pa}_{\hat{2}}{+}\hat{\pa}_{\hat{4}}$ and $\hat{\pa}_{0\dot{1}}\equiv\hat{\pa}_{\hat{1}}{-}\hat{\pa}_{\hat{3}},$ having defined $\hat{\pa}_{\hat{\mu}}\equiv
e_{\hat{\mu}}^{\nu}\pa_{\nu}$ by means of the tetrad $G^{\m\n}\equiv e^\m_{\hat\sigma}e^\n_{\hat\lambda}\eta^{\hat\sigma\hat\lambda}$. Both in the Leznov and in the Yang formulation of noncommutative SDYM all
tree level amplitudes beyond the three point functions vanish, as it should be for integrable theories. Indeed, this is in agreement with the string amplitudes computations for four dimensional $\mathcal N=2$
theory.

I will use these results to discuss the dimensional reduction from $(2+2)$--dimensional SDYM in order to get integrable noncommutative sine--Gordon generalization studied in the paper in collaboration with
O.~Lechtenfeld, S.~Penati, A.~Popov and L.~Tamassia \cite{LMPPT}.
\sectioncount{How to deform quantum field theories}\label{ncft}
The most natural generalization of field theories to noncommutative geometry is obtained by substituing the usual product with the noncommutative $\star$--product in the action. Other generalizations are of
course achievable, provided that in the $\theta\to0$ limit they reduce to the original commutative theory. In particular, I will explain in subsection \ref{integr nc ft} how to consistently deform
integrable quantum field theories. However, many typical features of noncommutative field theories can be deduced from the \emph{natural} generalization. I am going to illustrate this simple deformation and
apply it to some specific and meaningful examples.
\subsection{The free theory, interactions and Feynmann rules}
I now consider the action of an arbitrary field theory modified by the introduction of $\star$--product, replacing ordinary multiplication among fields.
\paragraph{Free and commutative}
Since $\star$--product is invariant under cyclic permutations of functions under integration and, in particular, $\star$--product of two generic functions is equivalent to usual commutative multiplication
between the two functions, all quadratic terms in the field theory action are not affected by noncommutativity. Hence, natural deformations of conventional free theories imply no differences with respect to
the ordinary theories.

An important consequence of the invariance of quadratic terms under noncommutative deformations is that free propagators in any noncommutative theory are equal to those evaluated when $\theta\to0$, i.e. in the
ordinary field theory.
\paragraph{Noncommutative interactions}
When an $n$--field interaction term is present in the field theory naturally generalized to NC geometry, the relevant action for this term gets modified by the phase factor coming from the $\star$--product.
In momentum space $\phi(x)=\int\frac{\intd^d k}{(2\p)^d}\ex^{\imm k\cd x}\phi(k)$, NC interactions are given by
\bea
\int\prod_{i=1}^n\frac{\intd^dk_i}{(2\p)^2}V(k_1,\dots,k_n)\phi(k_1)\cdots\phi(k_n)
\ena
with
\bea
V(k_1,\dots,k_n)=\d^{(d)}\left(\sum_{i=1}^n\right)\ex^{-\frac{\imm}{2}\sum_{i<j}k_i\times k_j}
\ena
I defined the antisymmetric product
\bea\label{times prod}
a\times b\equiv\theta^{\m\n}a_\m b_\n
\ena
As a consequence of the cyclicity of $\star$--product, the phase factor $V$ is invariant under cyclic permutations of momenta.
Feynmann rules get modified by the introduction of this phase factor, which is sensible to the ordering of momenta entering in each vertex.

The form of the phase factor distinguishes two different situations. In the simplest case, the phase is an overall factor depending only on the external momenta and hence do not modify the evaluation of the
internal loops. This means that the degree of divergence of the corresponding graph doesn't change. The \emph{planar} diagrams describe this situation. On the other hand, when the internal momenta
dependence is non trivial in $V$, the UV behavior of Feynmann diagrams may not be left unchanged with respect to the ordinary theory. This are referred to as the \emph{nonplanar} diagrams.
\subsection{Examples}
A couple of typical and illustrative examples of field theories generalized to noncommutative geometry are sketched below, following the rules of \emph{natural} deformation.
\subsubsection{The deformed $\phi^4$}
The action for the scalar $\phi^4$ noncommutative theory, following the rules just described, is given by
\bea
S\left[\phi\right]=\int\intd^4x\left[\frac{1}{2}\pa_{\m}\phi\pa^{\m}\phi-\frac{m^2}{2}\phi^2-\frac{\l}{4!}\phi\star\phi\star\phi\star\phi\right]
\ena
The interaction term written in the momentum space reads
\bea\label{feynman}
&&\frac{\l}{4!}\int\intd^4x\,\phi\star\phi\star\phi\star\phi=\non\\
&=&\frac{1}{(2\p)^{16}}\frac{\l}{4!}\int\intd^4k_1\,\intd^4k_2\,\intd^4k_3\,\intd^4k_4\,V(k_1,k_2,k_3,k_4)\,\phi(k_1)\phi(k_2)\phi(k_3)\phi(k_4)
\ena
with the phase factor yielding
\bea
V(k_1,k_2,k_3,k_4)&=&(2\p)^4\d^{(4)}\left(\sum_{i=1}^4 k_i\right)\left(\cos \frac{k_1\theta k_2}{2}\cos\frac{k_3\theta k_4}{2}+\right.\non\\
&&\left.+\cos\frac{k_1\theta k_3}{2}\cos\frac{k_2\theta k_4}{2}+\cos\frac{k_1\theta k_4}{2}\cos\frac{k_2\theta k_3}{2}\right)
\ena
An interesting calculation is the 2--point function evaluation at one loop. We know that it diverges in ordinary field theory and it determines the mass renormalization (at first order) in the ordinary case. At zeroth order there is no
change with respect to commutative 2--point function, since as I previously remarked the free propagator isn't involved in the noncommutative deformation. At one loop order, however, we get a nonplanar
contribution to the Feynmann diagrams, which amounts to
\bea
\int\frac{\intd^4k}{(2\p)^4}\frac{\ex^{\imm k\times p}}{k^2+m^2}
\ena
(up to the symmetry coefficient). The difference with respect to the planar contribution resides in the phase factor $\ex^{\imm k\times p}$ which is indeed absent in the planar graph --- the antisymmetric
product is defined in \refeq{times prod}. In addition, we define a product that is quadratic in the noncommutative parameter
\bea\label{circ product}
a\circ b\equiv-a_\m\theta^{\m\rho}\theta^\n_\rho b_\n
\ena
Then, an effective UV cutoff $\L_{\rm{eff}}^2=\oneover{p\circ p+1/\L^2}$, expressed in terms of the usual UV cutoff $\L$ and of the external momentum $p$ flowing in the two point function can be introduced.
Renormalization turns out to be formulated in terms of this effective cutoff.  Non planar Feynmann diagrams depend on the UV cutoff $\L$ only through $\L_{\rm eff}(p)$. This is crucial, since it shows the
typical mixing of ultraviolet and infrared divergences occurring in noncommutative field theories. In fact, the effective cutoff remains finite when $\L^2\to\infty$ if the external momentum is non zero.
However, if $p\to0$ (or in the commutative limit $\theta\to0$) then $\L_{\rm{eff}}\to\infty$ when the ordinary UV cutoff is taken to diverge, since $\L_{\rm{eff}}\to\L$. In other words, in the IR limit we
recover the usual UV divergence. The renormalized two point function at one loop yields
\bea
\G^{(2)}_R&=&p^2+m^2_R+\frac{g}{96\p^2}\oneover{p\circ p+\frac{1}{\L^2}}+\non\\
&&+\frac{g}{96\p^2}m^2_R\log\left[m^2_R\left(p\circ p+\oneover{\L^2}\right)\right]
\ena
with the renormalized mass equal to $m^2_R=m^2+\frac{g}{48\p^2}\L^2+\frac{g}{48\p^2}m^2\log\frac{m^2}{\L^2}$. We can take the two opposite limits $p\circ p\ll\frac{1}{\L^2}$ and $p\circ p\gg\oneover{\L^2}$.
In the first case the effective two point action reduces to the ordinary $\G^{(2)}_R=p^2+m'^2_R$ where the renormalized mass of the ordinary theory $m'_R$ is related to $m_R$ by
$m'^2_R=m^2_R+\frac{g}{96\p^2}\L^2+\frac{g}{96\p^2}m^2\log\frac{m^2}{\L^2}$. This agrees with the fact that IR limit captures the commutative UV divergence. In the UV limit instead, $p\circ
p\gg\oneover{\L^2}$, the 2--point function suffers from IR divergences for $p\circ p\to0$. We get a first order pole and a logarithmic divergence. These behaviors could be traced back to an effective action
including an extra (scalar) degree of freedom $\chi$ that suitably interacts with the scalar field $\phi$, yielding the correct two point amplitudes.

It is clear that in the $\theta\to0$ limit, the ordinary commutative theory is restored.
\subsubsection{Noncommutative gauge symmetries and Yang--Mills}
Ordinary gauge theories can be invariant under the action of gauge groups such as for instance $U(N)$, $SU(N)$, $SO(N)$ and $Sp(N)$. Noncommutative generalizations have a restricted gauge group choice, since
some of symmetry groups are not closed with respect to the Moyal product: $SU(N)$, $SO(N)$ and $Sp(N)$ for instance are not. Hence $U(N)$ gauge theories will be considered in the following.

Gauge transformations for the $N\times N$ gauge field $A_\m$ in commutative set--up read
\bea
\d_{\l}A_{\m}=\pa_{\m}\l+\imm[\l,A_{\m}]
\ena
where the gauge transformation parameter $\l$ is also a $N\times N$ matrix. The gauge invariant action is written in terms of the field strength
\bea\label{gaugeaction}
S=\int\intd^dx\; \traccia\left(F^{\m\n}F_{\m\n}\right)
\ena
(up to the coupling constant and topological terms). The field strength for the gauge field is as usual equal to
\bea
F_{\m\n}=\pa_{\m}A_{\n}-\pa_{\n}A_{\m}-\imm[A_{\m},A_{\n}]
\ena
The gauge transformation for $F_{\m\n}$ reads
\bea\label{gauge}
\d_{\l}F_{\m\n}=\imm[\l,F_{\m\n}]
\ena
Cyclicity of traces implies that \refeq{gaugeaction} is invariant under \refeq{gauge}.

Now, consider the \emph{natural} generalization to noncommutative geometry. The gauge invariant action gets changed into
\bea\label{actionnc}
S=\int\intd^dx\; \traccia\left(F^{\m\n}\star F_{\m\n}\right)
\ena
One can guess that gauge transformations are still a symmetry for the deformed action once the $\star$--product is introduced
\bea\label{gaugenc}
\d_{\l}A_{\m}&=&\pa_{\m}\l+\imm\l\star A_{\m}-\imm A_{\m}\star \l\cr
F_{\m\n}&=&\pa_{\m}A_{\n}-\pa_{\n}A_{\m}-\imm A_{\m}\star A_{\n}+\imm A_{\n}\star A_{\m}\cr
\d_{\l}F_{\m\n}&=&\imm\l\star F_{\m\n}-\imm F_{\m\n}\star \l
\ena
Here, the multiplication rules among matrix valued fields is understood to be a tensor product between the $\star$ and matrix products. The invariance of the action \refeq{actionnc} under \refeq{gaugenc} is
again ensured by the cyclicity of the trace, together with the ciclicity property of $\star$--product. Due to Moyal product, even the abelian case with the $U(1)$ gauge symmetry has non trivial transformation
rules, although the gauge transformation parameter $\l$ is a scalar function of coordinates. Indeed, from \refeq{gaugenc} one can see that $\d_\l F_{\m\n}\neq0$ and $\d_\l A_\m-\pa_\m\l\neq0$.

In the $\theta\to0$ limit, as usual, the commutative theory is recovered.
\subsubsection{The \emph{naturally} deformed sine--Gordon}
I showed that noncommutative $\phi^4$ theory would still be renormalizable (at one loop), if mixing of UV and IR divergences were not corrupting the correct behavior. Ordinary sine--Gordon (SG) is a
super--renormalizable theory (in the regime $\b^2<8\p$ for the coupling constant $\b$) and it would be interesting to see how natural noncommutative generalization could either spoil this property or not,
successively eventually considering the non--renormalizability regime ($\b^2\ge8\p$).

Indeed, it's easy to guess that even in the super--renormalizability regime noncommutativity will destroy usual renormalization, independently of IR/UV mixing. Let me write the deformed action
\bea\label{natural ncsg action}
S\left[\phi\right]=\int\intd^2x\left(\oneover{2}\pa^\m\phi\pa_\m\phi-\g\cos_\star\b\phi\right)
\ena
where I indicate with starred functions those whose power expansions are defined through the Moyal product. The key point in ordinary SG renormalization, as explained in subsection \ref{commutative
renormalization}, is that exactly the same coefficient appears in all diverging correlation functions, at each order in the $\b$ expansion, independently of the external propagators (which appear as an
overall factor). This very nice property allows to renormalize all correlation functions by defining the renormalized $\g_R$ alone. Unfortunately, due to $\star$--product, this feature is spoiled in the
noncommutative generalization. Coefficients get a non trivial and non factorized dependence on the number of external propagators. Moreover, the typical effective cutoff that usually arises in noncommutative
theories $\oneover{\L_{\rm{eff}}^2}=\oneover{\L^2}+\oneover{p\circ p}$ --- where $p$ is the external momentum or combinations of external momenta involved in the particular graph --- has to be introduced.

In other words, the divergences coming from ordinary graphs in the commutative theory are those of the planar diagrams in the noncommutative generalization. However, the symmetry coefficients are different,
since a single diagram in the ordinary model splits into planar and nonplanar parts. Nonplanar diagrams can be shown to be finite and destroy renormalizability. Furthermore, the phase factor coming from
$\star$--product gives origin to the effective cutoff, which is responsible of the IR/UV mixing. 

The main issue is that renormalization is impossible even in the super--renormalizability regime. This situation can only get worse in the non--renormalizability regime. Thus, natural deformation doesn't seem
to be a consistent noncommutative generalization of sine--Gordon theory, both with respect of quantum properties and for integrability. It has indeed been argued that integrability at classical level is not
preserved in this deformation, since there is no known method able to construct an infinite number of conserved currents for the theory \cite{CM}. Furthermore, scattering amplitudes for such a theory have
been computed and it has been found that particle production occurs due to the non vanishing of $2\to4$ amplitude. The purpose of my two papers, written in collaboration with M.~T.~Grisaru, S.~Penati,
L.~Tamassia, O.~Lechtenfeld and A.~Popov \cite{GMPT,LMPPT}, is to guarantee integrability in the noncommutative sine--Gordon and study S--matrix properties in order to be consistent with integrability
definition in two dimensions.
\subsubsection{Wess--Zumino--Witten model as a non \emph{natural} deformation}
Natural generalizations of field theories to noncommutative geometry are just one of many possible deformations that can be applied reducing to ordinary theories
in the commutative limit. I already remarked that there exists another procedure --- using the gauged bicomplex formulation --- which can be used to deform integrable field theories. It is interesting, and
this is what I will show in the following of this paragraph, to consider group valued fields and get \emph{nonablelian} like features (coming from the noncommutativity of Moyal product) even from ordinarily
abelian group $U(1)$. In other words, I exploit the fact that NC $U(1)$ displays nonabelian--like properties, but reduces to ordinary abelian group in the commutative limit.

For instance, the free scalar action in two dimensions can be restated as a principal chiral action for a field $g\in U(1)$. While introducing $\star$--product in the free field theory is armless, if we
consider the principal chiral action noncommutativity brings to a deformed (noncommutative) theory
\bea\label{pc action}
S_{\rm PC}=\int\intd^2 x\, \pa_\m g^{-1}\star\pa^\m g
\ena
In the noncommutative $U(1)$ theory the field $g$ can be written as $g(x)=\ex_\star^{\imm\a\phi(x)}$. The equations of motion are of the form
\bea
\pa_\m\left(g^{-1}\star\pa^\m g\right)=0
\ena
Both action and equations of motion reduce to trivial free theory action and equations for the free (massless) scalar $\phi$ in the $\theta\to0$ limit. Hence, \refeq{pc action} represents a non natural
deformation of free scalar field theory.

Another well known theory with a $U(1)$ symmetry can play the role of a noncommutative generalization of the free field theory, when formulated by means of Moyal product. This is the Wess--Zumino--Witten
(WZW) model, defined by the action
\bea
S_{\rm WZW}&=&\oneover{2}\int\intd^2x\,\pa_\m g^{-1}\pa^\m g-\oneover{3}\int d^2x\,\intd\l\,\e^{\m\n\sigma}\,\hat g^{-1}\pa_\m\hat g\star\hat g^{-1}\pa_\n\hat g\star\hat g^{-1}\pa_\sigma\hat g
\ena
The second term in the action, i.e. the Wess--Zumino (WZ) term, is a function of the homotopy path $\hat g(\l,x)$ satisfying $\hat g(0,x)\equiv1$ and $\hat g(1,x)\equiv g(x)$. As a consequence of the fact
that the variation of the WZ term is a total derivative in $\rho$, the equations of motion are two dimensional 
\bea
\bar\pa\left(g^{-1}\star\pa g\right)=0
\ena
where the derivatives with respect to the complex coordinates are as usual $\pa\equiv\pa_0 +\pa_1$ and $\bar \pa\equiv \pa_0-\pa_1$. The commutative and abelian WZW action reduces to free field theory for
$\phi$, since the WZ term vanishes in this limit. Noncommutativity acts as if the symmetry group were nonabelian and hence yields a non trivial deformation.

These simple examples are nice since they show different interesting non trivial ways to generalize the free scalar field theory to noncommutative geometry. WZW model will play an important role later on, in
the noncommutative sine--Gordon theories that I will examine in chapters \ref{sgI} and \ref{sgII}. It will come out to be indeed the noncommutative generalization of the kinetic part of the theory, as in the
example I just illustrated. 
\subsection{Infinite conserved currents and noncommutative deformations}\label{integr nc ft}
As I noted in the previous subsection, the natural deformation of sine--Gordon theory is not suitable for a quantum well--defined theory. The bicomplex approach guarantees in principle integrability also for
the noncommutative model. In fact, it preserves by construction the existence of an infinite number of local (in the sense on non integral functions) conserved currents. In ordinary geometry, this implies 
the factorization of the S--matrix, as I explained in subsection \ref{ordinary S matrix}. It is thus worthwhile to study the deformed bicomplex formulation of noncommutative generalizations and to test if the
integrability vs. factorization theorem is still valid in noncommutative geometry.

Some examples of integrable field theories generalization to noncommutative geometry by means of the deformed gauged bicomplex procedure have been considered in literature. Non linear Schr\"odinger
equations and the associated infinite symmetries have been considered in \cite{DMH}. The deformation through bicomplex is in this particular case equivalent to the deformation one would have obtained by
direct substitution of Moyal product in the equations of motion themselves. Other examples sharing the same property are the noncommutative Toda equations and noncommutative Kortweg--de--Vries equations,
studied in \cite{NCEOM,list2}. See also \cite{list1,murugan,hamanaka}.

Sine--Gordon model, on the contrary, gives different results depending on whether the deformation is operated in the gauged bicomplex equations or in the equations of motion for the scalar field. Direct
substitution at the level of the equations of motion yields
\bea\label{nc direct sg}
\pa\pab\phi=\g\sin_\star\phi
\ena
(in complex $z,\bar z$ coordinates in two dimensional space). This equation ensues from the noncommutative action of the natural deformation \refeq{natural ncsg action} (setting $\b=1$). Hence, the theory is
the previously discussed natural noncommutative version of sine--Gordon. Generalization to noncommutative geometry via bicomplex brings to the following equations of motion \cite{GP}
\bea\label{nc bicomplex sg}
\pab\left(\ex_\star^{-\frac{\imm}{2}\phi}\star\pa\ex_\star^{\frac{\imm}{2}\phi}-\ex_\star^{\frac{\imm}{2}\phi}\star\pa\ex_\star^{-\frac{\imm}{2}\phi}\right)=\imm\g\sin_\star\phi
\ena
where the $\star$--product have been introduced in the bicomplex definitions. The equation above (toghether with another constraint that reduces to free field equation in the commutative limit) is the
integrability condition associated to the Lax pair.  Equation \refeq{nc bicomplex sg}, as \refeq{nc direct sg}, relaxes to the usual sine--Gordon equation in the $\theta\to0$ limit. However, it is by
construction associated to an infinite number of conserved currents, so that integrability is ensured. As an equation of motion \refeq{nc bicomplex sg} looks more complicated than \refeq{nc direct sg}. The
action that produces this equation has been worked out in my publication with M.~Grisaru, S.~Penati and L.~Tamassia \cite{GMPT}. I will discuss it and its S--matrix in the next chapter. 

I note that \refeq{nc bicomplex sg} is not the only possible deformation for the sine--Gordon theory that can be obtained through the gauged bicomplex. Thinking of the sine--Gordon model as coming from
dimensional reduction of SDYM in four dimensions, having as an intermediate step the three dimensional modified sigma model with a WZW--like interaction term, a different set of equations are derived. This
has been done in collaboration with O.~Lechtenfeld, S.~Penati, A.~Popov and L.~Tamassia \cite{LMPPT}. In this case, the Moyal product is introduced in the sigma model linear system. Then, dimensional
reduction to two dimensions is applied. The resulting deformed equations can no longer be written in terms of a single scalar field playing the role of the sine--Gordon field $\phi$. Indeed, a new scalar is
non trivially interacting with $\phi$ and thus the equations of motion are no longer decoupled. In chapter \ref{sgII} I will explain how noncommutative conserved currents, solitons, action, and tree level
S--matrix have been constructed and analyzed.
\subsection{Phenomenological and field theoretical consequences of noncommutativity}\label{nc problems}
Since noncommutativity can be traced back to string theory in specific backgrounds and, in particular, noncommutative field theories arise as the low energy effective description of string theory in such
backgrounds, we would like to know something about phenomenological consequences or eventual quantum field theory inconsistencies of Moyal product. I already mentioned that $\star$--product provides an
affective cutoff that causes the mixing of infrared and ultraviolet divergences for noncommutative renormalized actions of quantum field theories. This and other characteristics are worth being discussed.
\paragraph{Lorentz invariance: the two dimensions}
In a space--time of dimensions lager that two, Lorentz invariance is not preserved by noncommutative generalization, due to the non tensorial nature of the noncommutativity parameter $\theta^{\m\n}$,
$\L^\m_\rho\L^\n_\s\theta^{\rho\s}\neq\theta^{\m\n}$. I recall that Lorentz invariance is not spoiled instead in two dimensions because of the specific form of Lorentz representation and noncommutativity
parameter matrices.

The failure of Lorentz invariance for four dimensional theories has been widely considered in literature as an important phenomenological issue \cite{Colla}. The scale for Lorentz symmetry breaking is set by
the noncommutativity parameter. Hence, eventual predictions on observations can be made.

Alternative solutions to the Lorentz violation problem have been proposed \cite{Morita}. The main idea is to change the nature of the noncommutativity parameter into tensor $\hat\theta^{\m\n}$. As a
consequence, Weyl formalism must be modified. The variable $\theta$ and its conjugate variable $\eta$ must be included in the Fourier transform definition and in the Weyl map $\hD$ that defines Weyl operators
--- as functions of noncommutative operators $\hx^\m$ and of the tensor $\hat\theta^{\m\n}$. The $\star$--product definition is left unchanged. This modified formalism allows to get
$\L^\m_\rho\L^\n_\s\hat\theta^{\rho\s}=\hat\theta^{\m\n}$, i.e.  it preserves Lorentz invariance.
\paragraph{Noncommutative violation of causality and unitarity}
Causality and unitarity are a fundamental concerns in quantum field theory. Violation of these two properties generally occurs when noncommutativity involves the time coordinate.

Acausality is a typical feature of time non locality, which comes from the infinite number of derivatives involved in the $\star$--product and from $\theta^{0i}$ being non zero for some $i$'s. An illustrative
and simple example displaying acausal behavior is the noncommutative $\phi^4$ theory \cite{SST,Bozkaya}. In the previous subsection I wrote the expression for the $\phi^4$ vertex with $\star$--product (eq.
\refeq{feynman}).  The four point amplitude can be written in terms of a wedge product defined through the time/space component of the noncommutativity parameter --- let me choose for instance
$\theta^{01}\equiv\theta$ ---
\bea
\imm{\cal M}\sim g[\cos(p_1\wedge p_2)\cos(p_3\wedge p_4)+(2\leftrightarrow3)+(2\leftrightarrow4)]
\label{ampli}
\ena
where
\bea
a\wedge b=\theta\left(a^0b^1-a^1b^0\right)
\ena
Momentum conservation rule yelds $p_1+p_2+p_3+p_4=0$. The expression \refeq{ampli} in the commutative limit $\theta\to0$ gives $\imm{\cal M}=-\imm g$. In the center of mass frame (characterized by the 
momentum $p$) the amplitude \refeq{ampli} takes the simple form
\bea
\imm{\cal M}\sim g[\cos(4p^2\theta)+2]
\ena
A sinusoidal behavior of the scattering amplitude is somehow usual in noncommutative field theories, even though the specific form depends on the matter content of the theory. From the amplitude
\refeq{ampli} one can deduce the wave function for the outgoing particles by computing the inverse Fourier transform of the wave packets. All in one, from an initial wave function concentrated in $p=\pm p_0$
\bea
\phi_{in}(p)\sim\left(\ex^{-\frac{1}{\l}(p-p_0)^2}+\ex^{-\frac{1}{\l}(p+p_0)^2}\right)
\ena
we obtain three outgoing packets described by
\bea\label{out wave}
\Phi_{out}(x)\sim g\left[F(x;-\theta,\l,p_0)+4\sqrt\l\ex^{-\l\frac{x^2}{4}}\ex^{i p_0 x}+F(x;\theta,\l,p_0)\right]+(p_0\rightarrow-p_0)\non\\
\ena
This expression is worked out in the approximation $p_0\gg\sqrt\l\gg\oneover{\theta p_0}$ and $\l\theta\gg1$.  The function $F(x;\theta,\l,p_0)$ gives a wave packet of width $8\sqrt\l\theta$ concentrated in
$x=-8\theta p_0$ 
\bea
F(x;\theta,\l,p_0)=\oneover{\sqrt{-4\imm\theta}}\ex^{-\frac{\left(x+8\theta p_0\right)^2}{64\theta^2\l}}\ex^{-\imm\frac{\left(x-\frac{p_0}{2\l^2\theta}\right)}{16\theta}}\ex^{\imm\frac{p_0^2}{4\l^2\theta}}
\ena
From \refeq{out wave} we conclude that the outgoing wave function is indeed made of three wave packets. Two contributions are centered respectively in $x=\pm8\theta p_0$, have the same width $8\sqrt\l\theta$
and oscillate with frequency $\exp\left\{\pm\imm\frac{(x\pm p_0/2\l^2\theta)^2}{16\theta}\right\}$, which approaches $p_0$ near the maxima. The remaining contribution is the wave packet arising in the
ordinary commutative theory, when $\theta=0$. It is centered at the origin and oscillates with frequency $p_0$. The two shifted wave packets represent respectively an advanced and delayed packet with respect
to the ordinary one. The acausal behavior is ascribed to the advanced packet, since it appears at $x=0$ before the incoming wave function hits the origin. The delayed packet doesn't affect causality, in
principle.

It is worth noticing that in string theory with an electric $B$--field, the acausal behavior, reflected by the advanced wave packet in the scattering process, is cancelled by a phase factor coming from
particular strings oscillations. Thus, string scattering (on the circle) is a causal process even with time/space components of the two form background field turned on, due to stringy modes.

Further works have considered noncommutativity implications about causality in field theories \cite{liao}. Some suggestions are related to a different definition for correlation functions in noncommutative
field theories. Causality should be preserved by a specific time ordering prescription in the computation of Green functions.

Another main problem in noncommutative field theories is unitarity of the S--matrix. One can see breaking of unitarity in a similar way as acausality. The authors of \cite{gm} showed that noncommutative
$\phi^3$ and $\phi^4$ theories do not obey to the cutting rule, which states
\bea\label{cutting rule}
2\Im{\cal M}_{ab}=\sum_n{\cal M}_{an}{\cal M}_{nb}
\ena
I indicate with ${\cal M}_{ab}$ the amplitude matrix between an initial state $a$ and a final state $b$ (on--shell). This rule is implied by unitarity, so unitarity is violated if it fails. The argument
that leads to non unitarity requires that $\circ$--product (defined in \refeq{circ product}) must be negative among external leg momenta --- $p\circ p<0$ for a two point function, for instance. This in turn
implies that non unitarity appears only for time/space noncommutativity, $\theta^{0i}\neq0$, and for space--like momenta. It is easy to see that cutting rule \refeq{cutting rule} can't be satisfied when these
conditions apply, by directly computing amplitudes in NC $\phi^4$ theory, for instance.

The issue of unitarity violation can be related to the impossibility of consistently formulating a low energy effective (time/space) noncommutative field theory description of string theory with electric
field backgrounds. I already remarked how electric backgrounds make impossible to take a decoupling limit of string theory without making the noncommutativity parameter vanish. This could suggest that
time/space noncommutative field theories are lacking of the degrees of freedom (namely massive string modes) that are instead present in noncommutative opens sting theory (with electric components of $B$
field turned on). The lack of massive modes eventually implies the loss of unitarity in field theory, as well as of causality.

As for causality, unitarity can be restored in the noncommutative field theories even with time/space noncommutativity, if a particular time--ordering is defined.

The issue of causality, together with the concern about integrability, has been considered in the two noncommutative generalizations of sine--Gordon model that we have studied. The first of the two
generalizations has been found to suffer from acausality behaviors, while the second model we constructed is perfectly causal at tree level. Since the second model has the advantage of being derived by
dimensional reduction from four dimensional SDYM, which in turn describes ${\cal N}=2$ RNS superstrings, one may think that the relation to string theory prevents causality to be violated in this case.
\paragraph{UV/IR mixing}
Let me make a brief summary about the mixing of UV and IR dynamics. In performing quantum calculations for \emph{naturally} deformed field theories, we found out new features of UV divergences. In
particular, nothing changes with respect to the ordinary theory for the planar contribution to the effective action. Instead, nonplanar graphs yield an effective cutoff which lowers the divergence degree
of the diagram, provided that we do not take the IR limit for small momenta. This is clear looking at the expression for the effective cutoff
\bea
\L_{\rm{eff}}^2=\oneover{\oneover{\L^2}+\oneover{\sqrt{p\circ p}}}
\ena
So, there is a stringent and unavoidable relation between UV and IR divergences, which prevents the theories to be conventionally renormalizable. 

The same situation happens in string theory. T--duality for instance relates high energy dynamics to small energy regions. In particular, the modular transformation $\t\to1/\t$ allows to interpret the UV
region of open string theory to the IR behavior of closed strings. If the closed string metric is identified by means of the noncommutativity parameter $g^{ij}\sim(\theta^2)^{ij}$ and the open string metric
is instead $G^{ij}\sim\delta^{ij}$, we precisely obtain the same correlation between UV and IR phenomena that is described by the noncommutative UV/IR mixing.

\chapter{Noncommutative sine--Gordon with non factorized S--matrix}\label{sgI}
The model we consider in this chapter is derived from a NC generalization of sine--Gordon, operated at the level of the gauged bicomplex formulation. As we have explained, \emph{natural} generalization of
sine--Gordon is not satisfactory nor from the quantum nor classical point of view, since both renormalizability and integrability are spoiled. The bicomplex approach seems more promising, since it guarantees
the existence of an infinite number of conserved currents. 

The equations of motion by means of this method were derived by M.~T.~Grisaru and S.~Penati in \cite{GP}. The authors also showed how to construct solitons perturbatively in the noncommutativity parameter
$\theta$. We will review in the next section how to obtain the deformed equations of motion.  Then, the corresponding action was calculated in my paper \cite{GMPT}, in collaboration with M.~T.~Grisaru,
S.~Penati and L.~Tamassia. This is explained in subsection \ref{section ncsg3 action}. Section \ref{section ncsg3 connections} illustrates the connections to SDYM and Thirring model. Finally, the relevant
properties of S--matrix are discussed in the following section.
\section{Noncommutative sine--Gordon from the bicomplex}
Here we follow the procedure we used in subsection \ref{ordinary bicomplex}, generalizing it to noncommutative geometry, in order to obtain an integrable noncommutative sine--Gordon --- in the sense of
preserving the infinite number of conserved currents. This method yields the equations of motion of the theory. It is not generally trivial to deduce the action generating these equations.
\subsection{The equations of motion}
The paper by M.~T.~Grisaru and S.~Penati \cite{GP} tackles the problem of classical integrability. The equations of motion they obtained, using the method of the bicomplex implemented in NC geometry, do not
resemble the ones of the \emph{natural} generalization that I previously discussed.

We consider the same linear space ${\cal M} = {\cal M}^0 \otimes \L$ as for the ordinary theory described in subsection \ref{ordinary bicomplex}, where ${\cal M}^0$ is the space of $2 \times 2$ matrices with
entries in ${\mathbb C}$, and $\L = \otimes_{r=0}^2 \L^r$ is a two dimensional graded vector space with the $\L^1$ basis $(\t,\s)$ satisfying $\t^2 = \s^2 = \t \s + \s \t =0$.  For any matrix function $f \in
{\cal M}^0$ the two linear maps are defined in analogy to the ordinary case 
\beq
\d\x=\pab\x\t-R\x\s\,,\qquad \intd\x=-S\x\t+\pa\x\s
\label{bicomplex}
\eeq
where $R,S$ are constant matrices with $[R,S]=0$. The bicomplex conditions $\d^2 = \intd^2 = (\intd\d + \d \intd) =0$ are trivially satisfied as we already know from commutative sine--Gordon.

The NC gauged bicomplex is obtained as in the ordinary theory by dressing one of the two differential operators
\beq
Df\equiv G^{-1}\star\intd(G\star\x)=-L\star\x\t+(\pa+M\star)\x\s
\eeq
where $G$ is a generic invertible ($G \star G^{-1} = G^{-1} \star G =I$) matrix in ${\cal M}^0$ and
\beq
L=G^{-1}\star SG\,, \qquad M=G^{-1}\star\pa G
\label{matrices}
\eeq
However, we introduced the NC $\star$--product in the above expressions, since we want to get a NC model. The condition $D^2=0$ implies $\pa L = [L, M]_{\star}$ which one can check to be still identically
satisfied (and represents the equation associated to the Lax pair of operators determining integrability of the sine--Gordon system in ordinary geometry, in the $\theta\to0$ limit).  The last condition $\{D,
\d\}=0$ gives instead the nontrivial equation
\beq
\pab M=[R, L]_{\star}
\label{matrix eq}
\eeq
We note that this equation is non trivial even in the commutative limit. 

In order to obtain a noncommutative version of the sine--Gordon equation we choose the $U(2)$ group valued fields
\bea
&&R=S=\sqrt{\g}\left(\begin{array}{cc}  0&0 \cr 0&1\end{array} \right)\nonumber \\
&&G=e_{\star}^{\frac{i}{2}\s_2\phi}=\left(\begin{array}{cc}\cos_{\star}{\frac{\phi}{2}}&\sin_{\star}{\frac{\phi}{2}}\cr-\sin_{\star}{\frac{\phi}{2}}&\cos_{\star}{\frac{\phi}{2}}\end{array}\right)
\label{sgdif}
\eea
which are the straightforward generalizations of the ordinary $SU(2)$ fields for commutative sine--Gordon. As a consequence, we get 
\bea
&& M = \frac{1}{2} \left( \begin{array}{cc} e_{\star}^{\frac{i}{2} \phi}  \star \pa 
\ex_{\star}^{-\frac{i}{2} \phi} +  \ex_{\star}^{-\frac{i}{2} \phi}  \star \pa 
\ex_{\star}^{\frac{i}{2} \phi}~&~  -i(\ex_{\star}^{-\frac{i}{2} \phi}  \star \pa 
\ex_{\star}^{\frac{i}{2} \phi} -  \ex_{\star}^{\frac{i}{2} \phi}  \star \pa 
\ex_{\star}^{-\frac{i}{2} \phi}) \cr
~~~& ~~~ \cr
i(e_{\star}^{-\frac{i}{2} \phi}  \star \pa 
e_{\star}^{\frac{i}{2} \phi} -  e_{\star}^{\frac{i}{2} \phi}  \star \pa 
e_{\star}^{-\frac{i}{2} \phi}) ~&~ e_{\star}^{\frac{i}{2} \phi} \star
\pa e_{\star}^{-\frac{i}{2} \phi} + e_{\star}^{-\frac{i}{2} \phi} \star
\pa e_{\star}^{\frac{i}{2} \phi} \end{array}\right)
\nonumber \\
&&~~~~~~~~~\nonumber \\
&&~~~~~~~~~\nonumber \\
&& L = \sqrt{\g} \left( \begin{array}{cc} \sin^2_{\star}{\frac{\phi}{2}}
& -\sin_{\star}{\frac{\phi}{2}} \star \cos_{\star}{\frac{\phi}{2}} \cr
-\cos_{\star}{\frac{\phi}{2}} \star \sin_{\star}{\frac{\phi}{2}} &
\cos^2_{\star}{\frac{\phi}{2}} \end{array}\right)
\eea
Computing $[R,L]_\star$ we obtain
\beq
[R,L] = \g \left( \begin{array}{cc} 0
& \sin_{\star}{\frac{\phi}{2}} \star \cos_{\star}{\frac{\phi}{2}} \cr
-\cos_{\star}{\frac{\phi}{2}} \star \sin_{\star}{\frac{\phi}{2}} & 0 \end{array}\right)
\eeq
Then, equation (\ref{matrix eq}) is a matrix equation in $U(2)$. In particular, the matrix $M$ develops a nontrivial trace part, as a consequence of the noncommutative nature of the $U(1)$ subgroup.
Therefore, writing eq. (\ref{matrix eq}) in components, leads to two nontrivial equations for the field $\phi$
\bea
\pab\left(\ex_{\star}^{\frac{i}{2} \phi}\star\pa\ex_{\star}^{-\frac{i}{2}\phi}+\ex_{\star}^{-\frac{i}{2} \phi}\star\pa\ex_{\star}^{\frac{i}{2}\phi}\right)&=&0 \nonumber \\
\pab\left(\ex_{\star}^{-\frac{i}{2}\phi}\star\pa\ex_{\star}^{\frac{i}{2}\phi}-\ex_{\star}^{\frac{i}{2} \phi}\star\pa\ex_{\star}^{-\frac{i}{2}\phi}\right)&=&\imm\g\sin_{\star}{\phi} \label{sg}
\eea
The first equation becomes trivial in the limit $\theta \to 0$, whereas the second reduces to the ordinary sine--Gordon equation \refeq{nc bicomplex sg}
\beq
\pa\pab\phi=\g\sin{\phi} 
\eeq
In the noncommutative case both  equations are meaningful and describe the dynamics of the field $\phi(z,\bar{z}, \theta)$. In particular, the first equation contains the potential term which is the
\emph{natural} generalization of the ordinary sine potential, whereas the other has the structure of a conservation law and can be interpreted as imposing an extra condition on the system. Both the
equations are in general complex and possess the $Z_2$ symmetry of the ordinary sine--Gordon (invariance under $\phi \to -\phi$). 

The reason why integrability seems to require two equations of motions can be traced back to the general structure of unitary groups in NC geometry.  In the bicomplex approach the ordinary equations are
obtained as zero curvature conditions for covariant derivatives defined in terms of $SU(2)$ gauge connections. If the same procedure is to be implemented in the noncommutative case, the group $SU(2)$, which is
known to be not closed in noncommutative geometry, has to be extended to a noncommutative $U(2)$ group and a NC $U(1)$ factor enters necessarily into the game.  The appearance of the second equation in (\ref{sg})
for our NC integrable version of sine--Gordon is then a consequence of the fact that the fields develop a nontrivial trace part. We note that the pattern of equations we have found seems to be quite general
and unavoidable if integrability is of concern. In fact, the same has been found in   \cite{CM} where a different but equivalent set of equations was proposed.  

The presence of two equations of motion is in principle very restrictive and one may wonder whether the class of solutions is empty. To show that this is not the case, in   \cite{GP} solitonic solutions
were constructed perturbatively which reduce to the ordinary solitons when we take the commutative limit. More generally, we observe that the second equation in (\ref{sg}) is automatically satisfied by any
chiral or antichiral function.  Therefore, we expect the class of solitonic solutions to be at least as large as the ordinary one. In the general case, instead, we expect the class of dynamical solutions to
be smaller than the ordinary one because of the presence of the nontrivial constraint. However, since the constraint equation is one order higher with respect to the dynamical equation, order by order in the
$\theta$--expansion a solution always exists. This means that a Seiberg--Witten map between the NC and the ordinary model does not exist as a mapping between physical configurations, but it might be
constructed as a mapping between equations of motion or conserved currents.  

The question which was left open in   \cite{GP} was the existence of an action for the set of equations (\ref{sg}). Here, we give an action and discuss the relation of our model with the NC
selfdual Yang--Mills theory and the NC Thirring model. Moreover, we discuss some properties of the corresponding S--matrix which, in spite of integrability, turns out to be acausal and not factorized. 
\subsection{The action} \label{section ncsg3 action}
We are now interested in the possibility of determining an action for the scalar field $\phi$ satisfying the system of eqs. (\ref{sg}). We are primarily motivated by the possibility to move on to a quantum
description of the system. 

In general, it is not easy to find an action for the dynamical equation (the first eq. in (\ref{sg})) since $\phi$ is constrained by the second one. One possibility could be to implement the constraint by the
use of a Lagrange multiplier.

We consider instead the equivalent set of equations (\ref{sg2}).  We rewrite them in the form
\bea
&&\pab(g^{-1}\star \pa g)=\frac{1}{4}\g\left(g^2-g^{-2}\right)\cr
&&\pab(g\star\pa g^{-1})=-\frac{1}{4}\g\left(g^2-g^{-2}\right)
\label{sg3}
\ena
where we have defined $g\equiv e_\star^{\frac{i}{2}\phi}$. Since $\phi$ is in general complex $g$ can be seen as an element of a noncommutative complexified $U(1)$. The gauge group valued function 
$\bar g\equiv (g^\dagger)^{-1}= e_\star^{\frac{i}{2}\phi^\dagger}$ 
is subject to the equations
\bea
&&\pab(\bar g \star\pa\bar g^{-1})=-\frac{1}{4}\g \left(\bar g^2-\bar g^{-2}\right)\cr
&&\pab(\bar g^{-1}\star\pa \bar g)=\frac{1}{4}\g \left(\bar g^2-\bar g^{-2}\right) \label{sg4}
\ena
obtained by taking the h.c. of (\ref{sg3}).

In order to determine the action it is convenient to concentrate on the first equation in (\ref{sg3}) and the second one in (\ref{sg4}) as the two independent complex equations of motion which describe the
dynamics of our system. 

We first note that the left--hand sides of equations (\ref{sg3}) and (\ref{sg4}) have the chiral structure which is well known to correspond to a NC version of the WZNW action \cite{NCWZ}. Therefore we are
led to consider the action 
\beq
S[g,\bar g]=S[g]+S[\bar g] \label{action g}
\eeq
where, introducing the homotopy path $\hat{g}(t)$ such that $\hat{g}(0)=1$, $\hat{g}(1) = g$ ($t$ is a commuting parameter) we have defined
\bea
&S[g]&=\int\intd^2z\left[\pa g\star\pab g^{-1}+\int_0^1\intd t~\hat{g}^{-1}\star\pa_{t}\hat{g}\star[\hat{g}^{-1}\star\pa\hat{g},\hat{g}^{-1}\star\pab\hat{g}]_{\star}-\frac{\g}{4}(g^2+g^{-2}-2)\right] \non\\
&&~~~~~~~~~~\label{actionWZNW}
\eea
and similarly for $S[\bar{g}]$. The first part of the action can be recognized as the NC generalization of a complexified $U(1)$ WZNW action \cite{NS}.

To prove that this generates the correct equations, we should take the variation with respect to the $\phi$ field ($g=\ex_\star^{\frac{i}{2}\phi}$) and deal with complications which follow from the fact that
in the NC case the variation of an exponential is not proportional to the exponential itself.  However, since the variation $\d \phi$ is arbitrary, we can forget about its $\theta$ dependence and write
$\frac{i}{2}\d \phi = g^{-1} \d g $, trading the variation with respect to $\phi$ with the variation with respect to $g$. Analogously, the variation with respect to $\phi^{\dagger}$ can be traded with the
variation with respect to $\bar{g}$.

It is then a simple calculation to show that
\beq
\d S[g] = \int \intd^2z ~2g^{-1} \d g \left[ \pab \left( g^{-1} \star \pa g \right) ~-~ \frac{\imm}{2} \g \sin_{\star}{\phi} \right] 
\eeq
from which we obtain the first equation in (\ref{sg3}). Treating $\bar g$ as an independent variable an analogous derivation gives the second equation in (\ref{sg4}) from $S[\bar g]$.

We note that, when $\phi$ is real, $g=\bar g$ and the action (\ref{action g}) reduces to $S_{\rm real}[g]=2S_{\rm WZW}[g]-\g(\cos_\star\phi -1)$.  In general, since the two equations (\ref{sg}) are complex it
would be inconsistent to restrict ourselves to real solutions. However, it is a matter of fact that the equations of motion become real when the field is real.  Perturbatively in $\theta$ this can be proved
order by order by direct inspection of the equations in   \cite{GP}. In particular, at a given order one can show that the imaginary part of the equations vanishes when the constraint and the equations of
motion at lower orders are satisfied. 

It would be interesting to obtain the action (\ref{action g}) from the dimensional reduction of the 4d SDYM action by generalizing to the NC case the procedure used in   \cite{SDYMsine}.
\section{Connections to strings and dualities}\label{section ncsg3 connections}
The NC sine--Gordon model we here propose can be related to string theory, via dimensional reduction from four dimensional NC SDYM. This is a NC of the ordinary dimensional reduction that gives sine--Gordon in
ordinary geometry from SDYM. However, we point out that the reduction in noncommutative geometry works only at the level of the equations of motion, but not for the action. On the other hand, bosonization of
NC massive and abelian Thirring model yields the action of our NC version of sine--Gordon, as we will show in subsection \ref{nc thirring boso}.
\subsection{Equations of motion from noncommutative selfdual Yang--Mills} 
The (anti--)selfdual Yang--Mills equation is well--known to describe a completely integrable classical system in four dimensions \cite{Y}.  In the ordinary case the equations of motion for many two
dimensional integrable systems, including sine--Gordon, can be obtained through dimensional reduction of the (A)SDYM equations \cite{SDYMred}.

A convenient description of the (A)SDYM system is the so called $J$--formulation, given in terms of a $SL(N,C)$ matrix--valued $J$ field satisfying
\beq
\pa_{\bar{y}}\left(J^{-1} \pa_y J\right)+\pa_{\bar{z}}\left(J^{-1}\pa_z J\right)=0 \label{selfdual}
\eeq
where $y$, $\bar{y}$, $z$, $\bar{z}$ are complex variables treated as formally independent.

In the ordinary case, the sine--Gordon equation can be obtained from (\ref{selfdual}) by taking $J$ in $SL(2, C)$ to be \cite{SDYMsine}
\beq
J=J(u,z,\bar{z})=\ex^{\frac{z}{2}\s_i}e^{\frac{\imm}{2}u\s_j} \ex^{-\frac{\bar{z}}{2}\s_i} \label{ansatz1}
\eeq
where $u=u(y,\bar y)$ depends on $y$ and $\bar y$ only and $\s_i$ are the Pauli matrices.

A noncommutative version of the (anti--)selfdual Yang--Mills system can be naturally obtained \cite{T} by promoting the variables $y$, $\bar y$, $z$ and $\bar z$ to be noncommutative thus extending the
ordinary products in (\ref{selfdual}) to $\star$--products. In this case the $J$ field lives in $GL(N,C)$. 

It has been shown \cite{LPS} that NC SDYM naturally emerges from open $N=2$ strings in a B--field background.  Moreover, in   \cite{T,Legare,hamanaka} examples of reductions to two dimensional NC
systems were given. It was also argued that  the NC deformation should preserve the integrability of the systems \cite{NCSDYM4}.  

We now show that our NC version of the sine--Gordon equations can be derived through dimensional reduction from the NC SDYM equations.  For this purpose we consider the NC version of equations (\ref{selfdual})
and choose $J_\star$ in $GL(2,C)$ as
\beq
J_\star=J_\star(u,z,\bar{z})=\ex_\star^{\frac{z}{2}\s_i}\star \ex_\star^{\frac{\imm}{2}u\s_j}\star \ex_\star^{-\frac{\bar{z}}{2}\s_i} \label{ansatz2}
\eeq
This leads to the matrix equation
\beq
\pa_{\bar{y}}a~ I ~+~ \imm\left(\pa_{\bar{y}}b+\frac{1}{2} \sin_\star{u} \right)\s_j=0 \label{sgmatrix}
\eeq
where $a$ and $b$ have been defined in (\ref{sg}).  Now, taking the trace we obtain $\pa_{\bar{y}} a = 0$  which is the constraint equation in (\ref{sg}). As a consequence, the term proportional to $\s_j$
gives rise to the dynamical equation in (\ref{sg}) for the particular choice $\g = -1$.  Therefore we have shown that the equations of motion of the NC version of sine--Gordon proposed in   \cite{GP} can
be obtained from a suitable reduction of the NC SDYM system as in the ordinary case. From this derivation the origin of the constraint appears even more clearly: it arises from setting to zero the trace part
which the matrices in $GL(2,C)$ naturally develop under $\star$--multiplication. 

Solving (\ref{sgmatrix}) for the particular choice $\s_j = \s_3$ we obtain the alternative set of equations
\bea
&& \pab \left(\ex_{\star}^{-\frac{\imm}{2} \phi}\star \pa \ex_{\star}^{\frac{\imm}{2} \phi}\right) ~=~\frac{\imm}{2} \g \sin_{\star}{\phi} 
\label{sg2} \\
&& \pab \left(\ex_{\star}^{\frac{\imm}{2} \phi}\star \pa \ex_{\star}^{-\frac{\imm}{2} \phi} \right) ~=~ -\frac{\imm}{2} \g \sin_{\star}{\phi} \nonumber
\eea
Order by order in the $\theta$--expansion the set of equations (\ref{sg}) and (\ref{sg2}) are equivalent. Therefore, the set (\ref{sg2}) is equally suitable for the description of an integrable noncommutative
generalization of sine--Gordon.

Since our NC generalization of sine--Gordon is integrable, the present result gives support to the arguments in favor of the integrability of NC SDYM system.

We note that our equations of motion, Wick rotated to Minkowski, can also be obtained by suitable reduction of the $(2+1)$ integrable noncommutative model studied in   \cite{LPS2,LP1}. 
\subsection{Noncommutative Thirring model and bosonization}\label{nc thirring boso}
In the ordinary case the equivalence between Thirring and  sine--Gordon models \cite{coleman} can be proven at the level of functional integrals by implementing the bosonization prescription \cite{witten,
bosonization} on the fermions.  The same procedure has been worked out in NC geometry \cite{MS,NOS}.  Starting from the NC version of Thirring described by 
\beq
S_T = \int d^2x \left[ \bar{\psi} \imm \g^\mu \pa_\mu \psi + m \bar{\psi} \psi - \frac{\l}{2}(\bar{\psi} \star \g^\mu \psi)(\bar{\psi} \star \g_\mu \psi)\right]
\label{Thirring}
\eeq
the bosonization prescription gives rise to the action for the bosonized NC massive Thirring model which turns out to be a NC WZNW action supplemented by a cosine potential term for the NC $U(1)$ group valued
field which enters the bosonization of the fermionic currents.  In particular, in the most recent paper in \cite{MS} it has been shown that working in Euclidean space the massless Thirring action corresponds
to the sum of two WZNW actions once a suitable choice for the regularization parameter is made. Moreover, in \cite{NOS} it was proven that the bosonization of the mass term in (\ref{Thirring}) gives rise to a
cosine potential for the scalar field with coupling constant proportional to $m$.

The main observation is that our action (\ref{action g}) is the sum of two NC WZNW actions plus cosine potential terms for the pair of $U(1)_C$ group valued fields $g$ and $\bar g$, considered as
independent.  Therefore, our action can be interpreted as coming from the bosonization of the massive NC Thirring model, in agreement with the results in \cite{MS,NOS}.

We have shown that even in the NC case the sine--Gordon field can be interpreted as the scalar field which enters the bosonization of the Thirring model, so proving that the equivalence between Thirring and
sine--Gordon can be maintained in NC generalizations of these models.  Moreover, the classical integrability of our NC version of sine--Gordon proven in   \cite{GP} should automatically guarantee the
integrability of the NC Thirring model.  

In the particular case of zero coupling ($\g = 0$), the equations (\ref{action g}) and (\ref{sg2}) correspond to the action and the equations of motion for a NC $U(1)$ WZNW model \cite{NCWZ}, respectively.
Again, we can use the results of   \cite{GP} to prove the classical integrability of the NC $U(1)$ WZNW model and construct explicitly its conserved currents.  
\section{Properties of S--matrix and integrability} 
It is well known that in integrable commutative field theories there is no particle production and the S--matrix factorizes. In the noncommutative case properties of the S--matrix have been investigated for
two specific models: The $\l \Phi^4$ theory in two dimensions \cite{seiberg} and the non integrable ``natural'' NC generalization the the sine--Gordon model \cite{CM}. In the first reference a very
pathological acausal behavior was observed due to the space and time noncommutativity. For an incoming wave packet the scattering produces an advanced wave which arrives at the origin before the incoming
wave. In the second model investigated it was found that particle production occurs. The tree level $2 \to 4$ amplitude does not vanish.   

It might be hoped that classical integrability would alleviate these pathologies. In the NC integrable sine--Gordon case, since we have an action, it is possible to investigate these issues. As described
below we have computed the scattering amplitude for the $2 \to 2 $ process and found that the acausality of   \cite{seiberg} is not cured by integrability.  We have also computed the production amplitudes
for the processes $2 \to 3$ and $2 \to 4$ and found that they don't vanish.

We started from our action (\ref{actionWZNW}) rewritten in terms of Minkowski space coordinates $x^0,x^1$ and real fields ($g=\ex_{\star}^{\frac{i}{2}\phi}$, $\hat{g}(t)=\ex_{\star}^{\frac{i}{2}t\phi}$ with
$\phi$ real)
\bea
&&S[g]=-\frac{1}{2}\int d^2x~g^{-1}\star\pa^\mu g\star g^{-1}\star\pa_\mu g-\frac{1}{3}\int d^3x~\e^{\m\n\rho}\hat{g}^{-1}\star\pa_\mu\hat{g}\star\hat{g}^{-1}\star\pa_\n\hat{g}\star\hat{g}^{-1}\star\pa_\rho\hat{g}\cr
&&~~~~~+\frac{\g}{4}\int d^2 x (g^2+g^{-2}-2) 
\ena
where $f \star g = f e^{\frac{i}{2} \theta \e^{\mu\nu} \overleftarrow{\pa}_\mu \overrightarrow{\pa}_\nu} g$, and we derived the following Feynman's rules
\begin{itemize}
\item The propagator
\beq
G(q)=\frac{4i}{q^2-2\g}
\eeq
\item
The vertices 
\bea
&& v_3(k_1,\dots,k_3)= \frac{2}{2^3 \cdot 3!}\e^{\m\n}k_{1\m}k_{2\n}F(k_1,\dots,k_3) \nonumber \\
&&v_4(k_1,\dots,k_4)= \imm \left(-\frac{1}{2^4 \cdot 4!}\left(k_1^2+3k_1\cdot k_3\right)+\frac{\g}{2 \cdot 4!}\right)F(k_1,\dots,k_4) \nonumber \\
&& v_5(k_1,\dots,k_5)= -\frac{2\e^{\m\n}}{2^5 \cdot 5!}\left(k_{1\m}k_{2\n}-k_{1\m}k_{3\n}+2k_{1\m}k_{4\n}\right)F(k_1,\dots,k_5)\nonumber \\
&&  v_6(k_1,\dots,k_6)= \imm\left[\frac{1}{2^6 \cdot 6!}\left(k_{1}^2+5k_{1}\cdot k_{3}-5k_{1} \cdot k_{4}+5k_{1} \cdot k_{5}\right)- \frac{\g}{2 \cdot 6!}\right] F(k_1,\dots,k_6)\nonumber \\
&&~~~~~~~~~\label{vertices}
\ena
where
\beq
F(k_1,\dots,k_n)=\exp\left(-\frac{\imm}{2}\sum_{i<j}k_i\times k_j\right)
\eeq
is the phase factor coming from the $\star$--products in the action (we have indicated $a\times b= \theta \e^{\m\n} a_\m b_\n$), $k_i$ are all incoming momenta and we used momentum conservation.
\end{itemize}

At tree level the $2 \to 2$ process is described by the diagrams with the topologies in Fig. 1.

\vskip 18pt
\noindent
%---------- FIGURE TOP ------------
\begin{minipage}{\textwidth}
\begin{center}
\includegraphics[width=0.60\textwidth]{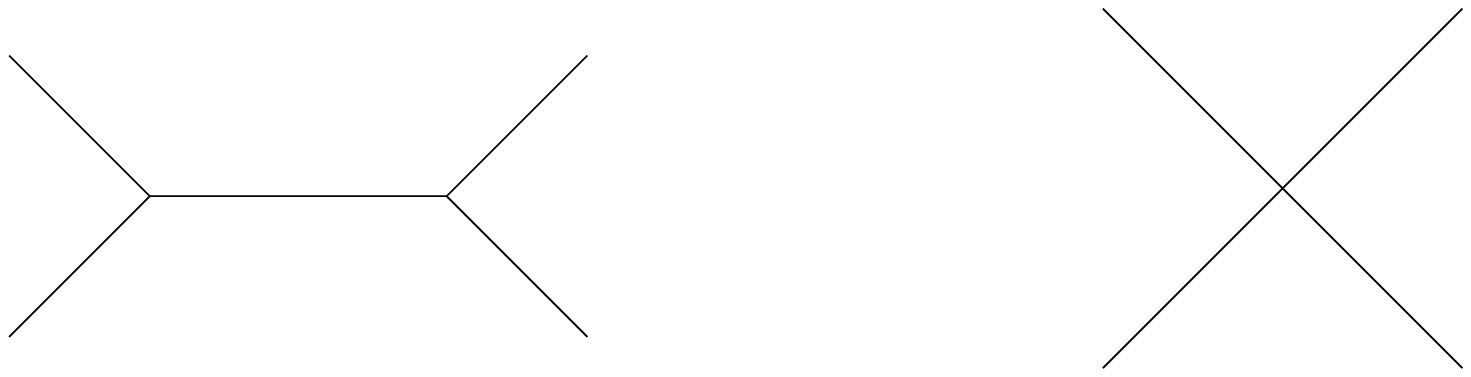}
\end{center}
\begin{center}
{\small{Figure 1: Tree level $2 \to 2$ amplitude}}
\end{center}
\end{minipage}
%---------- FIGURE END ------------

\vskip 20pt

Including contributions from the various channels and using the three point and four point vertices of eqs. (\ref{vertices}) we obtained for the scattering amplitude the expression  
\beq
-\frac{\imm}{2}E^2 p^2\left(\frac{1}{2E^2-\g}-\frac{1}{2p^2+\g}\right)\sin^2{(pE\theta)}~+~\frac{\imm\g}{2}\cos^2{(pE \theta)}\label{final}
\eeq
where $p$ is the center of mass momentum and $E = \sqrt{p^2 + 2\g}$.

For comparison with   \cite{seiberg} this should be multiplied by an incoming wave packet  
\beq
\Phi_{in}(p)\sim\left(\ex^{-\frac{(p-p_0)^2}{\l}}+\ex^{-\frac{(p+p_0)^2}{\l}}\right)
\eeq
and Fourier transformed with $\ex^{ipx}$.  We have not attempted to carry out the Fourier transform integration.  However, we note that for $p_0$ very large $E$ and $p$ are concentrated around large values and
the scattering amplitude assumes the form 
\beq
\imm\frac{\g}{4}\sin^2{(pE\theta)}~+~\imm\frac{\g}{2}\cos^2{(pE \theta)}
\eeq
which is equivalent to the result in   \cite{seiberg}, leading to the same acausal pathology \footnote{It is somewhat tantalizing that a change in the relative coefficient between the two terms would lead
to a removal of the trigonometric factors which are responsible for the acausal behavior.}. 

We describe now the computation of the production amplitudes $2 \to 3$ and $2 \to 4$. At tree level the contributions are drawn in Figures 2 and 3, respectively.

\vskip 18pt
\noindent
%---------- FIGURE TOP ------------
\begin{minipage}{\textwidth}
\begin{center}
\includegraphics[width=0.60\textwidth]{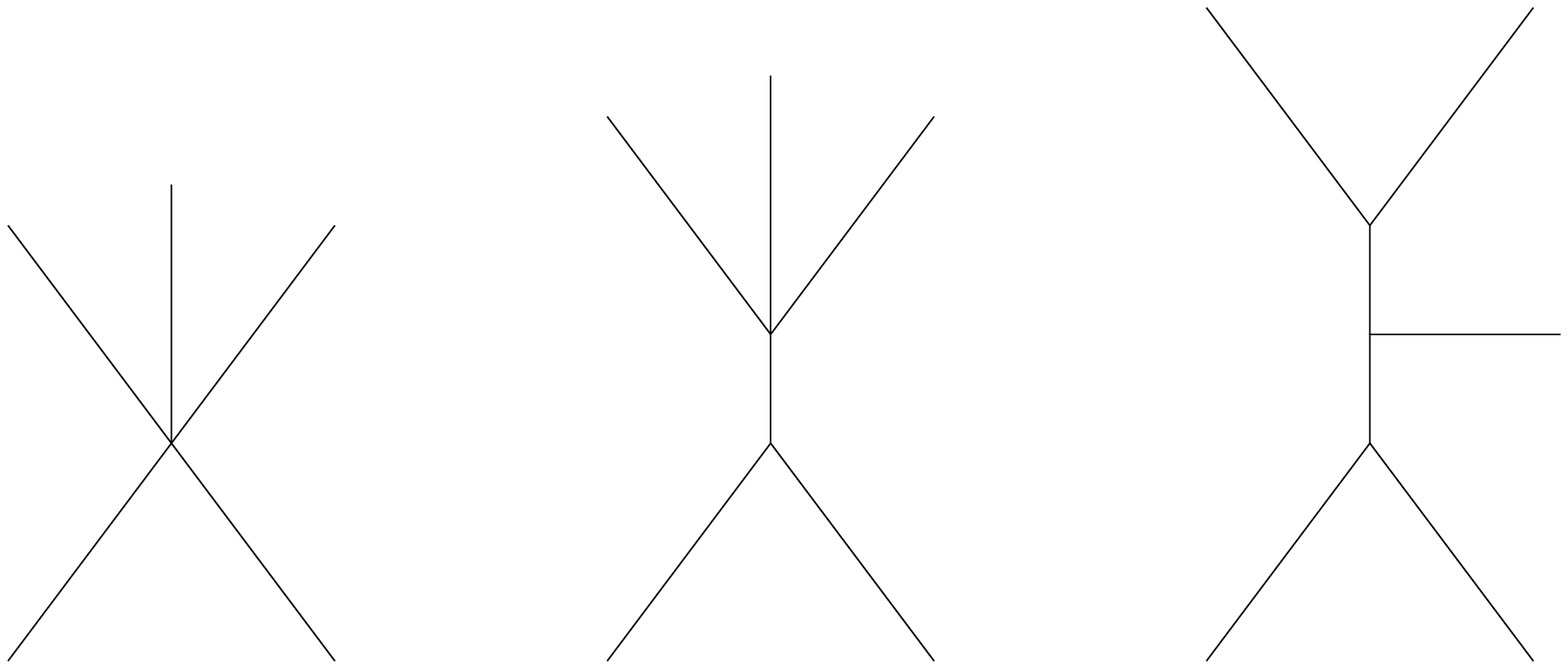}
\end{center}
\begin{center}
{\small{Figure 2: Tree level $2 \to 3$ amplitude}}
\end{center}
\end{minipage}
%---------- FIGURE END ------------

\vskip 18pt
\noindent
%---------- FIGURE TOP ------------
\begin{minipage}{\textwidth}
\begin{center}
\includegraphics[width=0.60\textwidth]{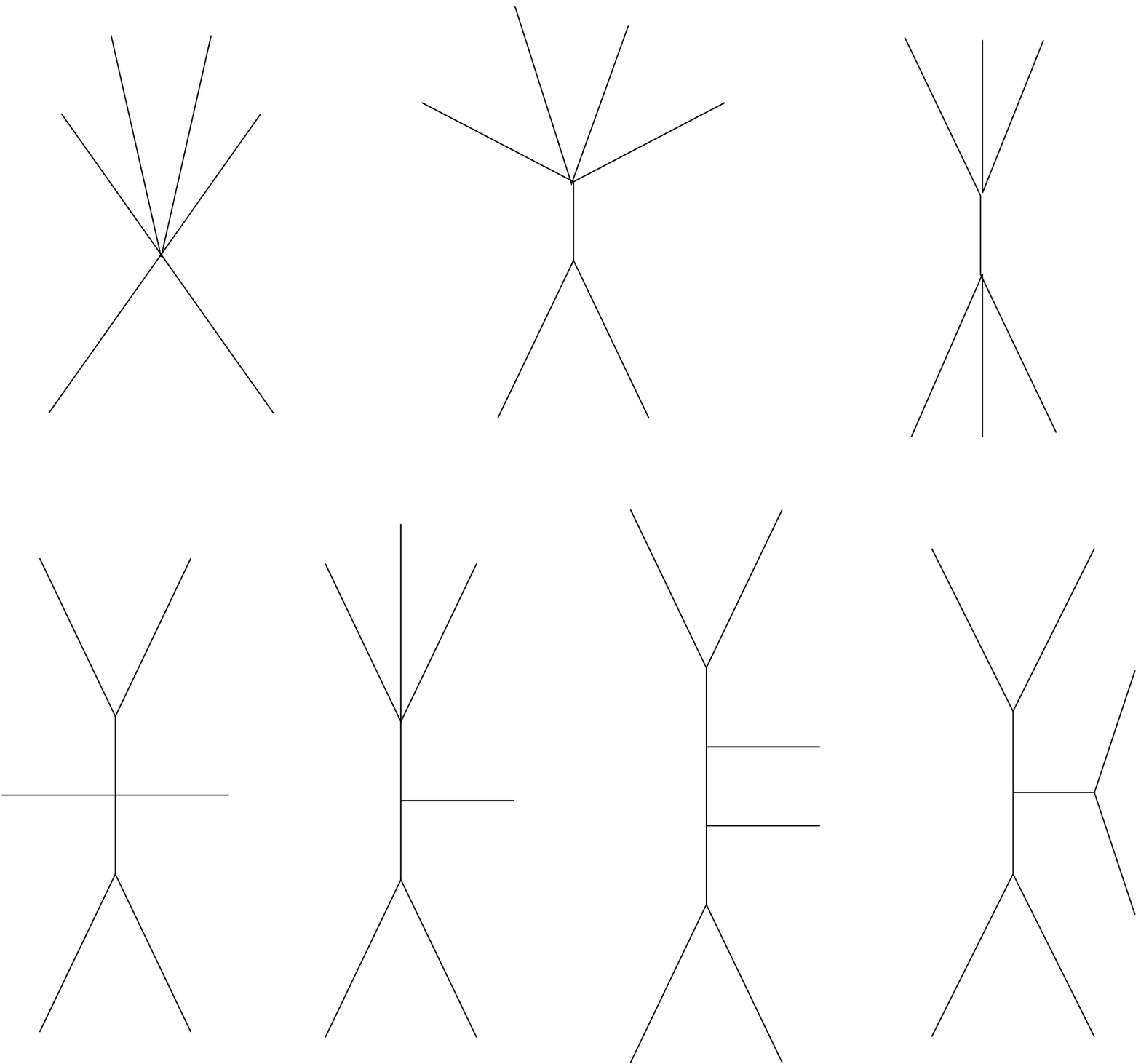}
\end{center}
\begin{center}
{\small{Figure 3: Tree level $2 \to 4$ amplitude}}
\end{center}
\end{minipage}
%---------- FIGURE END ------------

\vskip 20pt

For any topology the different possible channels must be taken into account.  This, as well as the complicated expressions for the vertices, has led us to use an algebraic manipulation program computer.  We
used {\em Mathematica}${}^{\copyright}$ to symmetrize completely the vertices (\ref{vertices}). This allows to take automatically into account the different diagrams obtained by exchanging momenta entering a
given vertex.  The contribution from each diagram was obtained as a product of the combinatorial factor, the relevant vertices and propagators.  Due to the length of the program it was impossible to handle
the calculation in a complete analytic way. Instead, the program was run with assigned values of the momenta and arbitrary  $\theta$ and $\g$. For both the  $2\rightarrow 3$ and $2\rightarrow 4$ processes the
result is non vanishing.  As a check of our calculation we mention that the production amplitudes vanish  when we set $\theta = 0$, for any value of the coupling and the momenta.

This investigation has been carried on in the particular case $\phi = \phi^\dag$. However, it can be easily proved that turning on a nontrivial imaginary part for the $\phi$ field does not cure the previous
pathologies.  
\section{Remarks on the (non)integrable noncommutative sine--Gordon} 
We have investigated some properties of the integrable NC sine--Gordon system proposed in \cite{GP} and further analyzed in my publication \cite{GMPT}. We succeeded in constructing an action which turned out to
be a WZNW action for a noncommutative, complexified $U(1)$ augmented by a cosine potential. We have shown that even in the NC case there is a duality relation between our integrable NC sine--Gordon model and
the NC Thirring. 

NC WZNW models have been shown to be one--loop renormalizable \cite{LFI}.  This suggests that the NC sine--Gordon model proposed in \cite{GP} is not only integrable but it might lead to a well--defined
quantized model, giving support to the existence of a possible relation between integrability and renormalizability. 

Armed with our action we investigated some properties of the S--matrix for elementary excitations.  However, in contradistinction to the commutative case, the S--matrix turned out to be acausal and
non factorizable \footnote{Other problems of the S--matrix have been discussed in \cite{gmbk,gm}.}. The reason for the acausal behavior has been discussed in \cite{seiberg}~ where it was pointed out
that noncommutativity induces a backward--in--time effect because of the presence of certain phase factors. It appears that in our case this effect is still present in spite of integrability. 

It is not clear why the presence of an infinite number of local conserved currents (local in the sense that they are not expressed as integrals of certain densities) does not guarantee factorization and
absence of production in the S--matrix as it does in the commutative case. The standard proofs of factorization use, among other assumptions, the mutual commutativity of the charges --- a property we have not
been able to check so far because of the complicated nature of the currents. But even if the charges were to commute the possibility of defining them as powers of momenta, as required in the proofs, could be
spoiled by acausal effects which prevent a clear distinction between incoming and outgoing particles.  Indeed, this may indicate some fundamental inconsistency of the model as a field theory describing
scattering in $(1+1)$--dimensional Minkowski space (simply going to Euclidean signature does not change the factorizability  properties of the S--matrix). However, it is conceivable that in  Euclidean space
the model could be used to describe some statistical mechanics system and this possibility might be worth investigating.

In a series of papers \cite{liao} a different approach to quantum NC theories has been proposed when the time variable is not commuting.  In particular, the way one computes Green's functions is different
there, leading to a modified definition of the S--matrix. It would be interesting to redo our calculations in that approach to see whether a well--defined factorized S--matrix for our model can be
constructed.  In this context it would be also interesting to investigate the scattering of solitons present in our model \cite{GP}. To this end, since our model is a reduction of the $(2+1)$ integrable model
studied in \cite{LP1,bieling,wolf}, it might be possible to exploit the results of those papers concerning multi--solitons and their scattering to investigate the same issues in our case.    
 
The model we have proposed here describes the propagation of two interacting scalar fields $\Re\phi$, $\Im\phi$ . The particular form of the interaction follows from the choice of the $U(2)$ matrices made in
\cite{GP} for the bicomplex formulation or, equivalently, from the particular ansatz (\ref{ansatz2}) in the reduction from NC SDYM (in the commutative limit and for $\Im\phi =0$ we are back to the ordinary
sine--Gordon).  In the commutative case the ansatz (\ref{ansatz1}) depends on a single real field and, independently of the choice of the Pauli matrices in the exponentials, we obtain the same equations of
motion.  In the NC case the lack of decoupling of the $U(1)$ subgroup requires the introduction of two fields. This implies that in general we can make an ansatz $J_\star=\ex_{\star}^{\frac{z}{2} \s_i} g(y,
\bar{y})\ex_{\star}^{-\frac{\bar{z}}{2} \s_i}$ where $g$ is a group valued field which depends on two scalar fields in such a way as to reduce to the ordinary ansatz (\ref{ansatz1}) in the commutative limit
and for a suitable identification of the two fields.  In principle there are different choices for $g$ as a function of the two fields satisfying this requirement. Different choices may be inequivalent and
describe different but still integrable dynamics for the two fields. Therefore, an interesting question  is whether an ansatz slightly different from (\ref{ansatz2}) exists which would give rise to an
integrable system described by a consistent, factorized S--matrix.  We might expect that if such an ansatz exists it should introduce a different interaction between the two fields and this might cure the
pathological behavior of the present scattering matrix. If such a possibility exists it would be interesting to compare the two different reductions to understand in the NC case what really drives the system
to be integrable in the sense of having a well--defined, factorized S--matrix since the existence of an infinite number of local conservation laws does not appear to be sufficient. These issues are discussed
in the next chapter, following the construction of the integral sine--Gordon model of my paper \cite{LMPPT}.

\chapter{Integrable noncommutative sine--Gordon}\label{sgII}
In the previous NC sine--Gordon model we proposed, the dimensional reduction from SDYM does not work at the level of the action, which turns out to be the sum of two WZW models augmented by a cosine
potential.  Evaluating tree--level scattering amplitudes we discovered that the S--matrix suffers from acausal behavior and is non factorized, meaning that particle production occurs. Integrability seems not
to translate in the nice properties of S--matrix, as one expects in two dimensions.

At this point it is important to note that the noncommutative deformation of an integrable equation is a priori not unique, because one may always add terms which vanish in the commutative limit. For the case
at hand, for example, different inequivalent ans\"atze for the $U(2)$ matrices entering the bicomplex construction \cite{DMH} are possible as long as they all reproduce the ordinary sine--Gordon equation in
the commutative limit. It is therefore conceivable that among these possibilities there exists an ansatz (different from the one in \cite{GP,GMPT}) which guarantees the classical integrability of the
corresponding noncommutative model. What is already certain is the necessity to introduce two real scalar fields instead of one, since in the noncommutative realm the $U(1)$ subgroup of $U(2)$ fails to
decouple. What has been missing is a guiding principle towards the ``correct'' field parametrization.

Since the sine--Gordon model can be obtained by dimensional reduction from $(2+2)$--dimensional SDYM theory via a $(2+1)$--dimensional integrable sigma model \cite{ward}, and because the latter's
noncommutative extension was shown to be integrable in \cite{LP1}, it seems a good idea to construct an integrable generalization of the sine--Gordon equation by starting from the linear system of this
integrable sigma model endowed with a time--space noncommutativity.  This is the key strategy of my paper \cite{LMPPT}, written in collaboration with O.~Lechtenfeld, A.~D.~Popov, S.~Penati and L.~Tamassia.
The reduction is performed on the equations of motion first, but it also works at the level of the action, so giving directly the $(1+1)$--dimensional action we are looking for. We interpret this success as an
indication that the new field parametrization proposed here is the proper one.

To be more precise, we propose three different parametrizations, by pairs of fields $(\phi_+,\phi_-)$, $(\rho,\varphi)$ and $(h_1,h_2)$, all related by nonlocal field redefinitions but all deriving
from the compatibility conditions of the underlying linear system \cite{LP1}. The first two appear in a Yang gauge \cite{Y}, while the third one arises in a Leznov gauge \cite{L}. For either field pair
in the Yang gauge, the nontrivial compatibility condition reduces to a pair of noncommutative sine--Gordon equations which in the commutative limit degenerates to the standard sine--Gordon equation for
$\sfrac12(\phi_+{+}\phi_-)$ or $\varphi$, respectively, while $\sfrac12(\phi_+{-}\phi_-)$ or $\rho$ decouple as free bosons.  The alternative Leznov formulation has the advantage of producing two polynomial
(actually, quadratic) equations of motion for $(h_1,h_2)$ but retains their coupling even in the commutative limit.

With the linear system comes a well--developed technology for generating solitonic solutions to the equations of motion. Here, we shall employ the dressing method \cite{dressing,faddeev} to explicitly outline
the construction of noncommutative sine--Gordon multi--solitons, directly in $(1+1)$ dimensions as well as by reducing plane--wave solutions of the $(2+1)$--dimensional integrable sigma model \cite{bieling}, as I
explained in section \ref{solitons}. We completely analyze the one--soliton sector where we recover the standard soliton solution as undeformed; noncommutativity becomes palpable only at the multi--soliton
level.

It was shown in \cite{LPS} that the tree--level $n$--point amplitudes of noncommutative $(2+2)$--dimensional SDYM vanish for $n>3$, consistent with the vanishing theorems for the $N{=}2$ string.  Therefore,
we may expect nice properties of the S--matrix to be inherited by our noncommutative sine--Gordon theory.  In fact, a direct evaluation of tree--level amplitudes reveals that, in the Yang as well as the
Leznov formulation, the S--matrix is {\em causal\/} and no particle production occurs.

In section \ref{ncsg sigma} we review the basic construction of the $(2+1)$--dimensional integrable sigma model of \cite{LP1} through a linear system, for the case of a noncommuting time coordinate.  In section
\ref{ncsg sg} we describe its dimensional reduction to the noncommutative integrable sine--Gordon model, both in the Yang and the Leznov formulation, mentioning the relation to the previously investigated NC
version of sine--Gordon.  Section \ref{ncsg nc solitons} is devoted to the construction of solitonic solutions for our model, by way of the iterative dressing approach.  The computation of scattering
amplitudes is finally described in section \ref{ncsg smatrix}, showing the nice results that we were expecting.
\section{Noncommutative integrable sigma model and the bicomplex}\label{ncsg sigma}
As has been known for some time, nonlinear sigma models in $(2+1)$ dimensions may be Lorentz--invariant or integrable but not both~\cite{ward}.  Since the integrable variant serves as our starting point for
the derivation of the sine--Gordon model and its soliton solutions, we shall present its noncommutative extension \cite{LP1} in some detail in the present section.
\paragraph{Noncommutative $\R^{2,1}$}
Classical field theory on noncommutative spaces may be realized by deforming the ordinary product of classical fields (or their components) to the noncommutative star product
\begin{equation}
(f \star g)(x)\ =\ f(x)\,\exp\,\bigl\{ \frac{\imm}{2}{\ld{\partial}}_a \,\theta^{ab}\, {\rd{\partial}}_b \bigr\}\,g(x)\quad,
\end{equation}
with a constant antisymmetric tensor~$\theta^{ab}$, where $a,b,\ldots=0,1,2$.  Specializing to $\R^{2,1}$, we shall use (real) coordinates $(x^a)=(t,x,y)$ in which the Minkowskian metric reads
$(\eta_{ab})=\textrm{diag}(-1,+1,+1)$.  For later use we introduce the light--cone coordinates 
\begin{equation} \label{lightcone}
u\ :=\ \sfrac{1}{2}(t+y)\quad,\qquad v\ :=\ \sfrac{1}{2}(t-y)\quad,\qquad \pa_u\ =\ \pa_t+\pa_y\quad,\qquad \pa_v\ =\ \pa_t-\pa_y \quad.
\end{equation} 
In view of the future reduction to $(1+1)$ dimensions, we choose the coordinate~$x$ to remain commutative, so that the only non--vanishing component of the noncommutativity tensor is 
\begin{equation}
\theta^{ty}\ =\ -\theta^{yt}\ =:\ \theta\ >\ 0 \quad.
\end{equation}

\paragraph{Linear system}
Consider on noncommutative $\R^{2,1}$ the following pair of linear differential equations~\cite{LP1},
\begin{equation}\label{linsys}
(\z \pa_x -\pa_u)\Psi\ =\ A\star\Psi \qquad\textrm{and}\qquad (\z \pa_v -\pa_x)\Psi\ =\ B\star\Psi \quad,
\end{equation}
where a spectral parameter~$\z\in\C P^1\cong S^2$ has been introduced.  The auxiliary field $\Psi$ takes values in U$(n)$ and depends on $(t,x,y,\z)$ or, equivalently, on $(x,u,v,\z)$.  The $u(n)$ matrices
$A$ and~$B$, in contrast, do not depend on~$\z$ but only on $(x,u,v)$.  Given a solution~$\Psi$, they can be reconstructed via\footnote{ Inverses are understood with respect to the star product,
i.e.~$\Psi^{-1}\star\Psi=\mbf{1}$.} 
\begin{equation} \label{ABfromPsi}
A \= \Psi\star(\pa_u-\z\pa_x)\Psi^{-1} \qquad\textrm{and}\qquad B \= \Psi\star(\pa_x-\z\pa_v)\Psi^{-1} \quad.
\end{equation}
It should be noted that the equations (\ref{linsys}) are not of first order but actually of infinite order in derivatives, due to the star products involved.  In addition, the matrix $\Psi$ is subject to the
following reality condition~\cite{ward}:
\begin{equation}\label{real}
\mbf{1} \= \Psi(t,x,y,\z)\,\star\,[\Psi(t,x,y,\bar{\z})]^{\dagger} \quad,
\end{equation}
where `$\dagger$' is hermitian conjugation.  The compatibility conditions for the linear system~(\ref{linsys}) read
\begin{align}
\pa_x B -\pa_v A\ =\ 0 \quad ,\label{comp1} \\[4pt]
\pa_x A -\pa_u B -A\star B +B\star A\ =\ 0 \quad . \label{comp2}
\end{align}
By detailing the behavior of~$\Psi$ at small~$\z$ and at large~$\z$ we shall now ``solve'' these equations in two different ways, each one leading to a single equation of motion for a particular field theory.
\paragraph{Yang--type solution}
We require that $\Psi$ is regular at $\z{=}0$~\cite{ivle1}, 
\begin{equation} \label{asymp1}
\Psi(t,x,y,\z\to0)\= \Phi^{-1}(t,x,y)\ +\ O(\z) \quad,
\end{equation}
which defines a U$(n)$--valued field $\Phi(t,x,y)$, i.e.~$\ \Phi^\dagger=\Phi^{-1}$.  Therewith, $A$ and~$B$ are quickly reconstructed via
\begin{equation} \label{ABfromPsi2}
A\=\Psi\star\pa_u\Psi^{-1}\big|_{\z=0} \=\Phi^{-1}\star\pa_u\Phi \qquad\textrm{and}\qquad B\=\Psi\star\pa_x\Psi^{-1}\big|_{\z=0} \=\Phi^{-1}\star\pa_x\Phi \quad.
\end{equation}
It is easy to see that compatibility equation (\ref{comp2}) is then automatic while the remaining equation~(\ref{comp1}) turns into~\cite{LP1}
\begin{equation} \label{yangtype}
\pa_x\,(\Phi^{-1}\star\pa_x\Phi)-\pa_v\,(\Phi^{-1}\star\pa_u\Phi)\ =\ 0 \quad.
\end{equation}
This Yang--type equation~\cite{Y} can be rewritten as
\begin{equation} \label{yangtype2}
(\eta^{ab}+v_c\,\eps^{cab})\,\pa_a (\Phi^{-1}\star\pa_b \Phi)\ =\ 0\quad,
\end{equation}
where $\eps^{abc}$ is the alternating tensor with $\eps^{012}{=}1$ and $(v_c)=(0,1,0)$ is a fixed space--like vector.  Clearly, this equation is not Lorentz--invariant but (deriving from a Lax pair) it is
integrable.

One can recognize (\ref{yangtype2}) as the field equation for (a noncommutative generalization of) a WZW--like modified U$(n)$ sigma model~\cite{ward,ioannidou} with the action\footnote{ which is obtainable
by dimensional reduction from the Nair--Schiff action~\cite{nair,moore} for SDYM in $(2+2)$ dimensions}
\begin{equation} \label{Yaction}
\begin{aligned}
S_{\textrm{Y}}\ & =\ -\sfrac12 \int\! \diff{t}\,\diff{x}\,\diff{y}\;\eta^{ab}\; \traccia\,\Bigl(\pa_a \Phi^{-1} \star\, \pa_b \Phi \Bigr) \\
&\quad\ -\sfrac13\int\!\diff{t}\,\diff{x}\,\diff{y} \int_0^1\!\diff{\l}\; \widetilde{v}_{\rho}\,\eps^{\rho\m\n\s}\;\traccia\,\Bigl(\Pht^{-1}\star\,\pa_{\m}\Pht\,\star\,\Pht^{-1}\star\,\pa_{\n}\Pht\,\star\,
\Pht^{-1}\star\,\pa_{\s}\Pht \Bigr) \quad,
\end{aligned}
\end{equation}
where Greek indices include the extra coordinate~$\l$, and $\eps^{\rho\m\n\s}$ denotes the totally antisymmetric tensor in~$\R^4$.  The field~$\Pht(t,x,y,\l)$ is an extension of~$\Phi(t,x,y)$, interpolating
between
\begin{equation}
\Pht(t,x,y,0)\ =\ \textrm{const} \qquad\quad\textrm{and}\qquad\quad \Pht(t,x,y,1)\ =\ \Phi(t,x,y) \quad,
\end{equation}
and `$\traccia$' implies the trace over the U$(n)$ group space.  Finally, $(\widetilde{v}_{\rho})=(v_c,0)$ is a constant vector in (extended) space--time. 
\paragraph{Leznov--type solution}
Finally, we also impose the asymptotic condition that $\ \lim_{\z\to\infty}\Psi=\Psi^0$ with some constant unitary (normalization) matrix~$\Psi^0$. The large~$\z$ behavior~\cite{ivle1}
\begin{equation} \label{asymp2}
\Psi(t,x,y,\z\to\infty)\= 
\bigl( \mbf{1}\ +\ \z^{-1}\Ups(t,x,y)\ +\ O(\z^{-2}) \bigr)\,\Psi^0
\end{equation}
then defines a $u(n)$--valued field $\Ups(t,x,y)$.  Again this allows one to reconstruct $A$ and~$B$ through
\begin{equation} \label{ABfromPsi3}
A\=-\lim_{\z\to\infty} \bigl(\z\,\Psi\star\pa_x\Psi^{-1}\bigr) \=\pa_x\Ups \qquad\textrm{and}\qquad B\=-\lim_{\z\to\infty} \bigl(\z\,\Psi\star\pa_v\Psi^{-1}\bigr) \=\pa_v\Ups
\quad.
\end{equation}
In this parametrization, compatibility equation (\ref{comp1}) becomes an identity but the second equation~(\ref{comp2}) turns into~\cite{LP1}
\begin{equation} \label{leznovtype}
\pa_x^2\Ups -\pa_u\pa_v\Ups - \pa_x \Ups \star \pa_v \Ups + \pa_v \Ups \star \pa_x \Ups \ =\ 0 \quad.
\end{equation}

This Leznov--type equation~\cite{L} can also be obtained by extremizing the action
\begin{equation} \label{Laction}
S_{\textrm{L}}\= \int\!\diff{t}\,\diff{x}\,\diff{y}\ \traccia\,\Bigl\{\sfrac12\,\eta^{ab}\,\pa_a \Ups \,\star\, \pa_b \Ups \ +\ \sfrac13\,\Ups \star 
\bigl( \pa_x \Ups\,\star\,\pa_v \Ups - \pa_v \Ups\,\star\,\pa_x \Ups \bigr) \Bigr\} \quad,
\end{equation}
which is merely cubic.

Obviously, the Leznov field $\Ups$ is related to the Yang field $\Phi$ through the non--local field redefinition
\begin{equation} \label{nonlocal}
\pa_x\Ups\=\Phi^{-1}\star\pa_u\Phi\qquad\textrm{and}\qquad \pa_v\Ups\=\Phi^{-1}\star\pa_x\Phi \quad.
\end{equation}
For each of the two fields $\Phi$ and~$\Ups$, one equation from the pair (\ref{comp1}, \ref{comp2}) represents the equation of motion, while the other one is a direct consequence of the parametrization
(\ref{ABfromPsi2}) or~(\ref{ABfromPsi3}).
\section{The noncommutative sine--Gordon action from dimensional reduction}\label{ncsg sg}
\paragraph{Algebraic reduction ansatz}
It is well known that the (commutative) sine--Gordon equation can be obtained from the self--duality equations for $SU(2)$ Yang--Mills upon appropriate reduction from $2{+}2$ to $(1+1)$ dimensions. In this process
the integrable sigma model of the previous section appears as an intermediate step in $(2+1)$ dimensions, and so we may take its noncommutative extension as our departure point, after enlarging the group to
$U(2)$.  In order to avoid cluttering the formulae we suppress the `$\star$' notation for noncommutative multiplication from now on: all products are assumed to be star products, and all functions are built on
them, i.e.~$\ex^{f(x)}$ stands for $\e_\star^{f(x)}$ and so on.

The dimensional reduction proceeds in two steps, firstly, a factorization of the coordinate dependence and, secondly, an algebraic restriction of the form of the $U(2)$ matrices involved.  In the language of
the linear system~(\ref{linsys}) the adequate ansatz for the auxiliary field~$\Psi$ reads
\begin{equation} \label{ansatzPsi}
\Psi(t,x,y,\z) \= V(x)\,\psi(u,v,\z)\,V^\+(x)\qquad\textrm{with}\qquad V(x) \= \Ecal\,\ex^{\imm\a\,x\,\s_1} \quad,
\end{equation}
where $\s_1=(\begin{smallmatrix} 0 & 1 \\ 1 & 0 \end{smallmatrix})$, $\Ecal$ denotes some constant unitary matrix (to be specified later) and $\a$ is a constant parameter. Under this factorization, the linear
system~(\ref{linsys}) simplifies to\footnote{ The adjoint action means $\mathrm{ad}\s_1\,(\psi)=[\s_1,\psi]$.}
\begin{equation}
(\pa_u -\imm\a\,\z\,\mathrm{ad}\s_1)\,\psi\=-a\,\psi \qquad\textrm{and}\qquad (\z\pa_v -\imm\a\,\mathrm{ad}\s_1)\,\psi\= b\,\psi 
\end{equation}
with $\ a=V^\+A\,V\ $ and $\ b=V^\+B\,V$.  Taking into account the asymptotic behavior (\ref{asymp1}, \ref{asymp2}), the ansatz~(\ref{ansatzPsi}) translates to the decompositions
\begin{align} \label{ansatzPhi}
\Phi(t,x,y) &\= V(x)\,g(u,v)\,V^\+(x)\qquad\textrm{with}\qquad g(u,v) \in U(2)\quad, \\[6pt] \label{ansatzUps}\Ups(t,x,y) &\= V(x)\,\chi(u,v)\,V^\+(x)\qquad\textrm{with}\qquad \chi(u,v) \in u(2) \quad.
\end{align}
To aim for the sine--Gordon equation, one imposes certain algebraic constraints on $a$ and $b$ (and therefore on~$\psi$).  Their precise form, however, is not needed, as we are ultimately interested only in
$g$ or~$\chi$. Therefore, we instead directly restrict $g(u,v)$ to the form
\begin{equation} \label{reducg}
g \= \Bigl(\begin{matrix} g_+ & 0 \\ 0 & g_- \end{matrix}\Bigr) \= g_+ P_+ + g_- P_-\qquad\textrm{with}\qquad g_+ \in \textrm{$U(1)$}_+ \quad\textrm{and}\quad    g_- \in \textrm{$U(1)$}_- 
\end{equation}
and with projectors $P_+=(\begin{smallmatrix}1&0\\0&0\end{smallmatrix})$ and $P_-=(\begin{smallmatrix}0&0\\0&1\end{smallmatrix})$.  This imbeds $g$ into a $U(1) \times U(1)$ subgroup of~$U(2)$.  Note that $g_+$
and $g_-$ do not commute, due to the implicit star product.  Invoking the field redefinition~(\ref{nonlocal}) we infer that the corresponding reduction for~$\chi(u,v)$ should be\footnote{ Complex conjugates of
scalar functions are denoted with a dagger to remind the reader of their noncommutativity.}
\begin{equation} \label{reduch}
\chi \= \imm \Bigl(\begin{matrix} 0 & h^\+ \\ h & 0 \end{matrix}\Bigr)\qquad\textrm{with}\qquad h \in \C \quad,
\end{equation}
with the ``bridge relations''
\begin{equation} \label{nonlocal2}
\begin{aligned}
\a\,(h-h^\+) &\= - g_+^\+ \pa_u g_+ \= g_-^\+ \pa_u g_- \quad,\\[6pt]
\sfrac{1}{\a}\,\pa_v h &\= g_-^\+ g_+ - \mbf{1} \qquad\textrm{and h.c.} \quad.
\end{aligned}
\end{equation}
In this way, the $u(2)$--matrix $\chi$ is restricted to be off--diagonal.

We now investigate in turn the consequences of the ans\"atze (\ref{ansatzPhi}, \ref{reducg}) and (\ref{ansatzUps}, \ref{reduch}) for the equations of motion (\ref{yangtype}) and (\ref{leznovtype}),
respectively.
\paragraph{Reduction of Yang--type equation}
Let us insert the ansatz~(\ref{ansatzPhi}) into the Yang--type equation of motion~(\ref{yangtype}).  After stripping off the $V$ factors one obtains
\begin{equation}
\pa_v (g^\+ \pa_u g) + \a^2 (\s_1 g^\+ \s_1 g - g^\+ \s_1 g \s_1) \= 0 \quad.
\end{equation}
Specializing with (\ref{reducg}) and employing the identities $\ \s_1 P_\pm \s_1 = P_\mp\ $ we arrive at $\ Y_+P_+ + Y_-P_- =0$, with
\begin{equation} \label{Yg}
\begin{aligned}
Y_+ &\ \equiv\ \pa_v (g_+^\+ \pa_u g_+) + \a^2 (g_-^\+ g_+ - g_+^\+ g_-) \= 0 \quad, \\[6pt]
Y_- &\ \equiv\ \pa_v (g_-^\+ \pa_u g_-) + \a^2 (g_+^\+ g_- - g_-^\+ g_+) \= 0 \quad.
\end{aligned}
\end{equation}
Since the brackets multiplying~$\a^2$ are equal and opposite, it is worthwhile to present the sum and the difference of the two equations:
\begin{equation} \label{Yg2}
\begin{aligned}
\pa_v\bigl( g_+^\+ \pa_u g_+ + g_-^\+ \pa_u g_- \bigr) &\= 0 \quad,\\[6pt]
\pa_v\bigl( g_+^\+ \pa_u g_+ - g_-^\+ \pa_u g_- \bigr) &\=2\a^2 \bigl( g_+^\+ g_- - g_-^\+ g_+ \bigr) \quad.
\end{aligned}
\end{equation}

It is natural to introduce the angle fields $\p_\pm(u,v)$ via
\begin{equation} \label{para1}
g \= \ex^{\frac{\imm}{2}\phi_+P_+}\,\ex^{-\frac{\imm}{2}\phi_-P_-}\qquad\Leftrightarrow\qquad g_+ \= \ex^{\frac{\imm}{2}\phi_+}\qquad\textrm{and}\qquad g_- \= \ex^{-\frac{\imm}{2}\phi_-} \quad.
\end{equation}
In terms of these, the equations~(\ref{Yg2}) read
\begin{equation} \label{Yphi}
\begin{aligned}
\pa_v\bigl( \ex^{-\frac{\imm}{2}\phi_+}\,\pa_u\ex^{\frac{\imm}{2}\phi_+} +\ex^{\frac{\imm}{2}\phi_-}\,\pa_u\ex^{-\frac{\imm}{2}\phi_-} \bigr) &\=0\quad,\\[6pt]
\pa_v\bigl( \ex^{-\frac{\imm}{2}\phi_+}\,\pa_u\ex^{\frac{\imm}{2}\phi_+} -\ex^{\frac{\imm}{2}\phi_-}\,\pa_u\ex^{-\frac{\imm}{2}\phi_-} \bigr) &\=
2\a^2\bigl( \ex^{-\frac{\imm}{2}\phi_+}\ex^{-\frac{\imm}{2}\phi_-} -\ex^{\frac{\imm}{2}\phi_-}\ex^{\frac{\imm}{2}\phi_+} \bigr) \quad.
\end{aligned}
\end{equation}
We propose to call these two equations ``the noncommutative sine--Gordon equations''.  Besides their integrability (see later sections for consequences) their form is quite convenient for studying the
commutative limit.  When $\theta\to0$, (\ref{Yphi}) simplifies to
\begin{equation} \label{Ycomm}
\pa_u\pa_v (\phi_+{-}\phi_-) \= 0 \qquad\textrm{and}\qquad\pa_u\pa_v (\phi_+{+}\phi_-) \= -8\a^2\,\sin\sfrac12(\phi_+{+}\phi_-) \quad.
\end{equation}
Because the equations have decoupled we may choose 
\begin{equation}
\phi_+ \= \phi_- \ =:\ \p  \qquad\Leftrightarrow\qquad g_+ \= g_-^\+ \qquad\Leftrightarrow\qquad g \in \textrm{$U(1)$}_{\textrm{A}}
\end{equation}
and reproduce the familiar sine--Gordon equation
\begin{equation} \label{sG}
(\pa_t^2 -\pa_y^2)\,\p \= -4\a^2\,\sin\p \quad.
\end{equation}
One learns that in the commutative case the reduction is $SU(2)\to U(1) _{\textrm{A}}$ since the $U(1) _{\textrm{V}}$ degree of freedom $\phi_+{-}\phi_-$ is not needed.  The deformed situation, however, requires
extending $SU(2)$ to $U(2)$, and so it is imperative here to keep both $U(1)$s and work with {\it two\/} scalar fields.

Inspired by the commutative decoupling, one may choose another distinguished parametrization of~$g$, namely
\begin{equation} \label{para2}
g_+ \= \ex^{\frac{\imm}{2}\rho}\,\ex^{\frac{\imm}{2}\varphi}\qquad\textrm{and}\qquad g_- \= \ex^{\frac{\imm}{2}\rho}\,\ex^{-\frac{\imm}{2}\varphi} \quad,
\end{equation}
which defines angles $\rho(u,v)$ and $\varphi(u,v)$ for the linear combinations $\textrm{U}(1)_{\textrm{V}}$ and $\textrm{U}(1)_{\textrm{A}}$, respectively.  Inserting this into (\ref{Yg}) one finds
\begin{equation} \label{Yrho}
\begin{aligned}
\pa_v\bigl(\ex^{-\frac{\imm}{2}\varphi}\,\pa_u\ex^{\frac{\imm}{2}\varphi}\bigr)+2\imm\a^2\,\sin\varphi &\=-\pa_v\bigl[\ex^{-\frac{\imm}{2}\varphi}\ex^{-\frac{\imm}{2}\rho}\,(\pa_u \ex^{\frac{\imm}{2}\rho})\ex^{\frac{\imm}{2}\varphi}\bigr] 
\quad,\\[6pt]
\pa_v\bigl(\ex^{\frac{\imm}{2}\varphi}\,\pa_u\ex^{-\frac{\imm}{2}\varphi} \bigr)-2\imm\a^2\,\sin\varphi&\=-\pa_v\bigl[\ex^{\frac{\imm}{2}\varphi}\ex^{-\frac{\imm}{2}\rho}\,(\pa_u\ex^{\frac{\imm}{2}\rho})\ex^{-\frac{\imm}{2}\varphi}\bigr]\quad.
\end{aligned}
\end{equation}
In the commutative limit, this system is easily decoupled to
\begin{equation} \label{sG2}
\pa_u \pa_v \rho \=0 \qquad\textrm{and}\qquad \pa_u \pa_v \varphi + 4\a^2\,\sin\varphi \= 0 \quad,
\end{equation}
revealing that $\ \rho\to\frac12(\phi_+{-}\phi_-)\ $ and $\ \varphi\to\frac12(\phi_+{+}\phi_-)=\p\ $ in this limit.

It is not difficult to write down an action for (\ref{Yg}) (and hence for (\ref{Yphi}) or (\ref{Yrho})). The relevant action may be computed by reducing (\ref{Yaction}) with the help of (\ref{ansatzPhi})
and~(\ref{reducg}). The result takes the form
\begin{equation} \label{gaction}
S[g_+,g_-] \= S_{\rm WZW}[g_+]\,+\,S_{\rm WZW}[g_-]\,+\,\alpha^2\int\!\diff{t}\,\diff{y}\; \bigl( g_+^{\dag} g_- + g_-^\dag g_+ - 2 \bigr) \quad,
\end{equation}
where $S_W$ is the abelian WZW action
\begin{equation} \label{WZWaction}
S_{\rm WZW}[f]\ \equiv\ -\sfrac12\int\!\intd t\,\intd y\; \pa_v f^{-1}\;\pa_u f\,
-\,\sfrac13\int\!\intd t\,\intd y\int_0^1\!\intd\l\;\eps^{\m\n\s}\,\hat f^{-1}\pa_\m\hat f\;\hat f^{-1}\pa_\n\hat f\;\hat f^{-1}\pa_\s\hat f\quad.
\end{equation}
Here $\hat{f}(\l)$ is a homotopy path satisfying the conditions $\hat{f}(0) = 1$ and $\hat{f}(1) = f$.  Parametrizing $g_{\pm}$ as in (\ref{para2}) and using the Polyakov--Wiegmann identity, the action for
$\rho$ and $\varphi$ reads
\begin{equation} \label{rhophiaction}
\begin{aligned}
S[\rho,\varphi] &\= 2 S_{\rm PC}\bigl[ \ex^{\frac{\imm}{2}\varphi}\bigr] \,+\,2\a^2\int\!\intd t\,\intd y\;\bigl( \cos{\varphi} -1 \bigr)\,+\,2 S_{\rm WZW}\bigl[ \ex^{\frac{\imm}{2}\rho}\bigr] \\ 
&\qquad - \int\!\intd t\,\intd y\;\ex^{-\frac{\imm}{2}\rho}\,\pa_v \ex^{\frac{\imm}{2}\rho}\bigl( \ex^{-\frac{\imm}{2}\varphi}\,\pa_u \ex^{\frac{\imm}{2}\varphi}
+ \ex^{\frac{\imm}{2}\varphi}\,\pa_u \ex^{-\frac{\imm}{2}\varphi}\bigr)\quad,
\end{aligned}
\end{equation}
where
\begin{equation}
S_{\rm PC}[f]\ \equiv\ -\sfrac12\int\!\intd t\,\intd y\;\pa_v f^{-1}\;\pa_u f\quad.
\end{equation}
In this parametrization the WZ term has apparently been shifted entirely to the $\rho$ field while the cosine--type self--interaction remains for the $\varphi$ field only. This fact has important consequences
for the scattering amplitudes.

It is well known \cite{witten, bosonization, bosonization2} that in ordinary commutative geometry the bosonization of $N$ free massless fermions in the fundamental representation of $SU(N)$ gives rise to a
WZW model for a scalar field in $SU(N)$ plus a free scalar field associated with the $U(1)$ invariance of the fermionic system.  In the noncommutative case the bosonization of a single massless Dirac fermion
produces a noncommutative $U(1)$ WZW model~\cite{MS}, which becomes free only in the commutative limit. Moreover, the $U(1)$ subgroup of U($N$) does no longer decouple~\cite{matsubara}, so that $N$ noncommuting
free massless fermions are related to a noncommutative WZW model for a scalar in U($N$).  On the other hand, giving a mass to the single Dirac fermion leads to a noncommutative cosine potential on the
bosonized side~\cite{mass}.

In contrast, the noncommutative sine--Gordon model we propose here is of a more general form.  The action~(\ref{gaction}) describes the propagation of a scalar field $g$ taking its value in $U(1) \times U(1)
\subset U(2)$. Therefore, we expect it to be a bosonized version of two fermions in some representation of $U(1) \times U(1)$. The absence of a WZ term for $\varphi$ and the lack of a cosine--type
self--interaction for~$\rho$ as well as the non--standard interaction term make the precise identification non--trivial however.
\paragraph{Reduction of Leznov--type equation}
Alternatively, if we insert the ansatz~(\ref{ansatzUps}) into the Leznov--type equation of motion~(\ref{leznovtype}) we get
\begin{equation}
\pa_u\pa_v\chi+2\a^2(\chi-\s_1\chi\s_1)+\imm\a\bigl[[\s_1,\chi],\pa_v\chi\bigr]\=0\quad.
\end{equation}
Specializing with (\ref{reduch}) this takes the form $\ Z\s_-+Z^\+\s_+=0\ $ with $\ \s_-=(\begin{smallmatrix}0&0\\1&0\end{smallmatrix})\ $ and $\ \s_+=(\begin{smallmatrix}0&1\\0&0\end{smallmatrix})$, where
\begin{equation} \label{Lh}
Z\ \equiv\ \pa_u\pa_v h + 2\a^2\,(h-h^\+)+ \a\,\bigl\{ \pa_v h\,,\,h-h^\+ \bigr\} \=0 \quad.
\end{equation}
The decomposition
\begin{equation} \label{para3}
\chi \= \imm(h_1\s_1 + h_2\s_2) \qquad\Leftrightarrow\qquad h \= h_1 + \imm h_2
\end{equation}
then yields
\begin{equation}
\begin{aligned} \label{Lh12}
\pa_u\pa_v h_1 -2\a\,\bigl\{ \pa_v h_2\,,\,h_2 \bigr\} &\=0 \quad, \\[6pt]
\pa_u\pa_v h_2 +4\a^2 h_2 +2\a\,\bigl\{ \pa_v h_1\,,\,h_2 \bigr\} &\=0 \quad.
\end{aligned}
\end{equation}
These two equations constitute an alternative description of the noncommutative sine--Gordon model; they are classically equivalent to the pair of~(\ref{Yg2}) or, to be more specific, to the pair
of~(\ref{Yrho}).  For the real fields the ``bridge relations''~(\ref{nonlocal2}) read
\begin{equation} \label{nonlocal3}
\begin{aligned}
& 2\imm\a\,h_2 \= -\ex^{-\frac{\imm}{2}\varphi}\ex^{-\frac{\imm}{2}\rho}\,\pa_u ( \ex^{\frac{\imm}{2}\rho}\ex^{\frac{\imm}{2}\varphi} )\= \ex^{\frac{\imm}{2}\varphi}\ex^{-\frac{\imm}{2}\rho}\,
\pa_u ( \ex^{\frac{\imm}{2}\rho}\ex^{-\frac{\imm}{2}\varphi} ) \quad, \\[6pt]
& \qquad\qquad \sfrac{1}{\a}\pa_v h_1 \= \cos\varphi-1 \qquad\textrm{and}\qquad\sfrac{1}{\a}\pa_v h_2 \= \sin\varphi \quad.
\end{aligned}
\end{equation}
One may ``solve'' one equation of~(\ref{Yrho}) by an appropriate field redefinition from~(\ref{nonlocal3}), which implies already one member of~(\ref{Lh12}). The second equation from~(\ref{Yrho}) then yields
the remaining ``bridge relations'' in~(\ref{nonlocal3}) as well as the other member of~(\ref{Lh12}). This procedure works as well in the opposite direction, from~(\ref{Lh12}) to (\ref{Yrho}).  The nonlocal
duality between $(\varphi,\rho)$ and $(h_1,h_2)$ is simply a consequence of the equivalence between (\ref{yangtype}) and (\ref{leznovtype}) which in turn follows from our linear system~(\ref{linsys}).

The ``$h$~description'' has the advantage of being polynomial.  It is instructive to expose the action for the system~(\ref{Lh12}).  Either by inspection or by reducing the Leznov action~(\ref{Laction}) one
obtains
\begin{equation}
S[h_1,h_2] \= \int\!\diff{t}\,\diff{y}\;\Bigl\{ \pa_u h_1 \pa_v h_1 + \pa_u h_2 \pa_v h_2 -4\a^2 h_2^2 -4\a\,h_2^2\,\pa_v h_1 \Bigr\} \quad.
\label{haction}
\end{equation}
\subsection{Connection to previous noncommutative sine--Gordon generalizations}
The noncommutative generalizations of the sine--Gordon model presented above are expected to possess an infinite number of conservation laws, as they originate from the reduction of an integrable model
\cite{LPS2}. It is worthwhile to point out their relation to previously proposed noncommutative sine--Gordon models which also feature an infinite number of local conserved currents.  

In \cite{GP} an alternative noncommutative version of the sine--Gordon model was proposed. Using the bicomplex approach the equations of motion were obtained as flatness conditions of a bidifferential
calculus,\footnote{This subsection switches to Euclidean space $\R^2$, where $\pa$ and $\bar\pa$ are derivatives with respect to complex coordinates.}
\beq
\bar{\pa} ( G^{-1} \star \pa G) \= [ R\,,\, G^{-1} \star S\,G ]_{\star} \quad,
\label{eq}
\eeq
where  
\beq
R \= S \= 2\alpha\, \Bigl(\begin{matrix} 0 & 0 \\ 0 & 1 \end{matrix}\Bigr) 
\eeq
and $G$ is a suitable matrix in $U(2)$ or, more generally, in complexified $U(2)$. In \cite{GP} the $G$ matrix was chosen as
\beq \label{sgdif}
G \= \e_{\star}^{\frac{\imm}{2} \s_2 \phi} \= \biggl( \begin{matrix}\phantom{-}\cos_{\star}{\frac{\phi}{2}} & \ \sin_{\star}{\frac{\phi}{2}}\\[4pt]
-\sin_{\star}{\frac{\phi}{2}} & \ \cos_{\star}{\frac{\phi}{2}}\end{matrix} \biggr)
\eeq
with $\phi$ being a complex scalar field. This choice produces the noncommutative equations (all the products are $\star$--products)
\bea
&& 
\bar{\pa} \bigl( \ex^{\frac{\imm}{2} \phi}  \pa \ex^{-\frac{\imm}{2} \phi}+ \ex^{-\frac{\imm}{2} \phi}   \pa \ex^{\frac{\imm}{2} \phi} \bigr)~=~ 0 \quad,\nonumber \\
&&
\bar{\pa} \bigl( \ex^{-\frac{\imm}{2} \phi}  \pa \ex^{\frac{\imm}{2} \phi}- \ex^{\frac{\imm}{2} \phi}   \pa \ex^{-\frac{\imm}{2} \phi} \bigr) ~=~ 4\imm\alpha^2 \sin{\phi} \quad.\label{sg3 II}
\eea
As shown in \cite{GMPT} these equations (or a linear combination of them) can be obtained as a dimensional reduction of the equations of motion for noncommutative $U(2)$ SDYM in $(2+2)$ dimensions.

The equations (\ref{sg3 II}) can also be derived from an action which consists of the sum of two WZW actions augmented by a cosine potential,
\beq
S[f,\bar f]\=S[f]+S[\bar f]  \qquad\text{with}\qquad S[f] \ \equiv\ S_W[f] -\alpha^2 \int\!\diff{t}\,\diff{y}\; \bigl( f^2+ f^{-2} -2 \bigr) \quad,\label{sg4 II}
\eeq
with $S_W[f]$ given in (\ref{WZWaction}) for $f\equiv\ex^{\frac{\imm}{2}\phi}$ in complexified $U(1)$.  However, this action cannot be obtained from the SDYM action in $(2+2)$ dimensions by performing the same field
parametrization which led to (\ref{sg3 II}). 

Comparing the actions (\ref{gaction}) and (\ref{sg4 II}) and considering $f$ and $\bar{f}$ as independent $U(1)$ group valued fields we are tempted to formally identify $f \equiv g_+$ and $\bar{f} \equiv g_-$.
Doing this, we immediately realize that the two models differ in their interaction term which generalizes the cosine potential.  While in (\ref{sg4 II}) the fields $f$ and $\bar{f}$ show only
self--interaction, the fields $g_+$ and $g_-$ in (\ref{gaction}) interact with each other.  As we will see in section \ref{ncsg smatrix} this makes a big difference when evaluating the S--matrix elements.

We close this section by observing that the equations of motion (\ref{Yphi}) can also be obtained directly in two dimensions by using the bicomplex approach described in \cite{GP}. In fact, if instead of
(\ref{sgdif}) we choose
\beq
G \= \biggl( \begin{matrix}\ex^{\frac{\imm}{2}\phi_+} + \ex^{-\frac{\imm}{2}\phi_-}& \ -\imm \ex^{\frac{\imm}{2}\phi_+} +\imm \ex^{-\frac{\imm}{2}\phi_-} \\[4pt]
\imm \ex^{\frac{\imm}{2}\phi_+} -\imm \ex^{-\frac{\imm}{2}\phi_-}& \phantom{-}\ \ex^{\frac{\imm}{2}\phi_+} + \ex^{-\frac{\imm}{2}\phi_-} \end{matrix} \biggr)
\eeq
it is easy to prove that (\ref{eq}) yields exactly the set of equations (\ref{Yphi}). Therefore, by exploiting the results in \cite{GP} it should be straightforward to construct the first nontrivial conserved
currents for the present model. 

\section{Noncommutative solitons}\label{ncsg nc solitons}
\subsection{Dressing approach in $(2+1)$ dimensions}
The existence of the linear system allows for powerful methods to systematically construct explicit solutions for $\Psi$ and hence for $\Phi^\+=\Psi|_{\z=0}$ or $\Ups$.  For our purposes the so--called
dressing method \cite{dressing,faddeev} proves to be most practical, and so we shall first present it here for our linear system~(\ref{linsys}), before reducing the results to solitonic solutions of the
noncommutative sine--Gordon equations.

The central idea is to demand analyticity in the spectral parameter~$\z$ for the linear system~(\ref{linsys}), which strongly restricts the possible form of~$\Psi$. The most elegant way to exploit this
constraint starts from the observation that the left hand sides of the differential relations (D):=(\ref{ABfromPsi}) as well as the reality condition (R):=(\ref{real}) do not depend on~$\z$ while their right
hand sides are expected to be nontrivial functions of~$\z$ (except for the trivial case $\Psi=\Psi^0$). More specifically, $\C P^1$ being compact, the matrix function~$\Psi(\z)$ cannot be holomorphic
everywhere but must possess some poles, and hence the right hand sides of (D) and (R) should display these (and complex conjugate) poles as well.  The resolution of this conundrum demands that the residues of
the right hand sides at any would--be pole in~$\z$ have to vanish.  We are now going to evaluate these conditions.

The dressing method builds a solution $\Psi_N(t,x,y,\z)$ featuring $N$~simple poles at positions $\m_1$, $\m_2,\ldots,\m_N$ by left--multiplying an $(N{-}1)$--pole solution $\Psi_{N-1}(t,x,y,\z)$ with a
single--pole factor of the form $\ \bigl(1+\frac{\m_N{-}\mb_N}{\z{-}\m_N}P_N(t,x,y)\bigr)$, where the $n{\times}n$ matrix function $P_N$ is yet to be determined.  In addition, we are free to right--multiply
$\Psi_{N-1}(t,x,y,\z)$ with some constant unitary matrix~$\Psh^0_N$.  Starting from $\Psi_0=\mbf{1}$, the iteration $\ \Psi_0\mapsto\Psi_1\mapsto\ldots\mapsto\Psi_N\ $ yields a multiplicative ansatz for
$\Psi_N$ which, via partial fraction decomposition, may be rewritten in an additive form (as a sum of simple pole terms). Let us trace this iterative procedure constructively.

In accord with the outline above, the one--pole ansatz must read
($\Psh^0_1=:\Psi^0_1$)
\begin{equation} \label{Psione}
\Psi_1 \= \Bigl(\mbf{1}\,+\,\frac{\m_1-\mb_1}{\z-\m_1}\,P_1\Bigr)\,\Psi^0_1\= \Bigl(\mbf{1} \,+\,\frac{\Lambda_{11}S_1^\+}{\z-\m_1}\Bigr)\,\Psi^0_1
\end{equation}
with some $n{\times}r_1$ matrix functions $\Lambda_{11}$ and~$S_1$ for some $1{\le}r_1{<}n$. The normalization matrix~$\Psi^0_1$ is constant and unitary. It is quickly checked that
\begin{equation} \label{R1}
\res_{\z=\mb_1} (R) =0 \qquad\Longrightarrow\qquad P_1^\+ \= P_1 \= P_1^2 \qquad\Longrightarrow\qquad P_1 \= T_1\,(T_1^\+ T_1)^{-1} T_1^\+ \quad,
\end{equation}
meaning that $P_1$ is a rank~$r_1$ projector built from an $n{\times}r_1$ matrix function~$T_1$. The columns of~$T_1$ span the image of~$P_1$ and obey $P_1T_1=T_1$. When using the second parametrization
of~$\Psi_1$ in~(\ref{Psione}) one finds that
\begin{equation} \label{R2}
\res_{\z=\mb_1} (R) =0 \qquad\Longrightarrow\qquad (\mbf{1}-P_1)\,S_1\Lambda_{11}^\+ \=0 \qquad\Longrightarrow\qquad T_1 \= S_1 \qquad\qquad\qquad{}
\end{equation}
modulo a freedom of normalization.  Finally, the differential relations yield
\begin{equation} \label{D1}
\res_{\z=\mb_1} (D) =0 \qquad\Longrightarrow\qquad (\mbf{1}-P_1)\,\Lb_1\,(S_1\Lambda_{11}^\+) \=0 \qquad\Longrightarrow\qquad \Lb_1\,S_1 \= S_1\,\G_1^{A,B}
\end{equation}
for some $r_1{\times}r_1$ matrices~$\G_1^A$ and $\G_1^B$, after having defined
\begin{equation}
\bar{L}_i^A\ :=\ \pa_u-\mb_i\pa_x \qquad\textrm{and}\qquad \bar{L}_i^B\ :=\ \m_i(\pa_x-\mb_i\pa_v) \qquad\textrm{for}\quad i=1,2,\ldots,N \quad.
\end{equation}
Because the $\Lb_i$ are linear differential operators it is easy to write down the general solution for~(\ref{D1}): Introduce ``co--moving coordinates''
\begin{equation} \label{comoving}
w_i \ :=\ x + \mb_i u + \mb_i^{-1} v \qquad\Longrightarrow\qquad \bar{w}_i \= x + \m_i u + \m_i^{-1} v \qquad\textrm{for}\quad i=1,2,\ldots,N 
\end{equation}
so that on functions of $(w_i,\bar{w}_i)$ alone the $\Lb_i$ act as
\begin{equation}
\bar{L}^A_i \= \bar{L}^B_i \= (\m_i{-}\mb_i)\frac{\pa}{\pa\bar{w}_i} \quad.
\end{equation}
Hence, (\ref{D1}) is solved by
\begin{equation}
S_1(t,x,y) \= \Sh_1(w_1)\,\ex^{\bar{w}_1 \G_1 /(\m_1-\mb_1)} \qquad\textrm{ for any $w_1$-holomorphic $n{\times}r_1$ matrix function $\Sh_1$}
\end{equation}
and $\G_1^A=\G_1^B=:\G_1$.  Appearing to the right of~$\Sh_1$, the exponential factor is seen to drop out in the formation of~$P_1$ via (\ref{R1}) and~(\ref{R2}). Thus, no generality is lost by taking
$\G_1=0$. We learn that any $w_1$--holomorphic $n{\times}r_1$ matrix $T_1$ is admissible to build a projector~$P_1$ which then yields a solution $\Psi_1$ (and thus~$\Phi$) via~(\ref{Psione}).  Note that
$\Lambda_{11}$ need not be determined separately but follows from our above result.  It is not necessary to also consider the residues at $\z{=}\m_1$ since their vanishing leads merely to the hermitian
conjugated conditions.

Let us proceed to the two--pole situation. The dressing ansatz takes the form ($\Psi^0_1\Psh^0_2=:\Psi^0_2$)
\begin{equation} \label{Psitwo}
\Psi_2 \= \Bigl(\mbf{1} \,+\,\frac{\m_2-\mb_2}{\z-\m_2}\,P_2\Bigr)\Bigl(\mbf{1} \,+\,\frac{\m_1-\mb_1}{\z-\m_1}\,P_1\Bigr) \,\Psi^0_2
\= \Bigl(\mbf{1} \,+\,\frac{\Lambda_{21}S_1^\+}{\z-\m_1}\,+\,\frac{\Lambda_{22}S_2^\+}{\z-\m_2}\Bigr) \,\Psi^0_2 \quad,
\end{equation}
where $P_2$ and $S_2$ are to be determined but $P_1$ and $S_1$ can be copied from above. Indeed, inspecting the residues of (R) and (D) at $\z=\mb_1$ simply confirms that
\begin{equation}
P_1 \= T_1\,(T_1^\+ T_1)^{-1} T_1^\+ \qquad\textrm{and}\qquad T_1\=S_1 \qquad\textrm{with}\qquad S_1 \= \Sh_1(w_1)
\end{equation}
is just carried over from the one--pole solution.  Relations for $P_2$ and $S_2$ arise from 
\begin{align}
\res_{\z=\mb_2} (R) =0 &\quad\Longrightarrow\quad (\mbf{1}{-}P_2)\,P_2\=0\quad\Longrightarrow\quad P_2 \= T_2\,(T_2^\+ T_2)^{-1} T_2^\+ \ ,\\[6pt]
\res_{\z=\mb_2} (R) =0 &\quad\Longrightarrow\quad \Psi_2(\mb_2)\,S_2\Lambda_{22}^\+ \= (\mbf{1}{-}P_2)(1-\sfrac{\m_1-\mb_1}{\m_1-\mb_2}P_1)\,S_2\Lambda_{22}^\+\=0\ ,\label{TnotS}
\end{align}
where the first equation makes use of the multiplicative form of the ansatz~(\ref{Psitwo}) while the second one exploits the additive version. We conclude that $P_2$ is again a hermitian projector (of some
rank~$r_2$) and thus built from an $n{\times}r_2$ matrix function~$T_2$. Furthermore, (\ref{TnotS}) reveals that $T_2$ cannot be identified with $S_2$ this time, but we rather have
\begin{equation} \label{T2fromS2}
T_2 \= \Bigl(1-\frac{\m_1{-}\mb_1}{\m_1{-}\mb_2}\,P_1\Bigr)\,S_2
\end{equation}
instead. Finally, we consider
\begin{equation}
\res_{\z=\mb_2} (D) =0 \qquad\Longrightarrow\qquad \Psi_2(\mb_2)\,\Lb_2\,(S_2\Lambda_{22}^\+) \=0 \qquad\Longrightarrow\qquad \Lb_2\,S_2 \= S_2\,\G_2^{A,B}
\end{equation}
which is solved by
\begin{equation}
S_2(t,x,y) \= \Sh_2(w_2)\,\ex^{\bar{w}_2 \G_2 /(\m_2-\mb_2)} \qquad\textrm{ for any $w_2$-holomorphic $n{\times}r_2$ matrix function $\Sh_2$}
\end{equation}
and $\G_2^A=\G_2^B=:\G_2$. Once more, we are entitled to put $\G_2=0$.  Hence, the second pole factor in (\ref{Psitwo}) is constructed in the same way as the first one, except for the small
complication~(\ref{T2fromS2}). Again, $\Lambda_{21}$ and $\Lambda_{22}$ can be read off the result if needed.

It is now clear how the iteration continues.  After $N$ steps the final result reads
\begin{equation}
\Psi_N \= \biggl\{ \prod_{\ell=0}^{N-1} \Bigl(\mbf{1} \,+\, \frac{\m_{N-\ell}-\mb_{N-\ell}}{\z-\m_{N-\ell}}\,P_{N-\ell} \Bigr)\biggr\}\,\Psi^0_N
\=\biggl\{\mbf{1}\,+\,\sum_{i=1}^N\frac{\Lambda_{Ni}S_i^\+}{\z-\m_i}\biggr\}\,\Psi^0_N \quad,
\end{equation}
featuring hermitian rank $r_i$ projectors~$P_i$ at $i=1,2,\ldots,N$, via
\begin{equation}
P_i \= T_i\,(T_i^\+ T_i)^{-1} T_i^\+ \qquad\textrm{with}\qquad T_i \= \biggl\{ \prod_{\ell=1}^{i-1} \Bigl(\mbf{1} \,-\,\frac{\m_{i-\ell}-\mb_{i-\ell}}{\m_{i-\ell}-\mb_i}\,P_{i-\ell}\Bigr)\biggl\}\,S_i \quad,
\end{equation}
where
\begin{equation}
S_i(t,x,y) \= \Sh_i(w_i)
\end{equation}
for arbitrary $w_i$--holomorphic $n{\times}r_i$ matrix functions $\Sh_i(w_i)$.  The corresponding classical Yang and Leznov fields are
\begin{align} \label{PhiN}
\Phi_N &\= \Psi_N^\+(\z{=}0) \= {\Psi^0_N}^\+\,\prod_{i=1}^N \bigl( \mbf{1}-\rho_i\,P_i \bigr)\qquad\textrm{with}\qquad \rho_i \= 1-\frac{\m_i}{\mb_i} \quad, \\[6pt]
\Ups_N &\=\lim_{\z\to\infty}\z\,\bigl(\Psi_N(\z)\,{\Psi^0_N}^\+-\mbf{1}\bigr) \= \sum_{i=1}^N (\m_i{-}\mb_i)\,P_i \quad. %\sum_{i=1}^N \Lambda_{Ni}\,S_i^\+ \quad.
\end{align}
The solution space constructed here is parametrized (slightly redundantly) by the set $\{\Sh_i\}_1^N$ of matrix--valued holomorphic functions and the pole positions~$\m_i$.  The so--constructed classical
configurations have solitonic character (meaning finite energy) when all these functions are algebraic.

The dressing technique as presented above is well known in the commutative theory; novel is only the realization that it carries over verbatim to the noncommutative situation by simply understanding all
products as star products (and likewise inverses, exponentials, etc.). Of course, it may be technically difficult to $\star$--invert some matrix, but one may always fall back on an expansion in powers
of~$\theta$.
\subsection{Solitons in noncommutative integrable sine--Gordon}
We should now be able to generate $N$--soliton solutions to the noncommutative sine--Gordon equations, say~(\ref{Yrho}), by applying the reduction from $(2+1)$ to $(1+1)$ dimensions (see section \ref{ncsg sg})
to the above strategy for the group~$U(2)$, i.e.~putting $n{=}2$.  In order to find nontrivial solutions, we specify the constant matrix~$\Ecal$ in the ansatz~(\ref{ansatzPsi}) for~$\Psi$ as
\begin{equation}
\Ecal \= \ex^{-\imm\frac{\pi}{4}\s_2} \= \sfrac{1}{\sqrt{2}} \Bigl(\begin{matrix} 1 & -1 \\ 1 & \phantom{-} 1 \end{matrix}\Bigr)
\end{equation}
which obeys the relations $\ \Ecal\s_3=\s_1\,\Ecal\ $ and $\ \Ecal\s_1=-\s_3\,\Ecal$.  Pushing $\Ecal$ beyond~$V$ we can write
\begin{equation} \label{Wdef} 
\Phi(t,x,y) \= W(x)\,\gt(u,v)\,W^\+(x) \qquad\textrm{with}\qquad W(x) \= \ex^{-\imm\a\,x\,\s_3}
\end{equation}
and
\begin{equation} \label{grot}
\gt(u,v) \= \Ecal\,g(u,v)\,\Ecal^\+ \= \Ecal\,\biggl( \begin{matrix} 
g_+ & \ 0 \\[4pt] 0 & \ g_- \end{matrix}\biggr)\,\Ecal^\+\= \sfrac12 \, \biggl(\begin{matrix}g_+{+}g_- & \ \ g_+{-}g_- \\[4pt] g_+{-}g_- & \ \ g_+{+}g_-\end{matrix}\biggr) \quad.
\end{equation}

With hindsight from the commutative case~\cite{faddeev} we choose
\begin{equation}
\Psh^0_i \= \s_3 \quad\forall i \qquad\Longleftrightarrow\qquad \Psi^0_N \=\s_3^N
\end{equation}
(which commutes with $W$) and restrict the poles of~$\Psi$ to the imaginary axis, $\m_i=\imm p_i\ $ with $\ p_i\in\R$.  Therewith, the co--moving coordinates~(\ref{comoving}) become
\begin{equation}
w_i \= x - \imm (p_i\,u - p_i^{-1} v) \ =:\ x - \imm\eta_i(u,v) \quad,
\end{equation}
defining $\eta_i$ as real linear functions of the light--cone coordinates.  Consequentially, from~(\ref{PhiN}) we get $\ \rho_i=2\ $ and find that
\begin{equation}
\gt_N(u,v) \= \s_3^N\,\prod_{i=1}^N \bigl( \mbf{1}-2\,\Pt_i(u,v) \bigr) \qquad\textrm{with}\qquad P_i \= W\,\Pt_i\,W^\+ \quad.
\end{equation}
Repeating the analysis of the previous subsection, one is again led to construct hermitian projectors
\begin{equation}
\Pt_i \= \Tt_i\,(\Tt_i^\+ \Tt_i)^{-1} \Tt_i^\+ \qquad\textrm{with}\qquad \Tt_i \= \prod_{\ell=1}^{i-1} \Bigl(\mbf{1} \,-\, \frac{2\,p_{i-\ell}}{p_{i-\ell}+p_i}\,\Pt_{i-\ell}\Bigr)\,\St_i \quad,
\end{equation}
where $2{\times}1$ matrix functions $\St_i(u,v)$ are subject to
\begin{equation} \label{Dred}
\Lbt_i\,\St_i \= \St_i\,\widetilde{\G}_i \qquad\textrm{for}\quad  i=1,2,\ldots,N
\end{equation}
and some numbers~$\widetilde\G_i$ (note that now rank $r_i{=}1$) which again we can put to zero.  On functions of the reduced co--moving coordinates~$\eta_i$ alone, 
\begin{equation}
\Lbt_i \= W^\+ \Lb_i W \= (\m_i{-}\mb_i)\, W^\+ \frac{\pa}{\pa\bar{w}_i} W \= p_i\,\Bigl( \frac{\pa}{\pa\eta_i} + \a \,\s_3 \Bigr)
\end{equation}
so that (\ref{Dred}) is solved by
\begin{equation}
\St_i(u,v) \= \widehat{\St}_i(\eta_i) \= \biggl(\begin{matrix}\g_{i1}\ex^{-\a\,\eta_i}\\[4pt] 
\imm\g_{i2}\ex^{+\a\,\eta_i}\end{matrix}\biggr) \=\ex^{-\a\,\eta_i\s_3}\,\biggl(\begin{matrix} \g_{i1} \\[4pt] \imm\g_{i2} \end{matrix}\biggr)\qquad\textrm{with}\quad \g_{i1}, \g_{i2} \in\C \quad.
\end{equation}
Furthermore, it is useful to rewrite
\begin{equation}
\g_{i1}\g_{i2} =: \l_i^2 \quad\textrm{and}\quad \g_{i2}/\g_{i1} =: \g_i^2 \qquad\Longleftrightarrow\qquad
\biggl(\begin{matrix} \g_{i1} \\[4pt] \imm\g_{i2} \end{matrix}\biggr)\=\l_i\, \biggl(\begin{matrix} \g_i^{-1} \\[4pt] \imm\g_i \end{matrix}\biggr)
\end{equation}
because then $|\g_i|$ may be absorbed into $\eta_i$ by shifting $\a\eta_i\mapsto\a\eta_i+\ln|\g_i|$. The multipliers~$\l_i$ drop out in the computation of~$\Pt_i$.  Finally, to make contact with the
form~(\ref{grot}) we restrict the constants $\g_i$ to be real.

Let us check the one--soliton solution, i.e.~put $N{=}1$.  Suppressing the indices momentarily, absorbing $\g$ into~$\eta$ and dropping~$\l$, we infer that
\begin{equation} \label{onesol}
\Tt\=\biggl(\begin{matrix} \ex^{-\a\eta} \\[4pt] \imm\ex^{\a\eta} \end{matrix}\biggr) \quad\Longrightarrow\quad
\Pt\=\frac{1}{2\,\ch 2\a\eta} \biggl(\begin{matrix} \ex^{-2\a\eta} & -\imm \\[4pt] \imm & \ex^{+2\a\eta} \end{matrix}\biggr) \quad\Longrightarrow\quad
\gt\= \Biggl(\begin{matrix} \tanh 2\a\eta & \ \frac{\imm}{\ch 2\a\eta}\\[6pt] \frac{\imm}{\ch 2\a\eta} & \ \tanh 2\a\eta \end{matrix}\Biggr)
\end{equation}
which has $\det\,\gt=1$.  Since here the entire coordinate dependence comes in the single combination~$\eta(u,v)$, all star products trivialize and the one--soliton configuration coincides with the commutative
one. Hence, the field~$\rho$ drops out, $\gt\in SU(2)$, and we find, comparing (\ref{onesol}) with 
(\ref{grot}), that
\begin{equation}
\sfrac12(g_+{+}g_-) \= \cos\sfrac{\varphi}{2} \= \tanh 2\a\eta \qquad\textrm{and}\qquad \sfrac{1}{2\imm}(g_+{-}g_-) \= \sin\sfrac{\varphi}{2} \= \sfrac{1}{\ch 2\a\eta}
\end{equation}
which implies
\begin{equation}
\tan\sfrac{\varphi}{4} \= \ex^{-2\a\eta} \qquad\Longrightarrow\qquad \varphi \= 4\,\arctan \ex^{-2\a\eta} \= -2\,\arcsin (\tanh 2\a\eta) \quad,
\end{equation}
reproducing the well known sine--Gordon soliton with mass $\ m=2\a$.  Its moduli parameters are the velocity $\ \n=\frac{1-p^2}{1+p^2}\ $ and the center of inertia $\ y_0=\frac{1}{\a}\sqrt{1{-}\n^2}\ln|\g|\ $
at zero time~\cite{faddeev}. In passing we note that in the ``$h$~description'' the soliton solution takes the form
\begin{equation}
h_1 \= p\,\tanh 2\a\eta \qquad\textrm{and}\qquad h_2 \= \sfrac{p}{\ch 2\a\eta} \qquad\Longrightarrow\qquad h \= p\,\tanh(\a\eta{+}\sfrac{\imm\pi}{4}) \= p\,\ex^{\frac{\imm}{2}\varphi} \quad.
\end{equation}

Noncommutativity becomes relevant for multi--solitons.  At $N{=}2$, for instance, one has
\begin{equation}
\begin{aligned}
& \gt_2 \= (1-2\Pt_1)\,(1-2\Pt_2) \qquad\textrm{with}\quad \Pt_1 = \Pt \quad\textrm{from (\ref{onesol})}\quad\textrm{and}\quad \Pt_2 \= \Tt_2\,(\Tt_2^\+ \Tt_2)^{-1} \Tt_2^\+ \\[6pt]
& \textrm{where}\qquad \Tt_2 \= \bigl(\mbf{1} - \sfrac{2 p_1}{p_1+p_2} \Pt_1\bigr)\,\widehat{\St}_2 \qquad\textrm{and}\qquad \widehat{\St}_2 \= \ex^{-\a\,\eta_2\s_3}\,
\bigl(\begin{smallmatrix} \g_2^{-1} \\ \imm\g_2 \end{smallmatrix}\bigr) \quad\ \textrm{with}\quad \g_2\in\R \quad.
\end{aligned}
\end{equation}
We refrain from writing down the lengthy explicit expression for~$\gt_2$ in terms of the noncommuting coordinates $\eta_1$ and~$\eta_2$, but one cannot expect to find a unit (star--)determinant for~$\gt_2$ except
in the commutative limit. This underscores the necessity of extending the matrices to~$U(2)$ and the inclusion of a nontrivial $\rho$ at the multi--soliton level.

It is not surprising that the just--constructed noncommutative sine--Gordon solitons themselves descend directly from BPS solutions of the $(2+1)$--dimensional integrable sigma model. Indeed, putting back the
$x$~dependence via~(\ref{Wdef}), the $(2+1)$--dimensional projectors~$P_i$ are built from $2{\times}1$ matrices
\begin{equation}
S_i \= W(x)\,\Sh_i(\eta_i) \= \ex^{-\imm\a\,w_i\s_3}\, \biggl(\begin{matrix} \g_i^{-1} \\[4pt] \imm\g_i \end{matrix}\biggr) \= 
\biggl(\begin{matrix} 1 \\[4pt] \imm\g_i^2\ex^{2\imm\a\,w_i}\end{matrix}\biggr)\, \g_i^{-1}\ex^{-\imm\a\,w_i} \quad.
\end{equation}
In the last expression the right factor drops out on the computation of projectors; the remaining column vector agrees with the standard conventions \cite{ward,LP1,faddeev,bieling}. Reassuringly, the
coordinate dependence has combined into~$w_i$. The ensueing $(2+1)$--dimensional configurations~$\Phi_N$ are nothing but noncommutative multi--plane--waves the simplest examples of which were already
investigated in~\cite{bieling}.
\section{Properties of S--matrix leading to integrability}\label{ncsg smatrix}
In this section we compute tree--level amplitudes for the noncommutative generalization of the sine--Gordon model proposed in section \ref{ncsg sg}, both in the Yang and the Leznov formulation.  In commutative
geometry the sine--Gordon S--matrix factorizes in two--particle processes and no particle production occurs, as a consequence of the existence of an infinite number of conservation laws.  In the
noncommutative case it is interesting to investigate whether the presence of an infinite number of conserved currents is still sufficient to guarantee the integrability of the system in the sense of having a
factorized S--matrix.

A previous noncommutative version of the sine--Gordon model with an infinite set of conserved currents was proposed in \cite{GP}, and its S--matrix was studied in \cite{GMPT}.  Despite the existence of an
infinite chain of conservation laws, it turned out that particle production occurs in this model and that the S--matrix is neither factorized nor causal.\footnote{ Acausal behavior in noncommutative field
theory was first observed in \cite{seiberg} and shown to be related to time--space noncommutativity.} As already stressed in section \ref{ncsg sg}, the noncommutative generalization of the sine--Gordon model
we propose in this chapter differs from the one studied in \cite{GP} in the generalization of the cosine potential. Therefore, both theories describe the dynamics of two real scalar fields, but the structure
of the interaction terms between the two fields is different.  We then expect the scattering amplitudes of the present theory to behave differently from those of the previous one. To this end we will compute
the amplitudes corresponding to $2\to 2$ processes for the fields $\rho$ and $\varphi$ in the $g$--model (Yang formulation) as well as for the fields $h_1$ and $h_2$ in the $h$--model (Leznov formulation).
In the $g$--model we will also compute $2\to 4$ and $3\to 3$ amplitudes for the massive field~$\varphi$. In both models the S--matrix will turn out to be {\em factorized\/} and {\em causal\/} in spite of their
time--space noncommutativity. 
\paragraph{Amplitudes in the ``$g$--model''. Feynman rules}
We parametrize the $g$--model with $(\rho, \varphi)$ as in (\ref{rhophiaction}) since in this parametrization the mass matrix turns out to be diagonal, with zero mass for $\rho$ and $m{=}2\alpha$ for
$\varphi$.  Expanding the action (\ref{rhophiaction}) up to the fourth order in the fields, we read off the following Feynman rules: 
\begin{itemize}
\item{The propagators 
\bea
\parbox{2cm}{\includegraphics[width=1.9cm]{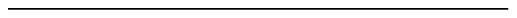}} &\ \equiv\ &\langle \varphi\varphi\rangle\=\frac{2\imm}{k^2-4\a^2}\quad,\\[4pt]
\parbox{2cm}{\includegraphics[width=1.9cm]{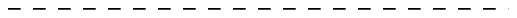}} &\ \equiv\ &\langle \rho\,\rho \rangle\=\frac{2\imm}{k^2}\quad.
\ena}
\item{The vertices (including a factor of ``i'' from the expansion of $\ex^{\imm S}$)
\bea
\parbox{2cm}{\includegraphics[width=1.9cm]{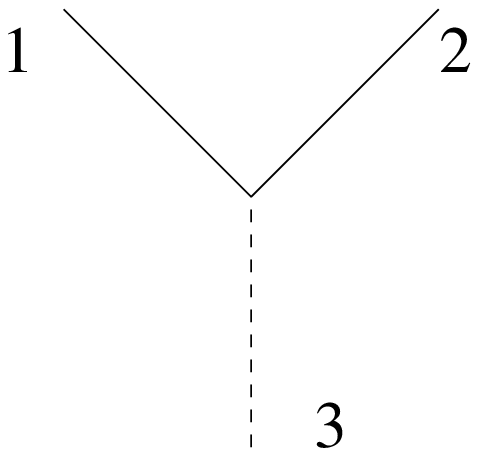}} &\=&-\frac{1}{2^3}(k_{2}^2-k_1^2-2k_1\wedge k_2) F(k_1,k_2,k_3)\quad,\\
\parbox{2cm}{\includegraphics[width=1.9cm]{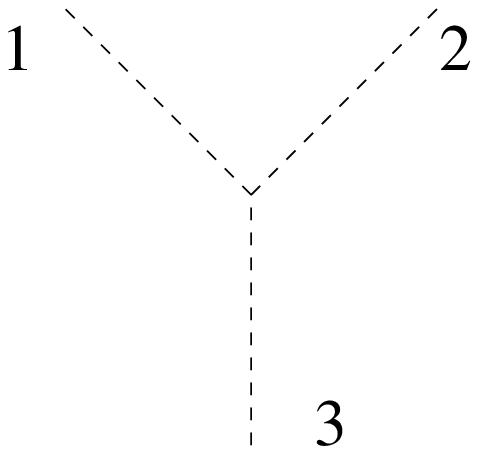}} &\=&\frac{1}{2\cdot 3!}\;k_{1}\wedge k_2\; F(k_1,k_2,k_3)\quad, 
\ena}
\bea
\parbox{2cm}{\includegraphics[width=1.9cm]{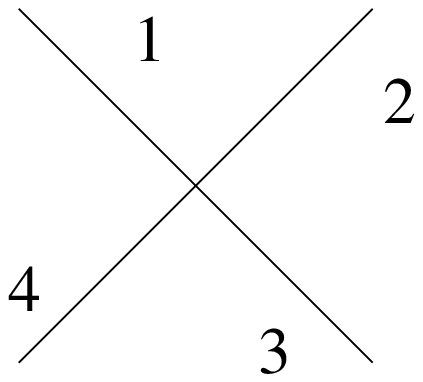}} &\=&\Bigl[-\frac{\imm}{2^3\cdot4!}\,(k_1^2+3k_1\cdot k_3)+\frac{2\imm\a^2}{4!}\Bigr]F(k_1,k_2,k_3,k_4)\quad,\\
\parbox{2cm}{\includegraphics[width=1.9cm]{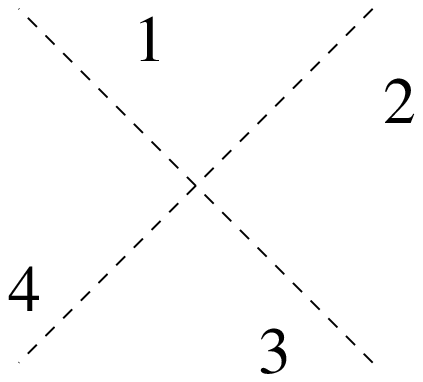}} &\=&-\frac{\imm}{2^3\cdot 4!}\,(k_1^2+3k_1\cdot k_3)\,F(k_1,k_2,k_3,k_4)\quad,\\
\parbox{2cm}{\includegraphics[width=1.9cm]{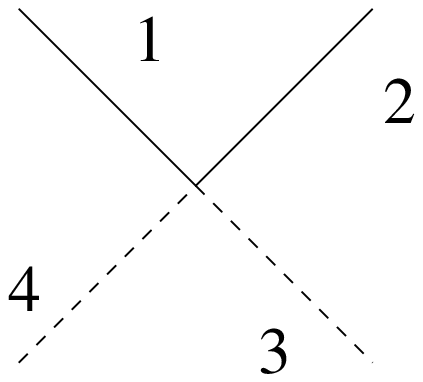}} &\=&-\frac{\imm}{2^5}\,(k_1^2-k_2^2+2 k_1\cdot k_3-2 k_2\cdot k_3 \non\\
&&\qquad\ +\,2k_1\wedge k_2+2k_1\wedge k_3+2k_3\wedge k_2)\,F(k_1,k_2,k_3,k_4) \quad,
\ena
\end{itemize}
where we used the conventions of section \ref{ncsg sigma} with the definitions
\beq
u\cdot v \= - \eta^{ab}\,u_a v_b \= u_tv_t - u_yv_y \qquad\text{and}\qquad u\wedge v\=   u_tv_y-u_yv_t \quad.
\eeq
Moreover, we have defined
\beq
F(k_1,\dots, k_n) \= \exp \bigl\{ -\sfrac{\imm}{2} \textstyle{\sum_{i<j}^n} k_i \wedge k_j \bigr\} \quad.
\eeq
and use the convention that all momentum lines are entering the vertex and energy--momentum conservation has been taken into account. 

We now compute the scattering amplitudes $\varphi\varphi \to \varphi\varphi$, $\rho\rho \to \rho\rho$ and $\varphi\rho\to \varphi\rho$ and the production amplitude $\varphi\varphi \to \rho\rho$.  We perform
the calculations in the center--of--mass frame. We assign the convention that particles with momenta $k_1$ and $k_2$ are incoming, while those with momenta $k_3$ and $k_4$ are outgoing.
\paragraph{Amplitude $\varphi\varphi\to \varphi\varphi$}
The four momenta are explicitly written as \bea\label{momenta aa-aa} k_1=(E,p)\ ,\quad k_2=(E,-p)\ ,\quad k_3=(-E,p)\ ,\quad k_4=(-E,-p)\ , \ena with the on--shell condition $E^2-p^2=4\a^2$.  There are two
topologies of diagrams contributing to this process.  Taking into account the leg permutations corresponding to the same particle at a single vertex, the contributions read\\
\begin{center}\begin{tabular}{r@{}clr@{}cl}
\parbox{1.9cm}{\includegraphics[width=1.9cm]{vertice1111.eps}} &=&$2\imm\a^2 \cos^2 (\theta Ep)\quad,$ & \parbox{2.4cm}{\includegraphics[width=2.4cm]{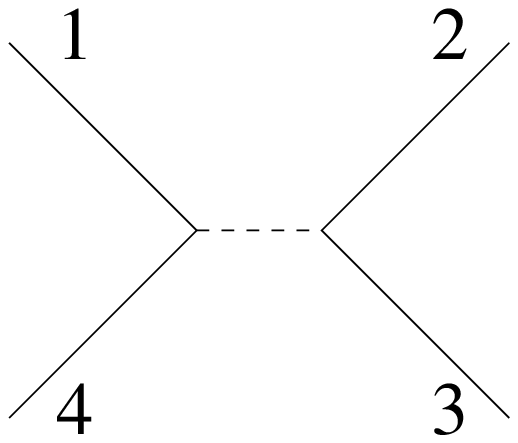}} &=&$0\quad,$\\
\parbox{1.9cm}{\includegraphics[width=1.9cm]{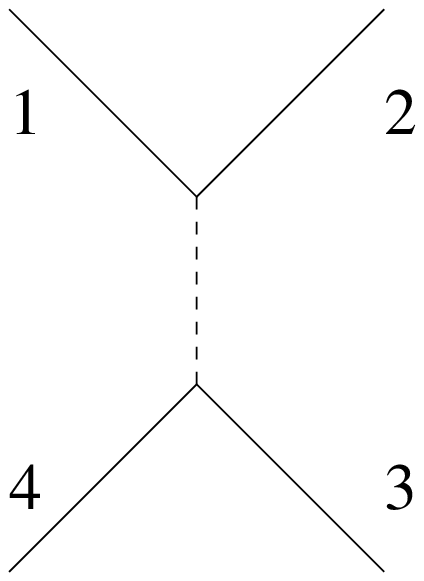}}&=&$-\frac{\imm}{2}p^2\sin^2(\theta Ep)\quad,$ &\parbox{2.4cm}{\includegraphics[width=2.4cm]{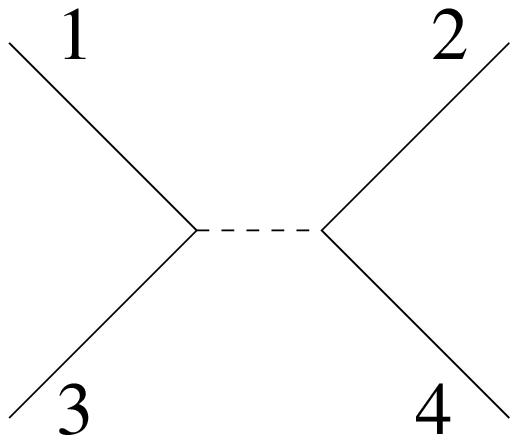}}&=&$ \frac{\imm}{2}E^2\sin^2(\theta Ep)\quad.$\end{tabular}\\ \end{center}
\noindent
The second diagram is actually affected by a collinear divergence since the total momentum $k_1+k_4$ for the internal massless particle is on--shell vanishing.  We regularize this divergence by temporarily
giving a small mass to the $\rho$ particle. It is easy to see that the amplitude is zero for any value of the small mass since the wedge products $k_1 \wedge k_4$ and $k_2 \wedge k_3$ from the two vertices
always vanish.  As an alternative procedure we can put one of the external particles slightly off--shell, so obtaining a finite result which vanishes in the on--shell limit. 

Summing all the contributions, for the $\varphi\varphi \to \varphi\varphi$ amplitude we arrive at 
\beq
A_{\varphi\varphi \to \varphi\varphi}\=2\imm\a^2 \quad,
\eeq
which perfectly describes a {\em causal\/} amplitude.

A non vanishing $\varphi\varphi \to \varphi\varphi$ amplitude appears also in the noncommutative sine--Gordon proposal of \cite{GP,GMPT}. However, there the amplitude has a nontrivial $\theta$--dependence
which is responsible for acausal behavior.  Comparing the present result with the result in \cite{GMPT}, we observe that the same kind of diagrams contribute. The main difference is that the exchanged
particle is now massless instead of massive.  This crucial difference leads to the cancellation of the $\theta$--dependent trigonometric behavior which in the previous case gave rise to acausality.
\paragraph{Amplitude $\rho\rho \to \rho\rho$}
In this case the center--of--mass momenta are given by
\beq
k_1=(E,E)\ ,\quad k_2=(E,-E)\ ,\quad k_3=(-E,E)\ ,\quad k_4=(-E,-E)\ ,
\label{rhomomenta}
\eeq  
where the on--shell condition $E^2-p^2=0$ has already been taken into account.  For this amplitude we have the following contributions\\
\begin{center}\begin{tabular}{r@{}clr@{}cl}
\parbox{2cm}{\includegraphics[width=1.9cm]{vertice2222.eps}} &=& $0\quad,$ & \parbox{2.5cm}{\includegraphics[width=2.4cm]{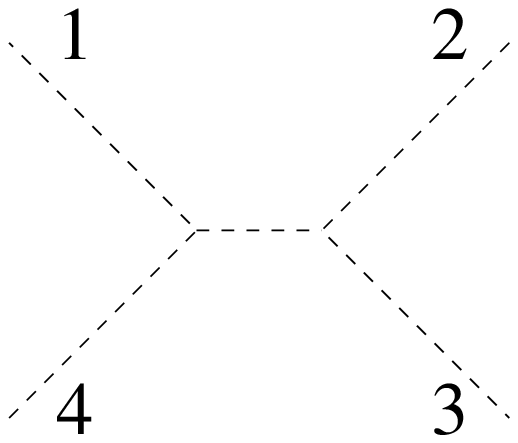}} &=& $0\quad,$ \\
\parbox{2cm}{\includegraphics[width=1.9cm]{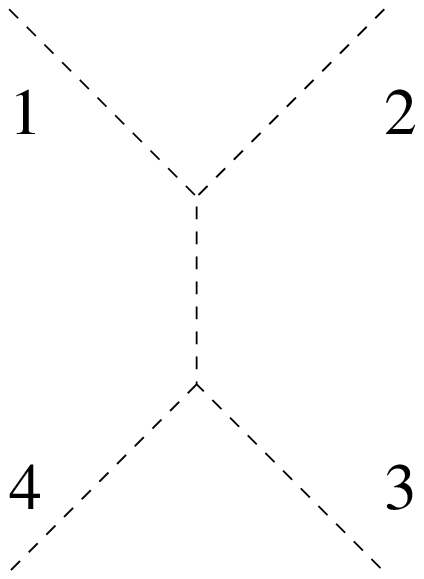}}&=&$-\frac{\imm}{2}E^2\sin^2(\theta E^2)\quad,$ &\parbox{2.5cm}{\includegraphics[width=2.4cm]{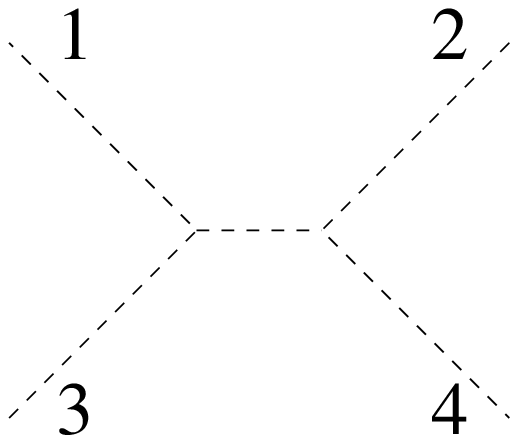}}&=& $\frac{\imm}{2}E^2\sin^2(\theta E^2)\quad.$
\end{tabular}\\ \end{center}
Again, a collinear divergence appears in the second diagram.  In order to regularize the divergence we can proceed as before by assigning a small mass to the $\rho$ particle. The main difference with respect
to the previous case is that now the $\rho$ particle also appears as an external particle, with the consequence that the on--shell momenta in (\ref{rhomomenta}) will get modified by the introduction of a
regulator mass.  A careful calculation shows that the amplitude is zero for any value of the regulator mass, due to the vanishing of the factors $k_1 \wedge k_4$ and $k_2 \wedge k_3$ from the vertices.

Therefore, the two non vanishing contributions add to
\beq
A_{\rho\rho \to \rho\rho}\=0 \quad.
\eeq

\paragraph{Amplitude $\varphi\rho \to \varphi\rho$}
There are two possible configurations of momenta in the center--of--mass frame, describing the scattering of the massive particle with either a left--moving or a right--moving massless one. In the left--moving
case the momenta are
\bea\label{momenta ab-ab 1} 
k_1=(E,p)\ ,\quad  k_2=(p,-p)\ ,\quad  k_3=(-E,p)\ , \quad  k_4=(-p,-p)\ ,
\ena
while in the right--moving case we have
\bea\label{momenta ab-ab 2} 
k_1=(E,-p)\ ,\quad k_2=(p,p)\ ,\quad k_3=(-E,p)\ , \quad  k_4=(-p,-p)\ .
\ena
For the left--moving case (\ref{momenta ab-ab 1}) the results are\\
\begin{center}
\begin{tabular}{r@{}cl@{\hspace{1cm}}r@{}cl} \parbox{2cm}{\includegraphics[width=1.9cm]{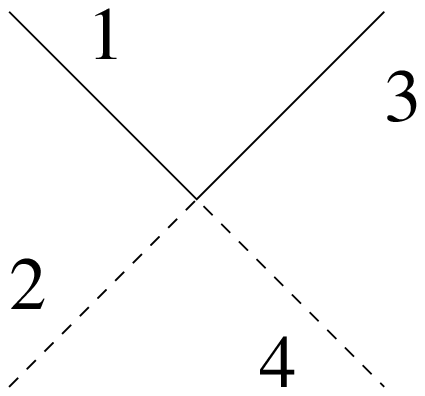}} &=&\multicolumn{4}{l}{$-\frac{\imm}{2}Ep\,\sin(\theta Ep)\,\sin(\theta p^2)\quad,$}\\
\parbox{2cm}{\includegraphics[width=1.9cm]{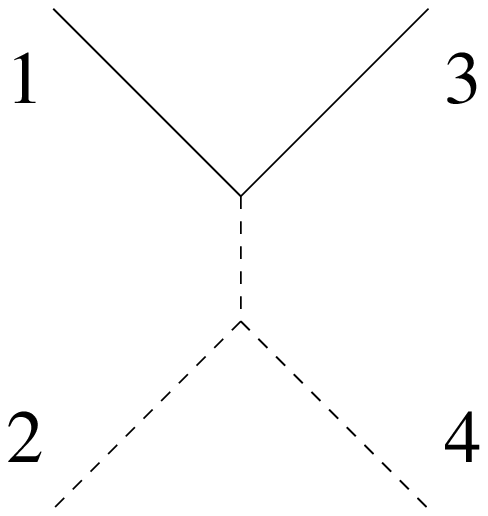}}&=&\multicolumn{4}{l}{$\frac{\imm}{2}Ep\,\sin(\theta Ep)\,\sin(\theta p^2)\quad,$}\\
\parbox{2.5cm}{\includegraphics[width=2.4cm]{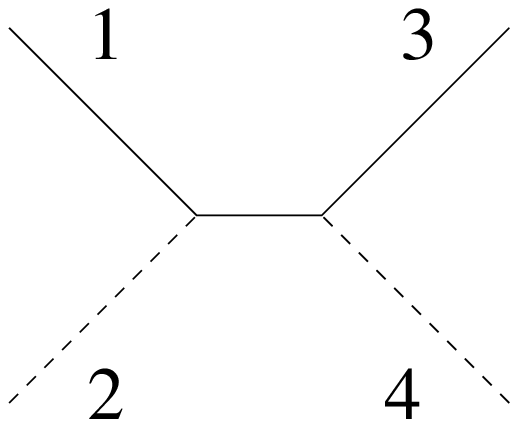}}&=&$0\quad,$ &\parbox{2.5cm}{\includegraphics[width=2.4cm]{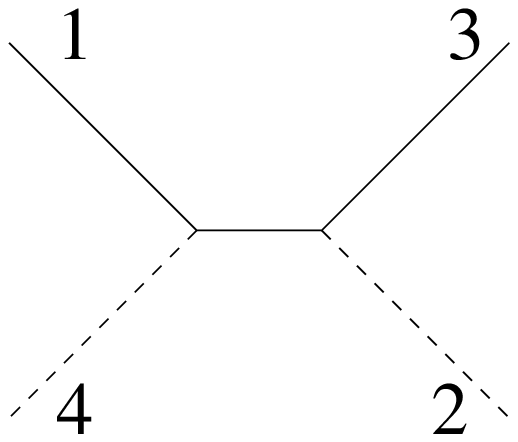}}&=&$0\quad.$\end{tabular}\\
\end{center}
For the right--moving choice (\ref{momenta ab-ab 2}), we obtain instead\\
\begin{center}
\begin{tabular}{r@{}cl@{\hspace{1cm}}r@{}cl}
\parbox{2cm}{\includegraphics[width=1.9cm]{varabab.eps}}&=&$0\quad,$ &\parbox{2cm}{\includegraphics[width=1.9cm]{varaapbb.eps}}&=&$0\quad,$ \\
\parbox{2.5cm}{\includegraphics[width=2.4cm]{varabpab.eps}}&=&$0\quad,$ &\parbox{2.5cm}{\includegraphics[width=2.4cm]{abpab3.eps}}&=&$0\quad.$\end{tabular}\\ 
\end{center}
In this second case an infrared divergence is present due to the massless propagator, but again it can be cured as described before.  In both cases the scattering amplitude vanishes,
\beq
A_{\varphi\rho \to \varphi\rho}\=0 \quad.
\eeq

\paragraph{Amplitude $\varphi\varphi \to \rho\rho$}
The momenta in the center--of--mass frame are given by
\bea\label{momenta aa-bb}
k_1=(E,p)\ ,\quad k_2=(E,-p)\ ,\quad k_3=(-E,E)\ ,\quad k_4=(-E,-E)\ .
\ena 
In this case we have three kinds of diagrams contributing.  The corresponding results are\\
\begin{center}
\begin{tabular}{r@{}cl@{\hspace{1cm}}r@{}cl}
\parbox{2cm}{\includegraphics[width=1.9cm]{vertice1122.eps}}&=& \multicolumn{4}{l}{$\frac{\imm}{2}Ep\,\sin(\theta Ep)\,\sin(\theta E^2)\quad,$} \\
\parbox{2cm}{\includegraphics[width=1.9cm]{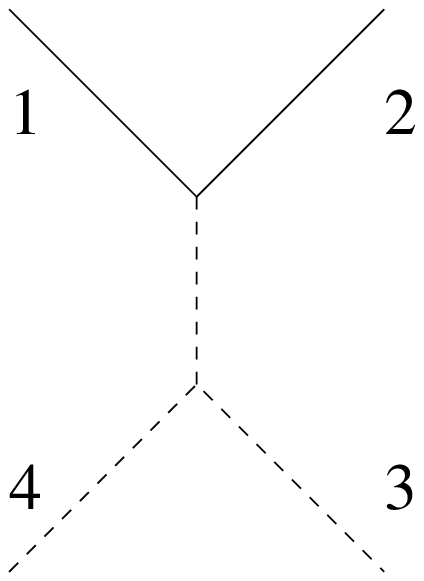}}&=& \multicolumn{4}{l}{$-\frac{\imm}{2}Ep\,\sin(\theta Ep)\,\sin(\theta E^2)\quad,$} \\
\parbox{2.5cm}{\includegraphics[width=2.4cm]{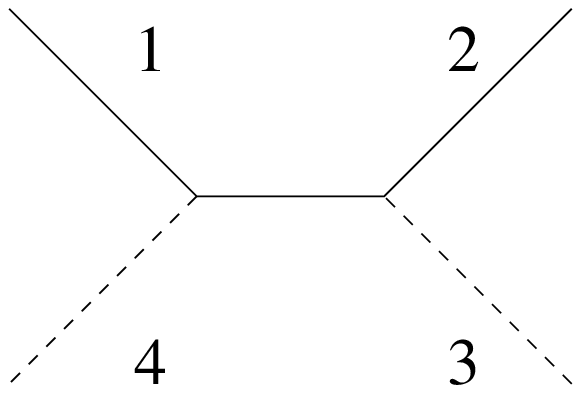}}&=&$0\quad,$ &\parbox{2.5cm}{\includegraphics[width=2.4cm]{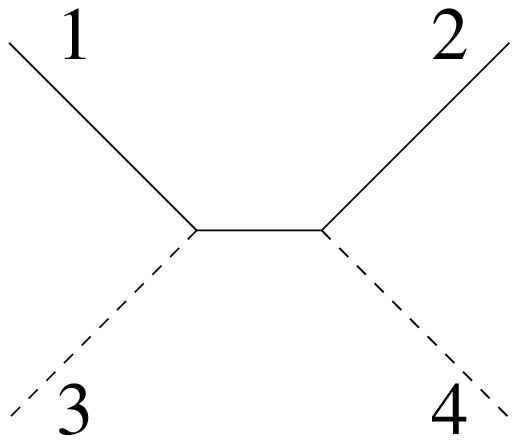}}&=&$0\quad.$\end{tabular}\\
\end{center}
\noindent
Summing the four contributions, we obtain
\beq
A_{\varphi\varphi \to \rho\rho}\=0
\eeq
as it should be expected for a production amplitude in an integrable model.  The same is true for the time--reversed production,
\beq
A_{\rho\rho \to \varphi\varphi}\=0 \quad.
\eeq

Summarizing, we have found that the only nonzero amplitude for tree--level $2 \to 2$ processes is the one describing the scattering among two of the massive excitations.  The result is constant, independent of
the momenta and so describes a perfectly {\em causal\/} process.  Since the result is independent of the noncommutation parameter $\theta$ it agrees with the four--point amplitude for the ordinary sine--Gordon
model.  Finally, we have found that the production amplitudes $\varphi \varphi \to \rho \rho$ and $\rho\rho \to \varphi\varphi$ vanish, as required for ordinary integrable theories.

As a further check of our calculation and an additional test of our model  we have computed the production amplitude $\varphi \varphi \to \varphi \varphi \varphi \varphi$ and the scattering amplitude $\varphi
\varphi \varphi \to \varphi \varphi \varphi$.  In both cases the topologies we have to consider are
\vspace{0.5cm}
\begin{center}
\begin{tabular}{c@{\hspace{0.8cm}}c@{\hspace{0.8cm}}c}
\parbox{2.5cm}{\includegraphics[width=2.4cm]{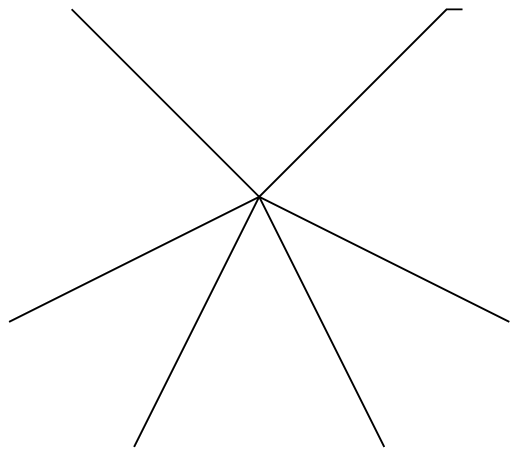}}&\parbox{2.3cm}{\includegraphics[width=2.2cm]{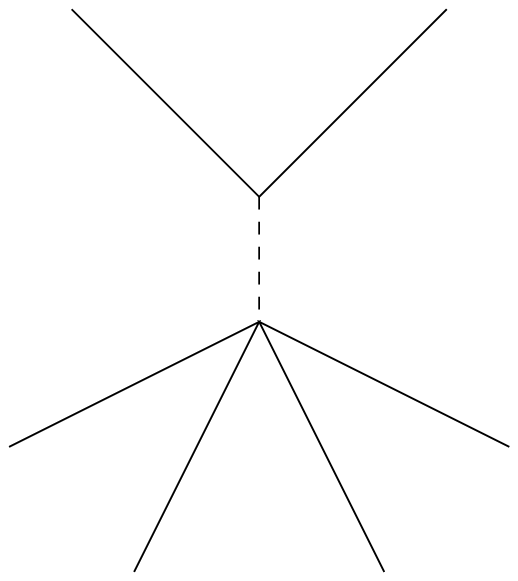}}&\parbox{2.0cm}{\includegraphics[width=1.9cm]{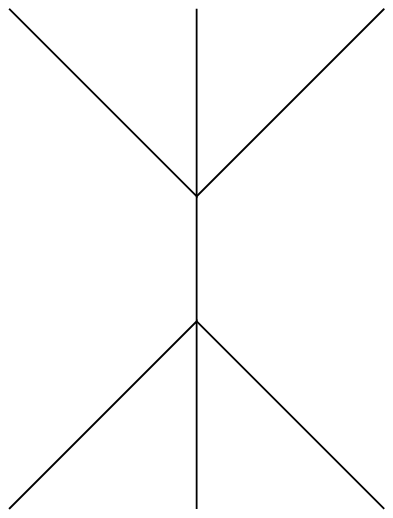}} \end{tabular} \\
\begin{tabular}{c@{\hspace{0.8cm}}c@{\hspace{0.8cm}}c@{\hspace{0.8cm}}c}
\parbox{2.0cm}{\includegraphics[width=1.9cm]{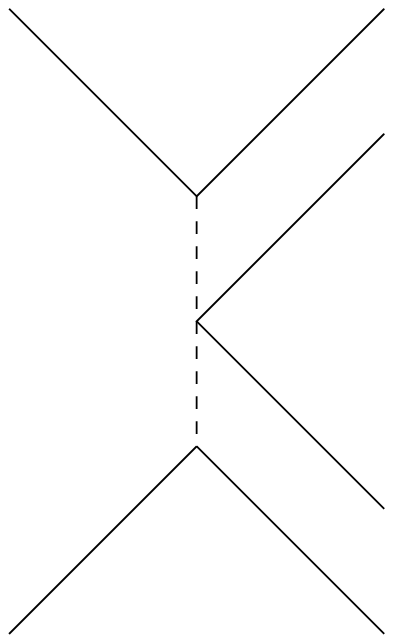}}&\parbox{2.0cm}{\includegraphics[width=1.9cm]{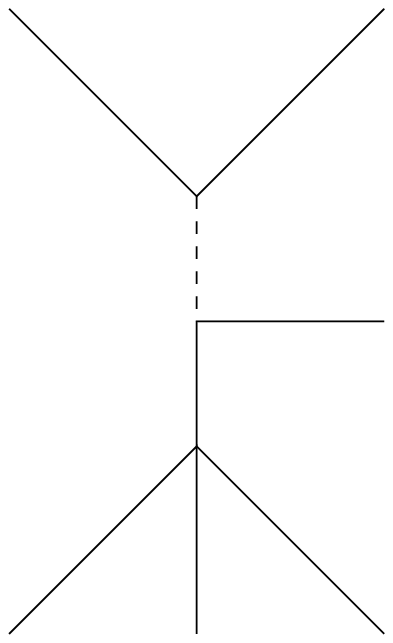}}&\parbox{2.0cm}{\includegraphics[width=1.9cm]{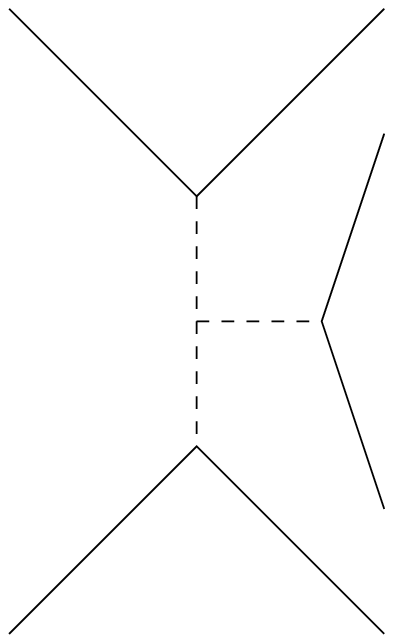}}&
\parbox{1.8cm}{\includegraphics[width=1.7cm]{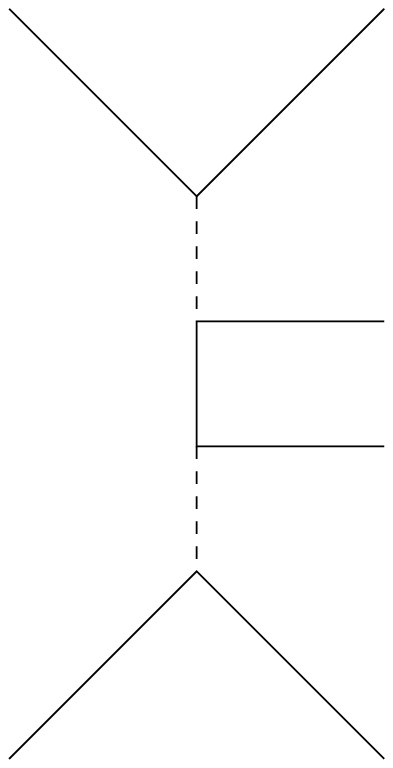}}\quad.\end{tabular}\\ \end{center}
\vspace{0.5cm}
\noindent
Due to the growing number of channels and ordering of vertices, it is no longer practical to perform the calculations by hand.  We have used {\it Mathematica}$^{\copyright}$ to symmetrize the vertices and
take automatically into account the different diagrams obtained by exchanging momenta entering a given vertex.  The computation has been performed with assigned values of the external momenta but arbitrary
values for $\alpha^2$ and $\theta$.  We have found a vanishing result for both the scattering and the production amplitude.  This is in agreement with the commutative sine--Gordon model results.
\paragraph{Amplitudes in the ``$h$--model''}
We now discuss the $2\to 2$ amplitudes in the Leznov formulation.  The theory is again described by two interacting fields, $h_1$ massless and $h_2$ massive. Referring to the action (\ref{haction}) we extract
the following Feynman rules,
\begin{itemize}
\item{The propagators
\bea
\parbox{2cm}{\includegraphics[width=1.9cm]{prop2.eps}}&\ \equiv\ &\langle h_1 h_1 \rangle\=\frac{\imm}{2k^2}\quad,\\
\parbox{2cm}{\includegraphics[width=1.9cm]{prop1.eps}}&\ \equiv\ &\langle h_2 h_2 \rangle\=\frac{\imm/2}{k^2-4\a^2}\quad.
\ena}
\item{The vertex
\bea
\parbox{2cm}{\includegraphics[width=1.9cm]{vertice112.eps}}
&\=&- 4\a\,(k_{3t}-k_{3y})\, F(k_1,k_2,k_3)\quad.
\ena}
\end{itemize}
Again, we compute scattering amplitudes in the center--of--mass frame.  Given the particular structure of the vertex, at tree level there is no $h_1 h_1 \to h_1 h_1$ scattering.  To find the $h_2 h_2 \to h_2
h_2$ amplitude we assign the momenta (\ref{momenta aa-aa}) to the external particles.  The contributions are \\
\begin{center}\begin{tabular}{r@{}clr@{}cl}
\parbox{2.3cm}{\includegraphics[width=2.2cm]{aapaa2.eps}}&=& $-16\imm\a^2\cos^2(\theta Ep)\quad,$ &\parbox{2cm}{\includegraphics[width=1.9cm]{11p11fig.eps}}&=& $16\imm\a^2\cos^2(\theta Ep)\quad,$ \\
\parbox{2.3cm}{\includegraphics[width=2.2cm]{aapaa3.eps}}&=&$0\quad.$\end{tabular}\\ 
\end{center}
\noindent
We note that a collinear divergence appears in the last diagram which can be regularized as described before.  Summing the two non vanishing contributions we obtain complete cancellation.

For the $h_2 h_2 \to h_1 h_1$ amplitude the center--of--mass momenta are given in (\ref{momenta aa-bb}).  The only topology contributing to this production amplitude has two channels, yielding \\
\begin{center}\begin{tabular}{r@{}clr@{}cl}
\parbox{2.5cm}{\includegraphics[width=2.4cm]{12p12fig.eps}}&=&$0\quad,$ &\parbox{2.3cm}{\includegraphics[width=2.2cm]{abpab2.eps}}&=& $0\quad,$\end{tabular}\\ 
\end{center}
\noindent 
which are both zero, so giving a vanishing result once more.  The same is true for the $h_1 h_1 \to h_2 h_2$ production process.

Finally, for the $h_1 h_2 \to h_1 h_2$ amplitude, we refer to the center--of--mass momenta defined in (\ref{momenta ab-ab 1}) and (\ref{momenta ab-ab 2}). In both cases the contributions are\\
\begin{center}\begin{tabular}{r@{}clr@{}cl}
\parbox{2.3cm}{\includegraphics[width=2.2cm]{varabpab.eps}}&=&$0\quad,$ &\parbox{2.3cm}{\includegraphics[width=2.2cm]{abpab3.eps}}&=&$0\quad,$\end{tabular}\\ \end{center}
\noindent 
and so we find that the sum of the two channels is always equal to zero.

Since all the $2 \to 2$ amplitudes vanish, the S--matrix is trivially causal and factorized.

Both in the ordinary and noncommutative cases the ``$h$--model''  is dual to the ``$g$--model''. In the commutative limit the ``$g$--model'' gives rise to a sine--Gordon model plus a free field which can be
set to zero.  In this limit our amplitudes exactly reproduce the  sine--Gordon amplitudes.  On the other hand, the amplitudes for the ``$h$--model'' all vanish. Therefore, in the commutative limit they do not
reproduce anything immediately recognizable as an ordinary sine--Gordon amplitude. This can be understood by observing that, both in the ordinary and in the noncommutative case, the Leznov formulation is an
alternative description of the sine--Gordon dynamics and obtained from the standard Yang formulation by the {\em nonlocal field redefinition\/} given in~(\ref{nonlocal3}). Therefore, it is expected that the
scattering amplitudes for the elementary excitations, which are different in the two formulations, do not resemble each other.
\section{Remarks on the noncommutative integrable sine--Gordon model}
We have proposed a novel noncommutative sine--Gordon system based on {\em two\/} scalar fields, which seems to retain all advantages of $(1+1)$--dimensional integrable models known from the commutative limit.
The rationale for introducing a second scalar field was provided by deriving the sine--Gordon equations and action through dimensional and algebraic reduction of an integrable $(2+1)$--dimensional sigma model:
In the noncommutative extension of this scheme it is natural to generalize the algebraic reduction of $SU(2)\to U(1)$ to one of $U(2) \to U(1) \times U(1)$.  We gave two Yang--type and one Leznov--type
parametrizations of the coupled system in (\ref{Yphi}), (\ref{Yrho}) and (\ref{Lh12}) and provided the actions for them, including a comparison with previous proposals.  It was then outlined how to explicitly
construct noncommutative sine--Gordon multi--solitons via the dressing method based on the underlying linear system.  We found that the one--soliton configuration agrees with the commutative one but already
the two--soliton solutions gets Moyal deformed.

What is the gain of doubling the field content as compared to the standard sine--Gordon system or its straightforward star deformation?  Usually, time--space noncommutativity adversely affects the causality
and unitarity of the S--matrix (see, e.g.~\cite{CM, GP, GMPT}), even in the presence of an infinite number of local conservation laws. In contrast, the model described here seems to possess an S--matrix which
is {\em causal\/} and {\em factorized\/}, as we checked for all tree--level $2\to 2$ processes both in the Yang and Leznov formulations. Furthermore, we verified the vanishing of some $3\to 3$ scattering
amplitudes and $2\to 4$ production amplitudes thus proving the absence of particle production.

It would be nice to understand what actually drives a system to be integrable in the noncommutative case. A hint in this direction might be that the model proposed in~\cite{GP} has been constructed directly
in two dimensions even if its equations of motion (but {\em not\/} the action) can be obtained by a suitable reduction of a four dimensional system (noncommutative self--dual Yang--Mills). The model proposed
here, instead, originates directly, already at the level of the action, from the reduction of noncommutative self--dual Yang--Mills theory which is known to be integrable and related to the $N{=}2$
string~\cite{LPS}. 

Several directions of future research are suggested by our results.  First, one might hope that our noncommutative two--field sine--Gordon model is equivalent to some two--fermion model via noncommutative
bosonization.  Second, it would be illuminating to derive the exact two--soliton solution and extract its scattering properties, either directly in our model or by reducing wave--like solutions of the
$(2+1)$--dimensional sigma model \cite{bieling,wolf}. Third, there is no obstruction against applying the ideas and techniques presented in \cite{LMPPT} to other $(1+1)$--dimensional noncommutative integrable
systems in order to cure their pathologies as well.

\part{Holography and cosmology}
\chapter{AdS/CFT correspondence and branes}\label{chapter adscft}
Strings emerged in the first part of my thesis because of their relation to noncommutative field theories. However, string theory is a much wider framework at the present stage in theoretical physics,
including the issue of noncommutativity that we previously analyzed. We found out that strings in the low energy limit can describe (either noncommutative or ordinary) gauge theories, which are the
fundamental theories that constitute physics detected by present experiments. Realistic gauge theories, though --- Standard Model, QCD --- are still quite far to be consistently embedded in the string
framework.  On the other hand, one of the main successes of string theory is the feature of representing a theory of quantum gravity, since gravitons are automatically included in the spectrum of strings
excitations. Quantization of (super)string theories is thoroughly studied, in order to investigate its various perturbative and non perturbative aspects. 

Making contact with the observable world implies the need of finding a suitable description for gauge theories and gravity in the string theory context. An early indication going in the direction of seeing
gauge theories as strings was proposed by 't Hooft in the 70s \cite{thooft:1973}. Stringy features arise in the large $N$ limit of $SU(N)$ gauge theories. The $1/N$ expansion, for $N\to\infty$, is a natural
expansion in the four dimensional $SU(N)$ Yang--Mills theories. In fact, there is no other free dimensionless parameter, since the gauge group coupling constant $g_{\rm YM}$ is to be related to the QCD scale
$\L_{QCD}$. More precisely, demanding $\L_{QCD}$ to remain constant in the large $N$ limit, we have to keep the 't Hooft parameter $\l_T\equiv g_{\rm YM}^2N$ fixed. This 't Hooft limit is consistent also if
matter in the adjoint representation is added in the theory lagrangian, provided that asymptotic freedom is preserved. Thus, in general, one can compute correlation functions in the large $N$ limit with
adjoint fields, by drawing Feynman diagrams in a double line notation, associating a single oriented line to the (anti)fundamental index in which the adjoint representation can bi--decompose. The factors $N$
in a particular diagram comes from the $N/\l_T$ corresponding to each vertex, the $\l_T/N$ of propagators and the $N$ for loops. Interestingly, Feynman graphs can be classified by the genus of the surface
they can be drawn on, just as string worldsheets in the perturbative expansion of string theories. I now illustrate this sentence in more detail. 

Each Feynman diagram contains powers of $N$ and $\l_T$ according to the following expression \bea N^{V-E-F}\l-T^{E-V}=N^\chi\l_T^{E-V} \ena $E$ denotes the propagator, or the number of edges in the surface
triangulation representation, $V$ are the vertices, $F$ are the loops, or the number of faces if we put diagrams on surfaces of genus $g=1-(V-E+F)/2$ and Euler character $\chi=V-E+F$. Hence, the perturbative
expansion in $1/N$ with fixed $\l_T$ resembles to the closed oriented strings perturbation theory loop expansion if one identifies the string coupling constant with $1/N$ \bea
\sum_{g=0}^{\infty}N^{2-2g}\sum_{n=0}^{\infty}c_{g,n}\l_T^{n}=\sum_{g=0}^\infty f_g(\l_T) \ena The dominating contribution comes from $N^2$ terms which are associated to Feynman diagrams that can be drawn on
a surface with the topology of a sphere (a genus zero surface, indeed). This is just like in string theory. The analogy can be pushed forward for every gauge invariant field correlation function. Moreover,
when matter in the fundamental representation is included, the corresponding propagators in the Feynman diagrams are drawn as single lines, giving rise to boundaries in the surface language. This indeed
corresponds to counting for surfaces with boundaries in the $1/N$ expansion, as it happens when open oriented strings are considered in addition to closed strings. Changing the gauge group would also imply a
change in the string theory that should be dual to the large $N$ gauge theory. In fact, in $SO(N)$ or $USp(N)$ theories the adjoint representation is no longer identified with the product of fundamental times
anti--fundamental, but just of two fundamentals, making the arrow disappear for the propagators in the diagram picture. For the surfaces on which diagrams are drawn, this imply that they are no longer
orientable. Consequently, the string theory underlying should contain unoriented strings.

The original conjecture by 't Hooft was based on four dimensional QCD on the gauge theory side, but it didn't give a precise form to the string theory dual. Even before the large $N$ evidence and before the
formulation of QCD, strings stemmed from the effort to explain the properties of strong interactions arising from experiments. Thinking hadrons as formed by string flux tubes yields results in agreement with
hadron spectrum and scattering processes, as well as with linear confinement and Regge trajectories. Perturbative hadronic features are now well known to be consistently contained in (perturbative) QCD, while
non perturbative calculations are up to now best evaluated putting QCD on a lattice. On the other hand, as I mentioned, there are strong reasons to look for a possible string description of QCD (or, more
generally, gauge theories), which could hopefully give a useful insight at the non perturbative level, for instance. 

It was suggested by Polyakov that a four dimensional gauge theory could be connected to a non critical string theory in (at least) five dimensions. The fifth direction is necessary in order to take account of
the extra field (Liouville field) appearing in the non critical part of the string worldsheet action, proportional to $(D-10)$ for the superstring (or $(D-26)$ for bosonic strings). The Liouville field
emerges because of the quantum breaking of conformal invariance for the worldsheet theory. Quantizing this theory may lead to a non consistent perturbative expansion, plagued by corrections larger than the
leading terms. However, non critical string theories are also widely considered in the context of the gauge/string duality \cite{Polyakovcave}.

A big step in the direction of finding the stringy origin of gauge theories was performed by Maldacena ten years ago.  The Maldacena conjecture \cite{Maldacena:1997re}, which has been now thoroughly checked
at least for the best known examples \cite{Aharony:1999ti}, gives the desired correspondence between specific gauge theories and critical string theory (or M theory). This duality can also be interpreted as
an avatar of the holographic principle, stating that the number of quantum gravity degrees of freedom in a region should not exceed the boundary area measured in Planck units. In other words, a theory of
quantum gravity should be described by the physics at the boundary, instead than in the whole space--time. Indeed, the gauge/string correspondence formulated by Maldacena illustrates the duality between a
string theory in some background and the lower dimensional gauge theory, that can be thought as living on the boundary of the string background --- more precisely, its degrees of freedom are sources localized
on the boundary for the supergravity theory. Holography in its original formulation was motivated by the Bekenstein bound on the maximal value that entropy can assume referring to some region in space--time.
This value should be determined by the area of the region boundary surface and not by its volume, in order not to violate the second law of thermodynamics. It is interesting to note that Bekenstein--Hawking
formula computation for entropy has represented one of the first concrete indications of the duality relating a specific D--brane (near horizon) solution in string theory (or supergravity in the low energy
limit) to the field theory living on the branes. Since the near horizon geometry of the string background turns out to be some AdS space (times compact spaces) when the field theory on the branes is
conformal, the correspondence is usually called the AdS/CFT. 

The early calculations, preceding the full AdS/CFT correspondence formulation, were performed in the background of the D1-D5 system. On the IIB supergravity side, compactified over a five dimensional space,
the near horizon solution corresponding to this brane configuration yields an $AdS_3\times S^3$ (times a four dimensional internal manifold $M^4$). The full background is a generalization of the
Reissner--Nordstrom black holes, charged under two $U(1)$ groups.  The field theory generated by the IIB open superstrings ending on the intersecting D1-D5 branes is a $(1+1)$--dimensional conformal field
theory, living on the branes intersection. Explicit calculations on the CFT side, corresponding to the entropy of extremal and near extremal black hole --- i.e. counting BPS states --- were indeed shown to be
in agreement with the Bekenstein--Hawking formula. There is an apparent contradiction related to the validity of the calculations in the two descriptions. In fact, supergravity and CFT are valid
approximations in opposite regimes of the string coupling constant.  However, the microstate counting in the field theory does not change even when we go to the CFT strong coupling regime, due to
supersymmetry. 

Another set--up has then been considered, which has been proven to be very useful for testing AdS/CFT duality, and has already appeared in my discussion related to noncommutative gauge theories.  This is the
IIB string theory with $N$ D3--branes, which yields a superconformal four dimensional Yang--Mills theory on the D3's, with $SU(N)$ gauge group. From the gravity solution viewpoint, the 3--branes infer an
$AdS_5\times S^5$ near horizon geometry --- where the radii of the AdS space and of the sphere are determined by $N$ ---, so that the (strongest version of) correspondence states in this case that $SU(N)$
${\cal N}=4$ SYM in four dimensions is equivalent to open string theory in an $AdS_5\times S^5$ background.  This is maybe the best celebrated example of the duality, due to the fact that much is known about
${\cal N}=4$ SYM, allowing to directly test the Maldacena conjecture.

Before going into the details of the concrete formulation of AdS/CFT correspondence in the next section, I would like to make a couple of remarks. First, I would like to stress the fundamental role that
branes play in the correspondence. In fact, gauge fields are the massless excitations of the open strings ending on the branes. This represents in general the main motivation to the brane--world idea, i.e.
the expectation that the observed universe is built out of branes, where the gauge theories that we experience are localized (I will discuss more on brane--worlds in chapter \ref{cosmo braneworld}). Black
branes are also important since they represent the other side of the duality. They are classical extended solutions of supergravity and hence sources for the supergravity fields. Their near horizon geometry
is the background for the string theory that constitutes the stringy side of the duality. As a consequence, the idea that black branes and D--branes should be two faces of the same object is an issue of the
AdS/CFT correspondence, already advocated by Polchinski \cite{Polchibrane}.
%The second observation is (summing on all AAdS spaces?) 
\sectioncount{The duality paradigm}\label{adscft} The general formulation of the Maldacena conjecture in its strongest version is that \emph{the full string theory (M theory) on AdS, with specific boundary
conditions, is dual to the CFT associated to open strings ending on the branes}. 

A powerful argument in favor of the AdS/CFT correspondence comes from the symmetries on the two sides. The isometries of AdS turn out to amount exactly to the superconformal group of the field theory.
Moreover, isometries of the internal space correspond to the R--symmetries in the supersymmetric theory living on the branes. The evaluation of absorption processes have also represented a great hint in the
direction of a proof for the duality. The absorption of particles from infinity translates into excitations in the gauge theory, in the D--branes (or M--branes) picture. Conversely, the tunneling of waves into the near horizon
region of the supergravity solution leads to excitations of the AdS geometry. The exact match of the absorption cross section of massless particles on both sides (in the small energy limit) thus confirms that
states in the conformal theory may be described by the supergravity geometry fluctuations.

The two famous examples of D3 and D1-D5 branes are examined in the first paper where the AdS/CFT duality is explicitly evinced \cite{Maldacena:1997re}. The near horizon geometries are respectively
$AdS_5\times S^5$ and $AdS_3\times S^3$, corresponding to 4D ${\cal N}=4$ SYM and (1+1)-dimensional (4,4) SCFT in the Higgs branch, respectively. Intuitively, the duality, let's say in the more immediate
D3--branes case, is clear when we take the decoupling limit. On the D--brane side, the open string modes described by the gauge theory on the D--branes are decoupled from closed string modes yielding gravity
in the bulk. The same kind of decoupling happens from the supergravity classical solution viewpoint. Namely, in the black brane geometry, there will be massless modes in the bulk and arbitrary modes near to
the horizon, since their energy gets redshifted due to the warped geometry. In the low energy limit, the two sets of modes decouple and bulk massless states describe gravity while it is indeed natural to
argue that near horizon excitations correspond to gauge theory states. This reasoning can be equivalently applied to other string theory configurations of D--branes or to M theory. I didn't make yet any
precise statement on what the low energy limit and the validity range for the two sides of the duality are. The small energy limit is taken sending $\a'\to0$. Then, the effective open string
coupling, due to the $N$ coincident D--branes configuration becomes $g_sN\sim g_{\rm YM}^2N$. In order to trust the perturbative gauge theory we then require $g_sN\ll1$. On the other hand, the supergravity
approximation giving AdS geometry is valid as long as the curvature radius is very large. Given that the radius of AdS is determined by $g_sN$, the right range to trust the supergravity solution is $g_sN\gg1$,
which is opposite to the gauge theory validity region. I will be more precise in discussing the specific examples in section \ref{D3 and M5}.

M theory configurations were also analyzed in \cite{Maldacena:1997re}, exemplifying the correspondence between the SCFT on the (eventually wrapped) M--branes and the AdS near horizon geometries arising as
11-dimensional supergravity solutions. It is particularly of my interest to cite the M5 case, which yields the duality between a (0,2) SCFT in six dimensions and the $AdS_7\times S^4$ supergravity background
and which will be shortly reviewed in subsection \ref{D3 and M5}. My work \cite{Mazzanti:2007dq} is indeed focused on a seven dimensional gravity set--up and its holographic dual in six dimensions.  The other
M theory dualities proposed in \cite{Maldacena:1997re} refer to M2--branes and M5--branes wrapped on four cycles of the six dimensional compactification manifold. The field theories are respectively
(2+1)--dimensional SCFT and 2--dimensional (0,4) SCFT, corresponding to M theory on $AdS_4\times S^7$ and $AdS_3\times S^2$ (times the compactification manifold $M^6$), respectively. A very large class of new
examples were consequently studied in the literature, including deformations of conformal field theories corresponding to changes in the internal manifold, RG flows which can be described as perturbations of
the AdS backgrounds (non--CFT's on the branes field theory side imply non--AdS gravity solutions), finite temperature theories. Furthermore, the interest in going towards a QCD gravity dual has produced much
work in more recent times.
\subsection{String theory fields vs. field theory operators}\label{field operator section}
We now need a general and precise prescription expressing the duality for physical quantities on the two sides. This has been formulated in \cite{Witten:1998qj,Gubser:1998bc} providing the identification of
the generating functional of correlation functions in the field theory to the string partition function with fixed boundary conditions on the supergravity (string) side. The field/operator identification can
be conceptually explained in the following way. Concerning the gravity description, waves tunneling in the throat region with AdS metric look like the effects of insertions at the boundary of AdS. Hence,
correlation functions in the gauge theory living on the boundary of AdS can be viewed as measuring the classical supergravity action as a function of the boundary conditions determined by the value of the
gauge field operator source. From a different point of view, adding source coupling terms in the field theory lagrangian amounts to setting the expectation value of the source field, which is in turn
determined by the boundary condition in the supergravity geometry. In formulae, this is expressed by
\bea
\ex^{-W_{\rm CFT}}=\langle\ex^{\int\intd^dx\phi_0(x){\cal O}(x)}\rangle_{\rm CFT}=Z_{\rm string}\left[\Phi(x,r)|_{\pa({\rm AdS})}=\phi_0(x)\right]\label{field operator correspondence}
\ena
The l.h.s. of \refeq{field operator correspondence} is the generating functional of correlation functions in the gauge theory on the boundary. The r.h.s. represents the string partition function with
appropriate boundary conditions, which reduces to the extremum of the supergravity action $\ex^{-I_{\rm sugra}}$ in the classical limit, where no stringy excitations are considered.

The metric of $AdS_{d+1}$ can be easily written in a $u$--frame or $r$--frame as % footnote
\footnote{This actually does not give the full AdS metric, but only the Poincar\'e patch. Indeed, it displays an horizon and analytically continuing beside the horizon recovers the entire AdS.}% footnote
\bea
\intd s^2=\frac{\ell^2}{u^2}\left(\eta_{\m\n}\intd x^\m\intd x^\n+\intd u^2\right)\label{u frame}\\
\intd s^2=\frac{r^2}{\ell^2}\eta_{\m\n}\intd x^\m\intd x^\n+\intd r^2\label{r frame}
\ena
where $u=\ell^2/r$. The boundary is located at $u=0$ or $r=\infty$, respectively, and we get an horizon at $u=\infty$ or $r=0$. The boundary conditions in \refeq{field operator correspondence} can be naively
thought to be determined at the true boundary. However, it will be in some cases necessary to impose an IR (large $r$) cutoff on the AdS space at $u=\epsilon$, which thus plays the role of a UV cutoff for the
CFT, due the the redshift $E_{\rm CFT}=E\ell/u=Er/\ell$ felt by a field theory observer placed at large $r$% footnote
~\footnote{A rigorous approach to the regularization of supergravity action on asymptotically AdS spaces will be reviewed in \ref{holographic renormalization}}.% footnote
This expresses the UV/IR relation inside the correspondence.

In a very general and sketchy way, one can use the correspondence \refeq{field operator correspondence} to infer the duality between supergravity fields of mass $m$ and field theory operators of conformal
dimension $\D$.  On the supergravity side, one can solve the wave equation for the massive scalar field (analytically continuing to Euclidean space). The resulting solution near the boundary $u\to0$, is a
linear combination of two contributions
\bea
\phi(x,u)\sim u^{d-\D}\left[\phi_0(x)+O(u^2)\right]+u^\D\left[A(x)+O(u^2)\right]
\ena
where % footnote
\footnote{The - root $\D=\frac{d}{2}-\sqrt{\frac{d^2}{4}+\ell^2m^2}$ can also be considered \cite{KW:1999}, where roughly the role of the $d-\D$ and $\D$ solutions to the wave equation are inverted in the
following argument.}% footnote
\bea\label{mass dimension correspondence}
\D=\frac{d}{2}+\sqrt{\frac{d^2}{4}+\ell^2m^2}
\ena
The boundary condition at the cutoff $u=\epsilon$ that we can substitute in \refeq{field operator correspondence} as $\epsilon\to0$ is thus
\bea
\phi(x,\epsilon)=\epsilon^{d-\D}\phi_0(x)\label{boundary condition massive scalar}
\ena
We can now deduce the conformal dimension of the associated operator ${\cal O}$ in \refeq{field operator correspondence}, by concluding from \refeq{boundary condition massive scalar} that $\phi_0$ has
dimension $d-\D$, so that ${\cal O}$ gets dimension $\D$. This evaluation can be derived with a more rigorous calculation by computing the two point function $\langle{\cal O}(x_1){\cal O}(x_2)\rangle$ on the
supergravity side, which should then give the  typical CFT result $\langle{\cal O}(x_1){\cal O}(x_2)\rangle\sim|x_1-x_2|^{-2\D}$.
\paragraph{The instructive two point correlation function example}
Let me consider the supergravity action for a massive scalar field of mass $m$ (Wick rotated to Euclidean space)
\bea\label{action massive scalar}
I_{\rm sugra}={1\over 2}\int\intd^{d}x\intd u\,u^{-d+1}\left[\left(\pa_u\phi\right)^2+\left(\pa_i\phi\right)^2+\frac{m^2}{u^2}\phi^2\right]
\ena
We should then evaluate this action on the classical solution to the wave equation, with boundary condition given by \refeq{boundary condition massive scalar}. First of all, the scalar field $\phi$ can be
determined as a function of its boundary value $\phi_0$, by means of the bulk--to--boundary Green function solution to the wave equation, that behaves as a delta function on the boundary
\bea
\left(-u^{d+1}\pa_uu^{-d+1}\pa_u+u^2\pa_i\pa^i+m^2\right)K(u,x,y)=0
\ena 
with
\bea
u^{\D-d}K(u,x,y)\stackrel{u\to0}{\longrightarrow}\d^{(d)}(x-y)
\ena
The Green function solution (for $\D>d/2$% footnote
~\footnote{The lower bound on $\D$ can be taken to be $\D>\frac{d}{2}-1$ \cite{KW:1999} by carefully considering the divergences of the action \refeq{action massive scalar}.}% footnote
) reads
\bea
K(u,x,y)=\frac{\G(\D)}{\p^{\frac{d}{2}}\G\left(\D-\frac{d}{2}\right)}\left(\frac{u}{u^2+|x-y|^2}\right)^\D
\ena
It can be used to determine the scalar field solution with boundary condition \refeq{boundary condition massive scalar}, which indeed yields
\bea\label{phi from K}
\phi(u,x)=\frac{\G(\D)}{\p^{\frac{d}{2}}\G\left(\D-\frac{d}{2}\right)}\int\intd^dy\left(\frac{u}{u^2+|x-y|^2}\right)^\D\phi_0(y)
\ena
Integrating by parts the action \refeq{action massive scalar} imposing the IR cutoff at $u=\epsilon$, one gets a boundary action for the scalar field, since the bulk part vanishes once we substitute the
classical equation of motion. Besides, using the expression for the derivative of $\phi$ w.r.t. $u$ at leading order in the small $u$ expansion from \refeq{phi from K}, one finally finds
\bea
I_{\rm sugra}=\frac{\D\G(\D)}{2\p^{\frac{d}{2}}\G\left(\D-\frac{d}{2}\right)}\int\intd^dx\intd^dy\frac{\phi_0(x)\phi_0(y)}{|x-y|^{2\D}}
\ena
Differentiating this expression twice w.r.t. the source $\phi_0$ yields the two point function, which has the expected dependence $\sim|x-y|^{-2\D}$ of a two point correlation function for an operator of
conformal dimension $\D$. Hence we roughly verified that the operator ${\cal O}$ associated via \refeq{field operator correspondence} to a scalar field of mass $m$, bears conformal dimension $\D$ determined
by \refeq{mass dimension correspondence}.

However, there are some subtleties about how to deal with IR divergences in the supergravity action $I_{\rm sugra}$, corresponding to the UV region in the dual CFT. A systematic approach that takes care of
regularization in the AdS/CFT context is holographic renormalization \cite{Henningson:1998gx,Skenderis:1999nb,deHaro:2000xn} which I will shortly review in section \ref{holographic renormalization}.
\subsection{The duality in practice}\label{D3 and M5}
It is interesting to explore in a more systematic way the most studied example of AdS/CFT correspondence: the ${\cal N}=4$ SYM and $AdS_5\times S^5$ background. Since the (0,2) SCFT on M5, dual to M theory on
$AdS_7\times S^4$ will be the basis of the main topic coming next in chapter \ref{7D RS dual}, i.e. 7D Randall--Sundrum holography, I will also briefly discuss it in the present subsection.
\subsubsection{The celebrated 4D ${\cal N}=4$ SYM example}
Since D3--branes configurations in type IIB string theory infer a four dimensional field theory, it represents one of the possibilities that allows to go towards a description of realistic gauge theories.
Moreover, the ${\cal N}=4$ SYM theory associated to the low energy dynamics of the open strings states is thoroughly studied in literature. For this reason, the D3 set--up is one of the clearest example of
the gauge/string duality.

We start from a stack of $N$ coincident D3--branes in type IIB string theory. At low energy, we must consider the dynamics of massless modes of both closed and open strings. Closed strings massless states are
described by the IIB supergravity in the ten dimensional bulk, since they give a gravity supermultiplet in ten dimensions. Open strings with ends attached on the D3's yield the ${\cal N}=4$ vector
supermultiplet in $(3+1)$ dimensions and the associated theory is indeed the four dimensional $SU(N)$ ${\cal N}=4$ SYM. Higher derivative corrections to the actions both on the branes and in the bulk, as well
as interaction terms between brane and bulk fields, are suppressed in the $\a'\to0$ decoupling limit. Hence, we get on the one hand ten dimensional supergravity and on the other hand ${\cal N}=4$ SYM,
decoupled from each other.

The supergravity dual, as conjectured by Maldacena, is represented by black 3--brane solution in classical supergravity. This is explicitly given by
\bea\label{black 3 brane solution}
\intd s^2&=&\oneover{f^{\oneover{2}}}\eta_{\m\n}\intd x^\m\intd x^\n+f^{\oneover{2}}\left(\intd\rho^2+\rho^2\intd\O_5^2\right)\non\\
F_5&=&(1+*)\intd t\intd x_1\intd x_2\intd x_3\intd f^{-1}\\
f&=&1+\frac{4\p g_s\a'^2N}{\rho^4}
\ena
It represents a solution for a three dimensional source of the supergravity four form, carrying $N$ units of electric charge.  The metric expression shows that energies get redshifted by the warp factor of
the geometry. If we compare the energy $E_{\rm YM}$ measured by an observer at infinity to the energy $E$ seen at a finite distance $\rho$, we obtain
\bea
E_{\rm YM}=\oneover{f^{\oneover{4}}}E
\ena
Hence, states in the near horizon region have vanishing energy as they approach the horizon $\rho=0$. The modes surviving in the low energy limit will thus be massless modes propagating in the bulk and
arbitrary states living close to the horizon. The low energy limit implies a decoupling of bulk and horizon modes, since the absorption cross section relative to the bulk modes becomes very small and near
horizon excitations have as well small probability to fall into the bulk region. In the $\a'\to0$ we thus get supergravity in the bulk and decoupled supergravity in the near horizon geometry determined
by the $\rho\to0$ limit in \refeq{black 3 brane solution}
\bea\label{black 3 brane near horizon}
\intd s^2&=&\frac{\rho^2}{\ell^2}\eta_{\m\n}\intd x^\m\intd x^\n+\frac{\ell^2}{\rho^2}\intd\rho^2+\ell^2\intd\O_5^2\\
\ell^4&=&4\p g_s\a'^2N\label{radius ads5}
\ena
This singles out an $AdS_5\times S^5$ near horizon geometry, namely a throat, with the radii of $AdS_5$ and $S^5$ both equal to $\ell$ \refeq{radius ads5}. In the exact low energy limit we have to keep
$\rho/\a'$ fixed, while $\a'\to0$, in order to have fixed energies $E$ measured in string units $\sqrt{\a'}E$ and fixed energies measured from infinity, which are the energies $E_{\rm YM}\sim\left(\sqrt{\a'}E\right)\left(\rho/\a'\right)$ seen by the dual gauge theory. Defining a new variable $U\equiv \rho/\a'$, the string parameter $\a'$ disappears from the background solution
\refeq{black 3 brane near horizon} except for an overall factor. 

The dynamics in the two low energy descriptions lead to the conjecture that $SU(N)$ ${\cal N}=4$ SYM is equivalent to string theory on $AdS_5\times S^5$, since we can identify the two ten dimensional free
supergravity theories. Now, as I already said, this is the strongest version of the duality. It is an extrapolation of what we can deduce from the above discussion. Indeed, the two descriptions we used holds
in different regimes for the effective string coupling $g_sN$, supposing that $g_s<1$ --- $g_s>1$ can be considered by performing an S--duality $g_s\to1/g_s$. Perturbative calculations in ${\cal N}=4$ SYM can
be performed only if the gauge coupling is small. We get that
\bea
\frac{4\p\imm}{g_{\rm YM}^2}+\frac{\theta}{2\p}=\frac{\imm}{g_s}+\frac{\chi}{2\p}
\ena
where $\theta$ is the angle associated to the topological part of the SYM action and $\chi$ is the expectation value of the RR scalar field. However, the effective gauge coupling that enters into the loop
expansion of the conformal field theory gets an additional factor $N$, so that perturbation theory is valid if
\bea
g_{
m YM}^2N\sim g_sN\ll1
\ena
On the contrary, we know that supergravity approximation of string theory can be trusted when the curvature of the geometry is much bigger than the string length. Hence, the radius of $AdS_5$ in units of
$\ell_s\sim\sqrt{\a'}$ must satisfy
\bea
\frac{\ell}{\ell_s}\sim g_sN\gg1
\ena
and $N\gg1$, since $g_s<1$. The correspondence between the two sides of the AdS/CFT duality relates theories being in opposite coupling regimes. When one side is weakly coupled, the other is strongly coupled
and viceversa. It is a common feature in gauge/string dualities and this makes the full correspondence highly non trivial.

I recall that an intuitive piece of evidence in favor of the correspondence is illustrated by the symmetries on the two sides. The superconformal algebra with 32 supercharges yields an $SO(4,2)$ conformal
group for the bosonic generators, which is exactly the same as the group of isometries in an $AdS_5$ background. Furthermore, ${\cal N}=4$ SYM fields carry R charges under the $SU(4)$ R--symmetry. This
corresponds to the isometries in the internal $S^5$, which is invariant under $SO(6)\sim SU(4)$ rotations.
\paragraph{The AdS/CFT tested}
Some of the most powerful tests of the AdS/CFT correspondence, relative to the case that I am here considering, are related to the matching of the supergravity and susy field theory spectra, as well as to the
CFT correlation functions calculated by means of the string partition function, and to the various possible deformations of ${\cal N}=4$ SYM corresponding to backgrounds different from $AdS_5\times S^5$.

I showed above that two point functions derived using the field/operator correspondence agree with the conformal field theory results. The matching provides a test of the full AdS/CFT for those fields that
don't get a renormalization of the conformal dimension $\D$, namely chiral primaries, proving that supergravity calculations indeed reproduce these specific properties of the dual field theory. Three and
four point function computations have also been achieved in some cases, supporting the duality (see \cite{Freedman:1998tz,Aharony:1999ti} and references therein).

The matching of chiral primary operators with the fields of IIB supergravity compactified on $AdS_5\times S^5$ nicely fits into the correspondence. It represents quite a strong test, since chiral primaries
are protected from quantum corrections. The field/theory correspondence is realized via the relation between masses and conformal dimensions, that for scalar fields in $AdS_5$ yields $\ell^2m^2=\D(\D-4)$.
Short chiral primaries are given by operators of the form ${\cal O}_n=\traccia\left(\phi^{i_1}\cdots\phi^{i_n}\right)$, where the indices of scalars are in the symmetric traceless products of the $\mathbf 6$
representations of the $SU(4)$ R--symmetry of the theory.  Each lowest dimension operator has conformal dimension $n$. One can then build the full chiral primary spectrum of higher dimension operators by
acting on the ${\cal O}_n$'s with the supercharge operators $Q$ and momentum operators $P$ of the ${\cal N}=4$ superconformal algebra.  On the other hand, one can also work out the spectrum of IIB
supergravity on $AdS_5\times S^5$ expanding supergravity fields in spherical harmonics, which correspond to the symmetric traceless products of $\mathbf 6$'s. Moreover, states arising from the dimensional
reduction of ten dimensional supergravity have helicities ranging from $-2$ to 2, indicating that they correspond to operators in small representation of the CFT superconformal algebra. Indeed, we find that
the graviton and four form spherical harmonics (their $S^5$ components) are associated to the lowest dimension scalar in the representations of weight $n$. Scalars of dimensions $n+1$ come from the two form
fields with indices in the $S^5$, while the symmetric tensor corresponds to the graviton components along the $AdS_5$. The complex scalar of weight $n+2$ is traced back to the dilaton scalar harmonics. The
matching of the spectra holds only for large $N$, due to the supergravity approximation.

Other interesting non trivial calculations include the Wilson loops, seen as strings stretching between a quark--antiquark pair, and anomalies --- I will devote subsection \ref{holographic conformal anomaly}
to the discussion on holographic conformal anomalies and consequent checks of AdS/CFT.
\subsubsection{About the (0,2) SCFT in six dimensions}
The $AdS_7$/${\rm SCFT}_6$ is an example of the AdS/CFT duality in M theory. In this case, the AdS space is seven dimensional. Hence the dual conformal theory lives in six dimensions and further
compactifications are required to define an effective four dimensional gauge theory, going in a more realistic direction. 

The low energy limit of M theory with $N$ coincident M5--branes yields a (0,2) SCFT on the worldvolume of the branes. We should also have M2--branes in the bulk, since they are electrically charged under the
three form for which M5--branes are magnetic sources. The ${\cal N}=(0,2)$ superconformal algebra in six dimensions has 32 supercharges, as the ${\cal N}=4$ four dimensional algebra. The irreducible massless
representations form a tensor multiplet, including five real scalars, a two form with self--dual field strength and fermions. The ${\rm SCFT}_6$ can be wrapped on two dimensional manifolds providing an
effective low energies $SU(N)$ (supersymmetric) gauge theory description, at energy lower that the inverse dimension of the internal space. 

On the classical supergravity side, the background solution reads
\bea\label{black 5 solution}
\intd s^2&=&\oneover{f^\oneover{3}}\eta_{\m\n}\intd x^\m\intd x^\n+f^{\frac{2}{3}}\left(\intd\rho+\rho^2\intd\O_4^2\right)\non\\
f&=&1+\frac{\p\ell_{Pl}^3N}{\rho^3}
\ena
where the $x^\m$ are the six directions along the 5--brane and the four form flux through the $S^4$ is quantized in units of $N$
\bea
\int_{S^4}F_4&=&N
\ena
As for the black 3--branes, the modes near the horizon $\rho\to0$ decouples from the bulk fields, in the low energy limit. In the near horizon region, the geometry \refeq{black 5 solution}
becomes $AdS_7\times S^4$
\bea
\intd s^2&=&\frac{2\rho^2}{\ell}\eta_{\m\n}\intd x^\m\intd x^\n+\frac{\ell^2}{\rho^2}\intd\rho^2+\frac{\ell^2}{4}\intd\O_4^2\label{black 5 near horizon}\\
\ell&=&2\ell_{Pl}\left(\p N\right)^{\oneover{3}}\label{radius ads7}
\ena
Since the radius $\ell$ of $AdS_7$ \refeq{radius ads7} is different from the radius $\ell/2$ of the four sphere, \refeq{black 5 near horizon} doesn't give an asymptotically flat space--time. 

The Maldacena conjecture in this case states that the ${\cal N}=(0,2)$ SCFT in six dimensions is holographically dual to M theory on $AdS_7\times S^4$ with $N$ units of four form flux on $S^4$. The
supergravity calculations can be trusted as usual when the curvature radius of the geometry is large with respect to the eleven dimensional Planck length $\ell_{Pl}$
\bea
\frac{\ell}{\ell_{Pl}}\sim N^{\frac{1}{3}}\gg1
\ena
A Lorentz invariant lagrangian formulation for the (0,2) SCFT is not as simple as for the ${\cal N}=4$ SYM theory. There are no dimensionless nor dimensionful parameter for this theory, since it represents
non trivial fixed points of the renormalization group in six dimensions. We expect a $1/N$ expansion, due to the $\ell_{Pl}^3/\ell\sim 1/N$ M theory corrections. 

As in the general AdS case, we can identify the $AdS_7$ isometries with the conformal symmetries of the field theory. In fact, these two symmetries are described by the same $SO(6,2)$ group. The (0,2) SCFT
scalar fields also carry R--symmetry charges and transform as the ${\bf 5}$ representation of $Sp(2)\sim SO(5)$. The $SO(5)$ group, on the other hand, is the isometry group of the four sphere in the
supergravity classical solution \refeq{black 5 near horizon}.
\paragraph{Checks and predictions}
The (0,2) SCFT is not well understood. However, a light--cone quantization procedure has been performed in \cite{Aharony:1997an} and it has been shown that the form of correlation functions of chiral
operators are constrained by the superconformal algebra. These results are in agreement with the calculations on the supergravity side, using the field/operator correspondence \refeq{field operator
correspondence} and evaluating the supergravity action on the classical solution to the equations of motion.

Moreover, the spectrum of protected chiral primaries in the $\mathcal N=(0,2)$ CFT is known. The lowest components are given by scalars in the symmetric traceless representations of $SO(5)$. It has been
compared to the spectrum of 11--dimensional supergravity compactified on $AdS_7\times S^4$, making use of the mass/dimension relation $\ell^2m^2=\D(\D-6)$. The matching between spherical harmonics on $S^4$
yielding a tower of KK massive modes with $m^2=4k(k-3)/\ell^2$ and the scalar chiral primary operators of conformal dimensions $\D=2k, k=2,3,\dots$ --- obtained by the light cone quantization --- is exact and
protected from quantum corrections on both sides of the duality. Higher dimension operators correspond to higher spin KK modes.

As in the ${\cal N}=4$ conformal theory in four dimensions, some other checks have be performed, such as the generalization of Wilson loops --- which are gauge invariant Wilson surfaces operators in the (0,2)
SCFT, related to a membrane bounded by the contour surface in the CFT description ---, or deformations of the field theory amounting to orbifolding the internal space. In particular, I will discuss the
conformal anomaly, which has not been evaluated in the (0,2) SCFT, and its computation by means of the AdS/CFT picture, in subsection \ref{holographic conformal anomaly}.
\sectioncount{Holographic renormalization: space--times with boundaries}\label{holographic renormalization}
In section \ref{field operator section} I illustrated the prescription that allows to perform calculations of physical quantities, as for instance CFT operators correlation functions, by means of the
field/operator correspondence \refeq{field operator correspondence}. To this scope, it is necessary to evaluate the on--shell supergravity action, in the supergravity approximated description of
strings on AdS spaces. The dual objects are correlation functions for the CFT operators. I showed that for instance supergravity results for two point function are in perfect agreement with the constraints
implied by conformal invariance, on the CFT side. Nevertheless, both descriptions suffer from divergences. The on--shell supergravity action gets infinite contributions from the large distances region, i.e.
IR divergences arise. On the other hand, the corresponding  gauge theory UV divergences must be subtracted in the CFT renormalization procedure. Hence we need to consistently obtain finite quantities to be
compared via the field/operator holographic correspondence.

In order to renormalize the supergravity action we have to introduce an IR cutoff $\epsilon$ on the space--time background, corresponding to an UV cutoff for the boundary field theory. The divergences will be
cured by adding suitable counterterms to the supergravity action, in order to keep it finite in the limit $\epsilon\to0$. The solutions to the equations of motion determining the value of the on--shell
action are set by the CFT data --- the boundary conditions ---, as it is established by the AdS/CFT duality. Viceversa, given a supergravity background, one can derive expectation values, correlation
functions and other relevant quantities in the gauge theory, from the renormalized supergravity action. 

In this section, I will explain how to construct this renormalized supergravity action following the holographic renormalization procedure \cite{Henningson:1998gx,Skenderis:1999nb,deHaro:2000xn}. I will
mainly focus on the pure gravity case, which is a very illustrative application allowing to calculate the stress--energy tensor of the dual CFT. In particular, I will derive the trace anomaly for conformal
field theories in the AdS/CFT context \cite{Henningson:1998gx}. Conformal anomaly will be fundamental for the analysis of holographic Randall--Sundrum cosmologies (see section \ref{4D RS holo cosmo} for the
4D case and chapter \ref{7D RS dual} four the seven dimensional set--up).
\subsection[Renormalized gravity action and boundary counterterms]{Renormalized gravity action and boundary counterterms in AAdS spaces}
The most useful result for the holographic analysis of the 7D RS model that I achieved (and I am going to discuss next) is conformal anomaly in an arbitrary number of dimensions for a CFT on a curved
asymptotically flat space. To this purpose, we shall disregard all supergravity fields except the metric and a negative cosmological constant $\L$, for the moment.  We consider the Einstein--Hilbert action of
a $(d+1)$--dimensional manifold $\cal M$ and the necessary Gibbons--Hawking term arising as a consequence of the presence of the boundary $\pa\cal M$
\bea \label{HSS action}
S_{\rm{gr}}={1 \over 16 \p G_N^{(d+1)}}\left[\int_{\cal M}\intd^{d+1}x\,\sqrt{-g}\,\left(R\left[g\right]-2\L\right)+\int_{\pa\cal M}\intd^d x\,\sqrt{-\g}\,2K\right],
\ena
Here $G_N^{(d+1)}$ is the $(d+1)$--dimensional Newton constant, $\L$ is the bulk cosmological constant, and $K$ is the trace of the boundary extrinsic curvature. The on--shell effective action is obtained by
evaluating \refeq{HSS action} on the solutions to the equations of motion, i.e. the Einstein equations. In this notation they read
\bea\label{HSS einstein eq}
G_{MN}=\L g_{MN}
\ena
Boundary conditions are associated to CFT data. However, the solutions to \refeq{HSS einstein eq} don't induce a metric on the boundary, but they despite induce a conformal class, which defines the metric on
the boundary up to conformal transformations. The reason for this is that $g_{MN}$ satisfying \refeq{HSS einstein eq} has a second order pole singularity at infinity $r\to\infty$ --- the asymptotics of the
metric is given in the expression \refeq{u frame} for the $u$--frame and \refeq{r frame} for the $r$--frame. Thus, the metric at the boundary $\hg_{(0)\m\n}$ can be defined by means of a function $u$, that
has a single pole and non zero derivative at $\pa\cal M$, $\hg_{(0)}=u^2g|_{\pa\cal M}$. This choice is not unique, since $\hg'_{(0)}=u'^2g|_{\pa\cal M}$, with $u'=u\ex^{\o}$ is still in the same conformal
class as $g_{(0)\m\n}$, determined by the classical solutions to \refeq{HSS einstein eq}. Clearly, the existence of an entire conformal class of solutions is related to the conformal invariance of the gauge
theory on the boundary.

Now, the aim is to evaluate the on--shell gravity action, given a conformal structure $\left[\hg_{(0)}\right]$ at the boundary, in order to compute the stress--energy tensor of the holographically dual CFT.
We can use the variable $\rho=u^2$ to expand the metric in powers of $\rho$, adopting the Fefferman and Graham \cite{Fefferman} coordinate system
\bea\label{HSS coord}
\intd s^2&=&g_{MN}\,\intd x^M\intd x^N=\ell^2\left[{\intd\rho^2\over4\rho^2}+{1\over\rho}\hg_{\m\n}(x,\rho)\intd x^\m\intd x^\n\right]\non\\
\hg(x,\rho)&=&\hg_{(0)}+\dots+\rho^{d/2}\hg_{(d)}+h_{(d)}\rho^{d/2}\log\rho+\dots 
\ena
Here the subscripts $(i)$ denote the number of derivatives contained in the relative term in the expansion. We remark that this metric parametrization reduces to exactly AdS when $g_{(0)\m\n}=\eta_{\m\n}$ and
all higher order terms vanish. Furthermore, the expansion \refeq{HSS coord} defines an asymptotically AdS (AAdS) space since
\bea
R_{MNRS}\left[g\right]=\left(g_{MR}\,g_{NS}-g_{MS}\,g_{NR}\right)+O(\rho)
\ena
The logarithmic term in the expansion \refeq{HSS coord} has to be introduced only when $d$ is even to find an order by order solution to the Einstein equations \refeq{HSS einstein eq}. For odd $d$'s $h_{(d)}$
is null. This property reflects the fact that conformal anomaly is absent in an odd number of dimensions. Indeed, $h_{(d)}$ will be later related to the trace anomaly. Let's proceed to the order by order
solution to \refeq{HSS einstein eq}, plugging the Fefferman--Graham parametrization \refeq{HSS coord}. The equations of motion for the $d$--dimensional metric $\hg_{\m\n}$ read
\bea
\rho\left[2 \hg''-2\hg'\hg^{-1}\hg'+\traccia(\hg ^{-1}\hg')\hg'\right]+{\rm Ric}\left[\hg\right]-(d-2)\hg'-\traccia\left(\hg ^{-1}\hg'\right)\hg&=&0\cr
\nabla_\m\traccia\left(\hg^{-1}\hg'\right)-\nabla^\n\hg_{\m\n}^\prime&=&0\cr
\traccia\left(\hg ^{-1}\hg''\right)-\frac{1}{2}\traccia\left(\hg^{-1}\hg'\hg ^{-1}\hg'\right)&=&0\label{HSS eqn}
\eea
where the prime denotes differentiation w.r.t. $\rho$. The coefficient $h_{(d)\m\n}$ of the logarithmic term in \refeq{HSS coord} must be traceless, in the sense that
$\traccia\left(\hg_{(0)}^{-1}h_{(d)}\right)=0$, and covariantly conserved $\nabla^\m h_{(d)\m\n}=0$, where the covariant derivative is built from $\hg_{\m\n}$. One can then derive the form of $\hg_{(i)}$ in
terms of $\hg_{(0)}$, for $i\neq d$, from the first of equations \refeq{HSS eqn}. The trace of each $\hg_{(i)}$ with odd $i$ vanishes, while for even $i$ can be calculated in terms of lowest order
$\hg_{(j)}$'s by means of the third equation in \refeq{HSS eqn}.  One can also get the expressions for the divergences of the $\hg_{(i)}$'s, which is null for $i$ odd and non trivial for even $i$. In
particular, for $d=2,4$ it gives a conservation equation of the form  
\bea
\nabla^\m\hg_{(d)\m\n}=\nabla^\m A_{(d)\m\n}
\ena
where $A_{(d)\m\n}$ is antisymmetric and determined by the $g_{(i)}$, $i<d$. As a result, we can obtain $\hg_{(d)\m\n}$ up to an integration constant $t_{\m\n}$, which remains undetermined (the trace and
divergence are known from the solution to the Einstein equations). The constant $t_{\m\n}$ is strictly related to the stress--energy tensor expectation value of the dual field theory on the boundary.  The
coefficient $h_{(d)}$ is also determined as a function of $\hg_{(0)},\dots,\hg_{(d)}$, for an even number of dimensions $d$. The explicit results, which can be extracted from \refeq{HSS eqn} by
differentiating w.r.t. $\rho$ $i$ times and then putting $\rho$ to zero, can be found in \cite{deHaro:2000xn}. However, $g_{(d)}$ is not completely determined by the conformal structure on the boundary and the boundary
stress--energy tensor evaluation is necessary to give a full description of the bulk metric expansion.
% results?

We now proceed to the computation of the on--shell gravity action. Since the action suffers from divergences, we regularize it imposing the IR cutoff at $\rho=\epsilon$
\bea\label{HSS regaction}
S_{\rm{gr,reg}}&=&{1\over16\p G_N^{(d+1)}}\left[\int_{\rho\geq\epsilon}\intd^{d+1}x\sqrt{-g}\left(R\left[g\right]-2\L\right)+\int_{\rho=\epsilon}\intd^dx\sqrt{-\g}2K\right]=\\
&=&{\ell^d\over16\p G_N^{(d+1)}}\int\intd^dx\left[\int_\epsilon\intd\rho{d\over\rho^{d/2+1}}\sqrt{-\hg}+\left.{1\over\rho^{d/2}}\left(-2d\sqrt{-\hg}+4\rho\partial_\rho\sqrt{-\hg}\right)
\right|_{\rho=\epsilon}\right] \non
\ena
When we substitute the Fefferman--Graham expansion in \refeq{HSS regaction} we can isolate the divergent part of the gravity action in the $\epsilon\to0$ limit. They arise as $n$--th order poles
$1/\epsilon^n$ whose coefficients can be determined by the coefficients of the metric expansion. In addition, for an even number of dimensions $d$, a logarithmic divergence, inherited from the metric
expansion, appears in the regularized action. The infinities then precisely read
\bea\label{HSS regaction1}
S_{\rm{gr,reg}}={\ell^{d-1}\over16\p G_N^{(d+1)}}\int\intd^dx\sqrt{\hg_{(0)}}\left(\epsilon^{-d/2}a_{(0)}+\dots+\epsilon^{-1}a_{(d-2)}-\log\epsilon\; a_{(d)}\right)+S_{\rm gr,fin}\non\\
\ena
All the coefficients $a_{(n)}$ are covariant, meaning that the pure gravity action can be renormalized in this framework by the subtraction of covariant counterterms 
\bea\label{HSS counteraction}
S_{\rm gr,count}&=&-{\ell^{d-1}\over16\p G_N^{(d+1)}}\int\intd^dx\sqrt{\hg_{(0)}}\left(\epsilon^{-d/2}a_{(0)}+\epsilon^{-d/2+1}a_{(2)}+\dots+\right.\non\\
&&\left.+\epsilon^{-1}a_{(d-2)}-\log\epsilon\; a_{(d)}\right)
\ena
The renormalized action can be written as
\bea
S_{\rm gr,ren}=\lim_{\epsilon\to0}\left(S_{\rm gr,reg}-S_{\rm gr,count}\right)
\ena
The gravity action evaluated on a classical metric solution to the Einstein equation corresponds, via the field/operator duality, to the CFT stress--energy tensor expectation value $\langle T^{\m\n}\rangle$
and to its correlation functions, upon differentiation w.r.t. the boundary metric $g_{(0)\m\n}$ (coupled to $T^{\m\n}$). Hence, this can be used both as a check of the conformal anomalies calculated in known
gauge theories and as a prediction for not yet worked out results --- I refer for instance to the (0,2) SCFT in six dimensions arising in the coincident M5 configuration dual to supergravity on $AdS_7$. 
\subsubsection{Conformal anomaly via holography}\label{holographic conformal anomaly}
It is interesting to remark that there is a straightforward argument leading to the evaluation of the anomaly, in terms of the $d$--th order counterterm. It can directly be shown that the last counterterm in
\refeq{HSS counteraction} is responsible for the anomaly just by using conformal symmetry arguments. In fact, the finite part of the gravity action $S_{\rm gr,fin}$ variation under conformal transformations is
associated to the trace anomaly
\bea\label{anomaly from variation}
\d S_{\rm gr,fin}=-\int_{\pa\cal M}\intd^dx\sqrt{-g_{(0)}}\d\s{\cal A}_{(d)}
\ena
Here $\d\s$ is the parameter for infinitesimal conformal transformations 
\bea\label{HSS infinitesimal conformal}
\d\hg_{(0)\m\n}=2\d\s\hg_{(0)\m\n}
\ena 
and ${\cal A}_{(d)}$ is the CFT conformal anomaly. On the other hand, from expressions \refeq{HSS regaction} and \refeq{HSS regaction1} one can deduce that the regulated action is invariant under
infinitesimal conformal transformations combined with a rescaling of the IR cutoff
\bea
\d\epsilon=2\d\s\,\epsilon
\ena 
Now, since all negative power terms in the $\epsilon$ expansion of the regularized action are individually invariant, the variation of the logarithmic contribution must be equal to \refeq{anomaly from
variation}. In fact, the covariant coefficients $a_{(n)}$ transform according to
\bea
\d a_{(n)}=-n\d\s a_{(n)}
\ena
Hence the variation of terms of the form $\sqrt{-\hg_{(0)}}\epsilon^{-(d+n)/2}a_{(n)}$ is null, while the logarithmic piece appearing for even $d$ yields a shift $\d\left(\log\epsilon\right)=2\d\s$ --- note
that $\sqrt{\hg_{(0)}}a_{(d)}$ is invariant by itself. This discussion demonstrates that
\bea\label{counterterm variation}
\d S_{\rm gr,count}={d\ell^{d-1}\over16\p G_N^{(d+1)}}\int_{\pa\cal M}\intd^dx\sqrt{-g_{(0)}}\d\s\, a_{(d)}
\ena
Finally, by comparing the two results \refeq{anomaly from variation} and \refeq{counterterm variation} and keeping in mind that $S_{\rm gr,reg}=S_{\rm gr,fin}-S_{\rm gr,count}$, we conclude that
\bea\label{HSS anomaly d}
{\cal A}_{(d)}=-{1\over16\p G_N^{(d+1)}}d\ell^{d-1}a_{(d)}
\ena
This is the main result that I will use in the following, in order to study the holographic cosmology evolution in Randall--Sundrum brane--worlds in four \cite{Kiritsis:2005bm} and six \cite{Mazzanti:2007dq}
dimensions (see section \ref{4D RS holo cosmo} and chapter \ref{7D RS dual}). 

Moreover, the full stress--energy tensor can be evaluated with some more involved computations.  If $\g_{\m\n}$ is the induced metric on the boundary of the regularized manifold, we get that the
stress--energy tensor expectation value in the boundary CFT must be given by
\bea\label{HSS tij1}
\langle T_{\m\n}\rangle={2\over\sqrt{-\hg_{(0)}}}{\pa S_{\rm{gr,ren}}\over\pa\hg_{(0)}^{\m\n}}=\lim_{\epsilon\to0}{2\over\sqrt{-\hg(\epsilon)}}{\pa S_{\rm{gr,ren}}\over\pa\hg^{\m\n}(\epsilon)}
=\lim_{\epsilon\to0}\left({1\over\epsilon^{d/2-1}}T_{\m\n}\left[\g\right]\right)
\ena
The last equality comes from the expression for the induced metric $\g_{\m\n}=\hg(\epsilon)/\epsilon$. Hence, we only need to know $T_{\m\n}\left[\g\right]$ up to order $\epsilon^{d/2-1}$. Plugging in
\bea
T_{\m\n}\left[\g\right]=T_{{\rm reg},\m\n}+T_{{\rm count},\m\n}
\ena
all the results determining $g_{(i)}$ $i\neq d$ and $a_{(n)}$ in terms of the boundary condition, i.e. the conformal class represented by $\hg_{(0)\m\n}$, one can derive the explicit expression for the
anomaly corresponding to any bulk solution $g_{MN}$ satisfying to the aforementioned boundary data. The general expressions for the two contributions are given by
\bea\label{HSS regT}
T_{\rm{reg},\m\n}&=&{1\over 8\p G_N^{(d+1)}}\left(K_{\m\n}-K\g_{\m\n}\right)\non\\
&=&-{1\over8\p G_N^{(d+1)}}\left\{-\pa_\epsilon\hg_{\m\n}(\epsilon)+\hg_{\m\n}(\epsilon) \traccia\left[\hg^{-1}(\epsilon)\pa_\epsilon\hg(\epsilon)\right]+{1-d\over\epsilon}\hg_{\m\n}(\epsilon)\right\}
% num ali
\ena
and
\bea\label{HSS countT}
T_{\rm{count},\m\n}&=&-{\ell^{-1}\over8\p G_N^{(d+1)}}\Bigg\{(d-1)\g_{\m\n}+{1\over(d-2)}\left(R_{\m\n}-\oneover{2}R\g_{\m\n}\right)\non\\
&&-{1\over(d-4)(d-2)^2}\bigg[-\nabla^2R_{\m\n}+2R_{\m\rho\n\s} R^{\rho\s}+{(d-2)\over2(d-1)}\nabla_\m\nabla_\n R\non\\
&&-{d\over2(d-1)}RR_{\m\n}-\oneover{2}\g_{\m\n}\left(R_{\rho\s}R^{\rho\s}-{d\over4(d-1)}R^2-{1\over(d-1)}\nabla^2R\right)\bigg]\non\\
&&-T_{(d)\m\n}\log\epsilon\Bigg\}
\ena
with 
\bea\label{T from a variation}
T_{(d)\m\n}=\oneover{\sqrt{-\g}}\frac{\pa}{\pa\g^{\m\n}}\int\intd^dx\sqrt{-\g}a_{(d)}
\ena
Note that the total stress--energy tensor $T_{\m\n}$ is covariantly conserved w.r.t. $\hg_{(0)}$, because $T_{{\rm reg},\m\n}$ and $T_{{\rm count},\m\n}$ are both conserved w.r.t. $\g_{\m\n}$ at the regulated
boundary. Furthermore, the expectation value of the total stress--energy tensor is given in terms of the undetermined integration constant $t_{\m\n}$. Since $\traccia\,t_{\m\n}$ is given as a function of the
boundary metric, the trace of stress--energy tensor is known too.

Making use of this formalism we can hence elegantly find the stress--energy tensor expressions for arbitrary AAdS spaces (with given boundary conformal class) using the AdS/CFT correspondence. The results
can be compared to known quantities, such as conformal anomalies in two and four dimensions. Namely, following \cite{Henningson:1998gx,Skenderis:1999nb,deHaro:2000xn} one comes to the traces
\bea
{\cal A}_{(2)}&=&\frac{\ell}{16\p G_N^{(3)}}R\\
{\cal A}_{(4)}&=&\frac{\ell^3}{2\p G_N^{(5)}}\left(E_{(4)}+I_{(4)}\right)
\ena
with
\bea
E_{(4)}&=&\frac{1}{64}\left(R^{\m\n\rho\s}R_{\m\n\rho\s}-4r^{\m\n}R_{\m\n}+R^2\right)\non\\
I_{(4)}&=&-\oneover{64}\left(R^{\m\n\rho\s}R_{\m\n\rho\s}-2R^{\m\n}R_{\m\n}+\oneover{3}R^3\right)
\ena
being the Euler density and the local conformal invariant in four dimensions.  The agreement with known CFT results is perfect. I will instead use the newly found expression for 6D conformal anomaly in
subsequent calculations (the form of the anomaly is explicitly written in appendix \ref{anomaly appendix}).

A further note. The equality \refeq{HSS anomaly d} uniquely relates the trace anomaly of CFT on curved AAdS spaces to the coefficient $a_{(d)}$ that can be computed following the holographic renormalization
procedure and the results given in \cite{deHaro:2000xn}. In turn, $a_{(d)}$ can be related to the coefficient of the logarithmic contribution to the metric expansion, written in terms of the Fefferman--Graham
parametrization \refeq{HSS coord}. More precisely, the stress--energy tensor contribution $T_{(d)\m\n}$ to $T_{{\rm count},\m\n}$ \refeq{T from a variation}, associated to the action
$\int\intd^dx\sqrt{-\g}a_{(d)}$, is indeed proportional to $h_{(d)}$. The correct expression is found comparing the logarithmic divergences that appear in the regularized stress--energy tensor computed on
one hand from the result \refeq{HSS regT} and, on the other hand, using \refeq{HSS countT}. The relation reads
\bea
h_{(d)\m\n}=-\frac{2}{d}T_{(d)\m\n}
\ena
Thus, the counterterm coefficient $a_{(d)}$, which determines the anomaly, is associated to a stress--energy tensor proportional to the coefficient $h_{(d)}$ of the logarithmic contribution in the metric
expansion.
% coefficients, signs and notations!
\subsection{Renormalized scalar action}
Introducing matter fields leads to a generalization of the previous procedure. One can consider, for instance, a scalar field $\phi(x,\rho)$ of mass $m$. For $\phi$ an analogous expansion in powers of $\rho$
plus the eventual logarithmic contribution will hold, as in the pure gravity case. Then, the scope is to evaluate the on--shell scalar field action, provided that we previously regularized the theory imposing
the IR cutoff. We will consequently derive the expectation value for the dual operator, which will get conformal dimension $\D=d/2\pm\sqrt{d^2/4+\ell^2m^2}$.

Now, the scalar field action is given by
\bea\label{HSS mataction}
S_{\rm mat}=\oneover{2}\int\intd^{d+1}x\sqrt{-g}\left(g^{MN}\pa_M\phi\pa_N\phi+m^2\phi^2\right)
\ena
Then, the field $\phi$ satisfies the following equations of motion order by order in $\rho$
\bea
\left(-\Box+m^2\right)\phi(x,\rho)&=&\non\\
\left[-(d-\D)\pa_\rho\log\hg\,\hat\phi+2(2\D-d-2)\pa_\rho\hat\phi-\hat\Box\hat\phi\right]+\rho\left[-2\pa_\rho\log\hg\,\pa_\rho\hat\phi-4\pa^2_\rho\hat\phi\right]&=&0% num ali
\ena
The scalar field can be expanded as
\bea
\phi=\rho^{(d-\D)/2}(\hat\phi_{(0)}+\rho\hat\phi_{(2)}+\dots)+\rho^{\D/2}(\phi_{(2\D-d)}+\log\rho\psi_{(2\D-d)}+\dots)
\ena
As in the gravity case, regularizing the action means imposing a cutoff at $\rho=\epsilon$ on the background, in order to compute the correct covariant counterterms. The regularized action reads
\bea
S_{\rm mat,reg}%&=&\half \int_{\r \geq \e} \d^{d+1} x\, \sqrt{G} \left( G^{\m \n} \pa_\m \F \pa_\n \F + m^2 \F^2 \right) \nonu
&=&-\int_{\rho=\epsilon}\intd^dx\,\sqrt{-\hg(\epsilon)}\epsilon^{-\D+d/2}\,\left[\oneover{2}\,(d-\D)\hat\phi^2(x,\epsilon)+\epsilon\,\hat\phi(x,\epsilon)\pa_\epsilon\hat\phi(x,\epsilon)\right]\non\\
&=&\int\intd^dx\,\sqrt{\hg_{(0)}}\,\left[\epsilon^{-\D+d/2}b_{(0)}+\epsilon^{-\D+d/2+1}b_{(2)}+\dots+\right.\non\\
&&\left.+\epsilon\,b_{(2\D-d+2)}-\log\epsilon\,a_{(2\D-d)}\right]+S_{\rm mat,fin}
\ena
One can proceed subtracting the counterterm action $S_{\rm mat,count}=S_{\rm mat,reg}-S_{\rm mat,fin}$, to the regularized action, achieving the holographic renormalization for the scalar field
\bea
S_{\rm mat,ren}&=&\lim_{\epsilon\to0}\Bigg\{S_{\rm mat,reg}-\int\intd^dx\,\sqrt{\hg_{(0)}}\,\bigg[\epsilon^{-\D+d/2}b_{(0)}+\epsilon^{-\D+d/2+1}b_{(2)}+\dots+\non\\
&&+\epsilon\,b_{(2\D-d+2)}-\log\epsilon\,a_{(2\D-d)}\bigg]\Bigg\}
\ena
Solving the equations of motion, leaves $\hat\phi_{(2\D-d)}$ undetermined, in analogy to the gravity case, where $\hg_{(d)}$ is not completely determined neither. Indeed, this coefficient is related to the
one point function of the dual operator $\cal O$ of conformal dimension $\D$
\bea
\langle{\cal O}\rangle=(2\D-d)\hat\phi_{(2\D-d)}+F\left(\hat\phi_{(i)},\psi_{(2\D-d)},\hg_{(j)}\right)\;,\qquad i<2\D-d
\ena
Furthermore, coupling the scalar field to gravity, allows to calculate the Ward identities that modify the conservation equation for the stress--energy tensor. The main result is that
\bea
\nabla^\n\langle T_{\m\n}\rangle=\langle\cal O\rangle\pa_\m\hat{\phi}_{(0)}
\ena
where $\hat\phi_{(0)}$ is the boundary condition CFT data for the scalar field. Explicit computations have been performed
\cite{Henningson:1998gx,Skenderis:1999nb,deHaro:2000xn,Nojiri:1998dh,Papadimitriou:2004ap,Papadimitriou:2004rz}, both for gravity and matter. The results comparable to known
quantities, such as matter conformal anomalies, agree with the CFT calculations. 

\vspace{0.5cm}
An alternative --- equivalent and more geometric --- formalism for holographic renormalization has been illustrated in \cite{Papadimitriou:2004ap}. It is an hamiltonian approach, which makes use of
Gauss--Codacci equations and ADM (Arnowitz--Deser--Misner) formalism to determine the $(d+1)$--dimensional Riemann tensor in terms of the intrinsic and extrinsic curvatures of the boundary hypersurface. This
allows to write the regularized action in such a way that expectation values of CFT operators are conceptually straightforward obtained as conjugated momenta w.r.t. the supergravity corresponding fields. When
renormalization takes place and divergences are covariantly subtracted, the only term in the conjugated momenta expansions that contributes to the one point functions is the one carrying conformal dimension
$\D$ --- where $\D$ is exactly the conformal dimension of the dual operator ($\D=d$ for the stress--energy tensor, for instance). The summarizing formula has the following form
\bea
\langle{\cal O}\rangle=\oneover{\sqrt{-\g}}\p_{(\D)}
\ena
(where $\p=\pa S_{\rm ren}/\pa\phi$ is the conjugated momenta w.r.t. the field $\phi$ associated to the operator $\cal O$). This is the paradigm from which one can get all the results previously reviewed,
both for stress--energy tensor and scalar fields.
\sectioncount{Randall--Sundrum and its holographic interpretation}\label{adscft in RS}
Holographic renormalization is fundamental in the derivation of the AdS/CFT dual to the background identified with a cutoff slice of AdS. I have barely deeply investigated the correspondence relating string
theory (or M theory) on $AdS_{(d+1)}$ to $d$--dimensional CFT, where operator sources are boundary conditions for string theory. The cases I discussed deal with the full $AdS_{(d+1)}$ space (or its Poincar\'e
patch).  Cutting the AdS is equivalent to introducing a UV regulator in the dual CFT, since AdS/CFT is a IR/UV correspondence, as I previously observed. Furthermore, regularized supergravity analysis in
the holographic duality context can be performed following the holographic renormalization prescription of the previous section.

Randall and Sundrum proposed a bulk gravity model in five dimensions that localizes the graviton modes on the four dimensional hypersurface cutting the warped space, i.e. on the 3--brane. I will explain the
features of Randall--Sundrum (RS) model in detail in section \ref{RS model}. The informations relevant to the AdS/CFT analysis, generalized to an arbitrary number of dimensions $(d+1)$, are the action and the
$\Zgr_2$ symmetry with fixed point defined by the location of the $d$--brane. The action is pure gravity in the bulk plus the localized energy source contribution from the brane
\bea
S_{\rm RS}=S_{\rm EH}+S_{\rm GH}+S_{b}
\ena
where
\bea
S_{\rm EH}&=&\oneover{16\p G_N^{(d+1)}}\int\intd^5x\sqrt{-g}\left(R\left[g\right]-2\L\right)\\
S_{\rm GH}&=&\oneover{8\p G_N^{(d+1)}}\int\intd^dx\sqrt{-g}\,K\\
S_{b}&\equiv&S_V+S_{\rm b,m}=\int\intd^4x\sqrt{-\g}\left(-V+\lagra_b\right)
\ena
Here $V$ denotes the tension of the $d$--brane and $\lagra_b$ is a generic matter lagrangian on the brane. The classical solution of the Einstein equations associated to the $S_{\rm RS}$ action variation
leads to a slice of $AdS_{(d+1)}$ with radius $\ell$, assuming the $\Zgr_{2}$ orbifold static ansatz for the metric. The static line element reads
\bea
\intd s^2=\ex^{-|y|/\ell}\intd x^\m\intd x_\m+\intd y^2
\ena
where $\ell$ is related to the bulk cosmological constant by $\L=d(d-1)/2\ell^2$. The solution also implies that the brane tension has to be fine--tuned in order to give zero effective cosmological constant
on the brane.

Now, AdS/CFT tells us that the partition function for gravity on a background with conformal structure $\left[\hg_{(0)}\right]$ at the boundary is equivalent to the generating functional of correlation
functions for a CFT on a space--time with metric $\hg_{(0)}$
\bea\label{field operator gravity}
Z_{\rm gr}\left[\hg_{(0)}\right]\equiv\int\intD g\,\ex^{-S_{\rm gr}\left[g\right]}=\int\intD\phi\,\ex^{-S_{\rm CFT}\left[\phi,\hg_{(0)}\right]}\equiv Z_{\rm CFT}\left[\hg_{(0)}\right]
\ena
This is a straightforward application of the field/operator correspondence \refeq{field operator correspondence}, in the case of pure gravity. However, due to the divergences appearing in \refeq{field
operator gravity}, we must carefully define the gravity action, adding the appropriate counterterms. Obviously, this is done by exploiting the holographic renormalization formalism illustrated in the previous
subsection, yielding
\bea
S_{\rm gr}=S_{\rm EH}+S_{\rm GH}-S_0-S_1+\dots-S_{d/2}
\ena
Here we explicitly write the first two counterterms $S_i,i=0,1$ (in chapter \ref{7D RS dual} the third will also be necessary and its explicit expression will be given there) which are of order $i$ in the
curvature $R=R\left[\g\right]$ ($\g_{\m\n}$ is as usual the induced metric on the boundary $\rho=\epsilon$)
\bea
S_0&=&\frac{(d-1)}{8\p G_N^{(d+1)}\ell}\int_{\rho=\epsilon}\intd^dx\sqrt{-\g}\\
S_1&=&-\frac{\ell}{16\p G_N^{(d+1)}(d-2)}\int_{\rho=\epsilon}\intd^dx\sqrt{-\g}R
\ena
I note that in any dimension, the brane tension action is precisely equal to twice the zeroth order counterterm with opposite sign $S_V=-2S_0$ --- it will be clear that this is due to the RS fine--tuning of
the brane tension to the bulk cosmological constant. 

Provided that \refeq{field operator gravity} holds, we wish to deduce the field/operator correspondence for a RS model
\bea
Z_{\rm RS}\left[\g\right]&=&\int_{{\cal R}_1\cup{\cal R}_2}\intD g\,\intD\phi\,\ex^{-S_{\rm RS}\left[\phi,g\right]}=\non\\
&=&\ex^{-2S_1\left[\g\right]}\int_{{\cal R}_1\cup{\cal R}_2}\intD g\,\intD\phi\,\ex^{-S_{\rm EH}\left[g\right]-S_{\rm GH}\left[g\right]-S_{
m b,m}\left[\phi,\g\right]}
\ena
where ${\cal R}_1$ and ${\cal R}_2$ are the two regions separated by the brane and are symmetric with respect to the $\Zgr_2$ orbifold. Using the expression for the pure gravity action and its dual
derivation, provided that the integral over ${\cal R}_1\cup{\cal R}_2$ can be decomposed into two independent integrals over ${\cal R}\equiv{\cal R}_1$ and ${\cal R}_2$ (which gives equal contributions), we
get
\bea
Z_{\rm RS}\left[\g\right]&=&\ex^{-2W_{\rm CFT}\left[\g\right]-2S_2\left[\g\right]+\dots-2S_{d/2}\left[\g\right]}\int_{\cal R}\intD\phi\,\ex^{-S_{
m b,m}\left[\phi,g\right]}
\ena
Therefore, the dual theory on the boundary corresponding the the RS background reads
\bea
S_{\widetilde{\rm RS}}=S_{\rm CFT}+S_{R}+S_{R^2}+\dots+S_{R^{d/2}}+S_{
m b,m}
\ena
where we have defined
\bea
S_{\rm CFT}\equiv2W_{\rm CFT} \,,\qquad S_{R^i}\equiv2S_i
\ena
Since $S_{R}$ is precisely the Einstein--Hilbert action in $d$ dimensions describing pure gravity on the boundary (with induced metric $\g_{\m\n}$ and Newton's constant $G_N^{(d)}=G_N^{(d+1)}/\ell$), the
AdS/CFT correspondence in the RS context takes the following form: \emph{gravity in a $(d+1)$--dimensional RS background is dual to a renormalized $d$--dimensional CFT plus $d$--dimensional gravity, plus
higher order corrections (and eventually the same matter on the brane)}.

The duality for RS backgrounds follows the discussions in \cite{Hawking:2000kj,Perez-Victoria:2001pa,Arkani-Hamed:2000ds,Nojiri:2000eb} and references therein. Further computations ensue from this application
of AdS/CFT, relating gravity quantities in RS set--up to the CFT coupled to gravity description. Since the holographic dual theory also includes gravity, cosmologies arise on both sides of the correspondence
and can be compared.  I will give a review and new results on the subject in section \ref{4D RS holo cosmo} and chapter \ref{7D RS dual}.

\chapter{Cosmology fundamentals and brane--worlds}\label{cosmo braneworld}
I am going to give an introduction about brane--world models, after an overview of the most salient features in (non)conventional cosmology. Brane--worlds are an attempt to incorporate the description of
conventional cosmology into stringy set--up. Most relevant questions in cosmology are of course the origin of and exit from primordial inflation and reheating. Problems related to current universe
description are above all finding an explanation for the evolution toward the present accelerated era and the cosmological constant problem (eventually associated to dark energy). The composition of the
universe is also a puzzle, since a consistent answer to the origin of dark matter and dark energy is still lacking. Some of these issues may be solved in the brane--worlds models I will consider.
\sectioncount{The conventional scenario of cosmological evolution}
The basic ingredients in the description of an expanding universe are the Friedmann--Robertson--Walker (FRW) metric and the Friedmann equations, coming indeed from Einstein equations in a FRW background. Nice
introductory lectures on Big Bang, inflation, dark energy --- among other issues --- and observations are \cite{Padmanabhan:2006kz}--\cite{Trodden:2004st}.
\subsubsection{Homogeneous and isotropic universe: FRW metric}
Homogeneity and isotropy are the two standard assumptions on the nature of our universe. Tests can be made: isotropy is confirmed by the measurements on the temperature of cosmic microwave background, which
is smooth in all directions. Invoking then the copernican principle which states that our location can't be distinguished as a special place in the universe, one comes to the conclusion that an everywhere
isotropic universe is homogeneous. 

A metric for three homogeneous and isotropic (visible) spatial dimensions plus one time direction for an expanding (or, in principle, contracting) universe is described by the most general ansatz
\bea\label{frwmetric}
ds^2=-\intd t^2+a^2(t)\left[\intd\rho^2+f^2(\rho)\left(\intd\theta^2+\sin^2\theta\,\intd\phi^2\right)\right]
\ena
where $f(\rho)$ labels the three possibilities for flat, hyperbolic and spherical space
\bea
f(\rho)=\left\{
\begin{array}{l@{$\quad\quad$}l}
\sin(\rho)&\rm{sphere}\\
\rho&\rm{plane}\\
\sinh(\rho)&\rm{hyperboloid}
\end{array}\right.\ena
The scale factor $a(t)$ can be used to define the Hubble parameter 
\bea
H(t)\equiv\frac{\dot a(t)}{a(t)}
\ena 
and the acceleration factor 
\bea
q(t)\equiv\frac{\ddot a(t)}{a(t)}
\ena
Furthermore, the time $t$ is a coordinate measured in the comoving frame. We define the conformal time $\t$ through the change of variables $a(\t)\intd\t=\intd t$ as
\begin{equation}\label{conformaltime}
\tau(t)\equiv\int_{t_0}^t \frac{\intd t'}{a(t')}
\end{equation}
The metric \refeq{frwmetric} can be rewritten explicitly parametrizing the spatial curvature by a constant $k=-1,0,+1$ (of mass dimension 2), which is positive, null or negative for locally spherical,
hyperbolic or flat spaces, respectively
\bea\label{frwmetric2}
ds^2=-\intd t^2 +a^2(t)\left[\frac{\intd \rho^2}{1-k\rho^2}+\rho^2\left(\intd\theta^2+\sin^2\theta\,\intd\phi^2\right)\right]
\ena
This is the form of FRW metric that I will mostly use in the following. Changing to conformal time we get
\bea\label{conffrwmetric2}
ds^2= a^2(\tau)\left[-\intd\tau^2+\frac{\intd \rho^2}{1-k\rho^2}+\rho^2\left(\intd\theta^2+\sin^2\theta\,\intd\phi^2\right)\right]
\ena
For flat spaces, $k=0$, the metric differs from Minkowski only by the conformal factor $a(\t)=a(t(\t))$. This is the reason why $\t$ is called the conformal time. In fact, the metric tensor can be written as
$g_{\m\n}=a^2(\t)\eta_{\m\n}$. Another useful parameter is the redshift $z$, defined as
\bea\label{redshift}
\frac{\l_0}{\l_e}\equiv1+z=\frac{a_0}{a(t_e)}
\ena
where 0 and $e$ subscripts denote respectively present and photon emission time. We are supposing that a photon was emitted at a time $t_e$ with wavelength $\l_e$ and is observed today at the time $t$ to have
wavelength $\l_0$. The relation between the two wavelengths is easily derived plugging the metric \refeq{conffrwmetric2} for flat space into the pure YM action describing photons. From \refeq{redshift} one
can consequently find the Hubble law in the form
\bea\label{hubble law}
z=H_0r
\ena
which is strictly valid for small redshifts, i.e. for photons emitted short time ago, $(t_0-t_e)\ll1$. In the small redshift approximation the difference between the emission and present time is equal to the
distance $r$ from the source of emission.
\subsubsection{The horizon}
Another useful element in cosmology is the maximum distance at which an observer can see signals (emitted at the initial time $t=0$). In other words, suppose that a light signal is emitted at zero time (or,
more generally, at whatever lower bound for the time variable) and propagates until a time $t$. At any time $t$ the observer receives signals from sources placed at a distance $r_{\rm{PH}}=r_{\rm PH}(t)$, in
the conformal frame. Hence, $r_{\rm PH}$ is the maximal distance of the observer from the source, such as at a given time $t$ he can probe signals emitted in the asymptotic past. This particle horizon is
precisely the conformal time
\bea
r_{\rm{PH}}=\t=\int_0^t\frac{\intd t'}{a(t')}
\ena
However, the physical distance (in the comoving frame) is given by
\bea
l_{\rm{PH}}=a(t)r_{\rm{PH}}=a(t)\int_0^t\frac{\intd t'}{a(t')}
\ena
Inversely, one can define an event horizon as the maximal distance $r_{\rm EH}=r_{\rm EH}(t)$ from the source at which an observer at the asymptotic future can be located in order to receive signals emitted
at time $t$ that propagate at the speed of light. The region of space--time lying outside the event horizon $r_{\rm EH}$ do not receive any signal from the source. The expression for the event horizon is 
\bea\label{event horizon}
r_{\rm{EH}}=\int_t^\infty\frac{\intd t'}{a(t')}
\ena
where we have supposed the time not to be bounded in the future. More generally one can replace the upper extremum of integration in \refeq{event horizon} with a finite time determining when the space closes.
A singularity $a(t_s)=0$ can for instance play this role. Similarly, the physical event horizon distance reads
\bea
l_{\rm{EH}}=a(t)r_{\rm{EH}}=a(t)\int_t^\infty\frac{\intd t'}{a(t')}
\ena
We note that knowing the scale factor $a(t)$ one could establish the age of the universe by evaluating $l_{\rm{PH}}(t_0)$ at the present time $t_0$.
\subsubsection{Einstein equations in the expanding universe: the Friedmann equations}
The gravitational dynamics in an expanding (homogeneous and isotropic) universe is governed by the Einstein equations evaluated on a FRW background. We have to plug the metric \refeq{frwmetric2} in the well
known equations
\bea\label{einstein}
R_{\m\n}-{1\over 2}Rg_{\m\n} =8\p G_NT_{\m\n}
\ena
Furthermore, one usually parametrizes the stress--energy tensor $T_{\m\n}$ with perfect fluid energy density $\rho$ and pressure $p$ in the following way
\bea\label{perfectfluid}
T_{\m\n} = (\rho+p)U_\m U_\n+pg_{\m\n}
\ena
Here $U_\m$ is the velocity of the perfect fluid which is normalized to $U^\m=(1,0,0,0)$ in the comoving frame. The general expression for the stress--energy tensor components is thus
\bea\label{stress--energy}
T^0_0=-\rho\,,\qquad T^i_j=pg^i_j
\ena
The relation between $\rho$ and $p=p(\rho)$ is the equation of state for the specific contribution to the Einstein equations. It is often approximated to a constant equation of state $p(t)=w\rho(t)$, with 
constant parameter $w$. 

Now, plugging the expression of the metric \refeq{frwmetric2} and the parametrization of the stress--energy tensor \refeq{stress--energy}, Einstein equations \refeq{einstein} become
\bea
\frac{\dot a^2}{a^2}&=&\frac{8\p G_N}{3}\rho-\frac{k}{a^2}  \label{friedmann intro}\\
2\frac{\ddot a}{a}+\frac{\dot a^2}{a^2}&=&-8\p G_N p-\frac{k}{a^2}  \label{evolution intro}
\ena
The first \refeq{friedmann intro} is the Friedmann equation and gives an algebraic relation between $H=\dot a/a$ and $\rho$ (and $k/a^2$ for non flat spaces). The second \refeq{evolution intro} tells us how
the Hubble parameter evolves in the FRW universe, since the l.h.s. is a function of $\dot H$ and $H^2$, while the r.h.s. in a function of $p$ (and again $k/a^2$ for non flat spaces). Putting the two
\refeq{friedmann intro} and \refeq{evolution intro} together we end up with an equation for the acceleration factor $q=\ddot a/a$
\bea
\frac{\ddot a}{a}=-\frac{4\p G_N}{3}\left(\rho+3p\right)
\ena
Of course we can generalize all the equations to the case of more than one contribution to the stress--energy tensor by substituting $\rho$ with $\sum_i\rho_i$ and $p$ with $\sum_ip_i$, with $i$ running over
all the different components. A critical energy density $\rho_c$ can be defined as
\bea\label{rho crit}
\frac{8\p G_N}{3}\rho_c=H^2
\ena
One can determine whether the space is positively or negatively curved, or flat, by means of the energy density evaluation. In fact \refeq{rho crit} defines an energy density for a spatially flat background.
The ratio of the measured density $\rho$ to the critical density $\rho_c$, $\O_{\rm{tot}}=\rho_{\rm{tot}}/\rho_c$, is then related to the spatial local geometry of the universe:
\bea
\O_{\rm tot}>1&\quad\Leftrightarrow\quad&k=+1\non\\
\O_{\rm tot}=1&\quad\Leftrightarrow\quad&k=0\\
\O_{\rm tot}<1&\quad\Leftrightarrow\quad&k=-1\non
\ena
Since at present $\O_{\rm tot}$ is measured to be $\O_{\rm tot}\simeq1$, we can deduce that our universe is approximately flat at the present stage of evolution.

It is well known that requiring energy conservation in General Relativity amounts to asking for a (covariantly) constant stress--energy tensor
\bea
\nabla^\m T_{\m\n}=0
\ena
(this equation can also be seen, in a more geometrical way, as the Bianchi identity for the Einstein tensor).  In turn, this implies the following conservation equation for the stress--energy tensor
parameters
\bea\label{conservation intro}
\dot\rho+3\frac{\dot a}{a}\left(\rho+p\right)=0
\ena
Using the constant equation of state $p=w\rho$, equation \refeq{conservation intro} further simplifies to
\bea\label{sol conservation}
\frac{\dot \rho}{\rho}+3(1+w)\frac{\dot a}{a}=0 \quad\Rightarrow\quad \rho(t)=\rho_0\left(\frac{a_0}{a(t)}\right)^{3(1+w)}
\ena
where $\rho_0$ and $a_0$ are the present values for the energy density and the scale factor, respectively. 

For flat universes, Friedmann equation \refeq{friedmann intro} is easily solved using \refeq{sol conservation}, yielding
\bea\label{scale sol}
a(t)=a_0\left[\frac{3(1+w)}{2}H_0 t\right]^{\frac{2}{3(1+w)}}
\ena
provided that $w>-1$. We furthermore supposed that $\rho_0=\rho_c$ which must be true for flat universes. In the last equality we have used the fact that in a flat universe, dominated by matter species
characterized by fixed $w>-1$, the age of the universe, i.e. $t_0$, can be simply evaluated by integration
\bea
t_0=\int_0^1\frac{\intd a}{a\,H(a)}=\frac{2}{3(1+w)}\oneover{H_0}
\ena
So that $H_0^{-1}\sim t_0$ is the Hubble time estimating the life--time of the universe. The initial value for the scale factor \refeq{scale sol} turns out to be $a\to0$, when $t\to0$. This means that the
space--time hits a singularity at the initial time $t=0$: the Big Bang. 

When $w=-1$ we instead get an exponential solution for the scale factor
\bea
a(t)=a_0\ex^{H_0(t-t_0)}
\ena
The universe exponentially expands in this case. 

An important remark is the dependence of the positive or negative acceleration on the equation of state parameter $w$. Differentiating the solution \refeq{scale sol} twice with respect to time, or
equivalently from equation \refeq{sol conservation}, one deduces that 
\bea\begin{array}{rcl@{\quad\Rightarrow\quad}rcl}
w&>&-1/3 & q&<&0\\
w&<&-1/3 & q&>&0
\end{array}\ena
Thus, the pressure of the perfect fluid must be negative in order to have a positive accelerated expansion of the universe.
\subsubsection{Survey on observations}
Observational data give quite a few pieces of information about what the universe used to look like at early times or at more recent ones. Main experiments are based on the observations of luminous matter
distribution, on measurements of star luminosity and supernovae and of cosmic microwave background (CMB) anisotropies.

First of all, isotropy and homogeneity at large scales can be tested by measuring distances and angular positions of galaxies and quasars. The distribution of luminous matter also shows the obvious
inhomogeneities at short scale: galaxies, stars, etc. Comparing the distribution at short scale with the structure formation simulations gives information about primordial density perturbations and about the
composition of the universe (namely dark matter, baryons, dark energy...). 

A big piece of information comes from CMB anisotropy measurements (recent data are from WMAP). Temperature anisotropies of the microwave background are evaluated. One can estimate as a consequence that the
curvature contribution to the Friedmann equation \refeq{friedmann intro} is very small and that our universe can be considered spatially flat to rather high precision. In other words, today total energy
density $\rho_{\rm tot}$ is equal to $\rho_c$ in good approximation --- or equivalently $\O_{\rm tot}\simeq1$. 

Both from CMB and from observation of abundance of light elements, where neutrons and protons combine, the important ratio $\eta=n_B/n_\g$ relating baryon density number to photons can be computed. The
primordial abundance of light atoms (from supernovae observations) can be traced back to the value of $\eta$ by considering the density number of neutrons at the time when their concentration freezed out. CMB
anisotropy can be connected to the number density of electrons which in turn must be equal to the number density of protons, by neutrality. Further insights on the composition of the universe might come from
the measurement of non baryonic dark matter --- i.e. non--relativistic matter that cannot be seen by direct observations --- in CMB experiments. On the other hand, one can evaluate the energy density of dark
matter from cluster formation simulations, comparing them with the observed distribution of luminous matter.

As a final point of this recapitulation, I would like to mention the feature of cosmological evolution represented by the observed alternating phases of acceleration and deceleration in our universe. From
type 1a supernovae we get that today's universe is accelerating, $q>0$. Supernovae observations give informations about large redshift $z$ (relatively small $z$ describes the late time evolution), when the
linear Hubble law \refeq{hubble law} is no more valid.  For higher redshifts, earlier in time, data imply that the universe underwent decelerated expansion. Late time acceleration has to be related to present
dark energy dominance, as I will explain next.

Absolute luminosity measurements at short distances and very small redshifts $z\ll1$ instead yield the present value for the Hubble constant $H_0=(71\pm3){\rm km/sec\,Mpc}$, using the Hubble law \refeq{hubble
law}.
\subsection{Big Bang! The early Universe}
Hot Big Bang theory describes the evolution of our universe at early times, extrapolating cosmological dynamics from known Standard Model and General Relativity physics. General Relativity tells us that an
homogeneous and isotropic universe should live in a FRW metric \refeq{frwmetric2} and obey to Friedmann equations \refeq{friedmann intro}--\refeq{evolution intro} (plus the conservation equation for the
energy density \refeq{conservation intro}). The point is now to understand what is the universe made of and how we shall describe its different components --- i.e. what are the relevant equations of state for
each species.

We start by considering particles at equilibrium as well as ``non matter'' sources for the stress--energy tensor (vacuum, curvature).
\subsubsection{Equilibrium}
The equation of state can be written as a very simple expression for non--relativistic matter as $p(t)\equiv0$ and for ultra--relativistic matter as $p(t)=\rho(t)/3$. We can thus more generally put the
equation of state in the form $p(t)=w\rho(t)$, where $w$ is a constant equal to $w_{\rm{urm}}=1/3$ for ultra--relativistic particles and $w_{\rm{nr}}=0$ for non--relativistic particles. This comes from the
distribution functions describing particles at equilibrium, i.e. particles whose interaction rate $\G$ is larger than the expansion rate, which is set by $H$. A species freezes, or decouples from the
background plasma, when $\G\ll H$. In words, before the decoupling time, the number density, energy density and pressure of a species at equilibrium is determined by the statistical distribution functions
(with Bose--Einstein or Fermi--Dirac statistics if we consider bosons or fermions, respectively). For ultra--relativistic and non--relativistic approximations the result is the aforementioned relation between
$p$ and $\rho$. 

It is important to note that ultra--relativistic matter, namely photons for instance, are still described by the same distribution function even when they decouple from the plasma. An effective temperature
scaling as
\bea\label{eff temp ultrarel}
T_{\rm ur}=T_{\rm ur,f}\left(\frac{a_{\rm f}}{a}\right)\sim\frac{1}{a}
\ena
must be introduced after photon freeze out.  Therefore, the temperature of the background radiation $T\sim1/a$ labels epochs of our universe, from \emph{recombination}. Indeed, recombination epoch denotes the
time at which protons and electrons formed hydrogen atoms. Soon after, the gas became neutral to photons (this roughly happened at $T\simeq10\,{\rm eV}$). In other words, photons freezed out. 

An analogous property, i.e. preserving the same statistic behavior in or out of equilibrium, also holds for non--relativistic matter. In this case, the distribution is redshifted such as
\bea
T_{\rm nr}=T_{\rm nr,f}\left(\frac{a_{\rm f}}{a}\right)^2\sim\frac{1}{a^2}
\ena
Using \refeq{eff temp ultrarel}, we know from quantum statistics that photons in equilibrium or free photons are characterized by a number density 
\bea
n_\g\sim T^3\sim \oneover{a^3}
\ena
On the other hand, baryonic matter, which is non--relativistic, behaves as
\bea
n_B\sim\oneover{a^3}
\ena
As a consequence, the ratio $\eta=n_B/n_\g$ is constant and is a very important parameter characterizing \emph{Big Bang nucleosynthesis} --- this indicates the epoch starting after the decoupling of neutrons,
when light nuclei formed (at about $T\simeq100 {\rm KeV}$) --- in agreement with experimental data.

What does the evolution for ultra-- and non--relativistic particles looks like in a FRW universe? Straightforward application of equations \refeq{sol conservation} and \refeq{scale sol} gives
\bea\begin{array}{r@{=}l@{\,\quad\Rightarrow\quad\,}r@{\,\sim\,}l@{\,\,,\quad\,}r@{\,\sim\,}l}
w&0 & a(t)&t^{2/3} & \rho(t)&a(t)^{-3}\sim t^{-2}\\
w&\frac{1}{3} & a(t)&t^{1/3} & \rho(t)&a(t)^{-4}\sim t^{-4/3}
\end{array}\ena
Hence radiation was dominating early in time. A transition occurred afterwards, from radiation to dust dominated universe, due to the densities scaling as $a^{-4}$ and $a^{-3}$, respectively. However, even
later in time dark energy gave a larger contribution to the total energy density, since, as we are going to see, it doesn't scale but stays constant.

The vacuum or cosmological equation of state directly follows from the stress--energy tensor associated to such a contribution
\bea
T_{\m\n}=-\frac{\L}{8\p G_N}g_{\m\n}
\ena
This shows that 
\bea
\rho_\L=\frac{\L}{8\p G_N}\,,\qquad p_\L=-\frac{\L}{8\p G_N}
\ena
Hence $w_\L=-1$.  Besides, since $\rho_\L$ is constant in time, this represents a vacuum energy. This is why I will interchange language between cosmological constant and vacuum energy in the context of
equations of state. The equation of state for vacuum energy implies the following evolution in a FRW universe
\bea
w=-1\quad\Rightarrow\quad a(t)\sim\ex^{H_\L t}\,,\quad \rho(t)=\rho_\L={\rm const}
\ena
The energy density and Hubble parameter are thus constant in time, meaning that at late times, when dust and radiation scale as some inverse power of $a$, it becomes the most relevant contribution.

There is another important contribution that we can usefully put in the constant equation of state form $p=w\rho$, with $w=-1/3$. This is the curvature $k$ for any non spatially flat background. In fact, the
curvature term can be isolated in the Einstein equations \refeq{einstein}. It gives a contribution to the stress--energy tensor such that one can define
\bea
\rho_{\rm{curv}}=-\frac{3k}{8\p G_N a^2}\,,\qquad p_{\rm{curv}}=\frac{k}{8\p G_N a^2}
\ena
As a consequence $w_{\rm{curv}}=-1/3$. If our universe was to be curved, we could however deduce that the corresponding contribution to energy density would have been less and less important back in time. In
fact, the curvature term scales as
\bea
w=-\oneover{3}\quad\Rightarrow\quad a(t)\sim t\,,\quad \rho(t)\sim a(t)^{-2}\sim t^{-2}
\ena
Curvature should then be important at times later that the matter dominated eras, but before dark energy dominance. Nonetheless, since today data reveal to good approximation a zero contribution from spatial
curvature, one gets that earlier in time this contribution has been more and more negligible. Hence, flatness of our universe is an expected property in Big Bang nucleosynthesis and earlier epochs.
\subsubsection{Early epochs}
Equilibrium thermodynamics can be applied to early epochs in the universe. Going backwards in time from \emph{recombination} epoch shows the different phases that the universe underwent according to the big
bang model, which gains considerable evidences from observations. At recombination ($T\simeq 10\,{\rm eV}$), the plasma of photons and electrons transformed into a neutral gas, where electrons combined with
protons to form hydrogen atoms transparent to electric radiation. Before recombination ($T\simeq 1\,{\rm MeV}$), neutrons decoupled and stopped being produced by electroweak reactions. Short after ($T\simeq
100\,{\rm KeV}$), light elements started to form --- while heavy elements could only be produced in other contexts, such as supernovae explosions. This is the \emph{nucleosynthesis} epoch. From the primordial
abundance of light elements, one can thus determine $\eta$ in agreement with the results independently derived by CMB observations, as I previously mentioned. At some time before nucleosynthesis ($T\simeq
1\,{\rm MeV}$) another species freezed out, i.e. \emph{neutrino decoupling} occurred. Since neutrinos are involved in the proton--to--neutron reactions, the temperature at which the decoupling happened is
relevant for determining the neutron--proton ratio before nucleosynthesis and, consequently, the light elements abundance. Further back in time, protons annihilated with electrons with the plasma.  Earlier,
also heavier particles such as muons and pions were in equilibrium in the plasma, before decoupling. However, at least two fundamental phase transitions occurred, which are predicted by the Standard Model
theory.  These are the QCD and electroweak transitions. Hadrons were formed from quarks and gluon during the \emph{QCD phase transition} (above $T\simeq 100\,{\rm MeV}$). Electroweak symmetry breaking,
instead, represents the \emph{electroweak phase transition} (above $T\simeq 100\,{\rm GeV}$). Theoretical expectations depend on the particular Standard Model extension one considers. At much earlier times
($T\simeq 10^{16}\,{\rm GeV}$), we may expect \emph{Grand Unification} to be recovered. Such high energies are hard to extrapolate, though.

Following the FRW evolution, we can moreover argue what was roughly the time corresponding to the transition from the matter dominated decelerated era to acceleration. This is simply achieved by applying the
Friedmann equation to a universe composed by dust and dark energy
\bea
\frac{{\dot a}^2}{a^2}=\frac{8\p G_N}{3\rho_c}\left(\O_{\rm nr}\frac{a_0^3}{a^3}+\O_\L\right)
\ena
Hence, we get that the transition occurred when $\ddot a=0$, i.e. when
\bea
\frac{a_0^3}{a^3}\equiv(1+z)^3=\frac{2\O_\L}{\O_{\rm nr}}
\ena
Substituting the experimental values for $\O_\L\simeq0.7$ and $\O_{\rm nr}\simeq0.3$ (I will discuss later in more detail the present composition of the universe, see subsection \ref{present era}), one
obtains $z\simeq0.7$, or else $T\simeq5\cd10^{-4}\,{\rm eV}$. Early universe could thus have been matter or radiation dominated. However, we can deduce from a similar sketchy calculation that primordial
nucleosynthesis took place in a radiation dominated universe. In fact, we know from equilibrium thermodynamics that
\bea
\frac{\rho_{\rm nr}}{\rho_{\rm ur}}=\frac{\rho_{{\rm nr},0}}{\rho_{{\rm ur},0}}\frac{a}{a_0}
\ena
Since phase transition from radiation to matter domination occurred approximately when $\rho_{\rm nr}\simeq\rho_{\rm ur}$, we find as a result the the corresponding redshift is
\bea
\frac{a_0}{a}\equiv(1+z)\simeq\frac{\O_{\rm nr}}{\O_{\rm ur}}
\ena
Once again, if we plug the experimental data $\O_{\rm nr}\simeq0.3$ and $\O_{\rm ur}\simeq10^{-4}$, we get $z\simeq3\cd10^3$, which means $T\simeq1\,{\rm eV}$. Since nucleosynthesis and earlier epochs were
contained in a radiation dominated era, the temperature can be quantitatively related to the age of the universe at the various steps of primordial evolution. Namely, Big Bang nucleosynthesis lasted from the
age of $t\simeq1\,{\rm sec}$ to $t\simeq3\,{\rm min}$, while QCD and electroweak transitions occurred approximately at $t\simeq3\cd10^{-5}\,{\rm sec}$ and $t\simeq10^{-10}\,{\rm sec}$. This calculations also
need species to be in thermal equilibrium with the plasma, which is a correct assumption at the early times associated with the different epochs.
\subsection{The importance of inflation}\label{inflation}
The Big Bang theory is very successful in predicting the physics that lead to the observed present universe, although initial conditions represent an unsolved issue in this framework. Indeed, there are some
main problems that seem to need further explanations. The presence of an inflationary era before Big Bang is able to give satisfactory answers to the initial condition open questions.
\paragraph{Horizon}
From CMB data we know that the universe is isotropic at large scales. The problem then stems from the computation of the maximal causally connected areas determined at the time of recombination. As I already
pointed out, photons freezed out at recombinations. Thus, causally connected regions today are derived from recombination horizon, expanded until actual time. The size of horizon at recombination epoch
measured today is indeed 
\bea
l_{{\rm H,rec},0}=\l_{\rm H,rec}(1+z_{\rm rec})\simeq300\,{\rm Mps}
\ena
Meanwhile, the present size of today horizon is given by $l_{{\rm H}0}\simeq10^4\,{\rm Mpc}$. CMB observations then imply isotropy over a much larger scale than the recombination horizon, meaning that non
causally connected patches of today horizon were subject to isotropic initial conditions, if we restrict to the Big Bang picture alone. These wouldn't be natural initial conditions.
\paragraph{Flatness}
The flatness measured in today universe implies an enormous fine--tuning of initial data. More precisely, we now observe an energy density from curvature contribution bounded by the small value $|\O_{\rm
curv}|<0.02$. Tracking this limiting value backwards to early epochs --- given that the curvature energy density scales as $1/a^2$ --- we get at nucleosynthesis $|\O_{\rm curv}|<10^{-16}$ and at electroweak
epoch $|\O_{\rm curv}|<10^{-26}$. The question is then why the universe had a such very large radius with respect to the inverse Hubble scale. This initial condition again is not a natural one, indeed.
\vspace{0.5cm}

There are other issues that lead to search for a new picture of the early universe, to be combined with the Big Bang theory. Entropy, for instance, should be given as an unnaturally large initial condition.
Also primordial perturbations must find an explanation, giving the initial conditions that yield the today measured anisotropies and structures.

Inflation is a useful mechanism that accomplishes to solve the initial value problems, since during this era the universe expands at large rate. A suitably large acceleration implies that small patches expand
to big sizes, so that the eventual initial curvature is smoothed, the horizon size increases putting in causal contact even distant regions of the CMB. To be more explicit, inflation is usually assimilated
with exponential acceleration 
\bea
a(t)\sim\ex^{\int\intd t\,H}
\ena
A long inflation thus allows to solve Big Bang initial condition problems. The number of e--folds 
\bea
N_{\rm efolds}(t)=\log\frac{a_{\rm end}}{a}=\int_t^{t_{\rm end}}\intd t\,H
\ena
is a basic parameter of inflation. Its value must be determined in order to give consistent data for the Big Bang theory. Precisely, a number of $N_{\rm efolds}\simeq60$ is necessary for a scenario of
primordial inflation to be consistent.
\subsubsection{Slow roll}
A very simple and illustrative model of inflation is described by a single field subject to slow rolling potential. Slow rolling inflation can be summarized as follows. Suppose that we have a scalar field
with action
\bea
S=\int\intd^4x\sqrt{-g}\left(\oneover{2}g^{\m\n}\pa_\m\phi\pa_\n\phi-V(\phi)\right)
\ena
subject to a power law potential $V(\phi)$. We assume that initially the scalar field is homogeneous over a large patch of the universe. Hence, we use FRW metric and a field $\phi=\phi(t)$ that doesn't depend
on space coordinates. The scalar field equation is
\bea
\ddot\phi+3\frac{\dot a}{a}\dot\phi=-V'(\phi)
\ena
The energy density and pressure associated to the scalar field are given by
\bea
\rho=\oneover{2}\dot\phi^2+V(\phi)\,,\qquad p=\oneover{2}\dot\phi^2-V(\phi)
\ena
and the Friedmann equation reads
\bea
\frac{\dot a^2}{a^2}=\frac{8\p}{3M_{Pl}^2}\left(\oneover{2}\dot\phi^2+V(\phi)\right)
\ena
The slow roll inflation takes place when we can discard the kinetic term for the scalar field with respect to its potential, in order to get a vacuum behavior $p=-\rho$. The potential is almost constant,
yielding approximately constant energy density. Usually, two parameters are defined and supposed to be very small
\bea
\epsilon&\equiv&\frac{M_{Pl}^2}{2}\frac{V'^2}{V}\ll1\\
|\eta|&\equiv&\left|M_{Pl}^2\frac{V''}{V}\right|\ll1
\ena
In the slow roll approximation, the equations for the inflaton scalar field become
\bea
\dot\phi&=&-\frac{V'(\phi)}{3H}\\
H^2&=&\frac{8\p}{3M_{Pl}^2}V(\phi)
\ena
In this particular model of inflation --- multi--field inflation can be obtained as a generalization of this set--up --- the number of e--folds is determined by the specific potential. It can be calculated
using the following expression
\bea
N_{\rm efolds}=\oneover{M_{Pl}^2}\int_{\phi}^{\phi_{\rm end}}\intd\phi\,\frac{V}{V'}
\ena

The great success of inflation is due to the fact that homogeneity initial condition is required only over a patch of the order of Planck scale or less, in order to solve the horizon problem. Many other
issues are related to inflation. One among the others is represented by density fluctuations, which can find their origin in the inflationary era. In fact, roughly speaking, small perturbations get extremely
enhanced during inflation and finally yield the observable structure formation (for a review, see \cite{Trodden:2004st}). 

I would still have to give some arguments about the transition from the end of inflation to the beginning of Big Bang. Indeed, during inflation everything gets diluted and redshifted, so that one has to find
a mechanism which recovers hot matter to get the Big Bang started. This mechanism is usually called \emph{reheating}. Reheating has been ascribed to inflaton decaying into other species.  Oscillations due to
this decay processes in some sense produce hot radiation which reaches an equilibrium at some reheating temperature. In more complex scenarios, however, reheating can be achieved with different mechanisms.
For instance, the potential may have some directions other than the almost flat direction followed by the inflaton during inflation. At the end of inflation, the other direction --- let's assume for
simplicity that there is only one extra direction --- may take the form of an unstable potential, letting the inflaton flowing towards a (meta)stable minimum of the theory.
\subsubsection{Conformal anomaly and inflation: Starobinsky model}
Starobinsky proposed a new model of inflation driven by quantum corrections to Einstein equation, and related to conformal anomaly. I here briefly summarize the mechanism leading to the inflationary phase and
to the graceful exit, without giving the details (the model is introduced in \cite{Starobinsky:1980te} and thoroughly discussed in \cite{Vilenkin:1985md}).

Einstein equations with some matter conformal field theory take the form
\bea\label{staro einstein}
G_{\m\n}=-8\p G_N\langle T_{\m\n}\rangle
\ena
They get quantum corrections from the evaluation of stress--energy tensor on curved manifolds. For free massless conformally invariant fields we obtain
\bea\label{staro stress}
8\p G_N\langle T_{\m\n}\rangle=\frac{H_{(3)\m\n}}{H_0^2}+\oneover{6}\frac{H_{(1)\m\n}}{M^2}
\ena
where
\bea\label{staro ano 1}
H_{(1)\m\n}=2\nabla_\m\nabla_\n R-2g_{\m\n}\nabla_\s\nabla^\s R+2R R_{\m\n}-\oneover{2}g_{\m\n}R^2\\
H_{(3)\m\n}=R^\s_\m R_{\n\s}-\frac{2}{3}rr_{\m\n}-\oneover{2}g_{\m\n}R^{\s\l}R_{\s\l}+\oneover{4}g_{\m\n}R^2 \label{staro ano 2}
\ena
The coefficient $H_0$ is determined by the matter content of the conformal theory
\bea
\left(8\p G_NH_0^2\right)^{-1}=\oneover{1440\p^2}\left(N_0+\frac{11}{2}N_{1/2}+31N_1\right)
\ena
where we denote with $N_0,N_{1/2},N_1$ the number of spin $0,1/2,1$ fields, respectively. The second coefficient, $M$, can be made arbitrary by adding a covariant counterterm to the action, which is necessary
to regulate the gravity action. Its arbitrary finite part changes the $1/M^2$ coefficient in the stress--energy tensor. Trace anomaly stems from \refeq{staro stress} plugging in \refeq{staro ano
1}--\refeq{staro ano 2}
\bea
8\p G_N\langle T^\m_\m\rangle=\oneover{H_0^2}\left(\oneover{3}R^2-R_{\m\n}R^{\m\n}\right)-\oneover{M^2}\Box R
\ena
Solving the Einstein equation \refeq{staro einstein} yields a de Sitter geometry fixed point for the FRW metric ansatz
\bea
a(t)=\left\{\begin{array}{l@{\,,\qquad}l}
\oneover{H_0}\cosh(H_0t)&k=+1\\
a_0\,\ex^{H_0t}&k=0\\
\oneover{H_0}\sinh(H_0t)&k=-1
\end{array}\right.\ena
Thus, this represents an inflationary era with exponential expansion. One first important point, relative to this de Sitter solution, is its instability. The linearized equation around the fixed point
solution can be analyzed and gives at least one negative root for the stability matrix eigenvalues, as expected for unstable points.

Successively, we can look at the slowly varying approximation $\dot H\ll H^2,\ddot H\ll\dot H$ for a spatially flat universe $k=0$ and $\left(H(t=0)-H_0\right)\ll1$ but still finite. The full evolution
equation can be written in terms of the Hubble parameter $H\equiv\dot a/a$ as
\bea
H^2\left(H^2-H_0^2\right)=\frac{H_0^2}{M^2}\left(2H\ddot H+6H^2\dot H-\dot H^2\right)
\ena
For slowly varying Hubble constant this reduces to the following solution
\bea\label{staro slow sol}
H(t)=H_0\tanh\left[\g-\frac{M^2t}{6H_0}\right]
\ena
with $\g=\log\left(\frac{2H_0}{|H(0)-H_0|}\right)/2$. The slowly varying approximation holds until $t\simeq t_*=6\g H_0/M^2$, when the Hubble parameter approaches the value $H\simeq M$. Besides, to obtain 
long inflation we must require $M^2\ll6H_0^2$, as we deduce from \refeq{staro slow sol}. Hence, the accelerated era terminates with a value for $H$ which is $H\ll H_0$. Let us investigate the fate of
evolution is this approximation. The Einstein equation now reads
\bea
2H\ddot H+6H^2\dot H-\dot H^2+M^2H^2=0
\ena
This equation implies the presence of damped oscillations that typically yield reheating. In fact, approximate solutions come first from ignoring the friction term $6H^2\dot H$, which is small. This simply
gives $H\propto\cos^2\left(Mt/2\right)$. Introducing friction we get the damping 
\bea
H=\frac{4}{3t}\cos^2\left(\oneover{2}Mt\right)\left(1-\frac{\sin Mt}{Mt}\right)+O(t^{-3})
\ena
The average scale factor over the period of oscillations is equal to $\overline{a(t)}\propto t^{2/3}$, describing non--relativistic matter. Furthermore, considering deviations from conformal invariance, one
can determine the temperature at which the oscillations are damped and thermalization occurs, leaving a radiation dominated universe.

Gravitational waves and quantum cosmology can also be considered in the Starobinsky model \cite{Vilenkin:1985md}. However, the feature I want to stress here is the role of conformal anomaly in driving an
inflationary era as well as reheating and thermalization. Indeed, conformal anomaly cosmologies, in the context of holographic duality, will be further analyzed in section \ref{4D RS holo cosmo} and chapter
\ref{7D RS dual}.
\subsection{Present era: dark energy and open questions}\label{present era}
The universe at early stages is well described by Big Bang theory, combined with inflation. However, observations also tell us how the universe looks like today. In particular, I will briefly focus on the
matter composition of our universe, which is related to equilibrium thermodynamics, and present acceleration. Open issues in cosmology will also be discussed.
\subsubsection{The composition}
The composition of current universe is due to relics left at the time of freeze out by the various species contributing to the energy density. Ultra--relativistic matter is characterized by the same
distribution function, both at equilibrium and after decoupling, provided that we define an effective temperature after freeze out, $T_{\rm ur}\sim1/a$. Its number density at the time it leaves equilibrium is
\bea
n_{\rm ur,f}\sim T^3_{\rm f}
\ena
This number gets then diluted by the factor $1/a^3$ until present epoch. The number density of relic photons, after recombination epoch, is $n_\g\simeq410\,{\rm cm}^{-1}$ and any ultra--relativistic relic
abundance should thus roughly be of the same order 
\bea
n_{\rm ur}\sim n_\g
\ena
Neutrinos for instance have number density $n_\n\simeq 115\,{\rm cm}^{-1}$ per species. 

Non--relativistic particles involve a less trivial analysis.  Number density of cold relics depend on their annihilation cross section $\s_0$ and on their mass $m_{\rm nr}$
\bea\label{nr relics}
n_{\rm nr}\sim\frac{n_\g}{\s_0m_{\rm nr}M_{Pl}}
\ena
From the number density, one can easily compute the energy density and the density parameter $\O_i$, relative to each species.

I already mentioned that our space is detected to be flat to good approximation. This means that
\bea
\sum_i\O_i\simeq1
\ena
From the number density for photons, we deduce their density parameter
\bea
\O_\g\simeq6\cd10^{-4}
\ena
Admitting a mass $m_{\n_e}<2.6{\rm eV}$ for neutrinos, we can calculate the upper bound for the associated density parameter, supposing that neutrinos are non--relativistic today. which is true. Since
$\rho_\n=\sum_\a m_{\n_\a}n_{\n_\a}$ we obtain $\O_\n<0.16$. Nevertheless, cosmological data measuring structures at small scales, which in turn are related to ultra--relativistic neutrino density, give
\bea
\O_\n<10^{-2}
\ena
yielding a stricter bound on the masses $\sum_\a m_{\n_\a}<0.42{\rm eV}$. 

Baryonic matter certainly contributes to density. Experiments teach us that
\bea
\O_B\simeq5\cd10^{-2}
\ena
Electrons contributes on the same footing as baryons to the total number density, because of neutrality. Nonetheless, since their mass is much smaller, their energy density is negligible with respect to
the baryon contribution, $\O_e\ll\O_B$. 

I summarized until now the stable visible particles that compose the universe. However, there is also some dark matter (cold dark matter) whose abundance can be estimated through indirect experiments --- as I
mentioned in the subsection dedicated to observations data --- and gives an important contribution to the total density parameter
\bea
\O_{\rm CDM}\simeq0.25
\ena
Finally, as we can deduce by summing all the contributions we summarized, $\O_\g+\O_\n+\O_B+\O_e+\O_{\rm CDM}\simeq0.3$, there should be another dark species that dominates the composition of our universe.
This is in agreement with observations revealing today acceleration, since nor usual matter nor radiation can lead to the present accelerated expansion. We instead need a negative pressure species, with
$w<-1/3$. Dark energy contribution thus reads
\bea
\O_\L\simeq0.7
\ena
This value is obtained assuming a constant equation of state where the parameter $w$ is constrained by supernovae observations to lay in the range $-1.2<w<0.8$. 
\subsubsection{Dark sector}
Finding a good candidate for dark energy is still an open problem. A cosmological constant would do the job, of course, but it is not in satisfactory agreement with data. One of the alternatives is dynamical
dark energy, which consists of some slowly varying field. I will not discuss this issue here, but the interested reader can find the topic reviewed in \cite{Trodden:2004st} and references therein. As for
inflaton, a slowly varying field causes a slowly varying equation of state parameter, approximately equal to $w\simeq-1$, implying acceleration. However, non standard FRW cosmology may develop future eternal
acceleration without the need of such a constraint on the equation of state parameter, as I will show in the context of brane--world cosmology, next in this chapter and in the following.

Cold dark matter, instead, must have been non--relativistic at freeze out, otherwise it would have influenced structure formation suppressing it (as neutrinos did). On the other hand it is obvious that
interactions of CDM with the other particles must be very weak, not to be detected by experiments. The energy density of CDM relics can be derived by \refeq{nr relics}. Assuming that CDM is composed by some
(anti)particles $Y,\bar Y$, their density parameter is independent of the mass $m_Y$, given that $\rho_Y=m_Yn_Y$. Thus, we obtain
\bea\label{density para CDM}
\O_Y\sim\frac{n_\g}{\s_0M_{Pl}\rho_c}
\ena
This expression ignores some negligible logarithmic factor $\log\left(\s_0M_{Pl}m_Y\right)$ and the effective number of massless degrees of freedom $g_*$ (which can however be easily included by defining an
effective Planck mass $M_{Pl*}=M_{Pl}/1.66\sqrt{g_*}$). The condition required for some particle $Y$ to be a consistent candidate for CDM is $\O_Y\simeq1/\s_0(10^9{\rm GeV}^2)\simeq1$. We deduce from
\refeq{density para CDM} that a weakly interacting massive particle (WIMP) with cross section $\s_0\simeq\a_WG_F\simeq10^{-9}{\rm GeV}^{-2}$ nicely fits into the picture. The lightest supersymmetric
particle (LSP) in Standard Model extensions is a good example of particle with such an annihilation cross section, indeed. LSP must be stable and furthermore supersymmetry must be broken at weak scale, in
this scenario. 
\vspace{0.5cm}

There are other fundamental issues in cosmology, that I will not discuss in detail. One is baryon asymmetry. This subject has to do with initial conditions. We need to find an explanation to the non equal
number of particles and antiparticles, namely quarks and antiquarks. This should not simply be an asymmetric initial condition that moreover doesn't fit into the inflation picture. Sakharov stated that three
conditions are necessary in oder to get the observed baryon--antibaryon number discrepancy: (i) violation of the baryon number, (ii) C and CP violation, (iii) deviation from thermal equilibrium. These three
conditions may be achieved independently of the specific particle model one considers. Many possibilities have been studied in literature --- Grand Unification mechanism, electroweak baryogenesis,
leptogenesis, for example. However, their description is beyond the scope of this work.

Another issue that I have not mentioned up to now, is the cosmological constant problem. It deals with the unnaturally small value of the cosmological constant, which is equivalent to fine--tuning vacuum
energy to a very small value. There are still no fully satisfactory answers to this problem. Most investigated attempts going in the direction of solving the cosmological constant problem are related to
supersymmetry. The landscape approach is also a possibility. Indeed, string compactification leads to a very large number of vacua, some of which may have the right vacuum energy property (in addition to the
other (Susy) Standard Model characteristics). However, the landscape still needs further investigations.

The main topic I focus on, in my work, is late time acceleration, driven by a brane--bulk energy exchange, and the possibility of finding a realistic evolution, including primordial inflation, in the context
of brane--world cosmology.
\sectioncount{Embedding cosmology in strings via brane--worlds}
String theory entails a certain number $n$ of extra dimensions with respect to the space--time that we experience through observations (see \cite{Csaki:2004ay} for interesting lectures on extra dimensions).
The number of extra dimensions is six for any critical superstring theory, since superstrings live in a 10--dimensional target space. In a large part of literature, the set--up involves exactly six extra
dimensions. However, there are other possibilities such as M theory, which yields an eleven dimensional target space, so that seven dimensions must be hidden. Non critical strings are not constrained by the
vanishing of the central charge --- $c\propto (D-10)$ --- to live in ten dimensions. Thus one can in principle consider both supercritical, $D>10$, or subcritical, $D<10$, string theories. The number $n$ of
extra dimensions in non critical backgrounds is as a consequence $n<6$ or $n>6$, respectively.

The question should be now why extra dimensions can be non visible to experimental observations. We can answer this question at different levels. The first aspect to consider is how to conveniently realize a
string theory with extra dimensions that are consistent with observations. We may have either tiny or large extra dimensions, either compact or warped or even flat, depending on the specific string theory
background. In what follows, I am going to review some of the possible set--up leading to an effective four dimensional world, namely brane--worlds. String theory indeed naturally provides such brane objects.
D$p$--branes are non perturbative membrane--like string configurations extended in $p$ spatial dimensions and in the time direction, which carry RR charges. $U(N)$ gauge fields typically live on a stack of
$N$ D$p$--brane.  Therefore, the idea of the Standard Model living on some (intersection of) stacks of D$p$--branes thrived. Other kinds of extended objects are present in string theory or M theory, such as
NS--branes (carrying NS--NS charges), M--branes (in eleven dimensional M theory), black $p$--branes (classical supergravity solutions with black hole--like thermodynamics), which can in some cases be related
among each other by string dualities.

On the other hand, there should be a profound reason why we live in a string theory vacuum with only four visible dimensions, where the dynamics of fields is described by Standard Model gauge theory and
General Relativity. There is much work going on addressing this question. The landscape of string vacua is studied in this context, in order to determine whether there are for instance statistical and/or
anthropic reasons that have driven the evolution of our universe towards the ``right'' vacuum. Although this is a fundamental issue that should be solved in the string theory picture, I am not going to cover
this topic here (see, for instance \cite{Douglas:2003um,Blumenhagen:2004xx}).

Instead, how the four dimensional world we observe is compatible with superstring theory and extra dimensions?
\subsection{Compatification of extra dimensions}
A first hypothesis on the structure of space--time embedding the known four dimensional world consists in assuming that all known fields (gauge and gravity) propagate in all dimensions. If the total number of
dimensions is $D=4+n$, we would like the $D$--dimensional gravity theory to contain the observed 4--dimensional one. In formulae, we must have the following form for the $D$--dimensional metric and curvature
\bea
\sqrt{g^{(D)}}=\ell^n\sqrt{g^{(4)}}\,,\qquad R^{(D)}[h]=R^{(4)}[h]
\ena
where the ansatz for the full metric is given by
\bea
\intd s^2\equiv g^{(D)}_{MN}\intd x^M\intd x^N=\left(\eta_{\m\n}+h_{\m\n}\right)\intd x^\m\intd x^\n-\ell^n \intd\O_{(n)}^2
\ena
Here $h_{\m\n}$ is the metric fluctuation around four dimensional Minkowski, $\ell$ is the radius of the extra dimensions and $\intd\O_{(n)}$ represents the line element of compact and locally flat extra
dimensions. The usual 4D Einstein--Hilbert action is thus contained in the higher dimensional theory
\bea
S^{(D)}=-M_{(D)}^{n+2}\int\intd^{D}x\sqrt{g^{(D)}}R^{(D)}=-M_{(D)}^{n+2}V_{(n)}\int\intd^{4}x\sqrt{g^{(4)}}R^{(4)}
\ena
The volume $V_{(n)}$ of the internal space is given by the integral $\int\intd\O_{(n)}\ell^n$. The relation between the four dimensional Planck mass $M_{Pl}$ and the $D$--dimensional scale $M_{(D)}$ is then
\bea\label{planck mass compact}
M_{Pl}^2= M_{(D)}^{n+2}V_{(n)}
\ena
The analogous comparison can be carried for the gauge couplings in four and $D$ dimensions --- $g_{\rm YM}$ and $g_{(D)}$, respectively. The result in this case gives
\bea
g_{\rm YM}^2=\frac{g_{(D)}^2}{V_{(n)}}
\ena
Putting all together, and assuming that the volume of the $n$--dimensional compact space scales as $V_{(n)}\simeq\ell^n$ (take as an example the $n$--torus: $V_{(n)}=(2\p\ell)^n$), the relation between four
dimensional Planck mass, four dimensional gauge coupling and radius of extra dimensions is straightforward obtained
\bea
\ell\sim\frac{\left(g_{\rm YM}\right)^{1+\frac{2}{n}}}{M_{Pl}}
\ena
Such tiny extra dimensions, scaling as the Planck length $\ell\sim M_{Pl}^{-1}$, are expected not to be observed in the relatively near future.

Large extra dimensions \cite{Arkani-Hamed:1998rs} can arise when the $D$--dimensional gravity scale is taken to be of the order of $M_{(D)}\gtrsim1{\rm TeV}$. This is in fact a lower bound determined by
experiments, since gravitational quantum effects haven't been observed yet. We can put the relation \refeq{planck mass compact} in the form
\bea\label{arkani radius}
\ell\sim\oneover{M_{(D)}}\left(\frac{M_{Pl}}{M_{(D)}}\right)^{\frac{2}{n}}\simeq10^{\frac{32}{n}}\, {\rm TeV}^{-1}
\ena
(with $M_{Pl}=10^{19}{\rm GeV}$).  The bound for the radius of extra dimensions is in turn determined by the distance that gravitational experiments on Newton law validity can probe. At the present state of
experiments, we must require $\ell$ to satisfy $\ell\lesssim0.1\,{\rm mm}$. The larger the number $n$ of extra dimensions is, the smaller extra dimensions are in this set--up. For $n=1$ equation \refeq{arkani
radius} gives $\ell\simeq2\cd10^{15}\,{\rm cm}$, which is much bigger than the experimental bound. If we instead admit the existence of two extra directions, the radius has a more reasonable value
$\ell\simeq2\,{\rm mm}$. Investigating the two extra dimension hypothesis, we also find that $\ell\lesssim0.2\,{\rm mm}$ changes the plausible bound for the $D$--dimensional gravity scale to
$M_{(D)}\gtrsim3\,{\rm TeV}$. More than two extra dimensions would imply such a tiny radius that they wouldn't be detected by any foreseeable future experiment.

This kind of model with large compact extra dimensions was proposed in order to solve the hierarchy problem, since it allows to define a fundamental gravity scale $M_{(D)}$ of the same order of the
electroweak interaction scale. Nevertheless, the hierarchy problem in this set--up doesn't disappear but rather translates in the problem of the unnatural large radius for the extra dimensions.
\subsection{The Randall--Sundrum alternative}\label{RS model}
Another possibility, which is the one exploited in recent brane--worlds models, is to relax the assumption of having all fields free to propagate in the whole space--time. Indeed, this can be done by
introducing a four dimensional defect, a brane, on which (some of the) Standard Model fields are constrained to live% footnote
~\footnote{The simplest example is to assume that \emph{all} Standard Model fields live on the same (stack of, intersection of) brane(s). However, more complicated and less naive configurations are possible
and have been widely investigated \cite{Gherghetta,Blumen,Blumenhagen:2004xx}, where the gravitational sector is free to propagate in the bulk and matter fields are localized at specific distances from the IR
brane. The only necessary requirement is that the Higgs boson must be placed on the IR brane hypersurface.}.% footnote

The goal of this introduction is mainly to discuss gravity/cosmology issues in the brane--world context. I won't generally go deeply into brane--world phenomenology, like Standard Model or QCD relevant
aspects. However, the search for a solution to the hierarchy problem has been the basic motivation for the Randall--Sundrum proposal, as an alternative to the large extra dimension models. Hierarchy gets
exponentially suppressed thanks to the fact that the extra dimension is a warped direction (which is essentially compact in the Randall--Sundrum model with two branes, RSI, and has infinite volume in the one
brane RSII). The feature of Randall--Sundrum model which is fundamental for cosmology is the localization of gravity on the four dimensional brane. Hence, even with an infinite extra dimension, gravitational
forces are effectively four dimensional (in the IR).

In the past, earlier warped extra dimension brane--worlds have been proposed \cite{Rubakov:1983bz}, with the idea of solving the cosmological constant problem \cite{selftune}. Indeed, balancing of the brane
tension with bulk cosmological constant gives, upon fine--tuning though, a zero effective cosmological constant on the brane representing the visible world. Hence, the problem of explaining such a small
observed vacuum energy in our universe is translated into a question about the fine--tuned value of the bulk vacuum energy which exactly cancels the Standard Model vacuum.

Let me now introduce the two Randall--Sundrum (RS) models \cite{Randall:1999ee} and the cosmologies that develop specially in the one brane RS set--up.
\subsubsection{RSI: the hierarchy solved}
The theory for this model is gravity in a five dimensional bulk, with five dimensional cosmological constant $\L_5$. The extra dimension is then compactified over a $S^1/\Zgr_2$ orbifold at whose fixed
points two sources of energy density, i.e. two co--dimension one branes, are located, in order to compensate for the bulk cosmological constant. The extra direction $y$ has hence a finite volume and the
boundaries of space--time are somehow replaced by the two 3--branes.  Einstein--Hilbert action, plus the branes tension contributions, plus arbitrary matter lagrangians localized on the branes (the specific
form is irrelevant to the Einstein equation results) take the form
\bea
S=S_{\rm EH}+S_{b_{\rm UV}}+S_{b_{\rm IR}}
\ena
with
\bea
S_{\rm EH}&=&\int\intd^5x\sqrt{-g}\left(M^3R-\L_5\right)\\
S_{b}&=&\int\intd^4x\sqrt{-\g}\left(-V+\lagra_b\right)
\ena
Here, the five dimensional gravitational scale is set by $M$ ($M^3\equiv1/16\p G_N^{(5)}$, $\L_5\equiv\L/8\p G_N^{(5)}$ to recover the notations of the previous chapter), $\g_{\rm UV}$ and $\g_{\rm IR}$
are the induced metrics on the IR and UV branes, respectively --- IR and UV indicate that energy scales are suppressed on the IR brane, as we will shortly deduce.  

The \emph{warp} ansatz for the background metric is essential in order to obtain the desired solution to the hierarchy problem. Indeed, we will see how energy scales are lowered by the warp factor
$\ex^{-2A(y)}$ which enters in the metric as
\bea\label{RS ansatz}
\intd s^2=\ex^{-2A(y)}\eta_{\m\n}\intd x^\m \intd x^\n+\intd y^2
\ena
The orbifold fixed points are $y_*=0$, where the UV brane sits and $y_*=y_{\rm IR}$. Standard Model (or at least the Higgs field) are thought as located on the IR brane (also called the TeV brane), while
gravity will show to be localized on the UV brane (or Planck brane). If the function $A(y)$ is linear $A(y)=y/\ell$, then the metric \refeq{RS ansatz} describes a slice of $AdS_5$ geometry with radius $\ell$
in the region between the two 3--branes. In these coordinates the Einstein tensor reads
\bea
G_{\m\n}&=&-3\left(A''-2A'^2\right)\non\\
G_{55}&=&6A'^2
\ena
The $55$ component of the Einstein equations
\bea
G_{MN}=\oneover{2M^3}T_{MN}
\ena
immediately gives --- keeping in mind the $\Zgr_2$ orbifold ---
\bea\label{RS sol A}
A=\sqrt{-\frac{\L_5}{24M^3}}|y|\equiv k|y|
\ena
where $k$ represents the scale of the $AdS_5$ geometry, $k=1/\ell$, and $\L_5$ is the negative cosmological constant of AdS space--time. The solution \refeq{RS sol A} tells us that the second derivative of $A$
has delta functions, and thus deltas are also appearing in the Einstein tensor. The contributions to the stress--energy tensor that compensate for these deltas are exactly the localized energy density sources
corresponding to the branes, provided that the brane tensions are respectively given by
\bea\label{RS tension contribution}
V_{b_{\rm UV}}=-V_{b_{\rm IR}}=24M^3k
\ena
In fact, the $\m\n$ components of the Einstein equations yield
\bea
A''=2k\left[\d(y)-\d(y-y_{\rm IR})\right]=\oneover{12M^3}\left[V_{b_{\rm UV}}\d(y)+V_{b_{\rm IR}}\d(y-y_{\rm IR})\right]
\ena
which is to be compared to the brane tension contributions \refeq{RS tension contribution}.  The RS metric can now be put in the simple form
\bea
\intd s^2=\ex^{-2k|y|}\eta_{\m\n}\intd x^\m \intd x^\n+\intd y^2
\ena

I now want to go towards the solution to the hierarchy problem. We need to analyze gravity in the RSI model and to find out which is the graviton zero mode and the associate KK tower. General fluctuations of
the metric should involve graviton, vector and scalar modes, as a consequence of the decomposition of 5D massless graviton in 4D. However, since the vector mode appears in the metric as $Y_\m\intd x^\m\intd
y$ it must vanish due to the $\Zgr_2$ symmetry $y\leftrightarrow-y$. The scalar mode has been carefully treated in \cite{Garriga,GKR,RSGR}, but doesn't bring relevant changes to the analysis I am reviewing.
Hence, let me consider the following fluctuation around the RS solution
\bea\label{fluctuations conformal}
\intd s^2=\left[\ex^{-2k|y|}\eta_{\m\n}+h_{\m\n}(x,z)\right]\intd x^\m \intd x^\n-\intd z^2
\ena
The linearized Einstein equations for $h_{\m\n}$ are derived by calculating the perturbed Einstein tensor and stress--energy tensor in the RS gauge $\pa^\n h_{\m\n}=h^\m_\m=0$. The following equation holds
\bea
\left\{\ex^{2k|y|}\Box+\pa_y^2+4k\left[\d(y)+\d(y-y_{\rm IR})\right]-4k^2\right\}h_{\m\n}=0
\ena
where $\Box$ is the four dimensional Minkowski d'Alambertian operator. I will always suppose all expressions to be even under the $\Zgr_2$ symmetry. Now, this can be put in the form of a Schr\"odinger
equation by redefining the fluctuations as $\hat h_{\m\n}=h_{\m\n}\ex^{k|y|/2}$ and performing a change of variables only involving the extra direction $ku={\rm sgn}(y)\left(\ex^{k|y|}-1\right)$. Defining in
addition the KK modes of mass $m$ by factorizing the fluctuation wave function as $\hat h_{\m\n}(x,u)=\psi_{\m\n}(u)\ex^{\imm p\cd x}$, we get the wave equation
\bea
\left[-\oneover{2}\pa^2_u+V(u)\right]\psi(u)=m^2\psi(u)
\ena
The mass of the KK modes must satisfy the relation to the KK momenta $p^2=m^2$ for each KK mode. The potential for this one dimensional quantum mechanical system is by construction given by
\bea
V(u)=\frac{15}{8}\frac{k^2}{\left(k|u|+1\right)^2}-\frac{3}{2}\frac{k}{\left(k|u|+1\right)}\left[\d(u)+\d(u-u_{\rm IR})\right]
\ena
In the $u$--frame, the orbifold fixed points are respectively $u_*=0$ and $u_*=u_{\rm IR}=\left(\ex^{ky_{\rm IR}}-1\right)/k$. The normalizable zero mode is then given by
\bea\label{zero mode finite}
\psi_0(u)=\frac{\hat N_0}{\left(k|u|+1\right)^{\frac{3}{2}}}
\ena
which is finite upon integration. The normalization constant $\hat N_0$ is set imposing $\int_{-u_{\rm IR}}^{u_{\rm IR}}\psi_0^2=1$. As one can deduce from the explicit form of the $\psi_0$ wave function
\refeq{zero mode finite}, the massless mode for the graviton is characterized by a peak at the location of the UV brane and it falls off away from it. The normalizability of the zero graviton mode is an
important point in the localization of gravity on the brane, also in the infinite volume extra dimension set--up. In fact, if the zero mode were not to be normalizable, we would not get effective 4D
gravitational interactions. The higher modes are linear combination of Bessel functions
\bea
\psi_m(u)=\sqrt{\frac{k|u|+1}{k}}\left[\hat N_{Jm}J_2\left(\frac{m}{k}(k|u|+1)\right)+\hat N_{Ym}Y_2\left(\frac{m}{k}(k|u|+1)\right)\right]
\ena
As usual, the wave functions can be normalized by requiring the unit integral, as for the zero mode. The boundary conditions determined by the delta functions in the potential --- $\pa_yh_{\m\n}=0$ on the
branes --- can be written as a solution to the determinant equation
\bea
\left|\begin{array}{cc} J_1\left(u=0\right)&Y_1\left(u=0\right)\\J_1\left(u=u_{\rm IR}\right)&Y_1\left(u=u_{\rm IR}\right)\end{array}\right|=0
\ena
They give a discrete spectrum for the KK states, labeled by
\bea
m_n=\x_nk\ex^{-ky_{\rm IR}}
\ena
Here the $\x_n$ denote the n--th root of $J_1$.

As a consequence of gravity localization on the UV brane and of the requirement for gauge fields to live on the IR brane, we can derive how matter field masses get suppressed by the warp factor with respect
to Planck mass. This would indeed solve the hierarchy between electroweak and Planck scales. The four dimensional Planck mass is obtained by writing the relevant effective action containing the four
dimensional Einstein--Hilbert term
\bea\label{RS effective EH action finite}
M^3\int\intd^4x\int_{-y_{\rm IR}}^{y_{\rm IR}}\intd y\sqrt{-g^{(4)}}R^{(4)}\ex^{-2k|y|}
\ena
This result comes from calculating the action that takes account of the massless graviton fluctuations of the metric, which we just computed. After integration over the $y$ variable, we get the value for the
effective four dimensional Planck mass
\bea
M_{Pl}^2=\frac{M^3}{k}\left(1-\ex^{-2ky_{\rm IR}}\right)
\ena
Until now I showed that fore reasonably large values of $y_{\rm IR}$ the Planck scale is barely unaffected by the warped geometry and remains of the order of the five dimensional Planck mass. On the contrary,
the Higgs vev gets lowered by an exponential factor. This can be seen by considering the scalar field action on the IR brane with Higgs potential
\bea
S_{\rm H}=\int\intd^4x\sqrt{\g_{\rm IR}}\left[\g^{\m\n}_{\rm IR}D_\m H^{\dag}D_\n H-\l\left(|H|^2-v^2\right)^2\right]
\ena
where the induced metric on the IR brane is given by $\g_{{\rm IR}\m\n}=\ex^{-2ky_{\rm IR}}g^{(4)}_{\m\n}$. Hence the Higgs action becomes the action for a non canonically normalized scalar
\bea
S_{\rm H}=\int\intd^4x\sqrt{g^{(4)}}\ex^{-4ky_{\rm IR}}\left[\ex^{2ky_{\rm IR}}g^{(4)\m\n}_{\rm IR}D_\m H^{\dag}D_\n H-\l\left(|H|^2-v^2\right)^2\right]
\ena
The scalar fields must be then rescaled to $H\to H\ex^{ky_{\rm IR}}$, yielding the following contribution to the effective action on the IR brane
\bea
S_{\rm H}=\int\intd^4x\sqrt{g^{(4)}}\left[g^{(4)\m\n}_{\rm IR}D_\m H^{\dag}D_\n H-\l\left(|H|^2-\ex^{-2ky_{\rm IR}}v^2\right)^2\right]
\ena
As a result, the vev of the Higgs field localized on the negative tension brane gets exponentially suppressed
\bea
v_{\rm IR}=\ex^{-ky_{\rm IR}}v
\ena
Since the Higgs vev determines all matter field masses, one obtains that its exponential suppression with respect to the barely unchanged Planck mass cancels the hierarchy between the electroweak and Planck
scales for sufficiently but reasonably large $ky_{\rm IR}$. Now the motivation of calling the UV brane \emph{Planck} brane and the IR brane \emph{TeV} brane becomes clear. On the Planck brane we indeed have
scales of the order of the Planck mass. This is also the reason why solving the hierarchy implies that at least the Higgs field of Standard Model should live on the IR brane. The main point is that far away
from the Planck brane the gravitational interaction is suppressed to the TeV scale, due to the localization of the graviton zero mode on the brane itself, while the other interactions have a strength of the
same order of the lowered Planck scale, i.e. of the Tev scale.

The problem of localizing all matter fields on the IR brane (which is addressed for instance in lectures \cite{Gherghetta}) is the suppression of higher dimensional operators (of dimension greater than four)
that are associates to proton decay, neutrino masses and flavor changing neutral currents. If fermions are also confined on the TeV brane, then these operators will conflict with experimental bounds on the
effects they are associates to. Moreover, Yukawa hierarchy cancellation is achieved only if Standard Model fermions are located at different distances from the TeV brane in the bulk, with the lightest (the
electron zero mode) being the furthest away from the negative tension brane (where the Higgs scalar sits), while the heaviest (the top zero mode) being the closest to the Higgs. Spreading fermions in the bulk
at degrees giving the correct Yukawa couplings allows as well to suppress the proton mass by a scale of an order larger than a TeV. Also flavor changing neutral currents are consistent with experimental data
in this scenario. Reasonable tiny neutrino masses, together with lepton number violation on the Planck branes, can in turn be obtained by placing the (left) right handed neutrino near to the (UV) IR brane.
What about gauge bosons? They equally couple to the TeV and Planck brane, i.e. their profile is flat, and they should be living in the bulk, since also fermions are.

Supersymmetric realization of Standard Models may also be implemented in the RSI scenario. Since supersymmetry automatically solves the hierarchy problem, translating it into the SUSY breaking scale problem
(which should be of the order of TeV to avoid fine--tuning), warped geometry has to provide the suppression of the scale of supersymmetry breaking, instead than of the electroweak scale. For a review on the
subject, I suggest \cite{Gherghetta}.
\subsubsection{RSII: gravity localization}
I should now make a remark on gravity localization. In the finite volume extra dimension case, the KK tower has a discrete spectrum and zero mode is peaked on the Planck brane. The continuum of KK modes that
appears in the $y_{\rm IR}\to\infty$ could instead spoil gravity localization. However we can see that this is not the case and the effective 4D gravitational potential acquires correction from the KK tower
which are not relevant on a large distance scale --- where here large means bigger than the inverse Planck mass. 

Calculations are analogous to those in the RSI case, when the boundary at $y_{\rm IR}$ is sent to infinity. Only one energy density source is left: the UV brane with positive tension, located at the origin
(in the $y$--frame). The action should be
\bea
S=S_{\rm EH}+S_{b_{\rm UV}}
\ena
The ansatz for the five dimensional metric is still given by the warped geometry \refeq{RS ansatz}, with the $\Zgr_2$ orbifold. The $S^1$ has degenerated, since its radius is taken to be infinite and hence
the coordinate $y$ spans from $-\infty$ to $+\infty$, preserving the symmetry $y\leftrightarrow -y$. The solutions to the Einstein equations are slightly modified by the absence of the delta contribution from
the missing IR brane
\bea\label{RSII sol A}
A=k|y|\,,\qquad A''=2k\d(y)
\ena
where $k$ is defined as before $k=\sqrt{-\L_5/24M^3}$ and the brane tension $V_{b_{\rm UV}}$ must be fine--tuned in order to compensate for the bulk cosmological constant
\bea
V_{b_{\rm UV}}=24M^3k=-\frac{\L_5}{k}
\ena
All together, this leads to a slice of $AdS_5$, cut by a 3--brane at the $y=0$ boundary. The positive $y$ part is reflected with respect to the position of the brane such that the metric reads
\bea
\intd s^2=\ex^{-2k|y|}\eta_{\m\n}\intd x^\m \intd x^\n-\intd y^2
\ena

It is now fundamental to find whether a normalizable graviton zero mode exists in this set--up, despite the infinite volume of extra direction. Following the calculations that brought to the KK spectrum and
zero mode in the compact extra dimension case, we come to the Schr\"odinger equation for the fluctuations modes $h_{\m\n}(u)\ex^{k|y|/2}=\psi_{\m\n}(u)\ex^{\imm p\cd x}$ (the change of variables is again
$ku={\rm sgn}(y)\left(\ex^{k|y|-1}\right)$)
\bea\label{RS schroedinger infinite}
\left[-\oneover{2}\pa^2_u+\frac{15}{8}\frac{k^2}{\left(k|u|+1\right)^2}-\frac{3}{2}\frac{k}{\left(k|u|+1\right)}\d(u)\right]\psi(u)=m^2\psi(u)
\ena
This equation is still solved by the zero mode
\bea\label{zero mode infinite}
\psi_0(u)=\frac{\hat N_0}{\left(k|u|+1\right)^{\frac{3}{2}}}
\ena
which is still normalizable, since its integral over the infinite $u$ direction is finite. We can thus demand the normalization constant to be defined through
\bea
\int_{-\infty}^{\infty}\intd u\;\psi_0^2(u)=1
\ena
So, the massless mode wave function \refeq{zero mode infinite} is large on the brane and rapidly drops as the distance from the brane increases. Gravity is still suppressed far away from the positive tension
brane. By noting that the 4D massless graviton gives the usual newtonian gravitational interaction, we could expect gravity to be effectively 4D even in the infinite volume extra dimension background.
However, the higher KK modes form a continuum without mass gap that could in principle recover the full 5D gravity. It is then necessary to explicitly consider those higher KK states in order to show that
this is not the case. The solutions to \refeq{RS schroedinger infinite}, for positive masses are again given by linear combinations of Bessel functions 
\bea
\psi_m(u)=\sqrt{\frac{k|u|+1}{k}}\left[\hat N_{J,m}J_2\left(\frac{m}{k}(k|u|+1)\right)+\hat N_{Y,m}Y_2\left(\frac{m}{k}(k|u|+1)\right)\right]
\ena
which are now normalized by requiring $\int_{-\infty}^{\infty}\intd u\psi_m\psi_{m'}=\d(m,m')$. Moreover, imposing the boundary conditions at the origin (due to the delta function in the potential) enforces a
relation between the two normalization parameters, leaving just $\hat N_m$ to be determined by the normalizability condition. If these modes are strongly suppressed by the potential in \refeq{RS
schroedinger infinite} they will not affect Newton's gravitational law at large distances.

The computation of the corrections to the gravitational potential is given by evaluating the KK wave function at the location of the brane $y=0$ (or equivalently $u=0$). Two massive particles, with masses
$m_1$ and $m_2$, placed at a distance $r$ on the Planck brane, are subject to a potential describing the exchange of the massless graviton mode plus the higher KK modes
\bea\label{RS gr potential infinite}
V_{\rm gr}(r)\sim G_N\frac{m_1m_2}{r}+\frac{1}{M^3}\int_0^\infty\intd m\frac{m_1m_2}{r}\,\ex^{-mr}\psi_m(0)
\ena
Imposing the right boundary conditions on the KK wave functions, one obtains $\psi_m(0)\sim\sqrt{m/k}$. Plugging this expression into the gravitational potential \refeq{RS gr potential infinite} and
integrating yields the following result
\bea
V_{\rm gr}(r)\sim G_N\frac{m_1m_2}{r}\left(1+\oneover{k^2r^2}\right)
\ena
I should mention that here $G_N\sim k/M^3$ from the relation between the effective 4D Planck scale $M_{Pl}$ and the higher dimensional Planck mass $M$
\bea
M_{Pl}^2=\frac{M^3}{k}
\ena
This directly comes from the evaluation of the relevant part of the effective 4D action when we consider the massless graviton fluctuations for the metric. The result is \refeq{RS effective EH action finite}
in the infinite extra direction limit $y_{\rm IR}\to\infty$.

The corrections to the Newton potential in the RS model with a non compact extra dimension are thus negligible at distances above the Planck length, since they are of order $1/k^2r^2$ with $k$ equal to the
scale of the $AdS_5$ (which is naturally of the order of the Planck mass $M$). So, gravity is effectively four dimensional in the RSII model, because of the localization of the graviton modes on the brane.
The model proposed by Randall and Sundrum hence represents a very interesting alternative to compactification, since it leads to 4D gravity in a higher dimensional bulk with warped but not necessarily compact
extra dimension.

A note on the RS relation to string theory is in order. The $AdS_5$ geometry is known to come from the near horizon type IIB string solution, in the supergravity decoupling limit, when the backreaction of a
(stack of $N$ coincident) D3--brane is taken into account. More precisely, the background metric is $AdS_5\times S^5$ and the background self--dual 5--form has quantized flux through the $S^5$ (in terms of
$N$). The theory supported by the D3--branes, which are indeed charged under the RR 5--form, is a superconformal ${\cal N}=4$ $SU(N)$ SYM gauge theory living in four dimensions. It is now established (on the
footing of various checks, mainly on non--renormalized operators) that 4D conformal field theory in the large $N$ approximation is dual to the string theory living on the background inferred by the D3--branes.
This constitute the AdS/CFT correspondence in the most well known case of $AdS_5\times S^5$. The rigorous results about gauge/gravity duality have been reviewed in section \ref{adscft}, while the relation to
RS models has been discussed in section \ref{adscft in RS}.

The calculations I have summarized for RSI and RSII may be extended to any number of dimensions with codimension--one branes. Higher codimensions have also been considered --- especially six dimensional
bulks with codimension--two 3--branes \cite{Cline:2003ak}. These set--up are more complicated since bulk curvature singularities enforced by the higher codimension branes are stronger than delta functions and
hence need to be regularized. Regularization can be achieved by the introduction of a finite thickness of the brane \cite{Kanno:2004nr}. Alternatively, one can add corrections to the gravitational action
which keep the singularities under control. Corrections may be gravity induced on the brane, or higher derivative terms such as string motivated Gauss--Bonnet corrections (which would give a RS model lus
higher order corrections \cite{Bostock:2003cv,Nojiri:2000gv,Nojiri:2006je}).
\subsubsection{7D RS: more KK and winding modes}\label{baolykken}
A case of codimension--one brane in a higher dimensional bulk has been analyzed in \cite{Bao:2005ni}. The set--up is seven dimensional RSI model with two branes. With the AdS warp ansatz for the metric, the
geometry is described by a slice of $AdS_7$. Now, $AdS_7\times S^4$ is a classical solution to eleven dimensional M theory in the supergravity limit, if we admit $N$ units of 4--form flux through the
4--sphere (where $N$ is the number of M5 branes that infer this geometry). The dual theory to this background --- via AdS/CFT --- is a $(0,2)$ SCFT in six dimensions (for more details, see section \ref{D3 and
M5}).

The warped geometry \`a la RS is obtained by solving Einstein equations imposing the presence of the 5--brane and the $S^1\times\Zgr_2$ orbifold. The RS bulk action plus the brane term take the form
\bea
S=S_{\rm EH}+S_{b_{\rm UV}}+S_{b_{\rm IR}}
\ena
where the Einstein--Hilbert contribution is seven dimensional and the energy density source is six dimensional
\bea
S_{\rm EH}&=&\int\intd^7x\sqrt{-g}\left(M^5R-\L_7\right)\\
S_{b}&=&\int\intd^6x\sqrt{-\g}\left(-V+\lagra_b\right)
\ena
$M$ is now the seven dimensional Planck mass, $\L_7$ is the bulk cosmological constant. The seven dimensional generalization of the RS metric ansatz is simply 
\bea\label{7D RS ansatz}
\intd s^2=\ex^{-2A(y)}\eta_{m n}\intd x^m \intd x^n+\intd y^2
\ena
with $m,n,\dots$ belonging to the 6--dimensional subspace and $y$ is as usual the coordinate transverse to the branes. Branes are located at $y_*=0$ and $y_*=y_{\rm IR}$. The solution to the Einstein
equations is analogous to the 5D case 
\bea\label{7D RS sol A}
A=\sqrt{-\frac{\L_7}{30M^5}}|y|\equiv k|y|
\ena
where now $k$ becomes the scale of the $AdS_7$ geometry and $\L_7$ has to be negative as in the 5D case. The (IR) UV 5--brane tension is (negative) positive and must satisfy the fine--tuning relation
\bea
V_{b_{\rm UV}}=-V_{b_{\rm IR}}=20M^3k=\sqrt{-\frac{40M^5\L_7}{3}}
\ena
However, our interest is now to get a four dimensional graviton zero mode and the relative KK tower, in analogy to the 5D RS model. We then compactify over a two dimensional internal space, taking the
following ansatz for the metric
\bea
\intd^s=a^2(y)\left[\eta_{\m\n}\intd x^\m\intd x^\n+R^2\left(\intd\theta^2_1+\intd\theta^2_2\right)\right]+\intd y^2
\ena
This identifies the internal space with a two torus of equal radii $R$ (a two sphere can also be considered). We used for simplicity the notation $a(y)\equiv\ex^{-2A(y)}$. The metric restricted to the two
dimensional compact space is defined as $\x_{ab}\intd\theta^a\intd\theta^b$, where $a,b,\dots$ run over the two compact directions. 

A complete analysis of the spectrum has been performed in \cite{Bao:2005ni}, I am not giving here a detailed derivation. I will study the solutions for scalar fields, since indeed graviton mode solutions are
obtained as a particular case of the scalar mode wave functions. In addition to the RS KK tower we have to consider the KK modes coming from the torus compactification and the winding modes associated to the
5--branes wrapping around the torus. The former spectrum is labeled by the momenta 
\bea
\x^{ab}k_ak_b=\frac{n_1^2+n_2^2}{a^2R^2}=\frac{\vec n^2}{a^2R^2}
\ena
with integers $n_a$, $a=1,2$. The winding modes arise as particle like states only if the branes are completely wrapped, with two directions around the torus and the remaining around some compact cycle in the
factored $S^4$. We can label them by the winding numbers
\bea
R^2\x_{ab}w^aw^b=R^2a^2\vec w^2
\ena
I have neglected the factors of eleven dimensional Planck length $\ell_{Pl}$ that enters into the contributions to the string world--sheet hamiltonian from the torus KK and winding modes. Winding mode masses
come with an extra $1/\ell_{Pl}^2$ factor, since the world--sheet action has an $\ell_{Pl}$ factor in front, while winding masses squared are quantized in terms of the compactification radius (in units of the
string length) $R^2/\ell_{Pl}$. Putting all together, we can immediately write the equations of motion for a scalar field $\phi(x,\theta,y)$ expanded in RS KK, toroidal compactification KK and winding modes
\bea
\phi_{m,\vec n,\vec w}(x,\theta,y)=\psi_{m,\vec n,\vec w}(y)\ex^{\imm p\cd x}\ex^{\imm k_a \theta^a}
\ena
The four dimensional momenta are on--shell, $p^2=m^2$. The massive scalar field equation then reads
\bea\label{7D RS eom scalar}
\left[\oneover{a^4}\pa_y\left(a^6\pa_y\right)+m^2-\frac{\vec n^2}{R^2}-a^2m_B^2-a^4\frac{R^2\vec w^2}{\ell_{Pl}^4}\right]\phi_{m,\vec n,\vec w}=0
\ena
where $m_B$ is the 7D bulk mass of the scalar field. Equation \refeq{7D RS eom scalar} takes the following form in the conformal frame, changing variables from $y$ to $z=1/ka(y)$ 
\bea\label{7D RS eom scalar z}
\psi''_{m,\vec n,\vec w}-\frac{5}{z}\psi'_{m,\vec n\vec w}+\left[m^2-\frac{\vec n^2}{R^2}\frac{m_B^2}{z^2}-\frac{R^2\vec w^2}{k^4\ell_{Pl}^4z^4}\right]\psi_{m,\vec n,\vec w}
\ena
We can notice in \refeq{7D RS eom scalar} that the warp factor contributes with different powers to the different higher modes. The original bulk squared mass gets an $a^2$ factor w.r.t. the four dimensional
RS mass $m^2$.  The torus KK modes don't carry any additional warping, while winding modes are multiplied by an $a^4$ compared to $m^2$. 

A fundamental remark is in order. Our main interest is to relate the graviton modes appearing in the 7D set--up with the ones we know from the original 5D RS model. The 7D spectrum will have toroidal KK
states and winding states on top of the RS like KK modes. The important result is that graviton states can be deduced from the scalar modes just by setting $m_B=0$ in \refeq{7D RS eom scalar}. It can be
shown that choosing for instance the harmonic gauge for gravitons, each of its components satisfies to the massless scalar field equation, due to the specific boundary conditions. Thus, all the results that
are found for massless scalar fields also apply to gravitons.

From equation \refeq{7D RS eom scalar z} with $\vec n=\vec w=0$ one can deduce that higher RS KK modes are differently spaced with respect to the 5D model. In fact, rewriting \refeq{7D RS eom scalar z}
with the appropriate rescaling $\psi(z)=k^3z^3\tilde\psi(z)$, zero winding and torus KK quantum numbers, we get
\bea
z^2\tilde\psi''_{m,0,0}+z\tilde\psi_{m,0,0}+\left(m^2z^2-\m^2\right)\tilde\psi_{m,0,0}=0
\ena
It is clear from the form of the equation that solutions will be linear combinations of Bessel functions $J_\m(mz)$ and $Y_\m(mz)$, where we defined $\m^2\equiv m_B^2+9$. Hence, graviton RS KK wave functions
will be determined by $J_3(mz)$ and $Y_3(mz)$, while in the 5D case we had $J_2$ and $Y_2$. The boundary conditions $\tilde\psi(z)=0$ on the branes then imply that the KK masses are quantized in terms of the
roots of $J_2=0$ rather than $J_1=0$.
\paragraph{KK modes from the torus}
We should put $\vec w=0$ in \refeq{7D RS eom scalar z} and solve it depending on the sign of the quantity $\s^2\equiv\left(m^2-\frac{\vec n^2}{R^2}\right)$. 

For $\s>0$, the modes represent the tower of toroidal KK states on top of the massive RS KK modes. Solutions are given in terms of the modified Bessel
functions $J_\m$ and $Y_\m$
\bea\label{KK solution 1}
\psi_{m,\vec n,0}(z)=\frac{k^3z^3}{N}\left[j_\m\left(z|\s|\right)+\hat N_{m,\vec n}Y_\m\left(z|\s|\right)\right]
\ena
where the constant $\hat N_{m,\vec n}$ is determined by the boundary conditions, which also give the mass spectrum. More precisely, the roots of $3J_\m(z_{\rm IR}|\s|)+(z_{\rm IR}|\s|J'_\m(z_{\rm IR}|\s|)=0$
label toroidal KK masses. When the toroidal compactification radius is comparable to the $S^1/\Zgr_2$ orbifold radius, $R\gtrsim y_{\rm IR}$, the separation among KK modes \refeq{KK solution 1} is comparable
to the one of RS KK solutions.

If $\s=0$, there is no solution, unless $\m=3$, i.e. $m_B=0$, due to boundary conditions. In this specific case, higher modes take the form
\bea
\psi_{m,mR,0}=\frac{k^3}{N}\left(1+\hat N_{m}z^6\right)
\ena
We then get the KK tower of massive modes over the zero mode. For negative $\s^2$ there is no solution at all. The excitations would be combinations of the $I_\m$ and $K_\m$ Bessel functions, but boundary
conditions yield an impossible equation.
\paragraph{Winding modes}
The excitations of the scalar field corresponding to the branes wrapping the two torus are found by putting $\vec n=0$ in \refeq{7D RS eom scalar z}. Moreover, the comments will be restricted to massless
scalar modes $m_B=0$ --- which is the relevant case to study graviton excitations. The higher wave functions can be calculated numerically. However, the zero mode equation is analytically solvable and gives a
linear combinations of $I_\m$ and $K_\m$ Bessel functions which are not compatible with boundary conditions. This means that for pure winding states there is no zero mode, i.e. no solution exists for $m=0$.
As for the toroidal KK modes, winding modes are new states with respect to the RS KK excitations. We can find a tower of winding modes sitting on each of the higher RS states, except for the zero RS mode. We
moreover remark that in spite of the fact that we should expect very heavy winding excitations when $R\sim y_{\rm IR}$, the double warping factor appearing in \refeq{7D RS eom scalar} in front of the winding
mode term suppresses the value of masses of these modes such that they really are of the order of $1/y_{\rm IR}$.
\vspace{0.5cm}

The 7D RS model that I just analyzed has been studied in my paper \cite{Mazzanti:2007dq} from the cosmological point of view (see chapter \ref{7D RS cosmo}) and also in the spirit of the AdS/CFT
correspondence (in chapter \ref{7D RS dual}). The set--up considered in \cite{Mazzanti:2007dq} is is analogous to the 7D RS proposed in \cite{Bao:2005ni}, except for the scale factors relative to the 2D
internal and 3D extended
space (the compact space is generic), which will be left different in general, and for the time dependence of the metric, due to the cosmological issue. Also, I will focus on the one brane RSII model, taking
the limit $y_{\rm IR}\to \infty$.
\section{Cosmologies in Randall--Sundrum}
Keeping in mind that gravity localization is a success of RS models, I wish to review here what kind of cosmologies arise in this context. The general bulk solution involves the bulk components of the
stress--energy tensor $T^5_5$ and $T^0_5$. However, a complete solution has been derived in \cite{Binetruy:1999ut} in the case of vanishing energy exchange between the brane and the bulk, i.e. $T^0_5=0$. If
we wish to study the implications of brane--bulk energy exchange, it's interesting to look at the cosmological evolution on the brane, i.e. solving the Einstein equations (or Friedmann--like equations)
evaluated on the boundary of the AdS space--time slice \cite{Kiritsis:2002zf}--\cite{Cai:2005qm}. Both of these procedures are summarized in what follows (for brane--world cosmology reviews, see for instance
\cite{Cline:2007yq,Langlois:2002bb,Maartens:2003tw,Brax:2004xh,Csaki:2004ay}).
\subsection{Non conventional vs. conventional cosmology}
Let me first mention some results precedent to RSII cosmology, which display non conventional features of brane--world evolution, with respect to the known General Relativity standard cosmology.
\subsubsection{Flat extra dimensions}
In the model inspired by the Horawa--Witten M theory solution \cite{Horava:1996ma} and proposed by Bin\'etruy, Deffayet and Langlois \cite{Binetruy:1999ut} the fifth dimension is compactified on a circle
which is then orbifolded as in RSI model. The Horawa--Witten solution corresponds to the $S^1/\Zgr_2$ orbifolding of the eleventh dimension in M theory at whose fixed points two 9--branes are located (with
all matter fields confined on these ten dimensional defects). The background in \cite{Binetruy:1999ut} is analogous but in five dimensions rather than eleven. Therefore, it is a RSI configuration with no
cosmological constant in the bulk and with matter placed on the UV brane. The relevant action is given by the five dimensional Einstein--Hilbert term and the four dimensional brane matter action
\bea\label{BDL action}
S=-M^3\int\intd^5x\sqrt{-g}R+\int\intd^4x\sqrt{-\g}\lagra_{b}
\ena
where $g_{MN}$ is the full 5D metric and $\g_{\m\n}$ is the induced metric on the brane. For a global solution, valid in the whole space--time, the presence of the IR brane carrying a non trivial
stress--energy tensor contribution will be explicitly considered, while for the purpose of finding the Friedmann type equation on the brane where matter is assumed to be localized only the action \refeq{BDL
action} is needed. Contrarily to RSI static set--up, the metric must exhibit a time dependence since the aim of the analysis is to compare the cosmological evolution arising in the Horawa--Witten--like model
to conventional cosmology. Thus, we can make the ansatz
\bea\label{BDL ansatz}
\intd s^2=g_{\m\n}\intd x^\m\intd x^\n+f^2\intd y^2=-n^2(t,y)\intd t^2+a^2(t,y)\d_{ij}\intd x^i\intd x^j+f^2(t,y)\intd y^2
\ena
This is a simple ansatz with flat 3D space. A straightforward generalization would be assuming isotropy and homogeneity for the 3D space, but eventually locally curved space. Since all matter fields are
confined on the 3--brane, the relevant part of the metric in order to compute physical quantities to be compared to observations is the induced metric on the brane, i.e. $\g_{\m\n}=g_{\m\n}(t,0)$. Matter can
be assimilated to a perfect fluid and its contribution to the total stress--energy tensor
\bea
T^M_N|_{\rm tot}=T^M_N|_{B}+T^M_N|_b
\ena
(subscript $B$ and $b$ respectively stand for bulk and brane contributions) is given in terms of the matter energy density $\rho$ and pressure $p$
\bea\label{BDL stress parametrization}
T^M_N|_b=\frac{\d(y)}{f}\diag(-\rho,p,p,p,0)
\ena
The bulk stress--energy tensor is generic $T^M_N|_{B}=T^M_N$. We would like to find the solutions to the five dimensional Einstein equations
\bea\label{einstein BDL}
G_{MN}=\oneover{2M^3}T_{MN}|_{\rm tot}
\ena
where the five dimensional Einstein tensor with the ansatz \refeq{BDL ansatz} takes the form
\bea
{G}_{00}&=&3\left\{\frac{\dot a}{a}\left(\frac{\dot a}{a}+\frac{\dot f}{f}\right)-\frac{n^2}{f^2}\left(\frac{a''}{a}+\frac{a'}{a}\left(\frac{a'}{a}-\frac{f'}{f}\right)\right)\right\}\label{ein00}\\
{G}_{ij}&=&\frac{a^2}{f^2}\d_{ij}\left\{\frac{a'}{a}\left(\frac{a'}{a}+2\frac{n'}{n}\right)-\frac{f'}{f}\left(\frac{n'}{n}+2\frac{a'}{a}\right)+2\frac{a''}{a}+\frac{n''}{n}\right\}\nonumber \\
&&+\frac{a^2}{n^2}\d_{ij}\left\{\frac{\dot a}{a}\left(-\frac{\dot a}{a}+2\frac{\dot n}{n}\right)-2\frac{\ddot a}{a}+\frac{\dot f}{f}\left(-2\frac{\dot a}{a}+\frac{\dot n}{n}\right)-\frac{\ddot f}{f}\right\}
\label{einij}\\
{G}_{05}&=&3\left(\frac{n'}{n}\frac{\dot a}{a}+\frac{a'}{a}\frac{\dot f}{f}-\frac{\dot{a}^{\prime}}{a}\right)\label{ein05}\\
{G}_{55}&=&3\left\{\frac{a'}{a}\left(\frac{a'}{a}+\frac{n'}{n}\right)-\frac{f^2}{n^2}\left(\frac{\dot a}{a}\left(\frac{\dot a}{a}-\frac{\dot n}{n}\right)+\frac{\ddot a}{a}\right)\right\}\label{ein55}
\ena
The conservation equation coming from the Bianchi identity for the Einstein tensor $\nabla_MG^M_N=0$, when using the Einstein equation \refeq{einstein BDL}, reads $\nabla_MT^M_N=0$. Therefore we get
\bea\label{BDL conservation}
\dot\rho+3\left(\rho+p\right)\frac{\dot a_o}{a_o}=0
\ena
where have used the parametrization \refeq{BDL stress parametrization}. The obtained equation is the usual four dimensional perfect fluid conservation equation in a FRW universe (the subscript $o$ is put on
quantities evaluated at the brane position $y=0$). From \refeq{einstein BDL}, plugging \refeq{ein00}--\refeq{ein55}, we get the jump equations from the $00$ and $ij$ components
\bea
a_{o^+}'=-a_{o^-}'&=&-\frac{f_o}{12M^3}a_o\rho\label{ap0}\\
n'_{o^+}=-n_{o^-}'&=&\frac{f_o}{12M^3}n_o\left(2\rho+3p\right)\label{np0}
\ena
Using these results, the $05$ component of the Einstein equations yields the conservation equation \refeq{BDL conservation}, while the $55$ component leads to the Friedmann--like equation on the brane
\bea\label{BDL modified Friedmann}
\frac{\ddot a_o}{a_o}+\frac{\dot a_o^2}{a_o^2}=-\oneover{6(12M^3)^2}\rho\left(\rho+3p\right)-\oneover{6M^3f_o^2}T^5_5
\ena
Here we chose the temporal gauge $n_o=1$. Equation \refeq{BDL modified Friedmann} clearly doesn't look like the Friedmann equation that governs conventional cosmology. In particular, we get a quadratic
dependence on the energy density which doesn't agree with the predictions of standard cosmology. We can say more about the scale factor assuming that $p=w\rho$ and neglecting the bulk term $T^5_5$ in
\refeq{BDL modified Friedmann} --- this is achieved in the approximation where the bulk energy density $\rho_B$ satisfies $M^3\rho_B\ll\rho^2$. Hence, one can easily derive from the conservation equation that
the matter energy density behaves as in the usual case
\bea
\rho\propto a_o^{-3(1+w)}
\ena
Nevertheless, the scale factor on the brane gets a different power dependence on time
\bea
a_o(t)\propto t^{\frac{1}{3(1+w)}}
\ena
in comparison to standard cosmology, which yields $a_o\propto t^{2/3(1+w)}$. The modified Friedmann equation \refeq{BDL modified Friedmann} can be written in a form which is more akin to Friedmann equation,
by using the conservation equation. As a result, we get
\bea
H^2=\oneover{6(12M^3)^2}\rho^2+\frac{\cal C}{a^4}
\ena
It thus produces a slower expansion of the universe with respect to what standard cosmology predicts, as we deduced from the comparison of scale factor power dependence.

As I anticipated, the search for a global solution implies that we must impose the presence of a second energy density source in the bulk, which we can assume to be the IR brane placed at the second orbifold
fixed point (the action \refeq{BDL action} is then modified with the addition of the second brane term, associated with a delta function $\d(y-y_{\rm IR})$)% footnote
~\footnote{It isn't necessary to locate the second brane at the orbifold point $y=y_{\rm IR}$. It could in principle be placed at an arbitrary point in the extra dimension, in order to have a consistent
solution to the Einstein equation. This point is here set to be the orbifold fixed point for simplicity. Furthermore, in the infinite volume extra dimension case, i.e. without the $S^1$ compactification,
there is no need to add a second energy source.}.% footnote
The stress--energy tensor contribution from this additional source is parametrized as in \refeq{BDL stress parametrization}, with energy density and pressure respectively $\rho_{\rm IR}$ and $p_{\rm IR}$
\bea\label{BDL stress parametrization IR}
T^M_N|_b=\frac{\d(y-y_{\rm IR})}{f}\diag(-\rho_{\rm IR},p_{\rm IR},p_{\rm IR},p_{\rm IR},0)
\ena
If we find a global solution to the Einstein equation everywhere in the bulk, we can learn how density and pressure on the IR brane are related to matter on the UV brane. Indeed, we make the simplest
ansatz for the scale factor, with static extra dimension and linear functions in $y$, compatible with the $S^1/\Zgr_2$ orbifold and jump equations,
\bea\label{BDL global sol}
a(t,y)&=&a_o(t)\left[1-\frac{f_o}{12M^3}\rho(t)|y|\right]\\
n(t,y)&=&n_o(t)\left[1-\frac{f_o}{12M^3}\left(2+3w\right)\rho(t)|y|\right]\\
f(t,y)&=&f_o \label{BDL global sol f}
\ena
One finds a specific relation between $\rho_{\rm IR}$ and $\rho$, and between $w_{\rm IR}\equiv\rho_{\rm IR}/p_{\rm IR}$ and $w\equiv\rho/p$. First of all, solving the remaining Einstein equations, one
determines $H_o(t)$ (while $n_o(t)$ is arbitrary and corresponds to some arbitrary temporal gauge choice), as well as the conservation equation associated to the UV brane
\bea
\frac{\dot a_o^2}{a_o^2}&=&\oneover{6(12M^3)^2}\rho^2\\
\dot\rho+3(1+w)\frac{\dot a_o}{a_o}\rho&=&0\label{BDL conservation UV}
\ena
The $00$ component of Einstein equations yields a ``topological constraint'' on the matter content of the second brane, once we plug the jump equation solutions. This is due to the compactness of the extra
dimension. The topological constraint reads
\bea
\rho_{\rm IR}&=&-\frac{\rho}{1-\frac{f_o}{24M^3}\rho}\\
\left(2+3w_{\rm IR}\right)&=&\frac{1-\frac{f_o}{24M^3}\rho}{1+\frac{f_o}{24M^3}(2+3w)\rho}(2+3w)
\ena
Substituting these equations into the conservation equation \refeq{BDL conservation UV}, the IR conservation equation automatically holds
\bea\label{BDL topological constraint}
\dot\rho_{\rm IR}+3(1+w_{\rm IR})\frac{\dot a_{\rm IR}}{a_{\rm IR}}\rho_{\rm IR}&=&0\label{BDL conservation IR}
\ena
More generally, admitting $f(t,y)$ to be linear in $y$ by making the ansatz $f=f_o+2|y|(f_{\rm IR}-f_o)$ assuming $f_o$ still to be constant and leaving $a,n$ undetermined, one gets to the following
constraints
\bea
\rho a_o&=&-\rho_{\rm IR}a_{\rm IR}\\
(2\rho+3p)n_o&=&-(2\rho_{\rm IR}+3p_{\rm IR})n_{\rm IR}
\ena
This reduces to \refeq{BDL topological constraint} in the static extra dimension limit $f=f_o$. The interesting aspect of the global solution \refeq{BDL global sol}--\refeq{BDL global sol f} is that it shows
how global behaviors of the bulk dynamics are determined by boundary conditions, i.e. by the metric and matter contents evaluated on the branes. This could be interpreted as a manifestation of the holographic
principle, which is illustrated in more detail in the context of the AdS/CFT correspondence. In particular, 5D RSII and 7D RSII are analyzed in section \ref{adscft in RS} and \ref{7D RS dual}, referring to my
publication \cite{Mazzanti:2007dq}. However, the computation suffers from instabilities, which I will point out later on.

A more generic covariant approach can be applied, allowing to split equations along the brane and transverse to the brane, for an arbitrary hypersurface. For a detailed derivation and specific examples one
can read \cite{Brax:2004xh,Maartens:2003tw}. The solution I just described can be derived specifying the covariant approach to the Bin\'etruy--Deffayet--Langlois (BDL) model. The basis of this geometrical
approach are Gauss--Codazzi equations. They relate the four dimensional curvature Riemann tensor to the five dimensional one and the extrinsic curvature, and determine the variation of the extrinsic
curvature itself along the direction transverse to the brane. Namely one has to define the induced metric on the hypersurface. The extrinsic curvature is a function of the five dimensional metric $g_{MN}$ and
of the unit normal vector $n^M$ as $\hat g_{MN}=g_{MN}-n_Mn_N$, $K_{MN}=\hat g_M^R\nabla_Rn_N$ (where $\nabla_M$ is the five dimensional covariant derivative and a hat denotes objects constructed from the
induced metric). The Gauss and Codazzi equations then read 
\bea
\hat R_{MNRS}&=&R_{ABCD}\hat g_M^A\hat g_N^B\hat g_R^C\hat g_S^D+2K_{M[R}K_{S]N}\\
\hat\nabla_NK^N_M-\hat\nabla_MK&=&R_{AB}\hat g_M^An^B
\ena
For arbitrary hypersurfaces they imply that the four dimensional Einstein tensor has the following expression
\bea\label{covariant 4D einstein}
\hat G_{MN}&=&\frac{2}{3}\left[G_{AB}\,\hat g_M^A\hat g_N^B+\left(G_{AB}\,n^An^B-\oneover{4}G\right)\hat g_{MN}\right]+\\
&&KK_{MN}-K^A_MK_{NB}-\oneover{2}\left(K^2-K^{AB}K_{AB}\right)\hat g_{MN}-E_{MN}
\ena
where $E_{MN}\equiv C_{MANB}\,n^Sn^B$ is the Weyl tensor projection orthogonal to $n^M$. Moreover, since the brane stress--energy tensor is proportional to a delta function $T_{MN}|_b=\hat T_{MN}\d(y)$, the
jump of the extrinsic curvature is determined by its value, according to
\bea\label{covariant 4D jump}
\left[K_{MN}\right]=-\oneover{2M^3}\left(\hat T_{MN}-\oneover{3}\hat T\,\hat g_{MN}\right)
\ena
with the jump function defined as $\left[f\right]\equiv\lim_{\eps\to0}\left[f(y+\eps)-f(y-\eps)\right]$. Furthermore, if there is no energy exchange between the bulk and the brane, the usual four dimensional
conservation equation holds $\hat\nabla^N\hat T_{MN}=0$. All the informations to get the 4D Einstein equation on the brane are contained in \refeq{covariant 4D einstein}. We have to plug the jump equation
\refeq{covariant 4D jump} and the 5D Einstein equations, which gives $G_{MN}$ in terms of the bulk and brane stress--energy tensors.
\subsubsection{Bulk cosmological constant and brane tension: RSI/RSII}
The geometric approach just described may be straightforward applied to a bulk with non null cosmological constant and branes with non zero tension. Introducing these cosmological terms opens the
possibility to restore conventional cosmology at late times, contrarily of the unavoidable non conventional behavior of the previously analyzed BDL model. We assume that the 5D Einstein equation
contains a 5D cosmological constant $\L_5$ and that the brane stress--energy tensor can be decomposed into a tension contribution and a matter stress tensor $\hat T_{MN}=\t_{MN}-V\hat g_{MN}$. Hence, the four
dimensional Einstein equation takes the simple form
\bea
\hat G_{MN}=\oneover{2M_{Pl}^2}\t_{MN}-\L_4\hat g_{MN}+\oneover{(2M^3)^2}\p_{MN}-E_{MN}
\ena
Here we defined
\bea
\p_{MN}&=&\oneover{12}\t\t_{MN}-\oneover{4}\t_{MA}\t^A_N+\oneover{8}\t_{AB}\t^{AB}\hat g_{MN}-\oneover{24}\t^2\hat g_{MN}\\
\L_4&=&\oneover{12M^3}\left(\L_5+\frac{V^2}{12M^3}\right) \label{BDEL 4D cosmo const}\\
M_{Pl}^2&=&\frac{24M^6}{V}
\ena

To be more specific, I will consider the scenario studied in \cite{Binetruy:1999ut}. The set--up is analogous to the BDL model, with the addition of a bulk contribution to the total stress--energy tensor
(namely a cosmological constant) and a brane tension contribution to the brane stress--energy tensor. In other words, I am going to describe the cosmology of RSI. The background metric is still given by the
ansatz \refeq{BDL ansatz}, the stress--energy tensor is parametrized as in \refeq{BDL stress parametrization} but now $\rho\to V+\rho$ and $p\to -V+p$ to account for the brane tension $V$. Indeed,
$T^M_N|_{b}=T^M_N|_{b,m}+T^M_N|_{b,v}$ with
\bea
T^M_N|_{b,m}&=&\frac{\d(y)}{f}\diag(-\rho,p,p,p,0)\\
T^M_N|_{b,v}&=&\frac{\d(y)}{f}\diag(-V,-V,-V,-V,0)
\ena
The bulk stress--energy tensor for a pure cosmological constant contribution is parametrized as 
\bea
T_N^M|_B=\diag(-\L_5,\dots,-\L_5)
\ena 
The $05$ component of the bulk stress--energy tensor, representing the brane--bulk energy exchange, is assumed to vanish. The Einstein tensor solving the five dimensional Einstein equation \refeq{einstein BDL}
has components \refeq{ein00}--\refeq{ein55}. A further assumption, which should however be considered more carefully% footnote
~\footnote{We will make some explicit remarks on the consequences of the staticity (or radion stabilization) assumption at the end of the derivation --- based on \cite{Csaki:1999mp}.}% footnote
, is the staticity of the extra direction $y$, i.e. $f(t,y)=f_o$ and with a gauge choice $f_o=1$.

The Friedmann--like equation that solves the Einstein equation can be obtained in analogy to the BDL solution \cite{Binetruy:1999ut}. We get the same junction equations \refeq{ap0}--\refeq{np0} and the following four
dimensional expression for the Hubble parameter on the brane
\bea\label{BDEL modified Friedmann}
H_o^2\equiv\frac{\dot a_o^2}{a_o^2}=\oneover{144M^6}\rho\left(\rho+2V\right)+\frac{\mathcal C}{a_o^4}+\l_{\rm RS}
\ena
Here $\l_{\rm RS}$ is the effective 4D cosmological constant on the brane and is in fact exactly defined as $\l_{\rm RS}=\L_4$, see \refeq{BDEL 4D cosmo const}. The fine--tuning required by consistency of RS
model makes $\l_{\rm RS}$ vanish. The constant $\mathcal C$ is an integration constant since the Friedmann equation comes from integrating a second order equation. We can easily implement the curvature term
$-k/a_o^2$ on the r.h.s. of \refeq{BDEL modified Friedmann} whenever the spatial curvature on the brane doesn't vanish. We now have to notice that conventional cosmology is restored in this scenario when 
matter energy density is small compared to brane tension $\rho\ll V$. In fact, we get that the non standard quadratic term in \refeq{BDEL modified Friedmann} can be neglected in this limit and linear
dependence of $H^2$ on $\rho$ alone is recovered. 

An explicit solution was found in \cite{Binetruy:1999ut}, putting $\mathcal C=0$ and using the usual conservation equation \refeq{BDL conservation} with constant equation of state $p=w\rho$. Integrating \refeq{BDEL
modified Friedmann} with $\l_{\rm RS}=0$ then yields
\bea
a_o(t)=a_i\left\{\frac{\rho_i}{2M^3}\left[\frac{(1+w)^2}{16M^3}Vt^2+\frac{(1+w)}{2}t\right]\right\}^{\frac{1}{3(1+w)}}
\ena
At early times, such a solution describes a non conventional evolution of the universe characterized by $a\sim t^{1/3(1+w)}$. Late time evolution follows the standard cosmology behavior, i.e. $a\sim
t^{2/3(1+w)}$, as we expected. If we demand $\l_{\rm RS}$ not to satisfy the RS fine--tuning, the solution displays three phases --- if $\l_{\rm RS}$ is small enough ---, since
\bea
a_o(t)=a_i\left\{\frac{\rho_i}{\tilde\L}\left[\frac{V}{\tilde\L}\left[\cosh\left(H_\L\right)-1\right]+\sinh\left(H_\L\right)\right]\right\}^{\oneover{3(1+w)}}
\ena
After the phase transition from non conventional to conventional cosmological evolution we get an exponential behavior weighted by $H_\L\equiv3(1+w)\tilde\L/2M^3$ with
$\tilde\L\equiv\sqrt{(12M^3)^{3}\l_{\rm RS}}$.  Thus the present acceleration of the universe is obtained at the expenses of a small violation of the fine--tuning for RS solution.
\paragraph{Radion stabilization and RSII}
In the procedure just described there is however an issue that cannot be ignored. As I anticipated, the hypothesis of static extra dimension has to be considered more carefully. Assuming $f$ to be constant
means that we are fixing the size of the extra dimension. There is nevertheless a problem in imposing staticity without a true mechanism to stabilize the radius of extra dimension. Of course, this problem
arises only when the extra dimension is compactified. In \cite{Csaki:1999mp} the authors show that independently of the radion stabilization mechanism that one uses, standard cosmology is restored (see
also \cite{CEHS,GW1,CEGH} for previous and subsequent discussions). The requirement of stabilizing the radion (i.e. the scalar field $f(t,x^\m,y)$), without having a suitable potential (the details are not
crucial --- Goldberg--Wise mechanism has been used as an example), forces the matter contents of the two branes to be related by some constraint (for instance, we derived the relations \refeq{BDL topological
constraint}). Indeed, the branes would tend to get infinitely separated, falling into the RSII configuration, if the size of the extra dimension was not fixed. Fixing the energy densities artificially
avoids this instability. However, when we introduce a potential for the radion there is no need to constrain the matter contents of the two branes and conventional cosmological evolution is restored. As I
mentioned, the RSII model avoids the radion stabilization problem and cosmology is given by the Friedmann equation obtained in the RSI set--up \refeq{BDEL modified Friedmann}, which is now fully justified
because we take the limit of infinite size for the extra dimension $y_{\rm IR}\to\infty$.
\subsection{Brane--bulk energy exchange and late time acceleration}\label{5D RS cosmo}
The analysis of RSII cosmology with the introduction of a non zero value for brane--bulk energy exchange is generally achieved by trading the possibility of studying the global solution to the Einstein
equation in the bulk in a rather simple way. Here I will proceed in studying the cosmological evolution of a RSII universe from the brane point of view with brane--bulk energy exchange --- mainly following
\cite{Kiritsis:2002zf,Kiritsis:2005bm}.

The set--up is RSII, with a bulk lagrangian ${\cal L}_{m,B}$ inserted
\bea\label{KKTTZ action}
S=\int\intd^5x\sqrt{-g}\left(M^3R-\L_5+{\cal L}_{m,B}\right)+\int\intd^4x\sqrt{-\hat g}\left(-V+{\cal L}_{m,b}\right)
\ena
Following my notation, $\g_{\m\n}$ is as usual the induced metric on the brane, $V$ is the positive brane tension (referring to the RS UV brane), ${\cal L}_{m,b}$ is the brane matter lagrangian which will
contribute to the matter perfect fluid stress--energy tensor. The bulk cosmological constant is $\L_5$ and $M$ is the five dimensional Planck mass. The metric is given by the time dependent ansatz
\bea
\intd s^{2}=-n^{2}(t,y)\intd t^{2}+a^{2}(t,y)\z_{ij}\intd x^{i}\intd x^{j}+f^{2}(t,y)\intd y^{2}
\ena
generalized to the case of an eventually non spatially flat brane. The three dimensional metric $\z_{ij}$ is maximally symmetric with spatial curvature $k=-1,0,1$, respectively denoting locally spherical,
flat or hyperbolic spaces. The brane is still placed at $y=0$ and the $\Zgr_2$ symmetry $y\leftrightarrow-y$ must be taken into account throughout the calculations. We wish to solve five dimensional Einstein
equations $G_{MN}=\oneover{2M^3}T_{MN}|_{\rm tot}$. The l.h.s. is given in terms of the metric by 
\bea
{G}_{00}&=&3\left\{\frac{\dot a}{a}\left(\frac{\dot a}{a}+\frac{\dot f}{f}\right)-\frac{n^2}{f^2}\left(\frac{a''}{a}+\frac{a'}{a}\left(\frac{a'}{a}-\frac{f'}{f}\right)\right)+k\frac{n^2}{a^2}\right\}
\label{kein00}\\
{G}_{ij}&=&\frac{a^2}{f^2}\z_{ij}\left\{\frac{a'}{a}\left(\frac{a'}{a}+2\frac{n'}{n}\right)-\frac{f'}{f}\left(\frac{n'}{n}+2\frac{a'}{a}\right)+2\frac{a''}{a}+\frac{n''}{n}\right\}\non\\
&&+\frac{a^2}{n^2}\z_{ij}\left\{\frac{\dot a}{a}\left(-\frac{\dot a}{a}+2\frac{\dot n}{n}\right)-2\frac{\ddot a}{a}+\frac{\dot f}{f}\left(-2\frac{\dot a}{a}+\frac{\dot n}{n}\right)-\frac{\ddot f}{f}\right\}
-k\z_{ij}\label{keinij}\\
{G}_{05}&=&3\left(\frac{n'}{n}\frac{\dot a}{a}+\frac{a'}{a}\frac{\dot f}{f}-\frac{\dot{a}^{\prime}}{a}\right),\label{kein05}\\
{G}_{55}&=&3\left\{\frac{a'}{a}\left(\frac{a'}{a}+\frac{n'}{n}\right)-\frac{f^2}{n^2}\left(\frac{\dot a}{a}\left(\frac{\dot a}{a}-\frac{\dot n}{n}\right)+\frac{\ddot a}{a}\right)-k\frac{f^2}{a^2}\right\}
\label{kein55}
\ena
The r.h.s. of Einstein equations is decomposed into the bulk $B$ and brane $b$, matter $m$ and vacuum $v$, contributions. We parametrize it as
\bea
T^M_N|_{\rm tot}&=&T^M_N|_{v,b}+T^M_N|_{v,B}+T^M_N|_{m,b}+T^M_N|_{m,B}\non\\
T^M_N|_{v,b}&=&\frac{\delta(y)}{f}\,\diag\left(-V,-V,-V,-V,0\right)\non\\
T^M_N|_{v,B}&=&\diag\left(-\L_5,-\L_5,-\L_5,-\L_5,-\L_5\right)\non\\
T^M_N|_{m,b}&=&\frac{\delta(y)}{f}\,\diag\left(-\rho,p,p,p,0\right)\non\\ 
T^M_N|_{m,B}&=&T^M_N 
\ena
Plugging all these pieces of information into the $00$ and $ij$ components of Einstein equations and integrating, we get the jump equations depending on bulk cosmological constant and brane tension
\bea\label{KKTTZ junction}
a'_+=-a'_-&=&-\frac{f}{12M^3}\left(V+\rho\right)\non\\
n'_+=-n'_-&=&-\frac{f}{12M^3}\left(-V+2\rho+3p\right)
\ena
where all the quantities are evaluated at the position of the brane $y=0$, but the $o$ subindices have been dropped to simplify notation (from now on all expressions will be on the brane). Then, the $05$ and
$55$ components of Einstein equations, once we substitute the junction equations \refeq{KKTTZ junction}, give
\bea\label{KKTTZ conservation}
\dot\rho+3\frac{\dot a}{a}\left(\rho+p\right)&=&\frac{2n^2}{f}T_{05}\\
\frac{\ddot a}{a}+\frac{\dot a^2}{a^2}-\frac{\dot a}{a}\frac{\dot n}{n}+\frac{k}{a^2}&=&-\frac{n^2}{144M^6}\left[\rho\left(\rho+3p\right)-V\left(\rho-3p\right)\right]+\non\\
&&+\frac{n^2}{6M^3}\left(\L_5+\frac{V^2}{12M^3}\right)-\frac{n^2}{6M^3}T^5_5\label{KKTTZ pre Friedmann}
\ena
Now we can go in a temporal gauge where $n=1$, simplifying the above equations. Moreover, since the model has an infinite volume extra dimension, there is no radion stabilization problem. We can thus set
$f=1$ to further simplify equations. The effective cosmological constant on the brane
\bea\label{4D RS fine tuning}
\l_{\rm RS}\equiv\oneover{12M^3}\left(\L_5+\frac{V^2}{12M^3}\right)
\ena
that appears in \refeq{KKTTZ pre Friedmann} must be fine--tuned to zero in a pure RS context, where also $T^0_5=T^5_5=0$. With these conditions, the RS vacuum is recovered since matter is absent in the bulk.

The scope of the analysis of this section is to describe the brane dynamics in the presence of energy exchange with the bulk. Hence, $T^0_5$ must be non zero. However, with a reasonable assumption on the bulk
matter contents, we can ignore the bulk ``self--interaction'', term $T^5_5$. If the ratio of bulk matter to bulk vacuum contribution is negligible with respect to brane matter relative to
brane vacuum energy, the bulk diagonal component $T^5_5$ is much smaller than the other terms in the r.h.s. of \refeq{KKTTZ pre Friedmann}. The assumption can be expressed in formulae by
\bea
\bigg|\frac{T^{(\mathrm{diag})}_{m,B}}{T^{(\mathrm{diag})}_{v,B}}\bigg|\ll\bigg|\frac{T^{(\mathrm{diag})}_{m,b}}{T^{(\mathrm{diag})}_{v,b}}\bigg|
\ena
At this point, the authors of \cite{Kiritsis:2002zf} introduce a mirage energy density $\chi$ determined by the bulk dynamics encoded in the parameter $T\equiv2T^0_5$. Integrating equation \refeq{KKTTZ pre
Friedmann} using the conservation equation \refeq{KKTTZ conservation} and suitably defining the mirage density one comes to the following set of equations
\bea
H^2&\equiv&\frac{\dot a}{a}=\oneover{144M^6}\rho^2+\frac{V}{72M^6}\left(\rho+\chi\right)-\frac{k}{a^2}+\l_{\rm RS}  \label{KKTTZ exchange Friedmann}\\
\dot\chi+4H\chi&=&\left(\frac{\rho}{V}+1\right)T  \label{KKTTZ exchange mirage}\\
\dot\rho+3H\left(\rho+p\right)&=&-T  \label{KKTTZ exchange conservation}
\ena
The first equation, eq. \refeq{KKTTZ exchange Friedmann} exactly reduces to the modified Friedmann equation derived in the RSI model \refeq{BDEL modified Friedmann} in the limit of zero energy exchange, when
$T\to0$. In this same limit, in fact, the mirage density obeys to a four dimensional free radiation conservation equation yielding $\chi=\chi_i/a^4$. If we rename the constant $\chi_i$ as ${\cal C}$, we then
precisely get \refeq{BDEL modified Friedmann}. Conservation equation for the matter energy density $\rho$ immediately gives the usual conservation equation in a FRW universe for $T=0$. The conventional
cosmology regime is restored at late times, if no energy exchange is present and $\chi_i=0$. The remaining of this section is devoted to the analysis of the consequences of non trivial brane--bulk energy
exchange in the brane cosmological evolution. 

I mention here that a physical desirable behavior of brane--bulk energy exchange parameter would imply a change of regime, from energy \emph{influx} to energy \emph{inflow}. We could eventually sketch
a scenario where energy is initially directed from the bulk onto the brane, if the brane comes with a low energy density initial condition. After a phase of influx, equilibrium should be reached for a
specific value of the matter energy density on the brane and energy should start to reverse its flow direction. Thus, energy outflow should start, causing a decrease in the brane matter energy density.
However, in most of the future discussions, the energy exchange parameter will be assumed of the form $T=A\rho^\n$, where $A$ is constant. Therefore, it won't be possible to directly detect any change of
regime, from negative to positive brane--bulk energy exchange. the analysis will be focused instead on the study of evolution phases where either influx or outflow takes place.
\subsubsection{Small densities}
When matter energy density is much smaller that the brane tension $\rho\ll V$ the set of equations \refeq{KKTTZ exchange Friedmann}--\refeq{KKTTZ exchange conservation} can be rewritten as
\bea
H^2&\equiv&\frac{\dot a}{a}=\frac{V}{72M^6}\left(\rho+\chi\right)-\frac{k}{a^2}+\l_{\rm RS}  \label{KKTTZ exchange Friedmann small}\\
\dot\chi+4H\chi&=&T  \label{KKTTZ exchange mirage small}\\
\dot\rho+3(1+w)H\rho&=&-T  \label{KKTTZ exchange conservation small}
\ena
where we also used the constant equation of state for the matter perfect fluid $p=w\rho$. 
\paragraph{Radiation dominated universe: explicit solutions}
An exact solution for $H^2$ can be found for \emph{radiation dominated} matter on the brane, $w=1/3$. In fact, the Hubble parameter can be written independently of $T$
\bea
H^2&=&\frac{V}{72M^6}\left(\rho_i+\chi_i\right)\frac{a_i^4}{a^4}-\frac{k}{a^2}  \label{KKTTZ Friedmann small}\\
\left(\rho+\chi\right)&=&\left(\rho_i+\chi_i\right)\frac{a_i^4}{a^4}
\ena
We note that we need three initial conditions in order to solve the system, differently from conventional cosmology scenario, since the additional condition is enforced by the presence of energy exchange. We
will consider RS fine--tuning to be valid $\l_{\rm RS}=0$, unless we explicitly specify it differently.

Although Friedmann equation \refeq{KKTTZ Friedmann small} follows the radiation dominated universe conventional evolution matter energy density evolution is modified, putting the initial condition $\chi_i=0$.
If we consider the case of energy \emph{outflow}, $T>0$, and parametrize energy exchange linearly in $\rho$ by $T=A\rho$, with $A>0$, it is easy to find
\bea
\rho&=&\rho_i\frac{a_i^4}{a^4}\ex^{-At}\\
\chi&=&\rho_i\frac{a_i^4}{a^4}\left(1-\ex^{-At}\right)
\ena
This solution indeed shows that the energy density $\rho$ decays more rapidly than $a^{-4}$ and that at late times the evolution is dominated by the mirage contribution
$\chi\stackrel{\scriptscriptstyle{t\to\infty}}{\sim}a^{-4}$. So, the radiation dominated behavior of the Hubble parameter comes from the balancing of matter and mirage energy densities, in which matter
density prevails at early times (remind that we imposed the initial condition $\chi_i=0$) while mirage density becomes important at late times. The acceleration factor for solution \refeq{KKTTZ Friedmann
small} is simply given by 
\bea
q\equiv\frac{\ddot a}{a}=-\frac{V}{72M^6}\left(\rho_i+\chi_i\right)\frac{a_i^4}{a^4}
\ena
Friedmann equation \refeq{KKTTZ Friedmann small} constraints $q$ to be negative when space is locally spherical or flat, $k\geq0$ so that we would only get decelerated expansion. However, if we consider a
locally hyperbolic space, we obtain $\left(\rho_i+\chi_i\right)<0$. Hence, positive acceleration and loitering universe, which is nevertheless a non realistic prediction for the eternal acceleration evolution
phase, is possibly achieved.

The effects of mirage energy, though, are present even in the zero energy exchange limit. These are manifest in equation \refeq{BDEL modified Friedmann}, where $T=0$ and ${\cal C}$ can be taken to be
different from zero, i.e. $\chi_i\neq0$. Then the Hubble parameter gets an additional radiation term $\mathcal C/a^4$, without brane--bulk energy exchange.  Analogous considerations on the consequences of the
mirage evolution can be made in the case of \emph{dust} dominated universe, $w=0$. With zero initial condition for $\chi$, $\chi_i=0$, we obtain the following integral expressions
\bea
\rho&=&\rho_i\frac{a_i^3}{a^3}\ex^{-At}\\
\chi&=&\rho_i\frac{a_i^4}{a^4}\left[A\int\intd t\frac{a(t)}{a_i}\ex^{-At}\right]
\ena
We still have that at late times mirage density radiation evolution dominates, since $\chi\stackrel{\scriptscriptstyle{t\to\infty}}{\sim}a^{-4}$. Initially only matter density is present, instead.

%\paragraph{Influx}
The Friedmann equation \refeq{KKTTZ Friedmann small} holds also in the case of energy \emph{influx}, as it is independent on the form of $T$. Moreover, conservation equation \refeq{KKTTZ exchange
conservation} takes the form
\bea\label{KKTTZ small influx}
\dot\rho+\frac{2}{t}\rho=-T(\rho)
\ena
when we use the solution for $H$ in a spatially flat universe. This immediately shows that, if we can neglect the linear term, matter energy density would grow unbounded for a negative energy exchange
parameter, until the small density approximation breaks up at $\rho\sim V$. The solution becomes unreliable and we should refer to the generic $\rho$ analysis. The general solution to \refeq{KKTTZ small
influx} depends on the specific form of the energy exchange parameter. If we suppose a power dependence, $T=A\rho^\n$, with $A<0$, we find the exact solution for $\n\neq1,3/2$
\bea
\rho^{1-\n}=\left(\rho_i\frac{t_i^2}{t^2}\right)^{1-\n}+\frac{1-\n}{3-2\n}\left(1-\frac{t_i^{3-2\n}}{t^{3-2\n}}\right)|A|t
\ena
while for $\n=1$ we get
\bea
\rho=\rho_i\frac{t_i^2}{t^2}\ex^{|A|(t-t_i)}
\ena
and for $\n=3/2$
\bea
\rho^{-\oneover{2}}=\rho_i^{-\oneover{2}}\frac{t_i^{-\oneover{2}}}{t^{-\oneover{2}}}-\sqrt{|A|}\oneover{2}t^{\oneover{2}}\log\frac{t}{t_i}
\ena
We can thus conclude that $\rho$ diverges at infinity for $0\leq\n<1$ as $\rho\sim t^{1/(1-\n)}$, while it exponentially diverges when $\n=1$. A divergence at finite time must occur for $1<\n<3/2$ at
\bea
t_\infty^{3-2\n}=t_i^{3-2\n}+\frac{3-2\n}{1-\n}\oneover{|A|}\left(\rho_it_i^2\right)^{1-\n}
\ena
as well as for $\n=3/2$ and for $\n>3/2$ if
\bea
\left(\rho_it_i^2\right)^{1-\n}-\frac{2\n-3}{\n-1}|A|t_i^{3-2\n}<0
\ena
Otherwise, in the range $\n>3/2$ with no divergence at finite time, the energy density is found to decrease at infinity, since $\rho\sim t^{-2}$ when $t\to\infty$.
\paragraph{Accelerated expansion as an attractor}
An insightful issue in the small density regime is the possibility of finding inflationary fixed point solutions to the system of equations \refeq{KKTTZ exchange Friedmann}--\refeq{KKTTZ exchange
conservation} which governs the cosmological evolution. Variables at the fixed point $\sH,\srho,\schi$ must solve
\bea\label{KKTTZ system small}
\sH^2&=&\frac{V}{72M^6}\left(\srho+\schi\right)\\
4\sH\schi&=&\sT\\
3(1+w)\sH\srho&=&-\sT
\ena
where $\sT=T(\srho)$. The solution to this system carries a $T\to-T$, $H\to-H$ symmetry, implying that influx fixed point solution gives expansion, while outflow yields contraction. We obtain the two roots
\bea\label{KKTTZ fix sol small}
\sH&=&\eta\sqrt{1-3w}\frac{V^{1/2}}{6M^3}\srho^{1/2}  \non\\
\schi&=&-\frac{3}{4}(1+w)\srho\\
\sT&=&-\eta(1+w)\sqrt{1-3w}\frac{V^{1/2}}{2M^3}\srho^{3/2}  \non
\ena
with $\eta=\pm1$. A real solution to these equations exists for $-1<w<1/3$. The $\eta=+1$ root describes an exponentially expanding universe, where the brane energy density dilution by expansion is
compensated by the energy influx onto the brane. In the cosmological model determined by the $\eta=-1$ root the situation is inverted, since the contraction of the universe is compensated by the energy
outflow and energy density on the brane remains constant. Of course, the trivial fixed point $\sH=\sT=0,\srho=-\schi$ is a solution to \refeq{KKTTZ system small}, which reduces to $\sH=\srho=\schi=0$ when we
assume $T=A\rho^\n$. 

The stability of the critical point solutions \refeq{KKTTZ fix sol small} is determined by only one parameter
\bea
\tn=\left.\frac{\intd\log|T|}{\intd\log\rho}\right|_\star
\ena
other than the equation of state parameter $w$. The linearized system for the perturbations $\d\rho,\d\chi$ has eigenvalues
\bea\label{KKTTZ eigen small}
\l_{1,2}=\oneover{6(1+w)}\frac{\sT}{\srho}\left[7-3\tn+3(1-\tn)\sqrt{\left(7-3\tn+3(1-\tn)\right)^2-24(3-2\tn)(1+w)}\right]\non\\
\ena
Clearly, in the case on energy influx at the critical point, both the eigenvalues have negative real part when $0\leq\tn<3/2$ in the region $-1<w<1/3$, implying that the solution is stable for those values of
$\tn$. We can also have spiral fixed points if the imaginary part of $\l_{1,2}$ doesn't vanish, within the range $-1/2<w<1/3$. Energy outflow always gives unstable or saddle points, as one can deduce by
plugging $\sT>0$ in \refeq{KKTTZ eigen small}. Moreover, the $w=-1$ fixed point is characterized by zero energy exchange and zero mirage density. It represents in fact the usual inflationary critical point
for a matter energy density being a root of $T(\rho)=0$. We note that supposing $T=A\rho^\n$ implies $\tn=\n$ in the formulae above.
\subsubsection{Generic densities}
We now consider the full set of equations \refeq{KKTTZ exchange Friedmann}--\refeq{KKTTZ exchange conservation}, with $\l_{\rm RS}=0$ and $p=w\rho$
\bea
H^2&\equiv&\frac{{\dot a}^2}{a^2}=\a\rho^2+2\a V\left(\rho+\chi\right)-\frac{k}{a^2}  \label{KKTTZ Friedmann gen}\\
\dot\chi+4H\chi&=&\left(\frac{\rho}{V}+1\right)T  \label{KKTTZ mirage gen}\\
\dot\rho+3\left(1+w\right)H\rho&=&-T  \label{KKTTZ conservation gen}
\ena
with $\a=1/144M^6$.  Some analytical comments can be made keeping $T$ generic, without specifying any dependence on the matter energy density. Since we eventually wish to find an realistic cosmological
evolution description, we have to look for acceleration, describing the present evolution of the universe detected by observations. 

I already observed that the system of equations \refeq{KKTTZ exchange Friedmann}--\refeq{KKTTZ exchange conservation} reduces to the modified Friedmann equation \refeq{BDEL modified Friedmann} plus standard
conservation equation arising in the model proposed and studied by Bin\'etruy, Deffayet, Ellwanger and Langlois \cite{Binetruy:1999ut}, in the zero energy exchange limit. This could lead to a late time accelerated era
for some negative ${\cal C}$, while at earlier times, however, the decelerated evolution is dominated by a $\rho^2$ dependence of the Hubble parameter $H^2$.

As a consequence of positive definiteness of r.h.s. of equation \refeq{KKTTZ Friedmann gen} we get that the acceleration factor $q$ is constrained to be limited by a parabola, as a function of
$\rho$. The quantity $q$ is independent of the actual form of energy exchange and can be written as
\bea
q=\dot H+H^2=-\a(2+3w)\rho^2-\a(1+3w)V\rho-2\a V\chi-\frac{k}{2a^2}
\ena
Using that $\a\left(\rho+2V\right)\rho+2\a V\chi-\left(k/a^2\right)>0$ we get that $q$ must lay below the curve
\bea
q<-\a(1+3w)\rho^2+\a(1-3w)V\rho-\frac{3k}{2a^2}
\ena
Hence, if $k\geq0$, $w>-1/3$ the universe is decelerating when $\rho/V>(1-3w)/(1+3w)$. Besides, if $w>1/3$ and the universe is spatially locally flat or spherical, we get deceleration at all times. In order
to get acceleration for a positively curved (or flat) space, supposing $w\ge-1/3$, we have to require negative mirage energy density $\chi<0$. For $k=0$ there is no other source of compensation for this
negative energy density than matter, so that we would remain left with a larger amount of dark matter than the one revealed by experimental data. In other words, dark plus visible matter density would be much
greater than the measured one. Indeed, since $\O_\chi<-1$, we would have to get $\O_{\rm CDM}+\O_{B}>2$. Only a locally hyperbolic space could eventually suppress the disagreement between these predictions
and observations, as negative curvature can play the role of an extra energy density source.

For a generic outflow configuration we immediately infer from the (non)conservation equation \refeq{KKTTZ conservation gen} that $\rho$ is a monotonically decreasing function in an expanding universe. This
agrees with the small energy density explicit solution previously described.

An accurate analysis is performed in \cite{Kiritsis:2002zf}. I won't discuss all the details here but give the main results and physical conclusions. As in the small energy density approximation, one can find all 
critical points of the differential system of equations \refeq{KKTTZ Friedmann gen}--\refeq{KKTTZ conservation gen}.
\paragraph{Accelerating fixed point solutions} 
There are non trivial critical points that solve the following algebraic set of equations in a spatially flat universe $k=0$
\bea\label{KKTTZ system gen}
\sH^2&=&\a\srho^2+\a V\left(\srho+\schi\right)\\
4\sH\schi&=&\sT\\
3(1+w)\sH\srho&=&-\sT
\ena
The symmetry relating expansion to influx and, viceversa, contraction to outflow, survives in the generic density case. Hence, inflationary solutions would necessarily imply energy flowing from the bulk onto
the brane. More precisely, static solutions are given in terms of $\srho$ as
\bea
\a(1+3w)\srho^2-\a(1-3w)V\srho+\frac{2|\sT|^2}{9(1+w)^2\srho^2}=0\label{KKTTZ critical rho gen}
\ena\bea
\sH&=&-\frac{\sT}{3(1+w)\srho} \quad\Rightarrow\quad \sqq=\frac{|\sT|^2}{9(1+w)^2\srho^2}\\
\schi&=&-\frac{3(1+w)}{4}\srho
\ena
The number of roots of equation \refeq{KKTTZ critical rho gen} depends on the specific form of the energy exchange. If we admit the validity of the ansatz $T=A\rho^\n$, we have a minimum number of roots for
$\n=2$, i.e. one non trivial root only, corresponding to a single non trivial critical point. For $\n=1$, meaning that $T$ depends linearly on $\rho$ --- two fixed points with interesting
cosmologies can arise. We see that for $\n=1$ the Hubble parameter evaluated at the fixed points is independent of the energy density $\sH=|A|/w$ (for energy influx).
Thus, all critical points in the phase portrait $q/\rho$ lay on a $q={\rm const}$ line and have the same positive acceleration.

The stability matrix associated to the linearized system of equations can be computed in the general case too. The significant $\n=1$ case is characterized by two fixed points with positive $\srho$ and $\sqq$
given by
\bea
\rho_{\star\pm}&=&\left(1\pm\sqrt{1-\frac{2}{9V^2}|A|}\right)V\\
\sqq&=&\frac{|A|^2}{9(1+w)^2}
\ena
for $|A|\leq3V\sqrt{2}$. The eigenvalues associated to the $+$ root have real part one opposite to the other. Therefore $\rho_{\star+}$ is always a saddle. The $-$ root gives instead a stable critical point,
which is in particular a stable node when $27V^2/8<|A|^2<9V^2/2$ and a stable counterclockwise spiral if $|A|^2<27V^2/8$. The small density limit critical point analysis with $\n=1$ is restored when we take
$|A|^2\ll4V^2$, implying indeed $\rho_{\star-}\ll V$. Hence $\rho_{\star-}$ represents the \refeq{KKTTZ fix sol small} fixed point, while $\rho_{\star+}\sim V$ lays outside the range of validity of the small
density approximation $\rho\ll V$ and thus isn't captured by the approximate analysis. The numerical phase portrait shows that three families of solutions exist. The $\rho_{\star-}$ critical point
attracts a wide range of trajectories, both starting with very small or very large energy density. There are however two families of solutions which are attracted to the saddle point $\rho_{\star+}$, which
represents new features with respect to a single critical point scenario. The first starts from very small energy density and travels very near to the limiting parabola which marks the upper bound for the
acceleration factor. Those trajectories then eventually flow towards a large $\rho$ and negative acceleration phase. The second family of solutions attracted by $\rho_{\star-}$ starts from large densities and
may or not reach a positive acceleration phase, to eventually end in eternal deceleration.

We can have some further analytical insight in the region of small $\rho$ manipulating the set of differential equations \refeq{KKTTZ Friedmann gen}--\refeq{KKTTZ conservation gen}. As it is explained in
\cite{Kiritsis:2002zf}, one finds that for $\n=1$ and \emph{influx} with $w>-1/3$ the trivial critical point $\srho=\sqq=0$ can never be attractive, as it is confirmed by the numerical calculations that illustrate the
phase portrait just described. In the case of \emph{outflow}, on the contrary, the Hubble parameter goes to zero when the energy density vanishes, in agreement with our general analytical considerations. In
this case all trajectories flow to the trivial fixed point but they can be distinguished into two families, one of which is decelerating at all times, while the other pass through an accelerated phase to 
decelerate again before reaching $\srho=\sqq=0$. Furthermore, it is worthwhile to note that the scale factor with the linear $\n=1$ assumption for energy exchange, behaves as in a radiation dominated universe
at late times.

In our analysis we mainly focused on flat 3D space. However, some remarks can also be made for locally hyperbolic or spherical spaces. If space can be considered to be locally a 3--sphere, the universe
eternally decelerates whenever energy flow from the brane into the bulk and $w\geq-1/3$. There may be an accelerated phase before turning to negative $q$. Interesting features appear for $k=-1$ and outflow.
Considering non--relativistic matter $w=0$ and $\n=1$ one can obtain eternal acceleration, after an initial decelerated era, while energy density goes monotonically to zero, as predicted by its (non)
conservation equation. Influx with the same conditions may imply an initially accelerating universe, which however decelerates at late times, with $\rho$ approaching a constant value.

This model shows a large range of possible cosmologies, displaying interesting features which may eventually lead to some realistic description. However, a full analytic treatment with generic energy exchange
is lacking. Still, we may get some intuition of how the cosmological evolution predicted in this context can agree with the known characteristics of inflation, Big Bang and data on present acceleration.
An analogous scenario has been the subject of my publication \cite{Mazzanti:2007dq}, where even a richer variety of cosmologies can be found, due to the additional degrees of freedom coming from the two more
compactified extra dimensions (see chapter \ref{7D RS cosmo}).
\section{Randall--Sundrum cosmology from the holographic point of view}\label{4D RS holo cosmo}
The 5D RS model just described has been investigated from the holographic point of view by Kiritsis in \cite{Kiritsis:2005bm}, relating the non conventional behavior of cosmological evolution in the gravity
set--up to non trivial dynamics in the dual theory, namely conformal anomaly. The aim here is to show how to explicitly construct the holographic dual, via AdS/CFT correspondence following the prescription of
section \ref{adscft in RS}.  Successively, I will discuss the cosmological feature on the CFT side and the generalization to non conformal and interacting case.
\subsection{The dual theory}
I briefly recall the holographic renormalization procedure, specializing to the 5D/4D case. From the AdS/CFT correspondence we learned that string theory on $AdS_5\times S^5$ with $N$ units of five form flux
and radii given in terms of $N$ is dual to the four dimensional $SU(N)$ ${\cal N}=4$ SYM theory. The correspondence is mathematically expressed by the equality
\bea
Z_{\rm string}\left[\Phi|_{\pa (AdS_5)}=\phi\right]=e^{-W_{\rm CFT}(\phi)}
\ena
Here we consider the classical limit on the supergravity side, when $N\gg1$ and $\l_T\equiv g_{\rm YM}^2N\sim g_sN\gg1$. The bulk action on $AdS_5$, once regulated following holographic renormalization, is
given by
\bea
S_{\rm gr}=S_{\rm EH}+S_{\rm GH}-S_0-S_1-S_2
\ena
where, in five dimensions, the different contributions read % footnote
\footnote{There is a subtlety here about the presence of the scheme dependent counterterm. However, we will note next that it will not contribute to the RS dual action derivation.}% footnote
\bea
S_{\rm EH}&=&M^3\int_{\rho\geq\epsilon}\intd^5x\sqrt{-g}\left(R\left[g\right]+\frac{12}{\ell^2}\right)\\
S_{\rm GH}&=&2M^3\int_{\rho=\epsilon}\intd^4x\sqrt{-g}\,K\\
S_0&=&6\frac{M^3}{\ell}\int_{\rho=\epsilon}\intd^4x\sqrt{-\g}\\
S_1&=&-\oneover{2}M^3\ell\int_{\rho=\epsilon}\intd^4x\sqrt{-\g}R\\
S_2&=&\frac{\log\epsilon}{4}M^3\ell^3\int_{\rho=\epsilon}\intd^4x\sqrt{-\g}\left(R_{\m\n}R^{\m\n}-\oneover{3}R^2\right)
\ena
As usual, we denote with $\g_{\m\n}$ the induced metric on the boundary, $R$ and $R_{\m\n}$ are the scalar and Ricci curvatures made out of $\g_{\m\n}$, $K$ is the trace of the extrinsic curvature on the
boundary $\rho=\epsilon$ and $\L_5$ is the negative cosmological constant $\L_5=-12M^3/\ell^2$. The radius $\ell$ of $AdS_5$ is determined by the supergravity classical solution to be $(4\p)^2M^3\ell^3=2N^2$.

The RS action in the five dimensional background takes the form
\bea
S_{\rm RS}=S_{\rm EH}+S_{\rm GH}-2S_0+S_{m,b}
\ena
I now apply the duality
\bea
\int\intD g\,\ex^{-S_{\rm EH}-S_{\rm GH}+S_0+S_1+S_2}=\int\intD\phi\,\ex^{-S_{\rm CFT}\left[\phi,\g\right]}
\ena
and the holographic renormalization calculations illustrated in section \ref{adscft in RS}. Feffermann--Gaham parametrization of the metric \refeq{HSS coord} must be used. We obtain as a result
\bea
S_{\widetilde{\rm RS}}=S_{\rm CFT}+S_R+S_{R^2}+S_{m,b}
\ena
with
\bea
S_{\rm CFT}\equiv2W_{\rm CFT} \,,\qquad S_{R}\equiv2S_1\,,\qquad S_{R^2}\equiv2S_2
\ena
This is the dual action that we expected for regularized 5D gravity on a slice of $AdS_5$ with $\Zgr_2$ orbifold: a renormalized CFT, $S_{\rm CFT}$, coupled to 4D gravity, $S_R$, plus higher order
corrections, $S_{R^2}$, in addition to the original matter action on the brane, $S_{b,m}$. The four dimensional Planck mass $M_{Pl}$ is given in terms of the five dimensional gravity mass scale $M$ as
$M_{Pl}^2=M^3\ell$.  The dual theory result is derived in a pure RS vacuum, i.e. with RS fine--tuning $\l_{\rm RS}=0$ (see eq. \refeq{4D RS fine tuning} for the definition of $\l_{\rm RS}$). This implies that
the effective cosmological constant on the brane vanishes. However, if we wish to take account of a non zero cosmological constant on the dual theory, we may consider the following generalization
\bea\label{4D RS dual action}
S_{\widetilde{\rm RS}}=S_{\rm CFT}+S_R+S_{R^2}+S_{m,b}+S_{\l}
\ena
with
\bea
S_{\l}=-2\l\int\intd^4x\sqrt{-\g}
\ena
More precisely, the RS background yields the following relations among dimensionful parameters
\bea
\ell=24\frac{M^3}{V}\,,\qquad \L_5=-\frac{V^2}{12M^3}
\ena
The 4D boundary parameters are related to the 5D bulk ones by
\bea
M_{Pl}=24\frac{M^6}{V}\,,\qquad c\equiv\left(\frac{N}{8\p}\right)^2=\left(12\frac{M^4}{V}\right)^3=\left(\oneover{2}M\ell\right)^3
\ena
($c$ will be related to the conformal anomaly coefficient).

We now have all the elements to study the holographic dual theory dynamics. In particular, it is interesting to derive the cosmological evolution in the dual theory, by making a time dependent ansatz for the
metric.
\subsection{Holographic cosmology in four dimensions}
Our set--up is now the four dimensional theory with action \refeq{4D RS dual action} and the four dimensional metric
\bea\label{4D RS metric}
\g^{\m\n}\intd x^\m\intd x^\n=-\intd t^2+a^2(t)\z_{ij}\intd x^i\intd^j
\ena
Here $\z_{ij}$ is a maximally symmetric three dimensional metric with spatial curvature $k$ and $a(t)$ is the scale factor associated to the Hubble parameter $H\equiv \dot a/a$. The Einstein equations get
contributions to the stress--energy tensors from the different terms in the action, yielding
\bea
\mpl G_{\m\n}+\l\g_{\m\n}=T_{\m\n}+W_{\m\n}+Z_{\m\n}
\ena
where we define
\bea
T_{\m\n}=\oneover{\sqrt{-\g}}\frac{\d S_{m,b}}{\d\g^{\m\n}}\,,\quad W_{\m\n}=\oneover{\sqrt{-\g}}\frac{\d S_{\rm CFT}}{\d\g^{\m\n}}\,,\quad Z_{\m\n}=\oneover{\sqrt{-\g}}\frac{\d S_{R^2}}{\d\g^{\m\n}}
\ena
All the contributions to the total stress--energy tensor are separately conserved
\bea
\nabla^\m T_{\m\n}=\nabla^\m W_{\m\n}=\nabla^\m Z_{\m\n}=0
\ena
Instead, they are not all traceless because of the conformal anomaly associated to the four dimensional CFT
\bea
W_\m^\m=-128\left(cE_{(4)}+aI_{(4)}\right)
\ena
The type A and B anomalies are given by the four dimensional Euler density $E_{(4)}$ and by the contraction of the Weyl tensor $I_{(4)}$, i.e. the unique conformal invariant with four derivatives (up to an
overall factor)
\bea
E_{(4)}&=&\oneover{64}\left(R^{\m\n\rho\s}R_{\m\n\rho\s}-4R^{\m\n}R_{\m\n}+R^2\right)\\
I_{(4)}&=&-\oneover{64}\left(R^{\m\n\rho\s}R_{\m\n\rho\s}-2R^{\m\n}R_{\m\n}+\oneover{3}R^2\right)
\ena
The coefficients $a$ and $c$ are determined as functions of the number of scalar, fermion and vector fields contained in the theory (see for example \cite{Kiritsis:2005bm} for the definition).
Actually, we ignored the scheme dependent contribution to anomaly, which takes the form $\nabla^\m J^{(4)}_\m=2b\Box R$ and is called divergence or D--type anomaly. Carrying on carefully the calculations
shows that this contribution gets cancelled, indeed, by the equal term appearing in the trace of $Z_{\m\n}$. This can be more clearly explained by saying that the scheme dependent counterterm that we
ignored, which should have been added in $S_{R^2}$, is exactly $S_b=-\frac{b}{3}\int\intd^4x\sqrt{-\g}R^2$. The stress--energy tensor associated to $S_b$ is characterized by a non vanishing trace, precisely
given by $Z^\m_\m=Z_b=-2b\Box R$, thus canceling $\nabla^\m J^{(4)}_\m$. We can argue that this happens in general for any number of (even) dimensions.

Since the second order counterterm stress--energy tensor is somehow related to the CFT, it is convenient to define the sum
\bea
V_{\m\n}=W_{\m\n}+Z_{\m\n}
\ena 
which satisfies the following conservation and trace equations
\bea
\nabla^\m V_{\m\n}=0\,,\qquad V^\m_\m=-128\left(cE_{(4)}+aI_{(4)}\right)
\ena
The matter tensor is traceless and conserved. We parametrized the two contributions form matter and CFT as perfect fluids
\bea\label{4D RS holo stress para}
T_{00}&=&\rho\,,\qquad T_{\ij}=p\,a^2\z_{ij}\\
V_{00}&=&\s\,,\qquad V_{ij}=\s_p\,a^2\z_{ij}
\ena

The Einstein equations evaluated on the metric \refeq{4D RS metric} yield the Friedmann equations
\bea
\mpl\frac{{\dot a}^2}{a^2}&=&\oneover{3}\left(\rho+\s+\l\right)-\frac{k}{a^2} \label{4D RS Friedmann 1}\\
\mpl\left(2\frac{\ddot a}{a}+\frac{{\dot a}^2}{a^2}\right)&=&-p-\s_p-\frac{k}{a^2}+\l \label{4D RS Friedmann 2}
\ena
If we plug \refeq{4D RS Friedmann 1} into \refeq{4D RS Friedmann 2} we get the expression for acceleration
\bea
\mpl\frac{\ddot a}{a}=-\oneover{6}\left(\rho+3p+\s+3\s_p\right)+\frac{\l}{3}
\ena
The conservation equations in terms of the perfect fluid energy densities and pressures read
\bea
\dot\rho+3\frac{\dot a}{a}\left(\rho+p\right)&=&0 \label{4D RS conservation rho}\\
\dot\s+3\frac{\dot a}{a}\left(\s+\s_p\right)&=&0 \label{4D RS conservation sigma}
\ena
Finally, the conformal anomaly equation can be written in the form
\bea\label{4D RS anomaly eq}
\s-3\s_p=48c\frac{\ddot a}{a}\left(\frac{{\dot a}^2}{a^2}+\frac{k}{a^2}\right)
\ena
where $c=N^2/(8\p)^2\gg1$ --- for the holographic derivation of the anomaly  and, in particular, of its coefficients, see for example \cite{Henningson:1998gx}. The solution to the Einstein equations can be
easily found in terms of the matter energy density $\rho$. In fact, we can solve the anomaly equation \refeq{4D RS anomaly eq} for $\s_p$ and plug the result into the conservation equation \refeq{4D RS
conservation sigma}, obtaining
\bea
\dot\s+4H\s-48cH\left(H^2+\frac{k}{a^2}\right)\left(\dot H+H^2\right)=0
\ena
This can be integrated, giving as a result
\bea
\s&=&\chi+12c\left(H^2+\frac{k}{a^2}\right)^2\\
\dot\chi+4H\chi&=&0 \quad\Rightarrow\quad \chi=\frac{\chi_0}{a^4}
\ena
Hence the mirage density $\chi$ is a free radiation energy density and it is introduced as the solution to the homogeneous equation associated to $\chi$. Using the expression for $\s$ to solve equation
\refeq{4D RS Friedmann 1} for $H^2$ one gets
\bea\label{4D RS holo sol}
H^2=\frac{\mpl}{8c}\left[1+\varepsilon\sqrt{1-\frac{16c}{3\Mpl}\left(\rho+\chi+\l\right)}\right]-\frac{k}{a^2}
\ena
Depending on the value for $\varepsilon=\pm1$ we get to distinguished roots for the Hubble parameter. Putting $\varepsilon=1$ and $\left(\rho+\chi+\l\right)=0$ we recover the Starobinsky solution described in
subsection \ref{inflation}, with
\bea
H=H_0\equiv\frac{M_{Pl}}{2\sqrt c}
\ena
However, inflation cannot generate damped oscillations as in the real Starobinsky model. These oscillations are due to the higher derivative scheme dependent contribution to the anomaly, which is absent in
the RS dual derivation --- since it is cancelled by the equal and opposite term in $S_{R^2}$. We can nevertheless think of other higher derivative terms that could eventually play the same role as conformal
anomaly, such as Gauss--Bonnet corrections (see the appendix in \cite{Kiritsis:2005bm} for the Gauss--Bonnet analysis in this context). We can now use dimensionless variables 
\bea
\tilde\rho&=&E_0\rho\,,\quad\tilde\chi=E_0\chi\,,\quad\tilde\l=E_0\l\,,\quad E_0=\frac{3\Mpl}{16c}\non\\
\tilde H&=&\frac{H}{H_0}\,,\quad\tilde k=H_0k\,,\quad \tilde t=\frac{t}{H_0}\non
\ena
to expand the solution for small densities. Expression \refeq{4D RS holo sol} becomes
\bea
\tilde H^2+\frac{\tilde k}{a^2}=\oneover{2}\left[1+\varepsilon\sqrt{1-\left(\tilde\rho+\tilde\chi+\tilde\l\right)}\right]
\ena
Its expansion for $\tilde\rho,\tilde\chi,\tilde\l\ll1$ immediately gives
\bea
\tilde H^2=\frac{1+\varepsilon}{2}-\frac{\varepsilon}{4}\left(\tilde\rho+\tilde\chi+\tilde\l\right)-\frac{\varepsilon}{16}\left(\tilde\rho+\tilde\chi+\tilde\l\right)^2-\frac{\tilde k}{a^2}+\dots
\ena
The constant term vanishes in the smooth branch $\varepsilon=-1$, while the coefficients for the linear and quadratic terms are positive. This matches with the 5D bulk description of cosmological evolution on
the brane, derived in the previous section. We can in particular match the 5D RS parameters to the 4D holographic theory description parameters
\bea\label{4D RS match}
V=2E_0\,,\quad M^3=\frac{E_0}{3H_0}\,,\quad \L_5=-E_0H_0
\ena
A discrepancy between the two descriptions arises when the densities are of order one. The 4D theory predicts an upper bound determined by $E_0$, while in the 5D set--up there is no visible constraint. In the
small density regime, however, the two sides of the duality yield the same evolution (at late times). 

On the gravity side, I have shown the effects of the presence of brane--bulk energy exchange. So, I now wish to establish the correspondence between this energy exchange and its dual object in the
generalization of the holographic theory, following \cite{Kiritsis:2005bm}.
\subsection{Generalization in the holographic description}
Intuitively, introducing an interaction term between matter and CFT fields corresponds to turning on the energy exchange parameter $T^0_5$ in the 5D description. All the same, deviations from
conformality correspond to the bulk self--interaction contributions to the 5D stress--energy tensor, $T^5_5$. Let's see this in more detail.

We both include interactions and remove conformal invariance, defining the interaction term $S_{\rm int}$ and substituing $S_{\rm CFT}$ with a strongly coupled gauge theory action $S_{\rm SCGT}$
\bea
S_{\tilde{\rm RS}*}=S_{\rm SCGT}+S_R+S_{R^2}+S_{\rm m,b}+S_{\rm int}
\ena
The strongly coupled fields can be integrated out allowing to replace the sum $S_{\rm SCGT}+S_{\rm int}$ with a non local functional of the metric and of the matter fields $W_{\rm SCGT}$. The action thus reads
\bea
S_{\tilde{\rm RS}*}=W_{\rm SCGT}+S_R+S_{R^2}+S_{b,m}
\ena
The Einstein equations can still be written as
\bea
\mpl G_{\m\n}=T_{\m\n}+V_{\m\n}
\ena
where
\bea
T_{\m\n}=\oneover{\sqrt{-\g}}\frac{\d S_{\rm m,b}}{\d\g^{\m\n}}\,,\quad V_{\m\n}=\oneover{\sqrt{-\g}}\frac{\d\left(W_{\rm SCGT}+S_{R^2}\right)}{\d\g^{\m\n}}
\ena
The conservation equations get modified by a term associated to the interactions, so that $T_{\m\n}$ and $V_{\m\n}$ are no longer separately conserved
\bea
\nabla^\m T_{\m\n}=-\nabla^\m V_{\m\n}=T
\ena
Trace anomaly may get an extra contribution from non conformally coupled scalar fields or from gravitons and gravitinos
\bea
V^\m_\m=-128\left(cE_{(4)}+aI_{(4)}\right)+D\,R^2
\ena
However, $D\ll c$ and we will consider in particular the case of vanishing $D$ (the interested reader can find the generalization to $D\neq0$ in the appendix of \cite{Kiritsis:2005bm}). The stress--energy
perfect fluid parametrizations \refeq{4D RS holo stress para} still hold. As a consequence, for zero effective cosmological constant on the brane $\l=0$, we obtain the Friedmann equations in the form
\bea
\mpl\frac{{\dot a}^2}{a^2}&=&\oneover{3}\left(\rho+\s\right)-\frac{k}{a^2} \label{4D RS Friedmann gen 1}\\
\mpl\left(2\frac{\ddot a}{a}+\frac{{\dot a}^2}{a^2}\right)&=&-p-\s_p-\frac{k}{a^2}+\l \label{4D RS Friedmann gen 2}
\ena
implying
\bea
\mpl\frac{\ddot a}{a}=-\oneover{6}\left(\rho+3p+\s+3\s_p\right)
\ena
Besides, the conservation equations are modified by the introduction of the non homogenous term $T$, depending on the form of the interactions,
\bea
\dot\rho+3\frac{\dot a}{a}\left(\rho+p\right)&=&-T \label{4D RS conservation rho gen}\\
\dot\s+3\frac{\dot a}{a}\left(\s+\s_p\right)&=&T \label{4D RS conservation sigma gen}
\ena
and the trace anomaly receives the extra contribution $X$ from conformal breaking
\bea\label{4D RS anomaly eq}
\s-3\s_p=48c\frac{\ddot a}{a}\left(\frac{{\dot a}^2}{a^2}+\frac{k}{a^2}\right)+X
\ena
The function $X$ contains $\b$--functions and masses that classically or quantum break conformal invariance. To be more specific, $X$ can be put in the following form before integrating out the strongly
coupled fields
\bea
X=\sum_{ij}\left(\b^{(1)}_{ij}+R\b^{(2)}_{ij}\right){\cal O}_i\O_j+\left(R_{\m\n}\b^{(3)}_{ij}+g_{\m\n}\b^{(4)}_{ij}\right){\cal O}^\m_i\O^\n_j
\ena
Here we have denoted with ${\cal O}_i$ the operators of the SCGT and with $\O_j$ the matter theory operators. After integration, the above expression becomes
\bea
X=\sum_{j}\left(B^{(1)}_{j}+B^{(2)}_{j}R\right)\O_j+\left(B^{(3)}_{j} R_{00}+B^{(4)}_{j}\right)\O_j^0
\ena
where $B_j^{(I)}$ contains the sum over strongly coupled fields expectation values $B_j^{(I)}=\sum_{i}\b_{ij}^{(I)}\langle{\cal O}_i\rangle$. We can now proceed integrating the equation for $\s$, as in the
conformal non interacting case
\bea
\s&=&\chi+12c\left(H^2+\frac{k}{a^2}\right)^2\\
\dot\chi+4H\chi&=&T+HX 
\ena
The system of Friedmann plus conservation equations can't be solved in general, without specifying the form for $X$ and $T$
\bea
3M_P^2\left(H^2+{k\over a^2}\right)-12c\left(H^2+{k\over a^2}\right)^2&=&\rho+\chi\\
\dot\chi+4H\chi&=&T+HX\\
\dot\rho+3\frac{\dot a}{a}\left(\rho+p\right)&=&-T
\ena
The solution for $H^2$ is still given by \refeq{4D RS holo sol}. From this solution we can also deduce the acceleration factor $q=\dot H+H^2$
\bea
q=-\varepsilon{-(\rho+3p)+X-2\left(\chi+12c\left[H^2+{k\over a^2}\right]^2\right)\over 6M_P^2\sqrt{1-{16c\over 3M_P^4}(\rho+\chi)}}
\ena
Looking at this expression we observe that we get acceleration in the smooth branch $\varepsilon=-1$ if
\bea
X>(\rho+3p)+2\left(\chi+12c\left[H^2+{k\over a^2}\right]^2\right)
\ena
In the Starobinsky branch, on the other hand, $q$ is positive when
\bea
X<(\rho+3p)+2\left(\chi+12c\left[H^2+{k\over a^2}\right]^2\right)
\ena
i.e. in the complementary region. This means that for each particular value of $X$ we can always choose a branch such that acceleration is positive.

We finally come to the issue of comparing the two descriptions in the generalized case. I recall the set of equations derived in the 5D gravity set--up
\bea
H^2&=&{1\over144M^6}\rho^2+{V\over72 M^6}\left(\rho+\chi\right)+\l\\
\dot{\chi}+4H\chi&=&\left({\rho\over V}+1\right)2T^0_5-24{M^3\over V}HT^5_5\\
\dot\rho+3H\left(\rho+p\right)&=&-2T^0_5
\ena
It matches with the smooth branch solution to the holographic Einstein equations as long as the parameters are related by \refeq{4D RS match} and
\bea
T\simeq2T^0_5\,,\qquad X\simeq24\frac{M^3}{V}T^5_5
\ena
The non exactness of equalities comes from the difference in the two descriptions arising when the $\rho^2$ and $\rho T^0_5$ terms are not negligible, i.e. when $\rho\simeq V$ (actually, one can redefine the
5D mirage density in such a way that it is possible to find an exact match for $X$, but it includes matter energy density and pressure and the Hubble parameter). Nevertheless, this are the relations we
expected, asserting the correspondence between the hidden/visible interactions and the brane--bulk energy exchange, on the one hand, and, on the other hand, between non conformality and bulk
self--interaction.

Further studies can be performed on this holographic model, including some specific deformations of the dual theory. Those can be analyzed following the prescription given in this section and combining it with
the known results of holographic renormalization. The scalar field example is analyzed in \cite{Kiritsis:2005bm}. I have here and in the previous section reviewed the basics of the 5D RS cosmology, of the 4D
RS holographic theory and its generalization, in the perspective of giving an illustrative picture for the more complicated 7D RS model. This indeed will be the subject of the next two chapters.

\chapter{Randall--Sundrum cosmology in seven dimensions}\label{7D RS cosmo}
As I reviewed in the previous chapter, RS model \cite{Randall:1999ee} is an alternative way of localizing gravity in four dimensions without compactifying the extra dimensions. This was achieved in RSII model
by assuming a warped extra direction, instead of a compact one (compact extra dimensions has been used \cite{Arkani-Hamed:1998rs,Kokorelis:2002qi} as an attempt of giving an explanation to the hierarchy problem, to which RSI
\cite{Randall:1999ee} represents an alternative way out). 
%The set--up of RSII is five dimensional gravity in a bulk space--time cut by a 3--brane. There is, in addition, a $\mathbb Z_2$ reflection of the extra dimension transverse to the brane, having as fixed
%point the location of the brane. The result is a bound state graviton mode localized on the brane and a tower of KK modes, without mass gap, that give negligible corrections to the effective 4D gravity
%description. 
The metric solving the equations of motion for the five dimensional action can be viewed as coming from the Type IIB string theory background for a stack of $N$ D3--branes, in the low energy effective field
theory approximation, which gives indeed an $AdS_5\times S^5$ near horizon geometry for a 3--brane supergravity solution. In RS analysis only gravity on $AdS_5$ is considered, since the $S^5$ is factored out
from the anti de Sitter space, giving KK modes. The truncation of $AdS_5$ with the 3--brane cuts out its boundary. 

In a recent work \cite{Bao:2005ni}, a different string background related to RS set--up has been considered. I illustrated in subsection \ref{baolykken} the analysis performed by Bao and Lykken
\cite{Bao:2005ni} in a seven dimensional RS anti de Sitter background rather than in the five dimensional original model. The background may come from the near horizon geometry of M5--branes in the eleven
dimensional M theory, which gives $AdS_7\times S^4$. As for the five dimensional model, the sphere is factored out and only the physics of gravity in $AdS_7$ plus KK modes are considered. A further step
performed in \cite{Bao:2005ni} is to reduce $AdS_7\rightarrow AdS_5\times \S^2$, where $\S^2$ is a two dimensional internal space (namely a two--sphere or a torus). In \cite{Bao:2005ni} the RS spectrum of KK
modes gets modified and supplemental KK and winding modes appear, because of the $\S^2$ compactification. In \cite{Avramis:2004cn} the 7D supergravity orbifold compactification on $S^1/\mathbb Z_2$ is
considered in the context of anomaly cancellation on the boundary on the background, showing that the matter contents of the theory cannot be completely generic.

On the cosmological side, RS models can give new descriptions of the cosmological evolution of our universe. A realistic model should be able to explain the existence of dark energy and the nature of dark
matter, early time inflation and eventually the exit from this phase, as well as late time acceleration coming from the present observations --- additional issues are related to the cosmological constant,
temperature anisotropies, etc. (see \cite{Padmanabhan:2006kz}--\cite{Trodden:2004st} for recent reviews on the observed cosmology). I reviewed in the previous chapter the RS cosmologies that have been studied
and found them to exhibit some of these features. Generally, brane--world cosmological models should take account of the energy exchange between brane and bulk that naturally arises because of the non
factorized extra direction.  The implications of this energy exchange has been analyzed in subsection \ref{5D RS cosmo}, following \cite{Kiritsis:2002zf}. The authors propose some scenarios describing the
cosmological evolution of a universe with two accelerating phases, as we expect from experimental data. Moreover, most of the fixed points were shown to be stable.  Earlier attempts include
\cite{Kiritsis:2001bc}.  Subsequent papers \cite{Hebecker:2001nv}--\cite{Cai:2005qm} have been written on the subject, also finding new solutions, some of which exact.

We now want to investigate the 7D RS cosmology with brane--bulk energy exchange and to explore (in the next chapter) the model from the holographic point of view, making an explicit comparison between the two
descriptions. Our starting point is gravity in the 7D bulk cut by a five--brane, with the usual RS $\mathbb Z_2$ reflection and generic matter term on the brane. In order to study the cosmological evolution
of the brane--world, the ansatz for the metric is time dependent. Besides, the direction transverse to the 5--brane is the warped direction characterizing RS models. Unlike in \cite{Bao:2005ni}, we have
different warp factors for the 3D extended space and for the two dimensional compact internal space. It is worth noticing that the gravitational coupling constant of the 4D space--time is dynamically related
to the 7D Newton constant, since the compact space volume is generally time dependent. Indeed, the 4D energy density and pressures are also dynamical functions of the density and the pressures defined on the
brane. I analyze the generic Friedmann and conservation equations, including the energy exchange terms, in order to get the expressions for the Hubble parameter of the 4D space--time as a function of the
density with the aim of describing realistic cosmologies. It will be interesting to study both analytically and numerically the system of Einstein equations. The analogous critical point analysis in the 5D
bulk was performed in \cite{Kiritsis:2002zf}. Some explicit solutions, derived with simplifying assumptions on the parameter of the internal space and on its geometry, are also illustrated in my work
\cite{Mazzanti:2007dq}. 

Interesting results for cosmologies with compactification emerge in the context of dynamical compactification \cite{Mohammedi:2002tv,Andrew:2006yt}. The compact space is treated with a different scale factor
(as in the approach that will be used in this paper). In particular, in the context of dynamical compactification, the scale factor for the internal space has an inverse power dependence on the scale factor
for the visible directions. The extra dimensions thus contract as the extended space expand. I also include some remarks on dynamical compactification applications in our set--up. Other attempts to reduce to
conventional cosmology and investigate issues such as the cosmological constant from models with arbitrary number of extra dimensions are given in \cite{Kamani:2006tv,Zhuk:2006kh}. In particular, cosmologies
in six dimensions are analyzed in \cite{Papantonopoulos:2006uj}, which could be related to the holographic description of my model --- analyzed in the next chapter.

Next section will describe the set--up of the seven dimensional RS model. Section \ref{bulk cosmology} will be focused on the cosmological evolution from the 7D point of view, admitting brane--bulk energy
exchange and particularly investigating the form of Einstein equations with some specific ansatz, while in section \ref{critical bulk} the critical point analysis will be illustrated, including numerical
phase space portrait and explicit solutions.
%%%%%%%%%%%%%%%%%%%%%%%%%%%%%%%%%%%%%%%%%%%%%%%%%%%%%%%%%%%%%%%%%%%%%%%%%%%%%%%%%%%%%%%%%%%%%%%%%%%%%%%%%%%%%%%%%%%%%%%%%%%%%%%%%%%%%%
%%%%%%%%%%%%%%%%%%%%%%%%%%%%%%%%%%%%%%%%%%%%%%%%%%%%%%%%%% set--up %%%%%%%%%%%%%%%%%%%%%%%%%%%%%%%%%%%%%%%%%%%%%%%%%%%%%%%%%%%%%%%%%%%
%%%%%%%%%%%%%%%%%%%%%%%%%%%%%%%%%%%%%%%%%%%%%%%%%%%%%%%%%%%%%%%%%%%%%%%%%%%%%%%%%%%%%%%%%%%%%%%%%%%%%%%%%%%%%%%%%%%%%%%%%%%%%%%%%%%%%%
\section{7D RS set--up}\label{setup} 
As I announced in the introduction, we will work in a seven dimensional bulk with a 5--brane located at the origin of the direction $z$ transverse to the brane itself and with a $z\rightarrow-z$ $\Zgr_2$
identification. In analogy to the 5D RS model, the action in this seven dimensional set--up is given by the sum of the Einstein--Hilbert action with 7D cosmological constant plus a contribution localized on
the brane that represents the brane tension. Besides, we also put a matter term both in the bulk and on the brane. In formula we thus have 
\bea\label{7D action}
S&=&S_{EH}+S_{m,B}+S_{tens}+S_{m,b}=\non\\&=&\int\intd^7 x\sqrt{-g}\left(M^5R-\L_7+\lagra_{m,B}\right)+\int\intd^6 x\sqrt{-\g}\left(-V+\lagra_{m,b}\right) 
\ena 
Here $V$ is the brane tension and we will call the associated contribution to the action $S_{tens}$. $S_{EH}$ is the usual Einstein--Hilbert action with the seven dimensional bulk cosmological constant
$\L_7$ and $\lagra_{m,B},\lagra_{m,b}$ are, respectively, the bulk and brane matter lagrangians. The action for matter fields in the bulk $S_{m,B}$ is an additional term with respect to the usual RS set--up,
whereas the matter on the brane contribution will be referred to as $S_{m,b}$. The metric $\g_{\m\n}$ is the induced metric on the brane. The brane tension is necessary in the RS models in order to
compensate for the presence of the cosmological constant in the bulk.

The classical solution of the equations of motion for the theory above, neglecting all the matter terms and with a static warped geometry of the kind $\intd s^2=\ex^{-W}\intd x^2+\intd z^2$ ($W=W(z)$ is the
warp factor and the 6D $x$--directions are flat), is the analogue of the solution described by Randall and Sundrum \cite{Randall:1999ee} for the 5D RSII model. The 7D solution gives as a result
$W(z)=2|z|\sqrt{-\frac{\L_7}{30M^5}}$, so that the space--time is a slice of $AdS_7$ with the typical $\Zgr_2$ reflection, where there has to exist a relation between the brane tension and the bulk
cosmological constant $3V^2=-40M^5\L_7$. 

The aim of this section and of the next one is to generalize the RS ansatz to a time dependent background and to wrap the 5--brane over a two dimensional internal space, ending up with an effective 4D
cosmology.  Taking account of the seventh warped extra dimension and of the compactification over the other two extra dimensions, giving two different warp factors to the 3D space and the internal 2D
space, the time dependent ansatz for the metric is of the form
\bea\label{7D metric}
\intd s^2=-n^2(t,z)\intd t^2+a^2(t,z)\z_{ij}\intd x^i\intd x^j+b^2(t,z)\xi_{ab}\intd y^a\intd y^b+f^2(t,z)\intd z^2
\ena
with the maximally symmetric $\z_{ij}$ background in three spatial dimensions (with spatial curvature $k$) and $\xi_{ab}$ for the 2D internal space (with spatial curvature $\k$). We use capital indices
$A,B,\dots$ to run over the seven dimensions, $i,j,\dots$ for the three spatial dimensions of the 4D space--time, $a,b,\dots$ for the two internal dimensions. In our notations $z$ represents the seventh
warped extra direction, the $y$ coordinates belong to the 2D internal space, while the $\{x_\m\}=\{t,x_i,y_a\}$ run over the 6D space--time on the brane. Summarizing, the structure of the bulk is thus made of
a time coordinate, three extended maximally symmetric spatial dimensions (that gives, together with the time, the visible 4D space--time), two compact dimensions and a warped direction. The 3D and 2D spaces
have two different scale factors $a(t,z)$ and $b(t,z)$ respectively, while a gauge choice will be made for the values of the $n(t,z)$ and $f(t,z)$ factors on the brane, i.e. when $z=0$.

A less physically meaningful background, but better understood, would be to have a five dimensional maximally symmetric space with some 5D metric $\tilde\z_{ij}$ and just one scale factor $\tilde a(t,z)$,
without compactifying on any two dimensional internal space. The solution to the equations of motion in this case is much simpler. The results related to this background will be briefly mentioned along with
the more realistic analysis with brane wrapping over the two dimensional internal space.
%%%%%%%%%%%%%%%%%%%%%%%%%%%%%%%%%%%%%%%%%%%%%%%%%%%%%%%%%%%%%%%%%%%%%%%%%%%%%%%%%%%%%%%%%%%%%%%%%%%%%%%%%%%%%%%%%%%%%%%%%%%%%%%%%%%%%%
%%%%%%%%%%%%%%%%%%%%%%%%%%%%%%%%%%%%%%%%%%%%%%%%%%% bulk cosmological evolution %%%%%%%%%%%%%%%%%%%%%%%%%%%%%%%%%%%%%%%%%%%%%%%%%%%%
%%%%%%%%%%%%%%%%%%%%%%%%%%%%%%%%%%%%%%%%%%%%%%%%%%%%%%%%%%%%%%%%%%%%%%%%%%%%%%%%%%%%%%%%%%%%%%%%%%%%%%%%%%%%%%%%%%%%%%%%%%%%%%%%%%%%%%
\section{Cosmological evolution in the bulk}\label{bulk cosmology}
In this section we will analyze some aspects of the cosmological evolution from the 7D bulk point of view. We write the equations of motion for the bulk action and solve them by making assumptions to
simplify their form and get explicit results evaluated on the brane.

Given the set--up described in the previous section, we parametrize all the contributions to the stress--energy tensor as 
\bea\label{stress energy para bulk}
T^A_C|_{v,b}&=&\frac{\delta(z)}{f}\,\diag\left(-V,-V,-V,-V,-V,-V,0\right)\non\\
T^A_C|_{v,B}&=&\diag\left(-\L_7,-\L_7,-\L_7,-\L_7,-\L_7,-\L_7,-\L_7\right)\non\\
T^A_C|_{m,b}&=&\frac{\delta(z)}{f}\,\diag\left(-\rho,p,p,p,\p,\p,0\right)\non\\ T^A_C|_{m,B}&=&T^A_C 
\ena 
with the subindices $v$ and $m$ indicating the vacuum and matter stress--energy tensors, while $b$ and $B$ stand for the brane and bulk contributions respectively.  A difference between this (4+2+1)D
background and the simpler (6+1)D analysis without the 2D compactification cited at the end of the previous section, is having in \refeq{stress energy para bulk} two different pressures in the 3D space and in
the 2D compact dimensions for the matter on the brane, while for the (6+1)D background we would put $\p=p$. This generalization is due to the fact that we don't assume homogeneity for the matter fluid in the
whole (3+2)--dimensional space, but only in the 3D and 2D spaces separately. 

Having calculated the Einstein tensor, we can put the explicit expression in the equation
\bea
G_{AC}=\oneover{2M^5}T_{AC}
\ena
evaluated on the brane (from now on all the functions are evaluated on the brane, i.e. at $z\rightarrow0$), in the specific background (\ref{7D metric}). As a consequence, for the $00$, $ij$ and $ab$
components we obtain the jump equations
\bea\label{jump eqs}
a'_+=-a'_-&=&-\frac{fa}{20M^5}\left(V+\rho+2p-2\p\right)\non\\
b'_+=-b'_-&=&-\frac{fb}{20M^5}\left(V+\rho-3p+3\p\right)\\
n'_+=-n'_-&=&\frac{fn}{20M^5}\left(-V+4\rho+3p+2\p\right)\non
\ena
These are the values of the warp factors in the limit $z\to0$, where the subscripts $+$ and $-$ distinguish the limit taken from below from the limit taken from above. The prime denotes the partial
derivative with respect to the $z$ coordinate, while the dot indicates the time derivative.  For the $07$ and $77$ components, substituting the expressions (\ref{jump eqs}) and choosing a gauge with $f(t,0)=1$
and $n(t,0)=1$, we get the (non)conservation equation
\bea\label{conservation 7D}
&&\dot\rho+3\frac{\dot a}{a}\left(\rho+p\right)+2\frac{\dot b}{b}\left(\rho+\p\right)=2T_{07}
\ena
and the Friedmann equation
\bea\label{Friedmann 7D}
&&3\frac{\ddot a}{a}+2\frac{\ddot b}{b}+3\frac{\dot a^2}{a^2}+\frac{\dot b^2}{b^2}+6\frac{\dot a}{a}\frac{\dot b}{b}+3\frac{k}{a^2}+\frac{\k}{b^2}=\non\\
&=&-\frac{5}{(20M^5)^2}\bigg[V\left(6p-\p-\rho\right)+\rho\left(6p-\p+\rho\right)+\rho^2+\non\\
&&\left.+\oneover{5}\left(p-\p\right)\left(V-19\rho-3p-7\p\right)\right]+15\l_{\rm RS}-\oneover{2M^5}T^7_7
\ena
We have defined the constant
\bea
\l_{\rm RS}=\frac{1}{30M^5}\left(\L_7+\frac{3}{40M^5}V^2\right)
\ena
which plays the role of an effective cosmological constant on the brane. These \refeq{conservation 7D}--\refeq{Friedmann 7D} are two equations in five variables $\{H,F,\rho,p,\p\}$. We will thus have to make
an ansatz for some of those variables. 

The pure RS system correspond to setting $T^0_7=T^7_7=0$, that means putting to zero the brane--bulk energy exchange and no cosmological term on the brane, i.e. $\l_{\rm RS}=0$, to restore RS fine--tuning.

We can now write a simplified version of the differential equations (\ref{conservation 7D})--(\ref{Friedmann 7D}), using the usual ansatz for the equation of state of the matter fluid on the brane, i.e.
\bea\label{ansatz pressures 7D}
p=w\rho,\qquad \p=w_\p\rho
\ena
The set of equations (\ref{conservation 7D})--(\ref{Friedmann 7D}), in terms of the Hubble parameters $H\equiv\frac{\dot a}{a},F\equiv\frac{\dot b}{b}$, becomes
\bea\label{Friedmann 7D ansatz}
3\dot H+2\dot
F+6H^2+6HF+3F^2+3\frac{k}{a^2}+\frac{\k}{b^2}
&=&-\oneover{M^{10}}\left(c_VV+c_\rho\rho\right)\rho+15\l_{\rm RS}-\frac{T^7_7}{2M^5}\non\\\\
\dot\rho+\left[3(1+w)H+2(1+w_\p)F\right]\rho&=&2T_{07}\label{conservation 7D ansatz}
\ena
with
\bea\label{cV crho}
c_V=\frac{31w-6w_\p-5}{400},\quad c_\rho=\frac{11w+14w_\p+10-(w-w_\p)(3w-7w_\p)}{400}
\ena

Looking at the definition of the two coefficients $c_V,c_\rho$, we can note that equation (\ref{Friedmann 7D ansatz}) gets further simplified when the two pressures $p$ and $\p$ are equal. This can be seen
also from the previous equation (\ref{Friedmann 7D}), where the ``non--standard'' term on the r.h.s. (standard with respect to the homogeneous background analysis) is proportional to $(p-\p)$. We will first
examine some cosmological solutions assuming $p=\p$ and then we will drop this equal pressure condition to find an expression for $H$, in terms of the energy density, in the particular limits of static
compact extra dimensions and equal scale factors.
%%%%%%%%%%%%%%%%%%%%%%%%%%%%%%%%%%%%%%%%%%%%%%%%%%%%%%%%%%%%%%%%%%%%%%%%%%%%%%%%%%%%%%%%%%%%%%%%%%%%%%%%%%%%%%%%%%%%%%%%%%%%%%%%%%%%%%
%%%%%%%%%%%%%%%%%%%%%%%%%%%%%%%%%%%%%%%%%%%%%%%%% equal pressures %%%%%%%%%%%%%%%%%%%%%%%%%%%%%%%%%%%%%%%%%%%%%%%%%%%%%%%%%%%%%%%%%%%%
%%%%%%%%%%%%%%%%%%%%%%%%%%%%%%%%%%%%%%%%%%%%%%%%%%%%%%%%%%%%%%%%%%%%%%%%%%%%%%%%%%%%%%%%%%%%%%%%%%%%%%%%%%%%%%%%%%%%%%%%%%%%%%%%%%%%%%
\subsection{Equal pressures in 3D and 2D compact space}\label{equal pressures}
We first try to find a solution by simplifying the computation assuming $\p=p$. In this case, equation (\ref{Friedmann 7D}) written in terms of the Hubble parameters of the 3D space and 2D extra dimensions,
defined as $H=\dot a/a$, $F=\dot b/b$ respectively, together with the (non)conservation equation (\ref{conservation 7D}), becomes
\bea
3\dot H+2\dot F+6H^2+6HF+3F^2+3\frac{k}{a^2}+\frac{\k}{b^2}&=&  \non\\
=-\oneover{80M^{10}}\left[V\left(5p-\rho\right)+\rho\left(5p+2\rho\right)\right]+15\l_{\rm RS}-\frac{T^7_7}{2M^5}&&  \label{Friedmann 7D equal pressures before}\\\non\\
\dot\rho+\left(3H+2F\right)\left(\rho+p\right)&=&2T_{07}  \label{conservation 7D equal pressures}
\ena
We note that the system of equations written above still contains three variables $H(t),F(t),\rho(t)$ but only two equations. So we are able to just determine the value of the 3D Hubble parameter $H(t)$ as
a function of the 2D one $F(t)$. Moreover, given the complicated form of this set of equations, we will make some assumptions on the internal space, such as flat compact extra dimensions ($\k=0$) or static
extra dimensions ($F(t)\equiv0$) in the following subsections.

Manipulating the system above, equation \refeq{Friedmann 7D equal pressures before} takes the form
\bea\label{Friedmann 7D equal pressures}
5\frac{\intd}{\intd t}\left(3H+2F\right)^2+6\left(3H+2F\right)^3+6\left(3H+2F\right)\left(H-F\right)^2&=&\non\\
=\frac{1}{8M^{10}}\left[5V\dot\rho+6\left(3H+2F\right)V\rho+5\rho\dot\rho+3\left(3H+2F\right)\rho^2\right]+150\left(3H+2F\right)\l_{\rm RS}+&&\non\\
+\frac{5}{8M^{10}}\left(V+\rho\right)2T^0_7-\frac{5}{M^5}\left(3H+2F\right)T^7_7-10\left(3H+2F\right)\left(3\frac{k}{a^2}+\frac{\k}{b^2}\right)&&
\ena
It is interesting to derive the first order ODEs from the second order one \refeq{Friedmann 7D equal pressures}, in order to find the expression for $H^2$ as a function of the localized matter energy density
$\rho$ and to perform the critical point analysis.
%%%%%%%%%%%%%%%%%%%%%%%%%%%%%%%%%%%%%%%%%%%%%%%%%%%%%%%%%%%%%%%%%%%%%%%%%%%%%%%%%%%%%%%%%%%%%%%%%%%%%%%%%%%%%%%%%%%%%%%%%%%%%%%%%%%%%%
%%%%%%%%%%%%%%%%%%%%%%%%%%%%%%%%%%%%%%%%%%%%%%%%%%% flat extra equal pressures %%%%%%%%%%%%%%%%%%%%%%%%%%%%%%%%%%%%%%%%%%%%%%%%%%%%%%%
%%%%%%%%%%%%%%%%%%%%%%%%%%%%%%%%%%%%%%%%%%%%%%%%%%%%%%%%%%%%%%%%%%%%%%%%%%%%%%%%%%%%%%%%%%%%%%%%%%%%%%%%%%%%%%%%%%%%%%%%%%%%%%%%%%%%%%
\subsubsection{Flat compact extra dimensions with equal pressures}\label{bulk flat}
In flat compact extra dimensions ($\k=0$) and flat 3D space ($k=0$), equation (\ref{Friedmann 7D equal pressures}) shows that we can deduce the solution for $(3H+2F)$ in terms of the localized energy density
$\rho$ and of a mirage density $\chi$ (that we will define below through a differential equation) in the limit in which the two Hubble parameters are almost equal. In fact, in this case the third term on the
l.h.s.  of (\ref{Friedmann 7D equal pressures}) is negligible, leaving a solvable differential equation.

The solution for $\left(H-F\right)\ll\left(3H+2F\right)$ and $k=\k=0$ is given by
\bea\label{7D flat Hubble}
\left(3H+2F\right)^2=\frac{1}{16M^{10}}\rho^2+\frac{V}{8M^{10}}\left(\rho+\chi\right)+25\l_{\rm RS}
\ena
It is written in terms of the mirage density $\chi$ and the localized energy density $\rho$. The mirage density must satisfy
\bea\label{7D flat chi}
\dot\chi+\frac{6}{5}\left(3H+2F\right)\chi=\left(\frac{\rho}{V}+1\right)2T^0_7-\frac{8M^5}{V}\left(3H+2F\right)T^7_7
\ena
and the equation for $\rho$ is the (non)conservation equation
\bea\label{7D flat rho}
\dot\rho+\left(3H+2F\right)\left(\rho+p\right)=2T_{07}
\ena

Here we get a linear and quadratic dependence of the Hubble parameter $H^2$ on $\rho$, as well as a dependence on the mirage density $\chi$ and on the hidden sector Hubble parameter $F$. The quadratic and
linear terms in $\rho$ are analogous to those in the 5D analysis \cite{Kiritsis:2005bm}, implying that for $\rho\ll V$ the cosmological evolution looks four dimensional, while it moves away from the 4D
behavior for $\rho\gg V$. The term in $\chi$ also already appears in the 5D model, as well as the $\l_{\rm RS}$ constant term. Besides, the mirage energy density dynamics are controlled by the bulk parameters
$T^0_7$ and $T^7_7$, that represent the brane--bulk energy exchange and bulk pressure, as in \cite{Kiritsis:2005bm}.  However, a new variable $F$, the internal dimension Hubble parameter, arises and remains
undetermined, unless we make an ansatz for the evolution of the two compact extra dimensions. We can argue that the solution (\ref{7D flat Hubble})--(\ref{7D flat chi}) is written in terms of a ``total''
Hubble parameter $\oneover{5}(3H+2F)$, that carries the same characteristics as the Hubble parameter $H$ in the 5D model, but also includes the dynamics of the evolution of the extra dimensions. For equal
scale factors, $F=H$, this ``total'' Hubble parameter reduces precisely to $H$, giving the exact analogue to the 5D RS cosmology in 7D. 
%%%%%%%%%%%%%%%%%%%%%%%%%%%%%%%%%%%%%%%%%%%%%%%%%%%%%%%%%%%%%%%%%%%%%%%%%%%%%%%%%%%%%%%%%%%%%%%%%%%%%%%%%%%%%%%%%%%%%%%%%%%%%%%%%%%%%%
%%%%%%%%%%%%%%%%%%%%%%%%%%%%%%%%%%%%%%%%%%%%%% equal scale factors equal pressures %%%%%%%%%%%%%%%%%%%%%%%%%%%%%%%%%%%%%%%%%%%%%%%%%%%
%%%%%%%%%%%%%%%%%%%%%%%%%%%%%%%%%%%%%%%%%%%%%%%%%%%%%%%%%%%%%%%%%%%%%%%%%%%%%%%%%%%%%%%%%%%%%%%%%%%%%%%%%%%%%%%%%%%%%%%%%%%%%%%%%%%%%%
\subsubsection{Equal scale factors with equal pressures}\label{bulk equal pressures equal scale factors} 
A special case in which the $\left(H-F\right)\ll\left(3H+2F\right)$ limit is valid is the equal scale factor case $F=H$. The results can directly be obtained from the previous subsection, yielding 
\bea\label{H 7D equal equal pressures}
H^2&=&\frac{1}{400M^{10}}\rho^2+\frac{V}{200M^{10}}\left(\rho+\chi\right)+\l_{\rm RS}-\oneover{10}\left(3\frac{k}{a^2}+\frac{\k}{a^2}\right)\\
\dot\chi+6H\chi&=&\left(\frac{\rho}{V}+1\right)2T^0_7-\frac{40M^5}{V}HT^7_7\\ \dot\rho+5H\left(\rho+p\right)&=&2T_{07} 
\ena 
We added the curvature contributions that can be computed exactly in this limit. This solution is particularly simple thanks to the simultaneous vanishing of the $(H-F)$ and $(p-\p)$ terms.  It shows the
quadratic dependence of $H^2$ on $\rho$ and the linear term in $\left(\rho+\chi\right)$. The mirage density reduces to free radiation in 6D space--time when we restrict to pure RS configuration, with no
energy exchange and bulk pressure. In this same limit, the localized matter energy density obeys to standard conservation equation in 6D. 

We will now drop the equal pressure ansatz and derive the expression for the Hubble parameter of the 3D space making particular assumptions on the internal space scale factor. We suppose from now on to live
in a spatially flat universe ($k=0$), where nevertheless the extra dimensions may be curved ($\k\neq0$ generally). 
%%%%%%%%%%%%%%%%%%%%%%%%%%%%%%%%%%%%%%%%%%%%%%%%%%%%%%%%%%%%%%%%%%%%%%%%%%%%%%%%%%%%%%%%%%%%%%%%%%%%%%%%%%%%%%%%%%%%%%%%%%%%%%%%%%%%%%
%%%%%%%%%%%%%%%%%%%%%%%%%%%%%%%%%%%%%%%%%%%%%% equal scale factors %%%%%%%%%%%%%%%%%%%%%%%%%%%%%%%%%%%%%%%%%%%%%%%%%%%%%%%%%%%%%%%%%%%
%%%%%%%%%%%%%%%%%%%%%%%%%%%%%%%%%%%%%%%%%%%%%%%%%%%%%%%%%%%%%%%%%%%%%%%%%%%%%%%%%%%%%%%%%%%%%%%%%%%%%%%%%%%%%%%%%%%%%%%%%%%%%%%%%%%%%%
\subsection{Equal scale factors (generic pressures)}\label{bulk equal scale factors}
For generic pressures, we use the generalization of the equation of state for a fluid with energy density $\rho$ and parametrize the pressures of both the non compact and internal dimensions $p,\p$ that we introduced in (\ref{ansatz pressures 7D}) by means of $w,w_\p$. 

We evaluate the Friedmann and (non)conservation differential equations (\ref{Friedmann 7D ansatz})--(\ref{conservation 7D ansatz}) $F=H$. The system (\ref{Friedmann 7D ansatz})--(\ref{conservation 7D ansatz})
takes the form
\bea
\frac{5}{2}\left(\dot{H^2}+6H^3\right)&=&-\oneover{M^{10}}\left(c_VV+c_\rho\rho\right)H\rho-\frac{T^7_7}{2M^5}+15H\l_{\rm RS}-\frac{\k}{a^2}H\label{Friedmann 7D equal}\\
\dot\rho+(3(1+w)+2(1+w_\p))H\rho&=&2T_{07}\label{conservation 7D equal}
\ena
with the coefficients $c_V,c_\rho$ still given by (\ref{cV crho}). With the help of (\ref{conservation 7D equal}), (\ref{Friedmann 7D equal}) can be brought in a form from which we can explicitly deduce $H$
as a function of $\rho$ and $\chi$
\bea\label{Friedmann 7D equal explicit}
\frac{5}{2}\left(\dot{H^2}+6H^3\right)&=&\frac{1}{M^{10}}\left[\tcveq V\left(\dot\rho+6H\rho\right)+\tcreq \left(\dot{\left(\rho^2\right)}+6H\rho^2\right)\right]+\non\\
&&+\frac{2T^0_7}{M^{10}}\left(\tcveq V+\tcreq \rho\right)-H\frac{T^7_7}{2M^5}+15H\l_{\rm RS}-\frac{\k}{a^2}H
\ena
yielding
\bea
H^2&=&\frac{\tcreq }{5M^{10}}\rho^2+\frac{2\tcveq V}{5M^{10}}\left(\rho+\chi\right)+\l_{\rm RS}-\oneover{10}\frac{\k}{a^2}\label{H 7D equal}\\
\dot\chi+6H\chi&=&2T^0_7\left(1+\frac{\tcreq }{\tcveq }\frac{\rho}{V}\right)-\frac{M^5}{2\tcveq V}HT^7_7  \label{chi 7D equal}\\
\dot\rho+(3(1+w)+2(1+w_\p))H\rho&=&2T_{07}\label{rho 7D equal}
\ena
The two constants $\tcveq ,\tcreq $ must be defined as
\bea\label{ctilde equal}
\tcveq =\frac{c_V}{3(1+w)+2(1+w_\p)-6},\qquad \tcreq =\frac{c_\rho}{3(1+w)+2(1+w_\p)-3}
\ena
in order to have the right coefficients in equation (\ref{Friedmann 7D equal explicit}). For some values of $w,w_\p$ the denominator of $\tcveq$ or $\tcreq$ may vanish. However we can fix $w_\p$ such that
$c_V$ (or $c_\rho$) is proportional to $3(1+w)+2(1+w_\p)-6$ (or $3(1+w)+2(1+w_\p)-3$ for $c_\rho$) and $\tcveq$ (or $\tcreq $) is finite. As an example consider $w_\p=w$ (equal pressure in the internal
space and 3D space, $\p=p$) and check that both $\tcveq$ and $\tcreq$ remains finite and equal to $1/80$. Clearly, when $\tcveq$ (or $\tcreq$) diverges we cannot write the Friedmann equation \refeq{Friedmann
7D equal} in the form
\refeq{H 7D equal}.  
%%%%%%%%%%%%%%%%%%%%%%%%%%%%%%%%%%%%%%%%%%%%%%%%%%%%%%%%%%%%%%%%%%%%%%%%%%%%%%%%%%%%%%%%%%%%%%%%%%%%%%%%%%%%%%%%%%%%%%%%%%%%%%%%%%%%%%
%%%%%%%%%%%%%%%%%%%%%%%%%%%%%%%%%%%%%%%%%%%%%%%%% static extra %%%%%%%%%%%%%%%%%%%%%%%%%%%%%%%%%%%%%%%%%%%%%%%%%%%%%%%%%%%%%%%%%%%%%%%
%%%%%%%%%%%%%%%%%%%%%%%%%%%%%%%%%%%%%%%%%%%%%%%%%%%%%%%%%%%%%%%%%%%%%%%%%%%%%%%%%%%%%%%%%%%%%%%%%%%%%%%%%%%%%%%%%%%%%%%%%%%%%%%%%%%%%%
\subsection{Static compact extra dimensions (generic pressures)}\label{bulk static extra}
We can follow the same procedure as in the equal scale factor limit for the case of static compact extra dimensions $F=0$. While in the previous subsection the two internal and observed spaces were evolving
according to the same dynamics, in this limit the extra dimensions do not evolve and remain static.

The two differential equations for this set--up become
\bea
\frac{3}{2}\left(\dot{H^2}+4H^3\right)&=&-\oneover{M^{10}}\left(c_VV+c_\rho\rho\right)H\rho-\frac{T^7_7}{2M^5}+15H\l_{\rm RS}-\frac{\k}{b_0^2}H\label{Friedmann 7D static}\\
\dot\rho+3(w+1)H\rho&=&2T_{07}\label{conservation 7D static}
\ena
where $c_V,c_\rho$ are as before \refeq{cV crho}. We introduce the new coefficients $\tcvst ,\tcrst $ defined by
\bea\label{ctilde static}
\tcvst =\frac{c_V}{3w-1},\qquad \tcrst =\frac{c_\rho}{3w+1}
\ena
After plugging (\ref{conservation 7D static}) into the Friedmann equation (\ref{Friedmann 7D static}) we come to the expressions for $H$ and $\chi$ 
\bea\label{H 7D static}
H^2&=&\frac{\tcrst }{3M^{10}}\rho^2+\frac{2\tcvst V}{3M^{10}}\left(\rho+\chi\right)+\frac{5\l_{\rm RS}}{2}-\oneover{6}\frac{\k}{b_0^2}\\
\dot\chi+4H\chi&=&\left(1+\frac{\tcrst }{\tcvst }\frac{\rho}{V}\right)2T^0_7-\frac{M^5}{2\tcvst V}HT^7_7  \label{chi 7D static}\\
\dot\rho+3(w+1)H\rho&=&2T_{07}\label{rho 7D static}
\ena
In analogy to the equal scale factor case, these expressions are valid as long as we don't have $w=1/3$ ($w=-1/3$) with $c_V\ne0$ ($c_\rho\ne0$).
%%%%%%%%%%%%%%%%%%%%%%%%%%%%%%%%%%%%%%%%%%%%%%%%%%%%%%%%%%%%%%%%%%%%%%%%%%%%%%%%%%%%%%%%%%%%%%%%%%%%%%%%%%%%%%%%%%%%%%%%%%%%%%%%%%%%%%
%%%%%%%%%%%%%%%%%%%%%%%%%%%%%%%%%%%%%%%%%%%%%%%%% proportional hubble %%%%%%%%%%%%%%%%%%%%%%%%%%%%%%%%%%%%%%%%%%%%%%%%%%%%%%%%%%%%%%%%
%%%%%%%%%%%%%%%%%%%%%%%%%%%%%%%%%%%%%%%%%%%%%%%%%%%%%%%%%%%%%%%%%%%%%%%%%%%%%%%%%%%%%%%%%%%%%%%%%%%%%%%%%%%%%%%%%%%%%%%%%%%%%%%%%%%%%%
\subsection{Proportional Hubble parameters}\label{proportional hubble}
We are going to combine the two limits of $a(t)=b(t)$ and $F=0$ in the same description, implying an equal cosmological evolution for the internal space and the 3D visible spatial
dimensions, in the first case, and, in the second case, the absence of evolution for the compact space. 

Both the two systems of differential equations (\ref{H 7D equal})--(\ref{rho 7D equal}) and (\ref{H 7D static})--(\ref{rho 7D static}) can be written in a unified
formulation that encloses them, defining some appropriate constant parameters. We introduce an ``effective'' number of dimensions $d$ that takes the values $d=6$ in the equal
scale factor limit and $d=4$ in the static compact extra dimensions. If we look at the equations (\ref{chi 7D equal}) and (\ref{chi 7D static}), we see that $d$ appears as the number of dimensions for which
the energy density $\chi$ satisfies the free radiation conservation equation in the limit of pure RS ($T^7_7=T^0_7=0$). In fact, the Friedmann equation plus the two (non)conservation equations can be
rewritten as
\bea
H^2&=&\frac{\tcrd }{(d-1)M^{10}}\rho^2+\frac{2\tcvd V}{(d-1)M^{10}}\left(\rho+\chi\right)-\oneover{2(d-1)}\frac{\k}{b^2}+\frac{30}{d(d-1)}\l_{\rm RS}  \label{bulk limits H}\\
\dot\chi+d H\chi&=&2T^0_7\left(1+\frac{\tcrd }{\tcvd }\frac{\rho}{V}\right)-\frac{M^5}{2\tcvd V}HT^7_7  \label{bulk limits chi}\\
\dot\rho+w_d H\rho&=&2T_{07}  \label{bulk limits rho}
\ena
We have in addition defined $w_d=3(1+w)+(d-4)(1+w_\p)$ and $\tcvd =c_V/(w_d-d),\tcrd =c_\rho/(w_d-d/2)$ where $c_V$ and $c_\rho$ are given in (\ref{cV crho}). We remind that in order to get an algebraic
equation for $H^2$ as a function of the energy densities $\rho$ and $\chi$ (\ref{bulk limits H}), we have to keep $\tcvd,\tcrd$ finite, i.e.  respectively $w_d\neq d,w_d\neq d/2$ unless $c_V=0,c_\rho=0$. For
example we cannot write $H$ in the form (\ref{bulk limits H}) if $w=1/3,w_\p=0$ in both the equal scale factor and the static compact extra dimension limit, since the linear term in $\rho$ has a diverging
coefficient. 

Moreover, we can further generalize this analysis introducing a parameter $\x$ such that $F=\xi H$. This description contains all the above studied limits. The analogous of the previous relations
(\ref{bulk limits H})--(\ref{bulk limits rho}) can be written as
\bea
H^2&=&\frac{\tcrx }{(3+2\xi)M^{10}}\rho^2+\frac{2\tcvx V}{(3+2\xi)M^{10}}\left(\rho+\chi\right)-\frac{1}{\x^2+3\x+6}\frac{\k}{a^{2\x}}+\frac{5}{\x^2+2\x+2}\l_{\rm RS} \non\\ \label{bulk limits H prop}\\
\dot\chi+d_\xi H\chi&=&2T^0_7\left(1+\frac{\tcrx }{\tcvx }\frac{\rho}{V}\right)-\frac{M^5}{2\tcvx V}HT^7_7  \label{bulk limits chi prop}\\
\dot\rho+w_\x H\rho&=&2T_{07}  \label{bulk limits rho prop}
\ena
where now $d_\xi$ is a more complicated function of the proportionality constant $\xi$ between the two Hubble parameters $d_\xi\equiv6\frac{\xi^2+2\xi+2}{3+2\xi}$ and it reduces to $d_\xi=6,d_\xi=4$ in the
two previously examined limits of equal scale factors and static compact extra dimensions ($\xi=1,\xi=0$). The constant $w_\x$ reduces to $w_d$ for $\x=0,\x=1$ and is defined by
$w_\x\equiv3(1+w)+2\x(1+w_\p)$.  The two coefficients $\tcvx,\tcrx$ are defined as $\tcvd,\tcrd$, with $w_d\to w_\x,d\to d_\x$. The result (\ref{bulk limits H prop}) is valid unless $\x=-3/2$. In that case
the equation for $H$ becomes algebraic --- though we still have the curvature term explicitly depending on the scale factor ---, thus
\bea\label{bulk limits H prop 3}
H^2&=&-\frac{\tilde c_{\rho,-\frac{3}{2}} }{33M^{10}}\rho^2-\frac{\tilde c_{V,-\frac{3}{2}} V}{33M^{10}}\rho-\frac{\k a^3}{33}+4\l_{\rm RS}\\
\dot\rho+3(w-w_\p)H\rho&=&2T_{07}\non
\ena
No mirage density appears and the Hubble parameter is a quadratic polynomial in the localized energy density $\rho$ alone. Besides, if the pressures are equal $w_\p=w$ in the pure RS set--up $T^0_7=0$, the
energy density is constant in time and so is $H^2+\frac{\k a^3}{33}$, for $\x=-3/2$. The set of equations \refeq{bulk limits H prop 3} also doesn't depend on $T^7_7$ at all. 

Again we have to restrict to $w_\x\ne d_\x,d_\x/2$ to keep $\tcvx,\tcrx$ finite (unless $c_V\propto w_\x-d_\x,c_\rho\propto w_\x-d_\x/2$). 

We remark that for the scale factors satisfying $b(t)=1/a(t)$, i.e. dynamical compactification with $\x=-1$, the equation for the mirage energy density $\chi$ in the pure RS set--up is still an effective 6D
free radiation conservation equation, as for $b(t)=a(t)$. In fact, the only solutions to $d_\x=6$ are $\x=\pm1$. To obtain a 4D free radiation equation for $\chi$ we have to require $\x=0$, since the second
solution to $d_\x=4$ is $\x=-3/2$, for which we don't define a mirage density \refeq{bulk limits H prop 3}.

However, this is not the end of the story. Introducing the effective 4D densities $\rhofo=V_{(2)}\rho$ and $\chifo=V_{(2)}\chi$ (where $V_{(2)}=v\,b^2(t)=v\,a^{2\x}(t)$ is the volume of the 2D internal
space), we have to replace the l.h.s. of equations \refeq{bulk limits chi prop}--\refeq{bulk limits rho prop} respectively by $\dot\chifo+(d_\x-2\x)H\chifo$ and $\dot\rhofo+(w_\x-2\x)H\rhofo$ (the r.h.s. are
also modified and we will explicitly write them at the end of subsection \ref{bulk critical points gen}). This tells us that the 4D mirage density is a free radiation energy density for pure RS in four
dimensions for $d_\x-2\x=4$, which has solutions $\x=0,\x=1$ --- i.e. static internal space or equal scale factors, justifying the study of these two limits, in particular.
%%%%%%%%%%%%%%%%%%%%%%%%%%%%%%%%%%%%%%%%%%%%%%%%%%%%%%%%%%%%%%%%%%%%%%%%%%%%%%%%%%%%%%%%%%%%%%%%%%%%%%%%%%%%%%%%%%%%%%%%%%%%%%%%%%%%%%
%%%%%%%%%%%%%%%%%%%%%%%%%%%%%%%%%%%%%%%%%%%%%%%%% comments %%%%%%%%%%%%%%%%%%%%%%%%%%%%%%%%%%%%%%%%%%%%%%%%%%%%%%%%%%%%%%%%%%%%%%%%%%%
%%%%%%%%%%%%%%%%%%%%%%%%%%%%%%%%%%%%%%%%%%%%%%%%%%%%%%%%%%%%%%%%%%%%%%%%%%%%%%%%%%%%%%%%%%%%%%%%%%%%%%%%%%%%%%%%%%%%%%%%%%%%%%%%%%%%%%
\subsection{Comments}
We here summarize some considerations about the bulk evolution equations derived in the previous subsections and, in particular, about the explicit expressions we have found for the Hubble parameters in the
discussed limits.
\begin{enumeratea}
\item
With the assumption of having the same pressure for the matter fluid in the two dimensional internal space and in the 3D visible space (i.e. $\p=p$), we found a form of the Friedmann equation that has the
advantage of keeping both the Hubble parameters not constrained by any particular ansatz. The Friedmann equation (\ref{Friedmann 7D equal pressures}) provides an expression for $(3H+2F)$ in terms of $\rho$
and $\chi$ (for spatially flat spaces). This solution, though, is satisfactory only in the limit of small $(H-F)$. When $(H-F)$ is not negligible w.r.t. $(3H+2F)$, the mirage density equation may be written
introducing an extra term independent of the bulk parameters $T^0_7,T^7_7$. This prevents $\chi$ to obey to a free radiation equation in the pure RS set--up ($T^0_7=T^7_7=0$), as it should instead be in the
context of the AdS/CFT correspondence (we will discuss the comparison between the bulk and the dual brane analysis in section \ref{examples}).
\item
In the simple limits of equal scale factors (\ref{H 7D equal}) and static compact extra dimensions (\ref{H 7D static}) we recovered an expression for $H^2$ containing a quadratic term in $\rho$ and a linear
term in $(\rho+\chi)$, where $\rho$ is the localized energy density and $\chi$ is an artificially introduced mirage density that accounts for the bulk dynamics. In fact it depends on the bulk parameters
$T^0_7,T^7_7$. When $T^0_7=T^7_7=0$, the mirage density obeys to 4D free radiation equation for the static compact extra dimension case and to 6D free radiation equation for the equal scale factor case. This
is in complete analogy to the 5D analysis \cite{Kiritsis:2002zf}, where the same dependence on $\rho$ and $\chi$ occurs and the mirage energy satisfies 4D free radiation for pure RS (i.e. $T^0_5=T^5_5=0$).
\item
When $w_\p=w$ (or equivalently $\p=p$) in the equal scale factor limit (\ref{H 7D equal}), we find the results given by the equal pressures subsection in the case $F=H$ (\ref{H 7D equal equal pressures}). The
two limits of equal pressures and equal scale factors then commute and the results are consistent.
\item
The description of section \ref{proportional hubble} encloses in a unifying way the results in the limits of static compact extra dimensions and equal scale factors. It moreover generalizes these results to
the case of evolutions governed by proportional Hubble parameters $F(t)=\x H(t)$. We will use the set of equations written in terms of the effective number of dimensions $d$ (that describes the two limits of
static internal space, with $d=4$, and equal scaling for the compactification space and the 3D space, with $d=6$) to study the corresponding cosmological evolution in the next section.  
\end{enumeratea}

We are now going to proceed to the analysis of the critical points for this seven dimensional universe in a 7D Randall--Sundrum set--up, including the energy exchange term. 
%%%%%%%%%%%%%%%%%%%%%%%%%%%%%%%%%%%%%%%%%%%%%%%%%%%%%%%%%%%%%%%%%%%%%%%%%%%%%%%%%%%%%%%%%%%%%%%%%%%%%%%%%%%%%%%%%%%%%%%%%%%%%%%%%%%%%%
%%%%%%%%%%%%%%%%%%%%%%%%%%%%%%%%%%%%%%%%%%%%%%%%% bulk fixed points %%%%%%%%%%%%%%%%%%%%%%%%%%%%%%%%%%%%%%%%%%%%%%%%%%%%%%%%%%%%%%%%%%
%%%%%%%%%%%%%%%%%%%%%%%%%%%%%%%%%%%%%%%%%%%%%%%%%%%%%%%%%%%%%%%%%%%%%%%%%%%%%%%%%%%%%%%%%%%%%%%%%%%%%%%%%%%%%%%%%%%%%%%%%%%%%%%%%%%%%%
\section{Bulk critical point analysis with energy exchange}\label{critical bulk}
We have until now transformed the second order differential equation (\ref{Friedmann 7D}) plus the (non)conservation equation (\ref{conservation 7D}) in a set of three linear differential equations
(\ref{bulk limits H})--(\ref{bulk limits rho}) for combined equal scale factor and static compact extra dimension limits, or more generally (\ref{bulk limits H prop})--(\ref{bulk limits rho prop}) for
proportional Hubble parameters. We have introduced the mirage density $\chi$ defined by its differential equation. In this section we will use the system of equations (\ref{bulk limits H})--(\ref{bulk
limits rho}), obtained to describe both the limit of equal scale factors and static compact extra dimensions, to find its fixed points and the corresponding stability. The critical point analysis will allow
us to study the cosmological evolution in the bulk description for $F=0$ or $F=H$. 

We make an assumption on the bulk components of the stress--energy tensor that appears in the differential equations for the energy densities $\rho$ and $\chi$. As in \cite{Kiritsis:2002zf}, we will take
the diagonal elements of the stress energy tensor to satisfy the relation
\bea\label{small diagonal components}
\bigg|\frac{T^{(\mathrm{diag})}_{m,B}}{T^{(\mathrm{diag})}_{v,B}}\bigg|\ll\bigg|\frac{T^{(\mathrm{diag})}_{m,b}}{T^{(\mathrm{diag})}_{v,b}}\bigg|
\ena 
This enforces the solution to the Friedmann equation to be reasonably independent of the bulk dynamics, since the $T^7_7$ term in (\ref{Friedmann 7D}) becomes negligible with respect to the first term on the
r.h.s. of the same equation. Imposing such a relation, $T^7_7$ disappears from the sets of linear differential equations, while we remain left with the $T^0_7$ component. For the future bulk calculations we
will define $T\equiv2T^0_7$ to simplify the notation.

Before starting the critical point analysis we note that when $T=0$ the system of equations \refeq{bulk limits H}--\refeq{bulk limits rho} have only trivial critical points characterized by zero visible
Hubble parameter when the internal space is flat. There are two of these critical points. One is given by $\sH^2=-\k/2(d-1)b^2$, $\srho=\schi=0$ (which is valid only if we are compactifying on hyperbolic or
flat spaces) and the other is $\sH=0$, $\tcrd \srho^2+2\tcvd V\left(\srho+\schi\right)=M^{10}\k/b$. 

We will first restrict to small density%%%%%%%%%%%%%%% footnote
\footnote{We are referring to the energy density localized on the 5--brane. The effective 4D density is $\rhofo=V_{(2)}\rho$, where $V_{(2)}$ is the volume of the internal compactification pace. Similar
relations are established for the mirage density and the pressures. We note that the volume of the 2D compact space varies in time, unless extra dimensions are static, since it is proportional to
$b^2(t)$: it contracts as the 4D visible space expands in the dynamical compactification approach \cite{Mohammedi:2002tv,Andrew:2006yt}}%%%%%%%%%%%%%%%% footnote
 $\rho\ll V$ and flat internal space $\k=0$ (remind that the 3D space is already supposed to be flat, having put $k=0$) and then go through the generic density
analysis. De Sitter stable solutions (for the 4D visible space--time) would represent the present accelerated era, while inflationary phases at early times may be associated to primordial inflation.
%%%%%%%%%%%%%%%%%%%%%%%%%%%%%%%%%%%%%%%%%%%%%%%%%%%%%%%%%%%%%%%%%%%%%%%%%%%%%%%%%%%%%%%%%%%%%%%%%%%%%%%%%%%%%%%%%%%%%%%%%%%%%%%%%%%%%%
%%%%%%%%%%%%%%%%%%%%%%%%%%%%%%%%%%%%%%%%%%%%%%%%% small densities %%%%%%%%%%%%%%%%%%%%%%%%%%%%%%%%%%%%%%%%%%%%%%%%%%%%%%%%%%%%%%%%%%%%
%%%%%%%%%%%%%%%%%%%%%%%%%%%%%%%%%%%%%%%%%%%%%%%%%%%%%%%%%%%%%%%%%%%%%%%%%%%%%%%%%%%%%%%%%%%%%%%%%%%%%%%%%%%%%%%%%%%%%%%%%%%%%%%%%%%%%%
\subsection{Small energy density and flat compact extra dimensions}
When the localized energy density is small and the internal space curvature vanishes, $\rho\ll V,\k=0$, the bulk Einstein equations (\ref{bulk limits H})--(\ref{bulk limits rho}) in terms of $H$, $\chi$ and
$\rho$ become
\bea
H^2&=&\frac{2\tcvd V}{(d-1)M^{10}}\left(\rho+\chi\right)  \label{bulk limits H small}\\
\dot\chi+d H\chi&=&T\\
\dot\rho+w_d H\rho&=&-T  \label{bulk limits rho small}
\ena
We note that in this approximation expression (\ref{bulk limits H small}) isn't valid for $w_d=d$, unless $c_V\propto(w_d-d)$ (this would for example determine a specific value for $w_\p$ as a function of
$w$). Nevertheless we will find in the rest of the section that $\tcvd$ always appears with the coefficient $(w_d-d)$ in the critical point analysis. We thus have to keep in mind that
$(w_d-d)\tcvd=c_V=(31w-6w_\p-5)/400$ always remains finite. 
\paragraph*{Fixed point solutions}
In the small density approximation, the fixed points in terms of the critical energy density can immediately be found (to have a full solution we have to make an ansatz on the form of the energy exchange
parameter). 

The first solution is given by
\bea\label{bulk fix small density}
\sH&=&-\frac{\sB}{w_d}\srho^{1/2}  \non\\
\schi&=&-\frac{w_d}{d}\srho\\
\sT&=&\sB\srho^{3/2}  \non
\ena
where $\sB\equiv-w_d\left(\frac{(d-w_d)2\tcvd V}{d(d-1)M^{10}}\right)^{1/2}$. This represents an inflationary critical point for the cosmological evolution, for $(w_d-d)\tcvd=c_V>0$, i.e. both $\tcvd>0$
and $d>w_d$ or $\tcvd<0$ and $d<w_d$% footnote
~\footnote{As examples of negative $\tcvd$ and $(d-w_d)$, both negativeness conditions can be satisfied if $w=1/3$ for $w_\p>8/9$ when $H=F$, but never when $F=0$. If $w=0$ we must have $w_\p>1/2$ in the
equal scale factors limit, while no solution can be found with static compact extra dimensions. We instead get positive $\tcv$ and $(d-w_d)$ if $w=0$ for $-5/6<w_\p<1/2$ with $H=F$ and for $w_\p>1/2$ with
$F=0$. No value of $w_\p$ satisfies the positiveness conditions if $w=1/3$.}% footnote
. The acceleration factor at the fixed point is simply given by $q_\star=\sH^2$. Since we assume $w_d$ to be positive, $\sB$ is negative and $\sH$ in \refeq{bulk fix small density} is positive. At this fixed
point the universe is thus expanding.

Another fixed point leaves $\schi$ unchanged, while $\sT$ and $\sH$ have switched signs with respect to \refeq{bulk fix small density}, meaning that $\sH$ is negative and the universe in contracting (there is
a symmetry $T\to-T,H\to-H$ in the whole system of equations)
\bea\label{bulk fix small density 2}
\sH&=&\frac{\sB}{w_d}\srho^{1/2}  \non\\
\schi&=&-\frac{w_d}{d}\srho\\
\sT&=&-\sB\srho^{3/2}  \non
\ena

The trivial critical point is characterized by mirage density equal and opposite to $\srho$, but zero Hubble parameter and energy exchange (in the case we admit for the energy exchange the form $T=A\rho^\n$
all the variables are zero at the trivial fixed point).

For positive critical energy densities $\srho$, we obtain a negative brane--bulk energy exchange parameter at the critical point if $\sH>0$ and, viceversa, we have positive critical energy exchange for a
contracting universe at the critical point. During the evolution, we can expect a change of regime going from negative to positive $T$ as the energy density localized on the brane grows, as for the 5D RS
critical point analysis with energy exchange carried in \cite{Kiritsis:2002zf}. Even though most of the analysis in this section will be performed supposing that the energy exchange parameter has fixed sign
determined by the sign of $A$ (since we will mainly assume $T=A\rho^\n$), we can argue that for small energy density $\rho$ the generic energy exchange is presumably negative, meaning that energy would be
transferred from the bulk onto the brane. In this hypothesis, an equilibrium can be reached, such that the energy density would have a large limiting value for which energy starts to flow back into the bulk
(with positive energy exchange).
\paragraph*{Stability analysis}
The real parts of the eigenvalues of the stability matrix for the $(\d\chi,\d\rho)$ linear perturbations corresponding to the critical point (\ref{bulk fix small density}) can have opposite signs or be
both negative. This depends on the value of $T$ as a function of the energy density $\rho$ at the fixed point \refeq{bulk fix small density} describing an expanding universe with energy influx. The explicit
form of the eigenvalues is
\bea
\l_\pm=\frac{\sB}{2w_d}\srho^{1/2}\left[d+(1-\tn)w_d\pm\sqrt{(d+(1-\tn)w_d)^2-2(3-2\tn)dw_d}\right]  \label{eigenval small}
\ena
where we have defined
\bea\label{def nu tilde}
\tn\equiv\frac{\pa\log|T|}{\pa\log\rho}\bigg|_\star
\ena
Expression \refeq{eigenval small} then shows that the two eigenvalues have negative real part when $\tn<3/2$. There is a second upper bound on $\tn$ derived from requiring negative real part for the
eigenvalues. Nonetheless, for the range of values $-1\le w,w_\p\le1$ and $d=4,6$ in which we are interested, this bound is always equal or greater than $3/2$. If $|T|$ is a decreasing or constant function of
$\rho$ near $\sT$, the non trivial inflationary critical point always is an attractor. Also for growing $|T|$ we can have stable inflationary fixed points, as long as $\tn<3/2$. In particular, the linear
case $\tn=1$ is included in the stable inflationary fixed point window and will be analyzed both solving the Einstein equations numerically, in the next subsection, and deriving an explicit solution, in
subsection \ref{explicit bulk section}. 

Besides, when 
\bea\label{spiral range small}
1-\sqrt{\frac{2d}{w_d}}-\frac{d}{w_d}<\tn<1+\sqrt{\frac{2d}{w_d}}-\frac{d}{w_d}
\ena
the eigenvalues have non zero imaginary part and the critical point is a stable spiral if in addition $\tn<3/2$. For values of $\tn$ out of the range \refeq{spiral range small}, we get a node. As an example,
let's assume the value $\tn=1$ in the equal scale factor background. This gives a stable spiral for $w_\p>1/2$ or $w_\p<1/2$ and $w>-2(1+w_\p)$ (considering $w,w_\p>-1$). This means that in the case
$w\simeq1/3$ and $w_\p=0$ the critical point is a stable spiral, while for both $w$ and $w_\p$ null we instead have a stable node.

For energy outflow $\sT>0$ (which goes along with contraction $\sH<0$), we get a minus sign overall modifying the eigenvalues \refeq{eigenval small} referring to the linearized system around the critical
point \refeq{bulk fix small density 2} (characterized indeed by energy outflow). The eigenvalues cannot be both negative in this case. In fact we should demand $w_d>d/(\tn-1)$ with $\tn>1$ but also $\tn<3/2$
to get a stable fixed point. Only the trivial point, as we will discuss later, can be attractive for energy outflow dynamics.
\paragraph*{Assumption $T=A\rho^\n$ and numerical solutions}
Assuming the brane--bulk energy exchange parameter to take the form $T=A\rho^\n$ (so that $\tn=\n$ referring to \refeq{def nu tilde}), we can rewrite the system of differential equations in term of
dimensionless quantities $\check\rho=\g^6\rho$, $\check\chi=\g^6\s$, $\check H=\g H$, $\check T=\g^7 T$, where we called $\g^4\equiv\frac{2V}{(d-1)M^{10}}$. The dimensionless variable $\rho/V$ used to perform
the small energy density expansion at the beginning of this section is related to the dimensionless variable $\crho$ by $\crho=\left(2\tcvd/(d-1)\right)^{3/2} \left(V/M^6\right)^{5/2}(\rho/V)$. So,
considering the small $\rho/V$ approximation is equivalent to considering small $\crho$ approximation if the brane tension $V$ satisfies $V\lesssim M^6$ with respect to the 7D Planck mass and $\tcvd$ is
reasonably of the order $\tcvd\lesssim1$. The complete set of fixed point solutions (discarding the trivial ones) can be calculated in terms of the parameters $A,\n$ characterizing the energy exchange and
$w_d,d$ denoting the background (static extra dimensions or equal scale factors) and the equations of state for both the 3D and internal spaces. 

The Einstein equations become
\bea\label{dimensionless ODEs small}
\cHH^2&=&\tcvd\left(\crho+\cchi\right)  \non\\
\dot\cchi+d\cHH\cchi&=&\cAA\crho^\n\\
\dot\crho+w_d\cHH\crho&=&-\cAA\crho^\n  \non
\ena
where $\cAA=\g^{1+6(1-\n)}A$. 

\begin{figure}[h!]
\begin{tabular}{cc}
\includegraphics[width=0.5\textwidth]{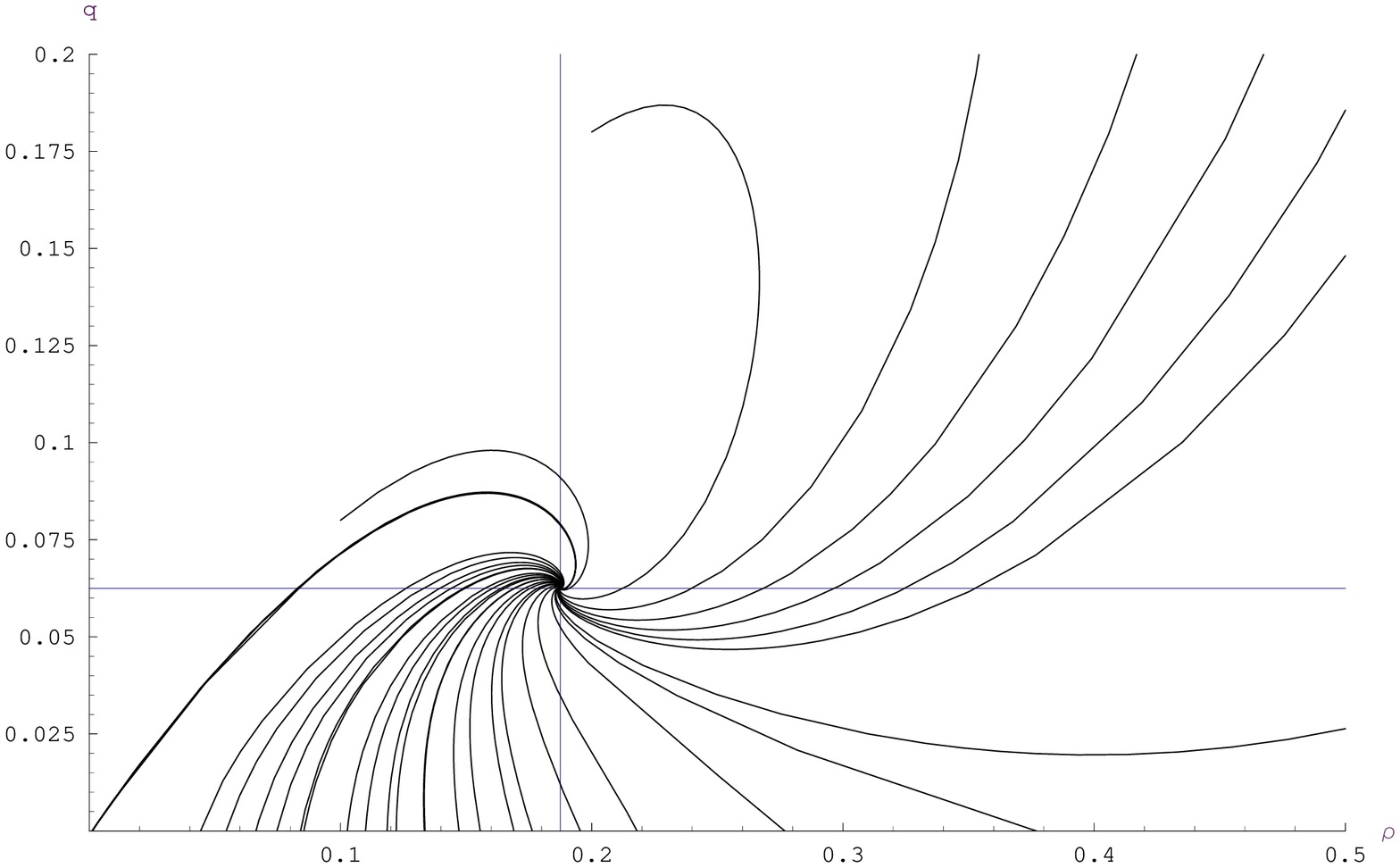}&
\includegraphics[width=0.5\textwidth]{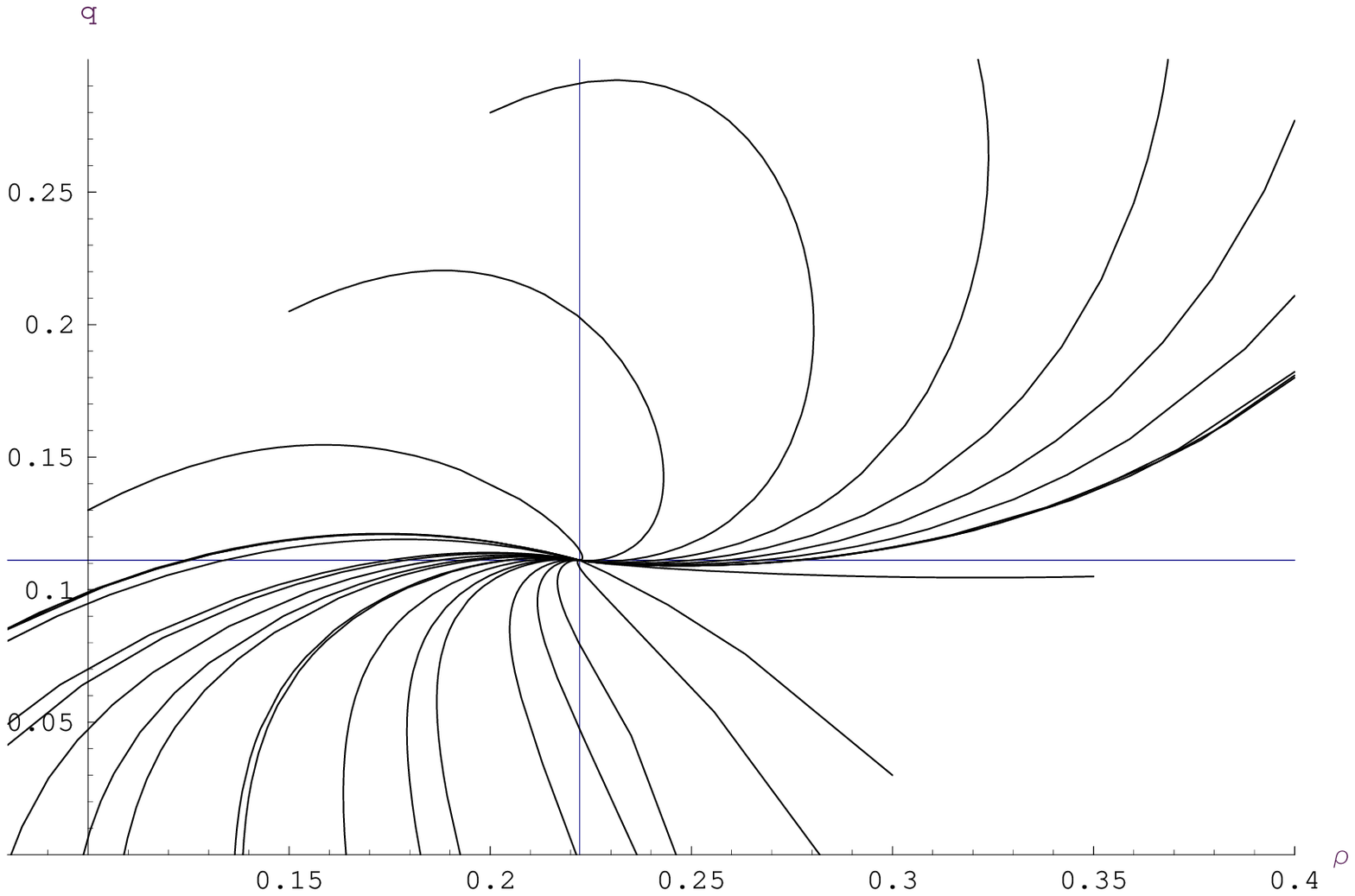}\\
(a)&(b)
\end{tabular}
\caption[Phase spaces $\cqq/\crho$ with influx of brane--bulk energy exchange in the small energy density $\rho$ approximation]{\label{phase space small} Phase spaces $\cqq/\crho$ for different initial
conditions $\crho_0$ and $\cqq_0$, with $\cAA=-1$, $\n=1$, $d=6$ (equal scale factors --- analogous pictures come from the static extra dimension case) and: (a) $w_d=4$ (for instance $w=0,w_\p=-1/2$ or
$w=-1/5,w_\p=0$) leading to a spiral--like stable critical point, (b) $w_d=3$ (for instance $w=0,w_\p=1/2$) determining a stable node. The extra blue grid lines intersection represents the fixed point.}
\end{figure}

The acceleration $\cqq$ can be evaluated independently of $\n$ and $\cAA$
\bea
\cqq=\left(1-\frac{w_d}{2}\right)\tcvd\,\crho+\left(1-\frac{d}{2}\right)\tcvd\,\cchi
\ena
as a function of the localized matter density and of the mirage density. Due to the positiveness constraint on $\tcvd\left(\crho+\cchi\right)$ coming from the first equation in \refeq{dimensionless ODEs
small}, the trajectories in the phase space must satisfy
\bea
\cqq\le (d-w_d)\tcvd\,\rho
\ena
as it is indeed showed in the numerical plots of figure \ref{phase space small}. For $w_d>2$ and $\tcvd>0$ we have positive acceleration only if the mirage density $\cchi$ is negative and smaller than
$-\crho(2-w_d)/(2-d)$.  On the other hand, $\cchi$ gets positive (suppose $\tcvd>0$) only if $\cqq<(2-w_d)\crho/2$.

The fixed points are given by
\bea
\check\sH^{3-2\n}&=&(-)^{3-2\n}\left(\frac{\tcvd(d-w_d)}{d}\right)^{1-\n}\frac{\cAA}{w_d}\\
\check\schi^{3-2\n}&=&(-)^{3-2\n}\frac{d^{2(\n-1)}\cAA^2}{\tcvd(d-w_d)w_d^{2\n-1}}\\
\check\srho^{3-2\n}&=&\frac{d\cAA^2}{w_d^2\tcvd(d-w_d)}\label{num fixed rho small}
\ena 
For $\n<3/2$, when the non trivial fixed point is stable, we have two roots of (\ref{num fixed rho small}) with opposite signs if $\n=1/2+m,m\in\mathbb Z$. Only one real root exists for integer $\n$ and it
carries the sign of the r.h.s. in \refeq{num fixed rho small}. Finally, for $\n=(2m+1)/4$, we have a positive root if the r.h.s. in \refeq{num fixed rho small} is negative. The two eigenvalues corresponding
to a negative $\csrho$ always have opposite real part, implying that this fixed point isn't be stable at linear order, it is a saddle. We further note that for $w_d>d$  and positive $\tcvd$ (or alternatively
$d>w_d$ and negative $\tcvd$) we can only have real and positive $\srho$ fixed point if $\n=(2m+1)/4,m\in\mathbb Z$. This implies that for $\n=1$ there is no non trivial fixed point with positive $\srho$,
when $(d-w_d)$ and $\tcvd$ have opposite signs. Moreover, these negative $\srho$ points are always characterized by negative $\sH$, so that they wouldn't be inflationary.

We can check the stability of the critical points by means of a numerical analysis of the differential system of equations (\ref{dimensionless ODEs small}). In the case of energy influx $\cAA<0$, putting
$\n=1$ and different values for the $d,w_d$ parameters, we get the phase spaces in figure \ref{phase space small}, plotting the acceleration factor $\check q(t)\equiv\ddot{\check{a}}(t)/\check a(t)$ as a
function of the energy density $\crho(t)$. We thus check that, solving the system of differential equations for variable initial conditions for $\cchi$ and $\crho$, including both positive and negative
initial $\cqq$, all the different trajectories converge to the non trivial fixed point, designated by the intersection of the two perpendicular lines in the picture. Besides, as we should expect from the
values of the parameters, in part \ref{phase space small}(a) they have a spiral behavior, while in the \ref{phase space small}(b) case they denote a node.

In the limit $w_d\to d$ we numerically recover the analytical solution discussed in subsection \ref{explicit bulk section}, neglecting the large density behavior.

For $\n>3/2$, i.e. when the non trivial fixed point is no more an attractor, we find that some of the trajectories go to the trivial critical point, while another branch of solutions to the Einstein equations
\refeq{dimensionless ODEs small} are characterized by diverging energy density $\crho$ (they become unreliable when $\crho^2\gtrsim(2\tcvd/(d-1))^{3}(V/M^6)^5$). This happens because, as it is suggested by
the integration of the third equation in \refeq{dimensionless ODEs small} with energy influx hypothesis, for $\crho$ big enough --- precisely for
$\crho^{\n-\frac{3}{2}}\gtrsim\left(w_d/A\right)\tcvd^{\frac{1}{2}}\left(1+\cchi/\crho\right)^\oneover{2}$ --- the function $\crho(t)$ starts growing, while it eventually goes to zero for small $\crho$.
Depending on the initial conditions we will have solutions ending in the trivial fixed point or diverging. 

The behavior of the system with energy outflow can be deduced analytically. Given the hypothesis $T>0$ it is clear that the trajectories can't be attracted by the fixed point solution $\sT=\sB\srho^{3/2}$.
They also can't flow to the critical point characterized by negative Hubble parameter and $\sT=-\sB\srho^{3/2}$. We already determined the non attractive nature of both these fixed points.  Indeed, from the
form of the Einstein equations \refeq{bulk limits rho small} we can conclude that all the trajectories in the phase space $\cqq/\crho$ go toward the trivial point, since for positive $T$ the density $\rho$ is
suppressed at late time. The way in which the trajectories go to the critical point depends on the positiveness of the function under the square root in (\ref{eigenval small}). We numerically checked that for
istance for $\n=1$ and $d=6,w_d=4$ the null fixed point is an attractor and, in particular, a stable node. 
%%%%%%%%%%%%%%%%%%%%%%%%%%%%%%%%%%%%%%%%%%%%%%%%%%%%%%%%%%%%%%%%%%%%%%%%%%%%%%%%%%%%%%%%%%%%%%%%%%%%%%%%%%%%%%%%%%%%%%%%%%%%%%%%%%%%%%
%%%%%%%%%%%%%%%%%%%%%%%%%%%%%%%%%%%%%%%%%%%%%%%%%%%% general density %%%%%%%%%%%%%%%%%%%%%%%%%%%%%%%%%%%%%%%%%%%%%%%%%%%%%%%%%%%%%%%%%
%%%%%%%%%%%%%%%%%%%%%%%%%%%%%%%%%%%%%%%%%%%%%%%%%%%%%%%%%%%%%%%%%%%%%%%%%%%%%%%%%%%%%%%%%%%%%%%%%%%%%%%%%%%%%%%%%%%%%%%%%%%%%%%%%%%%%%
\subsection{Critical points with general energy density}\label{bulk critical points gen}
Allowing $\k$ to be different from zero, not restricting the localized energy density to be small and with the assumptions \refeq{small diagonal components} on the bulk matter stress--energy tensor, we have
to solve the following set of equations
\bea
H^2&=&\frac{\tcrd }{(d-1)M^{10}}\rho^2+\frac{2\tcvd V}{(d-1)M^{10}}\left(\rho+\chi\right)-\oneover{2(d-1)}\frac{\k}{b^2}  \label{bulk limits H gen}\\
\dot\chi+d H\chi&=&T\left(1+\frac{\tcrd }{\tcvd }\frac{\rho}{V}\right)  \label{bulk limits chi gen}\\
\dot\rho+w_d H\rho&=&-T  \label{bulk limits rho gen}
\ena 

As in the small energy density regime, we can immediately notice that the analytical behavior for energy outflow is characterized by decreasing $\rho$ in time. The trajectories will thus be attracted to the
trivial fixed point characterized by vanishing $\srho$. 

We will now carry the general critical point analysis. Equation \refeq{bulk limits H gen} exhibits divergences if $w_d=d$ or $w_d=d/2$, unless $c_V\propto(w_d-d)$ or $c_\rho\propto(w_d-d/2)$. When the
divergence arises, the right system of equations is \refeq{Friedmann 7D ansatz}. Terms like $(w_d-d)\tcvd=c_V$ and $(w_d-d/2)\tcrd =c_\rho$ are always finite though (where we recall $c_V=31w-6w_\p-5$ and
$c_\rho=11w+14w_\p+10-(w-w\p)(3w-7w_\p)$). 
\paragraph*{Fixed point solutions}
If we demand $H$ to be positive, i.e. expanding universe, the solution for $\sH$ and $\sT$ is given by
\bea  \label{fixed line gen}
\sH&=&-\frac{\sB}{w_d}\srho^{1/2} \non\\
\schi&=&-\frac{w_d}{d}\left(1+\frac{\tcrd \srho}{\tcvd V}\right)\srho\\
\sT&=&\sB\srho^{3/2}\non
\ena
where $\sB=\sB\left(\srho\right)$ depends also on $\k$ and $\srho$ and is defined by
\bea\label{B star general}
\sB\left(\srho\right)=-w_d\left[\frac{(d-2w)\tcrd \srho+2(d-w)\tcvd V}{d(d-1)M^{10}}-\frac{\k}{2(d-1)b^2_\star\srho}\right]^{1/2}
\ena
We have a negative energy exchange parameter, as in the small density limit, corresponding to the positive Hubble parameter. 

The second fixed point solution is equal to the first except for the $\sH$ and $\sT$ signs switched (keep in mind the $H\to-H$, $T\to-T$ symmetry). $\sH$ is negative and we have energy outflow at the critical
point 
\bea  \label{fixed line gen}
\sH&=&\frac{\sB}{w_d}\srho^{1/2} \non\\
\schi&=&-\frac{w_d}{d}\left(1+\frac{\tcrd \srho}{\tcvd V}\right)\srho\\
\sT&=&-\sB\srho^{3/2}\non
\ena

The trivial critical point is characterized by vanishing $\sH$ and $\sT$, while the mirage density becomes 
\bea
\schi=\frac{M^{10}\k}{4\sbb^2\tcvd V}-\left(1+\frac{\tcrd \srho}{2\tcvd V}\right)\srho
\ena
If the energy exchange is supposed to be of the form $T=A\rho^\n$, the trivial fixed point is characterized by zero value for all the variables except for the mirage density $\chi$ which becomes
$\schi=M^{10}\k/4\sbb^2\tcvd V$ and is zero for flat compact spaces.

As in the limit of small energy density considered in the previous section, the constant $\sB$ is negative whenever the argument of the square root in (\ref{B star general}) is positive, i.e. when
\bea
\frac{\k}{2(d-1)\sbb^2}<\frac{(d-2w_d)\tcrd \srho+2(d-w)\tcvd V}{d(d-1)M^{10}}\srho
\ena
If the square root gives an immaginary number, we don't have any real valued fixed point except for the trivial one.
\paragraph*{Stability analysis} 
The positiveness of the eigenvalues of the stability matrix depends now on all the parameters and constants of the theory and not only on $\tn$, as for the small density, flat compact extra dimension simple
case.  We consider the situation in which the variation of $\k/b^2$ vanishes (this happens in the static compact extra dimension limit or for $\k=0$ in the equal scale factor background. Otherwise we would
have a linearized system of two differential equations plus one algebraic equation in the four variables $\d\k,\d\rho,\d\chi,\d H$. 

There are two conditions that must be satisfied, in order to get two negative eigenvalues and hence a stable fixed point. These conditions give two upper bounds for $\tn$ in terms of the constants $w_d,d$ and
$V,\srho,\k/b_\star^2,M$
\bea\label{bounds gen}
(\tn-1)&<&\frac{w_d}{d}\frac{w_d}{(d-1)\sB^2M^{10}}\left[(d-2w_d)\tcrd \srho+(d-w_d)\tcvd V\right]\\
(\tn-1)&<&\frac{d}{w_d}\non
\ena
The second bound in (\ref{bounds gen}) satisfies $d/w_d+1>3/2$ in the range $-1\le w,w_\p\le1$. Besides, the first bound reduces to $\n<3/2$ when we take the limit $\rho/V\ll1$ and put $\k=0$.  The results
are in agreement with the previous small density analysis. 

The bounds \refeq{bounds gen} depend on the fixed point value of $\rho$, which can't be determined without making any assumption on the form of $T$. However, we can make some remarks on the nature of the
fixed points. For values of $\tn$ in the range
\bea
1-\frac{d}{w_d}-R_\star<\tn<1-\frac{d}{w_d}+R_\star
\ena
where we defined
\bea
R_\star&\equiv&2\sqrt{\frac{w_d}{(d-1)\sB^2M^{10}}\left[(d-2w_d)\tcrd \srho+(d-w_d)\tcvd V\right]}\non
\ena
the stability matrix eigenvalues have non null immaginary part and the trajectories near to the critical point have a spiral--like behavior. When $\sR^2<0$ we always have node--like fixed points. In agreement
with the small density case (\ref{spiral range small}), when $\rho/V\ll1$ and $\k=0$ we get $\sR\to\sqrt{2d/w_d}$.

As an example, we assign the value $\tn=1$. Since the first bound (\ref{bounds gen}) can be rewritten as
\bea\label{spiral bound gen nu}
(\tn-1)<\frac{w_d}{d}\frac{\sR^2}{4}
\ena
this means that we should have $\sR^2>0$ to get stability. We also find that the fixed points have spiral shape when $\sR>d/w_d$ or $\sR<-d/w_d$, they will be nodes otherwise.
\paragraph*{Assumption $T=A\rho^\n$ and numerical solutions}
To do a more quantitative analysis we have to make an ansatz on the form of the energy exchange parameter $T$. As in the previous section, we suppose a power dependence on the energy density $\rho$ such that
$T=A\rho^\n$. The equations for generic energy densities and internal space curvature can be rewritten introducing dimensionless variables as in (\ref{dimensionless ODEs small})
\bea\label{dif sys gen num}
\cHH^2&=&\tcrd\,\a\crho^2+\tcvd\left(\crho+\cchi\right)-\ck   \non\\
\dot\cchi+d\cHH\cchi&=&\cAA\rho^\n\left(1+2\frac{\tcrd }{\tcvd}\a\crho\right)\\
\dot\crho+w_d\cHH\crho&=&-\cAA\crho^\n   \non
\ena
where $\a$ is a dimensionless constant defined by $\a^2\equiv\frac{(d-1)^3}{64}\left(\frac{M^6}{V}\right)^5$, $\ck$ is the dimensionless variable $\ck=\frac{\g^2\k}{2(d-1)b^2}$ --- we remind that we restrict
to constant $\ck$ approximation.

\begin{figure}[h!]
\begin{tabular}{cc}
%\parbox{0.5\textwidth}{\includegraphics[width=0.5\textwidth]{figures/phsp1g.eps}}&
%\parbox{0.5\textwidth}{\includegraphics[width=0.5\textwidth]{figures/phsp2g.eps}}\\
\includegraphics[width=0.5\textwidth]{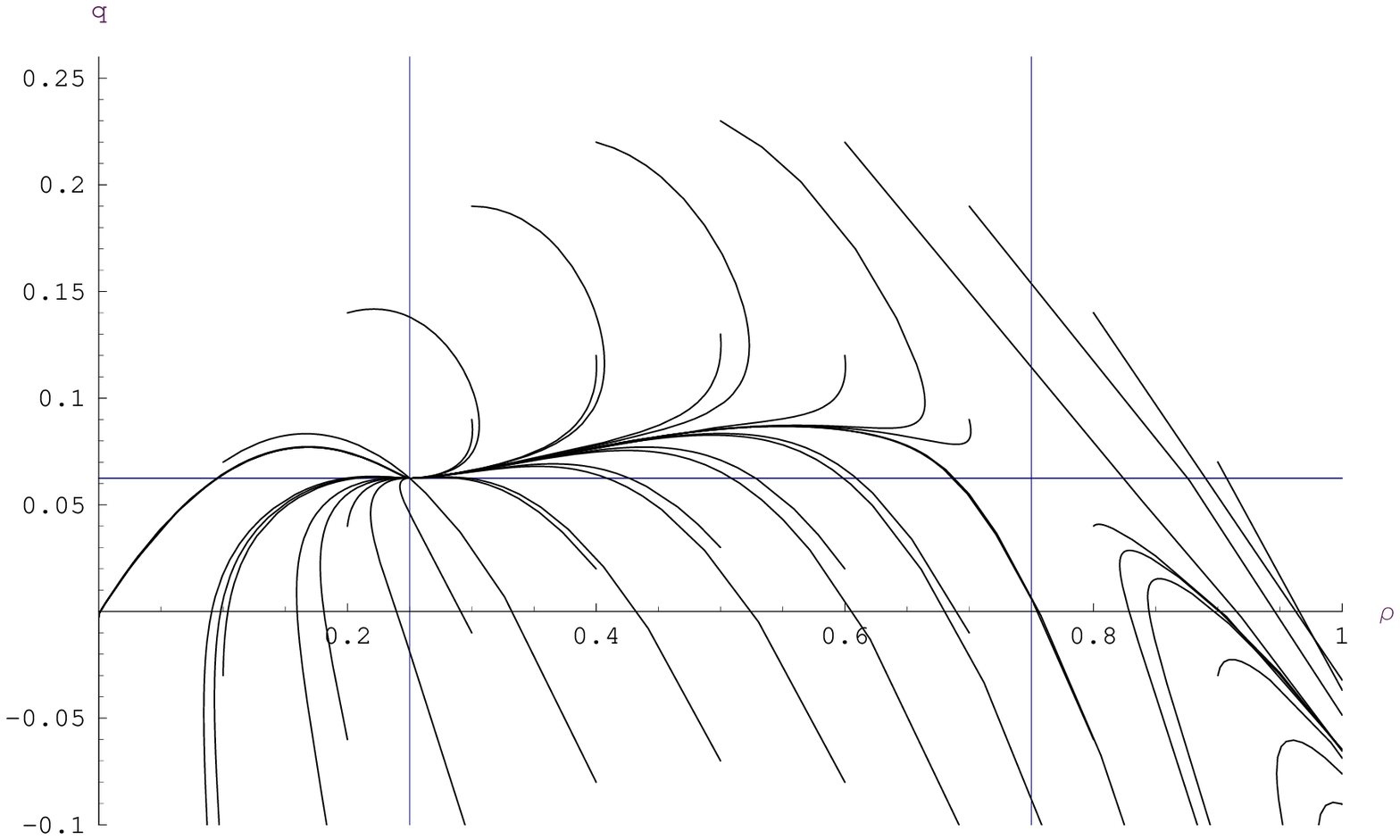}&
\includegraphics[width=0.5\textwidth]{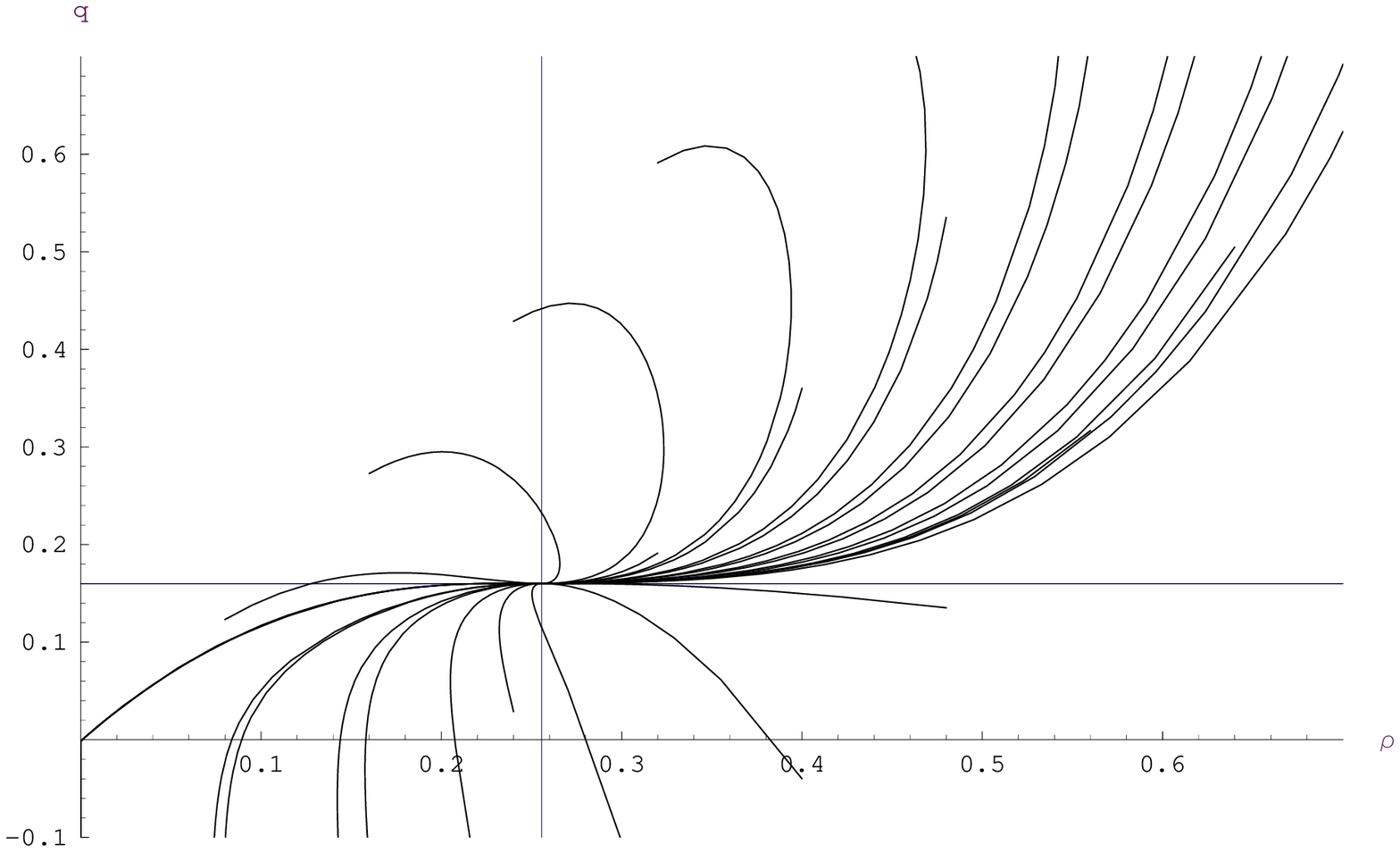}\\
(a)&(b)\\
%\parbox{0.5\textwidth}{\includegraphics[width=0.5\textwidth]{figures/phsp3g.eps}}&
%\parbox{0.5\textwidth}{\includegraphics[width=0.5\textwidth]{figures/phsp4g.eps}}\\
\includegraphics[width=0.5\textwidth]{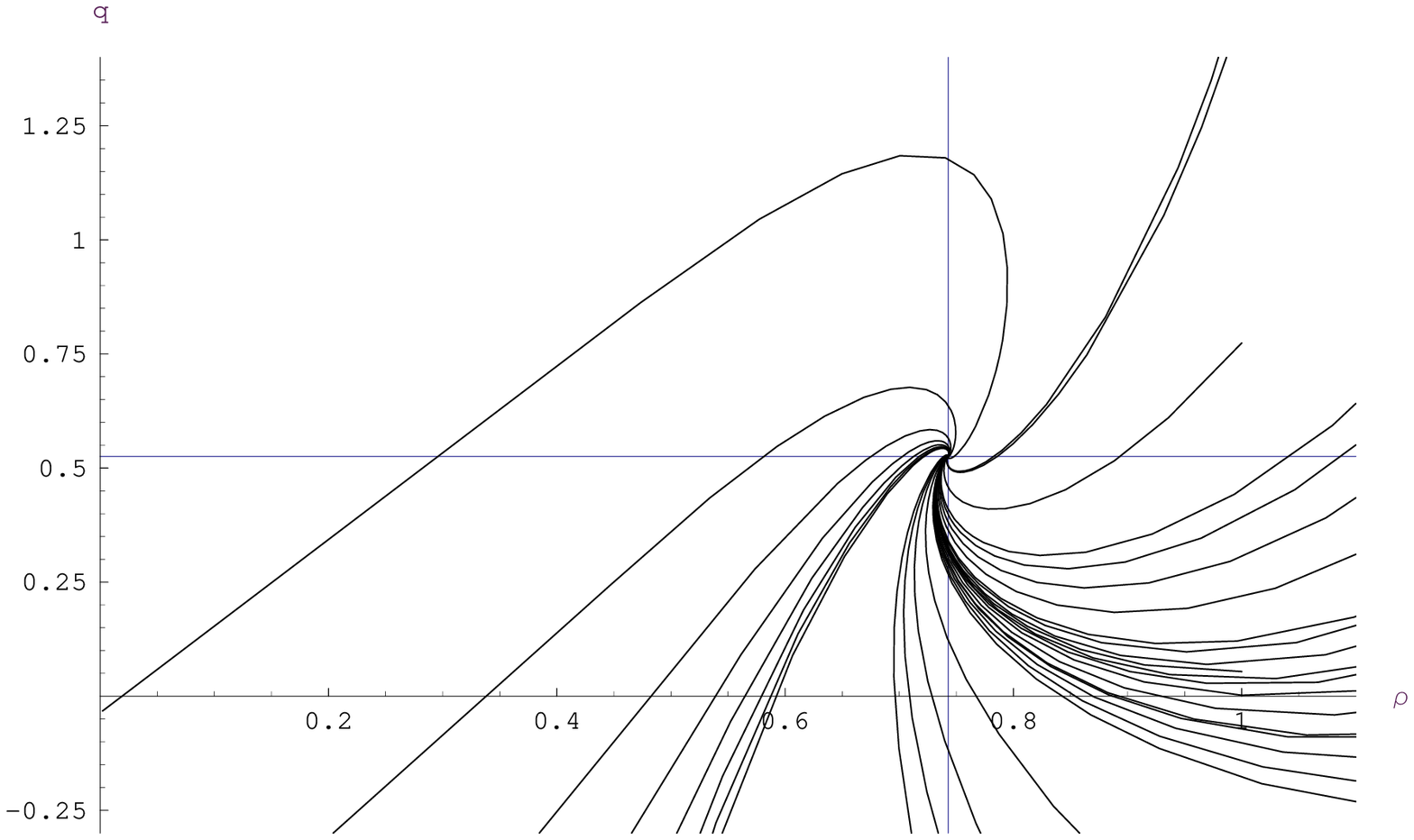}&
\includegraphics[width=0.5\textwidth]{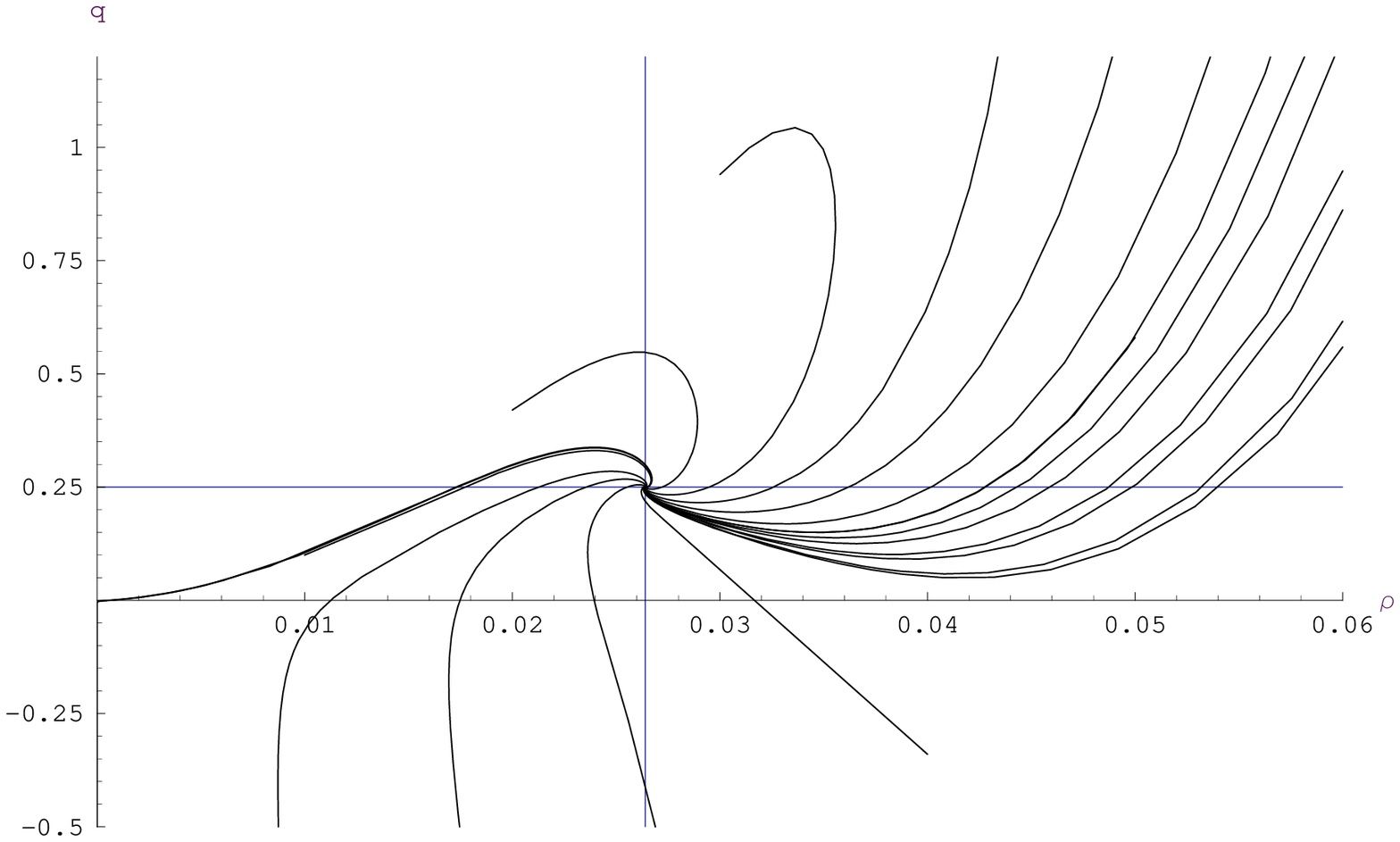}\\
(c)&(d)
\end{tabular}
\caption[Phase spaces $\cqq/\crho$ with influx of brane--bulk energy exchange and general energy density $\rho$]{\label{phase space gen}Trajectories in the phase spaces $\cqq/\crho$ with varying $\crho_0$ and
$\cchi_0$, $d=6,\cAA=-1$ (analogous pictures come from the static extra dimension case) and: (a) $w_d=4>d/2,\a=1,\n=1$ leading to stable node in $\csrho=1/4$ plus a second repulsive node in $\csrho=3/4$, (b)
$w_d=2.5<d/2,\a=1,\n=1$ determining a stable node, (c) $w_d=2.5<d/2,\a=1,\n=-1$ an example of a stable spiral, (d) $w_d=2<d/2,\a=10^3,\n=1$ a spiral behavior for large $\a$ (large $M^6/V$).}
\end{figure}

To obtain real observables, we have to restrict the possible values for $\crho$ and $\cchi$ such that $\tcrd\,\a\crho^2+\tcvd\left(\crho+\cchi\right)-\ck\ge0$. In fact, the plots show the presence of a
prohibited zone in the phase space --- in particular, in figure \ref{phase space gen}(a) it is clear that the region of the possible trajectories is delimited by a parabola. The relation that must be
satisfied, in terms of the acceleration parameter and the energy density, is
\bea  \label{acceleration parabola gen}
\cqq\le (d-2w_d)\tcrd\,\a\crho^2+(d-w_d)\tcvd\,\crho-d\k
\ena
In fact, the analytical expression for the acceleration $\cqq=\dot\cHH+\cHH^2$ can be written using (\ref{dif sys gen num}) in terms of the visible energy density and the mirage density. For any $\n$
\bea  \label{q dimensionless gen}
\cqq=(1-w_d)\tcrd\,\a\crho^2+\left(1-\frac{w_d}{2}\right)\tcvd\,\crho+\left(1-\frac{d}{2}\right)\tcvd\,\cchi-\ck
\ena
So, taking a specific value for $\crho$, we can have positive acceleration for our universe only if 
\bea
-\left(\tcrd\,\a\crho^2+\tcvd\,\crho-\ck\right)\le\tcvd\,\cchi<-\frac{2(w_d-1)\tcrd\,\a\crho^2+(w_d-2)\tcvd\,\crho+2\ck}{d-2}
\ena
and a necessary condition for this to be possible is a bound on the energy density $(d-2w_d)\tcvd\a\crho^2+(d-w_d)\tcvd\crho>d\ck$, as we can deduce from \refeq{acceleration parabola gen}. The mirage density
$\cchi$ has to be negative to get positive acceleration for $w_d\ge2,\k\ge0$. If instead $w_d\le1,\k\le0$, the mirage density is positive for negative $\cqq$.

Manipulating the set of equations \refeq{dif sys gen num}, we write the following differential equations in terms of the generic energy exchange parameter $T$
\bea
a\frac{\intd\cchi}{\intd a}&=&-d\cchi+\eta\cTT\left(1+2\frac{\tcrd }{\tcvd}\a\crho\right)\left[\tcrd\, \a\crho^2+\tcvd\left(\crho+\cchi\right)-\ck\right]^{-\frac{1}{2}}\\
a\frac{\intd\crho}{\intd a}&=&-w_d\crho-\eta\cTT\left[\tcrd\, \a\crho^2+\tcvd\left(\crho+\cchi\right)-\ck\right]^{-\frac{1}{2}}
\ena  
We thus come to the differential equation for the acceleration factor
\bea
\left(w_d\crho+\frac{\eta\cTT}{\cHH}\right)\frac{\intd\cqq}{\intd\crho}&=&-\frac{\eta\cTT}{2\cHH}\left(2\a c_\rho\crho+c_V\right)+\\
&&+\left[2\a(1-w_d)c_\rho\crho+\oneover{2}(2-w_d)c_V\right]\crho+d\cqq
\ena
where $\cHH=\sqrt{\left(2\a c_\rho\crho^2+c_V\crho+d\ck\right)/2+\cqq}$ and $\eta=\pm1$ denotes the two possible roots for $\cHH$. Again we note the presence of the symmetry $\cHH\to-\cHH$, $\cTT\to-\cTT$. We
have used the definitions $(w_d-d)\tcvd=c_V$ and $(w_d-d/2)\tcrd =c_\rho$.  From this equation we can infer that positive $\cqq$ implies growing $\cqq$ in an expanding universe ($\eta=+1$) with energy outflow
($\cTT>0$) if $c_V<0,c_\rho>0$ and $\crho<\frac{w_d-2}{w_d-1}\frac{|c_V|}{4\a c_\rho}\equiv\crho_{lim}$. For $\tcvd,\tcrd >0$, $c_V<0,c_\rho>0$ this is realized if $d/2<w_d<d$. In the case of energy influx
($\cTT<0$), we get increasing positive acceleration if $c_V,c_\rho>0$ and $w_d<1$ for all positive energy densities ($\tcvd,\tcrd $ has to be negative). Or else, $\cqq$ grows as $\crho$ grows if
$c_V>0,c_\rho<0$ and $w_d<1$, until the energy density reaches the bound $\crho_{lim}$.

The non trivial fixed points for energy influx are determined by the roots of the equation for $\csrho$
\bea\label{rho fixed gen}
\oneover{d}\left(d-2w_d\right)\tcrd \a\csrho^2+\oneover{d}\tcvd\left(d-w_d\right)\csrho-\frac{\cAA^2}{w_d^2}\csrho^{2(\n-1)}-\ck=0
\ena
while for $\cschi$ and $\csH$ we get the two functions of $\csrho$
\bea
\cschi=-\frac{w_d}{d}\left(1+2\frac{\tcrd }{\tcvd}\a\csrho\right)\csrho, \qquad \csH=-\frac{\cAA}{w_d}\csrho^{\n-1}
\ena
We thus have to fix a particular value for $\n$ in order to establish the precise number of roots and the explicit solution for the critical points. For integer and semi--integer $\n$ the number of
roots we can obtain, keeping $d\ne2w_d$, is
\bea
\n\ge2 \qquad &\Longrightarrow& \qquad \mbox{\# roots}=2(\n-1)\ge2  \non\\
1\le\n<2 \qquad &\Longrightarrow& \qquad \mbox{\# roots}=2  \non\\
\n<1 \qquad &\Longrightarrow& \qquad \mbox{\# roots}=2(2-\n)>2  \non
\ena 
These are all the roots of \refeq{rho fixed gen}, including trivial and complex roots. When $d=2w_d$ the critical value for the mirage energy density diverges due to the divergence of $\tcrd $, unless we fix
$w_\p$ to keep it finite. In this case, the number of roots changes to
\bea
\n>1 \qquad &\Longrightarrow& \qquad \mbox{\# roots}=2(\n-1)\ge1  \non\\
\n=1 \qquad &\Longrightarrow& \qquad \mbox{\# roots}=1  \non\\
\n<1 \qquad &\Longrightarrow& \qquad \mbox{\# roots}=3-2\n>1  \non
\ena 
For $\n>1$, one of the roots of equation (\ref{rho fixed gen}) is null if $\k=0$.

There are moreover two trivial fixed point solutions given by $\csrho=\csH=0,\cschi=\tcvd\,\ck$ and $\csrho=\cschi=0,\csH=\sqrt{-\ck}$, that reduce to a unique point with all vanishing variables when the
internal space is flat.

Let's study in more detail the case $\n=1$, since solutions can be written explicitly being the case with the minimum number of roots for \refeq{rho fixed gen}, together with the $\n=2$ case. As a result we
get a trivial fixed point solution with $\csH=\csrho=0,\cschi=\tcvd\,\ck$ and the trivial solution, acceptable only for negative and zero curvature, $\csH=\sqrt{-\ck},\csrho=\cschi=0$.  Finally, the two non
trivial solutions (for $d\ne2w_d$) are given by
\bea\label{fix gen nu}
\csH&=&-\frac{\cAA}{w_d} \non\\
\cschi&=&\frac{-\tcvd(d-w_d)-4(d-2w_d)\tcrd \,\a K^2\pm\sqrt{(d-w_d)^2\tcvd^2-4d(d-2w_d)\tcrd \,\a K^2}}{2(d-2w_d)\tcrd \,\a}\\
\csrho&=&\frac{-\tcvd(d-w_d)\pm\sqrt{(d-w_d)^2\tcvd^2-4d(d-2w_d)\tcrd \,\a K^2}}{2(d-2w_d)\tcrd\, \a}  \non
\ena
where $K$ corresponds to a shift and rescaling of $\cAA^2$ due to the non vanishing value of $\k$ and is defined by $K^2\equiv\cAA^2/w_d^2+\ck$. The two roots are both real only if the argument of the square
root in (\ref{fix gen nu}) is positive, i.e. when $w_d$ lies outside the two roots $-\tilde\a d\left(1\pm\sqrt{(\tilde\a+1)/\tilde\a}\right)$, with $\tilde\a\equiv4\tcrd\,\a K-1$. If $\tilde\a$ is in the
range bounded by $-1$ and $0$ the square root is always real, whatever $d,w_d$ we choose. If two or all among $\tcvd,\tcrd ,(d-w_d),(d-2w_d)$ have equal sign, one of the two solutions \refeq{fix gen nu}
always is characterized by a negative $\csrho$. We note that we can have at list a non trivial fixed point with positive energy density if $(d-2w_d)\tcrd\,\a>0,(d-w_d)\tcvd<0$ --- exactly two positive
$\csrho$ fixed points ---, or for $(d-2w_d)\tcrd\,\a<0$ --- only one critical point with positive energy density. 

The non trivial solution for $d=2w_d$ is 
\bea \label{fixed num gen 2}
\csH=-\frac{\cAA}{w_d}, \qquad  
\csrho=2\tcvd K^2,  \qquad  \cschi=-\tcvd K^2\left(4\tcrd  K^2\a+1\right)
\ena 
For this unique fixed point solution to be characterized by positive $\csrho$ we have to demand a positive $\tcvd K^2$. Moreover, we can derive from (\ref{spiral bound gen nu}) that for $\ck>\tcvd
w_d\csrho/4d$ the fixed point is a spiral, so that for instance, in a flat internal space, we always obtain a node since $\tcvd$ must be positive in order to have a positive $\csrho$ fixed point ($K=A$ in
this case).

The numerical analysis can now show some of the features that we commented for the cosmological evolution with generic density. The differential system of equations (\ref{dif sys gen num}) (substituting some
precise values for $\n$) can be solved numerically in order to check the existence of stable inflationary critical points. In figure \ref{phase space gen} we plot the dimensionless acceleration factor
$\cqq(t)$ as a function of the dimensionless energy density $\crho(t)$, as we did in the previous section for small densities.
In the plots, we consider for simplicity positive $\tcvd,\tcrd $ and flat compact extra dimensions $\k=0$ ($\k\ne0$ results in a shift for $\cAA^2/w_d^2$ in the critical point evaluation and in some scaling
of the suitable value of $\n$ in order to have stability).  

All the critical points we get are characterized by positive $\cqq$, i.e. they represent an inflationary point. This follows from $\cqq_star=\csH^2>0$. In the phase space portrait \ref{phase space gen}(a)
trajectories starting with positive acceleration factor and energy density lower than the critical one pass through an era of larger acceleration and then slow down to the fixed point, where inflation occurs.
There are then solutions starting with negative acceleration and going to the positive $\cqq$ critical point, eventually passing through a larger acceleration phase or through a smaller density phase. The
families of solutions that distinguish the diagram \ref{phase space gen}(a) from the others are characterized by both initial and final very high energy density, since they are repelled by the second non
attractive fixed point. Some start with high energy density and negative acceleration at late time, go through an era of larger acceleration (eventually positive) and then, while $\crho$ becomes very large,
they go to a region of large and negative $\cqq$. Other go from positive acceleration to large negative $\cqq$ and large $\crho$. The mirage density $\cchi$ can start from an initial condition smaller or
bigger than the critical value, has to be positive for $\cqq<-\left(3\crho+1\right)\crho$ --- as we deduce from \refeq{q dimensionless gen} plugging in the values for the parameters --- and approaches a
negative constant value. 

In diagram \ref{phase space gen}(b), trajectories starting with negative acceleration go to the positive $\cqq$ fixed point, eventually reaching a maximum $\cqq$ before ending into the critical point.
Positive acceleration initial condition lead to growing acceleration at very early times, when usually the energy density grows as well, then both $\cqq$ and $\crho$ decrease to reach the fixed point, passing
through a minimum for acceleration. The mirage density goes to the negative critical value being initially positive for the trajectories that come from negative acceleration conditions, with
$\cqq<-\left(4\crho+1\right)\crho/2$.  With the choice of parameters we used, we get $w_d<d/2$. We could also have used the fixed point solution \refeq{fixed num gen 2} if $w_d=d/2=3$ and the diagram for the
phase space would have been analogous to that in plot \ref{phase space gen}(c). 

In the phase space \ref{phase space gen}(c) all solutions converge to the fixed point with a spiral behavior. Energy density and acceleration parameter thus oscillate around the critical values. Here $\n=-1$
and $d<2w_d$. The number of roots corresponding to the critical point solutions are six, but four of them are complex roots and one is characterized by negative energy density. Only one real critical point
with positive energy density exists in this case and it has Hubble parameter. Since $\cqq$ can be negative at some time of the evolution, even if starting with a positive value, $\cchi$ can pass through a
positive phase, crossing zero before reaching the negative critical value, if $\cqq<-\left(4\crho+1\right)\crho/2$. A similar plot can be drawn also if $w_d=d/2$. There would be only five roots for $\csrho$,
four of which would be complex conjugated and the last would have positive energy density, representing the stable spiral.
 
Another stable spiral is represented in figure \ref{phase space gen}(d). Here, the dimensionless parameter $\a$, which is proportional to $M^6/V$, is large and $\n=1$. We find only one non trivial fixed point
with positive energy density and spiral behavior, so that trajectories has a shape analogous to the ones in \ref{phase space gen}(c).

For values of $\n$ different from $\n=1$ the number of fixed point roots may vary according to the previous discussion. Nonetheless, (as it is shown as an example in figure \ref{phase space gen}(c) for
$\n=-1$) some of the roots may be complex conjugated and thus not acceptable. Another simple case is $\n=2$, where we get two solutions to \refeq{rho fixed gen}, as with the $\n=1$ assumption. We will not
discuss this situation in detail since the phase spaces we can find are analogous to the $\n=1$ ones. 

The case of energy outflow is analogous to the small energy density analysis. In fact, the fixed point solution \refeq{fixed line gen} can't be characterized by positive energy exchange parameter $T$. We can
nevertheless have a critical point with energy outflowing from the brane into the bulk and negative Hubble parameter (as we can deduce from the expansion~$\to$~contraction, influx~$\to$~outflow symmetry). As
the differential equation for $\rho$ \refeq{bulk limits rho gen} shows, the energy density decreases and go to the trivial fixed point, meaning that the negative $H$ critical point isn't an attractor.
\paragraph{4D densities}
The 4D energy density in the static compact extra dimension case is just given by a constant rescaling of the 6D density. Thus, the phase portraits are given by the plots in figure \ref{phase space small} for
small energy density, and figure \ref{phase space gen} for generic density (up to constant rescaling). However, for equal scale factors $d=6$, the 4D effective energy density $\rhofo$ is dynamically
determined by the energy density localized on the 5--brane $\rho$ and the volume of the compact space $V_{(2)}$: $\rhofo=V_{(2)}\rho$. The 4D mirage density can be similarly defined as $\chifo=V_{(2)}\chi$.
We can single out the volume time dependence defining the dimensionful constant $v$ such that $V_{(2)}\equiv vb^2(t)$, where $b(t)$ is as usual the compact space scale factor --- in this particular case
$b(t)=a(t)$. The generic set of equations for the dimensionless variables $\cHH,\crhofo,\cchifo$ is
\bea
\cHH^2&=&\frac{\tcrsix\,\a}{v^2}\frac{\crhofo^2}{a^4}+\frac{\tcvsix}{v}\frac{\left(\crhofo+\cchifo\right)}{a^2}-\ck   \\
\dot\cchifo+4\cHH\cchifo&=&\cAAfo\,\frac{\rhofo^\n}{a^{2(\n-1)}}\left(1+2\frac{\tcrsix}{\tcvsix}\frac{\a}{v}\crhofo\right) \non\\
\dot\crhofo+(3(1+w)+2w_\p)\cHH\crhofo&=&-\cAAfo\,\frac{\crhofo^\n}{a^{2(\n-1)}}   \label{effective 4D equal gen sys dimensionless}
\ena
where $\Afo\equiv A/v^{\n-1}$. We first note that in the case of zero energy exchange $T=0$ ($\cAAfo=0$) the 4D mirage density satisfies the 4D free radiation equation, as in the static internal space
hypothesis. The 4D energy density $\rhofo$ does not have a definite behavior in the case of energy outflow. While in the static 2D compact space background it is clear that $\rhofo$, just as $\rho$, is
suppressed in time since $w_{\scriptscriptstyle{d=4}}>0$, here we may have a negative coefficient $w_{\scriptscriptstyle{d=6}}<0$ for the linear term in $\crhofo$ in \refeq{effective 4D equal gen sys
dimensionless}. If $w_\p<-3(1+w)/2$ --- which is possible only for $w<-1/3$ if $w_\p>-1$ --- we could have non trivial stable critical points, as in the energy influx context. This scenario would need further
investigations.
\paragraph{Dynamical compactification}
We can make some considerations regarding the more generic assumption on the relation between the Hubble parameters $H$ and $F$, $F=\x H$, of subsection \ref{proportional hubble}. For positive $\x$ the
qualitative behavior is analogous to what we deduced in the case of static compact extra dimensions and equal scale factors. When $\x$ is negative, meaning that we are using a dynamical compactification
approach \cite{Mohammedi:2002tv}, we could instead have some differences. In particular, it is worth noticing that $w_\x$ (appearing in the conservation equation for $\rho$) can become negative, implying a
diverging behavior for the localized energy density at late time in the case of energy influx. Thus, there won't be stable critical point in the dynamical compactification scenario with energy flowing from
the bulk onto the brane. 
%%%%%%%%%%%%%%%%%%%%%%%%%%%%%%%%%%%%%%%%%%%%%%%%%%%%%%%%%%%%%%%%%%%%%%%%%%%%%%%%%%%%%%%%%%%%%%%%%%%%%%%%%%%%%%%%%%%%%%%%%%%%%%%%%%%%%%
%%%%%%%%%%%%%%%%%%%%%%%%%%%%%%%%%%%%%%%%%%%%%%%%% simple solution with small density %%%%%%%%%%%%%%%%%%%%%%%%%%%%%%%%%%%%%%%%%%%%%%%%%
%%%%%%%%%%%%%%%%%%%%%%%%%%%%%%%%%%%%%%%%%%%%%%%%%%%%%%%%%%%%%%%%%%%%%%%%%%%%%%%%%%%%%%%%%%%%%%%%%%%%%%%%%%%%%%%%%%%%%%%%%%%%%%%%%%%%%%
\subsection{Small density and free radiation equation of state: an explicit solution}\label{explicit bulk section} We can write an explicit solution to the set of equations (\ref{bulk limits H})--(\ref{bulk
limits rho}) in the special limit of small localized energy density $\rho\ll V$ and when $w_d=d$. This last condition is realized if $w=1/3$ with static compact extra dimensions and if $w_\p=(1-3w)/2$ with
equal scale factors. We must be careful though, because in this limit $\tcvd$ generally diverges. It is thus important to keep it finite by imposing a priori a specific value for $w$ and $w_\p$. With these
assumptions, $H^2$ only depends on the sum $\left(\rho+\chi\right)$ and the equation for this sum can be easily integrated independently of the explicit form of $T$. To be more specific, we get the following
solution \bea\label{explicit bulk} H^2&=&\frac{2\tcvd V}{(d-1)M^{10}}\left(\rho_0+\chi_0\right)\frac{a_0^d}{a^d}-\oneover{2(d-1)}\frac{\k}{b^2}\\ \rho+\chi&=&\left(\rho_0+\chi_0\right)\frac{a_0^d}{a^d} \ena

As in the five dimensional RS model with energy exchange analyzed in \cite{Kiritsis:2002zf}, the evolution is determined by the initial value of the energy density (if we put, for example, $\chi_0=0$)
weighted by the expansion in a (effective six or four dimensional) radiation dominated era.

We deduce from \refeq{explicit bulk} that $|a(t)|$ must have an upper limiting value for $\k>0$. For small positive $a(t)$ the rate $\dot a(t)$ is a positive function and the universe expands until it
reaches the limiting value. If the compactification is over an hyperbolic space, the scale factor grows without bound. In particular, in a universe with extra dimensions evolving according to the same Hubble
parameter as for the observed space--time, the expansion rate goes to a constant positive value as the scale factor grows. In a static extra dimension set--up instead, $a(t)$ exponentially grows at infinity.
When $\k=0$ the explicit solutions for $a(t)$ reduce to $a(t)\sim t^{1/2}$ for static internal space, and $a(t)\sim t^{1/3}$ for equal scale factors. These represent exactly a radiation dominated flat
universe in four or six effective dimensions respectively. The evolution $a(t)\sim t^{1/3}$ can also be associated in 4D to the Friedmann equation for scalar field subject to a null potential.

If we further assume the energy exchange to be linear in the localized energy density $T=A\rho$, also imposing the initial condition $\chi_0=0$ and $w_d=d$, the integration of the $\chi$ and $\rho$ equations
yields 
\bea
\chi=\rho_0\frac{a_0^d}{a^d}\left(1-\ex^{-At}\right),\qquad  \rho=\rho_0\frac{a_0^d}{a^d}\,\ex^{-At}
\ena  
This solution shows that the initial amount of radiation energy density decays in favor of the mirage energy density, for energy outflow $A>0$. The late time evolution is thus governed by the mirage
density for outflow.

For equal scale factors it is interesting to write the explicit solution (remember $w_{\scriptscriptstyle{d=6}}=6$ in this case) in terms of the 4D densities $\rhofo,\chifo$ ($\rhofo=vb^2(t)\rho$,
$\chifo=vb^2(t)\chi$).  The ansatz implying static internal directions results in a constant rescaling of the 6D quantities only. The equal scale factor solution is given by
\bea
H^2&=&\frac{2\tcvsix\, V}{5M^{10}\, v}\left(\rhofz+\chifz\right)\frac{a_0^4}{a^6}-\oneover{10}\frac{\k}{a^2}\\
\rhofo+\chifo&=&\left(\rhofz+\chifz\right)\frac{a_0^4}{a^4}
\ena
The evolution is still weighted by the characteristic effective 6D radiation dominated era $1/a^6$. But we can see, exploring the solutions for $\rhofo$ and $\chifo$ in the case of energy exchange parameter
determined by $T=A\rho$ (with positive $A$), that the 4D localized energy density evolves as 4D radiation, exponentially suppressed in time (as for the static internal space background). In fact, assuming as
before$\chifz=0$, we get
\bea
\chifo=\rhofz\frac{a_0^4}{a^4}\left(1-\ex^{-At}\right),\qquad  \rhofo=\rhofz\frac{a_0^4}{a^4}\,\ex^{-At}
\ena
(in this case $\Afo=A$).

The scenario with energy influx is also analogous to the analysis in \cite{Kiritsis:2002zf}. For both static compact extra dimensions and equal scale factors we rewrite the conservation equation for $\rho$
(for flat internal space) as
\bea
\dot\rho+\frac{2}{t}\rho=-T
\ena 
where we have used the equation determining $H$ \refeq{explicit bulk}. This shows that for negative $T$, $\rho$ should increase without bounds at late time. Since we are in the low density approximation, we
can only rely on the generic $\rho$ analysis of the previous section for large $\rho$. However, assuming $T=A\rho^\n$, we can deduce that for $\n>3/2$ the energy density can flow to zero (for certain values
of the parameters). Indeed, $\n>3/2$ corresponds to non stable critical point in the small density analysis.

We remark that introducing the 4D density $\rhofo$ for a universe with equal scale factors brings to
\bea\label{effective 4D influx equal explicit}
\dot\rhofo+\frac{4}{3t}\rhofo=-T
\ena 
Equation \refeq{effective 4D influx equal explicit} tells us that the 4D localized energy density still grows unlimited at late time, if $T$ is linear in $\rhofo$ --- more general considerations are analogous
to the 6D density case. Again, we would need the full treatment for generic density.

The acceleration parameter $q\equiv\ddot a/a$ in this context is equal to
\bea
q=-\frac{(d-2)\tcvd V}{(d-1)M^{10}}\left(\rho_0+\chi_0\right)\frac{a_0^d}{a^d}
\ena
For non zero $\k$, the value of the acceleration can be either positive or negative. It has to be negative when $\k\ge0$, but may be positive for compactification on hyperbolic spaces ($\k<0$), giving as a
result a loitering universe.
%%%%%%%%%%%%%%%%%%%%%%%%%%%%%%%%%%%%%%%%%%%%%%%%%%%%%%%%%%%%%%%%%%%%%%%%%%%%%%%%%%%%%%%%%%%%%%%%%%%%%%%%%%%%%%%%%%%%%%%%%%%%%%%%%%%%%%%%%
%%%%%%%%%%%%%%%%%%%%%%%%%%%%%%%%%%%%%%%%%%%%%%%%%%%%%%%%% final remarks %%%%%%%%%%%%%%%%%%%%%%%%%%%%%%%%%%%%%%%%%%%%%%%%%%%%%%%%%%%%%%%%%
%%%%%%%%%%%%%%%%%%%%%%%%%%%%%%%%%%%%%%%%%%%%%%%%%%%%%%%%%%%%%%%%%%%%%%%%%%%%%%%%%%%%%%%%%%%%%%%%%%%%%%%%%%%%%%%%%%%%%%%%%%%%%%%%%%%%%%%%%
\section{Remarks on 7D RS cosmology}
In the context of the 7D RS proposal I studied the Friedmann--like equation that comes along with the introduction of a mirage energy density satisfying to the non homogeneous radiation equation in some effective number of
dimensions (which is six for equal scale factors in both 3D and 2D spaces and is four when the internal space is static). The bulk cosmological evolution is then determined by the Friedmann equation and by
the (non)conservation equations for the mirage density and the localized matter density on the 5--brane. Making use of some simple ansatz for the evolution of the 2D compactification space (such as putting
the corresponding Hubble parameter $F$ equal to the Hubble parameter of the visible space $H$ or to make it vanish) we found a wide spectrum of possible cosmologies that reduce to the RS vacuum in the absence
of matter (i.e. we imposed the RS fine--tuning $\l_{\rm RS}=0$).

Assuming small density approximation, I have described the explicit analytical solution in case of radiation dominated universe. The Hubble parameter evolves as in an effective 6D (4D) radiation dominated
era for equal scale factors (static compact extra dimensions), independently of the form of the brane--bulk energy exchange. The effective 4D mirage and matter energy densities obey to the 4D free radiation
equation in the absence of energy exchange. If energy flows from the brane into the bulk, the localized 4D energy density is suppressed in time, in favor of the mirage density, even with zero mirage initial
condition. For influx, the 4D matter energy density apparently grows unbounded (if $T$ is linear in $\rho$, otherwise energy density may go to zero for suitable power--like parametrizations),
eventually diverging at a finite time. The small density approximation must break down and the full analysis is needed. On the other hand, still for small densities but generic perfect fluid equation of state
(non necessarily pure vacuum energy) and energy influx, I found inflationary fixed point solutions that are stable for a large class of energy exchange parametrizations $T=A\rho^\n$. These thus represent
stable de Sitter solutions for our universe. I moreover argued that, differently than in the 5D RS approach \cite{Kiritsis:2002zf}, we may have stable de Sitter critical point solutions even for energy
outflowing to the bulk, in the case of equal scale factors with $w<-1/3$ (and $\n=1$). For dynamical compactification (i.e. $F=\x H$, $\x<0$) \cite{Mohammedi:2002tv} I could in principle also get an outflow
stable inflationary fixed point. I note that the 4D mirage energy density evolution without energy exchange is governed by the effective 4D free radiation equation only in the two limits of equal scale
factors $F=H$ and static compact extra dimensions $F=0$. 

Having dropped the small density approximation, more elaborate models of cosmologies developed. The number of possible inflationary critical point solutions can be larger than one, depending on the
parametrization for the brane--bulk energy exchange. For energy influx I showed the 6D picture of a scenario with two fixed points, where trajectories in the phase space can either always be characterized by
positive acceleration, either remain at all time with negative acceleration, or alternate acceleration and deceleration phases. The portraits are rigorously valid for the effective 4D energy density in the
case of static internal space (up to a constant rescaling). If we have equal scale factors, the evolution equations become much more complicated functions of the 4D densities and the computation is beyond the
scope of the paper. For $\n\ne1$ there seems to exist only the trivial critical point characterized by vanishing Hubble parameter, so that either the energy density grows without bounds as predicted by the
small density approximation, either it flows to zero. For energy outflow, the 6D energy density localized on the 5--brane decreases and the trajectories in the phase space go toward the trivial fixed point,
eventually passing through an accelerated era. The effective 4D picture may differ from this description in the equal scale factor case, since it would be possible in principle not to have decreasing density
at all times.

In the analysis of \cite{Mazzanti:2007dq}, there could be space to fit the cosmological evolution of the universe in this model, despite the fact that I didn't give a full cosmological description. Indeed,
one of the stable de Sitter critical point solutions I found could represent the actual accelerated era.  Besides, trajectories can end into the stable point first passing through a decelerated phase
representing the matter or radiation dominated universe. Another accelerated era may be present at early times, eventually corresponding primordial inflation. Still, there is no rigorous construction of such
a precise evolution.

\chapter{Holography in the seven dimensional Randall--Sundrum background}\label{7D RS dual}
I now derive the holographic dual theory of the 7D RS model proposed in my paper \cite{Mazzanti:2007dq} and illustrated in the previous chapter. This is done in analogy to the 5D/4D case, reviewed in section
\ref{4D RS holo cosmo}. Indeed, investigations on the 5D RS cosmology with brane--bulk energy exchange have been made from the holographic point of view \cite{Kiritsis:2005bm}, studying the dual theory in one
lower dimension. The gauge/gravity duality \cite{Maldacena:1997re,Witten:1998qj} that I reviewed in chapter \ref{chapter adscft} (see \cite{Aharony:1999ti} for a complete review) has undergone great
improvements over the last ten years and provides a new approach to the analysis of brane--world models. As it is explained in section \ref{adscft in RS} \cite{Hawking:2000kj}--\cite{Perez-Victoria:2001pa},
the truncation of the $AdS_{d+1}$ space is equivalent to introducing in the dual picture a UV cutoff for the $d$--dimensional gauge theory (conformal field theory). Earlier suggestions about this idea are
present in \cite{Nojiri:2000eb}. The presence of brane--bulk exchange corresponds to interactions between the gauge theory and the matter fields, while the bulk ``self--interaction'' is shown to be related to
the perturbation of the CFT (that becomes a strongly coupled gauge theory). In section \ref{4D RS holo cosmo} \cite{Kiritsis:2005bm} explicit examples of cosmological evolutions in the holographic 5D/4D
picture as well as comparison between the two dual theories have been discussed.  Other cosmological models have been analyzed in the context of the holographic correspondence
\cite{Hawking:2000bb}--\cite{Gubser:1999vj}.

Exploiting the AdS/CFT results, we build the holographic theory corresponding to the 7D RS background. The 7D RS dual theory is then a renormalized 6D CFT (the theory corresponding to the M5 system is an
anomalous \cite{Henningson:1998gx}--\cite{deHaro:2000xn} (0,2) SCFT, but any other six dimensional large--N CFT can be chosen) coupled to 6D gravity. See also \cite{Nojiri:1998dh} for other examples of
holographic Weyl anomaly derivations. The action also contains higher order corrections to gravity and the six dimensional matter fields. Higher derivative terms driven by conformal four dimensional anomaly
\cite{Deser:1993yx} were proved to lead to an inflationary critical point in the 4D Starobinsky model \cite{Starobinsky:1980te} and to a successive graceful exit from the long primordial inflation. As
illustrated in \cite{Vilenkin:1985md} and reviewed in subsection \ref{inflation} higher derivative contributions to the Einstein equations cause the universe to enter a matter dominated era where the scale
factor oscillates after inflation and to proceed through thermalization to a radiation dominated era. In our 6D holographic cosmological model we specially look for de Sitter fixed point solutions of the
equations of motion describing late time acceleration of our universe or critical points suitable for early time inflation, studying the associated stability matrix. A comparison with the 7D bulk analysis
results shows some peculiar features of the 7D/6D set--up. 

In section \ref{dual} and the following I will derive the 6D holographic dual to the 7D RS and the associated equations of motion. Section \ref{critical brane} summarizes the fixed points in the holographic
description and their stability. Some examples of the correspondence between the brane and bulk points of view will be given in section \ref{examples}. The generalization to non conformal and interacting
theory, corresponding to non vanishing brane--bulk energy exchange and bulk self interaction in the 7D approach, will be exposed in section \ref{general}.
%%%%%%%%%%%%%%%%%%%%%%%%%%%%%%%%%%%%%%%%%%%%%%%%%%%%%%%%%%%%%%%%%%%%%%%%%%%%%%%%%%%%%%%%%%%%%%%%%%%%%%%%%%%%%%%%%%%%%%%%%%%%%%%%%%%%%%
%%%%%%%%%%%%%%%%%%%%%%%%%%%%%%%%%%%%%%%%%%%%%%%%% holographic dual %%%%%%%%%%%%%%%%%%%%%%%%%%%%%%%%%%%%%%%%%%%%%%%%%%%%%%%%%%%%%%%%%%%
%%%%%%%%%%%%%%%%%%%%%%%%%%%%%%%%%%%%%%%%%%%%%%%%%%%%%%%%%%%%%%%%%%%%%%%%%%%%%%%%%%%%%%%%%%%%%%%%%%%%%%%%%%%%%%%%%%%%%%%%%%%%%%%%%%%%%%
\sectioncount{Construction of the holographic dual}\label{dual} 
The dual theory of the 7D RS model, via the AdS/CFT correspondence \cite{Maldacena:1997re}, will be derived in complete analogy to the 5D
set--up considered in \cite{Kiritsis:2005bm}. The RS model, with a time independent warped geometry, gives $AdS_7$ metric as a solution to the equations of motion for the gravity action in the bulk. It will
be useful to parametrize it according to Fefferman and Graham \cite{Fefferman} 
\bea\label{FG para} 
g_{MN}\intd x^M\intd x^N=\frac{\ell^2}{4}\rho^{-2}\intd \rho^2+\ell^2\rho^{-1} \hg_{\m\n}\intd x^\m\intd x^\n
\ena 
where the indices $M,N.\dots$ run over the 7D bulk space--time, $\m,\n,\dots$ span the 6D space--time on the 5--brane and $\rho$ is a reparametrization of the $z$ coordinate transverse to the brane.  The
location of the brane, translated to this new set of coordinates, is $\rho=0$ which represents the boundary of the background (\ref{FG para}). Generally, for all seven dimensional asymptotically AdS
space--times the 6D metric $g_{\m\n}$ can be expanded as \cite{Fefferman} 
\bea 
\hg=\hg_{(0)}+\rho \hg_{(2)}+\rho^2\hg_{(4)}+\rho^3\hg_{(6)}+\rho^3\log\rho\, h_{(6)}+\ogr(\rho^4) 
\ena 
The logarithmic piece appears only for space--times with an odd number of dimensions and is responsible for the cutoff dependent counterterm in the renormalized action.  As a consequence, it is also
responsible for the conformal anomaly of the holographic dual CFT living in an even number of dimensions \cite{Henningson:1998gx,deHaro:2000xn} --- in fact we don't have conformal anomaly in odd dimensional
CFT's. The subindices in the coefficients of the metric expansion stand for the number of derivatives contained in each term.

More precisely, RS background is a slice of $AdS_7$, where the boundary gets replaced by the 5--brane and the IR part is reflected, eliminating the UV slice. To describe seven dimensional gravity we will
take the usual Einstein--Hilbert action in the bulk, adding as usual a Gibbons--Hawking term \cite{Gibbons:1976ue} to take account of the boundary extrinsic curvature. Since the gravitational theory exhibits
divergences in a space--time with boundaries, we also have to regularize the Einstein--Hilbert plus Gibbons--Hawking action, cutting off the boundary of the space--time. I am now going to illustrate the
regularization and renormalization procedures.
%%%%%%%%%%%%%%%%%%%%%%%%%%%%%%%%%%%%%%%%%%%%%%%%%%%%%%%%%%%%%%%%%%%%%%%%%%%%%%%%%%%%%%%%%%%%%%%%%%%%%%%%%%%%%%%%%%%%%%%%%%%%%%%%%%%%%%
%%%%%%%%%%%%%%%%%%%%%%%%%%%%%%%%%%%%%%%%%%%%%% regularize gravity %%%%%%%%%%%%%%%%%%%%%%%%%%%%%%%%%%%%%%%%%%%%%%%%%%%%%%%%%%%%%%%%%%%%
%%%%%%%%%%%%%%%%%%%%%%%%%%%%%%%%%%%%%%%%%%%%%%%%%%%%%%%%%%%%%%%%%%%%%%%%%%%%%%%%%%%%%%%%%%%%%%%%%%%%%%%%%%%%%%%%%%%%%%%%%%%%%%%%%%%%%%
\subsection{Renormalization on the gravitational side} The renormalization for a gravitational theory in a background with boundaries has been explained in
\cite{Henningson:1998gx,Skenderis:1999nb,deHaro:2000xn} for a generic number of dimensions. I will apply those computations to the case of a seven dimensional bulk space--time.  

In general (see section \ref{holographic renormalization}), the bulk action  for gravity gets modified by the Gibbons--Hawking boundary term \cite{Gibbons:1976ue} and by some counterterms also localized
on the boundary 
\bea
S_{gr}=S_{EH}+S_{GH}-S_{count}
\ena
Using the Fefferman and Graham parametrization of the metric \refeq{FG para} and cutting off the boundary at $\rho=\epsilon$, we have
\bea
S_{EH}=M^5\int_{\rho\ge\epsilon}\intd^7x\sqrt{-g}\left(R[g]+\frac{30}{\ell^2}\right),\quad S_{GH}=2M^5\int_{\rho=\epsilon}\intd^6x\sqrt{-\g}K
\ena
where $R[g]$ is the bulk Ricci scalar, $K$ is the trace of the extrinsic curvature and $\g_{\m\n}$ is the induced metric on the boundary.  Putting the brane at $\rho=\epsilon$ corresponds to regularize the
gravity action. The counterterm contributions necessary to make it finite in the limit $\epsilon\rightarrow0$ are given by
\bea
S_{count}=S_0+S_1+S_2+S_3
\ena
$S_i$ are terms of order $i$ in the brane curvature $R\equiv R[\g]$ (the curvature of the induced metric $\g_{\m\n}$ on the boundary). In fact they can be written in terms of the induced metric $\g_{\m\n}$
and its Riemann tensor $R_{\m\n\rho\s}$, using the perturbative expansion relating $\g_{\m\n}$ to $\hg_{(0)\m\n}$ (see for instance \cite{deHaro:2000xn})
\bea\label{def count 0}
S_0&=&10\,\frac{M^5}{\ell}\int_{\rho=\epsilon}\intd^6x\sqrt{-\g}\\
S_1&=&-\oneover{4}M^5\ell\int_{\rho=\epsilon}\intd^6x\sqrt{-\g}R\\
S_2&=&\oneover{32}M^5\ell^3\int_{\rho=\epsilon}\intd^6x\sqrt{-\g}\left(R_{\m\n}R^{\m\n}-\frac{3}{10}R^2\right)\\
S_3&=&\frac{\log\epsilon}{64}M^5\ell^5\int_{\rho=\epsilon}\intd^6x\sqrt{-\g}\bigg(\oneover{2}RR_{\m\n}R^{\m\n}+\frac{3}{50}R^3+R^{\m\n}R^{\rho\s}R_{\m\rho\n\s}\non\\
&&+\oneover{5}R^{\m\n}\nabla_\m\nabla_\n R-\oneover{2}R^{\m\n}\Box R_{\m\n}\bigg)  \label{def count 3}
\ena
The third order term, $S_3$, depends on the cutoff $\epsilon$ and is thus responsible for the breaking of the scale invariance, i.e. it gives rise to the conformal anomaly for the dual 6D CFT in the context of
the AdS/CFT correspondence. We also note that the zeroth order term is related to the brane tension term of the RS model $S_{tens}$ in (\ref{7D action}) by $S_{tens}=-2S_0$, if we fine--tune $\l_{RS}=0$. In
fact, in the pure RS set--up, where the effective cosmological constant is null $\l_{RS}=0$, the brane tension is $V=20M^5/\ell$, since the bulk cosmological constant is given by $\L_7=-30M^5/\ell^2$ as a
function of the background length scale $\ell$ and the bulk Planck mass $M$, parametrizing the background metric by \refeq{FG para}. We will now use the AdS/CFT correspondence to compute the dual theory.
%%%%%%%%%%%%%%%%%%%%%%%%%%%%%%%%%%%%%%%%%%%%%%%%%%%%%%%%%%%%%%%%%%%%%%%%%%%%%%%%%%%%%%%%%%%%%%%%%%%%%%%%%%%%%%%%%%%%%%%%%%%%%%%%%%%%%%
%%%%%%%%%%%%%%%%%%%%%%%%%%%%%%%%%%%%%%%%%%%%%%%%%%%%% AdS/CFT %%%%%%%%%%%%%%%%%%%%%%%%%%%%%%%%%%%%%%%%%%%%%%%%%%%%%%%%%%%%%%%%%%%%%%%%
%%%%%%%%%%%%%%%%%%%%%%%%%%%%%%%%%%%%%%%%%%%%%%%%%%%%%%%%%%%%%%%%%%%%%%%%%%%%%%%%%%%%%%%%%%%%%%%%%%%%%%%%%%%%%%%%%%%%%%%%%%%%%%%%%%%%%%
\subsection{Gauge/gravity duality}
The AdS/CFT duality \cite{Maldacena:1997re} is realized between gravity (string theory or M theory decoupling limit) in a background with one or more stacks of some kind of branes and the gauge theory that
lives on the boundary of the near horizon geometry inferred by the branes \cite{Maldacena:1997re,Witten:1998qj}. In our particular case, $AdS_7\times S^4$ is the near horizon geometry of a system of $N$
parallel M5--branes in eleven--dimensional M theory. The radius of the AdS space is given in terms of the eleven dimensional Planck length $\ell_{Pl}$ and of the number $N$ of M5--branes
\bea
\ell=2(\p N)^{1/3}\ell_{Pl}
\ena
The radius of the four sphere is half the radius of $AdS_7$. The supergravity approximation for M theory is valid if $N\gg1$ and $\ell_{Pl}\sim N^{-1/3}\rightarrow0$, keeping the radius of the AdS large and
finite in units of $\ell_{Pl}$. The six dimensional theory that Maldacena \cite{Maldacena:1997re} conjectured to be dual to M theory in the background described above is a (0,2) SCFT. This theory is realized
as the open string theory in the world--volume of the M5--branes, in the low energy decoupling limit, and it does not contain dimensionless nor dimensionful parameters.  The $AdS_7\times S^4$ supergravity
background is characterized by a 4--form flux quantized in terms of the number of M5--branes and is not conformally flat, since the radii of the four--sphere and the AdS space are not coincident. 

The AdS/CFT correspondence relates the gravity (M theory) partition function for the bulk fields $\Phi_i$ (which is a function of the value of the fields on the boundary of $AdS_7$, $\phi_i$) to the
generating functional of correlation functions of the dual CFT operators with sources $\phi_i$
\bea
Z_{gr}\left[\phi_i\right]\equiv\int\intD\Phi_i\,\ex^{-S_{gr}}=\ex^{-W_{CFT}\left(\phi_i\right)}
\ena
Knowing that gravity on $AdS_7$ (the $S^4$ geometry can be factored out) corresponds to the specific CFT suggested by Maldacena \cite{Maldacena:1997re}, we can now obtain as a consequence the dual theory of
the 7D RS model, in analogy to \cite{Kiritsis:2005bm}. In fact, the action of the gravitational theory that I want to analyze via holography is 
\bea
S_{RS}=S_{EH}+S_{GH}+S_{tens}+S_m
\ena
We just add the Gibbons--Hawking term to (\ref{7D action}). We expect the hidden sector of the holographic theory to reflect the bulk dynamics when we go to the non conformal interacting generalization. 

We now have to keep in mind that the duality for gravity on $AdS_7$ can be stated as 
\bea\label{gravity duality}
Z_{gr}\left[\phi_i\right]\equiv\int_{\rho>\epsilon}\intD\Phi_i\,\ex^{-S_{EH}-S_{GH}+S_0+S_1+S_2+S_3}=\ex^{-W_{CFT}\left(\phi_i\right)}
\ena
Secondly, we have to remember that $S_{tens}=-2S_0$. Furthermore, we note that the integration in (\ref{gravity duality}) is over one half of the space--time appearing in the RS model, because of the
$\Zgr_2$ reflection along the $z$ direction. Since the integrals over the two specular regions are independent and equal we can write
\bea
Z_{RS}\left[\phi_i,\chi_i\right]&\equiv&\int_{\mathrm{all }\rho}\intD\Phi_i\intD\chi_i\ex^{-S_{EH}-S_{GH}+2S_0-S_m}\non\\
&=&\int_{\rho>\epsilon}\intD\Phi_i\intD\chi_i\,\ex^{-2S_{EH}-2S_{GH}+2S_0-S_m}
\ena
where $\chi_i$ are the matter fields on the brane.
Finally, putting all together, using equation (\ref{gravity duality}), we obtain
\bea
Z_{RS}\left[\phi_i,\chi_i\right]=\int_{\rho>\epsilon}\intD\Phi_i\intD\chi_i\,\ex^{-2W_{CFT}-2S_1-2S_2-2S_3-S_m}
\ena
The RS dual theory is
\bea\label{dual action RS}
S_{\widetilde{\rm RS}}=S_{CFT}+S_R+S_{R^2}+S_{R^3}+S_m
\ena
having defined
\bea\label{def RS dual}
S_{CFT}=2W_{CFT},\quad S_R=2S_1,\quad S_{R^2}=2S_2,\quad S_{R^3}=2S_3
\ena
The 6D Planck mass is thus given by $\Mpl=\frac{M^5\ell}{2}$ (the four dimensional Planck mass is therefore $M^2_{(4)}=V_{(2)}M^4_{Pl}$, where $V_{(2)}$ is the volume of the two dimensional internal
manifold).

We are now ready to calculate the equations of motion for the holographic 6D RS cosmology.
%%%%%%%%%%%%%%%%%%%%%%%%%%%%%%%%%%%%%%%%%%%%%%%%%%%%%%%%%%%%%%%%%%%%%%%%%%%%%%%%%%%%%%%%%%%%%%%%%%%%%%%%%%%%%%%%%%%%%%%%%%%%%%%%%%%%%%
%%%%%%%%%%%%%%%%%%%%%%%%%%%%%%%%%%%%%%%%%%%%%%%%% friedmann equations %%%%%%%%%%%%%%%%%%%%%%%%%%%%%%%%%%%%%%%%%%%%%%%%%%%%%%%%%%%%%%%%
%%%%%%%%%%%%%%%%%%%%%%%%%%%%%%%%%%%%%%%%%%%%%%%%%%%%%%%%%%%%%%%%%%%%%%%%%%%%%%%%%%%%%%%%%%%%%%%%%%%%%%%%%%%%%%%%%%%%%%%%%%%%%%%%%%%%%%
\sectioncount{Holographic cosmological evolution}\label{6D equations}
As we know, the RS classical solution in a 7D bulk with a warped geometry is $AdS_7$. Since our purpose is to study the cosmology associated to the 7D RS set--up, we have generalized the ansatz for the metric
to be time dependent in section \ref{setup} and \ref{bulk cosmology}. I have successively reviewed the notion of holographic dual theory in the previous section. What I want to do now is to describe the
cosmology of the seven dimensional RS model from the six dimensional holographic point of view, using the correspondence relation obtained in the previous section and generalizing the ansatz for the 6D
induced metric on the 5--brane to a time dependent geometry, as it has been done for the 7D bulk analysis.

We consider a 6D space--time, compactified on a 2D internal space, with a FRW metric for the four large dimensions. The induced metric tensor can be expressed as
\bea\label{metric 4+2}
\g_{\m\n}\intd x^\m\intd x^\n&=&
-\intd t^2+\frac{a^2(t)}{1-k r^2}\intd r^2+a^2(t)\,r^2\intd\th^2+a^2(t)\,r^2\sin^2\th\intd\phi^2+\non\\
&&+\frac{b^2(t)}{1-\k\rho^2}\intd\rho^2+b^2(t)\,\rho^2\intd\psi^2
\ena
where $k$ and $\k$, $a(t)$ and $b(t)$, $H(t)$ and $F(t)$ are, respectively, the curvatures, the scale factors, the Hubble parameters for the 3D and 2D spaces.

The action we are considering is 
\bea\label{action}
S_{\widetilde{\rm RS}}=S_{CFT}+S_{m,b}+S_{\l}+S_{R}+S_{R^2}+S_{R^3}
\ena
$S_{R}$, $S_{R^2}$, $S_{R^3}$ being respectively twice the first, second and third order terms in the curvature contributing to the counterterm action, as defined in (\ref{def count 0})--(\ref{def count 3}).
$S_{\l}$ is an effective cosmological term on the brane --- that represents a generalization to the case of a non exact RS fine--tuning with respect to the action (\ref{dual action RS}). $S_m$ and $S_{CFT}$
are the matter and (twice) the CFT action of the 6D description.

We want to solve the Friedmann equations, imposing the conservation and anomaly equations, defining the stress--energy tensor contributions as
\bea
T_{\m\n}&=&\oneover{\sqrt{-\g}}\frac{\d S_{m,b}}{\d \g^{\m\n}} \qquad\qquad
W_{\m\n}=\oneover{\sqrt{-\g}}\frac{\d S_{CFT}}{\d \g^{\m\n}} \\
Y_{\m\n}&=&\oneover{\sqrt{-\g}}\frac{\d S_{R^2}}{\d \g^{\m\n}} \qquad\qquad
Z_{\m\n}=\oneover{\sqrt{-\g}}\frac{\d S_{R^3}}{\d \g^{\m\n}} \non\\
\ena
and $V_{\m\n}=W_{\m\n}+Z_{\m\n}$. The equations of motion take the form
\bea\label{Fried}
M_{Pl}^4G_{\m\n}+\l \g_{\m\n}&=&T_{\m\n}+W_{\m\n}+Y_{\m\n}+Z_{\m\n}\\
\nabla^\n T_{\m\n}&=&0\non\\
\nabla^\n V_{\m\n}&=&0\non\\
V^\m_\m&=&\ano+Y\non
\ena
Here $\ano$ is the general anomaly for a 6D conformal theory \cite{Deser:1993yx,Bastianelli:2000rs} that is related to the $S_{3}$ contribution to the renormalized action
\cite{Henningson:1998gx}--\cite{deHaro:2000xn}%%%%%%%%%%%% footnote
\footnote{The scheme dependent contribution to the anomaly --- the type D anomaly --- gets cancelled by the equal and opposite scheme dependent contribution to $S^3$ --- which are local covariant counterterms
and can be derived from appendix C of \cite{Bastianelli:2000rs}. For an explicit example of how this happens, see \cite{Kiritsis:2005bm} in the four dimensional case. See also appendix \ref{anomaly appendix}
for further discussions}%%%%%%%%%%%%% footnote
, while $Y$ is the trace $Y^\m_\m$ of the variation of (twice) the second order counterterm action $S_{R^2}$. The trace of $Z_{\m\n}$ is null%%%%%%%%%%%%%%  footnote
\footnote{It is indeed proportional to the traceless tensor $h_{(6)}$ that appears in the Fefferman and Graham metric parametrization \refeq{metric 4+2} \cite{deHaro:2000xn}. I thank K. Skenderis for helpful
discussion on this subject.}%%%%%%%%%%%%  footnote
. The trace of $Y_{\m\n}$ is quadratic in the curvature of the metric \refeq{metric 4+2} 
\bea
Y=\oneover{32}M^5\ell\left(R^{\m\n}R_{\m\n}-\frac{3}{10}R^2\right)
\ena
The explicit form for the anomaly is a complicated expression of dimensions 6, cubic in the curvature, and is discussed in appendix \ref{anomaly appendix}. The effective cosmological constant on the brane is $\l$.
The stress--energy tensors are parametrized as
\bea\label{stress-en}\begin{array}{rclrclrcl}
T_{00}&=&\rho(t),\quad T_{ij}&=&p(t)\,\g_{ij}, \quad T_{ab}&=&\p(t)\,\g_{ab}\\
V_{00}&=&\s(t),\quad V_{ij}&=&\s_p(t)\,\g_{ij}, \quad V_{ab}&=&\s_\p(t)\,\g_{ab}\\
\end{array}\ena
where the indices $ij\dots$ parametrize the space part of the 4D FRW space--time and run from 1 to 3, while $ab\dots$ belong to the 2D internal space and take values in $(4,5)$.~%%%%%%%%%%% footnote
\footnote{The energy density $\rho(t)$ should not be confused with the 2D coordinate $\rho$, since the radius of the extra dimensions does not appear in the calculations.}%%%%%%%%%%%% footnote

Equations (\ref{Fried}) take the following form when we choose the metric (\ref{metric 4+2}) and the stress--energy tensors written in (\ref{stress-en}).  The Friedmann equations become
\bea\label{Friedmann eqs}
M_{Pl}^4\left(3H^2+6H\,F+F^2+3\frac{k}{a^2}+\frac{\k}{b^2}\right)&=&\rho+\s+\l\non\\
M_{Pl}^4\left(2\dot H+3H^2+4H\,F+2\dot F+3F^2+\frac{k}{a^2}+\frac{\k}{b^2}\right)&=&-p-\s_p+\l\non\\
M_{Pl}^4\left(3\dot H+6H^2+3H\,F+\dot F+F^2+\frac{k}{a^2}\right)&=&-\p-\s_\p+\l
\ena
then the conservation equations
\bea\label{conservation eqs}
\dot\s+3(\s+\s_p) H+2(\s+\s_\p)F&=&0\non\\
\dot\rho+3(\rho+p) H+2(\rho+\p)F&=&0
\ena
and finally the anomaly equation
\bea\label{anomaly eq}
&&\s-3\s_p-2\s_\p=\ano+Y
\ena
As I said, the anomaly comes from the cubic counterterm, so that it is cubic in the curvature. There are more precise statements about its form in appendix \ref{anomaly appendix}, where I also explicitely
give $Y$.
%%%%%%%%%%%%%%%%%%%%%%%%%%%%%%%%%%%%%%%%%%%%%%%%%%%%%%%%%%%%%%%%%%%%%%%%%%%%%%%%%%%%%%%%%%%%%%%%%%%%%%%%%%%%%%%%%%%%%%%%%%%%%%%%%%%%%%%%%%%%%%%%%
%%%%%%%%%%%%%%%%%%%%%%%%%%%%%%%%%%%%%%%%%%%%%%%%%%%%%%%%%%%%%%%% simplifying %%%%%%%%%%%%%%%%%%%%%%%%%%%%%%%%%%%%%%%%%%%%%%%%%%%%%%%%%%%%%%%%%%%%
%%%%%%%%%%%%%%%%%%%%%%%%%%%%%%%%%%%%%%%%%%%%%%%%%%%%%%%%%%%%%%%%%%%%%%%%%%%%%%%%%%%%%%%%%%%%%%%%%%%%%%%%%%%%%%%%%%%%%%%%%%%%%%%%%%%%%%%%%%%%%%%%%
\subsection{Simplifications and ansatz}\label{simplifications}
The set of equations (\ref{Friedmann eqs})--(\ref{anomaly eq}) does not contains six independent equations. Plugging the conservation in the first Friedmann equation derived w.r.t. time, we get a linear
combination of the other two Friedmann equations. So we will discard the last of \refeq{Friedmann eqs} from now on. We further note that that system contains only one algebraic equation: the Friedmann
equation. I will start by solving the anomaly equation in terms of one of the pressures coming from the hidden theory. 

Plugging the expression for $\s_p$ obtained evaluating (\ref{anomaly eq}) into the first of the conservation equations (\ref{conservation eqs}), we get a differential equation for $\s$ depending on $\s_\p$
\bea\label{conservation + anomaly}
\dot\s+2\left(2H+F\right)\s+2\left(H-F\right)\s_\p=\ano+Y
\ena
To obtain a solvable decoupled equation for $\s$ we can consider the limit in which the internal space has the same CFT pressure as the three large dimensions $\s_\p=\s_p$~%%%%%%%%%%%%%%%%%%%% footnote
\footnote{This limit comes from classical evaluation of stress--energy tensor derived from the action $S\propto\int\intd^6x\sqrt{-\g}\,H_{\m\n\rho}H^{\m\n\rho}$}%%%%%%%%%%%%%%% footnote
. Else, we can also consider the limit of zero pressure --- for the CFT --- in the internal space.  Putting these two limits together, we can try to solve the Friedmann equations imposing a more general
ansatz
\bea
\s_\p=\O\s_p
\ena
So that
\bea\label{manipulate anomaly eq}
\s-\oneover{\o}\s_p=\ano+Y
\ena
where $\o\equiv1/(3+2\O)$ ($\o$ is equal to $1/5,1/3$ in the two limits considered above). The differential equation for $\s$ becomes
\bea\label{conservation eq}
\dot\s+3\left[(1+\o)H+(1-\o)F\right]\s-\left[3\o\,H+(1-3\o)F\right]\left(\ano+Y\right)=0
\ena
We could now evaluate $\s$ solving the following integral
\bea\label{sigma anal sol}
\s&=&\chi+\oneover{a^{3(1+\o)}b^{3(1-\o)}}\int\intd t\;a^{3(1+\o)}b^{3(1-\o)}\left[3\o\frac{\dot a}{a}+(1-3\o)\frac{\dot b}{b}\right]\cdot\non\\
&&\cd \left[c_A\,E_{(6)}+c_B\,I_{(6)}+Y\right]
\ena
where $\chi$ is a solution for the homogeneous equation
\bea\label{homo eq}
\dot\chi+3\left[(1+\o)H+(1-\o)F\right]\chi=0\quad\Rightarrow\quad\chi=\frac{\chi_0}{a^{3(1+\o)}b^{3(1-\o)}}
\ena
We observe that generally (\ref{sigma anal sol}) is not explicitely integrable. In \refeq{sigma anal sol} we have written the anomaly in terms of its contributions that are the Euler density in six dimensions
$E_{(6)}$ and the local covariants included in $I_{(6)}$ (see appendix \ref{anomaly appendix} for further details).

The set of independent equations we finally have to solve, once we use (\ref{manipulate anomaly eq}) to eliminate $\s_p$ by means of
\bea\label{eliminate sigma_p}
\s_p=\o\s-\o\left(\ano+Y\right) 
\ena
is then
\bea\label{final system}
M_{Pl}^4\left(3H^2+6HF+F^2+3\frac{k}{a^2}+\frac{\k}{b^2}\right)&=&\rho+\s+\l\\
M_{Pl}^4\left(2\dot H+2\dot F+3H^2+4HF+3F^2+\frac{k}{a^2}+\frac{\k}{b^2}\right)&=&-w\rho-\o\sigma+\o\left(\ano+Y\right)+\l\\\non\\
\dot\s+3\left[(1+\o)H+(1-\o)F\right]\s&=&\left[3\o H+(1-3\o)F\right]\left(\ano+Y\right)\\
\dot\rho+\left[3(1+w)H+2(1+w_\p)F\right]\rho&=&0 \label{final system rho}
\ena
(the anomaly $\ano$ will be written explicitely --- in terms of $H$ and $F$ --- in the particular cases that I will take under examination in the following). We also use the three following ansatz relating
the pressures and the energy densities 
\bea\label{ansatz}
p&=&w\rho\non\\
\p&=&W\,p=w_\p\rho\non\\
\s_\p&=&\O\,\s_p \quad (3+2\O=1/\o)
\ena
Now we are left with a system of four equations (\ref{final system}) in four variables ($H,F,\rho,\s$). The other variables ($\s_p,\s_\p,p,\p$) are determined by the ansatz (\ref{ansatz}) and by the equation
(\ref{eliminate sigma_p}). In the next section, this system of differential equations will be studied restricting to some special limits, such as flat or static internal space, or equal scale factors. We
will find the critical point solutions and analyze the associated stability matrix. 
%%%%%%%%%%%%%%%%%%%%%%%%%%%%%%%%%%%%%%%%%%%%%%%%%%%%%%%%%%%%%%%%%%%%%%%%%%%%%%%%%%%%%%%%%%%%%%%%%%%%%%%%%%%%%%%%%%%%%%%%%%%%%%%%%%%%%%%
%%%%%%%%%%%%%%%%%%%%%%%%%%%%%%%%%%%%%%%%%%%%%%%%%% fixed points %%%%%%%%%%%%%%%%%%%%%%%%%%%%%%%%%%%%%%%%%%%%%%%%%%%%%%%%%%%%%%%%%%%%%%%
%%%%%%%%%%%%%%%%%%%%%%%%%%%%%%%%%%%%%%%%%%%%%%%%%%%%%%%%%%%%%%%%%%%%%%%%%%%%%%%%%%%%%%%%%%%%%%%%%%%%%%%%%%%%%%%%%%%%%%%%%%%%%%%%%%%%%%%
\sectioncount{Holographic critical point analysis}\label{critical brane}
The fixed points of the cosmological evolution of the universe we are considering may represent its inflationary eras --- for instance the early time or the late time acceleration ---, since the Hubble
parameters, just as the energy densities, are constant. If the constant value for the Hubble parameter is positive we have inflation. In this section, I am going to look for the existence of such
inflationary points for our specific holographic model and to find what kind of dependence they have on the parameters of the theory.

I will describe the fixed point solutions in the special limits of flat extra dimensions, curved static extra dimensions and equal scale factors for the internal and extended spaces. I will then study the
stability matrix associated with the critical points. Since the fixed points represents inflationary eras in the universe evolution, they could offer an explanation to the early inflation or to the late time
acceleration. In the first case they will have to be unstable or saddle points to allow the trajectory describing the cosmological evolution to flow away from inflation and exit from this phase. In the second
case the fixed points must be stable and act as attractors for the near trajectories. 

In what follows we will always suppose that the effective cosmological constant on the brane $\l$ is zero, unless I specify it differently.
%%%%%%%%%%%%%%%%%%%%%%%%%%%%%%%%%%%%%%%%%%%%%%%%%%%%%%%%%%%%%%%%%%%%%%%%%%%%%%%%%%%%%%%%%%%%%%%%%%%%%%%%%%%%%%%%%%%%%%%%%%%%%%%%%%%%%%%
%%%%%%%%%%%%%%%%%%%%%%%%%%%%%%%%%%%%%%%%%%%%%%%%%% flat extra %%%%%%%%%%%%%%%%%%%%%%%%%%%%%%%%%%%%%%%%%%%%%%%%%%%%%%%%%%%%%%%%%%%%%%%%%
%%%%%%%%%%%%%%%%%%%%%%%%%%%%%%%%%%%%%%%%%%%%%%%%%%%%%%%%%%%%%%%%%%%%%%%%%%%%%%%%%%%%%%%%%%%%%%%%%%%%%%%%%%%%%%%%%%%%%%%%%%%%%%%%%%%%%%%
\subsection{Flat compact extra dimensions}\label{brane critical flat}
In the limit of zero spatial curvature for the extra dimensions and for the extended space, the Friedmann plus conservation equations (\ref{final system}) take the form
\bea\label{final system flat}
M_{Pl}^4\left(3H^2+6HF+F^2\right)&=&\rho+\s+\l\\
M_{Pl}^4\left(2\dot H+2\dot F+3H^2+4HF+3F^2\right)&=&-w\rho-\o\sigma+\o\left(\ano+Y\right)+\l\\
\dot\s+3\left[(1+\o)H+(1-\o)F\right]\s&=&\left[3\o H+(1-3\o)F\right]\left(\ano+Y\right)\\
\dot\rho+\left[3(1+w)H+2(1+w_\p)F\right]\rho&=&0  \label{rho final system flat}
\ena
The fixed points of this system of differential equations and their stability are found with some further restrictions (see appendix \ref{fixed points appendix} in the flat extra dimension subsections for the
explicit calculations). As I point out in appendix \ref{anomaly appendix}, the anomaly $\ano$ generally depends on the Hubble parameters of the model, on their time derivatives up to the third order and on
the spatial curvatures. This remains true also for the flat extra dimension limit examined in this subsection.
\paragraph{Fixed point solutions}
With the assumption of flat internal space and zero curvature for the 3D space as well, we can find different fixed points depending on the value of the extra dimension Hubble parameter. They can be
summarized as follows.
\begin{enumeratei}
\item
As I illustrate in appendix \ref{fixed points appendix}, there are two non trivial time independent solution with $F$ non vanishing at the fixed point, $\sF\ne0$, one for $\o\ne1/5$ and one for $\o=1/5$.  The values
of the 3D Hubble parameter (the measurable Hubble parameter) are given in terms of the constants $\o,c_A,c_B,c_Y$ and the mass scale $M_{Pl}$. We note that the anomaly parameters $c_A,c_B$ are given by the
CFT, while $\o$ relates the hidden sector pressure of the internal space to the hidden pressure of the 3D space (\ref{ansatz}). In particular, the two Hubble parameters are related by the equality
$H_\star=(\cep +1)F_\star$ (where $\cep$ is a rather complicated function of $\o,c_A,c_B,c_Y\mpl$) when $\o\ne1/5$, while for $\o=1/5$ we have $H_\star=F_\star$. For $\o=1/5$, $F_\star\neq0$ and we can choose
it to be positive or negative, implying that the extra dimension scale exponentially at those points, with respectively either positive or negative velocity. Consequently $H$ would also describe either a
contracting or expanding universe.
\item
I also found a fixed point solution for which the extra dimensions are static, i.e. $F_\star=0$ ($\o\ne0,-1$ and $c_B\ne0$), meaning that, while our visible universe is exponentially growing or decreasing,
the internal space isn't expanding nor collapsing. The corresponding solution is given by 
\bea\label{fixed flat holo}
H_\star^2&=&-\frac{20}{3c_B}\frac{\o}{\o+1}M_{Pl}^2\left[48c_Y\pm\sqrt{6\left(384c_Y^2-c_B\frac{\o}{\o+1}\right)}\right] \\
\ssig&=&-\ssigpp=\frac{2\o}{3\o-1}\ssigpp=-\frac{20}{c_B}\frac{\o}{\o+1}\MPl\left[48c_Y\pm\sqrt{6\left(384c_Y^2-c_B\frac{\o}{\o+1}\right)}\right]\\
\srho&=&0\non
\ena
The roots are real if $384c_Y^2-c_B\o/(\o+1)>0$ and cannot be both positive. We never have a couple of fixed points in the phase space diagram.
\item
A third fixed point is characterized by zero extra dimension Hubble parameter and $\o=-1$. In this case the critical point exists only if the conformal field theory is characterized by a positive coefficient
for the type B anomaly. The solution is 
\bea\label{fixed flat omega holo}
H_\star^2=\frac{640c_Y}{c_B}\mpl, \quad \ssig=-\ssigp=\oneover{2}\ssigpp=\frac{640c_Y}{3c_B}\MPl,\quad \srho=0 
\ena
If $c_B$ is zero (i.e. the anomaly vanishes at the fixed point) we are left with the trivial fixed point alone. 
\item
For vanishing $\o$, $\l$ should be non zero to get the inflationary fixed point $H_\star^2=\frac{\l}{3\Mpl}$.
\item
There also exists a trivial fixed point with $\sH=\sF=0$, where the anomaly and the trace $Y^\m_\m$ become zero, and $\srho=\ssig=0$ if $\o\neq w$ or $\srho=-\ssig$ if $\o=w$.
\end{enumeratei}

In any case --- i.e. for every $\o$ and $\l$ --- the solution does not depend on $w,w_\p$ (in fact, the system of equations (\ref{Friedmann rho})--(\ref{rho}) doesn't contain $w,w_\p$). So, if a solution
exists for some $\o$ and $c_A,c_B,c_Y$, that solution always exists whatever values the two parameters relating matter pressures to energy density take. This marks a difference with the bulk analysis of
section \ref{critical bulk}, since here we don't have any bulk dynamics perturbing the conservation equations, being the matter theory conformal and non interacting.

All the critical points have zero localized energy density $\srho$ (except for the trivial point with $\o=w$). Also, when the Hubble parameter is non vanishing, we don't get a positive valued $\srho$.
\paragraph{Stability analysis}
In appendix \ref{fixed points appendix} I analyze the stability of the $\sF=0$ critical points linearizing the system of differential equations around the fixed point. I conclude that the studied fixed point
(\ref{fixed flat holo})--(\ref{fixed flat omega holo}), characterized by vanishing $\sF$ and $\o=-1$ or $\o\ne-1$ can both be saddles or attractors, depending on the value of the anomaly parameter $c_B$, of
$c_Y$ and of $\o$ (relating the two CFT pressures). The trivial fixed point cannot be analyzed at linear order, since its stability matrix is null. It is obvious from \refeq{rho final system flat} that for
positive Hubble parameters the energy density goes to zero at late time.

We can thus observe that, starting the cosmological evolution with a hidden energy density $\s$ different from zero and with a suitable value of the anomaly coefficient $c_B$, choosing $\o$ to be such that
the $\sF=0$ stability matrix has negative eigenvalues, the corresponding $\sF=0$ fixed point is a global attractor for the flat extra dimension universe. This critical point could eventually represent the
present accelerated era. However, it's a zero energy density critical point.
%%%%%%%%%%%%%%%%%%%%%%%%%%%%%%%%%%%%%%%%%%%%%%%%%%%%%%%%%%%%%%%%%%%%%%%%%%%%%%%%%%%%%%%%%%%%%%%%%%%%%%%%%%%%%%%%%%%%%%%%%%%%%%%%%%%%%%%
%%%%%%%%%%%%%%%%%%%%%%%%%%%%%%%%%%%%%%%%%%%%%%%%%% static extra %%%%%%%%%%%%%%%%%%%%%%%%%%%%%%%%%%%%%%%%%%%%%%%%%%%%%%%%%%%%%%%%%%%%%%%
%%%%%%%%%%%%%%%%%%%%%%%%%%%%%%%%%%%%%%%%%%%%%%%%%%%%%%%%%%%%%%%%%%%%%%%%%%%%%%%%%%%%%%%%%%%%%%%%%%%%%%%%%%%%%%%%%%%%%%%%%%%%%%%%%%%%%%%
\subsection{Static compact extra dimensions}\label{brane critical static}
The Einstein equations of motion (\ref{final system}) get simplified when we take the $b=\mathrm{const}$ ansatz for the internal space scale factor 
\bea\label{final system static}
M_{Pl}^4\left(3H^2+\frac{\k}{b_0^2}\right)&=&\rho+\s+\l\\
M_{Pl}^4\left(2\dot H+\frac{\k}{b_0^2}\right)&=&-w\rho-\o\sigma+\o\left(\ano+Y\right)+\l\\
\dot\s+3(1+\o)H\s&=&3\o H\left(\ano+Y\right)\\
\dot\rho+3(1+w)H\rho&=&0
\ena
From these equations we can deduce the corresponding fixed points and their criticality, following the calculations in appendix \ref{fixed points appendix}.
\paragraph{Fixed point solutions}
We find the inflationary fixed point for a universe with non evolving internal dimensions, i.e. with constant scale factor $b(t)\equiv b_0$. Besides the trivial fixed point $\sH=0$ there are other solutions.
\begin{enumeratei}
\item 
The existence of non trivial fixed points is determined by the values of the parameter of the specific conformal theory $c_A,c_B,c_Y$ and $\o$, but also by the mass scales $M_{Pl}$ and $\k/b_0$. An easy
critical point solution can be derived in the case of zero type B contribution to the anomaly and $\o=-1$
\bea\label{fixed static cA=cB}
H^2_\star&=&\frac{\k}{b_0^2}\left[9\pm\sqrt{90+\frac{5c_A}{192c_Y}\oneover{\mpl}\frac{\k}{b_0^2}}\right]^{-1}\\
\ssig&=&-\ssigp=2\ssigpp=\Mpl\left(3\left[9\pm\sqrt{90+\frac{5c_A}{192c_Y}\oneover{\mpl}\frac{\k}{b_0^2}}\right]^{-1}+1\right)\frac{\k}{b_0^2}\\
\srho&=&0\non
\ena
For $\k=0$ it reduces to the trivial fixed point. We have real roots for $\sH$ if $192c_Y^2\mpl>-c_A\k/45b_0^2$ and they will be both positive when $\k>0$ and $-192c_Y^2\mpl<c_A\k/45b_0^2<-192c_Y^2\mpl/252$, so
that $c_A$ must be negative.
\item
We also have a trivial critical point with $\sH=0$. In particular both $\srho$ and $\ssig$ can be different from zero, they are functions of $\k/b_0^2$ as we can see from the equation
$\srho+\ssig=\Mpl\k/b_0^2$, and are also related by 
\bea
(1+w)\srho+(1+\o)\ssig=\o\left(-\frac{c_B}{800}\frac{\k}{b_0^2}+\frac{4c_Y}{5}\mpl\right)\frac{\k^2}{b_0^4}
\ena
\item 
Other fixed points can be found, for example for $\o=0$. The internal space curvature $\k$ would have to be negative in those cases, since $H^2_\star=-\k/3b_0^2$ and we would have to compactify on
an hyperbolic space.
\end{enumeratei}
\paragraph{Stability analysis}
For the static extra dimension fixed points I found that we can get an attractor or a saddle, depending on the values of $c_A,c_B,cY\mpl$ and $\o$.

We can thus choose the hidden sector parameters such that we can get a stable fixed point. There exists however another critical point, i.e. the trivial one characterized by $\sH=0$ but generally non zero
$\srho$ and $\ssig$, which always is a saddle. So, trajectories may either be attracted by the non trivial critical point or flow away from the saddle point.
%%%%%%%%%%%%%%%%%%%%%%%%%%%%%%%%%%%%%%%%%%%%%%%%%%%%%%%%%%%%%%%%%%%%%%%%%%%%%%%%%%%%%%%%%%%%%%%%%%%%%%%%%%%%%%%%%%%%%%%%%%%%%%%%%%%%%%
%%%%%%%%%%%%%%%%%%%%%%%%%%%%%%%%%%%%%%%%%%%%%%%%% equal scale factors %%%%%%%%%%%%%%%%%%%%%%%%%%%%%%%%%%%%%%%%%%%%%%%%%%%%%%%%%%%%%%%%
%%%%%%%%%%%%%%%%%%%%%%%%%%%%%%%%%%%%%%%%%%%%%%%%%%%%%%%%%%%%%%%%%%%%%%%%%%%%%%%%%%%%%%%%%%%%%%%%%%%%%%%%%%%%%%%%%%%%%%%%%%%%%%%%%%%%%%
\subsection{Equal scale factors}\label{brane critical equal}
Another limit that simplifies some of the calculations is the equal scale factor assumption. In this case the Hubble parameters of the internal space and the 3D space are equal, $F=H$, and we are left with
the following set of equations for the variables $H,\rho,\s$
\bea\label{final system equal}
M_{Pl}^4\left(10H^2+\frac{\k}{b^2}\right)&=&\rho+\s+\l\\
M_{Pl}^4\left(4\dot H+10H^2+\frac{\k}{b^2}\right)&=&-w\rho-\o\sigma+\o\left(\ano+Y\right)+\l\\
\dot\s+6H\s&=&H\left(\ano+Y\right)\\
\dot\rho+(5+3w+2w_\p)H\rho&=&0
\ena
\paragraph{Fixed point solutions}
We find an inflationary fixed point where the internal space is also staying in an inflationary era, since the two Hubble parameters are equal (see appendix \ref{fixed points appendix} for calculations). The existence
of such fixed point depends on the values of $c_A,c_B$, $\k$ and $\o$.

Is is necessary to impose $\o=1/5$, implying that the two pressures characterizing the CFT stress--energy tensor must be equal $\s_\p=\s_p$, in order to obtain the following fixed point solutions.
\begin{enumeratei}
\item
We write the explicit solution for a flat internal space $\k=0$, where the critical value for the Hubble parameter is given in terms of the anomaly coefficients and of the only dimensionful parameter which is
the 6D Planck mass. This solution puts restrictions the possible values of $c_A$ and $c_B$, coming from the conformal field theory anomaly
\bea
\sH^2&=&-\frac{24}{c_A+2c_B}\mpl\left[24c_Y\pm\sqrt{576c_Y^2+(c_A+2c_B)}\right]\\  
\ssig&=&-\ssigp=-\ssigpp=-\frac{240}{c_A+2c_B}\MPl\left[24c_Y\pm\sqrt{576c_Y^2+(c_A+2c_B)}\right],  \qquad  \srho=0  \non
\ena
For $(24c_Y)^2>-(c_A+2c_B)$ the two roots are real. We must satisfy the condition $(c_A+2c_B)<0$ on the anomaly parameters in order to have two positive $\sH$ critical points (both denoted by zero
energy density $\srho$).
\item
For $\k\ne0$ there exists no fixed points, since $\left(\ano+Y\right)$ and $\left(10\sH^2+\frac{\k}{a^2}\right)$ cannot be both constant unless we enforce the staticity condition on the Hubble parameter
$H_\star=0$ and $\k=0$.
\end{enumeratei} 
\paragraph{Stability analysis}
The stability analysis in this last case of equal scale factors is carried in appendix \ref{fixed points appendix}. As a result, I found that the equal scale factor critical point with flat internal
space is an attractor. It could thus represent the eternal acceleration of the universe.
%%%%%%%%%%%%%%%%%%%%%%%%%%%%%%%%%%%%%%%%%%%%%%%%%%%%%%%%%%%%%%%%%%%%%%%%%%%%%%%%%%%%%%%%%%%%%%%%%%%%%%%%%%%%%%%%%%%%%%%%%%%%%%%%%%%%%%%
%%%%%%%%%%%%%%%%%%%%%%%%%%%%%%%%%%%%%%%%%%%%%%%%%%%% comments %%%%%%%%%%%%%%%%%%%%%%%%%%%%%%%%%%%%%%%%%%%%%%%%%%%%%%%%%%%%%%%%%%%%%%%%%
%%%%%%%%%%%%%%%%%%%%%%%%%%%%%%%%%%%%%%%%%%%%%%%%%%%%%%%%%%%%%%%%%%%%%%%%%%%%%%%%%%%%%%%%%%%%%%%%%%%%%%%%%%%%%%%%%%%%%%%%%%%%%%%%%%%%%%%
\subsection{Comments}
I want to summarize the interesting features of the critical analysis of sections \ref{brane critical flat}--\ref{brane critical equal}, in the perspective of the comparison with the 7D bulk gravitational
dual description that I illustrated in sections \ref{bulk cosmology} and \ref{critical bulk}.
\begin{enumeratea}
\item
All the critical point we can find in the brane description are characterized by exactly zero value for the localized matter energy density $\rho$ (except for some $\sH=0$ trivial points). To have a non
vanishing energy density it is necessary to introduce an interaction term between the matter fields and the hidden sector fields. The reason for this is that it modifies the conservation equation for $\rho$,
allowing for a non zero time--independent solution. This intuitively corresponds to turning on the brane--bulk energy exchange on the bulk gravity side.
\item
In most of the simple critical point solutions, the hidden sector pressure $\ssigp$ is related to the energy density $\ssig$ by $\ssigp=-\ssig$. This indicates a vacuum behavior for the equation of state of
the hidden sector of the holographic dual theory at the inflationary fixed points.
\item
There also appears to exist more than one critical point solution in some of the explicitely examined limits. Since the stability matrix analysis reveals that we can have either stable or saddle points
(depending on the CFT and counterterm parameters $c_A,c_B,c_Y$ and on the Planck mass), we can expect two kinds of behavior (if the number of critical point solution is two). Whenever one of the fixed points
is attractive and the other one is a saddle, we can generally depict a phase portrait such that some of the trajectories are attracted by the stable point, while others can be repulsed by the saddle and go
toward the large density region. If, on the other hand, we get two saddle points, trajectories bend near to the saddles and flow away. I note that the trivial critical point has undefined stability at linear
order in the perturbations, so that it may either attract or repel trajectories in its neighborhood. However, if $H>0$ and $w,w_\p>-1$, late time evolution is always described by $\rho\to0$. 
\item
Comparing these results with the bulk cosmology in the case of zero energy exchange (recall for instance \refeq{bulk limits H}--\refeq{bulk limits rho} with $T^7_7=T^0_7=0$) the main difference is that we have
non trivial fixed points in the CFT description, while the gravity cosmology is characterized only by critical points with zero Hubble parameter and zero acceleration. The origin of the discrepancy has to be
ascribed to the trace term coming from the CFT anomaly and to the counterterm trace that give a non trivial contribution to the conservation equation for the hidden sector energy density $\s$. The holographic
renormalization procedure, and in particular the quantum breaking of conformal invariance, is responsible for the presence of such terms.
\item
In the bulk description we could only find positive acceleration critical points since $q=H^2$. With the holographic approach, we could in principle also get non constant $H$ critical point solutions. In fact
we could solve the system of Einstein equations on the brane asking time--independence for $q=\dot H+H^2,\rho$ and $\s$. We would get a new system of first order (non linear) differential equations in $H$,
which in principle could have non trivial solutions.
\end{enumeratea}
%%%%%%%%%%%%%%%%%%%%%%%%%%%%%%%%%%%%%%%%%%%%%%%%%%%%%%%%%%%%%%%%%%%%%%%%%%%%%%%%%%%%%%%%%%%%%%%%%%%%%%%%%%%%%%%%%%%%%%%%%%%%%%%%%%%%%%%
%%%%%%%%%%%%%%%%%%%%%%%%%%%%%%%%%%%%%%%%%%%%%%%%%%%% correspondence %%%%%%%%%%%%%%%%%%%%%%%%%%%%%%%%%%%%%%%%%%%%%%%%%%%%%%%%%%%%%%%%%%%
%%%%%%%%%%%%%%%%%%%%%%%%%%%%%%%%%%%%%%%%%%%%%%%%%%%%%%%%%%%%%%%%%%%%%%%%%%%%%%%%%%%%%%%%%%%%%%%%%%%%%%%%%%%%%%%%%%%%%%%%%%%%%%%%%%%%%%%
\sectioncount{Brane/bulk correspondence at work}\label{examples}
Some solutions are now derived illustrating interesting aspects of the cosmological model we considered and in particular of the duality that relates the two descriptions. We will be able to
find explicit expressions by making special simplifying assumptions on the parameters of the holographic theory and on the space--time background. These examples allow us to make a comparison between the
results we will find in the holographic set--up and the expressions we derived in the bulk gravity theory. 

Since I am interested in comparing the two dual approaches, in the sense of AdS/CFT correspondence, I have to derive some expressions for $H$ in terms of the localized matter energy density $\rho$ and
of a mirage density $\chi$. In section \ref{bulk cosmology}, I performed the cosmological analysis in the 7D bulk description, expressing the 3D Hubble parameter $H$ in terms of the localized energy density
$\rho$, reducing the first order ODE in $H$ to an algebraic equation for $H^2$ plus a first order ODE for the mirage density $\chi$. In the holographic dual theory, the mirage density is identified with the
solution to the homogeneous equation associated to the conservation equation for the hidden sector density $\s$. It will thus have the property of obeying to the free radiation conservation equation in $d$
effective dimensions (where $d$ is the effective number of dimensions equal to $4$ in a static compact space background and to $6$ when $a(t)$ and $b(t)$ are equal).

To obtain the explicit result for $H^2$ in the brane dual description, however, it is necessary to integrate the differential equation \refeq{sigma anal sol} for the energy density $\s$. It is established
that the anomaly and the trace of the quadratic contribution to the variation of the dual theory action are highly non trivial functions of the Hubble parameters and the spatial curvatures and contain
derivatives of $H,F$ up to order three. So, an analytical integration of the $\s$ conservation equation is apparently unachievable in general. However, it is possible to neglect the $\ano$ and $Y$
contributions if we are in the slowly scaling approximation, which corresponds to a small curvature approximation. This is what I am going to discuss in the following.
%%%%%%%%%%%%%%%%%%%%%%%%%%%%%%%%%%%%%%%%%%%%%%%%%%%%%%%%%%%%%%%%%%%%%%%%%%%%%%%%%%%%%%%%%%%%%%%%%%%%%%%%%%%%%%%%%%%%%%%%%%%%%%%%%%%%%
%%%%%%%%%%%%%%%%%%%%%%%%%%%%%%%%%%%%%%%%%%%%%%%%%%%%%%% special 4 %%%%%%%%%%%%%%%%%%%%%%%%%%%%%%%%%%%%%%%%%%%%%%%%%%%%%%%%%%%%%%%%%%%
%%%%%%%%%%%%%%%%%%%%%%%%%%%%%%%%%%%%%%%%%%%%%%%%%%%%%%%%%%%%%%%%%%%%%%%%%%%%%%%%%%%%%%%%%%%%%%%%%%%%%%%%%%%%%%%%%%%%%%%%%%%%%%%%%%%%%
\subsection{Slowly scaling approximation}
I give a rough idea on how the correspondence between the brane and the bulk dual theories works. In fact, I will neglect all the higher order terms in the holographic description, which is equivalent to
ask that the Hubble parameter is negligible with respect to the Planck mass $H^2\ll\mpl$ (i.e. $\dot a\ll M_{Pl}\,a$). In this approximation, all the higher order curvature terms --- including the anomaly
and the trace $Y^\m_\m$ --- can be neglected in favor of contributions proportional to the Einstein tensor. The integration of the $\s$ equation would give terms of the order of $\mpl H^4$ (from $Y$) and
$H^6$ (from $\ano$). Once we plug the result for $\s$ in the Einstein equation \refeq{final system} these terms are suppressed, since the l.h.s. of \refeq{final system} is of order $\Mpl H^2$. 

I anticipate that, as a consequence of the small Hubble parameter approximation on the brane, we get a linear dependence of $H^2$ on the mirage plus visible matter energy densities (the hidden sector density
$\s$ is identified with the mirage density $\chi$). Higher order contributions due to the anomaly and to $Y$ would give rise to higher power dependence in a small density expansion for $H^2$. Since in the
holographic description we truncate to linear order in density, we also keep only linear terms in $\rho$ for the bulk gravity results. The bulk equations for $H$ and $\chi$ \refeq{bulk limits H}--\refeq{bulk
limits rho}, can be formulated independently of $w$ if we ignore higher (quadratic) order terms in $\rho$ and assign a specific value to $w_\p$. Neglecting the second order term in $\rho$ we will only have
one condition to determine $M$ and $V$ in terms of the brane parameter $M_{Pl}$, so that only the ratio $M^{10}/V$ will be identified.

Since in this approximation the quadratic and higher order dependence of $H^2$ on $\rho$ are absent, we won't capture the eventual Starobinsky \cite{Starobinsky:1980te} behavior of the solutions to the
Einstein equations \cite{Vilenkin:1985md}. The higher derivative terms are necessary in that case to calculate the exit from inflation to a matter dominated universe and the subsequent thermalization to
radiation dominated era. For Starobinsky, the higher derivative terms are represented by the type D conformal anomaly contribution to the trace of the stress--energy tensor. Nonetheless, in the 5D RS
holographic dual analysis of \cite{Kiritsis:2005bm}, where these terms are cancelled, stringy corrections like Gauss--Bonnet terms can play the same role. We note that in our set--up, the 6D conformal anomaly
contains suitable higher derivative terms, not only in the total derivative contributions but also in type B anomaly.

Here are the results to which the slowly scaling approximation leads in some particular limits.
%%%%%%%%%%%%%%%%%%%%%%%%%%%%%%%%%%%%%%%%%%%%%%%%%%%%%%%%%%%%%%%%%%%%%%%%%%%%%%%%%%%%%%%%%%%%%%%%%%%%%%%%%%%%%%%%%%%%%%%%%%%%%%%%%%%%%
%%%%%%%%%%%%%%%%%%%%%%%%%%%%%%%%%%%%%%%%%%%%%%%%%%%%%%% equal scale factors %%%%%%%%%%%%%%%%%%%%%%%%%%%%%%%%%%%%%%%%%%%%%%%%%%%%%%%%%
%%%%%%%%%%%%%%%%%%%%%%%%%%%%%%%%%%%%%%%%%%%%%%%%%%%%%%%%%%%%%%%%%%%%%%%%%%%%%%%%%%%%%%%%%%%%%%%%%%%%%%%%%%%%%%%%%%%%%%%%%%%%%%%%%%%%%
\subsubsection{Equal scale factors}\label{equal comparison}
We start looking at the Einstein, conservation and anomaly equations in the equal scale factor limit. The system of equations \refeq{final system}--\refeq{final system rho} for the theory on the brane takes
the form
\bea\label{final system equal compare}
\Mpl\left(10H^2+3\frac{k}{a^2}+\frac{\k}{a^2}\right)&=&\rho+\s+\l\\
\Mpl\left(4\dot H+10H^2+\frac{k}{a^2}+\frac{\k}{a^2}\right)&=&-w\rho-\o\sigma+\o\left(\ano+Y\right)+\l  \non
\ena
\bea
\dot\s+6H\s&=&H\left(\ano+Y\right)  \label{final system sigma equal compare}\\
\dot\rho+(3(1+w)+2(1+w_\p))H\rho&=&0 \label{final system rho equal compare}
\ena
We note that the set of equations is independent of the hidden sector parameter $\o$ and it is above all interesting that the homogeneous equation associated to \ref{final system sigma equal compare} is
indeed precisely the 6D free radiation equation, independently of the value for $\o$.  In particular, the two conservation equations can be written in the integral form
\bea
\s&=&\chi+a^{-6}\int\intd t\,a^6\,H\left(\ano+Y\right), \qquad \chi=\chi_0\,\left(\frac{a_0}{a}\right)^6\\
\rho&=&\rho_0\,\left(\frac{a_0}{a}\right)^{3(1+w)+2(1+w_\p)}\non
\ena
Plugging the result for $\s$ in the first equation on the system \refeq{final system equal compare} and neglecting the curvature higher order terms that come from the integration of $\left(\ano+Y\right)$,
we obtain the following expression for $H^2$, together with the $\rho$ and $\chi$ equations in their differential form
\bea
H^2+\oneover{10}\left(3\frac{k}{a^2}+\frac{\k}{a^2}\right)&=&\oneover{10\Mpl}\left(\rho+\chi\right)+\oneover{10\Mpl}\l\label{H 6D equal linear}\\
\dot\chi+6H\chi&=&0\\
\dot\rho+(3(1+w)+2(1+w_\p))H\rho&=&0\label{rho 6D equal linear}
\ena
It is now easy to compare \refeq{H 6D equal linear}--\refeq{rho 6D equal linear} with the corresponding system of equations in the bulk description of the equal scale factor universe, with zero brane--bulk
energy exchange $T^0_7$ and bulk ``self interaction'' $T^7_7$. The expression for $H^2$ can also be written in a $w$--independent way fixing $w_\p$ (in this particular case we could also include the quadratic
term in $\rho$ in the $w$--independent formulation).  Neglecting the second order term in the energy densities we obtain
\bea\label{H 7D equal linear}
H^2+\oneover{10}\left(3\frac{k}{a^2}+\frac{\k}{a^2}\right)&=&\frac{2\tcveq V}{5M^{10}}\left(\rho+\chi\right)+\l_{RS}\\
\dot\chi+6H\chi&=&0\\
\dot\rho+(3(1+w)+2(1+w_\p))H\rho&=&0\label{rho 7D equal linear}
\ena

The two systems of equations \refeq{H 6D equal linear}--\refeq{rho 6D equal linear} and \refeq{H 7D equal linear}--\refeq{rho 7D equal linear} perfectly agree at this order in the approximation. The matching
between the scales on the two sides of the duality is then
\bea\label{relation small density equal}
\frac{M^{10}}{V}=4\tcveq\Mpl  \qquad\stackrel{w_\p=w}{\longrightarrow}\qquad  \frac{M^{10}}{V}=\frac{\Mpl}{20}
\ena
As I announced, only the ratio $M^{10}/V$ can be determined, since we only have one condition to match the two descriptions. When higher order corrections are included in the brane description we would
generally find a matching for both $M$ and $V$, which in principle would depend on the particular CFT parameters ($c_A,c_B,c_Y$). We can guess that the ratio $M^{10}/V$ would not depend on them (indeed, for
the pure RS set--up we get $\Mpl V\propto M^{10}$). When $w_\p=w$ (i.e. when the pressures of the matter perfect fluid relative to the 2D internal space and the 3D space are equal $\p=p$) the coefficient
$\tcv$ becomes $\tcveq=1/80$. It is interesting to note that for $w=w_\p$ the matching exactly reduces to the RS condition for zero effective cosmological constant on the brane $\l_{RS}=0$. Since we are in
the limit $F=H$, it seems natural to have $\p=p$ too.
%%%%%%%%%%%%%%%%%%%%%%%%%%%%%%%%%%%%%%%%%%%%%%%%%%%%%%%%%%%%%%%%%%%%%%%%%%%%%%%%%%%%%%%%%%%%%%%%%%%%%%%%%%%%%%%%%%%%%%%%%%%%%%%%%%%%%
%%%%%%%%%%%%%%%%%%%%%%%%%%%%%%%%%%%%%%%%%%%%%%%%%%%%%%% static extra dimensions %%%%%%%%%%%%%%%%%%%%%%%%%%%%%%%%%%%%%%%%%%%%%%%%%%%%%
%%%%%%%%%%%%%%%%%%%%%%%%%%%%%%%%%%%%%%%%%%%%%%%%%%%%%%%%%%%%%%%%%%%%%%%%%%%%%%%%%%%%%%%%%%%%%%%%%%%%%%%%%%%%%%%%%%%%%%%%%%%%%%%%%%%%%
\subsubsection{Static compact extra dimensions}\label{static comparison}
We consider the static extra dimension limit $F=0$. The Einstein equations plus conservation and anomaly equations in this limit read
\bea\label{final system static compare}
\Mpl\left(3H^2+3\frac{k}{a^2}+\frac{\k}{b_0^2}\right)&=&\rho+\s+\l\\
\Mpl\left(2\dot H+3H^2+\frac{k}{a^2}+\frac{\k}{b_0^2}\right)&=&-w\rho-\o\sigma+\o\left(\ano+Y\right)+\l\\\non\\
\s+3(1+\o)H\s&=&3\o H\left(\ano+Y\right)  \label{final system sigma static compare}\\
\dot\rho+3(1+w)H\rho&=&0 \label{final system rho static compare}
\ena
The homogeneous equation associated to \refeq{final system sigma static compare} is the 4D free radiation equation only if $\o=1/3$, implying that the hidden sector pressure of the internal space $\s_\p$ must
be zero. With this assumption, the two conservation equations become
\bea\label{sigma integrate static compare}
\s&=&\chi+a^{-4}\int\intd t\,a^4\,H\left(\ano+Y\right), \qquad \chi=\chi_0\,\left(\frac{a_0}{a}\right)^4\\
\rho&=&\rho_0\,\left(\frac{a_0}{a}\right)^{3(1+w)}\non
\ena
The results for $\chi$ and $\rho$ agree with the bulk formulation for zero energy exchange. Plugging \refeq{sigma integrate static compare} into \refeq{final system static compare} and neglecting the
curvature higher order term, as we are in the slowly scaling approximation, we find (for $k=0$)
\bea
H^2+\oneover{3}\frac{\k}{b_0^2}&=&\oneover{3\Mpl}\left(\rho+\chi\right)\label{H 6D static linear}\\
\dot\chi+4H\chi&=&0\\
\dot\rho+3(1+w)H\rho&=&0
\ena
Although conservation equations agree, the Friedmann--like equation doesn't give the expected $1/6$ coefficient in front of the $\k/b_0^2$ term in \refeq{H 6D static linear}. In fact, the equations on the
bulk gravity side
\bea\label{H 7D static linear}
H^2+\oneover{6}\frac{\k}{b_0^2}&=&\frac{2\tcvst V}{3M^{10}}\left(\rho+\chi\right)\\
\dot\chi+4H\chi&=&0\\
\dot\rho+3(w+1)H\rho&=&0
\ena
The bulk equations are derived in the density linear approximation and for vanishing energy exchange $T^0_7$ and $T^7_7$. The coefficient $\tcvst$ can be written in a $w$--independent way if we fix $w_\p$.

As a consequence the ratio of the two bulk parameters can be identified with
\bea\label{relation small density static}
\frac{M^{10}}{V}=2\tcvst\Mpl  \qquad\stackrel{w_\p=\frac{w+5}{6}}{\longrightarrow}\qquad  \frac{M^{10}}{V}=\frac{\Mpl}{20}
\ena
The $\k/b_0^2$ terms differ in the two dual descriptions (in the static extra dimension background). The matching \refeq{relation small density static} gives a result that depends on the values of $w,w_\p$ in
a different way if compared to the equal scale factor limit \refeq{relation small density equal}. It is thus interesting to further examine how the matching varies according to the value of the internal space
Hubble parameter. I am indeed going to consider the proportionality ansatz $F=\x H$ to better understand this behavior. I note that in the limit $w_\p=(w+5)/6$ we recover in \refeq{relation small density
static} the RS fine--tuning determining zero effective cosmological constant on the brane $\l_{RS}=0$, since $\tcvst=1/40$.
%%%%%%%%%%%%%%%%%%%%%%%%%%%%%%%%%%%%%%%%%%%%%%%%%%%%%%%%%%%%%%%%%%%%%%%%%%%%%%%%%%%%%%%%%%%%%%%%%%%%%%%%%%%%%%%%%%%%%%%%%%%%%%%%%%%%%
%%%%%%%%%%%%%%%%%%%%%%%%%%%%%%%%%%%%%%%%%%%%%%%%%%%%% proportional Hubble %%%%%%%%%%%%%%%%%%%%%%%%%%%%%%%%%%%%%%%%%%%%%%%%%%%%%%%%%%%
%%%%%%%%%%%%%%%%%%%%%%%%%%%%%%%%%%%%%%%%%%%%%%%%%%%%%%%%%%%%%%%%%%%%%%%%%%%%%%%%%%%%%%%%%%%%%%%%%%%%%%%%%%%%%%%%%%%%%%%%%%%%%%%%%%%%%
\subsubsection{Proportional Hubble parameters}\label{proportional hubble comparison}
Following the computations in the last two sections and generalizing them, we derive the set of equations for $H^2,\chi$ and $\rho$ for proportional and small Hubble parameters. Since, as before, $\s=\chi$ if
we neglect higher order terms in the small curvature approximation, equations \refeq{final system}--\refeq{final system rho} lead to the following set
\bea
H^2+\oneover{(\x_b^2+6\x_b+3)}\frac{\k_b}{b^2}&=&\oneover{(\x_b^2+6\x_b+3)\Mpl}\left(\rho+\chi\right)\label{H 6D prop linear}\\
\dot\chi+d_{\x_b} H\chi&=&0\\
\dot\rho+w_{\x_b} H\rho&=&0
\ena
where $d_{\x_b}\equiv3(1+\o)+3\x_b(1-\o)$, $w_{\x_b}\equiv3(1+w)+2\x_b(1+w_\p)$ and $\x_b$ is the proportionality factor $F=\x_b H$ (or $b=a^{\x_b}$). The bulk equations where derived in \refeq{bulk limits H
prop}--\refeq{bulk limits rho prop}. For the moment we will keep two different proportionality factors so that we have in the bulk $F=\x_B H$. Following section \ref{proportional hubble} (putting zero $T^7_7$
and $T^0_7$) we get
\bea\label{H 7D prop linear}
H^2+\oneover{(\x_B^2+3\x_B+6)}\frac{\k_B}{b^2}&=&\frac{2\tcvx V}{(2\x_B+3)M^{10}}\left(\rho+\chi\right)\\
\dot\chi+d_{\x_B} H\chi&=&0\\
\dot\rho+w_{\x_B} H\rho&=&0
\ena
The following definitions have been used: $d_{\x_B}\equiv6(\x_B^2+2\x_B+2)/(2\x_B+3)$, $w_{\x_B}\equiv3(1+w)+2\x_B(1+w_\p)$ and recall that $\tcvx$ is $\tcvx\equiv c_V/(w_{\x B}-d_{\x B})$, where
$c_V=(31w-6w_\p-5)/400$.

In order for the two descriptions to be equivalent w.r.t. the $\chi$ and $\rho$ differential equations (which don't get any correction from higher order contributions), we have to put $\x_b=\x_B=\x$, as it
was expected. Besides, the parameter relating $\s_\p$ to $\s_p$ must be $\o=1/(2\x+3)$. However, assuming an equal proportionality relation on the two sides of the duality, the coefficients of the $\k$ terms
in \refeq{H 6D prop linear} and \refeq{H 7D prop linear} differ if the two curvatures in the bulk and brane descriptions are equal, unless $\x=1$. So, the only set--up that predicts the same effective
spatial curvature for the internal space for general backgrounds in the brane and bulk descriptions is the equal scale factor background (neglecting higher order corrections). However, we can determine an
effective spatial curvature for the internal space in the brane description, given by $\k_b=(\x^2+6\x+3)\k_B/(\x^2+3\x+6)$.

The matching for the scales of the two dual theories is given by
\bea\label{relation small density prop}
\frac{M^{10}}{V}=\frac{\x^2+6\x+3}{2\x+3}2\tcvx\Mpl  
\ena
It is always possible to choose a $w_\p$ such that the matching relation \refeq{relation small density prop} gives the RS fine--tuning condition $\Mpl=20M^{10}/V$. Otherwise, missing the fine--tuning would
amount to introducing a non vanishing effective cosmological constant on the RS brane.
%%%%%%%%%%%%%%%%%%%%%%%%%%%%%%%%%%%%%%%%%%%%%%%%%%%%%%%%%%%%%%%%%%%%%%%%%%%%%%%%%%%%%%%%%%%%%%%%%%%%%%%%%%%%%%%%%%%%%%%%%%%%%%%%%%%%%
%%%%%%%%%%%%%%%%%%%%%%%%%%%%%%%%%%%%%%%%%%%%%%%%%%%%% generalization %%%%%%%%%%%%%%%%%%%%%%%%%%%%%%%%%%%%%%%%%%%%%%%%%%%%%%%%%%%%%%%%
%%%%%%%%%%%%%%%%%%%%%%%%%%%%%%%%%%%%%%%%%%%%%%%%%%%%%%%%%%%%%%%%%%%%%%%%%%%%%%%%%%%%%%%%%%%%%%%%%%%%%%%%%%%%%%%%%%%%%%%%%%%%%%%%%%%%%
\sectioncount{Non conformal interacting generalization}\label{general}
To examine the general cosmological evolution that reflects presence of energy exchange between brane and bulk and of bulk pressure in the seven dimensional picture, we will drop the assumption of having a
conformal non interacting field theory living on the brane. Intuitively, a non vanishing $T^0_7$ in the bulk description corresponds to interactions between the gauge theory and the visible matter. The
diagonal $T^7_7$ component appears in the brane description as dual to a new trace term spoiling the conformal invariance. The generalization of the 6D RS dual action amounts to adding the new interaction
term and substituting the hidden sector CFT with a strongly coupled gauge theory (SCGT). The following analysis will be in analogy with
the 4D holographic dual generalization exposed in \cite{Kiritsis:2005bm}, corresponding to the 5D RS cosmology.

Using the notations of section \ref{dual} we write the generalized action as
\bea
S_{gen}=S_{SCGT}+S_R+S_{R^2}+S_{R^3}+S_{m,b}+S_{int}
\ena
where the new entry is the interaction term $S_{int}$ and $S_{CFT}$ has been changed into $S_{SCGT}$.
The strongly coupled fields can be integrated out, transforming the sum of the strongly coupled theory action plus the interaction term into an effective functional of the visible fields (and of the metric)
$W_{SCGT}$. As a result, the action becomes
\bea
S_{gen}=W_{SCGT}+S_R+S_{R^2}+S_{R^3}+S_{m,b}
\ena

As in the conformal non interacting case, we are now ready to calculate the general 6D equations of motion for the holographic generalized RS cosmology. 
%%%%%%%%%%%%%%%%%%%%%%%%%%%%%%%%%%%%%%%%%%%%%%%%%%%%%%%%%%%%%%%%%%%%%%%%%%%%%%%%%%%%%%%%%%%%%%%%%%%%%%%%%%%%%%%%%%%%%%%%%%%%%%%%%%%%%
%%%%%%%%%%%%%%%%%%%%%%%%%%%%%%%%%%%%%%%%%%%%%%%%% generalized equations %%%%%%%%%%%%%%%%%%%%%%%%%%%%%%%%%%%%%%%%%%%%%%%%%%%%%%%%%%%%%
%%%%%%%%%%%%%%%%%%%%%%%%%%%%%%%%%%%%%%%%%%%%%%%%%%%%%%%%%%%%%%%%%%%%%%%%%%%%%%%%%%%%%%%%%%%%%%%%%%%%%%%%%%%%%%%%%%%%%%%%%%%%%%%%%%%%%
\subsection{Generalized evolution equations}
The stress--energy tensors are defined in the following way
\bea
T_{\m\n}&=&\oneover{\sqrt{-\g}}\frac{\d S_m}{\d\g^{\m\n}},\quad W_{\m\n}=\oneover{\sqrt{-\g}}\frac{\d W_{SCGT}}{\d\g^{\m\n}}\\
Y_{\m\n}&=&\oneover{\sqrt{-\g}}\frac{\d S_{R^2}}{\d\g^{\m\n}},\quad Z_{\m\n}=\oneover{\sqrt{-\g}}\frac{\d S_{R^3}}{\d\g^{\m\n}}\non
\ena
\bea
V_{\m\n}&=&W_{\m\n}+Y_{\m\n}+Z_{\m\n}, \qquad Y^\m_\m=Y\non
\ena
The Einstein equation, the conservation conditions and the anomaly equation then read
\bea\label{Fried gen}
M_{Pl}^4G_{\m\n}&=&T_{\m\n}+W_{\m\n}+Y_{\m\n}+V_{\m\n}\non\\
\nabla^\n T_{\m\n}&=&T\non\\
\nabla^\n V_{\m\n}&=&-T\non\\
V^\m_\m&=&\ano+Y+X
\ena
The total stress--energy tensor is still conserved. Taking account of the interactions between the hidden theory and the matter generally amounts to having non separately conserved $V_{\m\n}$ and $T_{\m\n}$.
This will be reflected by the introduction of a non homogenous term in the conservation equations. The anomaly equation contains the general expression for the conformal anomaly in six dimensions $\ano$ and
the trace term $Y$.  Furthermore, it gets modified including an extra term $X$ that accounts for classical and quantum breaking of the conformal symmetry in a FRW plus compact space background. The
stress--energy tensors are parametrized as before
\bea\label{gen stress-en}\begin{array}{rclrclrcl}
T_{00}&=&\rho(t),\quad T_{ij}&=&p(t)\,\g_{ij}, \quad T_{ab}&=&\p(t)\,\g_{ab}\\
V_{00}&=&\s(t),\quad V_{ij}&=&\s_p(t)\,\g_{ij}, \quad V_{ab}&=&\s_\p(t)\,\g_{ab}\\
\end{array}\ena

The consequent changes in the equations written in terms of the Hubble parameters and of the energy densities and pressures are the following. The Friedmann equations remain the same
\bea\label{gen Friedmann eqs}
M_{Pl}^4\left(3H^2+6H\,F+F^2+3\frac{k}{a^2}+\frac{\k}{b^2}\right)&=&\rho+\s+\l\non\\
M_{Pl}^4\left(2\dot H+3H^2+4H\,F+2\dot F+3F^2+\frac{k}{a^2}+\frac{\k}{b^2}\right)&=&-p-\s_p+\l
\ena
the conservation equations now involve the quantity $T$
\bea\label{gen conservation eqs}
\dot\s+3(\s+\s_p) H+2(\s+\s_\p)F&=&T\non\\
\dot\rho+3(\rho+p) H+2(\rho+\p)F&=&-T
\ena
and the anomaly equation includes the conformal breaking term, as a consequence of the masses and $\b$--functions of the strongly coupled gauge theory
\bea\label{gen anomaly eq}
&&\s-3\s_p-2\s_\p=\ano+Y+X
\ena
$X$ has to be written in terms of the $\b$-functions and operators of the SCGT and matter theory.
Taking the same ansatz for the pressures as for the non interacting conformal theory 
\bea
\s_\p&=&\O\s_p \quad (\o^{-1}\equiv3+2\O)\non\\
p&=&w\rho,\qquad \p=w_\p\rho
\ena
and using the anomaly equation to eliminate $\s_p=\o\left(\s-\ano-Y-X\right)$ from the set of remaining equations, we get
\bea
M_{Pl}^4\left(3H^2+6H\,F+F^2+3\frac{k}{a^2}+\frac{\k}{b^2}\right)&=&\rho+\s+\l  \label{general H ODE 1}\\
M_{Pl}^4\left(2\dot H+3H^2+4H\,F+2\dot F+3F^2+\frac{k}{a^2}+\frac{\k}{b^2}\right)&=&-w\rho-\o\s+\l+\o\left(\ano+Y+X\right)  \non
\ena\bea
\dot\s+\left[3(1+\o)H+3(1-\o)F\right]\s&=&\left[3\o H+(1-3\o)F\right]\left(\ano+Y+X\right)+T  \label{general sigma ODE}\\
\dot\rho+\left[3(1+w)H+2(1-w_\p)F\right]\rho&=&-T  \label{general rho ODE}
\ena

The cosmological evolution described by these differential equations that include the non conformality (represented by the $X$ term) and the matter/hidden interactions (related to the $T$ term) could now be
investigated. In the spirit of the AdS/CFT correspondence, the CFT generalization amounts to introducing non trivial dynamics in the bulk and brane--bulk energy exchange (and bulk self--interaction) in the 7D
picture.
%%%%%%%%%%%%%%%%%%%%%%%%%%%%%%%%%%%%%%%%%%%%%%%%%%%%%%%%%%%%%%%%%%%%%%%%%%%%%%%%%%%%%%%%%%%%%%%%%%%%%%%%%%%%%%%%%%%%%%%%%%%%%%%%%%%%%
%%%%%%%%%%%%%%%%%%%%%%%%%%%%%%%%%%%%%%%%%%%%%%%%%%% fixed points and stability %%%%%%%%%%%%%%%%%%%%%%%%%%%%%%%%%%%%%%%%%%%%%%%%%%%%%%
%%%%%%%%%%%%%%%%%%%%%%%%%%%%%%%%%%%%%%%%%%%%%%%%%%%%%%%%%%%%%%%%%%%%%%%%%%%%%%%%%%%%%%%%%%%%%%%%%%%%%%%%%%%%%%%%%%%%%%%%%%%%%%%%%%%%%
\subsection{Critical points and stability}
The fixed points can be derived as I have done in appendix \ref{fixed points appendix} for the conformal non interacting theory and their stability can then be studied for specific theories. I won't discuss this
topic here. Since the new  deformation parameters $X,T$ depend on the 6D space--time curvature, they contain functions of the Hubble parameters and spatial curvatures and of the intrinsic energy scale of
the background, the $AdS_7$ radius (or $M_{Pl}$). They will thus in general modify the equations for the fixed points and their stability in a sensible way, depending on the specific generalization one wants
to consider. 

I will instead try to understand how the comparison with the bulk description gets changed when we go to the generalized scenario. This will be the subject of the next subsection.
%%%%%%%%%%%%%%%%%%%%%%%%%%%%%%%%%%%%%%%%%%%%%%%%%%%%%%%%%%%%%%%%%%%%%%%%%%%%%%%%%%%%%%%%%%%%%%%%%%%%%%%%%%%%%%%%%%%%%%%%%%%%%%%%%%%%%
%%%%%%%%%%%%%%%%%%%%%%%%%%%%%%%%%%%%%%%%%%%%%%% generalized correspondence %%%%%%%%%%%%%%%%%%%%%%%%%%%%%%%%%%%%%%%%%%%%%%%%%%%%%%%%%%
%%%%%%%%%%%%%%%%%%%%%%%%%%%%%%%%%%%%%%%%%%%%%%%%%%%%%%%%%%%%%%%%%%%%%%%%%%%%%%%%%%%%%%%%%%%%%%%%%%%%%%%%%%%%%%%%%%%%%%%%%%%%%%%%%%%%%
%\subsection{Comparison to 7D cosmology with energy exchange in slowly scaling regime}
\subsection{Correspondence: interactions and conformal breaking vs. energy exchange and bulk self--interaction}
As for $T=X=0$, we will make some assumptions simplifying the set of equations including Einstein, conservation and anomaly equations, with the aim of understanding some peculiar features of this cosmological
model and its two dual descriptions. First of all, we are going to neglect terms containing higher orders in the background curvature --- namely the anomaly and the trace contribution $Y$ coming from the
second order action $S_{R^2}$. 
\paragraph{Equal scale factors}
The correspondence works as in the conformal non interacting analysis of subsection \ref{equal comparison}. On the brane side (referring to eqs \refeq{general H ODE 1}--\refeq{general rho ODE}), the slowly
scaling approximation leads to the equations
\bea
H^2+\oneover{10}\left(3\frac{k}{a^2}+\frac{\k}{a^2}\right)&=&\oneover{10\Mpl}\left(\rho+\chi\right)+\oneover{10\Mpl}\l\label{H 6D equal linear gen}\\
\dot\chi+6H\chi&=&HX+T\\
\dot\rho+(3(1+w)+2(1+w_\p))H\rho&=&-T  \label{rho 6D equal linear gen}
\ena
To get the bulk description expressions for $H,\rho,\chi$ we truncate equations \refeq{bulk limits H}--\refeq{bulk limits rho} to first order in the density, neglecting $\rho/V$ w.r.t. order 1 terms 
\bea\label{H 7D equal linear gen}
H^2+\oneover{10}\left(3\frac{k}{a^2}+\frac{\k}{a^2}\right)&=&\frac{2\tcveq V}{5M^{10}}\left(\rho+\chi\right)+\l\\
\dot\chi+6H\chi&=&2T^0_7-\frac{40M^5}{V}HT^7_7\\
\dot\rho+(3(1+w)+2(1+w_\p))H\rho&=&-2T^0_7
\ena

The matching with the system of equations on the brane \refeq{H 6D equal linear gen}--\refeq{rho 6D equal linear gen} is exact if we have the following relations among the brane and bulk parameters
\bea
\Mpl&=&\frac{M^{10}}{2\tcveq V}  \quad\stackrel{w_\p=w}{\longrightarrow}\quad  \Mpl=20\frac{M^{10}}{V}\\
T&=&2T^0_7\\
X&=&-\frac{M^5}{2\tcveq V}T^7_7  \quad\Longrightarrow\quad  \frac{X}{\Mpl}=-\frac{T^7_7}{M^{5}} 
\ena
In the previous equations we have also explicitely evaluated the matching for $w_\p=w$, that gives the RS fine--tuning $\l_{RS}=0$.
\paragraph{Static compact extra dimensions}
The condition of static internal space $F=0$, together with the small Hubble parameter approximation, brings equations \refeq{general H ODE 1}--\refeq{general rho ODE}, for the brane description, and
equations \refeq{bulk limits H}--\refeq{bulk limits rho}, for the bulk description. Indeed, we get respectively
\bea
H^2+\oneover{3}\frac{\k}{b_0^2}&=&\oneover{3\Mpl}\left(\rho+\chi\right)\label{H 6D static linear gen}\\
\dot\chi+4H\chi&=&HX+T\\
\dot\rho+3(1+w)H\rho&=&-T
\ena
(where we assumed $\o=1/3$) and
\bea\label{H 7D static linear gen}
H^2+\oneover{6}\frac{\k}{b_0^2}&=&\frac{2\tcvst V}{3M^{10}}\left(\rho+\chi\right)\\
\dot\chi+4H\chi&=&2T^0_7-\frac{40M^5}{V}HT^7_7\\
\dot\rho+3(w+1)H\rho&=&-2T^0_7
\ena

The parameters in the gauge and gravity descriptions are thus related by the following expressions
\bea
\Mpl&=&\frac{M^{10}}{4\tcvst V}  \quad\stackrel{w_\p=\frac{w+5}{6}}{\longrightarrow}\quad  \Mpl=20\frac{M^{10}}{V}  \label{matching M equal general linear}\\
T&=&2T^0_7\\
X&=&-\frac{M^5}{2\tcvst V}T^7_7  \quad\Longrightarrow\quad  \frac{X}{\Mpl}=-2\frac{T^7_7}{M^{5}}  \label{matching X static general linear}
\ena
For $w_\p=(w+5)/6$ we get the zero effective cosmological constant on the RS brane.
\paragraph{Proportional Hubble parameters}
In the limit of proportional Hubble parameters or equivalently scale factors related by $b=a^{\x_b}$, we use the set of equations \refeq{general H ODE 1}--\refeq{general rho ODE}, substituting $F=\x_bH$ and
$\k\to\k_b$, and expanding in the slowly scaling approximation
\bea
H^2+\oneover{(\x_b^2+6\x_b+3)}\frac{\k_b}{b^2}&=&\oneover{(\x_b^2+6\x_b+3)\Mpl}\left(\rho+\chi\right)\label{H 6D prop linear gen}\\
\dot\chi+d_{\x_b} H\chi&=&\left(3\o+\x_b(1-3\o)\right)HX+T\\
\dot\rho+w_{\x_b} H\rho&=&-T
\ena
As before, $d_{\x_b}\equiv3(1+\o)+3\x_b(1-\o)$, $w_{\x_b}\equiv3(1+w)+2\x_b(1+w_\p)$. The bulk dynamics is described by \refeq{bulk limits H prop}--\refeq{bulk limits rho prop} with $F=\x_BH$ and $\k\to\k_B$
\bea\label{H 7D prop linear gen}
H^2+\oneover{(\x_B^2+3\x_B+6)}\frac{\k_B}{b^2}&=&\frac{2\tcvx V}{(2\x_B+3)M^{10}}\left(\rho+\chi\right)\\
\dot\chi+d_{\x_B} H\chi&=&2T^0_7-\frac{40M^5}{V}HT^7_7\\
\dot\rho+w_{\x_B} H\rho&=&-2T^0_7
\ena
with $d_{\x_B}\equiv6(\x_B^2+2\x_B+2)/(2\x_B+3)$, $w_{\x_B}\equiv3(1+w)+2\x_B(1+w_\p)$.

If $w$ and $w_\p$ are the same on both sides of the duality, then we must make the identification $\x_b=\x_B=\x$ to match $\rho$ equations. As a consequence $\o=1/(2\x+3)$ in order to have agreement for the
mirage density conservation equations and $\k_b=(\x^2+6\x+3)\k_B/(\x^2+3\x+6)$. With these conditions, the comparison between the two sets of equations thus gives the following matching relations
\bea
\Mpl&=&\frac{2\x+3}{\x^2+6\x+3}\frac{M^{10}}{2\tcvx V}  \\%\quad\stackrel{w_\p=\frac{w+5}{6}}{\longrightarrow}\quad  \Mpl=20\frac{M^{10}}{V}  \label{matching M equal general linear}\\
T&=&2T^0_7\\
X&=&-\frac{2\x+3}{2\x^2+3}\frac{M^5}{2\tcvx V}T^7_7  \quad\Longrightarrow\quad  \frac{X}{\Mpl}=-\frac{\x^2+6\x+3}{2\x^2+3}\frac{T^7_7}{M^{5}}  \label{matching X prop general linear}
\ena
%%%%%%%%%%%%%%%%%%%%%%%%%%%%%%%%%%%%%%%%%%%%%%%%%%%%%%%%%%%%%%%%%%%%%%%%%%%%%%%%%%%%%%%%%%%%%%%%%%%%%%%%%%%%%%%%%%%%%%%%%%%%%%%%%%%%%
%%%%%%%%%%%%%%%%%%%%%%%%%%%%%%%%%%%%%%%%%%%%%%%%%%%%% summary %%%%%%%%%%%%%%%%%%%%%%%%%%%%%%%%%%%%%%%%%%%%%%%%%%%%%%%%%%%%%%%%%%%%%%%
%%%%%%%%%%%%%%%%%%%%%%%%%%%%%%%%%%%%%%%%%%%%%%%%%%%%%%%%%%%%%%%%%%%%%%%%%%%%%%%%%%%%%%%%%%%%%%%%%%%%%%%%%%%%%%%%%%%%%%%%%%%%%%%%%%%%%
\sectioncount{Remarks on the 7D RS holographic description}\label{summary}
In the context of holographic cosmology, I have investigated the specific background of 7D RS gravity, including an energy exchange interaction between brane and bulk. Some novel features arise both on
the bulk side of the duality and in the conformal holographic theory on the brane. In particular, I found distinctive results in the comparison between the two descriptions that need a better understanding.
The originality with respect to the 5D/4D holographic cosmology \cite{Kiritsis:2002zf,Kiritsis:2005bm} is due to the compactification over a 2D internal space, around which we warp the 5--brane. The 6D
space--time filled by the brane acquires an inhomogeneity that distinguishes the 3D visible space from the 2D internal directions. Evolution can generally be different in the two spaces and pressures are
individually related to the energy density by the usual ansatz $p=w\rho$ and $\p=w_\p\rho$ ($p$ and $\p$ are respectively the 3D and 2D pressures), with $w_\p\neq w$ in general.

The rather detailed study of the various cosmologies emerging from the 7D RS model with energy exchange has been embodied in the context of the AdS/CFT correspondence. I have examined the role played in the
holographic critical point analysis by the 6D anomalous CFT coupled to 6D gravity --- with the addition of the higher order counterterms to the dual action. The dual theory on the brane is conformal
(classically) and non interacting (with the matter theory on the brane). This CFT would then correspond to the RS set--up with no energy exchange. Despite this fact, I may find inflationary critical point
solutions, depending on the anomaly parameters and on the coefficient of the second order counterterm (in the curvature). All this fixed points are characterized by zero matter energy density. Clearly,
neglecting all the higher order contributions (including the anomaly $\ano$ and the trace term $Y$) I recover the trivial fixed points of the pure RS gravity background.

The comparison between the two dual descriptions has been achieved in the approximation where all the higher order terms can be neglected, i.e. for small Hubble parameters. Since higher order terms are
truncated, we cannot access to the typical non conventional $\rho^2$ dependence in the expression for $H^2$ --- only linear terms are present in this approximation. 

Comparing the bulk Friedmann equation with the corresponding equation derived in the holographic description, we have to match the ratio $M^{10}/V$ ($M$ is the 7D Planck Mass and $V$ is the tension of the RS
brane) in terms of the 6D Planck mass $\Mpl$ (the 4D Planck mass $M_{(4)}$ is related to $M_{Pl}$ by $M_{(4)}^2=V_{(2)}\Mpl$). The RS fine--tuning $\Mpl=20M^{10}/V$ is restored when I recover homogeneity in
the background, imposing $F=H$ and $w_\p=w$. With these assumptions indeed, the matching is exact also with respect to the spatial curvature terms. As I move away from homogeneity, I have to define an
effective spatial curvature for the compact extra dimensions in the holographic description. The matching between the scales of the theories reflects the RS fine--tuning for a specific value of the matter
pressure in the internal space (determined by $w_\p$), depending on the proportionality factor $\x$ relating the two Hubble parameters $F(t)=\x H(t)$, or $b(t)=a^\x(t)$.  

I finally matched the evolution equations in the generalized holographic dual theory with the general bulk description. The interactions between hidden and visible sectors encodes the dynamics of the
brane--bulk energy exchange $T^0_7\ne0$ on the bulk gravity side while the breaking of conformal invariance (via non zero $\b$--functions or masses) amounts to turning on the bulk ``self--interaction''
$T^7_7\ne0$. 

In my paper \cite{Mazzanti:2007dq}, the 7D RS background has been quite accurately studied on the bulk side, though many profound cosmological aspects have not been explored. The holographic dual theory could
also give an interesting cosmological description of the brane--world. It would be interesting to exploit Starobinsky argument of graceful exit from primordial inflation via higher derivatives term in this
context.

\addcontentsline{toc}{chapter}{Summary and further considerations}
\chapter*{Summary and conclusive remarks}
\fancyhf{}
\fancyhead[RO]{\bf{Summary and conclusive remarks}}
\fancyhead[LE]{\bf{Summary and conclusive remarks}}
\fancyfoot[RO]{\thepage}
\fancyfoot[LE]{\thepage}
In this thesis I tackled two significant issues related to string theory. Noncommutative field theories, appearing in the low energy description of suitable string backgrounds, are the subject of the first
part of my work. The second part of my Ph.D. project deals with brane--world cosmology and holography, motivated by the demand of embedding cosmological data into the string framework.

% subject papers I,II
Noncommutative field theories are shown to exhibit special features, both at classical and quantum level. It is particularly interesting to investigate the faith of symmetries through generalization to
noncommutative geometry. In my two publications \cite{GMPT,LMPPT}, integrability of the sine--Gordon quantum field theory was considered. Integrable theories are of great interest, since they display nice
properties of the S--matrix that I discussed. Furthermore the S--matrix can be explicitely calculated in some cases --- in sine--Gordon model, for example. Quantum statistics applications are also very well
known, such as spin model analysis. Moreover, these remarkable theories play a rather relevant role in string theory, since they often arise as effective theories describing open string dynamics on branes.
In this context, their noncommutative generalization --- arising in noncommutative string backgrounds --- is an intriguing question that I addressed.

% bicomplex
Integrability is the consequence of the existence of infinitely many conserved currents in the classical theory. Thus, one may wish to deform a theory in such a way that conservation of these local currents
is automatically preserved. The existence of infinite deformed conserved currents is guaranteed if the noncommutativity parameter is introduced in the reformulation of the gauged bicomplex. Indeed, the
equations of motion for an integrable theory can be obtained as the compatibility conditions of a linear differential system. An infinite chain of differential one forms can then be constructed and yield the
conserved currents. Ordinary (euclidean) sine--Gordon follows from a specific matrix valued differential system, which produces a noncommutative version of the model upon the implementation of noncommutative
geometry in the matrix equation. Deformed currents are still conserved, so that the theory is formally classically integrable. The model derived by means of this procedure has been considered in my paper
\cite{GMPT}, written in collaboration with M.~Grisaru, S.~Penati and L.~Tamassia.

% S--matrix
However, a theorem by Zamolodchikov and Zamolodchikov states that integrability in two dimensions is equivalent to the introduction of restrictions on the S--matrix. Roughly, S--matrix is factorizable in two
particle scattering processes while production processes are absent. Despite the existence of the infinite number of conserved currents, the deformed theory generally do not exhibit the nice properties of the
S--matrix, expected in integrable models. This is indeed what happens in the aforementioned noncommutative sine--Gordon theory in \cite{GMPT}. We calculated the scattering amplitudes at tree level for
$2\to2$, $2\to3$ and $2\to4$ processes.  The results yield non zero value for production amplitudes, showing that Zamolodchikovs' theorem doesn't generally hold in noncommutative set--up. Moreover, we found
that causality was violated too, as it is usual in noncommutative field theories. In fact, causality and factorization of the S--matrix are two related issues, in the sense that the first condition seems to
be necessary in order for the second property to be satisfied.  We thus argued that an eventual causal noncommutative version of sine--Gordon could also be integrable, with respect to the full notion of
integrability implying S--matrix factorization.

% YM and bosonization
Moreover, the euclidean noncommutative sine--Gordon equations of motion construct\-ted in \cite{GP} can be traced back to four dimensional SDYM, via dimensional reduction. Nevertheless, the associated action
derived in \cite{GMPT} cannot be obtained by the same procedure, i.e. the dimensional reduction from SDYM doesn't work at the level of the action. On the other hand, we showed how this noncommutative
sine--Gordon model can be interpreted as the bosonization of noncommutative abelian massive Thirring model, naturally generalized to NC geometry. The abelianity is due to the fact that our sine--Gordon
contains two copies of the same WZW action plus the self--interaction for two fields $g$ and $\bar g$, belonging to the complexified $U(1)$. The appearance of two fields rather that one comes from the
noncommutative group structure. We deduced the equations of motion from a linear system defined in terms of a $U(2)$ gauge connection, which is the obliged generalization from the $SU(2)$ connection in
ordinary geometry --- since $SU(2)$ is no more closed under noncommutative product. Hence, an additional $U(1)$ factor pops out, so that we get two independent equations of motion and two abelian fields.

% groups
Since the results were not satisfactory in what concerns S--matrix properties, we were induced to consider a noncommutative generalization of sine--Gordon, with a different reduction from the stringy SDYM.
Indeed, we proposed in \cite{LMPPT} --- together with O.~Lechtenfeld, A.~D.~Popov, S.~Penati and L.~Tamassia --- a model which is a dimensional reduction from integrable noncommutative $(2+2)$--dimensional
SDYM. ${\cal N}=2$ superstrings, with $N$ coincident space filling D3--branes and constant NS--NS two form, are described at tree level by noncommutative $U(N)$ SDYM in four dimensions with signature
$(+,+,-,-)$. SDYM is known to be integrable also in its noncommutative version, since this can be proven by the vanishing of all amplitudes beyond three point functions. A first dimensional reduction yields
an intermediate $(2+1)$--dimensional modified sigma model. Further reduction leads to our integrable $(1+1)$--dimensional sine--Gordon model. However, this is done in two steps, first factorizing the
dependence on the extra coordinate and then restricting the form of the $U(2)$ fields appearing in the linear system. The second operation is crucial and causes the interaction terms in the action to mix the
two $U(1)$ fields $g_+$ and $g_-$. Indeed, the $U(2)$ group is broken to its diagonal $U(1)\times U(1)$ subgroup, which seems the most natural assumption.

% results II
Results are surprising, since S--matrix looks factorizable and causal. We computed all the $2\to2$, $2\to3$, $3\to3$ and $2\to4$ amplitudes at tree level, both in the Leznov and Yang gauges. Particle
production is absent and $2\to2$ processes are causal. Besides, Leznov formulation gives all zero amplitudes, in agreement with properties of SDYM in the Leznov gauge, coming from superstrings. I emphasize
that dimensional reduction works also at the level of the action, in this model. 

% solitons
Due to the mixed interaction term in the action, it is not trivial to relate the noncommutative integrable sine--Gordon to some fermionic noncommutative $U(1)\times U(1)$ Thirring--like model via
bosonization.  Indeed, this is still an open issue. However, an additional result we obtained is the construction of noncommutative multi--soliton solutions, which is achieved by means of the dressing method
in the gauged bicomplex approach --- again through dimensional reduction from the $(2+1)$--dimensional non linear sigma model.  \vspace{0.5cm}

% subject of paper III
I then moved to a more phenomenological application of string theory. This has to do with the idea that our world is somehow represented by an effectively four dimensional defect in ten dimensional
space--time, described by strings degrees of freedom (or eleven dimensional space--time if we consider M theory). The defect is naturally identified in string theory with a brane --- or a stack of coincident
branes, or else appropriate intersections of branes. The concept of brane--world solves the problem of finding an explanation to the number of visible dimensions (four) in comparison to the target space
dimension (ten) where critical superstrings are embedded. However, since gravity is free to propagate in the whole space--time, one still has to find a way to hide the extra dimensions, which are not detected
by experiments. This was usually achieved by compactification of extra dimensions. 

% RS
On the other hand, RS set--up leads to gravity localization on the brane, without the need of compactifying. Indeed, the massless graviton mode rapidly falls off as the distance from the brane increases.
Furthermore, massive modes are also localized and give negligible contributions to the gravitational potential at large distances (larger than the RS scale, defined in terms of the brane tension or,
equivalently, of the bulk cosmological constant). For small energies, Newton's law is thus recovered on the four dimensional brane in five dimensional bulk with warped extra dimension. The warp factor of RS
geometry is responsible for the localization of gravity and allows for an infinite volume extra direction. I indeed exploited the crucial characteristics of RS models in a higher dimensional set--up --- in
which compactification over an internal manifold is also involved --- in order to derive new cosmological evolution features in this context.

% cosmology
On general grounds, cosmology in RS brane--worlds appears to be non conventional, in the sense that standard Friedmann equations are not obtained. Evolution is hence not described by the usual formalism.
However, conventional cosmology is restored in the limit of small matter energy densities, at late times.  This is an interesting issue if we try to deal with the open questions of cosmology, as for instance
present time acceleration data, primordial inflation, dark energy.  Moreover, one can introduce an energy exchange between brane and bulk, yielding more intriguing and complicated cosmologies. These can
display non conventional behavior even at late times, hopefully matching some aspects of the observed cosmological evolution. The cosmological analysis with brane--bulk energy exchange has been first
performed in \cite{Kiritsis:2002zf,Kiritsis:2005bm} for the 5D RS model.

% results paper III/a
I proposed in my paper \cite{Mazzanti:2007dq} an analogous brane--world analysis for the 7D RS model with non zero energy exchange. Due to the two additional extra dimensions, a further compactification over
a two dimensional internal manifold was necessary in order to get effective four dimensional gravity at low energies. Two different scale factors $H$ and $F$ were introduced for the 3D and 2D spaces,
respectively.  I derived the Friedmann--like equations from Einstein equations by defining a mirage density through the associated differential equation. The mirage density encodes the bulk dynamics. It
reduces to 6D and 4D radiation in the equal scale factor $H=F$ and static extra dimensions $F=0$ background, respectively. However, the effective 4D mirage density is 4D free radiation in the RS vacuum for
both these configurations. 

I then considered the cosmological evolution of equal scale factors and static extra dimensions universes, showing interesting consequences of brane--bulk energy exchange in this model. In particular, when
influx is considered (energy flowing from the bulk onto the brane), I found stable accelerating fixed point solutions that can be interpreted as the present accelerated era. Scenarios can be roughly depicted,
where the evolution trajectory initially pass through an accelerated phase, to be identified with primordial inflation, starts decelerating after a change of regime from outflow to influx, and finally ends
into the inflationary critical point. The matching of the cosmological analysis to observational data would need a more rigorous formulation, indeed. I have moreover remarked the new features with respect to
the 5D model. For instance, the outflow case exhibits more complicated results, since it is not necessary for the 4D matter energy density to be attracted by the trivial fixed point, characterized by
vanishing acceleration and density. 

An explicit solution in the radiation dominated universe was found. It showed that if a particular linear ansatz on the outflowing energy exchange parameter is considered, the brane dynamics is dominated by
mirage radiation at late times, while matter radiation prevails at early times. The whole cosmological analysis is strongly dependent on the form of the energy exchange. It would be important to understand
more clearly how to eventually embed a realistic description of our universe in this context.

% AdS/CFT
It have already noted that RS brane somehow plays the role of a IR cutoff for gravity in the bulk. Moreover, since the RS static bulk is a slice of AdS space, it is natural to look for a holographic dual
description of RS models invoking the AdS/CFT correspondence. Holographic renormalization is a rigorous formulation of gauge/gravity duality which carefully cures the divergences that plague the two sides of
the duality. First of all, a regularization of the supergravity theory is needed. In RS models this is indeed provided by the cutoff brane. Then, AdS/CFT teaches us that the IR cutoff on the supergravity side
corresponds to a UV cutoff for the CFT. Hence, one obtains an holographic cutoff conformal theory. Besides, gravity gets renormalized with the addition of covariant boundary counterterms that appear in the
renormalized dual theory as the lower dimensional gravity plus higher order corrections.

% (holographic) cosmology
I constructed the holographic dual theory corresponding to the 7D RS model, using the holographic renormalization prescription specified to seven dimensions. The result is a 6D CFT coupled to 6D gravity with
higher derivative corrections. The Friedmann--like equations were calculated, together with conservation and anomaly equations. The mirage energy density is related to the energy density parametrizing the
stress--energy tensor contribution from CFT and higher order corrections, which are quadratic and cubic in the curvature. The anomaly is also cubic in the curvature and yields a very complicated expression in
terms of the Hubble parameters. A distinctive feature of this 7D/6D model is that the higher derivative terms contained in the anomaly may drive an exit from inflation as described in the Starobinsky model.
The absence of these contributions in the 5D/4D set--up is due to the fact that they are cancelled from the highest order counterterm appearing in the holographic action. In six dimensions some of the higher
derivative terms survive, since they are not uniquely contained in the scheme dependent part of the anomaly, which always gets cancelled. 

I performed the critical point analysis in equal scale factors and static extra dimensions backgrounds, finding non trivial inflationary points. However, an analytical expression of $H$ in terms of matter and
mirage energy densities has been obtained only working in the approximation where higher order terms are neglected. The Friedmann--like equation still displays non conventional behaviors, which have been
compared to the expression obtained in the 7D picture. Indeed, matching the dimensionful parameters yields the RS fine--tuning that implies zero effective cosmological constant on the brane. This result is
achieved in a homogeneous background, where the scale factors and pressures of the 3D and 2D spaces are equal. Deviations from this configuration lead to the generation of a non zero effective cosmological
constant on the brane.

% results paper III/b
Further insights about the duality ensue from the generalization of the field theory to a non conformal interacting strongly coupled gauge theory. Indeed, while the conformal theory corresponds to 7D gravity
bulk with no brane--bulk energy exchange, nor bulk self--interaction (which has been neglected in the cosmological analysis and represents the bulk diagonal component of stress--energy tensor associated to
matter in the bulk), its generalization holographically describes these two objects. More explicitely, we introduced a term in the trace equation representing the classical and quantum breaking of conformal
invariance. Conversely, non conservation of matter and hidden sector stress--energy tensors separately is related to the matter/hidden interactions. This is encoded in an additional factor appearing in the
conservation equations. As a result from the comparison of the two sides, I obtained that brane--bulk energy exchange and bulk self--interaction correspond through proportionality relations to the
matter/hidden interaction term and conformal breaking term, respectively.  Analogous relations appear in the 5D/4D set--up. This is an expected result, since the hidden sector in the holographic description
is associated to the bulk dynamics on the 7D gravity side. Hence, interactions between the visible and hidden fields are dual to the energy exchange between brane and bulk. Deformations of the hidden sector
theory instead amount to introducing some bulk self--interactions.

% conclusions on AdS/CFT and brane--worlds
The framework of AdS/CFT and brane--worlds offers a wide range of phenomenologically interesting subjects. I here considered a cosmological model leading to a rich description of the universe, eventually
giving some hints to solve open questions in cosmology --- from the holographic point of view as well. However, this model is not rigorously embedded in a string theory background. The first question to
address would then be how such a cosmological description can arise in string theory. 
%
%Nevertheless, another issue immediately breaks through, which I didn't discuss in my work: how do the matter and conformal field theories have to look like? Indeed, this opens an issue related to a great
%challenging branch in strings and holography, i.e. how to realize the gauge theories that we know to describe observed interactions.  
\vspace{0.5cm}

% conclusions of conclusions
To conclude, this thesis contains the detailed study of two distinct topics sharing the attribute of being recent applications of string theory. I investigated the properties of noncommutative field theories,
focusing on integrability, which is a field of great interest in string theory, as well as in AdS/CFT. My results show that sine--Gordon theory seems to be generalizable to noncommutative geometry preserving
integrability (at tree level). Hence, there is hope that integrable field theories coming from strings background inducing noncommutativity bear the same nice properties of the S--matrix as ordinary
integrable field theories do. Nonetheless, it is a challenge to determine whether noncommutative geometry arises at high energies as a quantum deformation of ordinary space--time.

I then began the analysis of a cosmological brane--world model \`a la Randall--Sundrum, which can be traced back to M5--branes in eleven dimensional M theory. Interesting results come both from the
cosmological analysis in the 7D set--up and from the construction and comparison of the 6D holographic dual theory. Cosmological scenarios are depicted, which give hints about possible descriptions of our
universe evolution. The matching of the two dual theories yields further intuitions on the AdS/CFT correspondence for RS brane--worlds with energy exchange. However, I walked the firsts steps in the holography
and cosmology of 7D RS, but a deeper understanding of the model is an open issue.

\addcontentsline{toc}{chapter}{Appendices}
\chapter*{Appendices}
\fancyhf{}
\fancyfoot[RO]{\thepage}
\fancyfoot[LE]{\thepage}
\appendix
\chapter{Hographic Weyl anomaly and critical point analysis}\label{appA}
\fancyhf{}
\fancyhead[RO]{\rightmark}
\fancyhead[LE]{\leftmark}
\fancyfoot[RO]{\thepage}
\fancyfoot[LE]{\thepage}
%%%%%%%%%%%%%%%%%%%%%%%%%%%%%%%%%%%%%%%%%%%%%%%%%%%%%%%%%%%%%%%%%%%%%%%%%%%%%%%%%%%%%%%%%%%%%%%%%%%%%%%%%%%%%%%%%%%%%%%%%%%%%%%%%%%%%%%%%%%%%
%%%%%%%%%%%%%%%%%%%%%%%%%%%%%%%%%%%%%%%%%%%%%%%%%%%%%%%%%%%%%% evaluation of the anomaly %%%%%%%%%%%%%%%%%%%%%%%%%%%%%%%%%%%%%%%%%%%%%%%%%%%%
%%%%%%%%%%%%%%%%%%%%%%%%%%%%%%%%%%%%%%%%%%%%%%%%%%%%%%%%%%%%%%%%%%%%%%%%%%%%%%%%%%%%%%%%%%%%%%%%%%%%%%%%%%%%%%%%%%%%%%%%%%%%%%%%%%%%%%%%%%%%%
\sectioncount{Conformal anomaly and traces in six dimensions}\label{anomaly appendix}
\paragraph{Conformal anomaly} The conformal anomaly for 6D theories has been studied in \cite{Deser:1993yx}. It can be derived using AdS/CFT and the gravitational renormalization procedure as in
\cite{Henningson:1998gx,Skenderis:1999nb,deHaro:2000xn}.

In our notations, the general expression for the trace anomaly in a six--dimensional CFT is 
\bea\label{gen anomaly} 
\ano=-\left(c_A E_{(6)}+c_B I_{(6)}+ \nabla_\m J^\m_{(5)}\right)
\ena 
$E_{(6)}$ is the Euler density in six dimensions (type A anomaly), $I_{(6)}$ is a fixed linear combination of three independent Weyl invariants of dimension six (type B anomaly) and $\nabla_iJ^\m_{(5)}$ is an
linear combinations of the Weyl variation of six independent local functionals (type D anomaly), so that at the end we have eight free coefficients in the general form of the anomaly depending on the specific
CFT. The type D anomaly is a trivial (it is a total derivative, indeed) scheme dependent term that can be cancelled by adding local covariant counterterms to the action \cite{Bastianelli:2000rs}. 

For our metric we obtain as a result that $E_{(6)}$ depends on the Hubble parameters and on their time derivatives up to order one (that is, up to the second time derivative of the scale factors). $I_{(6)}$
instead depends on $H$ and $F$ time derivatives up to order three, and so does the divergence term. To be more specific, $I_{(6)}$ is made up by three contributions $I_1,I_2,I_3$, with fixed coefficients; the
first two are two different contractions of three Weyl tensors (and contain only derivatives of the Hubble parameters up to order one), while in $I_3$ there are second order derivatives of the Weyl tensor
(i.e. third order time derivatives of the Hubble parameters). 

For a 6D FRW background the sum of type A plus type B anomalies depends on the Hubble parameters only up to the first time derivative, while the type D anomaly contains time derivatives up to order three.

In terms of the Riemann tensor, Ricci tensor and scalar curvature, the A, B and D contributions to the anomaly read \cite{Henningson:1998gx}
\bea
E_{(6)}&=&\frac{1}{6912}E_0\non\\
I_{(6)}&=&\frac{1}{1152}\left(-\frac{10}{3}I_1-\oneover{6}I_2+\oneover{10}I_3\right)\non\\
J^\m_{(5)}&=&-\frac{1}{1152}\left[-R^{\m\n\rho\s}\nabla^\t R_{\t\n\rho\s}+2\left(R_{\n\rho}\nabla^\m R^{\n\rho}-R_{\n\rho}\nabla^\n R^{\m\rho}\right)\right]+\non\\
&&-\oneover{2880}R^{\m\n}\nabla_\n R+\oneover{5760}R\nabla^\m R
\ena
where
\bea
E_0&=&K_1-12K_2+3K_3+16K_4-24K_5-24K_6+4K_7+8K_8\non\\
I_1&=&\frac{19}{800}K_1-\frac{57}{160}K_2+\frac{3}{40}K_3+\frac{7}{16}K_4-\frac{9}{8}K_5-\frac{3}{4}K_6+K_8\non\\
I_2&=&\frac{9}{200}K_1-\frac{27}{40}K_2+\frac{3}{10}K_3+\frac{5}{4}K_4-\frac{3}{2}K_5-3K_6+K_7\non\\
I_3&=&K_1-8K_2-2K_3+10K_4-10K_5-\frac{1}{2}K_9+5K_{10}-5K_{11}\non
\ena
and
\bea
(K_1,\dots,K_{11})&=&\left(R^3,RR_{\m\n}R^{\m\n},RR_{\m\n\rho\s}R^{\m\n\rho\s},R_\m{}^\n R_\n{}^\rho R_\rho{}^\m,R^{\m\n}R^{\rho\s}R_{\m\rho\s\n},\right.\cr
&&R_{\m\n}R^{\m\rho\s\t}R^\n{}_{\rho\s\t},R_{\m\n\rho\s}R^{\m\n\t\l}R^{\rho\s}{}_{\t n},R_{\m\n\rho\s}R^{\m\t\l\s}R^\n{}_{\t\l}{}^\rho,\cr
&&\left.R\Box R,R_{\m\n}\Box R^{\m\n},R_{\m\n\rho\s}\Box R^{\m\n\rho\s}\right)\non
\ena
In the analysis of the solutions to the Friedmann equations we plug in the specific expression for the Riemann tensor obtained considering our ansatz (\ref{metric 4+2}) for the metric. But before doing
this, we use the anomaly equation and some standard assumptions on the pressures that parametrize the stress--energy tensors to manipulate our system of differential equations.

To give an explicit result for the conformal anomaly in the specific case of 6D CFT on curved space--time, with the ansatz \ref{metric 4+2} for a 4D FRW plus a 2D compact internal space background, we write
the type A contribution, in terms of the 3D and 2D spaces Hubble parameters $H,F$ and spatial curvatures $k,\k$:
\bea
E_{(6)}&=&-\oneover{48}\bigg\{\kb\left(\dot H+H^2\right)\left(H^2+\ka\right)+F^2\left(\dot H+H^2\right)\left(3H^2+\ka\right)+\non\\
&&+2\left(\dot F+F^2\right)\left(H^2+\ka\right)\bigg\}
\ena
The type B and D contributions have a more complicated form and we write them when it is necessary, in the specific limits we consider throughout the paper. 

For the (0,2) SCFT dual to the $N$ M5 background, the anomaly coefficients are given by $c_A=c_B=4N^3/\p^3$ \cite{Henningson:1998gx}.
\paragraph{Counteterm traces}
The dual RS theory action contains the three counterterms $S_1$, $S_2$, $S_3$ written in \refeq{def count 0}. Varying these contributions w.r.t. the six dimensional induced metric $\g_{\m\n}$ on the brane
yields%%%%%%%%%%%%%% footnote
\footnote{Just for this formula we set $\ell=1$ to simplify the notation}%%%%%%%%%%%%%%% footnote
 \cite{deHaro:2000xn}
\bea \label{counterT}
T^{ct}_{\m\n}&=&-2M^5\left(5\g_{\m\n}+\frac{1}{4}\left(R_{\m\n}-\oneover{2}R\g_{\m\n}\right)\right.\non\\ 
&&\left.-\frac{1}{32}\left[-\Box R_{\m\n}+2R_{\m\s\n\rho}R^{\rho\s}+\frac{2}{5}\nabla_\m \nabla_\n R-\frac{3}{5}RR_{\m\n}\right.\right.\non\\
&&\left.\left.-\oneover2\g_{\m\n}\left(R_{\rho\s}R^{\rho\s}-\frac{3}{10}R^2-\frac{1}{5}\Box R\right)\right]-T^a_{\m\n}\log\e\right)
\ena
Where $T^a_{\m\n}$ is a traceless tensor of cubic order in the curvature. The trace of the conformal variation of $S_1$ (corresponding to the linear part of \refeq{counterT} in the curvature) gives a term
proportional to the Einstein tensor. The variation of the $S_3$ action (related to $T^a_{\m\n}$) is traceless because it is proportional \cite{deHaro:2000xn} to the traceless tensor $h_{(6)\m\n}$ that enters
into the parametrization of the metric \refeq{FG para} due to Fefferman and Graham. Finally, the trace of $S_2$ (equal to the trace of the quadratic contributions in \refeq{counterT}) is 
\bea
Y=\oneover{32}M^5\ell\left(R^{\m\n}R_{\m\n}-\frac{3}{10}R^2\right)
\ena
which can be expressed in terms of the Hubble parameters $H,F$ and of the spatial curvatures $k,\k$ of the (4D FRW + 2D compact space) background \refeq{metric 4+2} as
\bea
Y&=&-\frac{2M^5\ell}{160}\bigg\{-3\frac{k^2}{a^4}+6\frac{k}{a^2}\left(3\frac{\k}{b^2}+F^2+8FH+3H^2+6\dot F+4\dot H\right)+\non\\
&&+2\frac{\k}{b^4}+2\frac{\k}{b^2}\left[-\left(F-6H\right)\left(F+3H\right)+\dot F+9\dot H\right]+\non\\
&&-3F^4+48F^3H+6FH\left(21H^2+7\dot F+13\dot H\right)+F^2\left(111H^2-4\dot F+24\dot H\right)+\non\\
&&+3\left[6H^2-\left(\dot F-\dot H\right)^2+2H^2\left(7\dot F+3\dot H\right)\right]\bigg\}
\ena

We define $c_Y\equiv M^5\ell/32\mpl$ which is given as a function of the number $N$ of M5--branes in the gravity background by $c_Y=\sqrt{2N^3/\p^3}$.
%%%%%%%%%%%%%%%%%%%%%%%%%%%%%%%%%%%%%%%%%%%%%%%%%%%%%%%%%%%%%%%%%%%%%%%%%%%%%%%%%%%%%%%%%%%%%%%%%%%%%%%%%%%%%%%%%%%%%%%%%%%%%%%%%%%%%%%%%%%%%
%%%%%%%%%%%%%%%%%%%%%%%%%%%%%%%%%%%%%%%%%%%%%%%%%%%%%%% fixed points %%%%%%%%%%%%%%%%%%%%%%%%%%%%%%%%%%%%%%%%%%%%%%%%%%%%%%%%%%%%%%%%%%%%%%%%
%%%%%%%%%%%%%%%%%%%%%%%%%%%%%%%%%%%%%%%%%%%%%%%%%%%%%%%%%%%%%%%%%%%%%%%%%%%%%%%%%%%%%%%%%%%%%%%%%%%%%%%%%%%%%%%%%%%%%%%%%%%%%%%%%%%%%%%%%%%%%
\sectioncount{Fixed points in the holographic description}\label{fixed points appendix}
In this appendix, we are going to look for the existence of inflationary points for our specific holographic model and to find what kind of dependence they have on the parameters of the theory.  We will also
study the stability matrix determining --- in some special limits --- whether the critical points are stable or saddles.

In the calculations, we suppose that the effective cosmological constant on the brane $\l$ is null.
%%%%%%%%%%%%%%%%%%%%%%%%%%%%%%%%%%%%%%%%%%%%%%%%%%%%%%%%%%%%%%%%%%%%%%%%%%%%%%%%%%%%%%%%%%%%%%%%%%%%%%%%%%%%%%%%%%%%%%%%%%%%%%%%%%%%%%
%%%%%%%%%%%%%%%%%%%%%%%%%%%%%%%%%%%%%%%%%%%%%%%%% flat extra %%%%%%%%%%%%%%%%%%%%%%%%%%%%%%%%%%%%%%%%%%%%%%%%%%%%%%%%%%%%%%%%%%%%%%%%%
%%%%%%%%%%%%%%%%%%%%%%%%%%%%%%%%%%%%%%%%%%%%%%%%%%%%%%%%%%%%%%%%%%%%%%%%%%%%%%%%%%%%%%%%%%%%%%%%%%%%%%%%%%%%%%%%%%%%%%%%%%%%%%%%%%%%%%
\subsection{Flat compact extra dimensions}
We start by considering the case of flat internal space, which could be for example a two--torus. 
The three spatial dimensions of the 4D FRW are already supposed to be flat, so that the system of equations of motion simplify having dropped the terms proportional to the two spatial curvatures.

The general flat extra dimension fixed points ($F_\star\neq0$) are not easy to characterize. We choose to analyze the case in which the extra dimensions Hubble parameter is zero at the fixed point, meaning
that the fixed point represents a universe with static flat extra dimensions.   
%%%%%%%%%%%%%%%%%%%%%%%%%%%%%%%%%%%%%%%%%%%%%%%%%%%%%%%%%%%%%%%%%%%%%%%%%%%%%%%%%%%%%%%%%%%%%%%%%%%%%%%%%%%%%%%%%%%%%%%%%%%%%%%%%
%%%%%%%%%%%%%%%%%%%%%%%%%%%%%%%%%%%%%%%%%%%%%%%%% gen sol %%%%%%%%%%%%%%%%%%%%%%%%%%%%%%%%%%%%%%%%%%%%%%%%%%%%%%%%%%%%%%%%%%%%%%%
%%%%%%%%%%%%%%%%%%%%%%%%%%%%%%%%%%%%%%%%%%%%%%%%%%%%%%%%%%%%%%%%%%%%%%%%%%%%%%%%%%%%%%%%%%%%%%%%%%%%%%%%%%%%%%%%%%%%%%%%%%%%%%%%%
\subsubsection{$F_\star\ne0,\;\o\ne\frac{1}{5}$}
\paragraph{Fixed point solutions}
Looking for the solution of the Friedmann plus conservation set of equations with constant Hubble parameters ($H\equiv H_\star,F\equiv F_\star$) and zero curvatures ($k=\k=0$), we have to
consider the simplified system of equations (where we have already solved the equation for $\s$)
\bea
&&M_{Pl}^4\left(3H_\star^2+6H_\star\,F_\star+F_\star^2\right)-\left[3\o H_\star+\left(1-3\o\right)F_\star\right]\left(\stano+\stY\right)
=\l  \label{Friedmann rho}\\
&&M_{Pl}^4\left(3H_\star^2+4H_\star\,F_\star+3F_\star^2\right)-\o\left(3H_\star+2F_\star\right)\left(\stano+\stY\right)
=\l  \label{Friedmann p}\\
%%\ena
%%\bea
&&\s_\star=\left[3\o H_\star+\left(1-3\o\right)F_\star\right]\stano=\Mpl\left(3\sH^2+6\sH\sF+\sF^2\right)-\l  \label{sigma}\\
&&\s_{p\star}=-\o\left(3H_\star+2F_\star\right)\left(\stano+\stY\right)=-\Mpl\left(3\sH^2+4\sH\sF+3\sF^2\right)+\l  \label{sigma_p}\\
&&\rho_\star=0, \quad \chi_\star=0 \label{rho}
\ena
where the relation between $\O$ and $\o$ was defined to be $1/\o=2\O+3$ and the trace contributions $\sano,\sY$ take the form
\bea
\sano&=&\frac{c_A}{48}\left[2F_\star^3H_\star^3+3F_\star^2H_\star^4\right]+\frac{c_B}{4800}\left[12F_\star^6-128F_\star^5H_\star\right.+\non\\
&&\left.+291F_\star^4H_\star^2+184F_\star^3H_\star^3+557F_\star^2H_\star^4+138FH_\star^5-54H_\star^6\right]\\
\sY&=&\frac{6}{5}c_Y\mpl\left(2\sH^2+2\sH\sF+\sF^2\right)\left(3\sH^2+18\sH\sF-\sF^2\right)
\ena
and $\stano,\stY$ have been defined as 
\bea
\stano&\equiv&\sano/\left[3(1+\o)H_\star+3(1-\o)F_\star\right] \non\\
\stY&\equiv&\sY/\left[3(1+\o)H_\star+3(1-\o)F_\star\right]\non 
\ena
for $(1+\o)H_\star+(1-\o)F_\star\ne0$.

For $F_\star\ne0$ and $\o\ne1/5$, we can reformulate eqs (\ref{Friedmann rho}) and (\ref{Friedmann p}) in order to get 
\bea
&&(3-21\o)H_\star^2+(4-30)F_\star H_\star-(3-11\o)F_\star^2=(1-5\o)\,\frac{\l}{\Mpl}\\
&&\left(\stano+\stY\right)=2\Mpl\,\frac{H_\star-F_\star}{1-5\o}\label{eq ii+ii ano}
\ena
Imposing $\l=0$ (no effective constant on the brane) greatly simplifies the solution since $H_\star\propto F_\star$. Under that assumption, defining $\cep$ and $\dep$ --- as functions of $\o$ and
$\epsilon=\pm1$ ($\dep$ is a function of the anomaly parameters $c_A$ and $c_B$, of $c_Y$ and the Planck mass as well) --- such that $H_\star-F_\star=\cep F$ and $\left(\stano+\stY\right)=\dep F_\star^5$, the
solution takes the form
\bea
H_\star^2=M^2_{Pl}\left(\cep+1\right)^2\left[\frac{2\cep}{(1-5\o)\dep}\right]^{\frac{1}{2}}, \quad\label{sol H F}
F_\star^2=M^2_{Pl}\left[\frac{2\cep}{(1-5\o)\dep}\right]^\oneover{2} 
\ena
This solution exists for the values of $\o$ such that $\cep/\dep>0$ (for $\o<1/5$) or $\cep/\dep<0$ (for $\o>1/5$). 

The CFT energy density and pressures are then given by
\bea
&&\s_\star=M_{Pl}^6\left(1+3\o\cep\right)\dep\left[\frac{2\cep}{(1-5\o)\dep}\right]^\frac{3}{2}\label{sol sigma}\\
&&\s_{p\star}=\frac{2\o}{1-3\o}\s_\p=-M_{Pl}^6\,\o\left(5+3\cep\right)\dep\left[\frac{2\cep}{(1-5\o)\dep}\right]^\frac{3}{2},\quad \srho=0  \label{sol sigma_p}
\ena
(for $\o=1/3$ we have $\s_\p=0$).
%%%%%%%%%%%%%%%%%%%%%%%%%%%%%%%%%%%%%%%%%%%%%%%%%%%%%%%%%%%%%%%%%%%%%%%%%%%%%%%%%%%%%%%%%%%%%%%%%%%%%%%%%%%%%%%%%%%%%%%%%%%%%%%%%%%%%%
%%%%%%%%%%%%%%%%%%%%%%%%%%%%%%%%%%%%%%%%%%%%%%%%% omega=1/5 %%%%%%%%%%%%%%%%%%%%%%%%%%%%%%%%%%%%%%%%%%%%%%%%%%%%%%%%%%%%%%%%%%%%%%%%%%
%%%%%%%%%%%%%%%%%%%%%%%%%%%%%%%%%%%%%%%%%%%%%%%%%%%%%%%%%%%%%%%%%%%%%%%%%%%%%%%%%%%%%%%%%%%%%%%%%%%%%%%%%%%%%%%%%%%%%%%%%%%%%%%%%%%%%%
\subsubsection{$F_\star\ne0,\;\o=\frac{1}{5}$}
\paragraph{Fixed point solutions}
To analyze the case $\o=1/5$, it's better to reformulate equations (\ref{Friedmann rho}) and (\ref{Friedmann p}) in the following way:
\bea
&&2\Mpl \left(H_\star-F_\star\right)F_\star-(1-5\o)F_\star\left(\stano+\stY\right)=0\label{Fried no l}\\
&&M_{Pl}^4\left(3H_\star^2+6H_\star\,F_\star+F_\star^2\right)-\left[3\o H_\star+\left(1-3\o\right)F_\star\right]\left(\stano+\stY\right)=\l \label{Fried l}
\ena 
From the first equation we get $H_\star=F_\star$ and substituing in the second we find the equation for $H_\star$
\bea
-\frac{5}{288}(c_A+2c_B)H_\star^6-120c_Y\mpl \sH^4+10\Mpl H_\star^2=\l
\ena
For $\l=0$ it has a non trivial solution only if $(24c_Y)^2>(c_A+2c_B)$ 
\bea
H_\star^2=F_\star^2=-\frac{24}{c_A+2c_B}\mpl\left[24c_Y\pm\sqrt{576c_Y^2+(c_A+2c_B)}\right]
\ena
The energy density and pressures are equal to
\bea
\s_\star&=&-\ssigp=-\ssigpp=-\frac{240}{c_A+2c_B}\MPl\left[24c_Y\pm\sqrt{576c_Y^2+(c_A+2c_B)}\right] \\
\srho&=&0\non
\ena
%%%%%%%%%%%%%%%%%%%%%%%%%%%%%%%%%%%%%%%%%%%%%%%%%%%%%%%%%%%%%%%%%%%%%%%%%%%%%%%%%%%%%%%%%%%%%%%%%%%%%%%%%%%%%%%%%%%%%%%%%%%%%%%%%
%%%%%%%%%%%%%%%%%%%%%%%%%%%%%%%%%%%%%%%%%%%%%%%%% Fstar=0 %%%%%%%%%%%%%%%%%%%%%%%%%%%%%%%%%%%%%%%%%%%%%%%%%%%%%%%%%%%%%%%%%%%%%%%
%%%%%%%%%%%%%%%%%%%%%%%%%%%%%%%%%%%%%%%%%%%%%%%%%%%%%%%%%%%%%%%%%%%%%%%%%%%%%%%%%%%%%%%%%%%%%%%%%%%%%%%%%%%%%%%%%%%%%%%%%%%%%%%%%
\subsubsection{$F_\star=0$}
\paragraph{Fixed point solutions}
Supposing instead $F_\star=0$, the fixed point solution is 
\bea\label{F zero fix}
H_\star^2&=&-\frac{20}{3c_B}\frac{\o}{\o+1}M_{Pl}^2\left[48c_Y\pm\sqrt{6\left(384c_Y^2-c_B\frac{\o}{\o+1}\right)}\right] \\
\ssig&=&-\ssigpp=\frac{2\o}{3\o-1}\ssigpp=-\frac{20}{c_B}\frac{\o}{\o+1}\MPl\left[48c_Y\pm\sqrt{6\left(384c_Y^2-c_B\frac{\o}{\o+1}\right)}\right]\\
\srho=0\non
\ena
for $\o\ne-1$. This gives real Hubble parameter for $384c_Y^2-c_B\o/(\o+1)>0$. If $\o=-1$, we find 
\bea\label{F zero omega -1 fix}
H_\star^2=\frac{640c_Y}{c_B}\mpl, \quad \ssig=-\ssigp=\oneover{2}\ssigpp=\frac{640c_Y}{3c_B}\MPl,\quad \srho=0 
\ena
If the CFT is characterized by a positive $c_B$, there is no non trivial critical point. For vanishing $c_B$ the only fixed point with $\l\ne0$ is the trivial one.

When $F_\star=-H(1+\o)/(1-\o)$, there is only one possible solution, for which the parameters must have the values: $\o=-1$ (i.e. $\s_\p=-2\s_p$), $c_B=0$, $F_\star=0$ and the fixed point is thus the one in
(\ref{F zero omega -1 fix}).

\paragraph{Stability analysis}
For both fixed point characterized by the zero value of the extra dimension Hubble parameter, i.e. both for $\o\ne-1$ or $\o=-1$, we must find the eigenvalues of a $4\times4$ matrix. In fact we have a third
order linearized differential equation for the perturbation $\delta H(t)$ and a first order ODE for the energy density $\delta\rho$, while $\delta F(t)$ is found to be proportional to $\delta H(t)$ solving an
algebraic equation
\bea
\delta H^{(3)}&=&-a_2\,\delta\ddot H-a_1\,\delta\dot H-a_0\,\delta H+c_0\,\delta\rho\\
\delta\dot\rho&=&-3(1+w)H_\star\delta\rho\\
\delta F&=&\a\,\delta H
\ena
The coefficients in the differential equations are functions of the anomaly parameters $c_A,c_B$, of the trace parameter $c_Y$, of $\o$ and of $M_{Pl}$.

The eigenvalues $\l_1,\l_2,\l_3$ are then given by the roots of the third degree polynomial
\bea
\l^3+a_2\l^2+a_1\l+a_0=0
\ena
while $\l_4=-3(1+w)H_\star<0$. The coefficients $a_0,a_1,a_2$ are given by
\bea
a_0&=&\frac{12}{25}\frac{8000\Mpl(3+2\a+3(1+\a)\o)}{c_BH_\star(1-\a)\o}-\left(c_BH_\star^4(23\a-54)+144c_Y\mpl\sH^2\right)\o\non\\
a_1&=&\frac{1}{25}\frac{960000\Mpl(1+\a)}{c_BH_\star^2(1-\a)\o}-\left(c_BH_\star^4(137\a-222)+36c_Y\mpl\sH^2\right)\o\non\\
a_2&=&\frac{7-6\a}{1-\a}H_\star
\ena
where
\bea
\a=-3\,\frac{800\Mpl(1+\o)-\left(-3c_BH_\star^4+480c_Y\mpl\sH^2\right)\o}{800\Mpl(5-3\o)-\left(-3c_BH_\star^4+480c_Y\mpl\sH^2\right)(1-3\o)}
\ena

The sign of the eigenvalues $\l_1,\l_2,\l_3$ ($\l_3=\overline\l_2$ iff $27a_0^2+4a_1^3-18a_0a_1a_2-a_1^2a_2^2+4a_0a_2^3>0$, otherwise we get three real roots) determines the nature of the fixed point. Since
$\l_4<0$, we find that we can only have a completely stable fixed point or a saddle. In the case of one real and two complex conjugated roots the critical point can be attractive only if $a_2>0$ and
\bea
-\frac{2a_2}{3}<A+B<\frac{a_2}{3}
\ena
where
\bea
&&A=\sgn(R) \left(|R|+\sqrt{R^2-Q^3}\right)^{\frac{1}{3}},\quad B=\frac{Q}{A}\\
&&R\equiv\oneover{54}(2a_2^3-9a_1a_2+27a_0),\quad Q\equiv\oneover{9}(a_2^2-3a_1)
\ena
When the three roots are real, they are negative (corresponding to an attractive fixed point) iff $a_0,a_1,a_2>0$. For the other values of $a_0,a_1,a_2$ the critical point is a saddle.

The coefficients of the linearized differential equation don't depend on the anomaly parameter $c_A$ corresponding to the type A anomaly, so that only type B anomaly influence the characteristics of this
fixed point.
%%%%%%%%%%%%%%%%%%%%%%%%%%%%%%%%%%%%%%%%%%%%%%%%%%%%%%%%%%%%%%%%%%%%%%%%%%%%%%%%%%%%%%%%%%%%%%%%%%%%%%%%%%%%%%%%%%%%%%%%%%%%%%%%%%%%%%
%%%%%%%%%%%%%%%%%%%%%%%%%%%%%%%%%%%%%%%%%%%%%%%%% static extra %%%%%%%%%%%%%%%%%%%%%%%%%%%%%%%%%%%%%%%%%%%%%%%%%%%%%%%%%%%%%%%%%%%%%%%
%%%%%%%%%%%%%%%%%%%%%%%%%%%%%%%%%%%%%%%%%%%%%%%%%%%%%%%%%%%%%%%%%%%%%%%%%%%%%%%%%%%%%%%%%%%%%%%%%%%%%%%%%%%%%%%%%%%%%%%%%%%%%%%%%%%%%%
\subsection{Static compact extra dimensions}
We analyze the set of differential equations when the extra dimensions are compactified on a sphere, i.e. $\k>0$, supposing that the corresponding acceleration factor $b(t)$ remains constant, so that
$F(t)\equiv0$.

Beside the $H_\star=0$ fixed points, we only have two acceptable time--independent solutions to the Friedmann equations. 
The $H_\star=0,\k\ne0$ fixed points are always saddle points as we can conclude from the linear order analysis since the eigenvalues of the stability matrix (or their real parts) are one opposite to the
other.
%%%%%%%%%%%%%%%%%%%%%%%%%%%%%%%%%%%%%%%%%%%%%%%%%%%%%%%%%%%%%%%%%%%%%%%%%%%%%%%%%%%%%%%%%%%%%%%%%%%%%%%%%%%%%%%%%%%%%%%%%%%%%%%%%%%%%%
%%%%%%%%%%%%%%%%%%%%%%%%%%%%%%%%%%%%%%%%%%%%%%%%% omega neq 0 %%%%%%%%%%%%%%%%%%%%%%%%%%%%%%%%%%%%%%%%%%%%%%%%%%%%%%%%%%%%%%%%%%%%%%%%
%%%%%%%%%%%%%%%%%%%%%%%%%%%%%%%%%%%%%%%%%%%%%%%%%%%%%%%%%%%%%%%%%%%%%%%%%%%%%%%%%%%%%%%%%%%%%%%%%%%%%%%%%%%%%%%%%%%%%%%%%%%%%%%%%%%%%%
\subsubsection{$\o\ne0$}
\paragraph{Fixed point solutions}
The energy density $\rho_\star$ is zero, $H_\star$ and $\s_\star$ are then determined by
\bea
\left(\sano+\sY\right)&\equiv&\frac{c_A}{48}\frac{\k}{b_0^2} H^4_\star-\frac{c_B}{4800}\left(54H^6_\star-98\frac{\k}{b_0^2} H^4_\star+42\frac{\k^2}{b_0^4} H^2_\star-6\frac{\k^3}{b_0^6}\right)+ \non\\
&&+\frac{4c_Y}{5}\mpl\left(9\sH^4+18\frac{\k}{b_0^2}\frac{\k^2}{b_0^4}\right)= \non\\
&=&\frac{1+\o}{\o}\Mpl\left(3H^2_\star+\frac{\k}{b_0^2}\right)\label{ano=some}\\
\s_\star&=&\Mpl\left(3H^2_\star+\frac{\k}{b_0^2}\right)\\
\rho_\star&=&0
\ena
Restricting the possible values of $\o$, we can obtain at least one positive root $H_\star^2$ of eq (\ref{ano=some}), without having limitations on the anomaly coefficients $c_A,c_B$.

We can illustrate an example, choosing the simple case $c_B=0$ (i.e. there is no contribution from the conformal invariants in the anomaly) and also $\o=-1$, which simplifies the equation \refeq{ano=some}.
The fixed point is thus determined by
\bea
H^2_\star&=&\frac{\k}{b_0^2}\left[9\pm\sqrt{90+\frac{5c_A}{192c_Y}\oneover{\mpl}\frac{\k}{b_0^2}}\right]^{-1}\\
\ssig&=&-\ssigp=2\ssigpp=\Mpl\left(3\left[9\pm\sqrt{90-\frac{5c_A}{192c_Y}\oneover{\mpl}\frac{\k}{b_0^2}}\right]^{-1}+1\right)\frac{\k}{b_0^2}\\
\srho&=&0\non
\ena
which is real for $192c_Y^2\mpl>-c_A\k/45b_0^2$. We can moreover have two distinct positive $\sH$ fixed points if $5c_A\k/9b_0^2<-192c_Y^2\mpl$.

\paragraph{Stability analysis}
We now analyze the $H_\star\neq0$ fixed points behavior.

Regarding the fixed point determined by (\ref{ano=some}), we get a negative eigenvalue $\l_4=-3(1+w)H_\star$ (given that $w>-1,H_\star>0$) and the other three are the roots of the third degree polynomial
\bea\label{eigenval eq}
\l^3+a_2\l^2+a_1\l+a_0=0
\ena
where $a_{i}=\tilde a_{i}/\tilde a_3, i=0,1,2$ and
\bea
\tilde a_0&=&-\frac{c_A\o}{12}\frac{\k}{b_0^2} H^3_\star+\frac{c_B\o}{1200} H_\star\left(81H^4_\star-98\frac{\k}{b_0^2}H^2_\star+21\frac{\k^2}{b_0^4}\right)
+\frac{144c_Y\o}{5}\mpl\left(\sH^2+\frac{\k}{b_0^2}\right)+6\Mpl(1+\o)H_\star,\non\\
\tilde a_1&=&-\frac{c_A\o}{48}\frac{\k}{b_0^2} H^2_\star+\frac{c_B\o}{4800} \left(111H^4_\star-68\frac{\k}{b_0^2} H^2_\star+21\frac{\k^2}{b_0^4}\right)
+\frac{36c_Y\o}{5}\mpl\left(\sH^2+\frac{\k}{b_0^2}\right)+2\Mpl,\non\\
\tilde a_2&=&\frac{7c_B\o}{1920}H_\star\left(H^2_\star+\frac{\k}{b_0^2}\right),\quad \tilde a_3=\frac{c_B\o}{1920}\left(H^2_\star+\frac{\k}{b_0^2}\right)
\ena
We get a $4\times4$ stability matrix --- despite the fact that we should have only 2 variables ($H,\rho$) --- because the differential equations are of third order: the $\rho$ eigenvalue is $\l_4$, but $H$ is
a superposition of the four modes corresponding to the four eigevalues of the matrix.

As in the previous analysis for the flat extra dimensions, the solutions $\l_{1,2,3}$ of the equation (\ref{eigenval eq}) are such that $\l_1\in \mathbb{R}$, $\l_3=\overline \l_2$ or
$\l_1,\l_2,\l_3\in\mathbb{R}$. Besides, when we have the complex conjugated pair, there are only three possibilities:
\renewcommand{\theenumi}{\roman{enumi}}
\renewcommand{\labelenumi}{(\theenumi)}
\setcounter{enumi}{0}
\begin{enumerate}
\item $\l_1,\Re(\l_2)=\Re(\l_3),\l_4\le0$ $\Rightarrow$ the solution is stable (even if one of the eigenvalues are null, because that mode won't then contribute to the expression for $H$)
\item $\l_1,\l_4<0,\;\Re(\l_2)=\Re(\l_3)>0$ $\Rightarrow$ we get a saddle point
\item $\l_1>0,\;\Re(\l_2)=\Re(\l_3),\l_4<0$ $\Rightarrow$ in this case too, the fixed point is a saddle
\end{enumerate}
The equalities $\l_1=0$ and $\Re(\l_2)=\Re(\l_3)$ in the case (i) are possible, but not simultaneously.
When the roots are all real we can get a stable point iff $a_0,a_1,a_2>0$, which implies $\l_1,\l_2,\l_3<0$, and a saddle otherwise, with one negative two positive, or two negative one positive roots. 
%%%%%%%%%%%%%%%%%%%%%%%%%%%%%%%%%%%%%%%%%%%%%%%%%%%%%%%%%%%%%%%%%%%%%%%%%%%%%%%%%%%%%%%%%%%%%%%%%%%%%%%%%%%%%%%%%%%%%%%%%%%%%%%%%%%%%%
%%%%%%%%%%%%%%%%%%%%%%%%%%%%%%%%%%%%%%%%%%%%%%%%% equal scale factors %%%%%%%%%%%%%%%%%%%%%%%%%%%%%%%%%%%%%%%%%%%%%%%%%%%%%%%%%%%%%%%%
%%%%%%%%%%%%%%%%%%%%%%%%%%%%%%%%%%%%%%%%%%%%%%%%%%%%%%%%%%%%%%%%%%%%%%%%%%%%%%%%%%%%%%%%%%%%%%%%%%%%%%%%%%%%%%%%%%%%%%%%%%%%%%%%%%%%%%
\subsection{Equal scale factors}
Another limit that simplifies some of the calculations is the equal scale factor assumption. In this case the Hubble parameters of the internal space and the 3D space are equal, $F=H$, and we remain with a
set of equations for the variables $H,\rho,\s$, as in the static extra dimension limit.
%%%%%%%%%%%%%%%%%%%%%%%%%%%%%%%%%%%%%%%%%%%%%%%%%%%%%%%%%%%%%%%%%%%%%%%%%%%%%%%%%%%%%%%%%%%%%%%%%%%%%%%%%%%%%%%%%%%%%%%%%%%%%%%%%%%%%%
%%%%%%%%%%%%%%%%%%%%%%%%%%%%%%%%%%%%%%%%%%%%%%%%%%%%% \o\ne0 %%%%%%%%%%%%%%%%%%%%%%%%%%%%%%%%%%%%%%%%%%%%%%%%%%%%%%%%%%%%%%%%%%%%%%%%%
%%%%%%%%%%%%%%%%%%%%%%%%%%%%%%%%%%%%%%%%%%%%%%%%%%%%%%%%%%%%%%%%%%%%%%%%%%%%%%%%%%%%%%%%%%%%%%%%%%%%%%%%%%%%%%%%%%%%%%%%%%%%%%%%%%%%%%
\subsubsection{$\o\ne0$}
\paragraph{Fixed point solutions}
We observe that when $\o=0$ there is no acceptable solutions of the time--independent Einstein equations. So we want to find the fixed points in the $H=F$, $\o\ne0$ limit. The time--independent Friedmann plus
conservation equations lead to
\bea
\left(\sano+\sY\right)&\equiv&\frac{5}{48}(c_A+2c_B)\sH^4\left(\sH^2+\frac{\k}{a^2}\right)-\oneover{192}c_B\frac{\k^2}{a^4}\sH^2+\oneover{800}c_B\frac{\k^3}{a^6}+  \non\\
&&+\frac{4c_Y}{5}\mpl\left(150\sH^4+20\frac{\k}{b_0^2}\sH^2-\frac{\k^2}{b_0^4}\right)  \non\\
&=&6\Mpl\left(10\sH^2+\frac{\k}{a^2}\right)\\\label{H roots equal}
\ssig&=&\Mpl\left(10\sH^2+\frac{\k}{a^2}\right)\\
\srho&=&0
\ena
provided that $5+3w+2w_\p\neq0$%%%%%%%%%%%%%%%% footnote
\footnote{If the reasonable range of values for the pressures is given by $w,w_\p\ge-1$, this condition is satisfied whenever $w$ (or $w_\p$) is strictly greater than $-1$ and, since we are interested in
fixed points with a general $w\ne-1$, we will assume that this is the case.}%%%%%%%%%%%%%%%%% footnote
. As a consequence of solving the system of equations we also obtain a constraint on the CFT pressure of the hidden sector $\s_\p$, as we must impose $\o=1/5$, i.e. $\s_\p=\s_p$. The equation (\ref{H roots
equal}) yields the value of the Hubble parameter at the fixed point as a function of $c_A,c_B,c_Y,M_{Pl},\k$. 

In the case of flat extra dimensions $\k=0$ we immediately solve the system of equations finding (discarding the trivial $\sH=0$ solution)
\bea
\sH^2&=&-\frac{24}{c_A+2c_B}\mpl\left[24c_Y\pm\sqrt{576c_Y^2+(c_A+2c_B)}\right]\\  
\ssig&=&-\ssigp=-\ssigpp=-\frac{240}{c_A+2c_B}\MPl\left[24c_Y\pm\sqrt{576c_Y^2+(c_A+2c_B)}\right],  \qquad  \srho=0  \non
\ena
The solution is acceptable if $-(c_A-2c_B)<(24c_Y)^2$ and the two roots are both positive when $(c_A-c_B)<0$.
\paragraph{Stability analysis}
The last situation that we are going to consider is the case of equal scale factors, as in the fixed point analysis. In related subsection we found only one fixed point with $F=H$, that entails a relation
between the two CFT pressures $\s_\p=\s_p$ ($\o=1/5$). We calculate the stability matrix eigenvalues corresponding to this particular limit.

When the extra dimensions spatial curvature is zero $\k=0$, in addition to the vanishing 3D curvature ($k=0$), the stability matrix can be studied straightforward. All the eigenvalues are coincident since
$\d H\propto\d\s\propto\d\rho$. They are given by
\bea
\l=-(5+3w+2w_\p)\sH<0
\ena
For $\sH>0$ the fixed point is stable.

\addcontentsline{toc}{chapter}{References}
\bibliographystyle{unsrt}

\end{mainmatter}
\end{document}